\documentclass[12pt]{book}
\usepackage{latexsym, makeidx,amssymb,amsmath}
\usepackage{stmaryrd}

\begin{document}
\baselineskip=18pt

\newcommand{\la}{\langle}
\newcommand{\ra}{\rangle}
\newcommand{\psp}{\vspace{0.4cm}}
\newcommand{\pse}{\vspace{0.2cm}}
\newcommand{\ptl}{\partial}
\newcommand{\dlt}{\delta}
\newcommand{\sgm}{\sigma}
\newcommand{\al}{\alpha}
\newcommand{\be}{\beta}
\newcommand{\G}{\Gamma}
\newcommand{\gm}{\gamma}
\newcommand{\vs}{\varsigma}
\newcommand{\Lmd}{\Lambda}
\newcommand{\lmd}{\lambda}
\newcommand{\td}{\tilde}
\newcommand{\vf}{\varphi}
\newcommand{\yt}{Y^{\nu}}
\newcommand{\wt}{\mbox{wt}\:}
\newcommand{\der}{\mbox{Der}\:}
\newcommand{\ad}{\mbox{ad}\:}
\newcommand{\stl}{\stackrel}
\newcommand{\ol}{\overline}
\newcommand{\ul}{\underline}
\newcommand{\es}{\epsilon}
\newcommand{\dmd}{\diamond}
\newcommand{\clt}{\clubsuit}
\newcommand{\vt}{\vartheta}
\newcommand{\ves}{\varepsilon}
\newcommand{\dg}{\dagger}
\newcommand{\tr}{\mbox{Tr}\:}
\newcommand{\ga}{{\cal G}({\cal A})}
\newcommand{\hga}{\hat{\cal G}({\cal A})}
\newcommand{\Edo}{\mbox{End}\:}
\newcommand{\for}{\mbox{for}}
\newcommand{\kn}{\mbox{ker}\:}
\newcommand{\Dlt}{\Delta}
\newcommand{\rad}{\mbox{Rad}}
\newcommand{\rta}{\rightarrow}
\newcommand{\mbb}{\mathbb}
\newcommand{\rd}{\mbox{Res}\:}
\newcommand{\stc}{\stackrel{\circ}}
\newcommand{\stdt}{\stackrel{\bullet}}
\newcommand{\lra}{\Longrightarrow}
\newcommand{\llra}{\Longleftrightarrow}
\newcommand{\w}{\omega}
\newcommand{\sn}{\mbox{sn}\:}
\newcommand{\cn}{\mbox{cn}\:}
\newcommand{\dn}{\mbox{dn}\:}
\newcommand{\csch}{\mbox{csch}\:}
\newcommand{\sech}{\mbox{sech}\:}
\newcommand{\sta}{\theta}
\newcommand{\X}{{\cal X}}
\newcommand{\Y}{{\cal Y}}
\newcommand{\Z}{{\cal Z}}
\newcommand{\U}{{\cal U}}
\newcommand{\V}{{\cal V}}
\newcommand{\W}{{\cal W}}

\title{\LARGE \bf Algebraic Approaches to Partial \\ Differential Equations}
\vspace{6in}

\author{{\large \bf  Xiaoping Xu}\\ \\{\large Hua Loo-Keng Key
Mathematical Laboratory}\\ {\large Institute of Mathematics}
\\{\large Academy of Mathematics and System Sciences}\\{\large Chinese Academy of Sciences}
\\{\large 55 Zhongguancun Dong Lu}
\\{\large Beijing 100190, P. R. China}}
\vspace{2in}

\date{ 2012}

\thispagestyle{empty} \maketitle \enlargethispage*{1000pt}

\pagebreak

\thispagestyle{empty}

\pagebreak

\vspace*{8cm}

\begin{center}{\Huge \bf  Dedicated to My Wife Jing Jing }\end{center}
\thispagestyle{empty}

\newpage

\pagenumbering{roman}

\tableofcontents

\newpage

\chapter*{Preface}
\addcontentsline{toc}{chapter}{\numberline{}Preface}
\markboth{PREFACE}{PREFACE}

\texttt{}Partial differential equations are fundamental tools in
mathematics, sciences and engineering. For instance, the
electrodynamics is governed by the Maxwell equations, the
two-dimensional cubic nonlinear Schr\"{o}dinger equation is used to
describe the propagation of an intense laser beam through a medium
with Kerr nonlinearity and the Navier-Stokes equations are the
fundamental equations in  fluid dynamics. There are three major ways
of studying partial differential equations. The analytic way is to
study the existence and uniqueness of certain solutions of partial
differential equations and their mathematical properties. While the
numerical way is to find certain numerical solutions of partial
differential equations. In particular, physicists and engineers have
developed their own computational methods of finding physical and
practically useful numerical solutions, mostly motivated by
experiments. The algebraic way is to study symmetries, conservation
laws, exact solutions and complete integrability of partial
differential equations.

This book belongs to the third category. It is mainly an exposition
of the various algebraic techniques of solving partial differential
equations for exact solutions developed by the author in recent
years, with emphasis on physical equations such as: the
Calogero-Sutherland model of quantum many-body system in
one-dimension, the Maxwell equations, the free Dirac equations, the
generalized acoustic system, the Kortweg and de Vries (KdV)
equation, the Kadomtsev and Petviashvili (KP) equation, the equation
of transonic gas flows, the short-wave equation, the Khokhlov and
Zabolotskaya equation in nonlinear acoustics, the equation
 of geopotential  forecast, the nonlinear Schr\"{o}dinger equation and  coupled nonlinear Schr\"{o}dinger
 equations in optics, the Davey and Stewartson equations
of  three-dimensional packets of surface waves, the equation of the
dynamic convection in a sea,  the Boussinesq equations in
geophysics, the incompressible Navier-Stokes equations and the
classical boundary layer equations.

It is well known that most partial differential equations from
geometry are treated as the equations of elliptic type and most
partial differential equations from fluid dynamics are treated as
the equations of hyperbolic type. Analytically, partial differential
equations of elliptic type are easier than those of hyperbolic type.
Most of the nonlinear partial differential equations in this book
are from fluid dynamics. Our results show that algebraically,
partial differential equations of hyperbolic type are easier than
those of elliptic type in terms of exact solutions. Algebraic
approach and analytic approach have fundamental differences.

 This book was written based on the author's lecture notes on partial differential equations taught at
 the Graduate University of Chinese Academy of Sciences. It turned
 out that the course with the same title as the book was
 welcome not only by mathematical graduate students but also by
 physical and engineering students. Some engineering faculty members
 had also showed their interests in the course. The book is self-contained
 with the minimal prerequisite of calculus and linear algebra. It
 progresses according to the complexity of  equations and
 sophistication of the techniques involved.
 Indeed, it includes the basic algebraic techniques in
 ordinary differential equations and a brief introduction to special
 functions as the preparation for the main context.

 In
 linear partial differential equations, we focus on finding
 all the polynomial solutions and solving the initial-value
 problems. Intuitive derivations of  easily-using symmetry
 transformations of nonlinear  partial differential equations are
 given. These transformations generate sophisticated solutions with
 more parameters from relatively simple ones. They are also used to
 simplify  our process of finding exact solutions. We have
 extensively used moving frames, asymmetric conditions, stable
 ranges of nonlinear terms, special functions and linearizations in
 our approaches to nonlinear partial differential equations. The
 exact solutions we obtained usually contain multiple parameter
 functions and most of them are not of traveling-wave type.

   The book can serve as a research reference book for mathematicians,
   scientists and engineers. It can also be treated as a text book
   after a proper selection of materials
   for training students' mathematical skills and enriching their
knowledge.

\vspace{1cm}

\hfill Xiaoping Xu\\
Beijing, P. R. China\\
2012

\chapter*{Introduction}
\addcontentsline{toc}{chapter}{\numberline{}Introduction}
\markboth{INTRODUCTION}{INTRODUCTION}

In normal circumstances, the natural world operates according to
physical laws. Many of these laws were formulated in terms of
partial differential equations. For instance, the electromagnetic
fields in physics are governed by the well-known {\it Maxwell
equations}\index{Maxwell equations}
$$\ptl_t({\bf E})=\mbox{curl}\:{\bf B},\qquad\ptl_t({\bf B})=-\mbox{curl}\:{\bf
E}\eqno(0.1)$$ with
 $$\mbox{div}\:{\bf E}=f(x,y,z),
\qquad\mbox{div}\:{\bf B}=g(x,y,z),\eqno(0.2)$$ where the vector
function ${\bf E}$ stands for the electric field, the vector
function ${\bf B}$ stands for the magnetic field, the scalar
function $f$ is related to the charge density and the scalar
function $g$ is related to the magnetic potential. The {\it
two-dimensional cubic nonlinear Schr\"{o}dinger
equation}\index{Schr\"{o}dinger equation}
$$ i\psi_t+\kappa(\psi_{xx}+\psi_{yy})+\ves|\psi|^2\psi=0\eqno(0.3)$$
is used to describe the propagation of an intense laser beam through
a medium with Kerr nonlinearity, where $t$ is the distance in the
direction of propagation, $x$ and $y$ are the transverse spacial
coordinates,  $\psi$ is a complex valued function in $t,x,y$
standing for electric field amplitude, and $\kappa,\ves$ are nonzero
real constants. Moreover, the {\it coupled two-dimensional cubic
nonlinear Schr\"{o}dinger equations}\index{coupled Schr\"{o}dinger
equations}
$$ i\psi_t+\kappa_1(\psi_{xx}+\psi_{yy})+(\ves_1|\psi|^2+\es_1|\vf|^2)\psi=0,
\eqno(0.4)$$
$$ i\vf_t+\kappa_2(\vf_{xx}+\vf_{yy})+(\ves_2|\psi|^2+\es_2|\vf|^2)\vf=0
\eqno(0.5)$$ are used to describe the interaction of electromagnetic
waves with different polarizations in nonlinear optics, where
$\kappa_1,\kappa_2,\ves_1,\ves_2,\es_1$ and $\es_2$ are real
constants.

The most fundamental differential equations in the motion of
incompressible viscous fluids are the {\it Navier-Stokes
equations}\index{Navier-Stokes equations}
$$u_t+uu_x+vu_y+wu_z+\frac{1}{\rho}p_x=\nu (u_{xx}+u_{yy}+u_{zz})
 ,\eqno(0.6)$$
$$v_t+uv_x+vv_y+wv_z+\frac{1}{\rho}p_y=\nu (v_{xx}+v_{yy}+v_{zz})
 ,\eqno(0.7)$$
$$w_t+uw_x+vw_y+ww_z+\frac{1}{\rho}p_z=\nu (w_{xx}+w_{yy}+w_{zz})
 ,\eqno(0.8)$$
$$u_x+v_y+w_z=0,\eqno(0.9)$$
 where
$(u,v,w)$ stands for the velocity vector of the fluid, $p$ stands
for the pressure of the fluid, $\rho$ is the density constant and
$\nu$ is the coefficient constant of the kinematic viscosity.

Algebraic study of partial differential equations traces back to
Norwegian mathematician Sophus Lie [Lie], who invented the powerful
tool of continuous groups (known  as Lie groups) in 1874 in order to
study symmetry of differential equations. Lie's idea has been
carried on mainly by the mathematicians in the former states of
Soviet Union, East Europe and some mathematicians in North America.
Now it has become an important mathematical field known as ``group
analysis of differential equations," whose main objective is to find
symmetry group of partial differential equations, related
conservation laws and similarity solutions. The most influential
modern books on the subject may be the book ``Applications of Lie
Groups to Differential Equations" by Olver [Op] and the book ``Lie
Group Analysis of Differential Equations" by Ibragimov (cf. [In2,
In3]). In [X3], we found the complete set of functional generators
for the differential invariants of classical groups.

Soliton phenomenon was first observed by J. Scott Russel in 1834
 when he was riding on horseback
beside the narrow Union Canal near Edinburgh, Scotland. The
phenomenon had been theoretically studied by Russel, Airy (1845),
Stokes (1847), Boussinesq (1871, 1872) and Rayleigh (1876).   The
problem was finally solved by Kortweg and de Vries (1895) in terms
of the partial differential equation\index{KdV equation}
$$ u_t+6uu_x+u_{xxx}=0,\eqno(0.10)$$
 where $u$ is the
surface elevation of the wave above the equilibrium level, $x$ is
the distance from starting point and $t$ stands for time (later
people also realized that the above equation and its one-soliton
solution appeared in the Boussinesq's long paper [Bj]). However, it
was not until 1960 that any further application of the equation was
discovered. Gardner and Morikawa [GM] (1960) rediscovered the KdV
equation in the study of collision-free hydromagnetic waves.
Subsequently, the KdV equation has arisen in a number of other
physical contexts, such as, stratified internal waves, ion-acoustic
waves, plasma physics and lattice dynamics etc. Later a group led by
Kruskal [GGKM1, GGKM2, KMGZ, MGK] invented a special way of solving
the KdV equation (known as ``inverse scattering method") and
discovered infinite number of conservation laws of the equation.
Their works laid down the foundation  for the field of integrable
systems. We refer to the excellent book ``Solitons, Nonlinear
Evolution Equations and Inverse Scattering" by Ablowitz and Clarkson
[AC] for the details. Galaktionov and Svirshschevskii [GS] gave an
invariant-subspace approach to nonlinear partial differential
equations.

On the other hand, Gel'fand, Dikii and Dorfam [GDi1, GDi2,
GDo1-GDo3] introduced in 1970s a theory of Hamiltonian operators in
order to study the integrability of nonlinear evolution partial
differential equations (also cf. [Mf]). Our first experience with
partial differential equation was in the works [X1, X2, X4-X6] on
the structure of Hamiltonian operators and their supersymmetric
generalizations. In particular, we [X5] proved that vertex algebras
are equivalent to linear Hamiltonian operators as mathematical
structures. In this book, we are going to solve partial differential
equations directly based on the algebraic characteristics of
individual  equations. The tools we have employed are: the grading
technique from representation theory, the Campbell-Hausdorff-type
factorization of exponential differential operators, Fourier
expansions, matrix differential operators, stable-range of nonlinear
terms, generalized power series method, moving frames, classical
special functions in one variable and new multi-variable special
functions found by us, asymmetric conditions, symmetry
transformations and linearization techniques etc. The solved partial
differential equations are: flag partial differential equations
(including constant-coefficient linear equations), the
Calogero-Sutherland model of quantum many-body system in
one-dimension, the Maxwell equations, the free Dirac equations, the
generalized acoustic system,  the Kortweg and de Vries (KdV)
equation, the Kadomtsev and Petviashvili (KP) equation, the equation
of transonic gas flows, the short-wave equation,  the Khokhlov and
Zabolotskaya  equation in nonlinear acoustics, the equation
 of geopotential  forecast, the nonlinear Schr\"{o}dinger equation and  coupled nonlinear Schr\"{o}dinger
 equations in optics, the Davey and Stewartson equations
of  three-dimensional packets of surface waves, the equation of the
dynamic convection in a sea,  the Boussinesq equations in
geophysics, the Navier-Stokes equations and the classical boundary
layer equations.

The book consists of two parts. The first part is about basic
algebraic techniques of solving ordinary differential equations and
a brief introduction to special functions, most of which are
solutions of certain ordinary differential equations. This part
serves as a preparation for later solving partial differential
equations. It also makes the book accessible to the larger audience,
who may even not know what differential equation is about but have
the basic knowledge in calculus and linear algebra. The second part
is our main context, which consists of linear partial differential
equations, nonlinear scalar partial differential equations and
systems of nonlinear partial differential equations. Below we give
chapter-by-chapter detailed introductions.

In Chapter 1, we start with first-order linear ordinary differential
equations, and then turn to first-order separable equations,
homogenous equations and exact equations. Next we present the
methods of solving more special first-order ordinary differential
equations such as: the Bernoulli equations, the Darboux equations,
the Riccati equations, the Abel equations and the Clairaut's
equations.

Chapter 2 begins with solving  homogeneous linear ordinary
differential equations with constant coefficients by characteristic
equations. Then we solve the Euler equations and exact equations.
Moreover, the method of undetermined coefficients for solving
nonhomogeneous linear ordinary differential equations is presented.
Furthermore, we give the method of variation of parameters for
solving second-order nonhomogeneous linear ordinary differential
equations. In addition, we introduce the power series method to
solve variable-coefficient linear ordinary differential equations
and study the Bessel equation in detail.

Special functions are important objects both in mathematics and
physics. The problem of finding a function of continuous variable
$x$ that equals $n!$ when $x=n$ is a positive integer, was suggested
by Bernoulli and Goldbach, and was investigated by Euler in the late
1720s. In Chapter 3, we first introduce the gamma function $\G(z)$,
as a continuous generalization of $n!$. Then we prove the following
identities: (1) the {\it beta function}\index{beta function}
$B(x,y)=\int_0^1t^{x-1}(1-t)^{y-1}d=\G(x)\G(y)/\G(x+y)$; (2) {\it
Euler's reflection formula} $\G(z)\G(1-z)=\pi/\sin \pi z;$
\index{reflection formula}(3) the {\it product
formula}\index{product formula}
$$\G(z)\G\left(z+\frac{1}{n}\right)\G\left(z+\frac{2}{n}\right)\cdots\G\left(z+\frac{n-1}{n}\right)
=\frac{(2\pi)^{(n-1)/2}}{n^{nz-1/2}}\G(nz).\eqno(0.11)$$

In his thesis presented at G\"{o}ttingen in 1812, Gauss discovered
the one-variable function $_2F_1(\al,\be;\gm;z)$. We introduce it in
Chapter 3 as the power series solution of the {\it Gauss
hypergeometric equation}\index{Gauss hypergeometric equation}
$$z(1-z){y'}'+[\gm-(\al+\be+1)z]y'-\al\be y=0\eqno(0.12)$$
and prove the {\it Euler's integral representation}\index{Euler's
integral representation}
$$_2F_1(\al,\be;\gm;z)=\frac{\G(\gm)}{\G(\be)\G(\gm-\be)}\int_0^1t^{\be
-1}(1-t)^{\gm-\be-1}(1-zt)^{-\al}dt.\eqno(0.13)$$ Moreover, Jacobi
polynomials are introduced from the finite-sum cases of the Gauss
hypergeometric function and their orthogonality is proved. Legendre
orthogonal polynomials are discussed in detail.

Weierstrass's elliptic function $\wp(z)$ is a double-periodic
function with second-order poles, satisfying the nonlinear ordinary
differential equation
$${\wp'}^2(z)=4\wp^3(z)-g_2\wp(z)-g_3,\eqno(0.14)$$
whose consequence
$${\wp'}'(z)=6\wp^2(z)-\frac{g_2}{2}\eqno(0.15)$$
will be used later for solving nonlinear partial differential
equations. Weierstrass's zeta function $\zeta(z)$ is an integral of
$-\wp(z)$, that is, $\zeta'(z)=-\wp(z)$. Moreover, Weierstrass's
sigma function $\sgm(z)$ satisfies $\sgm'(z)/\sgm(z)=\zeta(z)$. We
discuss these functions and their properties in Chapter 3 to a
certain depth.

Finally in Chapter 3, we present Jacobi's elliptic functions
$\sn(z|m), \cn(z|m)$ and $\dn(z|m)$, and derive the nonlinear
ordinary differential equations satisfied by them. These functions
are also very useful in solving  nonlinear partial differential
equations.

Chapter 4 to Chapter 10 are the main contexts of this book. First in
Chapter 4. we derive the commonly used method of characteristic
lines for solving first-order quasilinear partial differential
equations, including boundary-value problems. Then we talk about
more sophisticated method of characteristic strip for solving
nonlinear first-order of partial differential equations. Exact
first-order partial differential equations are also handled.

 A {\it partial
differential equation of flag type}\index{flag partial differential
equation} is the linear differential equation of the form:
$$(d_1+f_1d_2+f_2d_3+\cdots+f_{n-1}d_n)(u)=0,\eqno(0.16)$$
where $d_1,d_2,...,d_n$ are certain commuting locally nilpotent
differential operators on the polynomial algebra
$\mbb{R}[x_1,x_2,...,x_n]$ and $f_1,...,f_{n-1}$ are polynomials
satisfying
$$d_l(f_j)=0\qquad\mbox{if}\;\;l>j.\eqno(0.17)$$
Many variable-coefficient (generalized) Laplace equations, wave
equations, Klein-Gordon equations, Helmholtz equations are
equivalent to the equations of this type. A general equation of this
type can not be solved by separation of variables. Flag partial
differential equations also naturally appear in the representation
theory of Lie algebras, in which the complete set of polynomial
solutions is crucial in determining the structure of many natural
representations. We use the grading technique from representation
theory to solve flag partial differential equations and find the
complete set of polynomial solutions. Our method also leads us to
obtain the solution of initial-value problem of the following type
of equations:
$$(\ptl_{x_1}^m-\sum_{r=1}^m\ptl_{x_1}^{m-r}f_r(\ptl_{x_2},...,\ptl_{x_n}))(u)=0,
\eqno(0.18)$$ where $m$ and $n>1$ are positive integers, and
$$f_r(\ptl_{x_2},...,\ptl_{x_n})\in\mbb{R}[\ptl_{x_2},...,\ptl_{x_n}].\eqno(0.19)$$
It turns out that the following family of new special functions
$${\cal Y}_\ell(y_1,...,y_m)=\sum_{\iota_1,...,\iota_m=0}^\infty {\iota_1+\cdots+\iota_m\choose
\iota_1,...,\iota_m}\frac{y_1 ^{\iota_1}y_2 ^{\iota_2}\cdots y_m
^{\iota_m}} {(\ell+\sum_{s=1}^ms\iota_s)!}\eqno(0.20)$$ play the key
roles, where $\ell$ is a nonnegative integer. In the case when all
$f_r=1$, we get that the functions
$$\vf_r(x)=x^r{\cal
Y}_r(b_1x,b_2x^2,...,b_mx^m)\;\;\mbox{with}\;\;
r=0,1,...,m-1\eqno(0.21)$$ form a fundamental set of solutions for
the ordinary differential equation
$$y^{(m)}-b_1y^{(m-1)}-\cdots-b_{m-1}y'-b_m=0.\eqno(0.22)$$
These results are  taken from our work [X11].

Barros-Neto and Gel'fand [BG1,BG2] (1998, 2002) studied  solutions
of the equation
$$u_{xx}+xu_{yy}=\dlt(x-x_0,y-y_0)\eqno(0.23)$$
related to the {\it Tricomi operator} $\ptl_x^2+x\ptl_y^2$.
\index{Tricomi operator}A natural generalization of the Tricomi
operator is
$\ptl_{x_1}^2+x_1\ptl_{x_2}^2+\cdots+x_{n-1}\ptl_{x_n}^2$. As
pointed out in [BG1, BG2], the Tricomi operator is an analogue of
the Laplace operator. So the equation
$$u_t=u_{x_1x_1}+x_1u_{x_2x_2}+\cdots+x_{n-1}u_{x_nx_n}\eqno(0.24)$$
is a natural analogue of heat conduction equation. In Chapter 4, we
use the method of characteristic lines to prove a
Campbell-Hausdorff-type factorization of exponential differential
operators and then solve the initial-value problem of the following
more general evolution equation
$$u_t=(\ptl_{x_1}^{m_1}+x_1\ptl_{x_2}^{m_2}+\cdots
+x_{n-1}\ptl_{x_n}^{m_n})(u)\eqno(0.25)$$ by
 Fourier expansions. Indeed we have solved analogous more general equations  related to tree diagrams.
We also use the Campbell-Hausdorff-type factorization to solve the
initial-value problem of analogous non-evolution flag partial
partial differential equations. The results are due to our work
[X7].

 The {\it Calogero-Sutherland model}\index{Calogero-Sutherland model} is an exactly solvable quantum many-body
system in one-dimension
 (cf. [Cf], [Sb]), whose Hamiltonian is given by
$$H_{CS}=\sum_{\iota=1}^n\ptl_{x_\iota}^2+K\sum_{1\leq p<q\leq n}\frac{1}{\sinh^2(x_p-x_q)},\eqno(0.26)$$
where $K$ is a constant. The model was used to study long-range
interactions of $n$ particles. Solving the model is equivalent to
find eigenfunctions and their corresponding eigenvalues of the
Hamiltonian $H_{CS}$ as a differential operator. We prove in Chapter
4 that the function
$$e^{2\mu_1(x_1+\cdots +x_n)}[\prod_{1\leq p<q\leq
n}(e^{2x_p}-e^{2x_q})]^{\mu_2}\eqno(0.27)$$ is a solution of the
Calogero-Sutherland model for any real numbers $\mu_1$ and $\mu_2$.
If $n=2$, we find a connection between the Calogero-Sutherland model
and the Gauss hypergeometric function. When $n>2$, a new class of
multi-variable hypergeometric functions are found based on Etingof's
work [Ep]. The results are taken from our work [X9]. Finally in
Chapter 4, we use matrix differential operators and Fourier
expansions to solve the Maxwell equations, the free Dirac equations
and the generalized acoustic system. The results come from our work
[X10].

Chapter 5 deals with nonlinear scalar (one dependent variable)
partial differential equations. First we do symmetry analysis on the
KdV equation (0.10), and obtain the {\it Galilean
boost}\index{Galilean boost} $G_c(u(t,x))=u(t,x+ct)-c/6$ for
$c\in\mbb{R}$. Solving the stationary equation $6uu_x+u_{xxx}=0$ and
using the Galilean boost $G_c$, we get the traveling-wave solutions
of the KdV equation in terms of the functions
$\wp(z),\tan^2z,\coth^2z$ and $\cn^2(z|m)$, respectively. In
particular, the soliton solution is obtained by taking $\lim_{m\rta
1}$ of a special case of the last solution. Moreover, we derive the
Hirota bilinear presentation of the KdV equation and use it to find
the two-soliton solution.

The {\it Kadomtsev and Petviashvili (KP) equation}\index{KP
equation}
$$(u_t+6uu_x+u_{xxx})_x+3\es u_{yy}=0\eqno(0.28)$$
with $\es=\pm 1$ is used to describe the evolution of long water
waves of small amplitude if they are weakly two-dimensional (cf.
[KP]). The choice of $\es$ depends on the relevant magnitude of
gravity and surface tension. The equation has also been proposed as
a model for surface waves and internal waves in straits or channels
of varying depth and width. The KP equation can be viewed as an
extension of the KdV equation (0.10). In Chapter 5, we have done the
symmetry analysis on the KP equation, and it possesses  the
following important symmetry transformation
$$T_{3,\al}(u(t,x,y))=u(t,x-\es\al'y/6,y+\al)+\es(2{\al'}'y-{\al'}^2)/72,
\eqno(0.29)$$ where $\al$ is any second-order differentiable
equation in $t$. Any solution of the KdV equation is obviously a
solution of the KP equation, and the above transformation
$T_{3,\al}$ maps such a solution independent of $y$ to a more
sophisticated solution of the KP equation that depends on $y$.
However, not all the interesting solutions of the KP equation are
obtained in this way. In fact, we  solve the KP equation for
solutions that are polynomial in $x$, and obtain many solutions that
can not be obtained from the solutions of the KdV equation; for
instance, we have the solution
$$u=-\frac{\es}{2} (x-\es
\al'y/6+\be)^2\wp(y+\al)+\frac{2{\al'}'y-{\al'}^2}{72\es}-\frac{\be'}{6},
\eqno(0.30)$$ where $\al$ and $\be$ are any functions in $t$ with
the above indicated differentiability. Furthermore, we find the
Hirota bilinear presentation of the KP equation and get the
following {\it lump solution} of the KP equation:\index{lump
solution}
$$u=2\ptl_x^2\ln((x-cy+3\es(b-c^2)t+a)^2+b(y+6\es ct)^2-\es/b^2),\eqno(0.31)$$
where $a,b,c\in\mbb{R}$ and $b\neq 0$. The above results in Chapter
5 are well-known (e.g., cf. [AC]) and we  reformulate them here just
for pedagogic purpose.

Lin, Reisner and Tsien [LRT] (1948) found the  equation
$$2u_{tx}+u_xu_{xx}-u_{yy}=0\eqno(0.32)$$\index{equation of
transonic gas flows}for two-dimensional non-steady motion of a
slender body in a compressible fluid, which was later called the
``equation of
 transonic gas flows" (cf. [Me1]). We derive in Chapter 5 the symmetry
 transformations of the above equation.  Using the stable range of the
nonlinear term $u_xu_{xx}$ and generalized power series method, we
find a family of singular solutions with seven arbitrary parameter
functions in $t$ and a family of analytic solutions with six
arbitrary parameter functions in $t$. For instance, we have the
solution
$$u=\frac{(x+\be'y+\al)^3}{3(y-\be)^2}
+({\be'}^2-2\al')x+2(\be'{\be'}'-{\al'}')y^2-2
{\be'}'xy-\frac{2{{\be'}'}'}{3}y^3+\mu\eqno(0.33)$$ which blows up
on a moving line $y=\be$,  where $\al,\be$ and $\mu$ are any
functions in $t$ with the above indicated differentiability. Such a
solution may reflect the phenomenon of abrupt high-speed wind. The
results are due to our work [X8].

Khristianovich and Rizhov [KR] (1958) discovered the {\it equations
of short waves}:\index{short-wave equations}
$$u_y-2v_t-2(v-x)v_x-2kv=0,\;\;v_y+u_x=0\eqno(0.34)$$
 in connection with the nonlinear reflection of weak
shock waves,  where $k$ is a real constant. Khokhlov and
Zabolotskaya [KZ] (1969) found the equation
$$2u_{tx}+(uu_x)_x-u_{yy}=0.\eqno(0.35)$$
for quasi-plane waves in nonlinear acoustics of bounded bundles.
More specifically,  the equation describes the propagation of a
diffraction sound beam in a nonlinear medium. The solutions of the
above equations similar to those of the equation (0.32) are derived
in Chapter 5 based on our work [X13].

In a book on short term weather forecast [Kt],  Kibel' (1954) used
the partial differential equation\index{equation of geopotential
forecast}
$$(H_{xx}+H_{yy})_t+H_x(H_{xx}+H_{yy})_y-H_y(H_{xx}+H_{yy})_x=k
H_x\eqno(0.36)$$  for geopotential forecast on a middle level in
earth sciences, where $k$ is a real constant. The  symmetry
transformations and two new families of exact solutions with
multiple parameter functions of the above equation are derived in
Chapter 5. The results are newly obtained by us.

In Chapter 6, we solve  the two-dimensional cubic nonlinear
Schr\"{o}dinger equation (0.3) and the coupled two-dimensional cubic
nonlinear Schr\"{o}dinger equations (0.4) and (0.5) by imposing a
quadratic condition on the related argument functions and using
their symmetry transformations. More complete families of exact
solutions of such type are obtained. The soliton solutions are
included. Many of them are the periodic, quasi-periodic, aperiodic
and singular solutions that may have practical significance. This
was our work [14].

 Davey and Stewartson [DS] (1974) used the method of multiple scales
 to derive the following system of nonlinear partial differential
 equations\index{Davey-Stewartson equation}
$$2iu_t+ \es_1u_{xx}+u_{yy}-2\es_2|u|^2u-2uv=0,\eqno(0.37)$$
$$v_{xx}-\es_1(v_{yy}+2(|u|^2)_{xx})=0\eqno(0.38)$$
that describe the long time evolution of  three-dimensional packets
of surface waves, where $u$ is a complex-valued function, $v$ is a
real valued function and $\es_1,\es_2=\pm 1$. In Chapter 6, we also
apply the above quadratic-argument approach to the Davey-Stewartson
equations and obtain four large families of solutions, including the
soliton solution. This part is a revision of our earlier preprint
[X18].

Both the atmospheric and oceanic flows are  influenced by the
rotation of the earth. In fact, the fast rotation and small aspect
ratio are two main characteristics of the large scale atmospheric
and oceanic flows. The small aspect ratio characteristic leads to
the primitive equations, and the fast rotation leads to the
quasi-geostropic equations. A main objective in climate dynamics and
in geophysical fluid dynamics is to understand and predict the
periodic, quasi-periodic, aperiodic, and fully turbulent
characteristics of the large scale atmospheric and oceanic flows.
The general model of atmospheric and oceanic flows is very
complicated.

Ovsiannikov (1967) introduced the following  equations in
geophysics:\index{equation of the dynamic convection}
$$u_x+v_y+w_z=0,\qquad \rho=p_z,\eqno(0.39)$$
$$\rho_t+u\rho_x+v\rho_y+w\rho_z=0,\eqno(0.40)$$
$$u_t+uu_x+vu_y+wu_z+v=-\frac{1}{\rho}p_x,\eqno(0.41)$$
$$v_t+uv_x+vv_y+wv_z-u=-\frac{1}{\rho}p_y\eqno(0.42)$$
to describe the dynamic convection in a sea, where $u,\:v$ and $w$
are components of velocity vector of relative motion of fluid in
Cartesian coordinates $(x,y,z)$, $\rho=\rho(x,y,z,t)$ is the density
of fluid and $p$ is the pressure (e.g., cf. Page 203 in [In3]).
Moreover, he determined the Lie point symmetries of the above
equations and found two very special solutions. In Chapter 7, we
give intuitive derivation of the symmetry transformations of the
above equations and solve them by the moving line, cylindrical
product and dimension reduction. This chapter is a revision of our
earlier preprint [X17].

The two-dimensional  Boussinesq equations for the incompressible
fluid in geophyics are\index{Boussinesq equations!two-dimensional}
$$u_t+uu_x+vu_y-\nu\Dlt u=-p_x,\qquad v_t+uv_x+vv_y-\nu\Dlt
v-\sta=-p_y,\eqno(0.43)$$
$$\sta_t+u\sta_x+v\sta_y-\kappa \Dlt\sta=0,\qquad
u_x+v_y=0,\eqno(0.44)$$ where $(u,v)$ is the velocity vector field,
$p$ is the scalar pressure, $\sta$ is the scalar temperature,
$\nu\geq 0$ is the viscosity and $\kappa\geq 0$ is the thermal
diffusivity. The above system is a simple model in atmospheric
sciences (e.g., cf. [Ma], [Cd]). By imposing asymmetric conditions
with respect to the spacial variables $x,y$ and using moving frame,
we find four families of multi-parameter solutions of the above
 Boussinesq equations in Chapter 8.

Another slightly simplified version of the system of  primitive
equations in geophysics is the three-dimensional stratified rotating
Boussinesq system (e.g., cf. [LTW1], [LTW2], [Ma],
[HMW]):\index{Boussinesq equations!three-dimensional}
$$u_t+uu_x+vu_y+wu_z-\frac{1}{R_0}v=\sgm(\Dlt u-p_x),\eqno(0.45)$$
$$v_t+uv_x+vv_y+wv_z+\frac{1}{R_0}u=\sgm(\Dlt v-p_y),\eqno(0.46)$$
$$w_t+uw_x+vw_y+ww_z-\sgm R T=\sgm(\Dlt w-p_z),\eqno(0.47)$$
$$T_t+uT_x+vT_y+wT_z=\Dlt T+w,\eqno(0.48)$$
$$ u_x+v_y+w_z=0,\eqno(0.49)$$
where $(u,v,w)$ is the velocity vector filed, $T$ is the temperature
function, $p$ is the pressure function, $\sgm$ is the Prandtle
number, $R$ is the thermal Rayleigh number and $R_0$ is the Rossby
number. Moreover, the vector $(1/R_0)(-v,u,0)$ represents the
Coriolis force and the term $w$ in (0.48) is derived using
stratification. By the similar method of solving the two-dimensional
equations, we derive in Chapter 8 five classes of multi-parameter
solutions of the equations (0.45)-(0.49). The results in Chapter 8
are reformulations of those in our work [X16].

In Chapter 9, we introduce a method of imposing asymmetric
conditions on the velocity vector with respect to independent
spacial variables and a method of moving frame for solving the three
dimensional Navier-Stokes equations (0.6)-(0.9). Seven families of
non-steady rotating asymmetric solutions with various parameters are
obtained. In particular, one family of solutions blow up  on a
moving plane, which may be used to study abrupt high-speed rotating
flows. Using Fourier expansion and two families of our solutions,
one can obtain discontinuous solutions that may be useful in study
of shock waves. Another family of solutions are partially
cylindrical invariant, containing two parameter functions in $t$,
which may be used to describe incompressible fluid in a nozzle. Most
of our solutions are globally analytic with respect to spacial
variables. The results are due to our work [X12].

In 1904, Prandtl observed that in the flow of slightly viscous fluid
past bodies, the frictional effects are confined to a thin layer of
fluid adjacent to the surface of the body. Moreover, he showed that
the motion of a small amount of fluid in this boundary layer decides
such important matters as the frictional drag, heat transfer, and
transfer of momentum between the body and the fluid. The
two-dimensional classical non-steady boundary layer
equations\index{boundary layer equations!two-dimensional}
$$u_t+uu_x+vu_y+p_x=u_{yy},\eqno(0.50)$$
$$p_y=0,\qquad u_x+v_y=0\eqno(0.51)$$
are used to describe the motion of a flat plate with the incident
flow parallel to the plate and directed to along the $x$-axis in the
Cartesian coordinates $(x,y)$, where $u$ and $v$ are the
longitudinal and the transverse components of the velocity, and $p$
is the pressure (e.g., cf. [In3]). The three-dimensional classical
non-steady boundary layer equations are:\index{boundary layer
equations!three-dimensional}
$$u_t+uu_x+vu_y+wu_z=-\frac{1}{\rho}p_x+\nu u_{yy},\eqno(0.52)$$
$$w_t+uw_x+vw_y+ww_z=-\frac{1}{\rho}p_z+\nu w_{yy},\eqno(0.53)$$
$$p_y=0,\qquad u_x+v_y+w_z=0,\eqno(0.54)$$
where $(u,v,w)$ denotes the velocity vector, $p$ stands for the
pressure,  $\rho$ is the density constant and $\nu$ is the
coefficient constant of the kinematic viscosity (e.g., cf. [In3]).

In Chapter 10, we introduce various schemes with multiple parameter
functions to solve these equations and obtain many families of new
explicit exact solutions with multiple parameter functions.
Moreover, symmetry transformations are used to simplify our
arguments.  The technique of moving frame is applied in the
three-dimensional case in order to capture the rotational properties
of the fluid. In particular, we obtain a family of solutions
singular on any moving surface, which may be used to study abrupt
high-speed rotating flows. Many other solutions are analytic related
to trigonometric and hyperbolic functions, which reflect various
wave characteristics of the fluid. Our solutions may also help
engineers to develop more effective algorithms to find physical
numeric solutions to practical models. The results are taken from
our work [X15]. Note that most of the nonlinear partial differential
equations  in this book are from fluid dynamics. Our results show
that algebraically, partial differential equations of hyperbolic
type are easier than those of elliptic type in terms of exact
solutions. The research in this book was partly supported  by the
National Natural Science Foundation of China (Grant No. 11171324).

\section*{Conventions}
\addcontentsline{toc}{section}{\numberline{}Notational Conventions}

$\mbb{C}$: the field of complex numbers.\\
$\ol{l,l+k}$: $\{l,l+1,i=l+2,...,l+k\}$, an index set.\\
$\dlt_{l,j}=1$ if $l=j$, and $0$ if $l\neq j$.\\
$\mbb{Z}$: the ring of integers.\\
$\mbb{N}$: $\{0,1,2,3,...\}$, the set of nonnegative integers\\
$i=\sqrt{-1}$: the imaginary number.\\
$\mbb{R}$: the field of real numbers.\\
$\ptl_x$: the operator of taking partial derivative with respect to
$x$.\\
$\bullet$ We assume that all partial differential derivatives can
change orders.\\
$\bullet$ We use prime $'$ to denote the derivative of a
one-variable function.\\
$\bullet$ When an expression appears, we always assume the
conditions that it makes sense; e.g., $\sqrt{a-b}\lra a\geq b$ if
$a,b\in\mbb{R}$.

\newpage

\pagenumbering{arabic}

\part{Ordinary Differential Equations}

\chapter{First-Order Ordinary Differential Equations}

In this chapter, we start with first-order linear ordinary
differential equations, and then turn to first-order separable
equations, homogenous equations and exact equations. Next we present
the methods of solving more special first-order ordinary
differential equations such as: the Bernoulli equations, the Darboux
equations, the Riccati equations, the Abel equations and the
Clairaut's equations.

\section{Basics}

In this section,  we deal with first-order linear ordinary
differential equations, separable equations, homogenous equations
and exact equations.

Let $y$ be a function of $t$. We use $y'=dy/dt$. A first-order
linear ordinary differential equation is written as
$$y'+f(t)y=g(t).\eqno(1.1.1)$$
To solve the equation, we multiply the integrating factor $e^{\int
f(t)dt}$ to the equation:
$$y'e^{\int
f(t)dt}+f(t)ye^{\int f(t)dt}=g(t)e^{\int f(t)dt},\eqno(1.1.2)$$
which can be rewritten as
$$(ye^{\int f(t)dt})'=g(t)e^{\int f(t)dt}.\eqno(1.1.3)$$
Thus
$$ye^{\int f(t)dt}=\int g(t)e^{\int f(t)dt}dt+c,\eqno(1.1.4)$$
where $c$ is an arbitrary constant. So we obtain the general
solution
$$y=e^{-\int f(t)dt}[\int g(t)e^{\int f(t)dt}dt+c].\eqno(1.1.5)$$
\pse

{\bf Example 1.1.1}. Solve the following initial-value problem:
$$ty'+2y=4t^2,\qquad\qquad y(1)=2\eqno(1.1.6)$$

{\it Solution}. Rewrite the equation in the standard form:
$$y'+{2\over t}y=4t.\eqno(1.1.7)$$
Then $f(t)=2/t$ and $g(t)=4t$. We calculate
$$e^{\int f(t)dt}=e^{\int (2/t)dt}\;\stackrel{\mbox{\small choose}}{=}\;e^{2\ln|t|}=e^{\ln t^2}=t^2.\eqno(1.1.8)$$
Thus the general solution is:
$$y={\int 4t\cdot t^2dt+c\over t^2}={t^4+c\over t^2}=t^2+ct^{-2}.\eqno(1.1.9)$$
The initial condition $y(1)=2$ implies
$$2=1+c\Longrightarrow c=1.\eqno(1.1.10)$$
The final solution is:
$$y=t^2+t^{-2}.\qquad\Box\eqno(1.1.11)$$
\pse

A first-order {\it separable}  ordinary differential equation is
written as $y'=f(t)g(y)$. \index{separable equation} The general
solution is given by
$$\int \frac{1}{g(y)}dy=\int f(t)dt+c.\eqno(1.1.12)$$

{\bf Example 1.1.2}. Solve
$$y'=\frac{ty^3}{\sqrt{1+t^2}},\qquad\qquad y(0)=1.\eqno(1.1.13)$$

{\it Solution}. We rewrite the equation as
$$\frac{2dy}{y^3}=\frac{2tdt}{\sqrt{1+t^2}}.\eqno(1.1.14)$$
So
$$-\int\frac{2dy}{y^3}=-\int\frac{2tdt}{\sqrt{1+t^2}}\lra\frac{1}{y^2}=c-2\sqrt{1+t^2}
\lra y=\pm\frac{1}{\sqrt{c-2\sqrt{1+t^2}}}.\eqno(1.1.15)$$ Since
$y(0)=1$, we choose positive sign and have
$$1=\frac{1}{\sqrt{c-2}}\lra c=3.\eqno(1.1.16)$$
Thus the final solution is
$$y=\frac{1}{\sqrt{3-2\sqrt{1+t^2}}}.\qquad\Box\eqno(1.1.17)$$
\pse

A first-order homogeneous ordinary differential equation is written
as $y'=f(y/t)$. To solve it, we change variable $u(t)=y(t)/t$. Then
$$y=tu\lra y'=u+tu'.\eqno(1.1.18)$$\index{homogeneous equation}
Thus the equation $y'=f(y/t)$ can be rewritten as
$$u+tu'=f(u)\lra u'=\frac{f(u)-u}{t},\eqno(1.1.19)$$
which is  a separable equation.\psp

{\bf Example 1.1.3}. Find the general solution of the following
homogeneous equation:
$$y'={2y^2-3t^2\over ty}.\eqno(1.1.20)$$

{\it Solution}. Rewrite
$$y'={2(y/t)^2-3\over y/t}.\eqno(1.1.21)$$
By changing variable $u(t)=y(t)/t$, we get
$$u+tu'={2u^2-3\over u}\Longrightarrow tu'= {2u^2-3\over u}-u={u^2-3\over u}.\eqno(1.1.22)$$
Thus
$${udu\over u^2-3}={dt\over t}\lra \int{2udu\over u^2-3}=\int{2dt\over t}\lra\ln|u^2-3|=\ln t^2+c_1.\eqno(1.1.23)$$
So $$u^2-3=ct^2\Longrightarrow u^2=3+ct^2.\eqno(1.1.24)$$ Hence
$$\left({y\over t}\right)^2=3+ct^2\Longrightarrow y^2=3t^2+ct^4.\qquad\Box\eqno(1.1.25)$$
\pse

 {\bf Example 1.1.4}. Solve the following equation
$$y'={t+y-2\over t-y+4}.\eqno(1.1.26)$$

{\it Solution}. In order to change the above equation to a
homogeneous equation, we change variable
$$\left\{\begin{array}{lll}T&=&t+k\\ Y&=&y+l,\end{array}\right.\eqno(1.1.27)$$
where $k$ and $l$ are constants to be determined. Since
$${t+y-2\over t-y+4}={T+Y-k-l-2\over T-Y-k+l+4},\eqno(1.1.28)$$
we let
$$\left\{\begin{array}{rrr}k+l+2&=&0\\ -k+l+4&=&0\end{array}\right.\lra \left\{\begin{array}{rrr}k+l&=&-2
\\ k-l&=&4\end{array}\right.\lra\left\{\begin{array}{rrr}k&=&1\\ l&=&-3.\end{array}\right.\eqno(1.1.29)$$
Hence
$$\left\{\begin{array}{lll}T&=&t+1\\ Y&=&y-3.\end{array}\right.\eqno(1.1.30)$$

The original equation changes to
$${dY\over dT}={T+Y\over T-Y}=\frac{1+\frac{Y}{T}}{1-\frac{Y}{T}}.\eqno(1.1.31)$$
Let
$$u=\frac{Y}{T}\lra {dY\over dT}=u+u'T.\eqno(1.1.32)$$
So
$$u+Tu'=\frac{1+u}{1-u}\lra u'T= \frac{1+u}{1-u}-u=\frac{1+u^2}{1-u}\eqno(1.1.33)$$
$$\lra \frac{1-u}{1+u^2}du=\frac{dT}{T}\lra \int\frac{1-u}{1+u^2}du=\int\frac{dT}{T}\eqno(1.1.34)$$
$$\lra \arctan u-\frac{1}{2}\ln (1+u^2)=\ln|T|+c_1.\eqno(1.1.35)$$
Thus
$$\frac{e^{\arctan u}}{\sqrt{1+u^2}}=c_2T,\eqno(1.1.36)$$
equivalently,
$$e^{\arctan u}=c_2T\sqrt{1+u^2}\lra e^{\arctan \frac{Y}{T}}=c_2T\sqrt{1+\frac{Y^2}{T^2}}\eqno(1.1.37)$$
$$\lra e^{\arctan \frac{Y}{T}}=\pm c_2\sqrt{T^2+Y^2}=c\sqrt{T^2+Y^2}.\eqno(1.1.38)$$
The final solution is
$$e^{\arctan
\frac{y-3}{t+1}}=c\sqrt{(t+1)^2+(y-3)^2}.\qquad\Box\eqno(1.1.39)$$\pse

A first-order {\it exact} ordinary differential equation has the
form
$$f(t,y)dt+g(t,y)dy=0,\qquad\mbox{where}\;\;\frac{\ptl f}{\ptl y}=
\frac{\ptl g}{\ptl t}.\eqno(1.1.40)$$ \index{exact equation} In this
case, the general solution is $U(t,y)=c$, where $U$ is a function
determined from
$$\frac{\ptl U}{\ptl t}=f,\qquad \frac{\ptl U}{\ptl
y}=g.\eqno(1.1.41)$$ Integrating the first equation yields $U=\int
f(t,y)dt+\psi(y)$, where $\psi(y)$ is a function to be determined.
In fact,
$$\psi'(y)=\frac{\ptl U}{\ptl
y}-\frac{\ptl \int f(t,y)dt}{\ptl y}=g-\frac{\ptl \int
f(t,y)dt}{\ptl y}.\eqno(1.1.42)$$

{\bf Example 1.1.5}. Solve the following exact equation:
$$(9t^2+y-1)dt-(4y-t)dy=0,\qquad\qquad y(1)=0.\eqno(1.1.43)$$

{\it Solution}. Let
$$U(t,y)=\int(9t^2+y-1)dt+\psi(y)=3t^3+(y-1)t+\psi(y).\eqno(1.1.44)$$
Taking partial derivative of the above equation with respect to $y$,
we have
$$U_y=t+\psi'(y)=-(4y-t).\eqno(1.1.45)$$
Thus
$$\psi'(y)=-4y.\qquad\mbox{Choose}\;\;\psi(y)=-2y^2.\eqno(1.1.46)$$
So $U=3t^3+(y-1)t-2y^2$ and the general solution is:
$$3t^3+(y-1)t-2y^2=c.\eqno(1.1.47)$$
When $y(1)=0$,
$$3-1=c\Longrightarrow c=2.\eqno(1.1.48)$$
The final solution is
$$3t^3+(y-1)t-2y^2=2.\qquad\Box\eqno(1.1.49)$$
\pse

An {\it integrating factor}\index{integrating factor} for the
equation $f(t,y)dt+g(t,y)dy=0$ is a function $\mu(t,y)$ such that
$$\mu(t,y)f(t,y)dt+\mu(t,y)g(t,y)dy=0\eqno(1.1.50)$$
is an exact equation, that is,
$$\frac{\ptl (\mu f)}{\ptl y}=
\frac{\ptl (\mu g)}{\ptl t}\lra g \frac{\ptl \mu}{\ptl t}-f
\frac{\ptl \mu}{\ptl y}=\left(\frac{\ptl f}{\ptl y}-\frac{\ptl
g}{\ptl t}\right)\mu\sim g\mu_t-f\mu_y=(f_y-g_t)\mu.\eqno(1.1.51)$$
The condition for $\mu$ to be a pure function in $t$ (i.e, $\ptl
\mu/\ptl y=0$) is $\mu_t/\mu=(f_y-g_t)/g$ is a pure function in
$t$.\psp

{\bf Example 1.1.6}. Solve the following equation by the method of
exact equations and integrating factors:
$$t(t^2+y^2+1)dt+ydy=0,\qquad\qquad y(0)=2.\eqno(1.1.52)$$

{\it Solution}. Note
$$f=t(t^2+y^2+t),\;\;g=y.\eqno(1.1.53)$$Moreover,
$$f_y=2ty,\;\;g_t=0.\eqno(1.1.54)$$
Since
$${f_y-g_t\over g}=2t,\eqno(1.1.55)$$
we look for an integrating factor $\mu(t)$. In this case,
$${\mu'\over \mu}=2t\;\;\stl{\mbox{\small choose}}{\lra}\;\;\mu=e^{t^2}.\eqno(1.1.56)$$
Thus the original equation is equivalent to the following exact
equation:
$$e^{t^2}t(t^2+y^2+1)dt+e^{t^2}ydy=0.\eqno(1.1.57)$$
Let \begin{eqnarray*}\qquad U(t,y)&=&\int
e^{t^2}t(t^2+y^2+1)dt+\psi(y)={1\over 2}\int
e^{t^2}(t^2+y^2+1)dt^2+\psi(y) \\ &=&{e^{t^2}(t^2+y^2)\over
2}+\psi(y).\hspace{8cm}(1.1.58)\end{eqnarray*}Then
$$U_y(t,y)=e^{t^2}y+\psi'(y)=e^{t^2}y\lra\psi'(y)=0\lra\psi\stackrel{\mbox{\small choose}}{=}0.\eqno(1.1.59)$$
Thus the general solution is:
$${e^{t^2}(t^2+y^2)\over 2}=c.\eqno(1.1.60)$$
Since $y(0)=2$, we have:
$$c={2^2\over 2}=2.\eqno(1.1.61)$$
Therefore, the final solution is:
$$e^{t^2}(t^2+y^2)=4.\qquad\Box\eqno(1.1.62)$$
\psp

If $(f_y-g_t)/f$ is a pure function in $y$, then we have the
integrating factor
$$\mu=\int\frac{g_t-f_y}{f}dy.\eqno(1.1.63)$$
 Let $\vf(z)$ be any one-variable function.
$$\mbox{If}\;\;f=y\vf(ty),\;g=t\vf(ty)\lra
\mu=\frac{1}{tf-yg}.\eqno(1.1.64)$$
$$\mbox{When}\;\;\frac{f_y-g_t}{g-f}=\vf(t+y)\lra
\mu=e^{\int \vf(z)dz},\;\;z=x+y.\eqno(1.1.65)$$
$$\mbox{If}\;\;\frac{f_y-g_t}{yg-tf}=\vf(ty)\lra
\mu=e^{\int \vf(z)dz},\;\;z=ty.\eqno(1.1.66)$$
$$\mbox{When}\;\;\frac{t^2(f_y-g_t)}{yg+tf}=\vf(y/t)\lra
\mu=e^{-\int \vf(z)dz},\;\;z=\frac{y}{t}.\eqno(1.1.67)$$
$$\mbox{If}\;\;\frac{f_y-g_t}{tg-yf}=\vf(t^2+y^2)\lra
\mu=e^{(1/2)\int \vf(z)dz},\;\;z=t^2+y^2.\eqno(1.1.68)$$ \psp

{\bf Excises 1.1}.\psp

1. Solve the equation:
$$ y'+y\tan t=t.$$

2. Find the general  solution of the equation:
$$y'=\frac{3t^2(1+e^{y^2})}{2y(1+t^3)}.$$

3. Solve the following equation
$$y'={t+2y-1\over 2t+3y+2}.$$

4. Find the general  solution of the equation:
$$y'=\frac{3t^2-y^2-7}{e^y+2ty+1}.$$

5. Solve the equation:
$$[3t^2\sin ty+y(t^3+3y+1)\cos ty]dt+[3\sin ty+t(t^3+3y+1)\cos ty]dy=0.$$

\section{Special Equations}

We present in this section the methods of solving the Bernoulli
equations, the Darboux equations, the Riccati equations, the Abel
equations and the Clairaut's equations.

A {\it Bernoulli equation}\index{bernoulli equation}  has the form
$$y'+f(t)y=g(t)y^a,\qquad a\neq 0,1.\eqno(1.2.1)$$
Changing variable $u(t)=y^{1-a}$, we get
$$u'=(1-a)y^{-a}y'\sim (1-a)y'=y^au'\eqno(1.2.2)$$
and (1.2.1) becomes
$$y^au'+(1-a)fy^a u=(1-a)gy^a\sim u'+(1-a)f u=(1-a)g.\eqno(1.2.3)$$
\pse

{\bf Example 1.2.1}. Solve the following Bernoulli equation :
$$y'-\frac{1}{t}y=y^3\sin t^3.\eqno(1.2.4)$$

{\it Solution}. Note that $y\equiv 0$ is an obvious solution.\\ We
assume that $y\not\equiv 0$. Rewrite the equation as:
$${y'\over y^3}-{1\over ty^2}=\sin t^3.\eqno(1.2.5)$$
Change variable:
$$u={1\over y^2};\qquad\;\;u'=-{2y'\over y^3}.\eqno(1.2.6)$$
Thus the original equation is equivalent to:
$$-{u'\over 2}-{u\over t}=\sin t^3,\eqno(1.2.7)$$equivalently,
$$u'+{2\over t}u=-2\sin t^3.\eqno(1.2.8)$$We calculate
$$e^{\int {2\over t}dt}\;\stackrel{\mbox{\small choose}}{=}\;e^{2\ln |t|}=e^{\ln t^2}=t^2.\eqno(1.2.9)$$
Thus
$$u={\int -2t^2\sin t^3dt +c\over t^2}={{2\over 3}\cos t^3+c\over t^2}={2\cos t^3+c_1\over 3t^2}.
\eqno(1.2.10)$$ Therefore,
$${1\over y^2}={2\cos t^3+c_1\over 3t^2}\lra y=\pm {\sqrt{3}t\over\sqrt{2\cos t^3+c_1}}.\qquad\Box\eqno(1.2.11)$$
\psp

A {\it Darboux equation}\index{Darboux equation} can be represented
as
$$(f(y/t)+t^a h(y/t))y'=g(y/t)+yt^{a-1}h(y/t).\eqno(1.2.12)$$
Using the substitution $y(t)=tz(t)$ and taking $z$ to be independent
variable, we have
$$\frac{d y}{dz}=y'\frac{dt}{dz}=t+z\frac{dt}{dz}.\eqno(1.2.13)$$
So (1.2.12) becomes
$$(f(z)+t^ah(z))y'\frac{dt}{dz}=(g(z)+zh(z)t^a)\frac{dt}{dz},\eqno(1.2.14)$$
equivalently,
$$(f(z)+t^ah(z))\left(t+z\frac{dt}{dz}\right)=(g(z)+zh(z)t^a)\frac{dt}{dz}.\eqno(1.2.15)$$
Thus
$$(zf(z)-g(z))\frac{dt}{dz}+f(z)t=-h(z)t^{a+1},\eqno(1.2.16)$$
which is a Bernoulli equation. \psp

A {\it Riccati equation}\index{Riccati equation} has the general
form
$$y'=f_2(t)y^2+f_1(t)y+f_0(t).\eqno(1.2.17)$$
If $f_2=0$, the equation is a linear equation. When $f_0=0$, it is a
Bernoulli equation. Changing variable
$$y=-\frac{u'(t)}{f_2(t)u(t)},\eqno(1.2.18)$$
we have
$$y'=\frac{f_2{u'}^2+f_2'uu'-f_2u{u'}'}{f_2^2u^2}\eqno(1.2.19)$$
and (1.2.17) becomes
$$\frac{f_2{u'}^2+f_2'uu'-f_2u{u'}'}{f_2^2u^2}=\frac{{u'}^2}{f_2u^2}
-\frac{f_1u'}{f_2u}+f_0\sim
{u'}'=\left(\frac{f_2'}{f_2}+f_1\right)u'-f_0f_2u=0,\eqno(1.2.20)$$
which is a second-order linear ordinary equation.\psp

{\bf Example 1.2.2}. Solve the Riccati equation:
$$y'=e^ty^2-y+e^{-t}.\eqno(1.2.21)$$

{\it Solution}. Now $f_2=e^t,\;f_1=-1$ and $f_0=e^{-t}$. Changing
variable
$$y(t)=-\frac{e^{-t}u'(t)}{u(t)},\eqno(1.2.22)$$
we get
$${u'}'=-u\eqno(1.2.23)$$
by (1.2.20). By a later method, the general solution of (1.2.23) is
$u=c_1\sin (t+c_2)$. Thus the general solution of (1.2.21) is
$$y=-e^{-t}\cot(t+c_2).\qquad\Box\eqno(1.2.24)$$\psp

Suppose that $y=\vf(t)$ is a particular solution of (1.2.17).
Changing variable $y(t)=\vf(t)+u(t)$, we reduce (1.2.17) to the
Bernoulli equation
$$u'=f_2u^2+(f_1+2f_2\vf)y.\eqno(1.2.25)$$
\pse

{\bf Example 1.2.3}. Solve the Riccati equation:
$$y'=y^2+\frac{t\tan t+2}{t}y+\frac{t\tan t+2}{t^2}.\eqno(1.2.26)$$

{\it Solution}. Observe that $y=-1/t$ is a particular solution of
(1.2.26). Changing variable $y(t)=u(t)-1/t$, we get
$$u'=u^2+\tan t\;u.\eqno(1.2.27)$$
Set $w=1/u$.  Then (1.2.27) becomes
$$w'=-1-\tan t\;w\lra w=\left[\frac{1}{2}\ln\frac{1-\sin t}{1+\sin t}+c\right]\cos t.\eqno(1.2.28)$$
So
$$u=\frac{\sec t}{\frac{1}{2}\ln\frac{1-\sin t}{1+\sin t}+c}\lra y
=\frac{\sec t }{\frac{1}{2}\ln\frac{1-\sin t}{1+\sin
t}+c}-\frac{1}{t}.\qquad\Box\eqno(1.2.29)$$\psp

An {\it Abel equation of the first kind}\index{Abel equation!first
kind} has the general form
$$y'=f_3(t)y^3+f_2(t)y^2+f_1(t)y+f_0(t),\qquad f_3(t)\not\equiv 0.\eqno(1.2.30)$$
The above equation is not integrable for arbitrary $f_n(t)$. We only
list two interesting special cases: \\
1. The Abel equation is generalized homogeneous:
$$y'=at^{2n+1}y^3+bt^ny^2+c\frac{y}{t}+dt^{-n-2}.\eqno(1.2.31)$$
Changing variable $y(t)=u(t)/t^{n+1}$, we obtain
$$y'=\frac{tu'-(n+1)u}{t^{n+2}}\eqno(1.2.32)$$
and
$$\frac{tu'-(n+1)u}{t^{n+2}}=a\frac{u^3}{t^{n+2}}+b\frac{u^2}{t^{n+2}}
+c\frac{u}{t^{n+2}}+d\frac{1}{t^{n+2}},\eqno(1.2.33)$$ equivalently,
$$tu'-(n+1)u=au^3+bu^2+cu+d\sim
tu'=au^3+bu^2+(c+n+1)u+d,\eqno(1.2.34)$$ which is a separable
equation.\\
2.  The Abel equation  has the form:
$$y'=at^{3n-m}y^3+bt^{2n}y^2+\frac{m-n}{t}y+dt^{2m}.\eqno(1.2.35)$$
Changing variable $y(t)=t^{m-n}u(t)$, we obtain
$$y'=t^{m-n}u'+(m-n)t^{m-n-1}u\eqno(1.2.36)$$
and
$$t^{m-n}u'+(m-n)t^{m-n-1}u=at^{2m}u^3+bt^{2m}u^2+(m-n)t^{m-n-1}u+dt^{2m},\eqno(1.2.37)$$
equivalently,
$$ t^{-m-n}u'=au^3+bu^2+d,\eqno(1.2.38)$$ which is a
separable equation.

From the above examples, we can try changing variable
$y=g_1(t)u(t)+g_0(t)$ to reduce the Abel equation to a separable
equation, where $g_0$ and $g_1$ are the functions to be determined.
\psp

An {\it Abel equation of the second kind}\index{Abel equation!second
kind} has the general form
$$(y+g(t))y'=f_2(t)y^2+f_1(t)y+f_0(t),\qquad g(t)\not\equiv 0.\eqno(1.2.39)$$
Again the above equation is not integrable for arbitrary $f_n(t)$.
We only list two interesting special cases: \\
1. The Abel equation of second kind is generalized homogeneous:
$$(y+kt^n)y'=a\frac{y^2}{t}+bt^{n-1}y+ct^{2n-1}.\eqno(1.2.40)$$
Changing variable $y(t)=t^nu(t)$, we obtain
$$y'=t^nu'+nt^{n-1}u\eqno(1.2.41)$$
and
$$(u+k)t^n(t^nu'+nt^{n-1}u)=at^{2n-1}u^2+bt^{2n-1}u+ct^{2n-1},\eqno(1.2.42)$$
equivalently,
$$(u+k)(tu'+nu)=au^2+bu+c\sim
t(u+k)u'=(a-n)u^2+(b-nk)u+c,\eqno(1.2.43)$$
 which is a separable equation.\\
2. The Abel equation of second kind has the form:
$$(y+g(t))y'=f_2(t)y^2+f_1(t)y+f_1(t)g(t)-f_2(t)g^2(t).\eqno(1.2.44)$$
Note that $y=-g(t)$ is a solution. Changing variable
$y(t)=u(t)-g(t)$, we obtain
$$u(u'-g')=f_2(u-g)^2+f_1(u-g)+f_1g-f_2g^2,\eqno(1.2.45)$$
equivalently, $$ uu'=f_2u^2+(f_1+g'-2gf_2)u\sim
u'=f_2u+f_1+g'-2gf_2,\eqno(1.2.46)$$ which is a first-order linear
equation.

From the above examples, we can again try changing variable
$y=g_1(t)u(t)+g_0(t)$ to reduce the Abel equation of second kind to
an integrable equation, where $g_0$ and $g_1$ are the functions to
be determined. \psp

A {\it Clairaut's equation}\index{Clairaut's equation} has the
general form
$$f(ty'-y)=g(y').\eqno(1.2.47)$$
Note that the linear function $y=at-b$ for which $f(b)=g(a)$ is a
solution. But the equation has more solutions in general.
Differentiating (1.2.47), we get
$${y'}'(tf'(ty'-y)-g'(y'))=0.\eqno(1.2.48)$$
Solving the system
$$f(ty'-y)=g(y'),\qquad tf'(ty'-y)=g'(y')\eqno(1.2.49)$$
 by viewing $y$ and $y'$ as variables, we get a singular solution of
 $y$.\psp

{\bf Example 1.2.4}. Solve the equation
$$(ty'-y)^2-{y'}^2-1=0.\eqno(1.2.50)$$
{\it Solution}. Rewrite the equation as $(ty'-y)^2={y'}^2+1$. Note
$f(z)=z^2$ and $g(z)=z^2+1$. Let
$$f(b)=g(a)\sim b^2=a^2+1\sim b=\pm\sqrt{a^2+1}.\eqno(1.2.51)$$
So we have the solution
$$y=at\pm\sqrt{a^2+1}.\eqno(1.2.52)$$
Now the second equation in (1.2.49) becomes
$$t(ty'-y)=y'\lra y'=\frac{ty}{t^2-1}.\eqno(1.2.53)$$
According to (1.2.50),
$$\frac{y^2}{(t^2-1)^2}-\frac{t^2y^2}{(t^2-1)^2}-1=0\sim
y^2+t^2=1.\qquad\Box\eqno(1.2.54)$$

We refer to [PZ] for more exact solutions of ordinary differential
equations.

\psp

{\bf Exercises 1.2}:\psp

1.  Solve the following Bernoulli equation
$$y'-\frac{1}{t}y=2y^2\tan t^2.$$

2. Solve the Riccati equation
$$y'=y^2+\frac{t\cot t+2}{t}y+\frac{t\cot t+2}{t^2}.$$

3. Solve the Abel equation  of the first kind:
$$y'=t^5y^3+t^2y^2-2\frac{y}{t}+\frac{1}{t^4}.$$

4. Solve the Abel equation of the first kind:
$$y'=t^3y^3-2t^4y^2+\frac{y}{t}+t^6.$$

5. Solve the Abel equation  of the second kind:
$$(y+5t^2)y'=5\frac{y^2}{t}+10ty+t^3.$$

6. Solve the Abel equation of the second kind:
$$(y+e^t)y'=-\frac{y^2}{t}+y\sin 2t+e^t\sin 2t+\frac{e^{2t}}{t}.$$

\chapter{Higher-Order Ordinary Differential Equations}

In this chapter, we  begin with  solving  homogeneous linear
ordinary differential equations with constant coefficients by
characteristic equations. Then we solve the Euler equations and
exact equations. Moreover, the method of undetermined coefficients
for solving nonhomogeneous linear ordinary differential equations is
presented. Furthermore, we give the method of variation of
parameters for solving second-order nonhomogeneous linear ordinary
differential equations. In addition, we introduce the power series
method to solve variable-coefficient linear ordinary differential
equations  and study the Bessel equation in detail.

\section{Basics}

This section deals with homogeneous linear ordinary differential
equation with constant coefficients, the Euler equations and exact
equations.

A second-order homogeneous linear ordinary differential equation
with constant coefficients is of the form
$$a{y'}'+by'+cy=0,\qquad a,b,c\in\mbb{R}.\eqno(2.1.1)$$
To find the general solution, we assume that $y=e^{\lmd t}$ is a
solution of (2.1.1), where $\lmd$ is a constant to be determined.
Substituting it into (2.1.1), we get
$$a\lmd^2e^{\lmd t}+b\lmd e^{\lmd t}+ce^{\lmd t}\sim
a\lmd^2+b\lmd+c=0,\eqno(2.1.2)$$ which is called the {\it
characteristic equation}\index{characteristic equation}  of (2.1.1).
 If the above equation has two
distinct real roots $\lmd_1$ and $\lmd_2$, then the general solution
of (2.1.1) is
$$y=c_1e^{\lmd_1 t}+c_2e^{\lmd_2 t},\eqno(2.1.3)$$
where $c_1$ and $c_2$ are arbitrary constants. When (2.1.2) has two
complex roots $r_1\pm r_2i$, then the real part and imaginary part
of $e^{(r_1+r_2i)t}$ are solutions of (2.1.1). So the general
solution of (2.1.1) is
$$y=c_1e^{r_1t}\sin r_2t+c_2e^{r_1t}\cos r_2t.\eqno(2.1.4)$$
In the case that (2.1.2) has repeated root $r$, the general solution
of (2.1.1) is
$$y=(c_1+c_2t)e^{r t}.\eqno(2.1.5)$$

\pse

{\bf Example 2.1.1}. The general solution of the equation
$${y'}'-2y'-3y=0\eqno(2.1.6)$$
is
$$y=c_1e^{3t}+c_2e^{-t}\eqno(2.1.7)$$
because $\lmd=3$ and $\lmd=-1$ are real roots of the characteristic
equation $\lmd^2-2\lmd-3=0$. Moreover, the general solution of the
equation
$${y'}'-4y'+13y=0\eqno(2.1.8)$$
is
$$y=c_1e^{2t}\sin 3t+c_2e^{2t}\cos 3t\eqno(2.1.9)$$
because $\lmd=2+3i$ and $\lmd=2-3i$ are  roots of the characteristic
equation
 $\lmd^2-4\lmd+13=0.$ Furthermore, the general solution of
the equation
$${y'}'+6y'+9y=0\eqno(2.1.10)$$
is
$$y=(c_1+c_2t)e^{-3t}.\qquad\Box\eqno(2.1.11)$$\pse

In general, the algebraic equation
$$b_n\lmd^n+b_{n-1}\lmd^{n-1}+\cdots+b_0=0\eqno(2.1.12)$$
is called the {\it characteristic equation} of the differential
equation
$$b_ny^{(n)}+b_{n-1}y^{(n-1)}+\cdots+b_0y=0,\qquad b_r\in\mbb{R}.\eqno(2.1.13)$$
If (2.1.12) has a real root $r$ with multiplicity $m$, then
$$(c_{m-1}t^{m-1}+\cdots+c_1t+c_0)e^{rt}\eqno(2.1.14)$$
is a solution of (2.1.13) for arbitrary
$c_0,c_1,...,c_{m-1}\in\mbb{R}$. When $r_1+r_2i$ is a complex root
of (2.1.12) with multiplicity $m$, then
$$(c_{m-1}t^{m-1}+\cdots+c_1t+c_0)e^{r_1t}\sin r_2t\eqno(2.1.15)$$
and
$$(a_{m-1}t^{m-1}+\cdots+a_1t+a_0)e^{r_1t}\cos r_2t\eqno(2.1.16)$$
are solutions of (2.1.13) for arbitrary $c_r,a_r\in\mbb{R}$. For
instance, if
$$(\lmd-1)(\lmd+2)^3(\lmd^2-4\lmd+13)^2=0\eqno(2.1.17)$$
is the characteristic equation of a differential equation of the
form (2.1.13), then the general solution of the differential
equation is
$$y=c_1e^t+(c_2t^2+c_3t+c_4)e^{-2t}+(c_5t+c_6)e^{2t}\sin 3t+
(c_7t+c_8)e^{2t}\cos 3t.\eqno(2.1.18)$$\pse

An {\it Euler ordinary differential equation}\index{Euler equation}
has the general form
$$b_nt^ny^{(n)}+b_{n-1}t^{n-1}y^{(n-1)}+\cdots+b_1ty'+b_0y=0,\qquad b_r\in\mbb{R}.\eqno(2.1.19)$$
We solve it by  changing variable $x=\ln t$. In fact,
$$y'=\frac{y_x}{t},\;\;{y'}'=\frac{y_{xx}-y_x}{t^2},\;\;{{y'}'}'=\frac{y_{xxx}-3y_{xx}+2y_x}{t^3}.
\eqno(2.1.20)$$ \pse

{\bf Example 2.1.2}. Solve the equation
$$t^2{y'}'-3ty'+5y=0.\eqno(2.1.21)$$

{\it Solution}. Changing variable $x=\ln t$, we get
$$y_{xx}-y_x-3y_x+5y=0\sim y_{xx}-4y_x+5y=0,\eqno(2.1.22)$$
whose characteristic equation is $\lmd^2-4\lmd+5=0$. The roots are
$\lmd=2\pm i$. So the general solution is
$$y=c_1e^{2x}\sin x+c_2e^{2x}\cos x=t^2(c_1\sin\ln t+c_2\cos\ln
t).\qquad\Box\eqno(2.1.23)$$ \pse

{\bf Example 2.1.3}. Solve the Euler equation
$$t^3{{y'}'}'-t^2{y'}'-2ty'-4y=0.\eqno(2.1.24)$$

{\it Solution}.  Changing variable $x=\ln t$, we get
$$y_{xxx}-3y_{xx}+2y_x-(y_{xx}-y_x)-2y_x-4y=0\sim
y_{xxx}-4y_{xx}+y_x-4y=0,\eqno(2.1.25)$$ whose characteristic
equation is
$$\lmd^3-4\lmd^2+\lmd-4=(\lmd-4)(\lmd^2+1)=0.\eqno(2.1.26)$$
Thus the general solution is
$$y=c_1e^{4x}+c_2\sin x+c_3\cos x=c_1t^4+c_2\sin\ln t+c_3\cos\ln
t.\qquad\Box\eqno(2.1.27)$$\pse

An $n$th-order ordinary differential equation is called an {\it
exact equation}\index{exact equation} if the equation can be
rewritten as
$$\frac{d\Phi(t,y,y',...,y^{(n-1)})}{dt}=0.\eqno(2.1.28)$$
We try to find $\Phi$  term by term.\psp

{\bf Example 2.1.4}. Solve the equation
$$ty{y'}'+t{y'}^2+yy'=0.\eqno(2.1.29)$$

{\it Solution}. Note that $\Phi=tyy'$. Thus (2.1.29) can rewritten
as $(tyy')'=0$. Thus
$$2tyy'=c_1\sim t(y^2)'=c_1\lra y^2=c_1\ln t+c_2.\qquad\Box\eqno(2.1.30)$$
\pse

{\bf Example 2.1.5}. Solve the equation
$$(1+t+t^2){{y'}'}'+(3+6t){y'}'+6y'=6t.\eqno(2.1.31)$$

{\it Solution}. We rewrite (2.1.31) as
$$(1+t+t^2){{y'}'}'+(1+2t){y'}'+(2+4t){y'}'+6y'-6t=0\eqno(2.1.32)$$
$$\lra[(1+t+t^2){y'}']'+(2+4t){y'}'+4y'+2y'-6t=0\eqno(2.1.33)$$
$$\lra[(1+t+t^2){y'}']'+[(2+4t)y']'+2y'-6t=0\eqno(2.1.34)$$
$$\lra[(1+t+t^2){y'}']'+[(2+4t)y']'+(2y)'-(3t^2)'=0\eqno(2.1.35)$$
$$\lra[(1+t+t^2){y'}'+(2+4t)y'+2y-3t^2]'=0\eqno(2.1.36)$$
$$\lra (1+t+t^2){y'}'+(2+4t)y'+2y-3t^2=2c_1\eqno(2.1.37)$$
$$\lra (1+t+t^2){y'}'+(1+2t)y'+(1+2t)y'+2y-3t^2=2c_1\eqno(2.1.38)$$
$$\lra[(1+t+t^2)y'+(1+2t)y-t^3]'=2c_1\eqno(2.1.39)$$
$$\lra(1+t+t^2)y'+(1+2t)y-t^3=2c_1t+c_2\eqno(2.1.40)$$
$$\lra[(1+t+t^2)y]'-t^3=2c_1t+c_2\eqno(2.1.41)$$
$$\lra(1+t+t^2)y-\frac{t^4}{4}=c_1t^2+c_2t+c_3.\qquad\Box\eqno(2.1.42)$$
\psp

{\bf Exercises 2.1}\psp

1. Find the general solution of the equation
$${y'}'-y'-6y=0.$$

2. Find the general solution of the equation
$${y'}'+6y'+13y=0.$$

3. Find the general solution of the equation
$$y^{(4)}+8{y'}'+16y=0.$$

4. Solve the Euler equation
$$t^3{{y'}'}'+3t^2{y'}'-2ty'+2y=0.$$

5. Solve the equation
$$ty{{y'}'}'+3ty'{y'}'+2y{y'}'+2{y'}^2=2\cos t-t\sin t.$$

\section{Method of
Undetermined Coefficients}

\index{method of undetermined coefficients}

In this section, we present the method of undetermined coefficients
for solving nonhomogeneous linear ordinary differential equations.

In order to solve linear nonhomogeneous ordinary differential
equation
$$f_n(t)y^{(n)}+f_{n-1}(t)y^{(n-1)}+\cdots+f_1(t)y=g(t),\eqno(2.2.1)$$
we find the general solution $\phi(t,c_1,...,c_n)$ of the
homogeneous equation
$$f_n(t)y^{(n)}+f_{n-1}(t)y^{(n-1)}+\cdots+f_1(t)y=0\eqno(2.2.2)$$
and a particular solution $y_0(t)$ of (2.2.1). Then the general
solution of (2.2.1) is $y=\phi(t,c_1,...,c_n)+y_0(t)$. It is often
that $y_0$ is obtained by guessing it of certain form with
undetermined coefficients based on the form of $g(t)$.\psp

{\bf Example 2.2.1}. Find the general solution of the equation
$${y'}'-\frac{2}{t^2}y=7t^4+3t^3.\eqno(2.2.3)$$

{\it Solution}. It is easy to see that $y=t^2$ and $y=1/t$ are
solutions of
$${y'}'-\frac{2}{t^2}y=0.\eqno(2.2.4)$$
So the general solution of (2.2.4) is
$$y=c_1t^2+\frac{c_2}{t}.\eqno(2.2.5)$$
Based on the form of (2.2.3), we guess a particular solution
$y_0(t)=at^6+bt^5$, where $a$ and $b$ are the constants to be
determined. Note
$$y_0'=6at^5+5bt^4\lra {y_0'}'=30at^4+20t^3.\eqno(2.2.6)$$
By (2.2.3),
$$30at^4+20t^3-2(at^4+bt^3)=7t^4+3t^3\sim 28a=7,\;18b=3\lra
a=\frac{1}{4},\;b=\frac{1}{6}.\eqno(2.2.7)$$ Thus $y_0=t^6/4+t^5/6$.
The general solution (2.2.3) is
$$y=c_1t^2+\frac{c_2}{t}+\frac{t^6}{4}+\frac{t^5}{6}.\qquad\Box\eqno(2.2.8)$$

{\bf Example 2.2.2}. Solve the equation
$${y'}'+3y'+2y=3\sin 2t.\eqno(2.2.9)$$

{\it Solution}. The general solution of ${y'}'+3y'+2y=0$ is
$y=c_1e^{-t}+c_2e^{-2t}$. We guess a particular solution of (2.2.9):
$$y_0=a\sin 2t+b\cos 2t.\eqno(2.2.10)$$
Then
$$ y_0'=2a\cos 2t-2b\sin 2t,\qquad {y_0'}'=-4a\sin 2t-4b\cos
2t.\eqno(2.2.11)$$ By (2.2.9),
$$-4a\sin 2t-4b\cos 2t+3(2a\cos 2t-2b\sin 2t)
+2(a\sin 2t+b\cos 2t)=3\sin 2t,\eqno(2.2.12)$$ equivalently,
$$-(2a+6b)\sin 2t+(6a-2b)\cos 2t=3\sin 2t.\eqno(2.2.13)$$
Hence
$$-(2a+6b)=3,\;\;6a-2b=0\lra
a=-\frac{3}{20},\;b=-\frac{9}{20}.\eqno(2.2.14)$$ So
$$y_0=-\frac{3}{20}\sin 2t-\frac{9}{20}\cos 2t\eqno(2.2.15)$$
and the general solution of (2.2.9) is
$$y=c_1e^{-t}+c_2e^{-2t}-\frac{3}{20}\sin 2t-\frac{9}{20}\cos 2t.\qquad\Box\eqno(2.2.16)$$

{\bf Example 2.2.3}. Find the solution of the following problem:
$$y''+y=2\cos t,\qquad\;\;y(0)=1,\;y'(0)=3.\eqno(2.2.17)$$
{\it Solution}. The general solution of the corresponding
homogeneous equation $y''+y=0$ is:
$$y=c_1\cos t+c_2\sin t.\eqno(2.2.18)$$
Thus we can not guess a particular solution $y_0=a\cos t+b\sin t$.
Instead, we guess that
$$y_0=at\cos t+bt\sin t\eqno(2.2.19)$$ is a
particular solution. Then
$$y_0'=(a+bt)\cos t+(b-at)\sin t,\eqno(2.2.20)$$
$$y_0''=(2b-at)\cos t-(2a+bt)\sin t.\eqno(2.2.21)$$
Substituting them into the equation in (2.2.17), we get
$$2b\cos t-2a\sin t=2\cos t.\eqno(2.2.22)$$So
$$a=0,\;b=1;\;\;y_0=t\sin t.\eqno(2.2.23)$$
Thus the general solution is:
$$y=c_1\cos t+(c_2+t)\sin t.\eqno(2.2.24)$$

Next
$$y'=(c_2+t)\cos t +(1-c_1)\sin t.\eqno(2.2.25)$$
Then $$y(0)=1\Longrightarrow c_1=1,\eqno(2.2.26)$$
$$y'(0)=3\Longrightarrow c_2=3.\eqno(2.2.27)$$
The final solution is:
$$y=\cos t+ (3+t)\sin t.\qquad\Box\eqno(2.2.28)$$

{\bf Example 2.2.4}. Find the solution of the following problem:
$$y''-4y'+4y=4(t^2+e^{2t}).\eqno(2.2.29)$$

{\it Solution}. The corresponding homogeneous equation is
$$y''-4y'+4y=0,\eqno(2.2.30)$$
whose characteristic equation is:
$$r^2-4r+4=0\Longrightarrow r=2\;\mbox{is a repeated root}.\eqno(2.2.31)$$
Thus the general solution is
$$y=(c_1+c_2t)e^{2t}.\eqno(2.2.32)$$

First we want to find a particular solution of the equation:
$$y''-4y'+4y=4t^2.\eqno(2.2.33)$$
Let
$$y_0=At^2+Bt+C\eqno(2.2.34)$$
be a particular solution. Then
$$y_0'=2At+B,\;\;y_0''=2A.\eqno(2.2.35)$$
Substitute them into the equation,
$$2A-4(2At+B)+4(At^2+Bt+C)=4t^2\eqno(2.2.36)$$
$$\Longrightarrow 4At^2+(4B-8A)t+2A-4B+4C=4t^2.\eqno(2.2.37)$$
$$4A=4,\;\;4B-8A=0,\;\;2A-4B+4C=0\Longrightarrow A=1,\;B=2,\;C={3\over 2}.\eqno(2.2.38)$$
So
$$y_0=t^2+2t+{3\over 2}.\eqno(2.2.39)$$

Next we want to find a particular solution of the equation:
$$y''-4y'+4y=4e^{2t}.\eqno(2.2.40)$$
Let
$$y_0=At^2e^{2t}\eqno(2.2.41)$$
be a particular solution. Then
$$y_0'=2A(t+t^2)e^{2t},\;\;y_0''=2A(1+4t+2t^2)e^{2t}.\eqno(2.2.42)$$
Substitute them into the equation,
$$2A(1+4t+2t^2)e^{2t}-8A(t+t^2)e^{2t}+4At^2e^{2t}=4e^{2t}\Longrightarrow 2Ae^{2t}=4e^{2t}.\eqno(2.2.43)$$
So $A=2$ and
$$y_0=2t^2e^{2t}.\eqno(2.2.44)$$

The final solution is
$$y=(c_1+c_2t+2t^2)e^{2t}+t^2+2t+{3\over 2}.\qquad\Box\eqno(2.2.45)$$
\pse

 {\bf Excises 2.2}. \psp

1. Find the general solution of the following equation:
$$y''+y'-2y=2t.$$

2. Solve the following initial value problem:
$$y''+2y'+5y=4e^{-x}\cos 2x,\qquad y(0)=1,\;\;y'(0)=0.$$

3.  Solve the following initial value problem:
$$y''-2y'-3y=\left\{\begin{array}{ll}3e^{-t}&\mbox{if}\;0\leq t\leq 1,\\ 2t^2&\mbox{if}\;t>1;\end{array}\right.
\qquad y(0)=0,\;\;y'(0)=1.$$

\section{Method of Variation of Parameters}

\index{method of variation of parameters}

 In this section, we give the method of variation of
parameters for solving second-order nonhomogeneous linear ordinary
differential equations.

Suppose that we know the fundamental solutions $y_1(t)$ and $y_2(t)$
of the linear homogeneous equation
$${y'}'+f_1(t)y'+f_0(t)y=0,\eqno(2.3.1)$$
that is, the general solution of (2.3.1) is $y=c_1y_1(t)+c_2y_2(t)$.
We want to solve the linear nonhomogeneous equation
$${y'}'+f_1(t)y'+f_0(t)y=g(t).\eqno(2.3.2)$$
Let $y=u_1(t)y_1+u_2(t)y_2$ be a solution of (2.3.2), where $u_1(t)$
and $u_2(t)$ are functions to be determined. Note
$$y'=u'_1y_1+u_2'y_2+u_1y_1'+u_2y_2'.\eqno(2.3.3)$$
In order to simplify the problem, we impose a condition
$$u'_1y_1+u_2'y_2=0.\eqno(2.3.4)$$
Then
$$y'=u_1y_1'+u_2y_2'\lra {y'}'=u_1{y_1'}'+u_2{y_2'}'+u_1'y_1'+u_2'y_2'
.\eqno(2.3.5)$$ According to (2.3.2),
$$u_1{y_1'}'+u_2{y_2'}'+u_1'y_1'+u_2'y_2'+f_1(u_1y_1'+u_2y_2')+f_0(u_1f_1+u_2f_2)=g(t)\eqno(2.3.6)$$
$$\lra
u_1({y_1'}'+f_1y_1+f_0y_1)+u_2({y_2'}'+f_2y_1+f_0y_2)+u_1'y_1'+u_2'y_2'=g(t),\eqno(2.3.7)$$
equivalently,
$$u_1'y_1'+u_2'y_2'=g(t)\eqno(2.3.8)$$
because $y_1$ and $y_2$ are solutions of (2.3.1).

The {\it Wronskian}\index{Wronskian} of the functions
$\{h_1,h_2,...,h_m\}$ is the determinant
$$W(h_1,h_2,...,h_m)=\left|\begin{array}{cccc}h_1&h_2&\cdots&h_m\\
h_1'&h_2'&\cdots&h_m'\\ \vdots& \vdots&
\vdots&\vdots\\h_1^{(m-1)}&h_2^{(m-1)}&\cdots&h_m^{(m-1)}\end{array}\right|.\eqno(2.3.9)$$
Solving the system (2.3.4) and (2.3.8) for $u_1'$ and $u_2'$ by
Crammer's rule, we get
$$u_1'=-\frac{g(t)y_2(t)}{W(y_1,y_2)},\qquad
u_2'=\frac{g(t)y_1(t)}{W(y_1,y_2)}.\eqno(2.3.10)$$ Thus
$$u_1=-\int\frac{g(t)y_2(t)}{W(y_1,y_2)}dt,\qquad
u_2=\int\frac{g(t)y_1(t)}{W(y_1,y_2)}dt.\eqno(2.3.11)$$ The final
solution is
$$y=-y_1(t)\int\frac{g(t)y_2(t)}{W(y_1,y_2)}dt+y_2(t)
\int\frac{g(t)y_1(t)}{W(y_1,y_2)}dt.\eqno(2.3.12)$$ \pse The above
method is called the {\it method of variation of parameters}.\psp

{\bf Example 2.3.1}. Find the general solution of the following
equation by the method of variation of parameters:
$$y''+4y={4\over \sin 2t},\qquad 0<t<{\pi\over 4}.\eqno(2.3.13)$$

{\it Solution}. The corresponding homogeneous equation is
$y''+4y=0,$ whose fundamental solutions are $y_1=\cos 2t$ and
$y_2=\sin 2t$. So
$$W(y_1,y_2)=\left|\begin{array}{cc}\cos 2t&\sin 2t\\ -2\sin 2t&2\cos
2t\end{array}\right|=2.\eqno(2.3.14)$$ Thus
$$u_1=-\int\frac{g(t)y_2(t)}{W(y_1,y_2)}dt=-2\int dt=c_1-2t,
\eqno(2.3.15)$$
$$u_2=\int\frac{g(t)y_1(t)}{W(y_1,y_2)}dt=\int \frac{2\cos 2t}{\sin 2t}dt=\ln \sin2t +c_2.\eqno(2.3.16)$$ The final solution is
$$y=(c_1-2t)\cos 2t+(c_2+\ln \sin 2t)\sin 2t.\qquad\Box\eqno(2.3.17)$$\pse

{\bf Example 2.3.2}. Solve the following initial value problem by
the method of variation of parameters:
$$y''-4y=g(t),\qquad\qquad y(0)=1,\;\;y'(0)=-1.\eqno(2.3.18)$$

{\it Solution}. First we solve the following initial value problem:
$$u''-4u=0,\qquad\qquad u(0)=1,\;\;u'(0)=-1.\eqno(2.3.19)$$
The general solution of the above equation is:
$$u=c_1e^{2t}+c_2 e^{-2t}.$$ So
$$u'=2(c_1e^{2t}-c_2e^{-2t}).$$
$$\left\{\begin{array}{l}u(0)=1\\ u'(0)=-1\end{array}\right.
\Longrightarrow \left\{\begin{array}{l}c_1+c_2=1\\
2(c_1-c_2)=-1\end{array}\right. \Longrightarrow
\left\{\begin{array}{l}c_1=1/4\\ c_2=3/4\end{array}\right.
\eqno(2.3.20)$$ The solution is:
$$u={1\over 4}(e^{2t}+3e^{-2t}).\eqno(2.3.21)$$

Next we want to solve the following problem:
$$v''-4v=g(t),\qquad\qquad v(0)=0,\;\;v'(0)=0.\eqno(2.3.22)$$
$$W(e^{2t},e^{-2t})=-4.$$
$$v=-e^{2t}\int_0^t{g(s)e^{-2s}\over -4}ds+e^{-2t}\int_0^t{g(s)e^{2s}\over -4}ds=
{1\over 2}\int_0^tg(s)\sinh 2(t-s)ds.\eqno(2.3.23)$$

The final solution is:
$$y=u+v={1\over 4}(e^{2t}+3e^{-2t})+{1\over 2}\int_0^tg(s)\sinh 2(t-s)ds.\qquad\Box\eqno(2.3.24)$$
\psp

If
$$v(t)=-y_1(t)\int^t_0\frac{g(s)y_2(s)}{W(y_1,y_2)(s)}ds+
y_2(t)\int^t_0\frac{g(s)y_1(s)}{W(y_1,y_2)(s)}ds,\eqno(2.3.25)$$
then
$$v'(t)=-y'_1(t)\int^t_0\frac{g(s)y_2(s)}{W(y_1,y_2)(s)}ds+
y'_2(t)\int^t_0\frac{g(s)y_1(s)}{W(y_1,y_2)(s)}ds.\eqno(2.3.26)$$
Thus we always have $v(0)=v'(0)=0$.\psp

{\bf Excises 2.3}. \psp

1. Solve the equation
$$y''+9y=\frac{9}{\cos 3t},\qquad 0<t<\frac{\pi}{6}.$$

2.  Solve the equation
$$y''-2y'+y=\frac{e^t}{1+t^2}.$$

3. Let $g(t)$ be a given function. Find the solution of the
following problem
$$y''-3y'-4y=g(t),\qquad\;\; y(0)=1,\;\;y'(0)=-1.$$

\section{Series Method and Bessel Functions}

In this section, we use power series to solve certain second-order
linear ordinary differential equations with variable coefficients:
$${y'}'+f_1(t)y'+f_0(t)y=0.\eqno(2.4.1)$$
Suppose that $f_1$ and $f_0$ are analytic at $t=0$. Around $t=0$,
$$f_0=\sum_{n=0}^\infty a_nt^n,\qquad f_1=\sum_{n=0}^\infty
b_nt^n,\qquad a_n,b_n\in\mbb{R}.\eqno(2.4.2)$$ We consider the
solution of the form
$$y=\sum_{n=0}^\infty c_nt^n,\qquad \mbox{where}\;c_n\;\mbox{are to be
determined}.\eqno(2.4.3)$$
$$y'=\sum_{n=1}^\infty nc_nt^{n-1},\qquad {y'}'=\sum_{n=2}^\infty
n(n-1)c_nt^{n-2}.\eqno(2.4.4)$$ Now (2.4.1) becomes
$$\sum_{n=0}^\infty
(n+1)(n+2)c_{n+2}t^n+(\sum_{n=0}^\infty b_nt^n)(\sum_{n=0}^\infty
(n+1)c_{n+1}t^n)+(\sum_{n=0}^\infty a_nt^n)(\sum_{n=0}^\infty
c_nt^n)=0.\eqno(2.4.5)$$
$$(n+1)(n+2)c_{n+2}=-\sum_{r=0}^n[(r+1)b_{n-r}c_{r+1}+a_{n-r}c_r].\eqno(2.4.6)$$
\pse

{\bf Example 2.4.1}. Solve the equation ${y'}'-t{y}'-y=0$.

{\it Solution}. Suppose that $y=\sum_{n=0}^\infty c_nt^n$ is a
solution. Note $a_r=-\dlt_{r,0}$ and $b_r=-\dlt_{r,1}$. Thus (2.4.6)
becomes
$$(n+1)(n+2)c_{n+2}=(n+1)c_n\sim c_{n+2}=\frac{c_n}{n+2}.\eqno(2.4.7)$$
Hence
$$y=c_0\sum_{n=0}^\infty \frac{t^{2n}}{(2n)!!}+c_1\sum_{n=0}^\infty
\frac{t^{2n+1}}{(2n+1)!!}.\qquad\Box\eqno(2.4.8)$$ \pse

Suppose
$$f_0=\sum_{n=-2}^\infty a_nt^n,\qquad f_1=\sum_{n=-1}^\infty
b_nt^n,\qquad a_n,b_n\in\mbb{R}.\eqno(2.4.9)$$ Assume that
$y=\sum_{n=0}^\infty c_nt^{n+\mu}$ is a solution of the equation
(2.4.1) with $c_0\neq 0$. Substituting it into (2.4.1), we find that
the coefficients of $t^{\mu-2}$ give
$$\mu(\mu-1)+\mu b_{-1}+a_{-2}=0\sim
\mu^2+(b_{-1}-1)\mu+a_{-2}=0,\eqno(2.4.10)$$ which is called the
{\it indicial equation}\index{indicial equation} of (2.4.1) with
(2.4.9). If (2.4.10) has two distinct real roots $\mu_1$ and $\mu_2$
such that $\mu_1-\mu_2$ is not an integer, then the equation (2.4.1)
has two linearly independent solutions of the forms:
$$y_1=t^{\mu_1}\sum_{n=0}^\infty c_nt^n,\qquad
y_2=t^{\mu_2}\sum_{n=0}^\infty d_nt^n.\eqno(2.4.11)$$ When (2.4.10)
has a repeated root $\mu$, then the equation (2.4.1) has two
linearly independent solutions of the forms:
$$y_1=t^{\mu}\sum_{n=0}^\infty c_nt^n,\qquad
y_2=y_1\ln t+t^{\mu}\sum_{n=0}^\infty d_nt^n.\eqno(2.4.12)$$
 If (2.4.10) has two
distinct real roots $\mu_1$ and $\mu_2$ such that $\mu_2-\mu_1$ is
an integer, then the equation (2.4.1) has two linearly independent
solutions of the forms:
$$y_1=t^{\mu_1}\sum_{n=0}^\infty c_nt^n,\qquad
y_2=ky_1\ln t+t^{\mu_2}\sum_{n=0}^\infty d_nt^n,\eqno(2.4.13)$$
where $k$ many be zero. \psp

{\bf Example 2.4.2}. Solve the following equation by power series:
$$t^2y''+3ty'+(1+t)y=0,\qquad\;\;t>0.\eqno(2.4.14)$$

{\it Solution}. Note that $t=0$ is a regular singular point. Let
$y=\sum_{n=0}^{\infty}c_nt^{n+\mu}$ be a solution with $c_0\neq 0$.
Then
$$y'=\sum_{n=0}^{\infty}(n+\mu)c_nt^{n+\mu-1},\;\;{y'}'=\sum_{n=0}^{\infty}(n+\mu)(n+\mu-1)c_nt^{n+\mu-2}.\eqno(2.4.15)$$
Substituting them into the equation, we have:
$$\sum_{n=0}^{\infty}(n+\mu)(n+\mu-1)c_nt^{n+\mu}+3\sum_{n=0}^{\infty}(n+\mu)c_nt^{n+\mu}
+(1+t)\sum_{n=0}^{\infty}c_nt^{n+\mu}=0,\eqno(2.4.16)$$
equivalently,
$$\sum_{n=0}^{\infty}(n+\mu)(n+\mu-1)c_nt^{n+\mu}+3\sum_{n=0}^{\infty}(n+\mu)c_nt^{n+\mu}
+\sum_{n=0}^{\infty}c_nt^{n+\mu}+\sum_{n=0}^{\infty}c_nt^{n+\mu+1}=0.\eqno(2.4.17)$$
So we have
$$[\mu(\mu-1)c_0+3\mu
c_0+c_0]t^\mu+\sum_{n=1}^{\infty}((n+\mu)(n+\mu-1)c_n+3(n+\mu)c_n+c_n+c_{n-1})t^{n+\mu}=0.\eqno(2.4.18)$$
Thus $\mu(\mu-1)c_0+3\mu c_0+c_0=0$ and for $n\geq 1$,
$$(n+\mu)(n+\mu-1)c_n+3(n+\mu)c_n+c_n+c_{n-1}=0\Longrightarrow (n+\mu+1)^2c_n=-c_{n-1}.\eqno(2.4.19)$$
$$c_n=-{c_{n-1}\over (n+\mu+1)^2}={(-1)^nc_0\over \prod_{j=1}^n(j+\mu+1)^2}.\eqno(2.4.20)$$
Denote
$$ b_n={(-1)^n\over \prod_{j=1}^n(j+\mu+1)^2}.\eqno(2.4.21)$$
Set
$$\varphi(\mu,t)=t^\mu(1+\sum_{n=1}^{\infty}b_nt^n).\eqno(2.4.22)$$
The indicial equation is
$$\mu(\mu-1)+3\mu+1=0\sim (\mu+1)^2=0\Longrightarrow \mu=-1\eqno(2.4.23)$$
is a double root. Then
$$y_1=\varphi(-1,t)=t^{-1}\left(1+\sum_{n=1}^{\infty}{(-1)^n\over \prod_{j=1}^nj^2}t^n\right)
=t^{-1}\left(1+\sum_{n=1}^{\infty}{(-1)^n\over(n!)^2}t^n\right)\eqno(2.4.24)$$
is a solution of (2.4.14).

 Observe
$$t^2\vf_{tt}+3t\vf_t+(1+t)\vf=t^\mu(\mu+1)^2\eqno(2.4.25)$$
(cf. the left hand side of (2.4.18) with $c_0=1$). Taking partial
derivative of (2.4.25) with respect to $\mu$, we get
$$t^2\vf_{tt\mu}+3t\vf_{t\mu}+(1+t)\vf_\mu=(\ln t)
t^\mu(\mu+1)^2+2t^\mu(\mu+1),\eqno(2.4.26)$$ equivalently,
$$t^2\vf_{\mu tt}+3t\vf_{\mu t}+(1+t)\vf_\mu=(2+(\mu+1)\ln t)
t^\mu(\mu+1).\eqno(2.4.27)$$ Taking $\mu=-1$ in the above equation,
we find
$$t^2\left(\frac{d}{dt}\right)^2\vf_\mu(-1,t)
+3t\frac{d}{dt}\vf_\mu(-1,t)+(1+t)\vf_\mu(-1,t)=0.\eqno(2.4.28)$$
Thus $y_2=\vf_\mu(-1,t)$ is another solution.  Note that for $n\geq
1$,
\begin{eqnarray*}\frac{d b_n}{d\mu}(-1)&=&\left({(-1)^n\over \prod_{j=1}^n(j+\mu+1)^2}\right)'|_{\mu=-1}
\\&=&\left({2(-1)^{n+1}\over \prod_{j=1}^n(j+\mu+1)^2}\right)\left(\sum_{j=1}^n{1\over j+\mu+1}\right)|_{\mu=-1}
\\&=&{2(-1)^{n+1}\over(n!)^2}\left(\sum_{j=1}^n{1\over j}\right).\hspace{8.4cm}(2.4.29)\end{eqnarray*}
Thus
$$y_2(t)=\varphi_\mu(-1,t)|_{r=-1}=y_1(t)\ln t+\sum_{n=1}^{\infty}{2(-1)^{n+1}\over(n!)^2}
\left(\sum_{j=1}^n{1\over j}\right)t^{n-1}.\eqno(2.4.30)$$ The
general solution is: $y=c_1y_1(t)+c_2y_2(t).\qquad\Box$\psp

The {\it Bessel equation}\index{Bessel equation} has the form
$${y'}'+t^{-1}y'+(1-\nu^2t^{-2})y=0,\eqno(2.4.31)$$
where $\nu$ is a constant called {\it order}.\index{order} The
indicial equation is
$$\mu^2-\nu^2=0\sim \mu=\pm\nu.\eqno(2.4.32)$$
We rewrite (2.4.31) as
$$t^2{y'}'+ty'+(t^2-\nu^2)y=0.\eqno(2.4.33)$$
Let $y=\sum_{n=0}^\infty c_nt^{n+\mu}$ be a solution of (2.4.33)
with $\mu=\pm\nu$ and $c_0\neq 0$. We have
$$ty'=\sum_{n=0}^\infty (n+\mu)c_nt^{n+\mu},\qquad
t^2{y'}'=\sum_{n=0}^\infty
(n+\mu)(n+\mu-1)c_nt^{n+\mu}.\eqno(2.4.34)$$ Denote by $\mbb{N}$ the
set of nonnegative integers. So (2.4.33) is equivalent to
$$c_1[(\mu+1)^2-\nu^2]=0,\;\;[(\mu+n+2)^2-\nu^2]c_{n+2}+c_n=0,\qquad
n\in\mbb{N}.\eqno(2.4.35)$$ Thus $c_{2r+1}=0$ for $r\in\mbb{N}$, and
$$c_{2n}=\frac{c_0}{\prod_{r=1}^n[\nu^2-(\mu+2r)^2]}=
\frac{(-1)^nc_0}{n!2^{2n}\prod_{r=1}^n(\mu+r)}.\eqno(2.4.36)$$ The
function
$$J_{\mu}(t)=\left(\frac{t}{2}\right)^{\mu}+\sum_{n=1}^\infty \frac{(-1)^n}
{n!\prod_{r=1}^n(\mu+r)}\left(\frac{t}{2}\right)^{2n+\mu}\eqno(2.4.37)$$
is called a {\it Bessel function of first kind}.\index{Bessel
function} If $\nu$ is not an integer, then the general solution of
(2.4.31) is
$$y=c_1J_\nu(t)+c_2J_{-\nu}(t).\eqno(2.4.38)$$

Note
$$\frac{d}{dt}(t^\mu
J_\mu)=\mu
t^\mu\left[\left(\frac{t}{2}\right)^{\mu-1}+\sum_{n=1}^\infty
\frac{(-1)^n}
{n!\prod_{r=1}^n(\mu+r-1)}\left(\frac{t}{2}\right)^{2n+\mu-1}\right]
=\mu t^\mu J_{\mu-1}\eqno(2.4.39)$$ and
$$\frac{d}{dt}(t^{-\mu}
J_\mu)= \sum_{n=1}^\infty \frac{(-1)^nt^{-\mu}}
{(\mu+1)(n-1)!\prod_{r=1}^{n-1}(\mu+r+1)}\left(\frac{t}{2}\right)^{2n+\mu-1}
=-\frac{ t^{-\mu} J_{\mu+1}}{\mu+1}.\eqno(2.4.40)$$ Thus
$$\frac{d}{tdt}(t^\mu J_\mu)=\mu t^{\mu-1} J_{\mu-1},\qquad
\frac{d}{tdt}(t^{-\mu} J_\mu)=-\frac{ t^{-\mu-1}
J_{\mu+1}}{\mu+1}.\eqno(2.4.41)$$ By induction,
$$\left(\frac{d}{tdt}\right)^m(t^\mu J_\mu)=[\prod_{r=0}^{m-1}(\mu-r)]t^{\mu-m}
J_{\mu-m}\eqno(2.4.42)$$ and
$$ \left(\frac{d}{tdt}\right)^m(t^{-\mu} J_\mu)=(-1)^m\frac{
t^{-\mu-m} J_{\mu+m}}{\prod_{r=1}^m(\mu+r)}.\eqno(2.4.43)$$

On the other hand, (2.4.39) gives
$$\mu t^{\mu-1}J_\mu+t^\mu J'_\mu=\mu t^\mu J_{\mu-1}\sim \mu J_\mu+t J'_\mu=\mu t
J_{\mu-1}\eqno(2.4.44)$$ and (2.4.40) yields
$$-\mu t^{-\mu-1}J_{\mu}+t^{-\mu}J_{\mu}'=-\frac{ t^{-\mu}
J_{\mu+1}}{\mu+1}\sim -\mu J_{\mu}+tJ_{\mu}'=-\frac{t
J_{\mu+1}}{\mu+1}.\eqno(2.4.45)$$ Thus
$$\mu
J_{\mu-1}+\frac{J_{\mu+1}}{\mu+1}=\frac{2\mu}{t} J_\mu,\qquad \mu
J_{\mu-1}-\frac{J_{\mu+1}}{\mu+1}=2\mu J_\mu'.\eqno(2.4.46)$$
Observe that
$$\left(\frac{d}{dt}\right)\frac{t^n}{n!}=\frac{t^{n-1}}{(n-1)!}\eqno(2.4.47)$$
for a positive integer $n$. If we have a continuous analogue of
$n!$, then we can simplify (2.4.42) and (2.4.43) by rescalling
$J_\mu$. Indeed, it is the spacial function $\G(s)$. \psp

When $\nu=n+1/2$ with $n\in\mbb{N}$, the indicial equation has two
roots $\mu_1=n+1/2$ and $\mu_2=-n-1/2$. Moreover, $\mu_1-\mu_2=2n+1$
is an integer. However, both $J_{n+1/2}(t)$ and $J_{-n-1/2}(t)$ are
well defined by (2.4.37). They form a set of fundamental solutions
of the Bessel equation. Suppose that $\nu=m$ is a positive integer.
the indicial equation has two roots $\mu_1=m$ and $\mu_2=-m$. The
function $J_m(t)$ is still well defined, but $J_{-m}(t)$ is not
defined. If $\mu=-m$,  by the second equation in (2.4.35) with
$n=2m-2$, we get
$$0=[(-m+2m-2+2)^2-m^2]c_{2m}=-c_{2m-2}=-\frac{c_0}{(m!)^2}\lra
c_0=0,\eqno(2.4.48)$$ which contradicts the assumption $c_0\neq 0$.
Thus we do not have a solution of the form $y=\sum_{n=0}^\infty
c_nt^{n-m}$. We look for another fundamental solution of the form
$$y=J_m(t)\ln t+\sum_{n=0}^\infty c_nt^{n-m},\eqno(2.4.49)$$
which is related to {\it Bessel functions of second kind}.\psp

{\bf Exercise 2.4}\psp

Solve the following equations by power series :

1. $(1-t^2)y''-ty'+16ty=0.$

2. $t^2y''+7ty'+(9-t)y=0,\qquad t>0.$

\chapter{Special Functions}

Special functions are important objects both in mathematics and
physics. This chapter is a brief introduction to them. The reader
may refer to [AAR] and [WG] for more extensive knowledge. First we
introduce the gamma function $\G(z)$ as a continuous generalization
of $n!$ and prove the beta-function identity, the Euler's reflection
formula and the product formula of the gamma function. Then we
introduce Gauss hypergeometric function as the power series solution
of the Gauss hypergeometric equation and prove the Euler's integral
representation. Moreover, Jacobi polynomials are introduced from the
finite-sum cases of the Gauss hypergeometric function and their
orthogonality is proved. Legendre orthogonal polynomials are
discussed in detail.

Weierstrass's elliptic function $\wp(z)$ is a double-periodic
function with second-order poles, which will be used later in
solving nonlinear partial differential equations. Weierstrass's zeta
function $\zeta(z)$ is an integral of $-\wp(z)$, that is,
$\zeta'(z)=-\wp(z)$. Moreover, Weierstrass's sigma function
$\sgm(z)$ satisfies $\sgm'(z)/\sgm(z)=\zeta(z)$. We discuss these
functions and their properties in this chapter to a certain depth.

Finally in this chapter, we present Jacobi's elliptic functions
$\sn(z|m), \cn(z|m)$ and $\dn(z|m)$, and derive the nonlinear
ordinary differential equations satisfied by them. These functions
are also very useful in solving  nonlinear partial differential
equations.

\section{Gamma and Beta Functions}

The problem of finding a function of continuous variable $x$ that
equals $n!$ when $x=n$ is a positive integer,  was suggested by
Bernoulli and Goldbach, and was investigated by Euler in the late
1720s.  For $a\in\mbb{C}$ and $n\in\mbb{N}+1$, we denote
$$(a)_n=a(a+1)\cdots (a+n-1),\qquad (a)_0=1.\eqno(3.1.1)$$
If $x$ and $n$ are positive integers, then
$$x!=\frac{(x+n)!}{(x+1)_n}=\frac{n!(n+1)_x}{(x+1)_n}=\frac{n!n^x}{(x+1)_n}
\frac{(n+1)_x}{n^x}.\eqno(3.1.2)$$ Since
$$\lim_{n\rta\infty}\frac{(n+1)_x}{n^x}=1,\eqno(3.1.3)$$
we have
$$x!=\lim_{n\rta\infty}\frac{n!n^x}{(x+1)_n}.\eqno(3.1.4)$$

Observe that for any $z\in\mbb{C}\setminus\{-\mbb{N}-1\}$,
\begin{eqnarray*}\qquad \qquad& &\left(\frac{n}{n+1}\right)^z\prod_{r=1}^n\left(1+\frac{z}{r}\right)^{-1}
\left(1+\frac{1}{r}\right)^z\\
&=&\left(\frac{n}{n+1}\right)^z\prod_{r=1}^n\left(\frac{z+r}{r}\right)^{-1}
\left(\frac{r+1}{r}\right)^z\\&=&\left(\frac{n}{n+1}\right)^z
\left(\frac{(z+1)_n}{n!}\right)^{-1}(n+1)^z=\frac{n!n^z}{(z+1)_n}.\hspace{3.9cm}(3.1.5)\end{eqnarray*}
Moreover, \begin{eqnarray*}& & \left(1+\frac{z}{r}\right)^{-1}
\left(1+\frac{1}{r}\right)^z\\ &=&
\left(1-\frac{z}{r}+\frac{z^2}{r^2}+O\left(\frac{1}{r^3}\right)\right)
\left(1+\frac{z}{r}+\frac{z(z-1)}{2r^2}+O\left(\frac{1}{r^3}\right)\right)\\
&=&1+\frac{z(z-1)}{2r^2}+O\left(\frac{1}{r^3}\right).\hspace{9.4cm}(3.1.6)\end{eqnarray*}
This shows that
$$\lim_{n\rta\infty}\prod_{r=1}^n\left(1+\frac{z}{r}\right)^{-1}
\left(1+\frac{1}{r}\right)^z\;\;\mbox{exists}.\eqno(3.1.7)$$ Thus we
have a function
$$\Pi(z)=\lim_{n\rta\infty}\frac{n!n^z}{(z+1)_n}=\prod_{r=1}^\infty\left(1+\frac{z}{r}\right)^{-1}
\left(1+\frac{1}{r}\right)^z\eqno(3.1.8)$$ and $\Pi(m)=m!$ for
$m\in\mbb{N}$ by (3.1.4). For notional convenience, we define the
gamma function
$$\G(z)=\Pi(z-1)=\lim_{n\rta\infty}\frac{n!n^{z-1}}{(z)_n}\qquad\for\;\;
z\in\mbb{C}\setminus\{-\mbb{N}-1\}.\eqno(3.1.9)$$ Then
\begin{eqnarray*}\qquad
\G(z+1)&=&\lim_{n\rta\infty}\frac{n!n^z}{(z+1)_n}=
z\lim_{n\rta\infty}\frac{n!n^z}{(z)_{n+1}}\\
&=&z\lim_{n\rta\infty}\frac{n}{z+n}\frac{n!n^{z-1}}{(z)_{n}}=
z\left(\lim_{n\rta\infty}\frac{n}{z+n}\right)\left(\lim_{n\rta\infty}\frac{n!n^{z-1}}{(z)_{n}}\right)
\\ &=&
z\lim_{n\rta\infty}\frac{n!n^{z-1}}{(z)_{n}}=z\G(z).\hspace{7.25cm}(3.1.10)\end{eqnarray*}
By (3.1.9), $\G(1)=1$. So $\G(m+1)=m!$ for $m\in\mbb{N}$.

For $x,y\in\mbb{C}$ with $\mbox{Re}\:x>0$ and $\mbox{Re}\:y>0$, we
define the beta function
$$B(x,y)=\int_0^1t^{x-1}(1-t)^{y-1}dt.\eqno(3.1.11)$$
\pse

{\bf Theorem 3.1.1}. {\it We have} $B(x,y)=\G(x)\G(y)/\G(x+y)$.

{\it Proof}. Note
$$B(x,y+1)=\int_0^1t^{x-1}(1-t)(1-t)^{y-1}dt=B(x,y)-B(x+1,y).\eqno(3.1.12)$$
On the other hand, integration by parts gives
\begin{eqnarray*}\qquad B(x,y+1)&=&\int_0^1t^{x-1}(1-t)^{y}dt\\ &=&\frac{t^x(1-t)^y}{x}|^1_0+\frac{y}{x}
\int_0^1t^{x}(1-t)^{y-1}dt=\frac{y}{x}B(x+1,y).\hspace{1.7cm}(3.1.13)\end{eqnarray*}
Thanks to the above two expressions, we have
$$B(x,y)-\frac{x}{y}B(x,y+1)=B(x,y+1)\lra
B(x,y)=\frac{x+y}{y}B(x,y+1) .\eqno(3.1.14)$$ By induction
$$B(x,y)=\frac{(x+y)_n}{(y)_n}B(x,y+n).\eqno(3.1.15)$$
Rewrite the above equation as
\begin{eqnarray*}\qquad
B(x,y)&=&\frac{(x+y)_n}{n!}\frac{n!}{(y)_n}\int_0^1t^{x-1}(1-t)^{y+n-1}dt\\
&\stl{t=s/n}{=}&\frac{(x+y)_n}{n!}\frac{n!}{(y)_n}\int_0^nn^{1-x}s^{x-1}\left(1-\frac{s}{n}\right)^{y+n-1}\frac{ds}{n}
\\
&=&\frac{(x+y)_n}{n!n^{x+y-1}}\frac{n!n^{y-1}}{(y)_n}\int_0^ns^{x-1}\left(1-\frac{s}{n}\right)^{y+n-1}ds\\
&=&\lim_{n\rta\infty}
\frac{(x+y)_n}{n!n^{x+y-1}}\frac{n!n^{y-1}}{(y)_n}\int_0^ns^{x-1}\left(1-\frac{s}{n}\right)^{y+n-1}ds
\\ &=&\frac{\G(y)}{\G(x+y)}\int_0^\infty
s^{x-1}e^{-s}ds.\hspace{6.6cm}(3.1.16)\end{eqnarray*} Taking $y=1$
in the above equation, we have
$$B(x,1)\G(x+1)=\int_0^\infty
s^{x-1}e^{-s}ds.\eqno(3.1.17)$$ Furthermore, (3.1.11) gives
$$B(x,1)=\int^1_0t^{x-1}dt=\frac{1}{x}.\eqno(3.1.18)$$
Thus
$$\G(x)=\frac{\G(x+1)}{x}=B(x,1)\G(x+1)=\int_0^\infty
s^{x-1}e^{-s}ds.\eqno(3.1.19)$$ Therefore,
$$B(x,y)=\frac{\G(y)}{\G(x+y)}\int_0^\infty
s^{x-1}e^{-s}ds=\frac{\G(x)\G(y)}{\G(x+y)}.\qquad\Box\eqno(3.1.20)$$

Recall the Euler's constant
$$\gm=\lim_{n\rta\infty}\left(\sum_{r=1}^n\frac{1}{r}-\ln
n\right).\eqno(3.1.21)$$ \pse

{\bf Theorem 3.1.2}. The following equation holds:
$$\frac{1}{\G(z)}=ze^{\gm z}\prod_{n=1}^\infty
\left(1+\frac{z}{n}\right)e^{-z/n}.\eqno(3.1.22)$$

{\it Proof}. Note
$$\left(1+\frac{z}{n}\right)e^{-z/n}=\left(1+\frac{z}{n}\right)\left(1-\frac{z}{n}+\frac{z^2}{2n^2}
+O\left(\frac{1}{n^3}\right)\right)=1-\frac{z^2}{2n^2}
+O\left(\frac{1}{n^3}\right).\eqno(3.1.23)$$ Thus the product in
(3.1.22) converges.  Moreover,
\begin{eqnarray*}\frac{1}{\G(z)}&=&\lim_{n\rta\infty}\frac{(z)_n}{n!n^{z-1}}
=\lim_{n\rta\infty}\frac{z(z+1)\cdots(z+n-1)}{(n-1)!n^z}
\\ &=&z\lim_{n\rta\infty}\left[\prod_{r=1}^{n-1}\left(1+\frac{z}{r}\right)\right]e^{-z\ln
n}\\ &=&z\lim_{n\rta\infty}e^{z[\sum_{r=1}^n1/r-\ln
n]}e^{-z/n}\prod_{r=1}^{n-1}\left(1+\frac{z}{r}\right)e^{-z/r}\\
&=&ze^{\gm
z}\prod_{r=1}^\infty\left(1+\frac{z}{r}\right)e^{-z/r}.\qquad\Box\hspace{7.5cm}(3.1.24)\end{eqnarray*}
\pse

{\bf Theorem 3.1.3}. {\it Euler's reflection formula:}
$$\G(z)\G(1-z)=\frac{\pi}{\sin \pi z}.\eqno(3.1.25)$$

{\it Proof}. From complex analysis,
$$\frac{\sin\pi
z}{\pi
z}=\prod_{n=1}^\infty\left(1-\frac{z^2}{n^2}\right).\eqno(3.1.26)$$
According to (3.1.22),
\begin{eqnarray*}\G(z)\G(-z)&=&\left[ze^{\gm z}\prod_{n=1}^\infty
\left(1+\frac{z}{n}\right)e^{-z/n}\right]^{-1}\left[-ze^{-\gm
z}\prod_{n=1}^\infty \left(1-\frac{z}{n}\right)e^{z/n}\right]^{-1}
\\ &=&-\frac{1}{z^2}\left[\prod_{n=1}^\infty
\left(1+\frac{z}{n}\right)\left(1-\frac{z}{n}\right)\right]^{-1}
\\ &=&-\frac{1}{z^2}\left[\prod_{n=1}^\infty
\left(1-\frac{z^2}{n^2}\right)\right]^{-1}=-\frac{\pi}{z\sin\pi
z}.\hspace{4.9cm}(3.1.27)\end{eqnarray*} Now (3.1.25) follows from
the fact $\G(1-z)=-z\G(-z).\qquad\Box$\psp

Letting $z=1/2$ in (3.1.25), we get $\G(1/2)=\sqrt{\pi}.$ Taking the
logarithm of (3.1.22), we have
$$-\ln\G(z)=\gm z+\ln z+\sum_{n=1}^\infty
\left[\ln\left(1+\frac{z}{n}\right)-\frac{z}{n}\right].\eqno(3.1.28)$$
Differentiating (3.1.28), we get
$$\psi(z)=\frac{\G'(z)}{\G(z)}=-\gm-\frac{1}{z}+\sum_{n=1}^\infty\left(\frac{1}{n}-\frac{1}{z+n}\right).\eqno(3.1.29)$$
In particular,
$$\psi'(z)=\sum_{n=0}^\infty\frac{1}{(z+n)^2}=\zeta(2,z),\eqno(3.1.30)$$
where the Riemman zeta function
$$\zeta(s,a)=\sum_{n=0}^\infty\frac{1}{(n+a)^s},\qquad\mbox{Re}\:s>1.\eqno(3.1.31)$$\pse

{\bf Theorem 3.1.3}. {\it The following product formula holds}:
$$\G(z)\G\left(z+\frac{1}{n}\right)\G\left(z+\frac{2}{n}\right)\cdots\G\left(z+\frac{n-1}{n}\right)
=\frac{(2\pi)^{(n-1)/2}}{n^{nz-1/2}}\G(nz).\eqno(3.1.32)$$

{\it Proof}.  Set
$$\phi(z)=\frac{n^{nz}}{n\G(nz)}\prod_{p=0}^{n-1}\G\left(z+\frac{p}{n}\right).\eqno(3.1.33)$$
Then (3.1.9) says
\begin{eqnarray*}
\qquad\phi(z)&=&\lim_{r\rta\infty}n^{nz-1}\frac{\prod_{p=0}^{n-1}\frac{r!r^{z+(p-n)/n}}{(z+p/n)_r}}
{\frac{(nr)!(nr)^{nz-1}}{(nz)_{nr}}}\\ &=&
\lim_{r\rta\infty}\frac{\prod_{p=0}^{n-1}\frac{r!r^{(p-n)/n}}{(z+p/n)_r}}
{\frac{(nr)!r^{-1}}{(nz)_{nr}}}=
\lim_{r\rta\infty}\frac{(r!)^nn^{rn}}
{(nr)!r^{(n+1)/2}}.\hspace{4.8cm}(3.1.34)
\end{eqnarray*}
Thus $\phi$ is a constant. Hence
$$\phi(z)=\phi(1/n)=\prod_{j=1}^{n-1}\G\left(\frac{j}{n}\right)=
\prod_{j=1}^{n-1}\G\left(1-\frac{j}{n}\right).\eqno(3.1.35)$$ So
$$\phi^2=\prod_{j=1}^{n-1}\G\left(\frac{j}{n}\right)\G\left(1-\frac{j}{n}\right)
=\prod_{j=1}^{n-1}\frac{\pi}{\sin j\pi/n}.\eqno(3.1.36)$$

Note
$$\sum_{r=0}^{n-1}z^r=\frac{z^n-1}{z-1}=\prod_{j=1}^{n-1}(z-e^{2j\pi
i/n}).\eqno(3.1.37)$$ Hence
\begin{eqnarray*}n&=&\prod_{j=1}^{n-1}(1-e^{2j\pi
i/n})=\prod_{j=1}^{n-1}e^{j\pi i/n}(e^{-j\pi i/n}-e^{j\pi i/n})\\
&=&e^{(n-1)\pi i/2}\prod_{j=1}^{n-1}(-2i\sin
j\pi/n)=2^{n-1}e^{(n-1)\pi i/2}(-i)^{n-1}\prod_{j=1}^{n-1}\sin
j\pi/n\\ &=&2^{n-1}e^{(n-1)\pi i/2}e^{3(n-1)\pi
i/2}\prod_{j=1}^{n-1}\sin j\pi/n=2^{n-1}\prod_{j=1}^{n-1}\sin
j\pi/n.\hspace{3.4cm}(3.1.38)\end{eqnarray*} By (3.1.36) and
(3.1.38),
$$\phi^2=\frac{(2\pi)^{n-1}}{n}\lra
\phi=\frac{(2\pi)^{(n-1)/2}}{\sqrt{n}}.\eqno(3.1.39)$$ Then (3.1.32)
follows from (3.1.33) and (3.1.39). $\qquad\Box$

\section{Gauss Hypergeometric Functions}

The term of ``hypergeometric'' was first used by Wallis in Oxford as
early as 1655 in his
    work {\it Arithmetrica Infinitorm} when referring  to any series which could be regarded as
     a generalization of the ordinary geometric series
$\sum_{n=0}^{\infty}z^n$. Nowadays a power series $\sum_{n=0}^\infty
c_nz^{n+\mu}$ is called a {\it hypergeometric function} if
$c_{n+1}/c_n$ is a rational function of $n$.  In this section, we
use $z$ to denote independent variable instead of $t$. The classical
hypergeometric equation is
$$z(1-z){y'}'+[\gm-(\al+\be+1)z]y'-\al\be y=0.\eqno(3.2.1)$$
We look for the solution of the form
$$y=\sum_{n=0}^\infty c_nz^{n+\mu},\eqno(3.2.2)$$
where $c_n$ and $\mu$ are constants to be determined. We calculate
$$y'=\sum_{n=0}^\infty (n+\mu)c_nz^{n+\mu-1},\qquad
{y'}'=\sum_{n=0}^\infty
(n+\mu)(n+\mu-1)c_nz^{n+\mu-2}.\eqno(3.2.3)$$ Substituting (3.2.3)
into (3.2.1), we get
$$z^\mu\sum_{n=0}^\infty
\{(n+\mu)(n+\mu-1)c_nz^{n-1}(1-z)+
(n+\mu)c_nz^{n-1}[\gm-(\al+\be+1)z]-\al\be c_nz^n\}=0,\eqno(3.2.4)$$
equivalently,
$$\mu(\mu-1+\gm)=0,\eqno(3.2.5)$$
$$(n+1+\mu)(n+\mu+\gm)c_{n+1} =[(n+\mu)(n+\mu+\al+\be)+\al\be]c_n\eqno(3.2.6)$$
for $n\in\mbb{N}$. We rewrite (3.2.6) as
$$(n+1+\mu)(n+\mu+\gm)c_{n+1} =(n+\mu+\al)(n+\mu+\be)c_n.\eqno(3.2.7)$$
By induction, we have
$$c_n=\frac{(\mu+\al)_n(\mu+\be)_n}{(\mu+1)_n(\mu+\gm)_n}c_0\qquad\for\;\;n\in\mbb{N}+1.\eqno(3.2.8)$$
Hence
$$y=c_0\sum_{n=0}^\infty
\frac{(\mu+\al)_n(\mu+\be)_n}{(\mu+1)_n(\mu+\gm)_n}z^{n+\mu}.\eqno(3.2.9)$$
According to (3.2.5), $\mu=0$ or $\mu=1-\gm$. Considering $\mu=0$,
we denote
$$_2F_1(\al,\be;\gm;z)=\sum_{n=0}^\infty
\frac{(\al)_n(\be)_n}{n!(\gm)_n}z^n,\eqno(3.2.10)$$
 which was introduced and studied  by Gauss
in his thesis presented at G\"{o}ttingen in 1812. We call it {\it
classical Gauss hypergeometric function}.\index{Gauss hypergeometric
function} Since
$$\lim_{n\rta\infty}\left[\frac{(\al)_{n+1}(\be)_{n+1}}{(n+1)!(\gm)_{n+1}}/\frac{(\al)_n(\be)_n}{n!(\gm)_n}\right]
=\lim_{n\rta\infty}\frac{(\al+n)(\be+n)}{(n+1)(\gm+n)}=1,\eqno(3.2.11)$$
the series in (3.2.10) converges absolutely when $|z|<1$. It can be
proved that $ _2F_1(\al,\be;\gm;z)$ has analytic extension on the
whole complex $z$ plane by complex analysis. Note that $
_2F_1(\al-\gm+1,\be-\gm+1;2-\gm;z)z^{1-\gm}$ is another solution of
(3.2.1) by (3.2.9).

Observe
$$_2F_1(\al,\be;\gm;0)=1.\eqno(3.2.12)$$
By (3.2.9), $ _2F_1(\al,\be;\gm;z)$ is the unique power series
solution of (3.2.1) satisfying (3.2.12). It has close relations with
elementary functions:
$$ _2F_1(-\al,\be;\be;-z)=\sum_{n=0}^\infty
\frac{(-1)^n(-\al)_n}{n!}z^n=\sum_{n=0}^\infty{\al\choose
n}z^n=(1+z)^\al,\eqno(3.2.13)$$
$$_2F_1(1,1;2;-z)z=\sum_{n=0}^\infty
\frac{n!n!}{n!(n+1)!}(-1)^nz^{n+1}=\ln (1+z),\eqno(3.2.14)$$
$$\lim_{\be\rta\infty}\: _2F_1(1,\be;1;z/\be)=\lim_{\be\rta\infty}\sum_{n=0}^\infty
\frac{n!(\be)_n}{n!n!\be^n}z^n=\sum_{n=0}^\infty
\frac{1}{n!}z^n=e^z,\eqno(3.2.15)$$
\begin{eqnarray*} \lim_{\al,\be\rta\infty}\:
_2F_1(\al,\be;3/2;-z^2/4\al\be)z&=&\lim_{\al,\be\rta\infty}\sum_{n=0}^\infty
\frac{(\al)_n(\be)_n}{n!(3/2)_n\al^n\be^n4^n}(-1)^nz^{2n+1}\\
&=&\sum_{n=0}^\infty \frac{(-1)^n}{(2n+1)!}z^{2n+1}=\sin
z,\hspace{3cm}(3.2.16)\end{eqnarray*}
\begin{eqnarray*} \lim_{\al,\be\rta\infty}\:
_2F_1(\al,\be;1/2;-z^2/4\al\be)&=&\lim_{\al,\be\rta\infty}\sum_{n=0}^\infty
\frac{(\al)_n(\be)_n}{n!(1/2)_n\al^n\be^n4^n}(-1)^nz^{2n}\\
&=&\sum_{n=0}^\infty \frac{(-1)^n}{(2n)!}z^{2n}=\cos
z.\hspace{4.2cm}(3.2.17)\end{eqnarray*} Less obviously,
$$ _2F_1(1/2,1/2;3/2;z^2)z=\arcsin z,\qquad
_2F_1(1/2,1;3/2;-z^2)z=\arctan z.\eqno(3.2.18)$$ In addition,
\begin{eqnarray*}\qquad\frac{d}{dz}\: _2F_1(\al,\be;\gm;z)&=&\sum_{n=0}^\infty
\frac{(\al)_n(\be)_n}{(n-1)!(\gm)_n}z^{n-1}\\ &=&\sum_{n=0}^\infty
\frac{(\al)_{n+1}(\be)_{n+1}}{n!(\gm)_{n+1}}z^n
\\ &=&\frac{\al\be}{\gm}\sum_{n=0}^\infty
\frac{(\al+1)_n(\be+1)_n}{n!(\gm+1)_n}z^n\\
&=&\frac{\al\be}{\gm}\:
_2F_1(\al+1,\be+1;\gm+1;z).\hspace{4.3cm}(3.2.19)\end{eqnarray*}
Furthermore, we have the following {\it Euler's Integral
Representation}.\psp

{\bf Theorem 3.2.1}. {\it If $\mbox{\it Re}\:\gm>\mbox{\it
Re}\:\be>0$, then
$$
_2F_1(\al,\be;\gm;z)=\frac{\G(\gm)}{\G(\be)\G(\gm-\be)}\int_0^1t^{\be
-1}(1-t)^{\gm-\be-1}(1-zt)^{-\al}dt\eqno(3.2.20)$$ in the $z$ plane
cut along the real axis from 1 to $\infty$.}

 {\it Proof}. First we
suppose $|z|<1$. We calculate
\begin{eqnarray*}\qquad & &\frac{\G(\gm)}{\G(\be)\G(\gm-\be)}\int_0^1t^{\be
-1}(1-t)^{\gm-\be-1}(1-zt)^{-\al}dt\\
&=&\frac{\G(\gm)}{\G(\be)\G(\gm-\be)}\sum_{n=0}^\infty(-1)^n{-\al\choose
n}z^n\int_0^1t^{\be+n-1}(1-t)^{\gm-\be-1}dt\\
&=&\frac{\G(\gm)}{\G(\be)\G(\gm-\be)}\sum_{n=0}^\infty\frac{(\al)_n}{n!}z^nB(\be+n,\gm-\be)
\\
&=&\frac{\G(\gm)}{\G(\be)\G(\gm-\be)}\sum_{n=0}^\infty\frac{(\al)_n}{n!}\frac{\G(\be+n)\G(\gm-\be)}{\G(\gm+n)}z^n
\\&=&\sum_{n=0}^\infty\frac{(\al)_n\G(\be+n)\G(\gm)}{n!\G(\be)\G(\gm+n)}z^n
\\&=&\sum_{n=0}^\infty\frac{(\al)_n(\be)_n}{n!(\gm)_n}z^n=\:_2F_1(\al,\be;\gm;z)\hspace{7cm}(3.2.21)\end{eqnarray*}
by (3.1.10), (3.1.11) and Theorem 3.1.1. So the theorem holds for
$|z|<1$.

Since the integral in (3.2.20) is analytic in the cut plane, the
theorem holds for $z$ in this region as well. $\qquad\Box$ \psp

{\bf Theorem 3.2.2 (Gauss (1812))}. {\it If $\mbox{\it
Re}(\gm-\al-\be)>0$, then}
$$
_2F_1(\al,\be;\gm;1)=\frac{\G(\gm)\G(\gm-\al-\be)}{\G(\gm-\al)\G(\gm-\be)}.\eqno(3.2.22)$$

{\it Proof}. By Abel's continuity theorem, (3.2.20) and Theorem
3.1.1,
\begin{eqnarray*} _2F_1(\al,\be;\gm;1)&=&\lim_{z\rta 1-}\frac{\G(\gm)}{\G(\be)\G(\gm-\be)}\int_0^1t^{\be
-1}(1-t)^{\gm-\be-1}(1-zt)^{-\al}dt\\
&=&\frac{\G(\gm)}{\G(\be)\G(\gm-\be)}\int_0^1t^{\be
-1}(1-t)^{\gm-\be-\al-1}dt\\
&=&\frac{\G(\gm)}{\G(\be)\G(\gm-\be)}B(\be,\gm-\be-\al)\\
&=&\frac{\G(\gm)\G(\gm-\al-\be)}{\G(\gm-\al)\G(\gm-\be)}
\hspace{7.9cm}(3.2.23)\end{eqnarray*} when $\mbox{ Re}\:\gm>\mbox{
Re}\:\be>0$ and $\mbox{Re}(\gm-\al-\be)>0$. The condition $\mbox{
Re}\:\gm>\mbox{ Re}\:\be>0$ can be removed in (3.2.22) by the
continuity in $\be$ and $\gm.\qquad\Box$\psp

By (3.1.10), we have:\psp

{\bf Corollary 3.2.3 (Chu-Vandermonde)}. {\it For $n\in\mbb{N}$,}
$$_2F_1(-n,\be;\gm;1)=\frac{(\gm-\be)_n}{(\gm)_n}.\eqno(3.2.24)$$

\section{Orthogonal Polynomials}

Let $k\in\mbb{N}$,
$$ _2F_1(-k,\be;\gm;z)=\sum_{n=0}^k
\frac{(-k)_n(\be)_n}{n!(\gm)_n}z^n=\sum_{n=0}^k {k\choose
n}\frac{(\be)_n}{(\gm)_n}(-z)^n \eqno(3.3.1)$$ is a polynomial. We
calculate the generating function
\begin{eqnarray*}& &\sum_{k=0}^\infty \frac{(\gm)_kx^k}{k!}\:
_2F_1(-k,\be;\gm;z) =\sum_{k=0}^\infty
\sum_{n=0}^k\frac{(\gm+n)_{k-n}(\be)_n}{n!(k-n)!}x^k(-z)^n
\\ &=&\sum_{k=0}^\infty \sum_{n=0}^k{\gm+k-1\choose k-n}{-\be\choose
n}x^kz^n =\sum_{m,n=0}^\infty {\gm+m+n-1\choose m}{-\be\choose
n}x^{m+n}z^n\\ &=&\sum_{n=0}^\infty(1-x)^{-\gm-n}{-\be\choose
n}x^nz^n = (1-x)^{-\gm}\left(1+\frac{xz}{1-x}\right)^{-\be}
\\&=& (1-x)^{\be-\gm}(1+(z-1)x)^{-\be}.\hspace{8.75cm}(3.3.2)\end{eqnarray*}

Set
$$w_k(\vt,\gm;z)=\:_2F_1(-k,\vt+k;\gm;z).\eqno(3.3.3)$$
According to (3.2.1),
$$z(1-z){w_k'}'+[\gm-(\vt+1)z]w_k'+k(\vt+k)w_k=0.\eqno(3.3.4)$$
Thus
$$\frac{d}{dz}[z^\gm (1-z)^{\vt-\gm+1}w'_k]+k(\vt+k)z^{\gm-1}
(1-z)^{\vt-\gm}w_k=0.\eqno(3.3.5)$$ Let $m,n\in\mbb{N}$ such that
$m\neq n$. Then
$$w_m\frac{d}{dz}[z^\gm (1-z)^{\vt-\gm+1}w'_n]+n(\vt+n)z^{\gm-1}
(1-z)^{\vt-\gm}w_mw_n=0\eqno(3.3.6)$$ and
$$w_n\frac{d}{dz}[z^\gm (1-z)^{\vt-\gm+1}w'_m]+m(\vt+m)z^{\gm-1}
(1-z)^{\vt-\gm}w_mw_n=0.\eqno(3.3.7)$$ Assume that
$\mbox{Re}\:\gm>0$, $\mbox{Re}\:(\vt-\gm)>-1$ and $\vt\not\in
-\mbb{N}-1$. Then
\begin{eqnarray*} & &\int_0^1z^{\gm-1}
(1-z)^{\vt-\gm}w_mw_ndz\\
&=&\frac{1}{(m-n)(m+n+\vt)}\int_0^1[m(\vt+m)-n(\vt+n)]z^{\gm-1}
(1-z)^{\vt-\gm}w_mw_ndz\\
&=&\frac{1}{(m-n)(m+n+\vt)}[\int_0^1w_m\frac{d}{dz}[z^\gm
(1-z)^{\vt-\gm+1}w'_n]dz\\ & &-\int_0^1w_n\frac{d}{dz}[z^\gm
(1-z)^{\vt-\gm+1}w'_m]dz]
\\ &=&\frac{z^\gm
(1-z)^{\vt-\gm+1}(w_mw_n'-w'_mw_n)
}{(m-n)(m+n+\vt)}|^1_0=0.\hspace{6.9cm}(3.3.8)\end{eqnarray*}

Let ${\cal C}_z$ be a loop around $z$. According to (3.3.2),
\begin{eqnarray*}\qquad& &\frac{(\gm)_k}{k!}\:
_2F_1(-k,\be;\gm;z)\\&=&\frac{1}{2\pi i}\int_{{\cal
C}_0}\frac{(1-x)^{\be-\gm}(1+(z-1)x)^{-\be}}{x^{k+1}}dx\\
&\stl{x=\frac{s-z}{s(1-z)}}{=}&\frac{1}{2\pi i}\int_{{\cal
C}_z}\frac{[z(1-s)/s(1-z)]^{\be-\gm}(z/s)^{-\be}}{[(s-z)/s(1-z)]^{k+1}}\frac{z}{s^2(1-z)}ds
\\ &=&\frac{z^{1-\gm}(1-z)^{\gm-\be+k}}{2\pi i}\int_{{\cal
C}_z}\frac{s^{\gm+k-1}(1-s)^{\be-\gm}}{(s-z)^{k+1}}ds
\\ &=&\frac{z^{1-\gm}(1-z)^{\gm-\be+k}}{2\pi i k!}\left(\frac{d}{dz}\right)^k\int_{{\cal
C}_z}\frac{s^{\gm+k-1}(1-s)^{\be-\gm}}{s-z}ds
\\ &=&\frac{z^{1-\gm}(1-z)^{\gm-\be+k}}{k!}\left(\frac{d}{dz}\right)^k[z^{\gm+k-1}(1-z)^{\be-\gm}].\hspace{4cm}(3.3.9)
\end{eqnarray*}
Hence
$$w_k(\vt,\gm;z)=\frac{z^{1-\gm}(1-z)^{\gm-\vt}}{(\gm)_k}\left(\frac{d}{dz}\right)^k[z^{\gm+k-1}(1-z)^{\vt-\gm+k}].
\eqno(3.3.10)$$ By (3.3.1),
$$\left(\frac{d}{dz}\right)^k(w_k)=\left(\frac{d}{dz}\right)^k(\sum_{n=0}^k {k\choose
n}\frac{(\vt+k)_n}{(\gm)_n}(-z)^n)=\frac{(-1)^kk!(\vt+k)_k}{(\gm)_k}.\eqno(3.3.11)$$
Thus
\begin{eqnarray*}\qquad& &\int_0^1z^{\gm-1}
(1-z)^{\vt-\gm}w_k^2dz\\
&=&\frac{1}{(\gm)_k}\int_0^1w_k\left(\frac{d}{dz}\right)^k[z^{\gm+k-1}(1-z)^{\vt-\gm+k}]dz
\\&=&\frac{(-1)^k}{(\gm)_k}\int_0^1\left(\frac{d}{dz}\right)^k(w_k)z^{\gm+k-1}(1-z)^{\vt-\gm+k}dz
\\&=&\frac{k!(\vt+k)_k}{((\gm)_k)^2}\int_0^1z^{\gm+k-1}(1-z)^{\vt-\gm+k}dz
\\&=&\frac{k!(\vt+k)_k\G(\gm+k)\G(\vt-\gm+k+1)}{((\gm)_k)^2\G(\vt+2k+1)}
\\&=&\frac{k!(\vt+k)_k\G(\gm)\G(\vt-\gm+k+1)}{(\gm)_k\G(\vt+2k+1)}
.\hspace{6.8cm}(3.3.12)
\end{eqnarray*}
Therefore $\{w_k(\vt,\gm;z)\mid k\in\mbb{N}\}$ forms a set of
orthogonal polynomials with respect to the weight $z^{\gm-1}
(1-z)^{\vt-\gm}$. The {\it Jacobi polynomials}\index{Jacobi
polynomial}
$$P^{(\al,\be)}_k(z)={\al+k\choose
k}w_k\left(\al+\be+1,\al+1;\frac{1-z}{2}\right).\eqno(3.3.13)$$
Indeed $\{P^{(\al,\be)}_k(z)\mid k\in\mbb{N}\}$ forms a complete set
of orthogonal functions on $[-1,1]$ with respect to the weight
$(1-z)^\al(1+z)^\be$. According (3.3.10),
$$P^{(\al,\be)}_k(z)=\frac{(-2)^{-k}}{k!}(1-z)^{-\al}(1+z)^{-\be}\left(\frac{d}{dz}\right)^k[(1-z)^{\al+k}(1+z)^{\be+k}].
\eqno(3.3.14)$$

The well known {\it Chebyshev polynomials of first
kind}\index{Chebyshev polynomials! of first kind}
$$T_k(z)=\frac{1}{{-1/2+k\choose
k}}P^{(-1/2,-1/2)}_k(z)=\frac{(-1)^k\sqrt{1-z^2}}{(2k-1)!!}
\left(\frac{d}{dz}\right)^k[(1-z^2)^{k-1/2}].\eqno(3.3.15)$$ The
well known {\it Chebyshev polynomials of second kind}\index{
Chebyshev polynomials! of second kind}
$$U_k(z)=\frac{(k+1)!}{{1/2+k\choose
k}}P^{(1/2,1/2)}_k(z)=\frac{(-1)^k(k+1)!}{(2k+1)!!\sqrt{1-z^2}}
\left(\frac{d}{dz}\right)^k[(1-z^2)^{k+1/2}].\eqno(3.3.16)$$ \pse

Equation
$$(1-z^2){y'}'-2zy'+\nu(\nu+1)y=0\eqno(3.3.17)$$
is called a {\it Legendre equation}, where $\nu$ is a constant.
Suppose that $y=\sum_{n=0}^\infty c_n z^n$ is a solution of
(3.3.17). Then
$$(1-z^2)(\sum_{n=2}^\infty n(n-1)c_nz^{n-2})
-2\sum_{n=1}^\infty nc_nz^n+\nu(\nu+1)\sum_{n=0}^\infty c_n
z^n=0,\eqno(3.3.18)$$ equivalently,
$$(n+2)(n+1)c_{n+2}+[\nu(\nu+1)-n(n+1)]c_n=0.\eqno(3.3.19)$$
Thus
$$c_{n+2}=\frac{(n-\nu)(n+1+\nu)}{(n+2)(n+1)}c_n.\eqno(3.3.20)$$
By induction,
$$c_{2n}=\frac{\prod_{i=0}^{n-1}(2i-\nu)(2i+1+\nu)}{(2n)!}c_0=
\frac{(-\nu/2)_n((1+\nu)/2)_n}{n!(1/2)_n}c_0,\eqno(3.3.21)$$
$$c_{2n+1}=\frac{\prod_{i=0}^{n-1}(2i+1-\nu)(2i+2+\nu)}{(2n+1)!}c_1=
\frac{((1-\nu)/2)_n((2+\nu)/2)_n}{n!(3/2)_n}c_1.\eqno(3.3.22)$$ Thus
for generic $\nu$, we have the fundamental solutions
$$\sum_{n=0}^\infty\frac{(-\nu/2)_n((1+\nu)/2)_n}{n!(1/2)_n}z^{2n}=\:_2F_1
\left(-\frac{\nu}{2},\frac{1+\nu}{2};\frac{1}{2};z^2\right)\eqno(3.3.23)$$
and
$$\sum_{n=0}^\infty\frac{((1-\nu)/2)_n((2+\nu)/2)_n}{n!(3/2)_n}z^{2n+1}=\;_2F_1
\left(\frac{1-\nu}{2},\frac{2+\nu}{2};\frac{3}{2};z^2\right)z,\eqno(3.3.24)$$
which are called {\it Legendre functions}.\index{Legendre function}
When $\nu=2k$ is nonnegative even integer, the first solution is a
polynomial and we denote the {\it Legendre
polynomial}\index{Legendre polynomial}
$$P_{2k}(z)=\frac{(-1)^k(1/2)_k}{k!}\:_2F_1
\left(-k,\frac{1}{2}+k;\frac{1}{2};z^2\right).\eqno(3.3.25)$$ If
$\nu=2k+1$ is an odd integer, the second solution is a polynomial
and we denote the {\it Legendre polynomial}
$$P_{2k+1}(z)=\frac{(-1)^k2(1/2)_{k+1}}{k!}\:_2F_1
\left(-k,\frac{3}{2}+k;\frac{3}{2};z^2\right)z.\eqno(3.3.26)$$ \pse

{\bf Theorem 3.3.1}. {\it For $n\in\mbb{N}$,}
$$P_n(z)=\frac{1}{2^nn!}\left(\frac{d}{dz}\right)^n[(z^2-1)^n].\eqno(3.3.27)$$

{\it Proof}. For convenience,  we set
$$\psi_n=\left(\frac{d}{dz}\right)^n[(z^2-1)^n].\eqno(3.3.28)$$
We want to prove
$$(1-z^2){\psi_n'}'-2z\psi_n'+n(n+1)\psi_n=0,\eqno(3.3.29)$$
which is equivalent to
$$[(1-z^2)\psi_n']'+n(n+1)\psi_n=0.\eqno(3.3.30)$$
Explicitly, (3.3.30) is
$$\left[(1-z^2)\left(\frac{d}{dz}\right)^{n+1}[(z^2-1)^n]
+n(n+1)\left(\frac{d}{dz}\right)^{n-1}[(z^2-1)^n]\right]'=0,\eqno(3.3.31)$$
equivalently,
$$(1-z^2)\left(\frac{d}{dz}\right)^{n+1}[(z^2-1)^n]
+n(n+1)\left(\frac{d}{dz}\right)^{n-1}[(z^2-1)^n]=0\eqno(3.3.32)$$
due to that both terms are equal to zero when $z=1$. Note
\begin{eqnarray*}& &(1-z^2)\left(\frac{d}{dz}\right)^{n+1}[(z^2-1)^n]\\
&=&(1-z^2)\left(\frac{d}{dz}\right)^{n+1}[(z-1)^n(z+1)^n]
\\ &=&-\sum_{s=0}^{n-1}{n+1\choose
s+1}[\prod_{p=0}^s(n-p)][\prod_{r=s+1}^nr](z-1)^{n-s}(z+1)^{s+1}\\
&=&-\sum_{s=0}^{n-1}\frac{(n+1)![\prod_{p=0}^s(n-p)][\prod_{r=s+1}^nr]}{(s+1)!(n-s)!}(z-1)^{n-s}(z+1)^{s+1}
\\
&=&-\sum_{s=0}^{n-1}\frac{(n+1)![\prod_{p=0}^{s-1}(n-p)][\prod_{r=s+2}^nr]}{s!(n-s-1)!}(z-1)^{n-s}(z+1)^{s+1}
\\ &=&-n(n+1)\sum_{s=0}^{n-1}\frac{(n-1)![\prod_{p=0}^{s-1}(n-p)][\prod_{r=s+2}^nr]}{s!(n-s-1)!}(z-1)^{n-s}(z+1)^{s+1}
\\ &=&-n(n+1)\sum_{s=0}^{n-1}{n-1\choose
s}[\prod_{p=0}^{s-1}(n-p)][\prod_{r=s+2}^nr](z-1)^{n-s}(z+1)^{s+1}
\\ &=&-n(n+1)\left(\frac{d}{dz}\right)^{n-1}[(z-1)^n(z+1)^n]
\\
&=&-n(n+1)\left(\frac{d}{dz}\right)^{n-1}[(z^2-1)^n],\hspace{7.8cm}(3.3.33)\end{eqnarray*}
that is (3.3.32) holds.

On the other hand,
$$\frac{1}{2^nn!}\psi_n=\left(\frac{d}{dz}\right)^n\sum_{r=0}^n\frac{(-1)^rz^{2n-2r}}{r!(n-r)!2^n}.\eqno(3.3.34)$$
Thus for $k\in\mbb{N}$,
$$\frac{1}{2^{2k}(2k)!}\psi_{2k}(z)|_{z=0}=\frac{(-1)^k(2k)!}{(k!)^22^{2k}}=\frac{(-1)^k(1/2)_k}{k!}\eqno(3.3.35)$$
and
$$\frac{1}{2^{2k+1}(2k+1)!z}\psi_{2k+1}(z)|_{z=0}=\frac{(-1)^k(2k+2)!}{k!(k+1)!2^{2k+1}}=
\frac{(-1)^k2(1/2)_{k+1}}{k!}.\eqno(3.3.36)$$ This shows that both
$\psi_n(z)/(2^nn!)$ and $P_n(z)$ are polynomial solutions of the
equation
$$(1-z^2){y'}'-2zy'+n(n+1)y=0\eqno(3.3.37)$$
with the same term of lowest degree. Observe that any power series
solution $y=\sum_{r=0}^\infty c_rz^r$ of (3.3.37) must be a linear
combination of (3.3.23) and (3.3.24), one of which is not
polynomial. Thus any two polynomial solutions of (3.3.37) must be
proportional. Hence $P_n(z)=\psi_n(z)/(2^nn!)$, that is, (3.3.27)
holds.$\qquad\Box$\psp

Let $m,n\in\mbb{N}$ such that $m\neq n$. Then
$$[(1-z^2)P_m'(z)]'P_n(z)+m(m+1)P_m(z)P_n(z)=0, \eqno(3.3.38)$$
$$ P_m(z)[(1-z^2)P_n'(z)]'+n(n+1)P_m(z)P_n(z)=0.\eqno(3.3.39)$$ Thus
\begin{eqnarray*}& &\int_{-1}^1P_m(z)P_n(z)dz\\
&=&\frac{1}{(m-n)(m+n+1)}\int_{-1}^1[m(m+1)-n(n+1)]P_m(z)P_n(z)dz
\\
&=&\frac{1}{(m-n)(m+n+1)}\left[\int_{-1}^1P_m(z)[(1-z^2)P_n'(z)]'dz-\int_{-1}^1[(1-z^2)P_m'(z)]'P_n(z)dz\right]
\\&=&\frac{1}{(m-n)(m+n+1)}(P_m(z)P_n'(z)-P_m'(z)P_n(z))(1-z^2)|^1_{-1}=0.\hspace{2.1cm}(3.3.40)\end{eqnarray*}
According to (3.3.34),
$$\left(\frac{d}{dz}\right)^n(P_n(z))=\frac{
(2n)!}{n!2^n}= (2n-1)!!.\eqno(3.3.41)$$ Hence
\begin{eqnarray*}\int_{-1}^1(P_n(z))^2dz&=&\frac{1}{n!2^n}\int_{-1}^1\left(\frac{d}{dz}\right)^n[(z^2-1)^n]P_n(z)dz
\\ &=&\frac{1}{n!2^n}\int_{-1}^1(-1)^n(z^2-1)^n\left(\frac{d}{dz}\right)^n(P_n(z))dz
\\
&=&\frac{(2n-1)!!}{n!2^n}\int_{-1}^1(1-z^2)^ndz=\frac{2(2n-1)!!}{n!2^n}\int_0^1(1-z^2)^ndz\\
\\&\stl{z=\sqrt{x}}{=}&\frac{(2n-1)!!}{n!2^n}\int_0^1x^{-1/2}(1-x)^ndx
=\frac{(2n-1)!!\G(1/2)\G(n+1)}{n!2^n\G(n+3/2)}\\
&=&\frac{2(2n-1)!!}{(2n+1)!!}=\frac{2}{2n+1}.\hspace{6.7cm}(3.3.42)\end{eqnarray*}
\pse

Legendre polynomials $\{P_k(z)\mid k\in\mbb{N}\}$ have been used to
solve the quantum two-body system.\psp

{\bf Exercise 3.3}\psp

 Find the differential equations satisfied by Jacobi polynomials
and prove that Chebyshev polynomials of each kind form a set of
orthogonal polynomials.

\section{Weierstrass's Elliptic Functions}

For two integers $m< n$, we denote
$$\ol{m,n}=\{m,m+1,....,n\},\;\;\ol{m,m}=\{m\},\;\ol{n,m}=\emptyset.\eqno(3.4.1)$$
Let $\w_1$ and $\w_2$ be two linearly independent elements in the
complex $z$-plane. Denote the lattice
$$L=\{m\w_1+n\w_2\mid m,n\in\mbb{Z}\},\qquad
L'=L\setminus\{0\}.\eqno(3.4.2)$$\pse

{\bf Lemma 3.4.1}. {\it For any $2<a\in\mbb{R}$, the series
$$\sum_{\w\in L'}\frac{1}{\w^a}\eqno(3.4.3)$$
converges absolutely}.

{\it Proof}. For $k\in\mbb{N}+1$, we denote
$$P_k=\{\pm k\w_1+r\w_2, r\w_1\pm k\w_2\mid
r\in\ol{-k,k}\},\eqno(3.4.4)$$ the set of the elements in $L$ lying
on the parallelogram with vertices $\{\pm k\w_1\pm k\w_2\}$. Denote
$$\dlt=\min\{|\w_1|,|\w_2|\}.\eqno(3.4.5)$$
Then
$$k\dlt\leq |\w|\;\;\mbox{for any}\;\w\in P_k.\eqno(3.4.6)$$
Moreover, the number of elements
$$|P_k|=8k.\eqno(3.4.7)$$

Now
$$\sum_{\w\in L'}\frac{1}{|\w|^a}=\sum_{k=1}^\infty\sum_{\w\in
P_k}\frac{1}{|\w|^a}<\sum_{k=1}^\infty\frac{8k}{(k\dlt)^a}=8\dlt^{-a}\sum_{k=1}^\infty\frac{1}{k^{a-1}},\eqno(3.4.8)$$
where the last series converges by calculus. $\qquad\Box$\psp

The {\it Weierstrass's Elliptic Function}\index{Weierstrass's
Elliptic Function}
$$\wp(z)=\frac{1}{z^2}+\sum_{\w\in L'}\left[\frac{1}{(z-\w)^2}-\frac{1}{\w^2}\right].\eqno(3.4.9)$$
For any $z\in\mbb{C}\setminus L$,
$$\lim_{|\w|\rta\infty}\frac{\left[\frac{1}{(z-\w)^2}-\frac{1}{\w^2}\right]}{\frac{1}{\w^3}}
=\lim_{|\w|\rta\infty}\frac{z\w
(2\w-z)}{(z-\w)^2}=2z.\eqno(3.4.10)$$ Since $\sum_{\w\in
L'}\frac{1}{\w^3}$ converges absolutely by Lemma 3.4.1, the series
in (3.4.9) converges absolutely. As $L'=-L'$, we have
\begin{eqnarray*}\qquad \wp(-z)&=&\frac{1}{(-z)^2}+\sum_{\w\in L'}\left[\frac{1}{(-z-\w)^2}-\frac{1}{\w^2}\right]
\\&=&\frac{1}{z^2}+\sum_{\w\in L'}\left[\frac{1}{(z+\w)^2}-\frac{1}{\w^2}\right]
\\&=&\frac{1}{z^2}+\sum_{-\w\in -L'}\left[\frac{1}{(z-(-\w))^2}-\frac{1}{(-\w)^2}\right]
\\&=&\frac{1}{z^2}+\sum_{\td\w\in
L'}\left[\frac{1}{(z-\td\w)^2}-\frac{1}{\td\w^2}\right]=\wp(z),\hspace{5.35cm}(3.4.11)\end{eqnarray*}
that is, $\wp(z)$ is an even function.

We calculate
$$\wp'(z)=-\frac{2}{z^3}-2\sum_{\w\in L'}\frac{1}{(z-\w)^3}=-2\sum_{\w\in
L}\frac{1}{(z-\w)^3},\eqno(3.4.12)$$ which converges absolutely for
any $z\in\mbb{C}\setminus L$. Since $L=-L$, $\wp'(z)$ is an odd
function by the similar argument as (3.4.11). For any $\w\in L$, we
have $L-\w=L$ and
\begin{eqnarray*}\qquad \wp'(z+\w)&=&-2\sum_{\w'\in
L}\frac{1}{(z+\w-\w')^3}=-2\sum_{\w'-\w\in
L-\w}\frac{1}{(z-(\w'-\w))^3}\\&=&-2\sum_{\td\w\in
L}\frac{1}{(z-\td\w)^3}=\wp'(z).\hspace{6.6cm}(3.4.13)\end{eqnarray*}
So the elements of $L$ are periods of $\wp'(z)$. Thus
$$\wp(z+\w)=\wp(z)+C\eqno(3.4.14)$$
for some constant $C$. Letting $z=-\w/2$ in (3.4.14), we have
$$\wp(\w/2)=\wp(-\w/2)+C\lra C=0\eqno(3.4.15)$$
by (3.4.11). Thus
$$\wp(z+\w)=\wp(z)\qquad\for\;\;\w\in L,\eqno(3.4.16)$$
that is, $\wp(z)$ is a {\it doubly periodic function}.\index{doubly
periodic function}

Note that the function
$$\wp_\ast(z)=\wp(z)-\frac{1}{z^2}=\sum_{\w\in
L'}\left[\frac{1}{(z-\w)^2}-\frac{1}{\w^2}\right]. \eqno(3.4.17)$$
is analytic at $z=0$. Moreover,
$$\wp_\ast^{(n)}(z)=(-1)^n(n+1)!\sum_{\w\in
L'}\frac{1}{(z-\w)^{n+2}}.\eqno(3.4.18)$$ In particular,
$$\wp_\ast^{(n)}(0)=(n+1)!\sum_{\w\in
L'}\frac{1}{\w^{n+2}}.\eqno(3.4.19)$$ For $m\in\mbb{N}$,
\begin{eqnarray*}\qquad\wp_\ast^{(2m+1)}(0)&=&(2m+2)!\sum_{\w\in
L'}\frac{1}{\w^{2m+3}}=-(2m+2)!\sum_{-\w\in
-L'}\frac{1}{(-\w)^{2m+3}}\\ &=&-(2m+2)!\sum_{\td\w\in
L'}\frac{1}{\td\w^{2m+3}}=-\wp_\ast^{(2m+1)}(0).\hspace{4cm}(3.4.20)\end{eqnarray*}
 Thus
$\wp_\ast^{(2m+1)}(0)=0.$ Thanks to (3.4.17), $\wp_\ast(0)=0$. Hence
$$\wp_\ast(z)=\sum_{m=1}^\infty c_{m+1}z^{2m}\eqno(3.4.21)$$
with
$$c_{m+1}=\frac{\wp_\ast^{(2m)}(0)}{(2m)!}=(2m+1)\sum_{\w\in
L'}\frac{1}{\w^{2m+2}}\eqno(3.4.22)$$ by (3.4.19).

Now
$$\wp(z)=\frac{1}{z^2}+\sum_{m=1}^\infty c_{m+1}z^{2m}\lra
\wp'(z)=-\frac{2}{z^3}+\sum_{m=1}^\infty
2mc_{m+1}z^{2m-1}.\eqno(3.4.23)$$ Moreover,
$$\wp^3(z)=\frac{1}{z^6}+\frac{3 c_2}{z^2}+3c_3+O(z),\eqno(3.4.24)$$
$${\wp'}^2(z)=\frac{4}{z^6}-\frac{8c_2}{z^2}-16c_3+O(z).\eqno(3.4.25)$$
Thus
$${\wp'}^2(z))-4\wp^3(z)=-\frac{20c_2}{z^2}-28c_3+O(z).\eqno(3.4.26)$$
Hence
$$\psi={\wp'}^2(z)-4\wp^3(z)+20c_2\wp(z)+28c_3\eqno(3.4.27)$$
is a function with periods in $L$ and only possible singular points
in $L$. Since $\psi(0)=0$, we have $\psi(\w)=\psi(0)=0$ for any
$\w\in L$. Hence $\psi$ is a holomorphic doubly periodic function.
So $\psi$ is bounded. Thus $\psi(z)\equiv\psi(0)=0$. This proves:
\psp

{\bf Theorem 3.4.2}. {\it For $z\in\mbb{C}\setminus L$,
$${\wp'}^2(z)=4\wp^3(z)-g_2\wp(z)-g_3\eqno(3.4.28)$$
with}
$$g_2=20c_2=60\sum_{\w\in
L'}\frac{1}{\w^4},\qquad g_3=28c_3=140\sum_{\w\in
L'}\frac{1}{\w^6}.\eqno(3.4.29)$$ \psp

Differentiating (3.4.28), we get
$$2\wp'(z){\wp'}'(z)=12\wp^2(z)\wp'(z)-g_2\wp'(z).\eqno(3.4.30)$$
Hence
$${\wp'}'(z)=6\wp^2(z)-\frac{g_2}{2},\eqno(3.4.31)$$
which is very important in solving nonlinear partial differential
equation.\psp

{\bf Remark 3.4.3}. Suppose $\mbox{Re}\:\w_1\neq 0$ and
$\mbox{Im}\:\w_1\neq 0$. Then $\w_1$ and its complex conjugate
$\ol{\w_1}$ are linearly independent. So we can take
$\w_2=\ol{\w_1}$. In this case, $\ol{L}=L$. If $z\in\mbb{R}$, then
$$\ol{\wp(z)}=\frac{1}{z^2}+\sum_{\w\in L'}\left[\frac{1}{(z-\ol{\w})^2}-\frac{1}{\ol{\w}^2}\right]
=\frac{1}{z^2}+\sum_{\td\w\in
\ol{L'}=L'}\left[\frac{1}{(z-\td\w)^2}-\frac{1}{\td\w^2}\right]=\wp(z).\eqno(3.4.32)$$
So $\wp(z)$ is a real-valued function on $\mbb{R}$. Similarly, $g_2$
and $g_3$ are real constants. Since $\w_1$ has two real freedom,
$g_2$ and $g_3$ can take any two real numbers such that
$g_2^3-27g_3^2\neq 0$ (the condition comes from ellipticity (cf.
[ARR, WG])).\psp

Observe
$$\frac{1}{z-\w}+\frac{1}{\w}+\frac{z}{\w^2}=\frac{1}{z-\w}+\frac{z+\w}{\w^2}=\frac{z^2}{\w^2(z-\w)}.\eqno(3.4.33)$$
Thus the series
$$\sum_{\w\in
L'}\left[\frac{1}{z-\w}+\frac{1}{\w}+\frac{z}{\w^2}\right]\eqno(3.4.34)$$
converges absolutely for any $z\in\mbb{C}\setminus L$. The {\it
Weierstrass's zeta function}:\index{Weierstrass's zeta function}
$$\zeta(z)=\frac{1}{z}+\sum_{\w\in
L'}\left[\frac{1}{z-\w}+\frac{1}{\w}+\frac{z}{\w^2}\right].
\eqno(3.4.35)$$ It is not the Riemann's zeta function! Obviously,
$$\zeta'(z)=-\wp(z).\eqno(3.4.36)$$
As the argument (3.4.11), $\zeta(z)$ is an odd function. Moreover,
$$\zeta'(z+\w)=-\wp(z+\w)=-\wp(z+\w)=\zeta'(\w)\qquad\for\;\;\w\in
L.\eqno(3.4.37)$$ In particular, this implies that
$$\zeta(z+\w_1)=\zeta(z)+2\eta_1,\qquad
\zeta(z+\w_2)=\zeta(z)+2\eta_2\eqno(3.4.38)$$ for some constants
$\eta_1,\eta_2\in\mbb{C}$. Taking $z=-\w_r/2$, we get
$$\zeta(\w_r/2)=\zeta(-\w_r/2)+2\eta_r.\eqno(3.4.39)$$
Hence
$$\eta_1=\zeta(\w_1/2),\qquad \eta_2=\zeta(\w_2/2).\eqno(3.4.40)$$

Now we assume
$$\mbox{Im}\;\frac{\w_2}{\w_1}>0.\eqno(3.4.41)$$
Let
$$A=-\frac{\w_1}{2}+\frac{\w_2}{2},\;\;B=\frac{\w_1}{2}+\frac{\w_2}{2},\;\;C=\frac{\w_1}{2}-\frac{\w_2}{2}
,\;\;D=-\frac{\w_1}{2}-\frac{\w_2}{2}.\eqno(3.4.42)$$ Denote by $XY$
the oriented segment from $X$ to $Y$ on the complex plane. Let
${\cal C}$ be the parallelogram $ABCD$ with counterclockwise
orientation. Since $z=0$ is the only pole of $\zeta(z)$ enclosed by
the parallelogram. We have
\begin{eqnarray*}\qquad2\pi i&=&\int_{\cal
C}\zeta(z)dz=\int_{DC}(\zeta(z)-\zeta(z+\w_2))dz\\
&
&+\int_{CB}(\zeta(z)-\zeta(z-\w_1)dz=-2\eta_2\w_1+2\eta_1\w_2.\hspace{3.9cm}(3.4.43)\end{eqnarray*}
 Thus $$\eta_1\w_2-\eta_2\w_1=\pi i.\eqno(3.4.44)$$

Note
\begin{eqnarray*}\qquad\qquad&
&\left(1-\frac{z}{\w}\right)e^{\frac{z}{\w}+\frac{z^2}{2\w^2}}\\
&=&\left(1-\frac{z}{\w}\right)\left(1+\frac{z}{\w}+\frac{z^2}{\w^2}+\frac{2z^3}{3\w^3}+O\left(\frac{z^4}{\w^4}\right)\right)
\\ &=&1-\frac{z^3}{3\w^3}+O\left(\frac{z^4}{\w^4}\right).\hspace{8.2cm}(3.4.45)\end{eqnarray*}
Since
$$\sum_{\w\in
L'}\left(\frac{Cz^4}{\w^4}-\frac{z^3}{3\w^3}\right)\eqno(3.4.46)$$
converges absolutely for any given $z$ and $C$, the product
$$\prod_{\w\in
L'}\left(1-\frac{z}{\w}\right)e^{\frac{z}{\w}+\frac{z^2}{2\w^2}}\;\;\mbox{converges
absolutely for any}\;z\in\mbb{C}\setminus L.\eqno(3.4.47)$$ We
define the {\it Weierstrass's sigma function}:\index{Weierstrass's
sigma function}
$$\sgm(z)=z\prod_{\w\in
L'}\left(1-\frac{z}{\w}\right)e^{\frac{z}{\w}+\frac{z^2}{2\w^2}}.\eqno(3.4.48)$$
Then
$$\ln\sgm(z)=\ln z+\sum_{\w\in
L'}\left[\ln\left(1-\frac{z}{\w}\right)+\frac{z}{\w}+\frac{z^2}{2\w^2}\right].\eqno(3.4.49)$$
Thus
$$\frac{\sgm'(z)}{\sgm(z)}=\frac{1}{z}+\sum_{\w\in
L'}\left[\frac{1}{z-\w}+\frac{1}{\w}+\frac{z}{\w^2}\right]=\zeta(z).\eqno(3.4.50)$$

By a similar argument as that of (3.4.11), $\sgm(z)$ is an odd
function. Moreover, (3.4.37) and (3.4.49) yield
$$\frac{\sgm'(z+\w_1)}{\sgm(z+\w_1)}-\frac{\sgm'(z)}{\sgm(z)}=2\eta_1,\;\;
\frac{\sgm'(z+\w_2)}{\sgm(z+\w_2)}-\frac{\sgm'(z)}{\sgm(z)}=2\eta_2.\eqno(3.4.51)$$
Thus
$$\frac{d}{dz}\ln\frac{\sgm(z+\w_r)}{\sgm(z)}=2\eta_r\lra\ln\frac{\sgm(z+\w_r)}{\sgm(z)}=2\eta_rz+C_r.\eqno(3.4.52)$$
So
$$\sgm(z+\w_r)=\sgm(z)e^{2\eta_rz+C_r}.\eqno(3.4.53)$$
Taking $z=-\w_r/2$ in (3.4.51), we get
$$\sgm(\w_r/2)=\sgm(-\w_r/2)e^{-\eta_r\w_r+C_r}\lra
e^{C_r}=-e^{\eta_r\w_r}.\eqno(3.4.54)$$ Therefore,
$$\sgm(z+\w_1)=-\sgm(z)e^{(2z+\w_1)\eta_1},\qquad
\sgm(z+\w_2)=-\sgm(z)e^{(2z+\w_2)\eta_2}.\eqno(3.4.55)$$\psp

Suppose $\mbox{Re}\:\w_1\neq 0$ and $\mbox{Im}\:\w_1<0$. Taking
$\w_2=\ol{\w_1}$, we get two real-valued functions $\zeta(z)$ and
$\sgm(z)$ for $z\in\mbb{R}$.

\section{Jacobian Elliptic Functions}

Let $0< m<1$ be a real constant. Jacobian elliptic function $\sn
(z|m)$ is the inverse function of the Legendre's {\it elliptic
integral of first kind}
$$z=\int_0^x\frac{dt}{\sqrt{(1-t^2)(1-m^2t^2)}},\eqno(3.5.1)$$
that is, $x=\sn (z|m)$. The number $m$ is the elliptic modulus of
$\sn (z|m)$. Moreover, we define
$$\cn (z|m)=\sqrt{1-\mbox{sn}^2(z|m)},\qquad \dn
(z|m)=\sqrt{1-m^2\mbox{sn}^2(z|m)}.\eqno(3.5.2)$$

Note
$$z=\lim_{m\rta
0}\int_0^x\frac{dt}{\sqrt{(1-t^2)(1-m^2t^2)}}=\int_0^x\frac{dt}{\sqrt{(1-t^2)}}=\arcsin
x.\eqno(3.5.3)$$ Thus
$$\lim_{m\rta 0}\sn (z|m)=\sin z,\qquad \lim_{m\rta 0}\cn (z|m)=\cos
z,\qquad \lim_{m\rta 0}\dn (z|m)=1.\eqno(3.5.4)$$ On the other hand,
$$z=\lim_{m\rta
1}\int_0^x\frac{dt}{\sqrt{(1-t^2)(1-m^2t^2)}}=\int_0^x\frac{dt}{1-t^2}=\frac{1}{2}\ln
\frac{1+x}{1-x},\eqno(3.5.5)$$ equivalently,
$$\frac{1+x}{1-x}=e^{2z}\sim \frac{2}{1-x}-1=e^{2z}\sim
1-x=\frac{2}{e^{2z}+1}\eqno(3.5.6)$$
$$\lra
x=1-\frac{2}{e^{2z}+1}=\frac{e^{2z}-1}{e^{2z}+1}=\frac{e^z-e^{-z}}{e^z+e^{-z}}=\tanh
z.\eqno(3.5.7)$$ Hence
$$\lim_{m\rta 1}\sn (z|m)=\tanh z,\qquad \lim_{m\rta 1}\cn (z|m)= \lim_{m\rta 1}\dn (z|m)=\mbox{sech}\: z.\eqno(3.5.8)$$

Taking derivative with respect to $z$ in (3.5.1), we get
$$1=\frac{1}{\sqrt{(1-x^2)(1-m^2x^2)}}\frac{dx}{dz}\sim
\frac{dx}{dz}=\sqrt{(1-x^2)(1-m^2x^2)}.\eqno(3.5.9)$$ So
$$\frac{d}{dz}\sn(z|m)=\sqrt{(1-\mbox{sn}^2(z|m))(1-m^2\mbox{sn}^2(z|m))}=\cn
(z|m)\;\dn(z|m).\eqno(3.5.10)$$ Moreover,
$$\frac{d}{dz}\cn(z|m)=-\frac{\sn
(z|m)}{\sqrt{1-\mbox{sn}^2(z|m)}}\frac{d}{dz}\sn(z|m)=-\sn(z|m)\;\dn(z|m),\eqno(3.5.11)$$
$$\frac{d}{dz}\dn(z|m)=-\frac{m^2\sn
(z|m)}{\sqrt{1-m^2\mbox{sn}^2(z|m)}}\frac{d}{dz}\sn(z|m)=-m^2\sn(z|m)\;\cn(z|m).\eqno(3.5.12)$$

Rewrite (3.5.2) as
$$\mbox{sn}^2(z|m)+\mbox{cn}^2(z|m)=1,\qquad
\mbox{dn}^2(z|m)+m^2\mbox{sn}^2(z|m)=1.\eqno(3.5.13)$$ Now
\begin{eqnarray*} \left(\frac{d}{dz}\right)^2\sn(z|m)
&=&\left(\frac{d}{dz}\cn(z|m)\right)\dn(z|m)+\cn(z|m)\left(\frac{d}{dz}\dn(z|m)\right)
\\&=&-\sn(z|m)\;\mbox{dn}^2(z|m)-m^2\sn(z|m)\;\mbox{cn}^2(z|m)\\ &=&
-\sn(z|m)\;(1-m^2\mbox{sn}^2(z|m))-m^2\sn(z|m)\;(1-\mbox{sn}^2(z|m))\\
&=&
2m^2\mbox{sn}^3(z|m)-(m^2+1)\sn(z|m),\hspace{4.6cm}(3.5.14)\end{eqnarray*}
\begin{eqnarray*} \left(\frac{d}{dz}\right)^2\cn(z|m)
&=&-\left(\frac{d}{dz}\sn(z|m)\right)\dn(z|m)-\sn(z|m)\left(\frac{d}{dz}\dn(z|m)\right)
\\&=&-\cn(z|m)\;\mbox{dn}^2(z|m)+m^2\cn(z|m)\;\mbox{sn}^2(z|m)\\ &=&
-\cn(z|m)\;(1-m^2+m^2\mbox{cn}^2(z|m))+m^2\cn(z|m)\;(1-\mbox{cn}^2(z|m))\\
&=&-2m^2\mbox{cn}^3(z|m)+(2m^2-1)\cn(z|m),\hspace{3.9cm}(3.5.15)\end{eqnarray*}
\begin{eqnarray*} \left(\frac{d}{dz}\right)^2\dn(z|m)
&=&-m^2\left(\frac{d}{dz}\sn(z|m)\right)\cn(z|m)-m^2\sn(z|m)\left(\frac{d}{dz}\cn(z|m)\right)
\\&=&-m^2\dn(z|m)\;\mbox{cn}^2(z|m)+m^2\dn(z|m)\;\mbox{sn}^2(z|m)\\ &=&
\dn(z|m)\;(1-m^2-\mbox{dn}^2(z|m))+\dn(z|m)\;(1-\mbox{dn}^2(z|m))\\
&=&-2\mbox{dn}^3(z|m)+(2-m^2)\dn(z|m).\hspace{4.7cm}(3.5.16)\end{eqnarray*}
The above three equations are very useful in solving nonlinear
partial differential equations such as nonlinear Schr\"{o}dinger
equations.

It is quite often to use (3.5.14)-(3.5.16) with similar equations
for trigonometric functions as follows:
$$\tan'z=\tan^2z+1,\qquad {\tan'}'z=2\tan^3z+2\tan z,\eqno(3.5.17)$$
$$\sec'z=\sec z\;\tan z,\qquad {\sec'}'z=2\sec^3 z-\sec
z,\eqno(3.5.18)$$
$$\coth'z=1-\coth^2z,\qquad {\coth'}' z=2\coth^3 z-2\coth
z,\eqno(3.5.19)$$
$$\csch'z=-\csch z\;\coth z,\qquad{\csch'}'(z)=2\csch^3z+\csch
z.\eqno(3.5.20)$$

\part{Partial Differential Equations}

\chapter{First-Order or Linear Equations}

First in this chapter, we derive the commonly used method of
characteristic lines for solving first-order quasilinear partial
differential equations, including boundary-value problems. Then we
talk about more sophisticated method of characteristic strip for
solving nonlinear first-order of partial differential equations.
Exact first-order partial differential equations are also handled.

Linear partial differential equations of flag type, including linear
equations with constant coefficients, appear in many areas of
mathematics and physics. A general equation of this type can not be
solved by separation of variables.  We use the grading technique
from representation theory to solve flag partial differential
equations and find the complete set of polynomial solutions. Our
method also leads us to find a family of new special functions by
which we are able to solve the initial-value problem of a large
class of linear equations with constant coefficients.

We use the method of characteristic lines to prove a
Campbell-Hausdorff-type factorization of exponential differential
operators and then solve the initial-value problem of flag evolution
partial differential equations. We also use the
Campbell-Hausdorff-type factorization to solve the initial-value
problem of generalized wave equations of flag type.

 The Calogero-Sutherland model is an exactly solvable quantum many-body
system in one-dimension
 (cf. [Cf], [Sb]). The model was used to study long-range
interactions of $n$ particles. We prove that a two-parameter
generalization of the Weyl function of type $A$ in representation
theory is a solution of the Calogero-Sutherland model. If $n=2$, we
find a connection between the Calogero-Sutherland model and the
Gauss hypergeometric function. When $n>2$, a new class of
multi-variable hypergeometric functions are found based on Etingof's
work [Ep]. Finally in Chapter 4, we use matrix differential
operators and Fourier expansions to solve the Maxwell equations, the
free Dirac equations and the generalized acoustic system.

\section{Method of Characteristics}

Let $n$ be a positive integer and let $x_1,x_2,...,x_n$ be $n$
independent variables. Denote
$$\vec x=(x_1,x_2,...,x_n).\eqno(4.1.1)$$
Suppose that $u(\vec x)=u(x_1,x_2,...,x_n)$ is a function in
$x_1,x_2,...,x_n$ determined by the quasilinear partial differential
equation
$$f_1(\vec x,u)u_{x_1}+f_2(\vec x,u)u_{x_2}+\cdots+f_n(\vec x,u)u_{x_n}=g(\vec
x,u)\eqno(4.1.2)$$ subject to the condition
$$\psi(\vec x,u)=0\qquad\mbox{on the surface}\;\;h(\vec
x)=0.\eqno(4.1.3)$$

Geometrically, the above problem is equivalent to find a hypersuface
$u=u(x_1,x_2,...,x_n)$ in the $(n+1)$-dimensional space of
$\{x_1,...,x_n,u\}$ passing through the codimension-2 boundary
(4.1.3) satisfying the equation (4.1.2). The idea of the {\it method
of characteristics} is to find all the lines on the hypersurface
passing through any point on the boundary (called {\it
characteristic lines}). Suppose that we have a line
$$x_1=x_1(s),\;\;x_2=x_2(s),\;\;...,\;\;x_n=x_n(s),\;\;u=u(s)\eqno(4.1.4)$$
passing through a point $(x_1,...,x_n,u)=(t_1,...,t_n,t_{n+1})$ on
the boundary (4.1.3). Since $u$ is a function of $x_1,...,x_n$
determining the hypersurface, we have
$$\frac{du}{ds}=u_{x_1}\frac{dx_1}{ds}+u_{x_2}\frac{dx_2}{ds}+\cdots+
u_{x_n}\frac{dx_n}{ds},\eqno(4.1.5)$$ equivalently,
$$(u_{x_1},...,u_{x_n},-1)\cdot
\left(\frac{dx_1}{ds},...,\frac{dx_n}{ds},\frac{du}{ds}\right)=0.\eqno(4.1.6)$$
On the other hand, (4.1.2) can be rewritten as
$$(u_{x_1},...,u_{x_n},-1)\cdot(f_1,...,f_n,g)=0.\eqno(4.1.7)$$
Comparing the above two equation, we find that original problem is
equivalent to solve the system of ordinary differential equations:
$$\frac{du}{ds}=g(\vec x,u),\;\;\frac{dx_r}{ds}=f_r(\vec x,u),\qquad
r\in\ol{1,n},\eqno(4.1.8)$$ subject to the initial conditions:
$$u|_{s=0}=t_{n+1},\;\;x_r|_{s=0}=t_r, \qquad
r\in\ol{1,n},\eqno(4.1.9)$$
$$\psi(t_1,...,t_n,t_{n+1})=0,\qquad h(t_1,...,t_n)=0.\eqno(4.1.10)$$
Solving (4.1.8) and (4.1.9), we find
$$u=\phi_{n+1}(s,t_1,...,t_{n+1}),\;\;x_r=\phi_r(s,t_1,...,t_{n+1}),\;\;r\in\ol{1,n}.
\eqno(4.1.11)$$ Eliminating possible variables in
$\{s,t_1,...,t_{n+1}\}$ by (4.1.10) and (4.1.11), we obtain the
solution of the original problem.\psp

{\bf Example 4.1.1}. Solve the equation $u_{x_1}-cu_{x_2}=0$ subject
to $u|_{x_1=0}=f(x_2)$, where $c$ is a constant and $f$ is a given
function.

{\it Solution}. The system of characteristic lines is:
$$\frac{du}{ds}=0,\;\;\frac{dx_1}{ds}=1,\;\;\frac{dx_2}{ds}=-c.\eqno(4.1.12)$$
Initial conditions are:
$$x_1|_{s=0}=t_1,\;\;x_2|_{s=0}=t_2,\;\;u|_{s=0}=t_3,\eqno(4.1.13)$$
$$t_3=f(t_2),\qquad t_1=0.\eqno(4.1.14)$$
The solution of (4.1.12) and (4.1.13) is
$$x_1=s,\;\;x_2=-cs+t_2,\;\;u=t_3.\eqno(4.1.15)$$
Thus $t_2=cx_1+x_2$ and the final solution is
$$u=f(cx_1+x_2).\qquad\Box\eqno(4.1.16)$$\pse

{\bf Example 4.1.2}. Solve the equation
$$u_x+x^2u_y=-yu\;\;\mbox{subject
to}\;\;u=f(y)\;\mbox{on}\;\;x=0.\eqno(4.1.17)$$

{\it Solution}. The system of characteristic lines is:
$$\frac{dx}{ds}=1,\;\;\frac{dy}{ds}=x^2,\;\;\frac{du}{ds}=-yu.\eqno(4.1.18)$$
Initial conditions are:
$$x|_{s=0}=t_1,\;\;y|_{s=0}=t_2,\;\;u|_{s=0}=t_3,\eqno(4.1.19)$$
$$t_3=f(t_2),\qquad t_1=0.\eqno(4.1.20)$$
The first equation in (4.1.18) gives $x=s$. Then the second equation
becomes
$$\frac{dy}{ds}=s^2\lra y=\frac{s^3}{3}+t_2.\eqno(4.1.21)$$
Now the third equation in (4.1.18) becomes
$$\frac{du}{ds}=-\left(\frac{s^3}{3}+t_2\right)u\sim \frac{du}{u}=
-\left(\frac{s^3}{3}+t_2\right)ds.\eqno(4.1.22)$$Thus
$$u=t_3e^{-s^4/12-t_2s}=f(t_2)e^{-s^4/12-t_2s}.\eqno(4.1.23)$$
Note $s=x$. So $t_2=y-x^3/3$. Thus the final solution is
$$u=f(y-x^3/3)e^{x^4/4-xy}.\qquad\Box\eqno(4.1.24)$$
\pse

{\bf Example 4.1.3}. Solve the the equation
$$u_x+u_y+xyu_z=u^2\;\;\mbox{subject
to}\;\;u=x^2\;\mbox{on}\;\;y=z.\eqno(4.1.25)$$

{\it Solutions}.  The system of characteristic lines is:
$$\frac{dx}{ds}=1,\;\;\frac{dy}{ds}=1,\;\;\frac{dz}{ds}=xy,\;\;\frac{du}{ds}=u^2.\eqno(4.1.26)$$
Initial conditions are:
$$x|_{s=0}=t_1,\;\;y|_{s=0}=t_2,\;\;z|_{s=0}=t_3,\;\;u|_{s=0}=t_1^2,\;\;t_2=t_3.\eqno(4.1.27)$$
The first two equations in (4.1.26) gives $x=s+t_1$ and $y=s+t_2$.
The third equation becomes
$$\frac{dz}{ds}=(s+t_1)(s+t_2)=s^2+(t_1+t_2)s+t_1t_2.\eqno(4.1.28)$$
Thus
$$z=\frac{s^3}{3}+\frac{t_1+t_2}{2}s^2+t_1t_2s+t_2.\eqno(4.1.29)$$
The last equation in (4.1.26) yields
$$\frac{1}{u}=-s+\frac{1}{t_1^2}\lra u=\frac{t_1^2}{1-st_1^2}.\eqno(4.1.30)$$
Note $t_1=x-s$ and $t_2=y-s$. Thus we obtained the parametric
solution
$$u=\frac{(x-s)^2}{1-s(x-s)^2},\;\;z=\frac{s^3}{3}-\frac{x+y}{2}s^2+xys+y-s.\qquad\Box\eqno(4.1.31)$$
\pse

{\bf Exercise 4.1} \psp

1. Solve the following problem
$$x^2u_x+2yu_y+4z^3u_z=0\;\;\mbox{subject
to}\;\;u=f(y,z)\;\;\mbox{on the plane}\;\;x=1.$$ \pse

2. Find the solution of the problem
$$u_x+2xu_y+3yu_z=4zu^3$$
subject to $$u^3=x^2+y+3\sin z\;\;\mbox{on the
surface}\;\;x=y^2+z^2.$$

\section{Characteristic Strip and Exact Equations}

Consider the partial differential equation
$$F(x,y,u,p,q)=0,\qquad p=u_x,\;q=u_y.\eqno(4.2.1)$$
We search for solution by solving the following system of {\it strip
equations}:
$$\frac{\ptl x}{\ptl s}=F_p,\;\;\frac{\ptl y}{\ptl s}=F_q,\;\;
\frac{\ptl u}{\ptl s}=pF_p+qF_q,\eqno(4.2.2)$$
$$\frac{\ptl p}{\ptl s}=-F_x-pF_u,\qquad \frac{\ptl q}{\ptl
s}=-F_y-qF_u,\eqno(4.2.3)$$ where we view $\{x,y,u,p,q\}$ as
functions of the two variables $\{s,t\}$, and $t$ is responsible for
the initial condition. The third equation in (4.2.2) is derived from
the first two via
$$\frac{\ptl u}{\ptl s}=u_x\frac{\ptl x}{\ptl s}+u_y\frac{\ptl y}{\ptl s}=pF_p+qF_q.\eqno(4.2.4)$$
Note $p_y=u_{xy}=u_{yx}=q_x$. Taking partial derivative of the first
equation in (4.2.1) with respect to $x$, we have
$$F_x+pF_u+p_xF_p+q_xF_q=0\sim
F_x+pF_u+p_xF_p+p_yF_q=0.\eqno(4.2.5)$$ Under the assumption the
first two equations in (4.2.2),
$$\frac{\ptl p}{\ptl s}=p_x\frac{\ptl x}{\ptl s}+p_y\frac{\ptl y}{\ptl s}=p_xF_p+q_xF_q=-F_x-pF_u,\eqno(4.2.6)$$
that is, the first equation in (4.2.3) holds. We can similarly
derive the second equation in (4.2.3). A solution of the system
(4.2.2) and (4.2.3) does give a characteristic line because
$$(u_x,u_y,-1)\cdot\left(\frac{\ptl x}{\ptl s},\frac{\ptl y}{\ptl
s},\frac{\ptl u}{\ptl
s}\right)=pF_p+qF_q-(pF_p+qF_q)=0.\eqno(4.2.7)$$

{\bf Example 4.2.1}. Solving the problem
$$u_xu_y-2u-x+2y=0\eqno(4.2.8)$$ subject
to $u=y^2$ on the line $x=0$.

{\it Solution}. Now $F=pq-2u-x+2y$. The strip equations are:
$$\frac{\ptl x}{\ptl s}=q,\qquad\frac{\ptl y}{\ptl s}=p,\qquad
\frac{\ptl u}{\ptl s}=2pq,\eqno(4.2.9)$$
$$\frac{\ptl p}{\ptl s}=1+2p\qquad \frac{\ptl q}{\ptl
s}=-2+2q.\eqno(4.2.10)$$ The initial conditions are given: when
$s=0$,
$$x=0,\qquad y=t,\qquad u=t^2.\eqno(4.2.11)$$
To find the condition for $p$ and $q$ when $s=0$, we calculate
$$\frac{du}{dt}=p\frac{dx}{dt}+q\frac{dy}{dt}\sim 2t=p\cdot 0+q\cdot
1\lra q=2t.\eqno(4.2.12)$$ On the other hand, when $s=0$, (4.2.8)
becomes
$$pq-2t^2+2t=0\lra p=t-1.\eqno(4.2.13)$$
According to (4.2.10), (4.2.12) and (4.2.13), we have
$$p=\frac{-1+(2t-1)e^{2s}}{2},\qquad q=1+(2t-1)e^{2s}.\eqno(4.2.14)$$

Next (4.2.9) becomes
$$ \frac{\ptl x}{\ptl s}=1+(2t-1)e^{2s},\;\;
\frac{\ptl y}{\ptl s}=\frac{-1+(2t-1)e^{2s}}{2},\eqno(4.2.15)$$
$$\frac{\ptl u}{\ptl
s}=(2t-1)^2e^{4s}-1.\eqno(4.2.16)$$ Thus
$$x=s+\frac{(2t-1)(e^{2s}-1)}{2},\qquad
y=-\frac{s}{2}+t+\frac{(2t-1)(e^{2s}-1)}{4},\eqno(4.2.17)$$
$$u=t^2-s+\frac{(2t-1)^2(e^{4s}-1)}{4}.\qquad\Box\eqno(4.2.18)$$
\pse

The equation
$$f(x,y,u)u_x=g(x,y,u)u_y\eqno(4.2.19)$$ is called {\it exact} if
$f_x=g_y$. For an exact equation, we look for a function
$\Psi(x,y,u)$ such that $\Psi_y=f$ and $\Psi_x=g$. Then
$\Psi(x,y,u)=0$ is a solution of (4.2.19). In fact, the equation
$\Psi(x,y,u)=0$ gives
$$\Psi_x+\Psi_uu_x=0,\qquad \Psi_y+\Psi_uu_y=0.\eqno(4.2.20)$$
Thus
$$u_x=-\frac{\Psi_x}{\Psi_u}=-\frac{g}{\Psi_u},\;\;
u_y=-\frac{\Psi_y}{\Psi_u}=-\frac{f}{\Psi_u},\eqno(4.2.21)$$ which
implies
$$fu_x=-f\frac{g}{\Psi_u}=-g\frac{f}{\Psi_u}=gu_y.\eqno(4.2.22)$$
\pse

{\bf Example 4.2.2}. Solve the equation
$$(x+\cos y+u)u_x=(y+e^x+u^2)u_y.\eqno(4.2.23)$$

{\it Solution}. Now $f=x+\cos y+u$ and $g=y+e^x+u^2$. Moreover,
$f_x=1=g_y$. The equation is exact. Let
$$\Psi=\int f(x,y,u)dy=\int(x+\cos y+u)dy=(x+u)y+\sin
y+\phi(x,u).\eqno(4.2.24)$$ Taking partial derivative of (4.2.24)
with respect to $x$, we get
$$y+\phi_x=\Psi_x=g=y+e^x+u^2\sim \phi_x=e^x+u^2.\eqno(4.2.25)$$
Hence
$$\phi=\int(e^x+u^2)dx=e^x+xu^2+h(u),\eqno(4.2.26)$$
where $h(u)$ is any differentiable function. The final answer is
$$(x+u)y+\sin y+e^x+xu^2+h(u)=0.\qquad\Box\eqno(4.2.27)$$

We refer to [Z] for more exact methods of solving differential
equations.\psp

{\bf Exercise 4.2}\psp

1. Find the solution of the following problem $u_xu_y-2u+2x=0$
subject to $u=x^2y$ on the line $x=y$.\psp

2. Solve the equation $(2xy+e^y)u_x=(y^2+x+\sin u)u_y.$

\section{Polynomial Solutions of Flag Equations}

A linear transformation $T$ on an infinite-dimensional vector space
$U$ is called {\it locally nilpotent} if for any $u\in U$, there
exists a positive integer $m$ (usually depends on $u$) such that
$T^m(u)=0$.

 A {\it partial differential equation of flag type} is the linear
differential equation of the form:\index{flag partial differential
equation}
$$(d_1+f_1d_2+f_2d_3+\cdots+f_{n-1}d_n)(u)=0,\eqno(4.3.1)$$
where $d_1,d_2,...,d_n$ are certain commuting locally nilpotent
differential operators on the polynomial algebra
$\mbb{R}[x_1,x_2,...,x_n]$ and $f_1,...,f_{n-1}$ are polynomials
satisfying
$$d_l(f_j)=0\qquad\mbox{if}\;\;l>j.\eqno(4.3.2)$$
Examples of such equations are: (1) Laplace equation
$$u_{x_1x_1}+u_{x_2x_2}+\cdots +u_{x_nx_n}=0;\eqno(4.3.3)$$
(2) heat conduction equation
$$u_t-u_{x_1x_1}-u_{x_2x_2}-\cdots -u_{x_nx_n}=0;\eqno(4.3.4)$$
(3) generalized  Laplace equation
$$u_{xx}+xu_{yy}+yu_{zz}=0.\eqno(4.3.5)$$
The aim of this section is to find all the polynomial solutions of
the equation (4.3.1). The contents are taken from the author's work
[X11].

Let $U$ be a vector space over $\mbb{R}$ and let $U_1$ be a subspace
of $U$. The quotient space
$$U/U_1=\{u+U_1\mid u\in U\}\eqno(4.3.6)$$ with linear operation
$$a(u_1+U_1)+b(u_2+U_1)=(au_1+bu_2)+U_1\qquad\for\;\;u_1,u_2\in
U,\;a,b\in\mbb{R},\eqno(4.3.7)$$ where the zero vector in $U/U_1$ is
$U_1$ and
$$u+v+U_1=u+U_1\qquad\for\;\;u\in U,\;v\in U_1.\eqno(4.3.8)$$
For instance, $U=\mbb{R}x+\mbb{R}y+\mbb{R}z$ and $U_1=\mbb{R}x$.
Then $U/U_1=\{by+cz+U_1\}\cong \mbb{R}y+\mbb{R}z$ and
$\{y+U_1,z+U_1\}$ forms a basis of $U/U_1$. Second example is
$U=\mbb{R}+\mbb{R}x+\mbb{R}x^2$ and $U_1=\mbb{R}(1+x+x^2)$. In this
case, $(1+U_1)+(x+U_1)+(x^2+U_1)=(1+x+x^2)+U_1=U_1$, the zero vector
in $U/U_1$. Thus $U/U_1=\{a+bx+U_1\mid
a,b\in\mbb{R}\}=\{ax+bx^2+U_1\mid a,b\in\mbb{R}\}$. Both
$\{1+U_1,x+U_1\}$ and $\{x+U_1,x^2+U_1\}$ are bases of $U/U_1$. But
we know $U/U_1\cong\mbb{R}^2$.

Recall that $\mbb{N}$ denotes the set of nonnegative integers. Let
$1\leq k<n$. Denote
$${\cal A}=\mbb{R}[x_1,x_2,...,x_n],\;\; {\cal
B}=\mbb{R}[x_1,x_2,...,x_k],\;\;V=\mbb{R}[x_{k+1},x_{k+2},...,x_n].\eqno(4.3.9)$$
Let $\{V_m\mid m\in\mbb{N}\}$ be a set of subspaces of $V$ such that
$$V_r\subset
V_{r+1}\;\;\for\;\;r\in\mbb{N}\;\;\mbox{and}\;\;V=\bigcup_{r=0}^\infty
V_r.\eqno(4.3.10)$$ For instance, we take $V_r=\{g\in
V\mid\mbox{deg}\:g\leq r\}$ in some special cases.\psp

{\bf Lemma 4.3.1}. {\it Let $T_1$ be a differential operator on
${\cal A}$ with a right inverse $T_1^-$ such that
$$T_1({\cal B}),\;T_1^-({\cal B})\subset{\cal B},\qquad
T_1(\eta_1\eta_2)=T_1(\eta_1)\eta_2,\qquad
T_1^-(\eta_1\eta_2)=T_1^-(\eta_1)\eta_2 \eqno(4.3.11)$$ for $\eta_1
\in {\cal B},\;\eta_2\in V$, and let $T_2$ be a differential
operator on ${\cal A}$ such that
$$ T_2(V_0)=\{0\},\;T_2(V_{r+1})\subset {\cal B}V_r,\;
T_2(f\zeta)=fT_2(\zeta) \;\;\for\;\; r\in\mbb{N},\;\;f\in{\cal
B},\;\zeta\in{\cal A}.\eqno(4.3.12)$$ Then we
have \begin{eqnarray*}\hspace{1cm}&&\{f\in{\cal A}\mid (T_1+T_2)(f)=0\}\\
& =&\mbox{Span}\{ \sum_{\iota=0}^\infty(-T_1^-T_2)^\iota(hg)\mid
g\in V,\;h\in {\cal B};\;T_1(h)=0\},
\hspace{3.3cm}(4.3.13)\end{eqnarray*} where the summation is finite.
Moreover, the operator $\sum_{\iota=0}^\infty(-T_1^-T_2)^\iota
T_1^-$ is a right inverse of $T_1+T_2$}.

{\it Proof}. For $h\in {\cal B}$ such that $T_1(h)=0$ and $g\in V$,
we have
\begin{eqnarray*}& &(T_1+T_2)(\sum_{\iota=0}^\infty(-T_1^-T_2)^\iota(hg))\\
&=&T_1(hg)-\sum_{\iota=1}^\infty
T_1[T_1^-T_2(-T_1^-T_2)^{\iota-1}(hg)]+
\sum_{\iota=0}^\infty T_2[(-T_1^-)^\iota(hg)]\\
&=&T_1(h)g-\sum_{\iota=1}^\infty
(T_1T_1^-)T_2(-T_1^-T_2)^{\iota-1}(hg)+
\sum_{\iota=0}^\infty T_2(-T_1^-T_2)^\iota(hg)\\
&=&-\sum_{\iota=1}^\infty T_2(-T_1^-T_2)^{\iota-1}(hg)+
\sum_{\iota=0}^\infty T_2(-T_1^-T_2)^\iota(hg)=0\hspace{4.3cm}
(4.3.14)\end{eqnarray*} by (4.3.11). Set $V_{-1}=\{0\}$. For
$j\in\mbb{N}$, we take $\{\psi_{j,r}\mid r\in I_j\}\subset V_j$ such
that
$$\{\psi_{j,r}+V_{j-1}\mid r\in I_j\}\;\;\mbox{forms a basis
of}\;\;V_j/V_{j-1},\eqno(4.3.15)$$ where $I_j$ is an index set.
 Let
$${\cal A}^{(m)}={\cal B}V_m=\sum_{s=0}^m\;\sum_{r\in I_s}{\cal
B}\psi_{s,r}.\eqno(4.3.16)$$ Obviously,
$$T_1({\cal
A}^{(m)}),\;T_1^-({\cal A}^{(m)}),\; T_2({\cal A}^{(m+1)})\subset
{\cal A}^{(m)}\qquad\for\;\;m\in\mbb{N}\eqno(4.3.17)$$ by (4.3.11)
and (4.3.12), and
$${\cal A}=\bigcup_{m=0}^\infty {\cal A}^{(m)}.\eqno(4.3.18)$$

  Suppose $\phi\in {\cal
A}^{(m)}$ such that $(T_1+T_2)(\phi)=0$. If $m=0$, then
$$\phi=\sum_{r\in I_0}h_r\psi_{0,r},\qquad h_r\in {\cal
B}.\eqno(4.3.19)$$ Now
$$0=(T_1+T_2)(\phi)=\sum_{r\in I_0}T_1(h_r)\psi_{0,r}+\sum_{r\in
I_0}h_r T_2(\psi_{0,r})=\sum_{r\in
I_0}T_1(h_r)\psi_{0,r},\eqno(4.3.20)$$ Since $T_1(h_r)\in {\cal B}$
by (4.3.11), (4.3.20) gives $T_1(h_r)=0$ for $r\in I_0$. Denote by
${\cal S}$ the right hand side of the equation (4.3.13). Then
$$\phi=\sum_{r\in I_0}\;\sum_{m=0}^\infty(-T_1^-T_2)^m(h_r\psi_{0,r})\in {\cal
S}.\eqno(4.3.21)$$

Suppose $m>0$. We write
$$\phi=\sum_{r\in I_m}h_r\psi_{m,r}+\phi',\qquad h_r\in{\cal
B},\;\phi'\in {\cal A}^{(m-1)}.\eqno(4.3.22)$$ Then
$$0=(T_1+T_2)(\phi)=\sum_{r\in
I_m}T_1(h_r)\psi_{m,r}+T_1(\phi')+T_2(\phi).\eqno(4.3.23)$$ Since
$T_1(\phi')+T_2(\phi)\in {\cal A}^{(m-1)}$, we have $T_1(h_r)=0$ for
$r\in I_m$. Now
$$\phi-\sum_{r\in
I_m}\sum_{j=0}^\infty(-T_1^-T_2)^j(h\psi_{m,r})=\phi'-\sum_{r\in
I_m}\sum_{j=1}^\infty(-T_1^-T_2)^j(h_r\psi_{m,r})\in {\cal
A}^{(m-1)}\eqno(4.3.24)$$ and (4.3.14) implies
$$(T_1+T_2)(\phi-\sum_{r\in
I_m}\sum_{j=0}^\infty(-T_1^-T_2)^j(h_r\psi_{m,r}))=0.\eqno(4.3.25)$$
By induction on $m$, $$\phi-\sum_{r\in
I_m}\sum_{j=0}^\infty(-T_1^-T_2)^j(h_r\psi_{m,r})\in {\cal
S}.\eqno(4.3.26)$$ Therefore, $\phi\in {\cal S}.$

For any $f\in {\cal A}$, we have:
\begin{eqnarray*} \hspace{1cm} & &(T_1+T_2)(\sum_{\iota=0}^\infty(-T_1^-T_2)^\iota T_1^-)(f)
\\ &=&f-\sum_{\iota=1}^\infty
T_2(-T_1^-T_2)^{\iota-1}T_1^-(f)+\sum_{\iota=0}^\infty
T_2(-T_1^-T_2)^\iota
T_1^-(f)=f.\hspace{2.1cm}(4.3.27)\end{eqnarray*} Thus
 the operator
$\sum_{\iota=0}^\infty(-T_1^-T_2)^\iota T_1^-$ is a right inverse of
$T_1+T_2.\qquad\Box$\psp

 We remark that the above operators $T_1$ and $T_2$ may not
 commute. The assumption $T_2(V_{r+1})\subset{\cal B}V_r$ instead of
 $T_2(V_{r+1})\subset V_r$ because we want our lemma working for a
 special case like $T_1=\ptl_{x_1}^2,\;T_2=x_1\ptl_{x_2}^2,\;{\cal
 B}=\mbb{R}[x_1]$ and $V=\mbb{R}[x_2]$.

Define
$$x^\al=x_1^{\al_1}x_2^{\al_2}\cdots
x_n^{\al_n}\;\;\for\;\;\al=(\al_1,...,\al_n)\in\mbb{N}^{\:n}.\eqno(4.3.28)$$
Moreover, we denote
$$\es_\iota=(0,...,0,\stl{\iota}{1},0,...,0)\in \mbb{N}^{\:n}.\eqno(4.3.29)$$
 For each
$\iota\in\ol{1,n}$, we define the linear operator $\int_{(x_\iota)}$
on ${\cal A}$ by:
$$\int_{(x_\iota)}(x^\al)=\frac{x^{\al+\es_\iota}}{\al_\iota+1}\;\;\for\;\;\al\in
\mbb{N}^{\:n}.\eqno(4.3.30)$$ Furthermore, we let
$$\int_{(x_\iota)}^{(0)}=1,\qquad\int_{(x_\iota)}^{(m)}=\stl{m}{\overbrace{\int_{(x_\iota)}\cdots\int_{(x_\iota)}}}
\qquad\for\; \;0<m\in\mbb{Z}\eqno(4.3.31)$$ and denote
$$\ptl^{\al}=\ptl_{x_1}^{\al_1}\ptl_{x_2}^{\al_2}\cdots
\ptl_{x_n}^{\al_n},\;\;
\int^{(\al)}=\int_{(x_1)}^{(\al_1)}\int_{(x_2)}^{(\al_2)}\cdots
\int_{(x_n)}^{(\al_n)}\qquad\for\;\;\al\in
\mbb{N}^{\:n}.\eqno(4.3.32)$$ Obviously, $\int^{(\al)}$ is a right
inverse of $\ptl^\al$ for $\al\in \mbb{N}^{\:n}.$ We remark that
$\int^{(\al)}\ptl^\al\neq 1$ if $\al\neq 0$ due to
$\ptl^\al(1)=0$.\psp

{\bf Example 4.3.1}. Find polynomial solutions of the heat
conduction equation $u_t=u_{xx}$.

{\it Solution}. In this case,
$${\cal A}=\mbb{R}[t,x],\;\;{\cal
B}=\mbb{R}[t],\;\;V=\mbb{R}[x],\;\;V_r=\{g\in V\mid\deg g\leq
r\}.\eqno(4.3.33)$$ The equation can be written as
$(\ptl_t-\ptl_x^2)(u)=0$. So we take
$$T_1=\ptl_t,\qquad T_1^-=\int_{(t)},\qquad
T_2=-\ptl_x^2.\eqno(4.3.34)$$ It can be verified that the conditions
in Lemma 4.3.1 are satisfied. Note that
$$\{f\in {\cal B}\mid T_1(f)=0\}=\{f\in \mbb{R}[t]\mid
\ptl_t(f)=0\}=\mbb{R}.\eqno(4.3.35)$$

We calculate
$$(-T_1T_2)^\iota(x^k)=(\int_{(t)}\ptl_x^2)^\iota(x^k)=\int_{(t)}^\iota(1)\ptl_x^{2\iota}(x^k)
=\frac{[\prod_{s=0}^{2\iota-1}(k-s)]t^{\iota}x^{k-2\iota}}{\iota!}.\eqno(4.3.36)$$
 Thus the space of the polynomial solutions is
$$\mbox{Span}\left\{\sum_{\iota=0}^{\llbracket k/2
\rrbracket}\frac{[\prod_{s=0}^{2\iota-1}(k-s)]t^{\iota}x^{k-2\iota}}{\iota!}\mid
k\in\mbb{N}\right\}.\qquad\Box\eqno(4.3.37)$$\pse

{\bf Example 4.3.2}. Find  polynomial solutions of the Laplace
equation $u_{xx}+u_{yy}=0$.

{\it Solution}. In this case,
$${\cal A}=\mbb{R}[x,y],\;\;{\cal
B}=\mbb{R}[x],\;\;V=\mbb{R}[y],\;\;V_r=\{g\in V\mid\mbox{deg}\:g\leq
r\}.\eqno(4.3.38)$$ Moreover, we take
$$T_1=\ptl_x^2,\qquad T_1^-=\int_{(x)}^2,\qquad
T_2=\ptl_y^2.\eqno(4.3.39)$$ It can be verified that the conditions
in Lemma 4.3.1 are satisfied. Note that
$$\{f\in {\cal B}\mid T_1(f)=0\}=\{f\in \mbb{R}[x]\mid
\ptl_x^2(f)=0\}=\mbb{R}+\mbb{R}x.\eqno(4.3.40)$$

We calculate
\begin{eqnarray*}\qquad(-T_1T_2)^\iota(y^k)&=&(-\int_{(x)}^2\ptl_y^2)^\iota
(y^k)=(-1)^\iota\int_{(x)}^{2\iota}(1)\ptl_y^{2\iota}(y^k)
\\
&=&\frac{[\prod_{s=0}^{2\iota-1}(k-s)](-x^2)^{\iota}y^{k-2\iota}}{(2\iota)!},
\hspace{5.6cm}(4.3.41)\end{eqnarray*}
\begin{eqnarray*}\qquad(-T_1T_2)^\iota(xy^k)&=&(-\int_{(x)}^2\ptl_y^2)^\iota
(xy^k)=(-1)^\iota\int_{(x)}^{2\iota}(x)\ptl_y^{2\iota}(y^k)
\\
&=&\frac{(-1)^\iota[\prod_{s=0}^{2\iota-1}(k-s)]x^{2\iota+1}y^{k-2\iota}}{(2\iota+1)!}.
\hspace{4.7cm}(4.3.42)\end{eqnarray*}
 Thus the space of the polynomial solutions is
\begin{eqnarray*}\qquad\qquad& &\mbox{Span}\big\{\sum_{\iota=0}^{\llbracket k/2
\rrbracket}\frac{[\prod_{s=0}^{2\iota-1}(k-s)](-x^2)^{\iota}y^{k-2\iota}}{(2\iota)!},\\
& & \sum_{\iota=0}^{\llbracket k/2
\rrbracket}\frac{(-1)^\iota[\prod_{s=0}^{2\iota-1}(k-s)]x^{2\iota+1}y^{k-2\iota}}{(2\iota+1)!}
\mid
k\in\mbb{N}\big\}.\qquad\Box\hspace{3.2cm}(4.3.43)\end{eqnarray*}\pse

Consider the wave equation in Riemannian space with a
nontrivial\index{wave equation in Riemannian space} conformal group:
$$u_{tt}-u_{x_1x_1}-\sum_{\iota,j=2}^ng_{\iota,j}(x_1-t)u_{x_\iota x_j}=0,\eqno(4.3.44)$$
where we assume that $g_{\iota,j}(z)$ are one-variable polynomials.
Change variables:
$$z_0=x_1+t,\qquad z_1=x_1-t.\eqno(4.3.45)$$
Then
$$\ptl_t^2=(\ptl_{z_0}-\ptl_{z_1})^2,\qquad
\ptl_{x_1}^2=(\ptl_{z_0}+\ptl_{z_1})^2.\eqno(4.3.46)$$ So the
equation (4.3.44) changes to:
$$2\ptl_{z_0}\ptl_{z_1}+
\sum_{\iota,j=2}^ng_{\iota,j}(z_1)u_{x_\iota x_j}=0.\eqno(4.3.47)$$
Denote
$$T_1=2\ptl_{z_0}\ptl_{z_1},\qquad
T_2=\sum_{\iota,j=2}^ng_{\iota,j}(z_1)\ptl_{x_\iota}\ptl_{x_j}.\eqno(4.3.48)$$
Take $T_1^-=\frac{1}{2}\int_{(z_0)}\int_{(z_1)}$, and
$${\cal B}=\mbb{R}[z_0,z_1],\qquad V=\mbb{R}[x_2,...,x_n],\qquad \qquad
V_r=\{f\in V\mid\mbox{deg}\;f\leq r\}.\eqno(4.3.49)$$ Then the
conditions in Lemma 4.3.1 hold. Thus we have:
 \psp

{\bf Theorem 4.3.2}. {\it The space of all polynomial solutions for
the equation (4.3.44) is:
\begin{eqnarray*} \hspace{2cm}& &\mbox{Span}\:\{\sum_{m=0}^\infty(-2)^{-m}
(\sum_{\iota,j=2}^n\int_{(z_0)}\int_{(z_1)}g_{\iota,j}(z_1)\ptl_{x_\iota}\ptl_{x_j})^m(f_0g_0+f_1g_1)\\
& & \mid
f_0\in\mbb{R}[z_0],\;f_1\in\mbb{R}[z_1],\;g_0,g_1\in\mbb{R}[x_2,...,x_n]\}
\hspace{3.8cm}(4.3.50)\end{eqnarray*} with $z_0,z_1$ defined in
(4.3.45).}\psp

Let $m_1,m_2,...,m_n$ be positive integers. According to Lemma
4.3.1, the set
\begin{eqnarray*}\hspace{1.9cm}& &\{\sum_{k_2,...,k_n=0}^\infty(-1)^{k_2+\cdots+k_n}{k_2+\cdots+k_k\choose
k_2,...,k_n} \int_{(x_1)}^{((k_2+\cdots +k_n)m_1)}(x_1^{\ell_1})\\
& &\times\ptl_{x_2}^{k_2m_2}(x_2^{\ell_2})\cdots
\ptl_{x_n}^{k_nm_n}(x_n^{\ell_n})\mid
\ell_1\in\ol{0,m_1-1},\;\ell_2,...,\ell_n\in\mbb{N}\}\hspace{1.7cm}
(4.3.51)\end{eqnarray*} forms a basis of the space of polynomial
solutions for the equation
$$(\ptl_{x_1}^{m_1}+\ptl_{x_2}^{m_2}+\cdots+\ptl_{x_n}^{m_n})(u)=0\eqno(4.3.52)$$
in  ${\cal A}$.

The above results can theoretically generalized as follows. Let
$$f_\iota\in\mbb{R}[x_1,...,x_\iota]\qquad \for\;\;\iota\in\ol{1,n-1}.\eqno(4.3.53)$$
Consider the equation:
$$(\ptl_{x_1}^{m_1}+f_1\ptl_{x_2}^{m_2}+\cdots+f_{n-1}\ptl_{x_n}^{m_n})(u)=0
\eqno(4.3.54)$$
 Denote
$$d_1=\ptl_{x_1}^{m_1},\;\;
d_r=\ptl_{x_1}^{m_1}+f_1\ptl_{x_2}^{m_2}+\cdots+f_{r-1}\ptl_{x_r}^{m_r}\qquad\for\;\;r
\in\ol{2,n}.\eqno(4.3.55)$$ We will apply Lemma 4.3.1 with
$T_1=d_r,\;T_2=\sum_{\iota=r}^{n-1}f_\iota\ptl_{x_{\iota+1}}^{m_{\iota+1}}$
and ${\cal B}=\mbb{R}[x_1,...,x_r],\;V=\mbb{R}[x_{r+1},...,x_n]$,
$$V_k=\mbox{Span}\:\{x_{r+1}^{\ell_{r+1}}\cdots x_n^{\ell_n}\mid
\ell_s\in\mbb{N},\;\ell_{r+1}+\sum_{\iota=r+2}^n\ell_\iota(\mbox{deg}\:f_{r+1}+1)\cdots
(\mbox{deg}\:f_{\iota-1}+1)\leq k\}.\eqno(4.3.56)$$ The motivation
of the above definition can be shown by the spacial example
$T_2=x_1\ptl_{x_2}+x_2^3\ptl_{x_3}$ and $V=\mbb{R}[x_2,x_3]$. In
this example, $T_2$ does not reduce the usual degree of the
polynomials in $V$. If we define new degree by $\mbox{deg}\;x_2^m=m$
and $\mbox{deg}\;x_3^m=4m$, then $T_2$ does  reduce the new degree
of the polynomials in $V$. Since $T_2(V_0)=\{0\}$ and
$T_2(V_{r+1})\subset {\cal B}V_r$ for $r\in\mbb{N}$, this gives a
proof that $T_2$ is locally nilpotent.

 Take a right inverse
$d_1^-=\int_{(x_1)}^{(m_1)}$.  Suppose that we have found a right
inverse $d_s^-$ of $d_s$ for some $s\in\ol{1,n-1}$ such that
$$x_\iota d_s^-=d_s^-x_\iota,\;\;
\ptl_{x_\iota}d_s^-=d_s^-\ptl_{x_\iota}\qquad\for\;\;\iota\in\ol{s+1,n}.\eqno(4.3.57)$$
Lemma 4.3.1 enable us to take
$$d_{s+1}^-=\sum_{\iota=0}^\infty(-d_s^-f_s)^\iota d_s^-\ptl_{x_{s+1}}^{\iota m_{s+1}}\eqno(4.3.58)$$
as a right inverse of $d_{s+1}$.  Obviously, $$ x_\iota
d_{s+1}^-=d_{s+1}^-x_\iota,\;\;
\ptl_{x_\iota}d_{s+1}^-=d_{s+1}^-\ptl_{x_\iota}\qquad\for\;\;\iota\in\ol{s+2,n}
\eqno(4.3.59)$$ according to (4.3.55). By induction, we have found a
right inverse $d_s^-$ of $d_s$ such that (4.3.57) holds for each
$s\in\ol{1,n}$.

We set
$${\cal S}_r=\{g\in \mbb{R}[x_1,...,x_r]\mid
d_r(g)=0\}\qquad\for\;\;r\in\ol{1,k}.\eqno(4.3.60)$$ By (4.3.55),
$${\cal S}_1=\sum_{i=0}^{m_1-1}\mbb{R}x_1^i.\eqno(4.3.61)$$
Suppose that we have found ${\cal S}_r$ for some $r\in \ol{1,n-1}$.
Given $h\in {\cal S}_r$ and $\ell\in \mbb{N}$, we define
$$\sgm_{r+1,\ell}(h)=\sum_{s=0}^\infty
(-d_r^-f_r)^s(h)\ptl_{x_{r+1}}^{sm_{r+1}}(x_{r+1}^{\ell}),\eqno(4.3.62)$$
which is actually a finite summation. Lemma 4.3.1 says
$${\cal S}_{r+1}=\sum_{\ell=0}^\infty \sgm_{r+1,\ell}({\cal
S}_r).\eqno(4.3.63)$$ By  induction, we obtain:\psp

{\bf Theorem 4.3.3}. {\it The set
$$\{\sgm_{n,\ell_n}\sgm_{n-1,\ell_{n-1}}\cdots\sgm_{2,\ell_2}(x_1^{\ell_1})\mid
\ell_1\in\ol{0,m_1-1},\;\ell_2,...,\ell_n\in\mbb{N}\}\eqno(4.3.64)$$
forms a basis of the polynomial solution space ${\cal S}_n$ of the
partial differential equation (4.3.54).}\psp

{\bf Example 4.3.3}. Let $m_1,m_2,n$ be positive integers. Consider
the following equations
$$\ptl_{x}^{m_1}(u)+x^n\ptl_{y}^{m_2}(u)=0\eqno(4.3.65)$$
Now
$$d_1=\ptl_{x}^{m_1},\;\;d_1^-=\int_{(x)}^{(m_1)}.\eqno(4.3.66)$$Then
\begin{eqnarray*}
\sgm_{2,\ell_2}(x^{\ell_1})&=&\sum_{r=0}^\infty(-\int_{(x)}^{(m_1)}x^n)^r
(x^{\ell_1})\ptl_{y}^{rm_2}(y^{\ell_2})\\
&=&x^{\ell_1}y^{\ell_2}+\sum_{r=1}^{\llbracket \ell_2/m_2
\rrbracket}\frac{(-1)^r[\prod_{s=0}^{rm_2-1}(\ell_2-s)]x^{r(n+m_1)+\ell_1}y^{\ell_2-rm_2}}
{\prod_{\iota=1}^{m_1}\prod_{j=1}^r(jn+(j-1)m_1+\iota+\ell_1)}.\hspace{1.9cm}
(4.3.67)\end{eqnarray*} The polynomial solution space of (4.3.65)
has a basis
$\{\sgm_{2,\ell_2}(x^{\ell_1})\mid\ell_1\in\ol{0,m_1-1},\;\ell_2\in\mbb{N}\}$.
\psp

In some practical problem, people found the linear wave equation
with dissipation:
$$u_{tt}+u_t-u_{x_1x_1}-u_{x_2x_2}-\cdots-u_{x_nx_n}=0.\eqno(4.3.68)$$
In order to find the polynomial solutions for the equations of the
above type pivoting at the variable $t$, we need the following
lemma. \psp

{\bf Lemma 4.3.4}. {\it Let $d=a\ptl_t+\ptl_t^2$ with $0\neq
a\in\mbb{R}$. Take a right inverse
$$d^-=\int_{(t)}\sum_{r=0}^\infty a^{-r-1}(-\ptl_t)^r\eqno(4.3.69)$$
of $d$. Then}
$$(d^-)^\iota(1)=\frac{t^\iota}{\iota!a^\iota}-\frac{t^{\iota-1}}{(\iota-2)!a^{\iota+1}}+\sum_{r=2}^{\iota-1}\frac{(-1)^r\prod_{s=1}^{r-1}(\iota+s)}
{(\iota-r-1)!r!a^{r+\iota}}t^{\iota-r}.\eqno(4.3.70)$$

{\it Proof}. For
$$f(t)=\sum_{\iota=1}^mb_\iota t^\iota\in\mbb{R}[t]t,\eqno(4.3.71)$$
we have
$$d(f(t))=amb_mt^{m-1}+\sum_{\iota=1}^{m-1}\iota(ab_\iota+(\iota+1)b_{\iota+1})t^{\iota-1}.\eqno(4.3.72)$$
Thus $d(f(t))=0$ if and only if $f(t)\equiv 0$. So for any given
$g(t)\in\mbb{R}[t]$, there exists a unique $f(t)\in \mbb{R}[t]t$
such that $d(f(t))=g(t)$.

Set
$$\xi_{a,\iota}(t)=\frac{t^\iota}{\iota!a^\iota}-\frac{t^{\iota-1}}{(\iota-2)!a^{\iota+1}}+\sum_{r=2}^{\iota-1}\frac{(-1)^r\prod_{s=1}^{r-1}(\iota+s)}
{(\iota-r-1)!r!a^{r+\iota}}t^{\iota-r},\eqno(4.3.73)$$ where we
treat
$$\xi_{a,0}(t)=1,\;\;\xi_{a,1}(t)=\frac{t}{a},\;\;\xi_{a,2}(t)=\frac{t^2}{2a^2}-\frac{t}{a^3}.
\eqno(4.3.74)$$ Easily verify
$d(\xi_{a,\iota}(t))=\xi_{a,\iota-1}(t)$ for $\iota=1,2$.

 Assume $\iota>2$. We have
\begin{eqnarray*} &&d(\xi_{a,\iota}(t))\\ &=&(a\ptl_t+\ptl_t^2)\left(\frac{t^\iota}{\iota!a^\iota}-
\frac{t^{\iota-1}}{(\iota-2)!a^{\iota+1}}+\sum_{r=2}^{\iota-1}\frac{(-1)^r\prod_{s=1}^{r-1}(\iota+s)}
{(\iota-r-1)!r!a^{r+\iota}}t^{\iota-r}\right)\\
&=&\frac{t^{\iota-1}}{(\iota-1)!a^{\iota-1}}-\frac{(\iota-1)t^{\iota-2}}{(\iota-2)!a^\iota}+
\sum_{r=2}^{\iota-1}\frac{(-1)^r(\iota-r)\prod_{s=1}^{r-1}(\iota+s)}
{(\iota-r-1)!r!a^{r+\iota-1}}t^{\iota-r-1}\\ &
&+\frac{t^{\iota-2}}{(\iota-2)!a^\iota}-\frac{(\iota-1)t^{\iota-3}}{(\iota-3)!a^{\iota+1}}+\sum_{r=2}^{\iota-1}
\frac{(-1)^r(\iota-r)\prod_{s=1}^{r-1}(\iota+s)}
{(\iota-r-2)!r!a^{r+\iota}}t^{\iota-r-2}\\
&=&\frac{t^{\iota-1}}{(\iota-1)!a^{\iota-1}}-\frac{t^{\iota-2}}{(\iota-3)!a^\iota}+
\frac{(\iota-2)(\iota+1)}
{(\iota-3)!2!a^{\iota+1}}t^{\iota-3}-\frac{(\iota-1)t^{\iota-3}}{(\iota-3)!a^{\iota+1}}\\
& &+
\sum_{r=3}^{\iota-1}(-1)^r\left[\frac{(\iota-r)\prod_{s=1}^{r-1}(\iota+s)}
{r!}-\frac{(\iota-r+1)\prod_{s=1}^{r-2}(\iota+s)}
{(r-1)!}\right]\frac{t^{\iota-r-1}}{(\iota-r-1)!a^{r+\iota-1}}\\ &=&
\frac{t^{\iota-1}}{(\iota-1)!a^{\iota-1}}-\frac{t^{\iota-2}}{(\iota-3)!a^\iota}+
\frac{(\iota-2)(\iota+1)-2(\iota-1)}{(\iota-3)!2!a^{\iota+1}}t^{\iota-3}
\\ & &+\sum_{r=3}^{\iota-1}(-1)^r\frac{[(\iota-r)(\iota+r-1)-r(\iota-r+1)]\prod_{s=1}^{r-2}(\iota+s)}
{(\iota-r-1)!r!a^{r+\iota-1}}t^{\iota-r-1}
\\ &=&
\frac{t^{\iota-1}}{(\iota-1)!a^{\iota-1}}-\frac{t^{\iota-2}}{(\iota-3)!a^\iota}+
\frac{\iota(\iota-3)}{(\iota-3)!2!a^{\iota+1}}t^{\iota-3}
\\ & &+\sum_{r=3}^{\iota-1}(-1)^r\frac{\iota(\iota-1-r)\prod_{s=1}^{r-2}(\iota+s)}
{(\iota-r-1)!r!a^{r+\iota-1}}t^{\iota-r-1}\\&=&
\frac{t^{\iota-1}}{(\iota-1)!a^{\iota-1}}-\frac{t^{\iota-2}}{(\iota-3)!a^\iota}+
\frac{\iota}{(\iota-4)!2!a^{\iota+1}}t^{\iota-3}
+\sum_{r=3}^{\iota-2}(-1)^r\frac{\iota\prod_{s=1}^{r-2}(\iota+s)}
{(\iota-r-2)!r!a^{r+\iota-1}}t^{\iota-r-1}\\&=&\frac{t^{\iota-1}}{(\iota-1)!a^{\iota-1}}-\frac{t^{\iota-2}}{(\iota-3)!a^\iota}+
\sum_{r=2}^{\iota-2}(-1)^r\frac{\prod_{s=1}^{r-1}(\iota-1+s)}
{(\iota-r-2)!r!a^{r+\iota-1}}t^{\iota-r-1}\\ &
=&\xi_{a,\iota-1}(t).\hspace{12.1cm}(4.3.75)\end{eqnarray*} Since
$(d^-)^0(1)=1,\;(d^-)^\iota(1)\in\mbb{R}[t]t$ by (4.3.69) and
$d[(d^-)^\iota(1)]=(d^-)^{\iota-1}(1)$ for $\iota\in\mbb{N}+1$, we
have $(d^-)^r(1)=\xi_{a,r}(t)$ for $r\in\mbb{N}$ by the uniqueness,
that is, (4.3.70) holds.$\qquad\Box$\psp

By Lemma 4.3.1 and the above lemma, we obtain:\psp

{\bf Theorem 4.3.5}. {\it The set
\begin{eqnarray*} \hspace{2cm}& &\{\sum_{r_1,...,r_n=0}^{\infty}{r_1+\cdots+r_n\choose
r_1,...,r_n}\left[\prod_{\iota=1}^n(2r_\iota)!{\ell_\iota\choose
2r_\iota}\right]
\\ & &\times\xi_{1,r_1+\cdots +r_n}(t)x_1^{\ell_1-2r_1}\cdots
x_n^{\ell_n-2r_n} \mid
\ell_1,...,\ell_n\in\mbb{N}\}\hspace{3.4cm}(4.3.76)\end{eqnarray*}forms
a basis of the polynomial solution space of the equation (4.3.68)}.
\psp

 Consider the Klein-Gordan equation:\index{Klein-Gordan equation}
$$u_{tt}-u_{xx}-u_{yy}-u_{zz}+a^2u=0,\eqno(4.5.77)$$ where $a$ is a nonzero
real number. Changing variable $u=e^{ai t}v$, we get
$$v_{tt}+2aiv_t-v_{xx}-v_{yy}-v_{zz}=0.\eqno(4.3.78)$$ We write
$$\xi_{2a i,\iota}=\zeta_{\iota,0}(t)+\zeta_{\iota,1}(t)i,\eqno(4.3.79)$$
where $\zeta_{\iota,0}(t)$ and $\zeta_{\iota,1}(t)$ are real
functions. According to (4.3.73),
$$\zeta_{2\iota,0}(t)=(-1)^\iota\left[\frac{t^{2\iota}}{(2\iota)!(2a)^{2\iota}}
+\sum_{r=1}^{\iota-1}\frac{(-1)^r\prod_{s=1}^{2r-1}(2\iota+s)}{(2r)!(2(\iota-r)-1)!(2
a)^{2(\iota+r)}}t^{2(\iota-r)}\right],\eqno(4.3.80)$$
\begin{eqnarray*}\hspace{2cm}& &\zeta_{2\iota,1}(t)=(-1)^\iota[\frac{t^{2\iota-1}}{(2\iota-2)!
(2a)^{2\iota+1}}\\ & &
+\sum_{r=1}^{\iota-1}\frac{(-1)^r\prod_{s=1}^{2r}(2\iota+s)}{(2r+1)!
[2(\iota-r-1)]!(2a)^{2\iota+2r+1}}t^{2\iota-2r-1}],\hspace{3.1cm}(4.3.81)\end{eqnarray*}
\begin{eqnarray*}\hspace{2cm}& &
\zeta_{2\iota+1,0}(t)=(-1)^\iota[\frac{t^{2\iota}}{(2\iota-1)! (2a)^{2(\iota+1)}}\\
& &
+\sum_{r=1}^{\iota-1}\frac{(-1)^r\prod_{s=1}^{2r}(2\iota+s+1)}{(2r+1)!(2\iota-2r-1)!
(2a)^{2(\iota+r+1)}}t^{2(\iota-r)}],\hspace{3.3cm}(4.3.82)\end{eqnarray*}
$$
\zeta_{2\iota+1,1}(t)=(-1)^{\iota+1}
\left[\frac{t^{2\iota+1}}{(2\iota+1)!(2a)^{2\iota+1}}+
\sum_{r=1}^\iota\frac{(-1)^r\prod_{s=1}^{2r-1}(2\iota+s+1)}{(2r)!(2\iota-2r)!(2
a)^{2\iota+2r+1}}t^{2\iota-2r+1}\right].\eqno(4.3.83)$$ Recall the
three-dimensional Laplace operator
$\Dlt=\ptl_x^2+\ptl_y^2+\ptl_z^2$. By Lemma 4.3.1 and Lemma 4.3.4,
$$\{e^{ait}\sum_{r=0}^\infty
\xi_{2ai,r}(t)\Dlt^r(x^\ell_1y^{\ell_2}z^{\ell_3})\mid
\ell_1,\ell_2,\ell_3\in\mbb{N}\}\eqno(4.3.84)$$ are complex
solutions of the Klein-Gordan equation (4.3.77). Taking real parts
of (4.3.84), we get \psp

{\bf Theorem 4.3.6}. {\it The Klein-Gordan equation (4.3.77) has the
following set of linearly independent trigonometric-polynomial
solution}:
\begin{eqnarray*} & &\{\sum_{r_1,r_2,r_3=0}^{\infty}{r_1+r_2+r_3\choose
r_1,r_2,r_3}\left[\prod_{s=1}^3(2r_s)!{\ell_s\choose 2r_s}\right]
(\zeta_{r_1+r_2+r_3,0}(t)\cos at-\zeta_{r_1+r_2+r_3,1}(t)\sin at)\\
& &\times
x^{\ell_1-2r_1}y^{\ell_2-2r_2}z^{\ell_3-2r_3},\;\sum_{r_1,r_2,r_3=0}^{\infty}{r_1+r_2+r_3\choose
r_1,r_2,r_3}\left[\prod_{s=1}^3(2r_s)!{\ell_s\choose 2r_s}\right]
(\zeta_{r_1+r_2+r_3,0}(t)\sin at\\ & &+\zeta_{r_1+r_2+r_3,1}(t)\cos
at)x^{\ell_1-2r_1}y^{\ell_2-2r_2}z^{\ell_3-2r_3},\mid
\ell_1,\ell_2,\ell_3\in\mbb{N}\}\hspace{3.8cm}(4.3.85)\end{eqnarray*}
\pse

The following lemmas will be used to handle some special cases when
the operator $T_1$  in Lemma 4.3.1 does not have a right inverse. We
again use the settings in (4.3.9) and (4.3.10).\psp

 {\bf Lemma 4.3.7}. {\it Let $T_0$
be a differential operator on ${\cal A}$ with right inverse $T_0^-$
such that
$$T_0({\cal B}),T_0^-({\cal B})\subset {\cal B},\;\;T_0(\eta_1\eta_2)=
T_0(\eta_1)\eta_2\qquad\for\;\;\eta_1 \in {\cal B},\;\eta_2\in
V,\eqno(4.3.86)$$ and let $T_1,...,T_m$ be commuting differential
operators on ${\cal A}$ such that $T_\iota(V)\subset V$,
$$T_0T_\iota=T_\iota T_0,\qquad T_\iota(f\zeta)=fT_\iota(\zeta)
\qquad\mbox{\it for}\;\;\iota\in\ol{1,m},\;f\in{\cal
B},\;\zeta\in{\cal A}.\eqno(4.3.87)$$ If $T_0^m(h)=0$ with
$h\in{\cal B}$ and $g\in V$, then
\begin{eqnarray*}\qquad u&=&\sum_{\iota=0}^\infty(\sum_{s=1}^m(T_0^-)^sT_s)^\iota(hg)=
\sum_{\iota_1,...,\iota_m=0}^\infty {\iota_1+\cdots+\iota_m\choose \iota_1,...,\iota_m}\\
& &\times
(T_0^-)^{\sum_{s=1}^ms\iota_s}(h)(\prod_{r=1}^mT_r^{\iota_r})(g)\hspace{7.5cm}(4.3.88)\end{eqnarray*}
is a solution of the equation:
$$(T_0^m-\sum_{r=1}^m T_0^{m-\iota}T_\iota)(u)=0.\eqno(4.3.89)$$
Suppose
$$T_\iota(V_r)\subset V_{r-1}\qquad\mbox{\it
for}\;\;\iota\in\ol{1,m},\;r\in\mbb{N},\eqno(4.3.90)$$ where
$V_{-1}=\{0\}$. Then any polynomial solution of (4.3.89) is a linear
combinations of the solutions of the form (4.3.88).}

{\it Proof}. Note that
$$T_0^{m-\iota}=T_0^m
(T_0^-)^\iota\qquad\for\;\;\iota\in\ol{1,m}\eqno(4.3.91)$$ and
\begin{eqnarray*} \hspace{2cm}& &\sum_{\iota_1+\cdots+\iota_m=\iota+1}{\iota+1\choose
\iota_1,...,\iota_m}y_1^{\iota_1}\cdots y_m^{\iota_m}=(y_1+\cdots+y_m)^{\iota+1}\\
&=&\sum_{r=1}^m\sum_{\iota_1+\cdots+\iota_m=\iota}{\iota\choose
\iota_1,...,\iota_m}y_ry_1^{\iota_1}\cdots
y_m^{\iota_m}.\hspace{4.9cm}(4.3.92)\end{eqnarray*}
 Thus
\begin{eqnarray*} & &(T_0^m-\sum_{p=1}^m T_0^{m-p}T_p)\left[\sum_{\iota_1,...,\iota_m=0}^\infty
{\iota_1+\cdots+\iota_m\choose
\iota_1,...,\iota_m}(T_0^-)^{\sum_{s=1}^ms\iota_s}(h)(\prod_{r=1}^mT_r^{\iota_r})(g)\right]\\
&=&\sum_{\iota_1,...,\iota_m\in\mbb{N};\:\iota_1+\cdots+\iota_m>0}
{\iota_1+\cdots+\iota_m\choose
\iota_1,...,\iota_m}T_0^m(T_0^-)^{\sum_{s=1}^ms\iota_s}(h)(\prod_{r=1}^mT_r^{\iota_r})(g)\\
& &- \sum_{\iota_1,...,\iota_m=0}^\infty\sum_{p=1}^m
{\iota_1+\cdots+\iota_m\choose
\iota_1,...,\iota_m}T_0^m(T_0^-)^{\iota_p+\sum_{s=1}^ms\iota_s}(h)(T_p\prod_{r=1}^mT_r^{\iota_r})(g)=0.
\hspace{1cm}(4.3.93)\end{eqnarray*}

Suppose that (4.3.90) holds.  Let  $u\in {\cal B}V_k\setminus{\cal
B}V_{k-1}$ be a solution of (4.3.89). Take a basis
$\{\phi_\iota+V_{k-1}\mid \iota\in I\}$ of $V_k/V_{k-1}$. Write
$$u=\sum_{\iota\in I}h_\iota\phi_\iota+u',\qquad h_\iota\in{\cal B},\;u'\in{\cal
B}V_{k-1}.\eqno(4.3.94)$$ Since
$$T_r(\phi_\iota)\in V_{k-1}\qquad\for\;\;\iota\in
I,\;r\in\ol{1,m}\eqno(4.3.95)$$ by (4.3.90), we have
$$(T_0^m-\sum_{r=1}^m T_0^{m-\iota}T_\iota)(u)\equiv \sum_{\iota\in
I}T_0^m(h_\iota)\phi_\iota\equiv 0\;\;(\mbox{mod}\;{\cal
B}V_{k-1}).\eqno(4.3.96)$$ Hence
$$T_0^m(h_\iota)=0\qquad\for\;\;\iota\in I.\eqno(4.3.97)$$
Now
$$u-\sum_{j\in I}\sum_{\iota_1,...,\iota_m=0}^\infty
{\iota_1+\cdots+\iota_m\choose
\iota_1,...,\iota_m}(T_0^-)^{\sum_{s=1}^ms\iota_s}(h_j)(\prod_{r=1}^mT_r^{\iota_r})(\phi_j)\in
{\cal B}V_{k-1}\eqno(4.3.98)$$ is a solution of (4.3.89). By
induction on $k$, $u$ is a linear combinations of the solutions of
the form (4.3.88).$\qquad\Box$\psp

We remark that the above lemma does not imply Lemma 4.3.1 because
$T_1$ and $T_2$ in Lemma 4.3.1 may not commute.

 Let $d_1$ be a differential operator on
$\mbb{R}[x_1,x_2,...,x_r]$ and let $d_2$ be a locally nilpotent
differential operator on $V=\mbb{R}[x_{r+1},...,x_n]$. Set
$$V_m=\{f\in V\mid d_2^{m+1}(f)=0\}\qquad\for\;\;m\in\mbb{N}.\eqno(4.3.99)$$
Then $V=\bigcup_{m=0}^\infty V_m$ because $d_2$ is locally
nilpotent. We treat $V_{-1}=\{0\}$. Take a subset $\{\psi_{m,j}\mid
m\in\mbb{N},\;j\in I_m\}$ of $V$ such
 that $\{\psi_{m,j}+V_{m-1}\mid j\in I_m\}$ forms a basis of
 $V_m/V_{m-1}$ for $m\in\mbb{N}$. In particular, $\{\psi_{m,j}\mid
 m\in\mbb{N},\;j\in I_m\}$ forms a basis of $V$.
 Fix $h\in \mbb{R}[x_1,...,x_r]$.\psp

{\bf Lemma 4.3.8}. {\it Let $m$ be a positive integer. Suppose that
$$u=\sum_{j\in I_m}f_j\psi_{m,j}+u'\in
\mbb{R}[x_1,x_2,...,x_n]\eqno(4.3.100)$$ with
$f_j\in\mbb{R}[x_1,...,x_r]$ and $d_2^m(u')=0$ is a solution of the
equation:
$$(d_1-hd_2)(u)=0.\eqno(4.3.101)$$
Then  $d_1(f_j)=0$ for $j\in I_m$ and the system
$$\xi_0=f_j,\;\;d_1(\xi_{s+1})=h\xi_s\qquad
\for\;\;s\in\ol{0,m-1}\eqno(4.3.102)$$ has a solution
$\xi_1,...,\xi_m\in \mbb{R}[x_1,...,x_r]$ for each $j\in I_m$.}

{\it Proof}. Observe that if $\{g_j+V_p\mid j\in J\}$  is a linearly
independent subset of $V_{p+1}/V_p$, then $\{d_2^s(g_j)+V_{p-s}\mid
j\in J\}$ is a linearly independent subset of $V_{p-s+1}/V_{p-s}$
for $s\in\ol{1,p+1}$ by (4.3.99). By induction, we take a
subset$\{\phi_{m-s,j}\mid j\in J_{m-s}\}$ of $V_{m-s}$ for each
$s\in\ol{1,m}$ such that
$$\{d^s_2(\psi_{m,j_1})+V_{m-s-1},d_2^{s-p}(\phi_{m-p,j_2})+V_{m-s-1}\mid
p\in\ol{1,s},\;j_1\in I_m,\;j_2\in J_{m-p}\}\eqno(4.3.103)$$ forms a
basis of $V_{m-s}/V_{m-s-1}$ for $s\in\ol{1,m}$. Denote
$${\cal
U}=\sum_{s=1}^m\sum_{p=0}^{m-s}\sum_{j\in
J_{m-s}}\mbb{R}[x_1,...,x_r]d_2^p(\phi_{m-s,j}).\eqno(4.3.104)$$

Now we write
$$u=\sum_{j\in
I_m}[f_j\psi_{m,j}+\sum_{s=1}^m f_{s,j}d_2^s(\psi_{m,j})]+v,\qquad
v\in{\cal U},\;f_{s,j}\in\mbb{R}[x_1,...,x_r].\eqno(4.3.105)$$ Then
(4.3.101) becomes
\begin{eqnarray*}& &\sum_{j\in
I_m}[d_1(f_j)\psi_{m,j}+(d_1(f_{1,j})-hf_j)
d_2(\psi_{m,j})+\sum_{s=2}^m(d_1(f_{s,j})-hf_{s-1,j})d^s(\psi_{m,j})]
\\ & &+(d_1-hd_2)(v)=0.\hspace{10.1cm}(4.3.106)\end{eqnarray*} Since $(d_1-hd_2)(v)\in{\cal U}$,
we have:
$$d_1(f_j)=0,\;\;d_1(f_{1,j})=hf_j,\;\;d_1(f_{s,j})=hf_{s-1,j}\eqno(4.3.107)$$
for $j\in I_m$ and $s\in\ol{2,m}$. So (4.3.102) has a solution
$\xi_1,...,\xi_m\in \mbb{R}[x_1,...,x_r]$ for each $j\in
I_m.\hspace{0.7cm}\Box$\psp

We remark that our above lemma implies that if (4.3.102) does not
have a solution for some $j$, then the equation (4.3.101) does not
have a solution of the form (4.3.100). Set
$${\cal S}_0=\{f\in \mbb{R}[x_1,...,x_r]\mid
d_1(f)=0\}\eqno(4.3.108)$$ and
$${\cal S}_m=\{f_0\in {\cal S}_0\mid d_1(f_s)=hf_{s-1}\;\mbox{for
some}\;\;f_1,...,f_m\in\mbb{R}[x_1,...,x_r]\}\eqno(4.3.109)$$ for
$m\in\mbb{N}+1$. For each $m\in\mbb{N}+1$ and $f\in {\cal S}_m$, we
fix $\{\sgm_1(f),...,\sgm_m(f)\}\subset\mbb{R}[x_1,...,x_r]$ such
that
$$d_1(\sgm_1(f))=hf,\;\;d_1(\sgm_s(f))=h\sgm_{s-1}(f)\qquad\for\;\;s\in\ol{2,m}.
\eqno(4.3.110)$$ Denote $\sgm_0(f)=f$.  \psp

{\bf Lemma 4.3.9}. {\it The set
$${\cal S}=\sum_{m=0}^\infty\sum_{j\in
I_m}\sum_{f\in{\cal
S}_m}\mbb{R}(\sum_{s=0}^m\sgm_s(f)d^s_2(\psi_{m,j}))\eqno(4.3.111)$$
is the solution space of the equation (4.3.101) in
$\mbb{R}[x_1,x_2,...,x_n]$.}\psp

{\it Proof}. For $f\in {\cal S}_m$,
\begin{eqnarray*} \qquad & &(d_1-hd_2)(\sum_{s=0}^m\sgm_s(f)d^s_2(\psi_{m,j}))\\
&=&
\sum_{s=1}^mh\sgm_{s-1}(f)d^s_2(\psi_{m,j})-\sum_{s=0}^{m-1}h\sgm_s(f)d^{s+1}_2(\psi_{m,j})=0.
\hspace{3.5cm}(4.3.112)\end{eqnarray*} Thus
$\sum_{s=0}^m\sgm_s(f)d^s_2(\psi_{m,j})$ is a solution of (4.3.101).

Suppose that $u$ is a solution (4.3.101). Then $u$ can be written as
(4.3.100) such that $f_j\neq 0$ for some $j\in I_m$ due to
$V=\bigcup_{m=0}^\infty V_m$. If $m=0$, then $u\in {\cal S}$
naturally. Assume that $u\in {\cal S}$ if $m<\ell$. Consider
$m=\ell$. According to Lemma 4.3.8, $f_j\in {\cal S}_m$ for any
$j\in I_m$ (cf. (4.3.109)). Thus $\sum_{j\in
I_m}\sum_{s=0}^m\sgm_s(f_j)d_2^s(\psi_{m,j})$ is a solution of the
equation (4.3.101). Hence $u-\sum_{j\in
I_m}\sum_{s=0}^m\sgm_s(f_j)d_2^s(\psi_{m,j})$ is a solution of
(4.3.101) and
$$d_2^m(u-\sum_{j\in
I_m}\sum_{s=0}^m\sgm_s(f_j)d_2^s(\psi_{m,j}))=0.\eqno(4.3.113)$$ So
$$u-\sum_{j\in
I_m}\sum_{s=0}^m\sgm_s(f_j)d_2^s(\psi_{m,j})\in
\mbb{R}[x_1,...,x_r]V_{m-1}.\eqno(4.3.114)$$ By assumption,
$$u-\sum_{j\in
I_m}\sum_{s=0}^m\sgm_s(f_j)d_2^s(\psi_{m,j})\in {\cal
S}.\eqno(4.3.115)$$ Since $\sum_{j\in
I_m}\sum_{s=0}^m\sgm_s(f_j)d_2^s(\psi_{m,j})\in {\cal S}$, we have
$u\in{\cal S}$. By induction, $u\in{\cal S}$ for any solution of
(4.3.101). $\qquad\Box$\psp

 Let $\es\in\{1,-1\}$ and let $\lmd$ be a nonzero real
number. Next we want to find all the polynomial solutions of the
equation:
$$u_{tt}+\frac{\lmd}{t}u_t-\es(u_{x_1x_1}+u_{x_2x_2}+\cdots+u_{x_nx_n})=0,
\eqno(4.3.116)$$ which is the {\it generalized anisymmetrical
Laplace equation}\index{anisymmetrical Laplace equation} if
$\es=-1$. Rewrite the above equation as:
$$tu_{tt}+\lmd u_t-\es t(u_{x_1x_1}+u_{x_2x_2}+\cdots+u_{x_nx_n})=0.
\eqno(4.3.117)$$
 Set
$$d_1=t\ptl_t^2+\lmd\ptl_t,\qquad d_2=\Dlt_n=\sum_{r=1}^n\ptl_{x_r}^n,\;\;h=\es t.\eqno(4.3.118)$$
Denote
$${\cal S}=\{f\in\mbb{R}[t]\mid d_1(f)=0\}.\eqno(4.3.119)$$
Note that
$$d_1(t^m)=m(\lmd+m-1)t^{m-1}\qquad\for\;\;m\in\mbb{N}.\eqno(4.3.120)$$
So
$${\cal
S}=\left\{\begin{array}{ll}\mbb{R}&\mbox{if}\;\lmd\not\in-(\mbb{N}+1),\\
\mbb{R}+\mbb{R}t^{-\lmd+1}&\mbox{if}\;\lmd\in-(\mbb{N}+1).\end{array}\right.
\eqno(4.3.121)$$ In particular, $t^{-\lmd}\not\in d_1(\mbb{R}[t])$
and so $d_1$ does not have a right inverse when $\lmd$ is a negative
integer. Otherwise $t^{-\lmd}=d_1(d_1^-(t^{-\lmd}))\in
d_1(\mbb{R}[t])$.

 Set
$$\phi_0(t)=1,\;\;\phi_m(t)=\frac{\es^mt^{2m}}{m!2^m\prod_{r=0}^{m-1}(\lmd+2r+1)}
\eqno(4.3.122)$$ for $m\in\mbb{N}+1$ and $\lmd\neq
-1,-3,...,-(2m-1)$.  Then $d(\phi_{r+1}(t))=\es t\phi_r(t)$ for
$r\in\ol{0,m-1}$. If $\lmd=-2k-1$, there does not exist a function
$\phi(t)\in\mbb{R}[t]$ such that $d_1(\phi(t))=\es t\phi_k(t)$
because $d_1(t^{2k+2})=(2k+2)(2k+1+\lmd)t^{2k+1}=0$. When $\lmd\in
-(\mbb{N}+1)$, we set
$$\psi_0=t^{1-\lmd},\;\;\psi_m=\frac{\es^mt^{2m+1-\lmd}}{2^m m!\prod_{r=1}^m(2r+1-\lmd)}\qquad
\for\;\;m\in\mbb{N}+1.\eqno(4.3.123)$$ It can be verified that
$d_1(\psi_{r+1}(t))=\es t\psi_r(t)$ for $r\in\mbb{N}$. Define
$$V=\mbb{R}[x_1,x_2,...,x_n],\;\;\Dlt_{2,n}=
\sum_{s=2}^n\ptl_{x_s}^2 \eqno(4.3.124)$$ and
$$V_m=\{f\in V\mid
\Dlt_n^{m+1}(f)=0\}\qquad\for\;\;m\in\mbb{N}.\eqno(4.3.125)$$
Observe
\begin{eqnarray*}\hspace{1cm}&
&\sum_{j_1,...,j_\ell=0}^\infty(-1)^{j_1+\cdots+j_\ell}{j_1+\cdots+j_\ell\choose
j_1,...,j_\ell}\prod_{r=1}^\ell \left[{\ell\choose
r}t^r\right]^{j_r}=
\sum_{p=0}^\infty\left(-\sum_{s=1}^\ell{\ell\choose s}t^s\right)^p\\
&=& \frac{1}{(1+t)^\ell}=\sum_{r=0}^\infty (-1)^r{\ell+r-1\choose
r}t^r\hspace{6.2cm}(4.3.126)\end{eqnarray*} for $|t|<1.$
 Applying Lemma 4.3.7 to
$\Dlt_n^{m+1}=\sum_{r=0}^{m+1}{m+1\choose
r}\ptl_{x_1}^{2(m+1-r)}\Dlt_{2,n}^r$, $T_0=\ptl_{x_1}^2$ and
$T_r=-{m+1\choose r}\Dlt_{2,n}^r$ for $r\in\ol{1,m+1}$, we get a
basis
$$\left\{\sum_{r=0}^\infty (-1)^r{m+r\choose
r}\frac{x_1^{\ell_1+2r}}{(\ell_1+2r)!}\Dlt_{2,n}^r(x_2^{\ell_2}\cdots
x_n^{\ell_n})\mid
\ell_1\in\ol{0,2m+1},\;\ell_2,...,\ell_n\in\mbb{N}\right\}\eqno(4.3.127)$$
 of $V_m$. Hence we obtain:\psp

{\bf Theorem 4.3.10}. {\it If $\lmd\not\in -(\mbb{N}+1)$, then the
set
$$\{\sum_{r=0}^\infty
\phi_r(t)\Dlt_n^r(x_1^{\ell_1}\cdots x_n^{\ell_n})\mid
\ell_1,...,\ell_n\in\mbb{N}\}\eqno(4.3.128)$$ forms a basis of the
space of the polynomial solutions for the equation (4.3.117). When
$\lmd$ is a negative even integer, the set $$\{\sum_{r=0}^\infty
\phi_r(t)\Dlt_n^r(x_1^{\ell_1}\cdots x_n^{\ell_n}),\sum_{r=0}^\infty
\psi_r(t)\Dlt_n^r(x_1^{\ell_1}\cdots x_n^{\ell_n})\mid
\ell_1,...,\ell_n\in\mbb{N}\}\eqno(4.3.129)$$ forms a basis of the
space of the polynomial solutions for the equation (4.3.117). Assume
that $\lmd=-2k-1$ is a negative odd integer. The set
\begin{eqnarray*}\hspace{1cm} & &\{\sum_{s=0}^k\sum_{r=0}^\infty
(-1)^r {k+r\choose r}\phi_s(t)\Dlt_n^s
\left[\frac{x_1^{\ell_1+2r}}{(\ell_1+2r)!}\Dlt_{2,n}^r(x_2^{\ell_2}\cdots
x_n^{\ell_n})\right],\\ & &\sum_{r=0}^\infty
\psi_r(t)\Dlt_n^r(x_1^{\ell_1'}x_2^{\ell_2}\cdots x_n^{\ell_n}) \mid
\ell_1\in\ol{0,2k+1},\;\ell_1',\ell_2,...,\ell_n\in\mbb{N}\}\hspace{2cm}
(4.3.130)\end{eqnarray*} is a basis of the space of the polynomial
solutions for the equation (4.3.117).}\psp

Finally, we consider the {\it special Euler-Poisson-Darboux
equation}: \index{special Euler-Poisson-Darboux equation}
$$u_{tt}-u_{x_1x_1}-u_{x_2x_2}-\cdots -u_{x_nx_n}
-\frac{m(m+1)}{t^2}u=0\eqno(4.3.131)$$ with $m\neq -1,0$. Change the
equations to:
$$t^2u_{tt}-t^2(u_{x_1x_1}+u_{x_2x_2}+\cdots +u_{x_nx_n})-m(m+1)u=0.\eqno(4.3.132)$$
Letting $u=t^{m+1}v$, we have:
$$t^2u_{tt}=m(m+1)t^{m+1}v+2(m+1)t^{m+2}v_t+t^{m+3}v_{tt}.\eqno(4.3.133)$$
Substituting (4.3.133) into (4.2.132), we get
$$tv_{tt}+2(m+1)v_t-t(v_{x_1x_1}+v_{x_2x_2}+\cdots +v_{x_nx_n})=0.
\eqno(4.3.134)$$ If we change variable $u=t^{-m}v$, then the
equation (4.3.132) becomes
$$tv_{tt}-2mv_t-t(v_{x_1x_1}+v_{x_2x_2}+\cdots+_{x_nx_n})=0
. \eqno(4.3.135)$$ Equations (4.3.134) and (4.3.135) are special
cases of the equation (4.3.117) with $\es=1$, and $\lmd=2(m+1)$ and
$\lmd=-2m$, respectively. \psp

{\bf Exercise 4.3} \psp

Find a basis of the polynomial solution space of the generalized
Laplace equation
$$u_{xx}+xu_{yy}+yu_{zz}=0.$$

\section{ Use of Fourier Expansion I}

In this section, we mainly use  Fourier expansion to solve
constant-coefficient linear partial differential equations. Let us
first look at three simple examples which are commonly used in
engineering mathematics. Kovalevskaya Theorem says that their
solutions are unique.

\psp

 {\bf Example 4.4.1}. Solve the following {\it heat conduction
 equation}
 \index{heat conduction equation}
$$u_t=u_{xx}\;\;\;\mbox{subject
to}\;\;u(t,-\pi)=u(t,\pi)\;\;\mbox{and}\;\;u(0,x)=g(x)\;\;\mbox{for}\;\;x\in[-\pi,\pi],\eqno(4.4.1)$$
where $g(x)$ is a given continuous function.

{\it Solution}. We assume the separation of variables
$u=\eta(t)\xi(x).$ Then the equation becomes
$$\eta'(t)\xi(x)=\eta(t){\xi'}'(x)\lra {\eta'(t)\over\eta(t)}={\xi''(x)\over \xi(x)}=\lmd\eqno(4.4.2)$$
is a constant. Solving the problem
$$\xi''=\lmd\xi,\;\;\;\xi(-\pi)=\xi(\pi)=0,\eqno(4.4.3)$$
we take $\lmd=-n^2$ for some $n\in\mbb{N}$ and $\xi=C_1\cos
nx+C_2\sin nx$. Moreover, $\eta'(t)=-n^2\eta(t)\lra \eta=C_3e^{-n^2
t}$. Thus
$$u=e^{-n^2t}(a\cos nx+b\sin nx)\eqno(4.4.4)$$
is a solution of the problem:
$$u_t=u_{xx},\;\;u(t,-\pi)=u(t,\pi).\eqno(4.4.5)$$
By {\it superposition principle} (additivity of solutions for
homogeneous linear equations),\index{superposition principle} we
have more general solutions of (4.4.5):
$$u(t,x)=\sum_{n=0}^\infty e^{-n^2t}(a_n\cos nx+b_n\sin
nx),\eqno(4.4.6)$$ where $a_n$ and $b_n$ are constants to be
determined. To satisfy the last condition in (4.4.1), we require
$$ \sum_{n=0}^\infty (a_n\cos
nx+ b_n\sin nx)=u(0,x)=g(x).\eqno(4.4.7)$$ According to the theory
of Fourier expansion,
$$a_0=\frac{1}{2\pi}\int_{-\pi}^\pi g(s)ds,\;\;a_n=\frac{1}{\pi}\int_{-\pi}^\pi g(s)\cos ns\:ds,\;\; b_n=\frac{1}{\pi}\int_{-\pi}^\pi g(s)\sin
ns\:ds\eqno(4.4.8)$$ for $n\geq 1$. So the final solution of (4.1)
is
\begin{eqnarray*}& &u(t,x)=\frac{1}{2\pi}\int_{-\pi}^\pi g(s)ds\\ & &+\frac{1}{\pi}\sum_{n=1}^\infty e^{-n^2t}\left[\cos nx\:\int_{-\pi}^\pi g(s)\cos
ns\:ds+\sin nx\:\int_{-\pi}^\pi g(s)\sin ns\:ds\right]
\\ &=&\frac{1}{2\pi}\int_{-\pi}^\pi g(s)ds+\frac{1}{\pi}\sum_{n=1}^\infty e^{-n^2t}\int_{-\pi}^\pi g(s)(\cos nx\:\cos+\sin nx\:\sin
ns)\:ds\\ &=&\frac{1}{2\pi}\int_{-\pi}^\pi
g(s)ds+\frac{1}{\pi}\sum_{n=1}^\infty e^{-n^2t}\int_{-\pi}^\pi
g(s)\cos n(x-s)\:ds.\qquad\Box\hspace{3.3cm}(4.4.9)\end{eqnarray*}
\pse

{\bf Example 4.4.2}. Solve the following {\it wave equation}
\index{wave equation}
$$u_{tt}=u_{xx}\;\;\;\mbox{subject
to}\;\;u(t,-\pi)=u(t,\pi)\eqno(4.4.10)$$ and
$$u(0,x)=g_1(x),\;u_t(0,x)=g_2(x)\;\;\for\;\;x\in[-\pi,\pi],\eqno(4.4.11)$$ where
$g_1(x)$ and $g_2(x)$ are given continuous functions.

{\it Solution}. We assume the {\it separation of
variables}\index{separation of variables} $u=\eta(t)\xi(x).$ Then
the equation becomes
$${\eta'}'(t)\xi(x)=\eta(t){\xi'}'(x)\lra {{\eta'}'(t)\over\eta(t)}={\xi''(x)\over \xi(x)}=\lmd\eqno(4.4.12)$$
is a constant. As the above example, we find the general solution of
(4.4.10) is
$$u(t,x)=\sum_{n=0}^\infty \cos nt\;(a_n\cos nx+b_n\sin
nx)+\sum_{n=1}^\infty\sin nt\;(\hat a_n\cos nx+\hat b_n\sin nx)+\hat
a_0t,\eqno(4.4.13)$$ where $a_n,b_n,\hat a_n,\hat b_n\in\mbb{R}$.
Note
$$u(0,x)=\sum_{n=0}^\infty(a_n\cos nx+b_n\sin
nx)=g_1(x).\eqno(4.4.14)$$ Since
$$u_t(t,x)=\sum_{n=1}^\infty n[ \sin nt\;(a_n\cos nx+b_n\sin
nx)+\cos nt\;(\hat a_n\cos nx+\hat b_n\sin nx)]+\hat
a_0,\eqno(4.4.15)$$ we have
$$u_t(0,x)=\sum_{n=1}^\infty n(\hat a_n\cos nx+\hat b_n\sin
nx)+\hat a_0=g_2(x).\eqno(4.4.16)$$ As (4.4.6)-(4.4.9), the final
solution is \begin{eqnarray*}u(t,x)&=&\frac{1}{2\pi}\int_{-\pi}^\pi
(g_1(s)+tg_2(s))ds+\frac{1}{\pi}\sum_{n=1}^\infty \cos nt\;\int_{-\pi}^\pi g_1(s)\cos n(x-s)\:ds\\
& &+\frac{1}{\pi}\sum_{n=1}^\infty\frac{\sin nt}{n}\int_{-\pi}^\pi
g_2(s)\cos
n(x-s)\:ds.\qquad\Box\hspace{4.3cm}(4.4.17)\end{eqnarray*} \pse

{\bf Example 4.4.3}. Solve the following {\it Laplace
equation}\index{Lapalce equation}
$$u_{xx}+u_{yy}=0\;\;\;\mbox{subject
to}\;\;u(x,-\pi)=u(x,\pi)\eqno(4.4.18)$$ and
$$u(0,y)=g_1(y),\;u_x(0,y)=g_2(y)\;\;\for\;\;y\in[-\pi,\pi],\eqno(4.4.19)$$ where
$g_1(y)$ and $g_2(y)$ are given continuous functions.

{\it Solution}. We assume the separation of variables
$u=\eta(x)\xi(y).$ Then the equation becomes
$${\eta'}'(x)\xi(y)=-\eta(x){\xi'}'(y)\lra -{{\eta'}'(x)\over\eta(x)}={\xi''(y)\over \xi(y)}=\lmd\eqno(4.4.20)$$
is a constant. As Example 4.4.1, we find the general solution of
(4.4.18) is
\begin{eqnarray*}\qquad u(x,y)&=&\sum_{n=0}^\infty \cosh nx\;(a_n\cos
ny+b_n\sin ny)+\hat a_0x\\ & &+\sum_{n=1}^\infty\sinh nx\;(\hat
a_n\cos ny+\hat b_n\sin ny),\hspace{5.05cm}(4.4.21)\end{eqnarray*}
where $a_n,b_n,\hat a_n,\hat b_n\in\mbb{R}$. As (4.4.14)-(4.4.17),
we get the final solution
 \begin{eqnarray*}\qquad u(x,y)&=&\frac{1}{2\pi}\int_{-\pi}^\pi
(g_1(s)+xg_2(s))ds+\frac{1}{\pi} \sum_{n=1}^\infty \cosh
nx\;\int_{-\pi}^\pi g_1(s)\cos n(y-s)\:ds\\ &
&+\frac{1}{\pi}\sum_{n=1}^\infty\frac{\sinh nx}{n}\int_{-\pi}^\pi
g_2(s)\cos
n(y-s)\:ds.\qquad\Box\hspace{3.05cm}(4.4.22)\end{eqnarray*} \pse

 The rest of this section is taken from the author's
work [X11].

Let $m$ and $n>1$ be positive integers and let
$$f_r(\ptl_{x_2},...,\ptl_{x_n})\in\mbb{R}[\ptl_{x_2},...,\ptl_{x_n}]
\qquad\for\;\;r\in\ol{1,m}.\eqno(4.4.23)$$ We want to solve the
equation:
$$(\ptl_{x_1}^m-\sum_{r=1}^m\ptl_{x_1}^{m-r}f_r(\ptl_{x_2},...,\ptl_{x_n}))(u)=0
\eqno(4.4.24)$$ with $x_1\in\mbb{R}$ and $x_r\in[-a_r,a_r]$ for
$r\in\ol{2,n}$, subject to the condition
$$\ptl_{x_1}^s(u)(0,x_2,...,x_n)=g_s(x_2,...,x_n)\qquad\for\;\;s\in\ol{0,m-1},
\eqno(4.4.25)$$ where $a_2,...,a_n$ are positive real numbers and
$g_0,...,g_{m-1}$ are continuous functions. For convenience, we
denote
$$k^\dg_\iota=\frac{k_\iota}{a_\iota},\;\;\vec
k^\dg=(k^\dg_2,...,k_n^\dg)\qquad\for\;\;\vec
k=(k_2,...,k_n)\in\mbb{Z}^{n-1}.\eqno(4.4.26)$$ Set
$$e^{\pi (\vec k^\dg\cdot\vec x)i}=e^{\sum_{r=2}^n\pi
k^\dg_rx_ri}.\eqno(4.4.27)$$

For $r\in\ol{0,m-1}$, Lemma 4.3.7 with $T_0=\ptl_{x_1}$ and
$T_q=f_q(\ptl_{x_2},...,\ptl_{x_n})$ gives that
\begin{eqnarray*} & &
\frac{1}{r!}\sum_{\iota_1,...,\iota_m=0}^\infty
{\iota_1+\cdots+\iota_m\choose
\iota_1,...,\iota_m}\int_{(x_1)}^{(\sum_{s=1}^ms\iota_s)}(x_1^r)(\prod_{p=1}^mf_p(\ptl_{x_2},...,
\ptl_{x_n})^{\iota_p})(e^{\pi (\vec k^\dg\cdot\vec x)i})\\
&=&\sum_{\iota_1,...,\iota_m=0}^\infty
{\iota_1+\cdots+\iota_m\choose
\iota_1,...,\iota_m}\frac{x_1^{r+\sum_{s=1}^ms\iota_s}}{(r+\sum_{s=1}^ms\iota_s)!}
\\ & &\times\left[\prod_{p=1}^mf_p(k_2^\dg\pi i,...,k_n^\dg\pi i)^{\iota_p}\right]e^{\pi
(\vec k^\dg\cdot\vec x)i}\hspace{7.4cm}(4.4.28)\end{eqnarray*} is a
complex solution of the equation (4.4.24) for any $\vec
k\in\mbb{Z}^{\:n-1}$. We write
\begin{eqnarray*}& &\sum_{\iota_1,...,\iota_m=0}^\infty {\iota_1+\cdots+\iota_m\choose
\iota_1,...,\iota_m}\frac{x_1^r\prod_{p=1}^m(x_1^pf_p(k_2^\dg\pi
i,...,k_n^\dg\pi i)) ^{\iota_p}} {(r+\sum_{s=1}^ms\iota_s)!}
\\&=&\phi_r(x_1,\vec k)+\psi_r(x_1,\vec
k)i,\hspace{9.6cm}(4.4.29)\end{eqnarray*} where $\phi_r(x_1,\vec k)$
and $\psi_r(x_1,\vec k)$ are real functions. Moreover,
$$\ptl_{x_1}^s(\phi_r)(0,\vec k)=\dlt_{r,s},\;\;\ptl_{x_1}^s(\psi_r)(0,\vec
k)=0\qquad\for\;\;s\in\ol{0,r}.\eqno(4.4.30)$$ We define $\vec
0\prec \vec k$ if its first nonzero coordinate is a positive
integer. By superposition principle and Fourier expansions, we
get:\psp

{\bf Theorem 4.4.1}. {\it The solution of the equation (4.4.24)
subject to the condition (4.4.25) is:
\begin{eqnarray*}u&=&\sum_{r=0}^{m-1}\sum_{\vec 0\preceq\vec
k\in\mbb{Z}^{\:n-1}}[b_r(\vec k)(\phi_r(x_1,\vec k^\dg)\cos \pi(\vec
k^\dg\cdot\vec x)-\psi_r(x_1,\vec k^\dg)\sin \pi(\vec k^\dg\cdot\vec
x))\\ & &+c_r(\vec k)(\phi_r(x_1,\vec k^\dg)\sin \pi(\vec
k^\dg\cdot\vec x)+\psi_r(x_1,\vec k^\dg)\cos \pi(\vec k^\dg\cdot\vec
x))],\hspace{3.4cm}(4.4.31)\end{eqnarray*}with
\begin{eqnarray*}\hspace{1.6cm}b_r(\vec k)&=&\frac{1}{2^{n-2+\dlt_{\vec k,\vec 0}}a_2\cdots
a_n}\int_{-a_2}^{a_2}\cdots \int_{-a_n}^{a_n}g_r(x_2,...,x_n)\cos
\pi (\vec k^\dg\cdot\vec x)\:dx_n\cdots dx_2\\&
&-\sum_{s=0}^{r-1}(b_s(\vec k)\ptl_{x_1}^r(\phi_s)(0,\vec
k)+c_s(\vec k)\ptl_{x_1}^r(\psi_s)(0,\vec k))
\hspace{3.2cm}(4.4.32)\end{eqnarray*}
\begin{eqnarray*}\hspace{1cm}c_r(\vec k)&=&\frac{1}{2^{n-2}a_2\cdots
a_n}\int_{-a_2}^{a_2}\cdots \int_{-a_n}^{a_n}g_r(x_2,...,x_n)\sin
\pi (\vec k^\dg\cdot\vec x)\:dx_n\cdots dx_2\\
&&-\sum_{s=0}^{r-1}(c_s(\vec k) \ptl_{x_1}^r(\phi_s)(0,\vec
k)-b_s(\vec k) \ptl_{x_1}^r(\psi_s)(0,\vec
k)).\hspace{3.7cm}(4.4.33)\end{eqnarray*} The convergence of the
series (4.4.31) is guaranteed  by the Kovalevskaya Theorem on the
existence and uniqueness of the solution of linear partial
differential equations when the functions in (4.4.25) are analytic.}

\pse

{\bf Remark 4.4.2}. (1) If we take $f_\iota=b_\iota$ with
$\iota\in\ol{1,m}$ to be constant functions and $\vec k=\vec 0$ in
(4.4.29), we get $m$ fundamental solutions
$$\vf_r(x)=\sum_{\iota_1,...,\iota_m=0}^\infty {\iota_1+\cdots+\iota_m\choose
\iota_1,...,\iota_m}\frac{x^r\prod_{p=1}^m(b_px^p) ^{\iota_p}}
{(r+\sum_{s=1}^ms\iota_s)!},
 \qquad r\in\ol{0,m-1},\eqno(4.4.34)$$ of the constant-coefficient ordinary
differential equation
$$y^{(m)}-b_1y^{(m-1)}-\cdots-b_{m-1}y'-b_m=0.\eqno(4.4.35)$$
Given the initial conditions:
$$y^{(r)}(0)=c_r\qquad\for\;\;r\in\ol{0,m-1},\eqno(4.4.36)$$
we define $a_0=c_0$ and
$$a_r=c_r-\sum_{s=0}^{r-1}\sum_{\iota_1,...,\iota_{r-s}\in\mbb{N};\:\sum_{p=1}^rp\iota_p=r-s}{r-s\choose
\iota_1,...,\iota_{r-s}}a_sb_1^{\iota_1}\cdots
b_{r-s}^{\iota_{r-s}}\eqno(4.4.37)$$ by induction on
$r\in\ol{1,m-1}$. Now the solution of (4.4.35) subject to the
condition (4.4.36) is exactly
$$y=\sum_{r=0}^{m-1}a_r\vf_r(x).\eqno(4.4.38)$$
From the above results, it seems that the following functions
$${\cal Y}_r(y_1,...,y_m)=\sum_{\iota_1,...,\iota_m=0}^\infty {\iota_1+\cdots+\iota_m\choose
\iota_1,...,\iota_m}\frac{y_1 ^{\iota_1}y_2 ^{\iota_2}\cdots y_m
^{\iota_m}} {(r+\sum_{s=1}^ms\iota_s)!}\qquad\for\;\;r\in\mbb{N}
\eqno(4.4.39)$$\index{Special function ${\cal Y}_r(y_1,...,y_m)$}
are important natural functions. Indeed,
$${\cal Y}_0(x)=e^x,\;\;{\cal Y}_0(0,-x^2)=\cos x,\;\;{\cal
Y}_1(0,-x^2)=\frac{\sin x}{x},\eqno(4.4.40)$$
$$\vf_r(x)=x^r{\cal
Y}_r(b_1x,b_2x^2,...,b_mx^m)\eqno(4.4.41)$$ and $$\phi_r(x_1,\vec
x)+\psi_r(x_1,\vec x)i=x_1^r{\cal Y}_r(x_1f_1(k_2^\dg\pi
i,...,k_n^\dg\pi i)),..., x_1^mf_m(k_2^\dg\pi i,...,k_n^\dg\pi
i))\eqno(4.4.42)$$ for  $r\in\ol{0,m}$.

(2) We can solve the  initial value problem (4.4.24) and (4.4.25)
with the constant-coefficient differential operators
$f_\iota(\ptl_2,...,\ptl_n)$ replaced by variable-coefficient
differential operators
$\phi_\iota(\ptl_2,...,\ptl_{n_1})\psi_\iota(x_{n_1+1},...,x_n)$ for
some $2<n_1<n$ , where $\phi_\iota(\ptl_2,...,\ptl_{n_1})$ are
polynomials in $\ptl_2,...,\ptl_{n_1}$ and
$\psi_\iota(x_{n_1+1},...,x_n)$ are polynomials in
$x_{n_1+1},...,x_n$. \psp

{\bf Exercise 4.4} \psp

1. Solve the following heat conduction problem: $u_t=2u_{xx}$
subject to $u_x(t,0)=0,\;\; u_x(t,3)=0$ and $u(0,x)=2x-1.$

2. Find the solution of the wave equation $u_{tt}=3u_{xx}$ subject
to $u(t,0)=u(t,4)=0$ and $u(0,x)=2-x,\;\;u_t(0,x)=|x-2|.$

3. Find the solution of the equation
$$u_{xxx}-u_{xxy}-u_{xz}-u_{zz}=0$$
with $x\in\mbb{R}$ and $y,z\in [-2,2]$ subject to
$$u(0,y,z)=y+z,\;\;u_x(0,y,z)=y-z,\;\;u_{xx}(0,y,z)=yz.$$

\section{ Use of Fourier Expansion II}

In this section, we mainly use  Fourier expansion to solve the
evolution equations and generalized wave equations of flag type
subject to initial conditions. The results in this section are taken
from the author's work [X7].

Barros-Neto and Gel'fand [BG1,BG2] (1998, 2002) studied solutions of
the equation
$$u_{xx}+xu_{yy}=\dlt(x-x_0,y-y_0)\eqno(4.5.1)$$
related to the {\it Tricomi operator} $\ptl_x^2+x\ptl_y^2$.
\index{Tricomi operator} A natural generalization of the Tricomi
operator is
$\ptl_{x_1}^2+x_1\ptl_{x_2}^2+\cdots+x_{n-1}\ptl_{x_n}^2$. The
equation
$$u_t=u_{x_1x_1}+u_{x_2x_2}+\cdots+u_{x_nx_n}\eqno(4.5.2)$$
is a well known classical heat conduction equation related to the
Laplacian operator $\ptl_{x_1}^2+\ptl_{x_2}^2+\cdots+\ptl_{x_n}^2$.
As pointed out in [BG1, BG2], the Tricomi operator is an analogue of
the Laplacian operator. An immediate  analogue of heat conduction
equation is
$$u_t=u_{x_1x_1}+x_1u_{x_2x_2}+x_2u_{x_3x_3}+\cdots+x_{n-1}u_{x_nx_n}.\eqno(4.5.3)$$
Another related well-known equation is the wave equation
$$u_{tt}=u_{x_1x_1}+u_{x_2x_2}+\cdots+u_{x_nx_n}.\eqno(4.5.4)$$
Similarly, we have the following analogue of wave equation:
$$u_{tt}=u_{x_1x_1}+x_1u_{x_2x_2}+x_2u_{x_3x_3}+\cdots+x_{n-1}u_{x_nx_n}.\eqno(4.5.5)$$
The purpose of this section is to give the methods of solving linear
partial differential equations of the above types subject to initial
conditions.

 Graphically, the
above equation are related to the {\it Dynkin diagram}\index{Dynkin
diagram} of the special linear Lie algebra:

\setlength{\unitlength}{3pt}

\begin{picture}(70,8)\put(2,0){${\cal T}_{A_n}$:}
\put(21,0){\circle{2}}\put(22,0){\line(1,0){12}}\put(21,-5){1}
\put(35,0){\circle{2}}\put(35,-5){2} \put(36,0){\line(1,0){12}}
\put(49,0){\circle{2}}\put(49,-5){3}
\put(57,0){...}\put(67,0){\circle{2}} \put(67,-5){n-1}
\put(68,0){\line(1,0){12}}\put(81,0){\circle{2}}\put(81,-5){n}
\end{picture}
\vspace{0.8cm}

\noindent Naturally, we should also consider similar equations
related to the graph:

\begin{picture}(108,25)\put(2,0){${\cal T}_{E^{n_0}_{n_1,n_2}}$:}
\put(21,0){\circle{2}}\put(21,-5){1}\put(22,0){\line(1,0){12}}
\put(36,0){\circle{2}}\put(36,-5){2}
\put(44,0){...}\put(52,0){\circle{2}}\put(52,-5){$n_0-1$}\put(53,0){\line(1,0){12}}
\put(66,0){\circle{2}}\put(66,3){$n_0$}\put(66.6,0.5)
{\line(3,1){12}}\put(80,4.5){\circle{2}}\put(80,0){$n_0+1$}\put(84,5.8){.}
\put(86,6.46){.}\put(88,7.14){.}\put(92,8.5){\circle{2}}\put(92,4){$n_0+2n_1-3$}
\put(92.6,8.8){\line(3,1){13.6}}
\put(107,13.7){\circle{2}}\put(107,8.3){$n_0+2n_1-1$}
\put(66.6,-0.5)
{\line(3,-1){12}}\put(80,-4.5){\circle{2}}\put(72,-9){$n_0+2$}\put(84,-5.8){.}
\put(86,-6.46){.}\put(88,-7.14){.}\put(92,-8.5){\circle{2}}\put(74,-13.5){$n_0+2n_2-2$}
\put(92.6,-8.8){\line(3,-1){13.6}}
\put(107,-13.7){\circle{2}}\put(100,-18.7){$n_0+2n_2$}
\end{picture}
\vspace{2.5cm}

\noindent which is the Dynkin diagram of an orthogonal Lie algebra
$o(2n)$ when $n_1=n_2=1$, and the Dynkin diagram of the simple Lie
algebra of types $E_6,\;E_7,\;E_8$ if
$(n_0,n_1,n_2)=(3,1,2),\;(3,1,3),\;(3,1,4)$, respectively. When
$(n_0,n_1,n_2)=(3,2,2),\;(4,1,3),\;(6,1,2)$, it is also the Dynkin
diagram of the affine Kac-Moody Lie algebra of types
$E_6^{(1)},\;E_7^{(1)},\;E_8^{(1)}$,  respectively (cf. [Kv]). These
diagrams are special examples of trees in graph theory.

A {\it tree}\index{tree} ${\cal T}$ consists of a finite set of {\it
nodes} ${\cal N}= \{\iota_1,\iota_2,...,\iota_n\}$ and a set of {\it
edges}
$${\cal E}\subset\{(\iota_p,\iota_q)\mid 1\leq p<q\leq n\}\eqno(4.5.6)$$
such that for each node $\iota_q\in{\cal N}$, there exists a unique
sequence $\{\iota_{q_1},\iota_{q_2},...,\iota_{q_r}\}$ of nodes with
$1=q_1<q_2<\cdots<q_{r-1}<q_r=q$ for which
$$(\iota_{q_1},\iota_{q_2}),(\iota_{q_2},\iota_{q_3}),...,(\iota_{q_{r-2}},\iota_{q_{r-1}}),
(\iota_{q_{r-1}},\iota_{q_r})\in{\cal E}.\eqno(4.5.7)$$  We also
denote the tree ${\cal T}=({\cal N},{\cal E})$. We  identify a tree
${\cal T}=({\cal N},{\cal E})$ with a graph by depicting a small
circle for each node in ${\cal N}$ and a segment connecting $r$th
circle to $j$th circle for the edge $(\iota_r,\iota_j)\in{\cal E}$
(cf. the above dynkin diagrams of type $A$ and $E$).

 For a tree ${\cal T}=({\cal N},{\cal E})$, we call the
differential operator
$$d_{\cal T}=\ptl_{x_1}^2+\sum_{(\iota_p,\iota_q)\in{\cal
E}}x_p\ptl_{x_q}^2\eqno(4.5.8)$$ a {\it generalized Tricomi operator
of type}\index{Tricomi operator!generalized} ${\cal T}$. Moreover,
we call the partial differential equation
$$u_t=d_{\cal T}(u)\eqno(4.5.9)$$
a {\it generalized heat conduction  equation associated with the
tree} ${\cal T}$, \index{heat conduction equation!generalized} where
$u$ is a function in $t,x_1,x_2,...,x_n$. For instance, the
generalized heat equation of type ${\cal T}_{E^{n_0}_{n_1,n_2}}$ is:
\begin{eqnarray*}\hspace{1.5cm}u_t&=&(\ptl_{x_1}^2+\sum_{q=1}^{n_0-1}x_q\ptl_{x_{q+1}}^2
+\sum_{r=0}^{n_2-1}x_{n_0+2r}\ptl_{x_{n_0+2r+2}}^2\\ &&
+x_{n_0}\ptl_{x_{n_0+1}}^2 +\sum_{p=1}^{n_1-1}x_{n_0+2p-1}
\ptl_{x_{n_0+2p+1}}^2)(u).\hspace{4.7cm}(4.5.10)\end{eqnarray*}
Similarly, we have the generalized wave equation associated with the
tree ${\cal T}$:
$$u_{tt}=d_{\cal T}(u)\eqno(4.5.11)$$

Let $m_0,m_1,m_2,...,m_n$ be $n+1$ positive integers. The difficulty
of solving the equations (4.5.9) and (4.5.10) is the same as that of
solving the following more general partial differential equation:
$$\ptl^{m_0}_t(u)=(\ptl_{x_1}^{m_1}+\sum_{(\iota_p,\iota_q)\in{\cal
E}}x_p\ptl_{x_q}^{m_q})(u).\eqno(4.5.12)$$ Obviously, we want to use
the operator $\sum_{\iota=0}^\infty(-T_1^-T_2)^\iota$ in Lemma
4.3.1. Then the main difficulty turns out to be how to calculate the
powers of the operator
$\ptl_{x_1}^{m_1}+\sum_{(\iota_p,\iota_q)\in{\cal
E}}x_p\ptl_{x_q}^{m_j}$. This essentially involves the
Campbell-Hausdorff formula, whose simplest nontrivial case
$e^{t(\ptl_{x_1}+x_1\ptl_{x_2})}=e^{tx_1\ptl_{x_2}}e^{t\ptl_{x_1}}e^{t\ptl_{x_2}/2}$
has been extensively used by physicists. \psp

{\bf Lemma 4.5.1}. {\it Let $f(x)$ be a smooth function and let $b$
be a constant. Then}
$$e^{b\frac{d}{dx}}(f(x))=f(x+b).\eqno(4.5.13)$$

{\it Proof}. Note
$$e^{b\frac{d}{dx}}(f(x))=\sum_{n=0}^\infty\frac{f^{(n)}(x)}{n!}b^n=f(x+b)
\eqno(4.5.14)$$ by Tayler's expansion.$\qquad\Box$\psp

For $n-1$ positive integers $m_1,m_2,...,m_{n-1}$, we denote
$$D=t(\ptl_{x_1}+x_1^{m_1}\ptl_{x_2}+x_2^{m_2}\ptl_{x_3}+\cdots+
x_{n-1}^{m_{n-1}}\ptl_{x_n})\eqno(4.5.15)$$ and set $\eta_1=t$,
$$\eta_\iota=\int_0^t(x_{\iota-1}+\int^{y_{\iota-1}}_0(x_{\iota-2}+...+\int_0^{y_2}(x_1+y_1)^{m_1}dy_1...)^{m_{\iota-2}}dy_{\iota-2})^{m_{\iota-1}
}dy_{\iota-1}\eqno(4.5.16)$$ for $\iota\in\ol{2,n}$.\psp

{\bf Lemma 4.5.2}. {\it We have the following
Campbell-Hausdorff-type factorizaton}:
$$e^D=e^{\eta_n\ptl_{x_n}}e^{\eta_{n-1}\ptl_{x_{n-1}}}\cdots
e^{\eta_1\ptl_{x_1}}.\eqno(4.5.17)$$\index{Campbell-Hausdorff-type
factorization}

{\it Proof}. Let $f(x_1,x_2,...,x_n)$ be any given smooth function.
We want to solve the equation
$$u_t-u_{x_1}-x_1^{m_1}u_{x_2}-x_2^{m_2}u_{x_3}-\cdots
-x_{n-1}^{m_{n-1}}u_{x_n}=0\eqno(4.5.18)$$ subject to
$u(0,x_1,...,x_n)=f(x_1,...,x_n)$. According to the method of
characteristic lines in Section 4.1, we solve the following problem:
$$\frac{dt}{ds}=1,\;\;\frac{dx_1}{ds}=-1,\;\;\frac{dx_{r+1}}{ds}=-x_r^{m_r},\;\;r\in\ol{1,n-1},\;\;\frac{d u}{ds}=0,\eqno(4.5.19)$$
subject to
$$t|_{s=0}=0,\;x_p|_{s=0}=t_p,\;\;p\in\ol{1,n},\;\;u|_{s=0}=t_{n+1},\;\;t_{n+1}=f(t_1,...,t_n).\eqno(4.5.20)$$
We find
$$u=t_{n+1},\;\;t=s,\;\;x_1=-s+t_1,\;\;x_2=t_2-\int_0^s(t_1-s_1)^{m_1}ds_1,\eqno(4.5.21)$$
$$x_3 =t_3-\int_0^s(t_2-\int_0^{s_2}(t_1-s_1)^{m_1}ds_1)^{m_2}ds_2,...,\eqno(4.5.22)$$
$$x_{r+1}=t_{r+1}-\int_0^s(t_r-\int_0^{s_r}(t_{r-1}-\cdots-\int_0^{s_2}(t_1-s_1)^{m_1}
ds_1\cdots)^{m_{r-1}}ds_{r-1})^{m_r}ds_r.\eqno(4.5.23)$$ Note that
$t_1=x_1+t=x_1+\eta_1,$
\begin{eqnarray*}\qquad t_2&=&x_2+\int_0^s(t_1-s_1)^{m_1}ds_1=
x_2+\int_0^t(x_1+t-s_1)^{m_1}ds_1\\
&\stl{y_1=t-s_1}{=}&x_2-\int_t^0(x_1+y_1)^{m_1}dy_1=
x_2+\int_0^t(x_1+y_1)^{m_1}dy_1=x_2+\eta_2,\hspace{1cm}(4.5.24)\end{eqnarray*}
\begin{eqnarray*}\qquad t_3&=&x_3+\int_0^t(t_2-\int_0^{s_2}(t_1-s_1)^{m_1}ds_1)^{m_2}ds_2
\\ &=&x_3+\int_0^t(x_2+\int_0^t(x_1+y_1)^{m_1}dy_1-\int_0^{s_2}(x_1+t-s_1)^{m_1}ds_1)^{m_2}ds_2
\\&=&x_3+\int_0^t(x_2+\int_0^t(x_1+y_1)^{m_1}dy_1+\int_t^{t-s_2}(x_1+y_1)^{m_1}dy_1)^{m_2}ds_2
\\&=&x_3+\int_0^t(x_2+\int_0^{t-s_2}(x_1+y_1)^{m_1}dy_1)^{m_2}ds_2
\\&\stl{y_2=t-s_2}{=}&x_3-\int_t^0(x_2+\int_0^{y_2}(x_1+y_1)^{m_1}dy_1)^{m_2}dy_2
\\
&=&x_3+\int_0^t(x_2+\int_0^{y_2}(x_1+y_1)^{m_1}dy_1)^{m_2}dy_2=x_3+\eta_3.\hspace{2.5cm}(4.5.25)
\end{eqnarray*}
This gives us a pattern of find general $t_p$. In the above, we have
also proved
$$t_2-\int_0^{s_2}(t_1-s_1)^{m_1}ds_1=x_2+\int_0^{t-s_2}(x_1+y_1)^{m_1}dy_1.\eqno(4.5.26)$$
Suppose that $t_r=x_r+\eta_r$ and
\begin{eqnarray*}\qquad& &t_{r-1}-\int_0^{s_{r-1}}(t_{r-2}-\cdots-\int_0^{s_2}(t_1-s_1)^{m_1}
ds_1\cdots)^{m_{r-2}}ds_{r-2}\\
&=&x_{r-1}+\int_0^{t-s_{r-1}}(x_{r-2}+\cdots+\int_0^{y_2}(x_1+y_1)^{m_1}dy_1...)
^{m_{r-2}}dy_{r-2}.\hspace{2cm}(4.5.27)
\end{eqnarray*}
Then we have
\begin{eqnarray*}&
&t_r-\int_0^{s_r}(t_{r-1}-\cdots-\int_0^{s_2}(t_1-s_1)^{m_1}
ds_1\cdots)^{m_{r-1}}ds_{r-1}
\\ &=&x_r+\int_0^t(x_{r-1}+\int^{y_{r-1}}_0(x_{r-2}+...+\int_0^{y_2}(x_1+y_1)^{m_1}dy_1...)^{m_{r-2}}dy_{r-2})^{m_{r-1}
}dy_{r-1}
\\ & &-\int_0^{s_r}(x_{r-1}+\int_0^{t-s_{r-1}}(x_{r-2}+\cdots+\int_0^{y_2}(x_1+y_1)^{m_1}dy_1...)
^{m_{r-2}}dy_{r-2})d_{s_{r-1}}\\
&=&x_r+\int_0^t(x_{r-1}+\int^{y_{r-1}}_0(x_{r-2}+...+\int_0^{y_2}(x_1+y_1)^{m_1}dy_1...)^{m_{r-2}}dy_{r-2})^{m_{r-1}
}dy_{r-1}
\\ & &+\int_t^{t-s_r}(x_{r-1}+\int_0^{y_{r-1}}(x_{r-2}+\cdots+\int_0^{y_2}(x_1+y_1)^{m_1}dy_1...)
^{m_{r-2}}dy_{r-2})dy_{r-1}
\\&=&x_r+\int_0^{t-s_r}(x_{r-1}+...+\int_0^{y_2}(x_1+y_1)^{m_1}dy_1...)^{m_{r-1}
}dy_{r-1}\hspace{3.6cm}(4.5.28)
\end{eqnarray*}
\begin{eqnarray*}t_{r+1}&=&x_{r+1}+\int_0^t(t_r-\int_0^{s_r}(t_{r-1}-\cdots-\int_0^{s_2}(t_1-s_1)^{m_1}
ds_1\cdots)^{m_{r-1}}ds_{r-1})^{m_r}ds_r
\\ &=&x_{r+1}+\int_0^t(x_r+\int_0^{t-s_r}(x_{r-1}+...+\int_0^{y_2}(x_1+y_1)^{m_1}dy_1...)^{m_{r-1}
}dy_{r-1} )^{m_r}ds_r
\\ &\stl{y=t-s_r}{=}&x_{r+1}-\int_t^0(x_r+\int_0^{y_r}(x_{r-1}+...+\int_0^{y_2}(x_1+y_1)^{m_1}dy_1...)^{m_{r-1}
}dy_{r-1} )^{m_r}dy_r\\
&=&x_{r+1}+\int_0^t(x_r+\int_0^{y_r}(x_{r-1}+...+\int_0^{y_2}(x_1+y_1)^{m_1}dy_1...)^{m_{r-1}
}dy_{r-1} )^{m_r}dy_r\\
&=&x_{r+1}+\eta_{r+1}.\hspace{9.9cm}(4.5.29)\end{eqnarray*} By
induction, we have
$$t_p=x_p+\eta_p\qquad\for\;\;p\in\ol{1,n}.\eqno(4.5.30)$$
Thus the solution
\begin{eqnarray*}\qquad u&=&t_{n+1}=f(t_1,t_2,...,t_n)\\
&=&f(x_1+\eta_1,x_2+\eta_3,...,x_n+\eta_n)\\&=&e^{\eta_n\ptl_{x_n}}e^{\eta_{n-1}\ptl_{x_{n-1}}}\cdots
e^{\eta_1\ptl_{x_1}}(f(x_1,...,x_{n-1},x_n)).\hspace{4.6cm}(4.5.31)\end{eqnarray*}

According to (4.5.15),
$$\ptl_t(e^{D}(f))=(\ptl_{x_1}+x_1^{m_1}\ptl_{x_2}+x_2^{m_2}\ptl_{x_3}+\cdots+
x_{n-1}^{m_{n-1}}\ptl_{x_n})e^{D}(f)\eqno(4.5.32)$$ and
$$e^{D}(f)|_{t=0}=e^0(f)=f.\eqno(4.5.33)$$
Hence
$$e^{D}(f)=u=e^{\eta_n\ptl_{x_n}}e^{\eta_{n-1}\ptl_{x_{n-1}}}\cdots
e^{\eta_1\ptl_{x_1}}(f).\eqno(4.5.34)$$ Since $f$ is an arbitrary
smooth function in $x_1,x_2,...,x_n$, (4.5.34) implies (4.5.17).
$\qquad\Box$\psp

We remark that the above lemma was proved pure algebraically in [X7]
by the Campbell-Hausdorff formula. The above result can be
generalized as follows. Recall the definition of a tree given in the
paragraph of (4.5.6) and (4.5.7).   We define a tree diagram ${\cal
T}^d$ to be a tree ${\cal T}=({\cal N},{\cal E})$ with a weight map
$d:{\cal E}\rta \mbb{N}+1$, denoted as ${\cal T}^d=({\cal N},{\cal
E},d)$. Set
$$D_{{\cal T}^d}=t(\ptl_{x_1}+\sum_{(\iota_p,\iota_q)\in{\cal
E}}x_p^{d[(\iota_p,\iota_q)]}\ptl_{x_q}).\eqno(4.5.35)$$ In order to
factorize $e^{D_{{\cal T}^d}}$, we need a new notion. For a node
$\iota_q$ in a tree ${\cal T}$, the unique sequence
$${\cal C}_q=\{\iota_{q_1},\iota_{q_2},...,\iota_{q_r}\}\eqno(4.5.36)$$ of nodes with
$1=q_1<q_2<\cdots<q_{r-1}<q_r=q$ satisfying
$(\iota_{q_k},\iota_{q_{k+1}})\in{\cal E}$ for $k\in\ol{1,r-1}$ is
called the {\it clan} of the node $\iota_q$.

Again we define $\eta_1^{{\cal T}^d}=t$. For any $q\in\ol{2,n}$ with
the clan (4.5.36), we define
\begin{eqnarray*}\hspace{2cm}\eta_q^{{\cal T}^d}&=&\int_0^t(x_{q_{r-1}}+\int^{y_{q_{r-1}}}_0(x_{q_{r-2}}+...
+\int_0^{y_{q_2}} (x_1+y_1)^{d[(\iota_{q_1},\iota_{q_2})]}\\
&&dy_1...
)^{d[(\iota_{q_{r-2}},\iota_{q_{r-1}})]}dy_{q_{r-2}})^{d[(\iota_{q_{r-1}},\iota_{q_r})]}
dy_{q_{r-1}}.\hspace{3.6cm}(4.5.37)\end{eqnarray*} \psp

{\bf Corollary 4.5.3}. {\it For a tree diagram ${\cal T}^d$ with $n$
nodes, we have
$$e^{D_{{\cal T}^d}}=e^{\eta_n^{{\cal T}^d}\ptl_{x_n}}e^{\eta_{n-1}^{{\cal T}^d}\ptl_{x_{n-1}}}\cdots
e^{\eta_1^{{\cal T}^d}\ptl_{x_1}}.\eqno(4.5.38)$$ In particular,
$u=g(x_1+\eta_1^{{\cal T}^d},x_2+\eta_2^{{\cal
T}^d},...,x_n+\eta_n^{{\cal T}^d})$ is the solution of the evolution
equation
$$u_t=(\ptl_{x_1}+\sum_{(\iota_p,\iota_q)\in{\cal
E}}x_p^{d[(\iota_p,\iota_q)]}\ptl_{x_q})(u)\eqno(4.5.39)$$ subject
to $u(0,x_1,...,x_n)=g(x_1,...,x_n).$}\psp

 Since $d:{\cal E}\rta \mbb{N}+1$ is an arbitrary
map, we can solve more general problem of replacing monomial
functions by any first-order differentiable functions. Let $\vec
h=\{h_{p,q}(x)\mid (\iota_p,\iota_q)\in{\cal E}\}$ be a set of
first-order differentiable functions. Suppose ${\cal
C}_q=\{\iota_{q_1},\iota_{q_2},...,\iota_{q_r}\}$. We define
\begin{eqnarray*}\qquad\qquad\eta_q^{\vec h}&=&\int_0^th_{q_{r-1},q_r}(x_{q_{r-1}}+
\int^{y_{q_{r-1}}}_0h_{q_{r-2},q_{r-1}}(x_{q_{r-2}}\\
&&+...+\int_0^{y_{q_2}}h_{q_1,q_2}
(x_1+y_1)dy_{q_1}...)dy_{q_{r-2}})
dy_{q_{r-1}}.\hspace{3.2cm}(4.5.40)\end{eqnarray*} Set
$$D_{\vec h}=t(\ptl_{x_1}+\sum_{(\iota_p,\iota_q)\in{\cal
E}}h_{p,q}(x_p)\ptl_{x_q}).\eqno(4.5.41)$$\pse

{\bf Corollary 4.5.4}. {\it We have the factorization:
$$e^{D_{\vec h}}=e^{\eta_n^{\vec h}\ptl_{x_n}}e^{\eta_{n-1}^{\vec h}\ptl_{x_{n-1}}}\cdots
e^{\eta_1^{\vec h}\ptl_{x_1}}.\eqno(4.5.42)$$ In particular,
$u=g(x_1+\eta_1^{\vec h},x_2+\eta_2^{\vec h},...,x_n+\eta_n^{\vec
h})$ is the solution of the evolution equation
$$u_t=(\ptl_{x_1}+\sum_{(\iota_p,\iota_q)\in{\cal
E}}h_{p,q}(x_p)\ptl_{x_q})(u)\eqno(4.5.43)$$ subject to the
condition $u(0,x_1,...,x_n)=g(x_1,...,x_n).$ }\psp

Next we consider
$$\hat D=t(\ptl_{x_1}^{m_1}+x_1\ptl_{x_2}^{m_2}+\cdots
+x_{n-1}\ptl_{x_n}^{m_n}).\eqno(4.5.44)$$ To study the factorization
of $e^{\hat D}$, we need the following preparations. Denote
$\hat{\cal A}=\mbb{R}[x_0,x_1,...,x_n]$. We denote
$$x^\al=\prod_{r=0}^nx_r^{\al_r},\qquad \ptl^\al=\prod_{r=0}^n
\ptl_{x_r}^{\al_r}\qquad\for\;\;\al=(\al_0,\al_1,...,\al_n)\in\mbb{N}^{n+1}.\eqno(4.5.45)$$
For $\al,\be\in\mbb{N}^{n+1}$, we define
$$\be\preceq\al\qquad
\mbox{if}\;\;\be_r\leq\al_r\;\;\for\;\;r\in\ol{0,n}\eqno(4.5.46)$$
and in this case,
$${\al\choose \be}=\prod_{r=0}^n{\al_r\choose
\be_r},\;\;\gm!=\prod_{r=0}^n\gm_r!\qquad\for\;\;\be\in
\mbb{N}^{n+1}.\eqno(4.5.47)$$ Set
$$\mbb{A}=\mbox{Span}\:\{x^\al\ptl^\be\mid\al,\be\in\mbb{N}^{n+1}\},\eqno(4.5.48)$$
the space of all algebraic differential operators on $\hat{\cal A}$.
For $T_1,T_2\in\mbb{A}$, the multiplication $T_2\cdot T_2$ is
defined by
$$(T_1\cdot T_2)(f)=T_1(T_2(f))\qquad\for\;\;f\in\hat{\cal
A}.\eqno(4.5.49)$$ Note that for $f,g_1,g_2\in\hat{\cal A}$ and
$\al,\be\in\mbb{N}^{n+1}$,
$$(g_1\ptl^\al\cdot
g_2\ptl^\be)(f)=\sum_{\al\succeq\gm\in\mbb{N}^{n+1}}{\al\choose
\gm}g_1\ptl^\gm(g_2)\ptl^{\be+\gm}(f). \eqno(4.5.50)$$ Thus
$$g_1\ptl^\al\cdot
g_2\ptl^\be=\sum_{\al\succeq\gm\in\mbb{N}^{n+1}}{\al\choose
\gm}g_1\ptl^\gm(g_2)\ptl^{\al+\be-\gm}.\eqno(4.5.51)$$ So
$(\mbb{A},\cdot)$ forms an associative algebra.

Define a linear transformation $\tau:\mbb{A}\rta \mbb{A}$ by
$$\tau(x^\al\ptl^\be)=x^\be\ptl^\al\qquad\for\;\;\al,\be\in\mbb{N}^{n+1}.\eqno(4.5.52)$$
\pse

{\bf Lemma 4.5.5}. {\it We have $\tau(T_1\cdot T_2)=\tau(T_2)\cdot
\tau(T_1)$.}

{\it Proof}. For $\al,\al',\be,\be'\in\mbb{N}^{n+1}$, we have
\begin{eqnarray*}\qquad\tau(x^\al\ptl^{\al'}\cdot
x^\be\ptl^{\be'})&=&\sum_{\al'\succeq\gm\in\mbb{N}^{n+1}}\gm!{\al'\choose
\gm}{\be\choose \gm}\tau(x^{\al+\be-\gm}\ptl^{\al'+\be'-\gm})
\\&=&\sum_{\be\succeq\gm\in\mbb{N}^{n+1}}\gm!{\be\choose
\gm}{\al'\choose \gm}x^{\al'+\be'-\gm}\ptl^{\al+\be-\gm}
\\&=& x^{\be'}\ptl^\be\cdot x^{\al'}\ptl^\al=\tau(
x^\be\ptl^{\be'})\cdot
\tau(x^\al\ptl^{\al'}).\qquad\Box\hspace{2.5cm}(4.5.53)
\end{eqnarray*}

Denote
$$\td D=t(x_1^{m_1}\ptl_{x_0}+x_2^{m_2}\ptl_{x_1}+\cdots
+x_{n-1}^{m_{n-1}}\ptl_{x_{n-2}}+x_n^{m_n}\ptl_{x_{n-1}}).\eqno(4.5.54)$$
Changing variables
$$z_r=\frac{x_r}{x_n^{\prod_{p=r+1}^nm_p}}\qquad\for\;\;r\in\ol{0,n-1}.\eqno(4.5.55)$$
Then
$$\ptl_{z_r}=x_n^{\prod_{p=r+1}^nm_p}\ptl_{x_r}\qquad\for\;\;r\in\ol{0,n-1}.\eqno(4.5.56)$$
Moreover,
$$\td D=t(\ptl_{z_{n-1}}+z_{n-1}^{m_{n-1}}\ptl_{z_{n-2}}+\cdots
+z_2^{m_2}\ptl_{z_1}+z_1^{m_1}\ptl_{z_0}).\eqno(4.5.57)$$ According
to (4.5.15)-(4.3.17), we define $\td\eta_{n-1}=t$ and
$$\td\eta_r=\int_0^t(z_{r+1}+\int^{y_{r+1}}_0(z_{r+2}+...+\int_0^{y_{n-2}}(z_{n-1}+y_{n-1})^{m_{n-1}}
dy_{n-1}...)^{m_{r+2}}dy_{r+2})^{m_{r+1} }dy_{r+1}\eqno(4.5.58)$$
for $r\in\ol{0,n-2}$. By Lemma 4.5.2,
$$e^{\td D}=e^{\td \eta_0\ptl_{z_0}}e^{\td \eta_1\ptl_{z_1}}\cdots
e^{\td\eta_{n-1}\ptl_{z_{n-1}}}.\eqno(4.5.59)$$ Note
\begin{eqnarray*} \eta^\ast_r&=&x_n^{\prod_{p=r+1}^nm_p}\td\eta_r=
\int_0^t(x_n^{\prod_{p=r+2}^nm_p}z_{r+1}+\int^{y_{r+1}}_0(x_n^{\prod_{p=r+2}^nm_p}z_{r+2}+\\
& &...+
\int_0^{y_{n-2}}(x_n^{m_n}z_{n-1}+x_n^{m_n}y_{n-1})^{m_{n-1}}
dy_{n-1}...)^{m_{r+2}}dy_{r+2})^{m_{r+1} }dy_{r+1}
\\&=&\int_0^t(x_{r+1}+\int^{y_{r+1}}_0(x_{r+2}+...+
\int_0^{y_{n-2}}(x_{n-1}\\ & &+x_n^{m_n}y_{n-1})^{m_{n-1}}
dy_{n-1}...)^{m_{r+2}}dy_{r+2})^{m_{r+1}
}dy_{r+1}\hspace{5.35cm}(4.5.60)\end{eqnarray*} for $r\in\ol{0,n-2}$
and let $\eta_{n-1}^\ast=tx_n^{m_n}$. By (4.5.56), we find
$$e^{\td D}=e^{\eta_0^\ast\ptl_{x_0}}e^{\eta_1^\ast\ptl_{x_1}}\cdots
e^{\eta_{n-1}^\ast\ptl_{x_{n-1}}}.\eqno(4.5.61)$$

According to Lemma 4.5.5,
\begin{eqnarray*}\qquad\qquad& &e^{t(x_0\ptl_{x_1}^{m_1}+x_1\ptl_{x_2}^{m_2}+\cdots
+x_{n-1}\ptl_{x_n}^{m_n})}\\ &=&e^{\tau(\td D)}=\tau(e^{\td D})=
\tau[e^{\eta_0^\ast\ptl_{x_0}}e^{\eta_1^\ast\ptl_{x_1}}\cdots
e^{\eta_{n-1}^\ast\ptl_{x_{n-1}}}]
\\ &=&e^{\tau(\eta_{n-1}^\ast\ptl_{x_{n-1}})}\cdots e^{\tau(\eta_1^\ast\ptl_{x_1})}e^{\tau(\eta_0^\ast\ptl_{x_0})}
\\&=&e^{x_{n-1}\tau(\eta_{n-1}^\ast)}\cdots
e^{x_1\tau(\eta_1^\ast))}e^{x_0\tau(\eta_0^\ast)}.
\hspace{6.7cm}(4.5.62)\end{eqnarray*} Denote
$\hat\eta_{n-1}=\tau(\eta^\ast_{n-1})=t\ptl_{x_n}^{m_n}$ and
\begin{eqnarray*}\hat\eta_r=\tau(\eta^\ast_r)&=&\int_0^t(\ptl_{x_{r+1}}+\int^{y_{r+1}}_0(\ptl_{x_{r+2}}+...+
\int_0^{y_{n-2}}(\ptl_{x_{n-1}}\\ &
&+\ptl_{x_n}^{m_n}y_{n-1})^{m_{n-1}}
dy_{n-1}...)^{m_{r+2}}dy_{r+2})^{m_{r+1}
}dy_{r+1}\hspace{3.9cm}(4.5.63)\end{eqnarray*} for $r\in\ol{0,n-2}$.
\psp

{\bf Theorem 4.5.6}. {\it We have the following factorization:}
$$e^{\hat D}=e^{t(\ptl_{x_1}^{m_1}+x_1\ptl_{x_2}^{m_2}+\cdots
+x_{n-1}\ptl_{x_n}^{m_n})}=e^{x_{n-1}\hat\eta_{n-1}}\cdots
e^{x_1\hat\eta_1}e^{\hat\eta_0}.\eqno(4.5.64)$$\pse

Next we want to solve the evolution equation
$$u_t=(\ptl_{x_1}^{m_1}+x_1\ptl_{x_2}^{m_2}+\cdots
+x_{n-1}\ptl_{x_n}^{m_n})(u)\eqno(4.5.65)$$ subject to the initial
condition:
$$u(0,x_1,...,x_n)=f(x_1,x_2,...,x_n)\qquad\for\;\;x_r\in[-a_r,a_r],\eqno(4.5.66)$$
where $f$ is a continuous function in $x_1,...,x_n$. For
convenience, we denote
$$k^\dg_r=\frac{k_r}{a_r},\;\;\vec
k^\dg=(k^\dg_1,...,k_n^\dg)\qquad\for\;\;\vec
k=(k_1,...,k_n)\in\mbb{Z}^{\:n}.\eqno(4.5.67)$$ Set
$$e^{\pi (\vec k^\dg\cdot\vec x)i}=e^{\sum_{r=1}^n\pi
k^\dg_rx_ri}.\eqno(4.5.68)$$ Note that $\hat\eta_r$ is a polynomial
in $t,\ptl_{x_{r+1}},...,\ptl_{x_n}$. So we denote
$$\hat\eta_r
=\hat\eta_r(t,\ptl_{x_{r+1}},...,\ptl_{x_n}).\eqno(4.5.69)$$

Observe that
$$e^{\hat D}(e^{\pi (\vec k^\dg\cdot\vec x)i})
=e^{x_{n-1}\eta_{n-1}(t,\pi k^\dg_ni)}\cdots e^{x_1\hat\eta(t,\pi
k^\dg_2i,...,\pi k^\dg_ni)} e^{\hat\eta(t,\pi k^\dg_1i,...,\pi
k^\dg_ni)}e^{\pi (\vec k^\dg\cdot\vec x)i}\eqno(4.5.70)$$ is a
solution of (4.5.65) for any $\vec k=(k_1,...,k_n)\in\mbb{Z}^{\:n}$.
Denote the right hand of (4.5.70) as $\phi_{\vec
k}(t,x_1,...,x_n)+\psi_{\vec k}(t,x_1,...,x_n)i$, where $\phi_{\vec
k}$ and $\psi_{\vec k}$ are real-valued functions. Then
$$\phi_{\vec k}(0,x_1,...,x_n)=\cos \pi (\vec
k^\dg\cdot\vec x),\;\;\psi_{\vec k}(0,x_1,...,x_n)=\sin \pi (\vec
k^\dg\cdot\vec x).\eqno(4.5.71)$$ We define $0\prec \vec k$ if its
first nonzero coordinate is a positive integer.

By Fourier expansion theory, we get:\psp

{\bf Theorem 4.5.7}. {\it The solution of the equation (4.5.65)
subject to (4.5.66) is
$$u=\sum_{0\preceq\vec k\in\mbb{Z}^{\:n}}(b_{\vec k}\phi_{\vec
k}(t,x_1,...,x_n)+c_{\vec k}\psi_{\vec
k}(t,x_1,...,x_n))\eqno(4.5.72)$$ with
$$b_{\vec k}=\frac{1}{2^{n-1+\dlt_{\vec k,\vec 0}}a_1a_2\cdots a_n}\int_{-a_1}^{a_1}\cdots
\int_{-a_n}^{a_n}f(x_1,...,x_n)\cos \pi (\vec k^\dg\cdot\vec
x)\:dx_n\cdots dx_1\eqno(4.5.73)$$ and}
$$c_{\vec k}=\frac{1}{2^{n-1}a_1a_2\cdots a_n}\int_{-a_1}^{a_1}\cdots
\int_{-a_n}^{a_n}f(x_1,...,x_n)\sin \pi (\vec k^\dg\cdot\vec
x)\:dx_n\cdots dx_1.\eqno(4.5.74)$$ \vspace{0.1cm}

{\bf Example 4.5.1}. Consider the case $n=2$, $m_1=m_2=2$ and
$a_1=a_2=\pi$. So the problem becomes
$$u_t=u_{x_1x_1}+x_1u_{x_2x_2}\eqno(4.5.75)$$
subject to
$$u(0,x_1,x_2)=f(x_1,x_2)\qquad\for\;\;x_1,x_2\in[-\pi,\pi].\eqno(4.5.76)$$
In this case,
\begin{eqnarray*}\hspace{2cm}\hat\eta_0(t,\ptl_{x_1},\ptl_{x_2})&=&
\int_0^t(\ptl_{x_1}+y_1\ptl^2_{x_2})^2dy_1\\
&=&\int_0^t(\ptl_{x_1}^2+2y_1\ptl_{x_1}\ptl^2_{x_2}+y_1^2\ptl_{x_2}^4)dy_1\\
&=&t\ptl_{x_1}^2+t^2\ptl_{x_1}\ptl^2_{x_2}+\frac{t^3\ptl_{x_2}^4}{3}
\hspace{5.1cm}(4.5.77)\end{eqnarray*} and
$\hat\eta_1(t,\ptl_{x_2})=t\ptl_{x_2}^2$. Thus
\begin{eqnarray*}\hspace{2cm}&
&e^{tx_1\ptl_{x_2}^2}e^{t\ptl_{x_1}^2+t^2\ptl_{x_1}\ptl^2_{x_2}+t^3\ptl_{x_2}^4/3}(e^{(k_1
x_1+k_2 x_2)i})\\
&=&e^{k_2^4t^3/3-k_2^2tx_1-k_1^2t}e^{(k_1 x_1+k_2 x_2-k_1k_2^2t^2)i}
.\hspace{5.6cm}(4.5.78)\end{eqnarray*} Hence $$\phi_{\vec
k}(t,x_1,x_2)=e^{k_2^4t^3/3-k_2^2tx_1-k_1^2t}\cos(k_1 x_1+k_2
x_2-k_1k_2^2t^2),\eqno(4.5.79)$$
$$\psi_{\vec
k}(t,x_1,x_2)=e^{k_2^4t^3/3-k_2^2tx_1-k_1^2t}\sin(k_1 x_1+k_2
x_2-k_1k_2^2t^2).\eqno(4.5.80)$$ The final solution of (4.5.75) and
(4.5.76) is \begin{eqnarray*}
u&=&\frac{1}{4\pi^2}\int_{-\pi}^\pi\int_{-\pi}^\pi
f(s_1,s_2)d_{s_1}d_{s_2}+\frac{1}{2\pi^2}
\sum_{0\prec(k_1,k_2)\in\mbb{Z}^2}
e^{k_2^4t^3/3-k_2^2tx_1-k_1^2t}\int_{-\pi}^\pi\int_{-\pi}^\pi
f(s_1,s_2)\\
& &\times
\cos[k_1(x_1-s_1)+k_2(x_2-s_2)-k_1k_2^2t^2]\:ds_1ds_2.\hspace{4.8cm}(4.5.81)\end{eqnarray*}
 \pse

 Theorem 4.5.6 gives a way of how to calculate the powers
of $\ptl_{x_1}^{m_1}+x_1\ptl_{x_2}^{m_2}+\cdots
+x_{n-1}\ptl_{x_n}^{m_n}$. Then we can use the powers to solve the
equation
$$\ptl_t^{m_0}(u)=(\ptl_{x_1}^{m_1}+x_1\ptl_{x_2}^{m_2}+\cdots
+x_{n-1}\ptl_{x_n}^{m_n})(u).\eqno(4.5.82)$$\psp

 {\bf Example
4.5.2}. Find the solution of the problem
$$u_{tt}=u_{x_1x_1}+x_1u_{x_2x_2}\eqno(4.5.83)$$
subject to
$$u(0,x_1,x_2)=f_1(x_1,x_2),\;\;u_t(0,x_1,x_2)=f_2(x_1,x_2)\qquad\for\;\;x_1,x_2\in[-\pi,\pi].\eqno(4.5.84)$$

{\it Solution}. According to the above example,
$$(\ptl_{x_1}^2+x_1\ptl_{x_2}^2)^m=\sum_{n_r\in\mbb{N};n_0+n_1+2n_2+3n_3=m}
\frac{m!}{n_0!n_1!n_2!n_3!3^{n_3}}
 x_1^{n_0}\ptl_{x_2}^{2(n_0+n_2+2n_3)}\ptl_{x_1}^{2n_1+n_2}
.\eqno(4.5.85)$$ By Lemma 4.3.1 with
$T_1=\ptl_t^2,\;T_1^-=\int_{(t)}^2$ (cf. (4.3.31)) and
$T_2=-(\ptl_{x_1}^2+x_1\ptl_{x_2}^2)$, we have the complex solutions
\begin{eqnarray*}&
&\sum_{m=0}^\infty (-T_1^-T_2)^m(e^{(k_1 x_1+k_2 x_2)i})=
\sum_{m=0}^\infty\frac{t^{2m}(\ptl_{x_1}^2+x_1\ptl_{x_2}^2)^m}{(2m)!}
(e^{(k_1 x_1+k_2 x_2)i})\\
&=&\sum_{m=0}^\infty\sum_{n_r\in\mbb{N};n_0+n_1+2n_2+3n_3=m}
\frac{t^{2m}
x_1^{n_0}\ptl_{x_2}^{2(n_0+n_2+2n_3)}\ptl_{x_1}^{2n_1+n_2}}{(2m-1)!!n_0!n_1!n_2!n_3!2^m3^{n_3}}(e^{(k_1
x_1+k_2 x_2)i})\\
&=&\sum_{m=0}^\infty\sum_{n_r\in\mbb{N};n_0+n_1+2n_2+3n_3=m}
\frac{(-1)^{n_0+n_1+n_2}i^{n_2}t^{2m}}{(2m-1)!!n_0!n_1!n_2!n_3!2^m3^{n_3}}\\
& &\times x_1^{n_0}k_2^{2(n_0+n_2+2n_3)}k_1^{2n_1+n_2}(e^{(k_1
x_1+k_2 x_2)i})\hspace{6.8cm}(4.5.86)\end{eqnarray*} and
\begin{eqnarray*}&
&\sum_{m=0}^\infty (-T_1^-T_2)^m(te^{(k_1 x_1+k_2
x_2)i})=\sum_{m=0}^\infty\frac{t^{2m+1}(\ptl_{x_1}^2+x_1\ptl_{x_2}^2)^m}{(2m+1)!}
(e^{(k_1 x_1+k_2 x_2)i})
\\
&=&\sum_{m=0}^\infty\sum_{n_r\in\mbb{N};n_0+n_1+2n_2+3n_3=m}
\frac{(-1)^{n_0+n_1+n_2}i^{n_2}t^{2m+1}}{(2m+1)!!n_0!n_1!n_2!n_3!2^m3^{n_3}}\\
& &\times x_1^{n_0}k_2^{2(n_0+n_2+2n_3)}k_1^{2n_1+n_2}(e^{(k_1
x_1+k_2 x_2)i})\hspace{6.8cm}(4.5.87)
\end{eqnarray*}
of (4.5.83). Thus we have the following real solutions
\begin{eqnarray*}&
&\phi_{k_1,k_2}(t,x_1,x_2)=\sum_{m=0}^\infty\sum_{n_r\in\mbb{N};n_0+n_1+4n_2+3n_3=m}(-1)^{n_0+n_1+n_2}k_1^{2(n_1+n_2)}k_2^{2(n_0+2n_2+2n_3)}
\\ & &\frac{t^{2m}x_1^{n_0}}{n_0!n_1!n_3!2^m3^{n_3}}
\left[ \frac{\cos
(k_1x+k_2x)}{(2m-1)!!(2n_2)!}+\frac{k_1k_2^2t^2\sin
(k_1x+k_2x)}{4(2m+3)!!(2n_2+1)!}
 \right], \hspace{3cm}(4.5.88)
\end{eqnarray*}
\begin{eqnarray*}&
&\psi_{k_1,k_2}(t,x_1,x_2)=\sum_{m=0}^\infty\sum_{n_r\in\mbb{N};n_0+n_1+4n_2+3n_3=m}
(-1)^{n_0+n_1+n_2}k_1^{2(n_1+n_2)}k_2^{2(n_0+2n_2+2n_3)}
\\ & &\frac{t^{2m}x_1^{n_0}}{n_0!n_1!n_3!2^m3^{n_3}}
\left[ \frac{\sin
(k_1x+k_2x)}{(2m-1)!!(2n_2)!}-\frac{k_1k_2^2t^2\cos
(k_1x+k_2x)}{4(2m+3)!!(2n_2+1)!}
 \right], \hspace{3cm}(4.5.89)
\end{eqnarray*}
\begin{eqnarray*}&
&\hat\phi_{k_1,k_2}(t,x_1,x_2)=\sum_{m=0}^\infty\sum_{n_r\in\mbb{N};n_0+n_1+4n_2+3n_3=m}
(-1)^{n_0+n_1+n_2}k_1^{2(n_1+n_2)}k_2^{2(n_0+2n_2+2n_3)}
\\ & &\frac{t^{2m+1}x_1^{n_0}}{n_0!n_1!n_3!2^m3^{n_3}}
\left[ \frac{\cos
(k_1x+k_2x)}{(2m+1)!!(2n_2)!}+\frac{k_1k_2^2t^2\sin
(k_1x+k_2x)}{4(2m+5)!!(2n_2+1)!}
 \right], \hspace{3cm}(4.5.90)
\end{eqnarray*}
\begin{eqnarray*}&
&\hat\psi_{k_1,k_2}(t,x_1,x_2)=\sum_{m=0}^\infty\sum_{n_r\in\mbb{N};n_0+n_1+4n_2+3n_3=m}
(-1)^{n_0+n_1+n_2}k_1^{2(n_1+n_2)}k_2^{2(n_0+2n_2+2n_3)}\\ &
&\frac{t^{2m+1}x_1^{n_0}}{n_0!n_1!n_3!2^m3^{n_3}} \left[ \frac{\sin
(k_1x+k_2x)}{(2m+1)!!(2n_2)!}-\frac{k_1k_2^2t^2\cos
(k_1x+k_2x)}{4(2m+5)!!(2n_2+1)!}
 \right]. \hspace{3.1cm}(4.5.91)
\end{eqnarray*}
Moreover,
$$\phi_{k_1,k_2}(0,x_1,x_2)=\frac{\ptl \hat\phi_{k_1,k_2}}{\ptl t}(0,x_1,x_2)=\cos
(k_1x+k_2x),\eqno(4.5.92)$$ $$\psi_{k_1,k_2}(0,x_1,x_2)=\frac{\ptl
\hat\psi_{k_1,k_2}}{\ptl
t}(0,x_1,x_2)=\sin(k_1x+k_2x),\eqno(4.5.93)$$
$$\frac{\ptl \phi_{k_1,k_2}}{\ptl t}(0,x_1,x_2)=\frac{\ptl \psi_{k_1,k_2}}{\ptl t}(0,x_1,x_2)=
\hat\phi_{k_1,k_2}(0,x_1,x_2)=\hat\psi_{k_1,k_2}(0,x_1,x_2)=
0.\eqno(4.5.94)$$ Thus the solution of the problem (4.5.83) and
(4.5.84) is
\begin{eqnarray*}\qquad
u&=&\sum_{0\preceq(k_1,k_2)\in\mbb{Z}^2}[a_{_{k_1,k_2}}\phi_{k_1,k_2}(t,x_1,x_2)
+c_{_{k_1,k_2}}\psi_{k_1,k_2}(t,x_1,x_2)\\ & &\qquad \qquad+\hat
a_{_{k_1,k_2}}\hat\phi_{k_1,k_2}(t,x_1,x_2)+ \hat
c_{_{k_1,k_2}}\hat\psi_{k_1,k_2}(t,x_1,x_2)],\hspace{2.8cm}(4.5.95)\end{eqnarray*}
where
$$a_{_{k_1,k_2}}=\frac{1}{2^{1+\dlt_{k_1,0}\dlt_{k_2,0}}\pi^2}\int_{-\pi}^\pi\int_{-\pi}^\pi
f_1(s_1,s_2)\cos(k_1s+k_2s)\;ds,\eqno\eqno(4.5.96)$$
$$c_{_{k_1,k_2}}=\frac{1}{2\pi^2}\int_{-\pi}^\pi\int_{-\pi}^\pi
f_1(s_1,s_2)\sin(k_1s+k_2s)\;ds,\eqno\eqno(4.5.97)$$
$$\hat a_{_{k_1,k_2}}=\frac{1}{2^{1+\dlt_{k_1,0}\dlt_{k_2,0}}\pi^2}\int_{-\pi}^\pi\int_{-\pi}^\pi
f_2(s_1,s_2)\cos(k_1s+k_2s)\;ds,\eqno\eqno(4.5.98)$$
$$\hat c_{_{k_1,k_2}}=\frac{1}{2\pi^2}\int_{-\pi}^\pi\int_{-\pi}^\pi
f_2(s_1,s_2)\sin(k_1s+k_2s)\;ds.\qquad\Box\eqno\eqno(4.5.99)$$ \pse

The above results can be generalized as follows. Recall the
definition of a tree given in the paragraph of (4.5.6) and (4.5.7).
A tree diagram ${\cal T}^d$ is a tree ${\cal T}=({\cal N},{\cal E})$
with a weight map $d:{\cal E}\rta \mbb{N}+1$. A node $\iota_q$ of a
tree ${\cal T}$ is called a {\it tip} if there does not exist $q\leq
p\leq n$ such that $(\iota_q,\iota_p)\in{\cal E}$. Set
$$\Psi=\{q\mid \iota_q\;\mbox{is a tip of}\;{\cal T}\}.\eqno(4.5.100)$$
Take  a tree diagram ${\cal T}^d$ with $n$ nodes and  a set
$\Psi^\dg=\{m_q\mid q\in\Psi\}$ of a positive integers. From
(4.5.63) and (4.5.64), we have to generalize the operator $\hat D$
in (4.5.44) in reverse order and set
$$ D^\dg=t(\sum_{(\iota_p,\iota_q)\in{\cal
E}}x_q\ptl_{x_p}^{d[(\iota_p,\iota_q)]}+\sum_{r\in\Psi}\ptl_{x_r}^{m_r}).\eqno(4.5.101)$$
Recall the definition of clan around (4.5.36). Given $q\in\ol{2,n}$,
we have the clan ${\cal
C}_q=\{\iota_{q_1},\iota_{q_2},...,\iota_{q_r}\}$ of the node
$\iota_q$ with $1=q_1<q_2<\cdots<q_{r-1}<q_r=q$. If $r=2$, we define
$\eta^\dg_q=t\ptl_{x_1}^{d[(q_1,q_2)]}$. When $r>2$, we define
\begin{eqnarray*}\qquad\eta_q^\dg&=&\int_0^t(\ptl_{x_{q_{r-1}}}+\int^{y_{q_{r-1}}}_0(\ptl_{x_{q_{r-2}}}+...+
\int_0^{y_3}(\ptl_{x_{q_2}}\\ &
&+\ptl_{x_1}^{d[(q_1,q_2)]}y_2)^{d[(q_2,q_3)]}
dy_2)...)^{d[(q_{r-2},q_{r-1})]}dy_{r-2})^{d[(q_{r-1},q_r)]
}dy_{r-1}\hspace{2cm}(4.5.102)\end{eqnarray*} For $q\in\Psi$, we
also define
\begin{eqnarray*}\eta_q^\clt&=&\int_0^t(\ptl_{x_q}+\int_0^{y_{q_r}}(\ptl_{x_{q_{r-1}}}+\int^{y_{q_{r-1}}}_0(\ptl_{x_{q_{r-2}}}+...+
\int_0^{y_3}(\ptl_{x_{q_2}}\\ &
&+\ptl_{x_1}^{d[(q_1,q_2)]}y_2)^{d[(q_2,q_3)]}
dy_2)...)^{d[(q_{r-2},q_{r-1})]}dy_{r-2})^{d[(q_{r-1},q_r)]
}dy_{r-1})^{m_q}dy_{q_r}\hspace{1.4cm}(4.5.103)\end{eqnarray*} By
Theorem 4.5.6, we have the following conclusion.\psp

{\bf Proposition 4.5.8}. {\it The following factorization holds}:
$$e^{D^\dg}=e^{x_2\eta^\dg_2}e^{x_3\eta^\dg_3}\cdots
e^{x_n\eta^\dg_n}\prod_{q\in\Psi}e^{\eta^\clt_q}.\eqno(4.5.104)$$
\pse

As Theorem 4.5.7, the above factorization can be used to solve the
evolution equation
$$u_t=(\sum_{(\iota_p,\iota_q)\in{\cal
E}}x_q\ptl_{x_p}^{d[(\iota_p,\iota_q)]}+\sum_{r\in\Psi}\ptl_{x_r}^{m_r})(u)\eqno(4.5.105)$$
subject to $u(0,x_1,...,x_n)=f(x_1,...,x_n)$.

Since the weight map $d$ is arbitrary and $\Psi^\dg$ can vary, we
can do further generalization as follows. Take two sets
$\{h_{p,q}(x)\mid (\iota_p,\iota_q)\in{\cal E}\}$ and $\{h_q(x)\mid
q\in\Psi\}$ of polynomials in $x$. We generalize (4.5.101) to
$$ D^\dg=t(\sum_{(\iota_p,\iota_q)\in{\cal
E}}x_qh_{p,q}(\ptl_{x_p})+\sum_{r\in\Psi}h_r(\ptl_{x_r})).\eqno(4.5.106)$$
 Given $q\in\ol{2,n}$,
we have the clan ${\cal
C}_q=\{\iota_{q_1},\iota_{q_2},...,\iota_{q_r}\}$ of the node
$\iota_q$ with $1=q_1<q_2<\cdots<q_{r-1}<q_r=q$. If $r=2$, we define
$\eta^\dg_q=th_{q_1,q_2}(\ptl_{x_1})$. When $r>2$, we define
\begin{eqnarray*}\qquad\eta_q^\dg&=&\int_0^th_{q_{r-1},q_r}(\ptl_{x_{q_{r-1}}}+\int^{y_{q_{r-1}}}_0
h_{q_{r-2},q_{r-1}}(\ptl_{x_{q_{r-2}}}+...\\
& &+
\int_0^{y_3}h_{q_2,q_3}(\ptl_{x_{q_2}}+h_{q_1,q_2}(\ptl_{x_1})y_2)
dy_2)...)dy_{r-2})dy_{r-1}\hspace{3.1cm}(4.5.107)\end{eqnarray*} For
$q\in\Psi$, we also define
\begin{eqnarray*}\eta_q^\clt&=&\int_0^th_q(\ptl_{x_q}+\int_0^{y_{q_r}}h_{q_{r-1},q_r}(\ptl_{x_{q_{r-1}}}
+\int^{y_{q_{r-1}}}_0h_{q_{r-2},q_{r-1}}(\ptl_{x_{q_{r-2}}}+...\\ &
&+
\int_0^{y_3}h_{q_2,q_3}(\ptl_{x_{q_2}}+h_{q_1,q_2}(\ptl_{x_1})y_2)
dy_2)...)dy_{r-2})dy_{r-1})dy_{q_r}.\hspace{2.8cm}(4.5.108)\end{eqnarray*}
Then (4.5.104) still holds.\psp

{\bf Exercise 4.5}\psp

1. Find the solution of the equation $u_{xx}+xu_{yy}=0$ for
$x\in\mbb{R}$ and $y\in[-\pi,\pi]$ subject to $u(0,y)=f_1(y)$ and
$u_x(0,y)=f_2(y)$, where $f_1(y)$ and $f_2(y)$ are continuous
functions on $[-\pi,\pi]$ (cf. Example 4.5.2).\pse

2. Solve the problem $u_t=u_{xx}+xu_{yy}+yu_{zz}$ for $t\in\mbb{R}$
and $x,y,z\in[-\pi,\pi]$ subject to $u(0,x,y,z)=g(x,y,z)$, where
$g(x,y,z)$ is a continuous function for $x,y,z\in[-\pi,\pi]$.\pse

3. Use (4.5.104) to solve the problem
$$u_t=(y\ptl_x^3+z\ptl_x^2+\ptl_y^2+\ptl_z^2)(u)$$
for $t\in\mbb{R}$ and $x,y,z\in[-\pi,\pi]$ subject to
$u(0,x,y,z)=g(x,y,z)$, where $g(x,y,z)$ is a continuous function for
$x,y,z\in[-\pi,\pi]$.

\section{ Calogero-Sutherland Model}

The Calogero-Sutherland model\index{Calogero-Sutherland model} is an
exactly solvable quantum many-body system in one-dimension
 (cf. [Cf], [Sg]), whose Hamiltonian is given by
$$H_{CS}=\sum_{\iota=1}^n\ptl_{x_\iota}^2+K\sum_{1\leq p<q\leq n}\frac{1}{\sinh^2(x_p-x_q)},\eqno(4.6.1)$$
where $K$ is a constant. The model was used to study long-range
interactions of $n$ particles. Solving the model is equivalent to
find eigenfunctions and their corresponding eigenvalues of the
Hamiltonian $H_{CS}$ as a differential operator:
$$H_{CS}(f(x_1,...,x_n))=\nu f(x_1,...,x_n)\eqno(4.6.2)$$
with $\nu\in\mbb{R}$. In other words, the above is the equation of
motion for the Calogero-Sutherland model.

In this section, we prove that a two-parameter generalization of the
Weyl function of type $A$ in representation theory is a solution of
the Calogero-Sutherland model. If $n=2$, we find a connection
between the Calogero-Sutherland model and the Gauss hypergeometric
function. When $n>2$, a new class of multi-variable hypergeometric
functions are found based on Etingof's work [Ep]. The results in
this section are taken from the author's work [X9]

Change variables
$$z_\iota=e^{2x_\iota}\qquad\for\;\;\iota\in\ol{1,n}.\eqno(4.6.3)$$
Then
$$\ptl_{x_\iota}=2e^{x_\iota}\ptl_{z_\iota}=2z_\iota\ptl_{z_\iota}\qquad\for\;\;\iota\in\ol{1,n}\eqno(4.6.4)$$
by the chain rule of taking partial derivatives. Moreover,
$$\frac{1}{\sinh^2(x_p-x_q)}=\frac{4}{(e^{x_p-x_q}-e^{x_q-x_p})^2}=
\frac{4}{[e^{-x_p-x_q}(e^{2x_p}-e^{2x_q})]^2}=\frac{4z_pz_q}{(z_p-z_q)^2}.\eqno(4.6.5)$$
So the Hamiltonian
$$H_{CS}=4\left[\sum_{\iota=1}^n(z_\iota\ptl_{z_\iota})^2+K\sum_{1\leq p<q\leq
n}\frac{z_pz_q}{(z_p-z_q)^2}\right]. \eqno(4.6.6)$$ Replacing $\nu$
by $4\nu$ and $f$ by $\Psi(z_1,...,z_n)$, we get the new equation of
motion for the Calogero-Sutherland model:
$$\sum_{\iota=1}^n(z_\iota\ptl_{z_\iota})^2(\Psi)+K\left(\sum_{1\leq p<q\leq n}\frac{z_pz_q}{(z_p-z_q)^2}
\right)\Psi=\nu \Psi.\eqno(4.6.7)$$ First we will introduce some
simple but nontrivial solutions.

Let $\{f_{p,q}(z)\mid p,q\in\ol{1,n}\}$ be a set of one-variable
differentiable functions and
 let $d_\iota$ be a one-variable differential operator in $z_\iota$ for $\iota\in\ol{1,n}$. It is easy to
 verified the following lemma:
\psp

{\bf Lemma 4.6.1}. {\it We have the following equation on
differentiation of determinants}:
\begin{eqnarray*}\qquad & &(\sum_{\iota=1}^nd_\iota)\left(\left|\begin{array}{cccc}f_{1,1}(z_1)&
f_{1,2}(z_2)&\cdots & f_{1,n}(z_n)\\ f_{2,1}(z_1)& f_{2,2}(z_2)&\cdots & f_{2,n}(z_n)\\
 \vdots&\vdots&\vdots&\vdots\\ f_{n,1}(z_1)& f_{n,2}(z_2)&\cdots &f_{n,n}(z_n)\end{array}
 \right|\right)\\&=&\sum_{\iota=1}^n\left|\begin{array}{cccc}f_{1,1}(z_1)& f_{1,2}(z_2)&
\cdots & f_{1,n}(z_n)\\ \vdots&\vdots&\vdots&\vdots\\
f_{\iota-1,1}(z_1)& f_{\iota-1,2}(z_2)&\cdots & f_{\iota-1,n}(z_n)\\
d_1(f_{\iota,1}(z_1))& d_2(f_{\iota,2}(z_2))&\cdots &
d_n(f_{\iota,n}(z_n))
\\ f_{\iota+1,1}(z_1)& f_{\iota+1,2}(z_2)&\cdots & f_{\iota+1,n}(z_n)\\ \vdots&\vdots&\vdots&\vdots\\
 f_{n,1}(z_1)& f_{n,2}(z_2)&\cdots & f_{n,n}(z_n)\end{array}\right|.\hspace{3.9cm}(4.6.8)
 \end{eqnarray*}

Denote the {\it Vandermonde determinant}\index{Vandermonde
determinant}
$${\cal W}(z_1,z_2,...,z_n)=\left|\begin{array}{cccc}1&1&\cdots &1\\ z_n&z_{n-1}&\cdots &z_1\\ z_n^2&z_{n-1}^2
&\cdots &z_1^2 \\ \vdots&\vdots&\vdots&\vdots\\
z_n^{n-1}&z_{n-1}^{n-1}&\cdots &z_1^{n-1}
\end{array}\right|=\prod_{1\leq p<q\leq n}(z_p-z_q).\eqno(4.6.9)$$
According to the last expression,
\begin{eqnarray*} & &(\sum_{\iota=1}^n(z_\iota\ptl_{z_\iota})^2)({\cal W})=(\sum_{r=1}^n
(z_r\ptl_{z_r})^2)(\prod_{1\leq p<q\leq n}(z_p-z_q))\\
&=&\sum_{r=1}^nz_r[\sum_{r\neq
s\in\ol{1,n}}(z_r\ptl_{z_r})^2(z_r-z_s)\cdot\frac{{\cal
W}}{z_r-z_s}\\ & &+2\sum_{1\leq s_1<s_2\leq n;s_1,s_2\neq
r}z_r\ptl_{z_r}(z_{s_1}-z_r)\cdot
z_r\ptl_{z_r}(z_{s_2}-z_r)\cdot\frac{{\cal
W}}{(z_{s_1}-z_r)(z_{s_2}-z_r)}]
\\ &=&\sum_{r=1}^nz_r\left[\sum_{r\neq
s\in\ol{1,n}}\frac{z_r{\cal W}}{z_r-z_s}+2\sum_{1\leq s_1<s_2\leq
n;s_1,s_2\neq r}\frac{z_r^2{\cal
W}}{(z_{s_1}-z_r)(z_{s_2}-z_r)}\right]
\\ &=& \sum_{1\leq r<s\leq n}\frac{(z_r-z_s){\cal
W}}{z_r-z_s}+2\sum_{r=1}^n\sum_{1\leq s_1<s_2\leq n;s_1,s_2\neq
r}\frac{z_r^2{\cal W}(z_1,z_2,...,z_n)}{(z_{s_1}-z_r)(z_{s_2}-z_r)}
\\
&=&\left(\frac{n(n-1)}{2}+2 \sum_{r=1}^n\sum_{1\leq s_1<s_2\leq
n;\:s_1,s_2\neq r}\frac{z_r^2}{(z_{s_1}-z_r)(z_{s_2}-z_r)}
\right){\cal W}.\hspace{2.8cm}(4.6.10)\end{eqnarray*} On the other
hand, Lemma 4.6.1 implies
$$ (\sum_{\iota=1}^n(z_\iota\ptl_{z_\iota})^2)({\cal
W})=(\sum_{\iota=1}^{n-1}\iota^2) {\cal W}
=\frac{(n-1)n(2n-1)}{6}{\cal W}.\eqno(4.6.11)$$
 Thus
(4.6.10) and (4.6.11) yield
\begin{eqnarray*}\qquad\qquad& &\sum_{r=1}^n\sum_{1\leq s_1<s_2\leq n;\:s_1,s_2\neq r}
\frac{z_r^2}{(z_{s_1}-z_r)(z_{s_2}-z_r)}\\
&=&\frac{1}{2}\left[\frac{(n-1)n(2n-1)}{6}
-\frac{n(n-1)}{2}\right]\\
&=&\frac{(n-1)n(n-2)}{6}={n\choose
3}.\hspace{7.5cm}(4.6.12)\end{eqnarray*}

Let
$$\phi_{\mu_1,\mu_2}(z_1,...,z_n)=(z_1z_2\cdots z_n)^{\mu_1}{\cal W}^{\mu_2}(z_1,z_2,...,z_n)\qquad\for\;\;
\mu_1,\mu_2\in\mbb{R},\eqno(4.6.13)$$ where the special case
$\phi_{(1-n)/2,1}$ is the {\it Weyl function}\index{Weyl function}
of type-$A_{n-1}$ simple Lie algebra.  Then
$$z_r\ptl_{z_r}(\phi_{\mu_1,\mu_2})=\left(\mu_1+\mu_2\sum_{r\neq s\in\ol{1,n}}\frac{z_r}{z_r-z_s}
\right)\phi_{\mu_1,\mu_2}\eqno(4.6.14)$$ for $r\in\ol{1,n}$. Hence
\begin{eqnarray*}& &\sum_{r=1}^n(z_r\ptl_{z_r})^2(\phi_{\mu_1,\mu_2})\\ &=&\sum_{r=1}^n
[\mu_1^2+\sum_{r\neq
s\in\ol{1,n}}\left(2\mu_1\mu_2\frac{z_r}{z_r-z_s}
-\mu_2\frac{z_sz_r}{(z_s-z_r)^2}+\mu_2^2\frac{z_r^2}{(z_s-z_r)^2}\right)\\
& &+2\mu_2^2 \sum_{1\leq s_1<s_2\leq n;\:s_1,s_2\neq
r}\frac{z_r^2}{(z_{s_1}-z_r)(z_{s_2}-z_r)}] \phi_{\mu_1,\mu_2}\\
&=&[n\mu_1^2+2\mu_1\mu_2\sum_{1\leq r<s\leq
n}\frac{z_r-z_s}{z_r-z_s}+2\mu_2^2 {n\choose 3}-2\mu_2\sum_{1\leq
r<s\leq n} \frac{z_sz_r}{(z_s-z_r)^2}\\ & &+\mu_2^2\sum_{1\leq
r<s\leq n}\frac{z_r^2+z_s^2}
{(z_s-z_r)^2}]\phi_{\mu_1,\mu_2}\\&=&[n\mu_1^2+n(n-1)\mu_1\mu_2+2\mu_2^2
{n\choose 3}-2\mu_2\sum_{1\leq r<s\leq n}
\frac{z_sz_r}{(z_s-z_r)^2}\\ & &+\mu_2^2\sum_{1\leq r<s\leq
n}\frac{z_r^2+z_s^2-2z_rz_s+2z_rz_s}{(z_s-z_r)^2}]\phi_{\mu_1,\mu_2}\\
&=&[n\mu_1^2+n(n-1)(\mu_1+\mu_2/2)\mu_2+2{n\choose 3}\mu_2^2\\ &
&+2\mu_2(\mu_2-1)\sum_{1\leq r<s\leq
n}\frac{z_sz_r}{(z_s-z_r)^2}]\phi_{\mu_1,\mu_2}\hspace{6.8cm}(4.6.15)\end{eqnarray*}
by (4.6.12) and (4.6.14). Therefore, we have: \psp

{\bf Theorem 6.2}. {\it The function $\phi_{\mu_1,\mu_2}$ satisfies:
\begin{eqnarray*}\qquad& &\sum_{r=1}^n(z_r\ptl_{z_r})^2(\phi_{\mu_1,\mu_2})+2\mu_2(1-\mu_2)
\left(\sum_{1\leq l<j\leq n}\frac{z_lz_j}{(z_l-z_j)^2}\right)\phi_{\mu_1,\mu_2}\\
&=&\left[n\mu_1^2+n(n-1)(\mu_1+\mu_2/2)\mu_2+2{n\choose
3}\mu_2^2\right]\phi_{\mu_1,\mu_2},\hspace{3.8cm}(4.6.16)\end{eqnarray*}
equivalently, $\phi_{\mu_1,\mu_2}(e^{2x_1},...,e^{2x_n})$ is a
solution of the Calogero-Sutherland model with $K=2\mu_2(1-\mu_2)$
and the corresponding eigenvalue is
$2n[2\mu_1(\mu_1+(n-1)\mu_2)+(n-1)(2n-1)\mu_2^2/3]$.}
 \psp

The above theorem for generic $\mu_1$ and $\mu_2$ was proved by us
in [X9], and it was  known when $\mu_1=\mu_2$ or $\mu_1=0$ before
our work [X9]. Next we will explore the connection between the
Calogero-Sutherland model and hypergeometric functions.

We first consider the case $n=2$. Now (4.6.16) becomes
\begin{eqnarray*}\qquad& &[(z_1\ptl_{z_1})^2+(z_2\ptl_{z_2})^2](\phi_{\mu_1,\mu_2})+2\mu_2(1-\mu_2)\frac{z_1z_2}
{(z_1-z_2)^2}\phi_{\mu_1,\mu_2}\\
&=&(2\mu_1^2+2\mu_1\mu_2+\mu_2^2)
\phi_{\mu_1,\mu_2}.\hspace{8.2cm}(4.6.17)\end{eqnarray*} Let $g(z)$
be a differentiable function. Denote
$$\xi=\frac{z_2}{z_2-z_1}.\eqno(4.6.18)$$
Then
$$z_1\ptl_{z_1}(g(\xi))=-z_2\ptl_{z_2}(g(\xi))=\frac{z_1z_2}{(z_2-z_1)^2}
g'(\xi).\eqno(4.6.19)$$ Moreover,
$$(z_1\ptl_{z_1})^2(g(\xi))=(z_2\ptl_{z_2})^2(g(\xi))=\frac{z_1^2z_2^2}
{(z_1-z_2)^4}{g'}'(\xi)+\frac{z_1z_2(z_1+z_2)}{(z_2-z_1)^3}
g'(\xi).\eqno(4.6.20)$$ According to (4.6.14),
$$z_1\ptl_{z_1}(\phi_{\mu_1,\mu_2})=\left(\mu_1
+\mu_2\frac{z_1}{z_1-z_2}\right)\phi_{\mu_1,\mu_2},\eqno(4.6.21)$$
$$z_2\ptl_{z_2}(\phi_{\mu_1,\mu_2})=\left(\mu_1-\mu_2\frac{z_2}{z_1-z_2}
\right)\phi_{\mu_1,\mu_2}.\eqno(4.6.22)$$

By (4.6.19)-(4.6.22), we have
\begin{eqnarray*}& &[(z_1\ptl_{z_1})^2+(z_2\ptl_{z_2})^2](\phi_{\mu_1,\mu_2}
g(\xi))\\
&=&\phi_{\mu_1,\mu_2}[\left(
2\mu_2(\mu_2-1)\frac{z_1z_2}{(z_1-z_2)^2}+(2\mu_1^2+2\mu_1\mu_2+\mu_2^2)\right)
g(\xi)\\
& &+
2\left(\mu_1+\mu_2\frac{z_1}{z_1-z_2}\right)\frac{z_1z_2}{(z_1-z_2)^2}g'(\xi)
-2\left(\mu_1-\mu_2\frac{z_2}{z_1-z_2}\right)\frac{z_1z_2}{(z_1-z_2)^2}g'(\xi)\\
&&+2\frac{z_1^2z_2^2}{(z_1-z_2)^4}{g'}'(\xi)+2\frac{z_1z_2(z_1+z_2)}
{(z_2-z_1)^3}g'(\xi)]\\
&=& \phi_{\mu_1,\mu_2}[\left(
2\mu_2(\mu_2-1)\frac{z_1z_2}{(z_1-z_2)^2}+(2\mu_1^2+2\mu_1\mu_2+\mu_2^2)\right)g
(\xi)\\ & &+2(1-\mu_2)\frac{z_1z_2(z_1+z_2)}{(z_2-z_1)^3}
g'(\xi)+\frac{2z_1^2z_2^2}{(z_1-z_2)^4}{g'}'(\xi)].\hspace{4.9cm}(4.6.23)\end{eqnarray*}
Observe that
$$\frac{z_1+z_2}{z_2-z_1}=2\frac{z_2}{z_2-z_1}-1=2\xi-1,\eqno(4.6.24)$$
$$\frac{z_1z_2}{(z_1-z_2)^2}=\frac{z_1z_2}{(z_2-z_1)^2}=
\frac{z_2^2}{(z_2-z_1)^2}-\frac{z_2}{z_2-z_1}=\xi(\xi-1).\eqno(4.6.25)$$
Thus
\begin{eqnarray*}& &2(1-\mu_2)\frac{z_1z_2(z_1+z_2)}{(z_2-z_1)^3}
g'(\xi)+\frac{2z_1^2z_2^2}{(z_1-z_2)^4}{g'}'(\xi)\\
&=&-\frac{2z_1z_2}{(z_1-z_2)^2}\left[(1-\mu_2)\frac{z_1+z_2}{z_1-z_2}
g'(\xi) -\frac{z_1z_2}{(z_1-z_2)^2}{g'}'(\xi)\right]\\
&=&-\frac{2z_1z_2}{(z_1-z_2)^2}[\xi(1-\xi){g'}'(\xi)+(1-\mu_2)(1-2\xi)g'(\xi)].\hspace{4.5cm}(4.6.26)
\end{eqnarray*}

Recall the classical Gauss hypergeometric equation
$$z(1-z){y'}'+[\gm-(\al+\be+1)z]y'-\al\be y=0.\eqno(4.6.27)$$
We take $\gm=1-\mu_2$ and
$$\al+\be+1=2(1-\mu_2)\lra \be=1-2\mu_2-\al,\eqno(4.6.28)$$
where $\al$ is arbitrary.\psp

{\bf Theorem 4.6.3}. {\it Let $\al,\mu_1,\mu_2\in\mbb{R}$. If $g(z)$
is a nonzero function satisfying the following classical Gauss
hypergeometric equation
$$z(1-z){g'}'+(1-\mu_2)(1-2z)g'-\al(1-\al-2\mu_2)g=0,\eqno(4.6.29)$$
then the function
$$\psi=(z_1z_2)^{\mu_1}(z_1-z_2)^{\mu_2}g\left(\frac{z_2}{z_2-z_1}\right)\eqno(4.6.30)$$
satisfies the equation for the Calogero-Sutherland model
\begin{eqnarray*}\qquad& &[(z_1\ptl_{z_1})^2+(z_2\ptl_{z_2})^2](\psi)+2\mu_2(1-\mu_2)\frac{z_1z_2}
{(z_1-z_2)^2}\psi\\
&=&[2\mu_1^2+2\mu_1\mu_2+\mu_2^2+2\al(\al+2\mu_2-1)]
\psi\hspace{5.6cm}(4.6.31)\end{eqnarray*} with $K=2\mu_2(1-\mu_2)$
and the eigenvalue
$\nu=2\mu_1^2+2\mu_1\mu_2+\mu_2^2+2\al(\al+2\mu_2-1)$.} \psp

Suppose that $\mu_2$ is not an integer, then the fundamental
solutions of the equation (4.6.29) are
$_2F_1(\al,1-\al-2\mu_2;1-\mu_2;z)$ and
$_2F_1(\al+\mu_2,1-\mu_2-\al;1+\mu_2;z)z^{\mu_2}$ (cf. (3.2.10)).

Next we consider $n>2$. Let
$$\G_A=\sum_{1\leq p<q\leq n}\mbb{N}\es_{q,p}\eqno(4.6.32)$$
be the additive semigroup  of  rank $n(n-1)/2$ with $\es_{q,p}$ as
base elements.
 For $\al=\sum_{1\leq p<q\leq n}\al_{q,p}\es_{q,p}\in \G_A$, we denote
$$\al_{1\ast}=\al_n^\ast=0,\;\;\al_{k\ast}=\sum_{r=1}^{k-1}\al_{k,r},\;\;
\al_l^\ast=\sum_{s=l+1}^n\al_{s,l}\eqno(4.6.33)$$
 Given $\vt\in\mbb{C}\setminus\{-\mbb{N}\}$ and $\tau_r\in\mbb{C}$ with $r\in\ol{1,n}$,
 we define our $(n(n-1)/2)$-variable {\it hypergeometric function of type A}
 by \index{hypergeometric function of type A}
$${\cal X}_A(\tau_1,..,\tau_n;\vt)\{z_{j,k}\}=\sum_{\be\in\G_A}\frac{\left[\prod_{s=1}^{n-1}
(\tau_s-\be_{s\ast})_{_{\be_s^\ast}}\right](\tau_n)_{_{\be_{n\ast}}}}
{\be!(\vt)_{_{\be_{n\ast}}}}z^\be,\eqno(4.6.34)$$ where
$$\be!=\prod_{1\leq k<j\leq n}\be_{j,k}!,\qquad z^\be=\prod_{1\leq k<j\leq n}
z_{j,k}^{\be_{j,k}}.\eqno(4.6.35)$$
 Set
$$\xi_{r_2,r_1}^A=\prod_{s=r_1}^{r_2-1}\frac{z_{r_2}}{z_{r_2}-z_s}\qquad
 \for\;\;1\leq r_1<r_2\leq n.\eqno(4.6.36)$$
Take $(\lmd_1,...,\lmd_n)\in\mbb{C}^n$  such that
$$\lmd_1-\lmd_2=\cdots
=\lmd_{n-2}-\lmd_{n-1}=\mu\;\;\mbox{and}\;\;\lmd_{n-1}-\lmd_n=\sgm
\not\in\mbb{N}, \eqno(4.6.37)$$ for some constants $\mu$ and $\sgm$.
Then we have the following result which was proved by representation
theory: \psp

{\bf Theorem 4.6.4}. {\it The function
$$\prod_{r=1}^nz_r^{\lmd_r+(n+1)/2-r}
{\cal
X}_A(\mu+1,..,\mu+1,-\mu;-\sgm)\{\xi_{r_2,r_1}^A\}\eqno(4.6.38)$$ is
a solution of the equation (4.6.7).}\psp

Below we want to show that the functions ${\cal
X}_A(\tau_1,..,\tau_n;\vt)\{z_{j,k}\}$ are  indeed natural
generalizations of the Gauss hypergeometric function
$_2F_1(\al,\be;\gm;z)$. Note
$$D=z\frac{d}{dz}\lra
D^2=z^2\frac{d^2}{dz^2}+z\frac{d}{dz}.\eqno(4.6.39)$$ Then the
classical hypergeometric equation (3.2.1) can be rewritten as
$$(\gm+D)\frac{d}{dz}(y)=(\al+D)(\be+D)(y).\eqno(4.6.40)$$

Denote
$${\cal D}_{p\ast}=\sum_{r=1}^{p-1}z_{p,r}\ptl_{z_{p,r}},\qquad {\cal D}_{q}^\ast=
\sum_{s=q+1}^nz_{s,q}\ptl_{z_{s,q}}\qquad\for\;\;p\in\ol{2,n},\;q\in\ol{1,n-1}.\eqno(4.6.41)$$
The following result was proved by author. \psp

{\bf Theorem 4.6.5}. {\it We have:
$$(\tau_{r_2}-1-{\cal D}_{r_2\ast}+{\cal D}_{r_2}^\ast)\ptl_{z_{r_2,r_1}}({\cal X}_A)
=(\tau_{r_2}-1-{\cal D}_{r_2\ast})(\tau_{r_1}-{\cal
D}_{r_1\ast}+{\cal D}_{r_1}^\ast) ({\cal X}_A)\eqno(4.6.42)$$ for
$1\leq r_1<r_2\leq n-1$ and
$$(\vt+{\cal D}_{n\ast})\ptl_{z_{n,r}}({\cal X}_A)=(\tau_n+{\cal D}_{n\ast})
(\tau_r-{\cal D}_{r\ast}+{\cal D}_r^\ast)({\cal X}_A)\eqno(4.6.43)$$
for $r\in\ol{1,n-1}$.}\psp

Recall the differentiation property
$$\frac{d}{dz}\:
_2F_1(\al,\be;\gm;z)=\frac{\al\be}{\gm}\:
_2F_1(\al+1,\be+1;\gm+1;z)\eqno(4.6.44)$$ (cf. (3.2.19)). For two
positive integers $k_1$ and $k_2$ such that $k_1<k_2$, a {\it path}
from  $k_1$ to
 $k_2$ is a sequence $(m_0,m_1....,m_r)$ of positive integers such that
$$k_1=m_0<m_1<m_2<\cdots <m_{r-1}<m_r=k_2.\eqno(4.6.45)$$
One can imagine a path from $k_1$ to $k_2$ is a way of a super man
going from $k_1$th floor to $k_2$th floor through a stairway. Let
$${\cal P}_{k_1}^{k_2}=\mbox{the set of all paths from}\;k_1\;\mbox{to}\;k_2.\eqno(4.6.46)$$
The {\it path polynomial} from $k_1$ to $k_2$ is defined as
$$P_{[k_1,k_2]}=\sum_{(m_0,m_1,...,m_r)\in {\cal P}_{k_1}^{k_2}}(-1)^rz_{m_1,m_0}z_{m_2,m_1}
\cdots z_{m_{r-1},m_{r-2}}z_{m_r,m_{r-1}}.\eqno(4.6.47)$$ In fact
$$\left(\begin{array}{ccccc}1&0&0&\cdots&0\\
P_{[1,2]}&
1&0&\cdots&0\\ P_{[1,3]}&P_{[2,3]}&1&\ddots&\vdots\\
\vdots&\vdots&\ddots&\ddots&0\\
P_{[1,n]}&P_{[2,n]}&\cdots&P_{[n-1,n]}&
1\end{array}\right)=\left(\begin{array}{ccccc}1&0&0&\cdots&0\\
z_{2,1}&
1&0&\cdots&0\\ z_{3,1}&z_{3,2}&1&\ddots&\vdots\\
\vdots&\vdots&\ddots&\ddots&0\\ z_{n,1}&z_{n,2}&\cdots&z_{n,n-1}&
1\end{array}\right)^{-1}.\eqno(4.6.48)$$ For convenience, we simply
denote
$$P_{[k,k]}=1\qquad\for\;\;0<k\in\mbb{N},\eqno(4.6.49)$$
$${\cal X}_A={\cal X}_A(\tau_1,..,\tau_n;\vt)\{z_{j,k}\},\eqno(4.6.50)$$
$${\cal X}_A[l,j]={\cal X}_A(\tau_1,...,\tau_l+1,...,\tau_j-1,...\tau_n;\vt)\{z_{r_2,r_1}\}
\eqno(4.6.51)$$ obtained from ${\cal X}_A$ by changing $\tau_l$ to
$\tau_l+1$ and $\tau_j$ to $\tau_j-1$ for $1\leq i<j\leq n-1$ and
$${\cal X}_A[k,n]={\cal X}_A(\tau_1,..,\tau_k+1,...,\tau_n+1;\vt+1)\{z_{r_2,r_1}\}\eqno(4.6.52)$$
obtained from ${\cal X}_A$ by changing $\tau_k$ to $\tau_k+1$,
$\tau_n$ to $\tau_n+1$ and $\vt$ to $\vt+1$ for $k\in\ol{1,n-1}$.
The following result was proved by the author. \psp

{\bf Theorem 4.6.6}. {\it For $1\leq r_1<r_2\leq n-1$ and $r\in
\ol{1,n-1}$, we have
$$\ptl_{z_{r_2,r_1}}({\cal X}_A)=\sum_{s=1}^{r_1}\tau_sP_{[s,r_1]}{\cal X}_A[s,r_2],\eqno(4.6.53)$$
$$\ptl_{z_{n,r}}({\cal X}_A)=\frac{\tau_n}{\vt}\sum_{s=1}^r\tau_sP_{[s,r]}{\cal X}_A[s,n].
\eqno(4.6.54)$$}\pse

Recall the integral representation
$$
_2F_1(\al,\be;\gm;z)=\frac{\G(\gm)}{\G(\be)\G(\gm-\be)}\int_0^1t^{\be
-1}(1-t)^{\gm-\be-1}(1-zt)^{-\al}dt\eqno(4.6.55)$$ (cf. Theorem
3.2.1). We have the following integral representation: \psp

{\bf Theorem 4.6.7}. {\it Suppose $\mbox{\it Re}\:\tau_n>0$ and
$\mbox{\it Re}\:(\vt-\tau_n)>0$.
 We have}
$${\cal X}_A=\frac{\G(\vt)}{\G(\vt-\tau_n)\G(\tau_n)}\int_0^1\left[\prod_{r=1}^{n-1}
(\sum_{s=r}^{n-1}P_{[r,s]}+tP_{[r,n]})^{-\tau_r}\right]t^{\tau_n-1}(1-t)^{\vt-\tau_n-1}dt
\eqno(4.6.56)$$ on the region
$P_{[r,n]}/(\sum_{s=r}^{n-1}P_{[r,s]})\not\in (-\infty,-1)$ for
$r\in\ol{1,n-1}$.\psp

Heckman and Opdam [HO, Hg1-Hg3, Oe1-Oe5, BO] introduced
hypergeometric equations related to root systems and analogous to
(4.6.7). They proved the existence of
 solutions (hypergeometric functions) of their equations.
 Gel'fand and Graev studied analogues of classical hypergeometric functions
  (so called GG-functions) by generalizing the differential property of the classical
  hypergeometric functions (e.g. cf. [GG]).

\section{Maxwell Equations}

\index{Maxwell equations} The electromagnetic fields in physics are
governed by the well-known Maxwell equations (e.g., cf. [In3]):
$$\ptl_t({\bf E})=\mbox{curl}\:{\bf B},\qquad\ptl_t({\bf B})=-\mbox{curl}\:{\bf
E}\eqno(4.7.1)$$ with
 $$\mbox{div}\:{\bf E}=f(x,y,z),
\qquad\mbox{div}\:{\bf B}=g(x,y,z),\eqno(4.7.2)$$ where the vector
function ${\bf E}$ stands for the electric field, the vector
function ${\bf B}$ stands for the magnetic field, the scalar
function $f$ is related to the charge density and the scalar
function $g$ is related to the magnetic potential. We want to  use
matrix-differential-operators and Fourier expansion to solve the
Maxwell equations (4.7.1) subject to the following initial
condition:
$${\bf E}(0,x,y,z)={\bf E}_0(x,y,z),\;\;{\bf B}(0,x,y,z)={\bf
B}_0(x,y,z)\eqno(4.7.3)$$ for $
x\in[-a_1,a_1],\;y\in[-a_2,a_2],\;z\in[-a_3,a_3],$ where ${\bf
E}_0(x,y,z)$ and ${\bf B}_0(x,y,z)$ are given real vector-valued
functions satisfying (4.7.2), and $a_r$ are positive real constants.
We denote
$${\bf B}=\left(\begin{array}{c}B_1\\ B_2\\
B_3\end{array}\right),\qquad {\bf E}=\left(\begin{array}{c}E_1\\ E_2\\
E_3\end{array}\right).\eqno(4.7.4)$$
 Then the Maxwell equations becomes
$$\ptl_t({\bf E})=\left(\begin{array}{c}\ptl_y(B_3)-\ptl_z(B_2)\\
\ptl_z(B_1)-\ptl_x(B_3)\\
\ptl_x(B_2)-\ptl_y(B_1)\end{array}\right)=\left(\begin{array}{ccc}0
&-\ptl_z& \ptl_y\\ \ptl_z&0&-\ptl_x\\
-\ptl_y&\ptl_x&0\end{array}\right){\bf B},\eqno(4.7.5)$$
$$\ptl_t({\bf B})=-\left(\begin{array}{c}\ptl_y(E_3)-\ptl_z(E_2)\\
\ptl_z(E_1)-\ptl_x(E_3)\\
\ptl_x(E_2)-\ptl_y(E_1)\end{array}\right)=-\left(\begin{array}{ccc}0
&-\ptl_z& \ptl_y\\ \ptl_z&0&-\ptl_x\\
-\ptl_y&\ptl_x&0\end{array}\right){\bf E}.\eqno(4.7.6)$$ Set
$$\mbb{D}=\left(\begin{array}{ccc}0
&-\ptl_z& \ptl_y\\ \ptl_z&0&-\ptl_x\\
-\ptl_y&\ptl_x&0\end{array}\right).\eqno(4.7.7)$$ Then we can
combine  the two equations in (4.7.1) into one equation:
$$\ptl_t\left(\begin{array}{c}{\bf E}\\ {\bf
B}\end{array}\right)=\left(\begin{array}{cc}0&\mbb{D}\\
-\mbb{D}&0\end{array}\right)\left(\begin{array}{c}{\bf E}\\ {\bf
B}\end{array}\right).\eqno(4.7.8)$$ Thus the solution is given by
$$\left(\begin{array}{c}{\bf E}\\ {\bf
B}\end{array}\right)=\left[\exp t\left(\begin{array}{cc}0&\mbb{D}\\
-\mbb{D}&0\end{array}\right)\right]\left(\begin{array}{c}{\bf E}_0\\
{\bf B}_0\end{array}\right)=\left(\begin{array}{cc}\cos
t\mbb{D}&\sin t\mbb{D}\\ -\sin t\mbb{D}&\cos
t\mbb{D}\end{array}\right)\left(\begin{array}{c}{\bf E}_0\\ {\bf
B}_0\end{array}\right),\eqno(4.7.9)$$
 where
$$\left(\begin{array}{c}{\bf E}_0\\
{\bf B}_0\end{array}\right)=\left(\begin{array}{c}{\bf E}\\ {\bf
B}\end{array}\right)|_{t=0}\eqno(4.7.10)$$ is a given first-order
differentiable field in $x,y,z$ satisfying the constraint (4.7.2).

Now the key point is how to calculate $\cos t\mbb{D}$ and $\sin
t\mbb{D}$. In order to do this, we consider the $3\times 3$
skew-symmetric matrix:
$$A=\left(\begin{array}{rrr}0&-a&-b\\ a&0&-c\\
b&c&0\end{array}\right),\qquad 0\neq a,b,c\in\mbb{R},\eqno(4.7.11)$$
where $\mbb{R}$ is the field of real numbers. Note that
$$A^2=\left(\begin{array}{rrr}0&-a&-b\\ a&0&-c\\
b&c&0\end{array}\right)\left(\begin{array}{rrr}0&-a&-b\\ a&0&-c\\
b&c&0\end{array}\right)=-\left(\begin{array}{ccc} a^2+b^2&bc &-ac\\
bc&a^2+c^2&ab\\ -ac &ab&b^2+c^2\end{array}\right).\eqno(4.7.12)$$
Moreover,
\begin{eqnarray*}\hspace{1cm}A^3&=&-\left(\begin{array}{ccc} a^2+b^2&bc &-ac\\
bc&a^2+c^2&ab\\ -ac &ab&b^2+c^2\end{array}\right)\left(\begin{array}{rrr}0&-a&-b\\ a&0&-c\\
b&c&0\end{array}\right)\\ &=&-(a^2+b^2+c^2)\left(\begin{array}{rrr}0&-a&-b\\ a&0&-c\\
b&c&0\end{array}\right)=-(a^2+b^2+c^2)A,\hspace{2.4cm}(4.7.13)\end{eqnarray*}
which implies
$$A^{2k+1}=[-(a^2+b^2+c^2)]^kA,\qquad
A^{2k+2}=[-(a^2+b^2+c^2)]^kA^2\qquad\for\;\;k\in\mbb{N},\eqno(4.7.14)$$
where $\mbb{N}$ stands for the set of nonnegative integers. Thus
$$\sin tA=\left(\sum_{k=0}^\infty
\frac{(a^2+b^2+c^2)^kt^{2k+1}}{(2k+1)!}\right)A,\eqno(4.7.15)$$ $$
\cos tA=I_3-\left(\sum_{k=0}^\infty
\frac{(a^2+b^2+c^2)^kt^{2k+2}}{(2k+2)!}\right)A^2,\eqno(4.7.16)$$
where $I_3$ is the $3\times 3$ identity matrix.

Denote
$$\Dlt=\ptl_x^2+\ptl_y^2+\ptl_z^2.\eqno(4.7.17)$$
By (4.7.7), (4.7.15) and (4.7.16), we have:
$$\sin t\mbb{D}=\left(\sum_{k=0}^\infty
\frac{\Dlt^kt^{2k+1}}{(2k+1)!}\right) \left(\begin{array}{ccc}0
&-\ptl_z& \ptl_y\\ \ptl_z&0&-\ptl_x\\
-\ptl_y&\ptl_x&0\end{array}\right)\eqno(4.7.18)$$ and
$$\cos t\mbb{D}=I_3+\left(\sum_{k=0}^\infty
\frac{\Dlt^kt^{2k+2}}{(2k+2)!}\right)
\left(\begin{array}{ccc}\ptl_y^2+\ptl_z^2 &-\ptl_x\ptl_y&
-\ptl_x\ptl_z\\ -\ptl_x\ptl_y&\ptl_x^2+\ptl_z^2
&-\ptl_y\ptl_z\\
-\ptl_x\ptl_z&-\ptl_y\ptl_z&\ptl_x^2+\ptl_y^2\end{array}\right).\eqno(4.7.19)$$

As operators,
$$\mbox{div}\circ\mbox{curl}=0.\eqno(4.7.20)$$
This shows
$$\ptl_t(\mbox{div}\:{\bf E})=\mbox{div}(\ptl_t{\bf
E})=\mbox{div}(\mbox{curl}\:{\bf B})=0,\eqno(4.7.21)$$
$$\ptl_t(\mbox{div}\:{\bf B})=\mbox{div}(\ptl_t{\bf
B})=-\mbox{div}(\mbox{curl}\:{\bf E})=0.\eqno(4.7.22)$$ Thus the
constraint (4.7.2) is satisfied if the initial field ${\bf E}_0$ and
${\bf B}_0$ satisfy it. Solving (4.7.2), we get
$${\bf E}_0=\left(\begin{array}{c}\int_0^x
f(s,y,z)ds-\ptl_y(f_1(x,y,z))\\
\ptl_x(f_1(x,y,z))-\ptl_z(f_2(x,y,z))\\
\ptl_y(f_2(x,y,z))\end{array}\right), \eqno(4.7.23)$$ $$ {\bf
B}_0=\left(\begin{array}{c}\int_0^x
g(s,y,z)ds-\ptl_y(g_1(x,y,z))\\
\ptl_x(g_1(x,y,z))-\ptl_z(g_2(x,y,z))\\
\ptl_y(g_2(x,y,z))\end{array}\right),\eqno(4.7.24)$$ which imply
that ${\bf E}_0$ is completely determined by two second-order
differentiable functions $f_1$ and $f_2$, and ${\bf B}_0$ is
completely determined by two second-order differentiable functions
$g_1$ and $g_2$. In other words, giving initial fields ${\bf E}_0$
and ${\bf B}_0$ is equivalent to giving four second-order
differentiable functions $f_1,\;g_1,\;f_2,\;g_2$.

For convenience, we denote
$$k^\dg_r=\frac{k_r}{a_r},\;\;\vec
k^\dg=(k^\dg_1,k_2^\dg,k_3^\dg)\qquad\for\;\;\vec
k=(k_1,k_2,k_3)\in\mbb{Z}^{\:3}.\eqno(4.7.25)$$Moreover, we write
$$\vec x=\left(\begin{array}{c}x\\ y\\
z\end{array}\right)\eqno(4.7.26)$$ and
$$\vec k^\dg\cdot \vec x=k_1^\dg x+k_2^\dg y+k_3^\dg
z.\eqno(4.7.27)$$ Set
$$|\vec k^\dg|=\sqrt{(k_1^\dg)^2+(k_2^\dg)^2+(k_3^\dg)^2}.\eqno(4.7.28)$$
Observe that
$$-\sum_{s=0}^\infty
\frac{(-1)^sx^{2s}(\pi t)^{2s+2}}{(2s+2)!}=\frac{\cos \pi
xt-1}{x^2},\;\;\sum_{s=0}^\infty \frac{(-1)^sx^{2s}(\pi
t)^{2s+1}}{(2s+1)!}=\frac{\sin \pi xt}{x}.\eqno(4.7.29)$$ Moreover,
we treat
$$\frac{\cos \pi
xt-1}{x^2}|_{x=0}=-\frac{\pi^2t^2}{2},\qquad \frac{\sin \pi
xt}{x}|_{x=0}=\pi t.\eqno(4.7.30)$$

 For $\lmd_1,\lmd_2,\lmd_3\in\mbb{R}$ and $\vec k\in\mbb{Z}^3$,
\begin{eqnarray*}& &\cos t\mbb{D}\:\left(\begin{array}{c}\lmd_1e^{\pi(\vec k^\dg\cdot
\vec x)i}\\ \lmd_2e^{\pi(\vec k^\dg\cdot \vec x)i}\\
\lmd_3e^{\pi(\vec k^\dg\cdot \vec
x)i}\end{array}\right)\\
&=&\left[I_3+\left(\sum_{s=0}^\infty
\frac{\Dlt^st^{2s+2}}{(2s+2)!}\right)
\left(\begin{array}{ccc}\ptl_y^2+\ptl_z^2 &-\ptl_x\ptl_y&
-\ptl_x\ptl_z\\ -\ptl_x\ptl_y&\ptl_x^2+\ptl_z^2
&-\ptl_y\ptl_z\\
-\ptl_x\ptl_z&-\ptl_y\ptl_z&\ptl_x^2+\ptl_y^2\end{array}\right)\right]\left(\begin{array}{c}
\lmd_1e^{\pi(\vec k^\dg\cdot \vec x)i}\\ \lmd_2e^{\pi(\vec
k^\dg\cdot \vec x)i}\\ \lmd_3e^{\pi(\vec k^\dg\cdot \vec
x)i}\end{array}\right)
\\
&=&\mbb{K}(\vec k,t)\left(\begin{array}{c} \lmd_1e^{\pi(\vec
k^\dg\cdot \vec x)i}\\ \lmd_2e^{\pi(\vec k^\dg\cdot \vec x)i}\\
\lmd_3e^{\pi(\vec k^\dg\cdot \vec
x)i}\end{array}\right)\hspace{9.5cm}(4.7.31)\end{eqnarray*} with
\begin{eqnarray*}\qquad& &\mbb{K}(\vec k,t)=I_3+\frac{\cos \pi t |\vec
k^\dg|-1}{|\vec
k^\dg|^2} \\& &\times \left(\begin{array}{ccc}(k_2^\dg)^2+(k_3^\dg)^2 &-k_1^\dg k_2^\dg& -k_1^\dg k_3^\dg\\
-k_1^\dg k_2^\dg&(k_1^\dg)^2+(k_3^\dg)^2
&-k_2^\dg k_3^\dg\\
-k_1^\dg k_3^\dg&-k_2^\dg
k_3^\dg&(k_1^\dg)^2+(k_2^\dg)^2\end{array}\right),\hspace{3.9cm}(4.7.32)\end{eqnarray*}
and
\begin{eqnarray*}& &\sin t\mbb{D}\:\left(\begin{array}{c}\lmd_1e^{\pi(\vec k^\dg\cdot
\vec x)i}\\ \lmd_2e^{\pi(\vec k^\dg\cdot \vec x)i}\\
\lmd_3e^{\pi(\vec k^\dg\cdot \vec
x)i}\end{array}\right)\\
&=&\left(\sum_{s=0}^\infty \frac{\Dlt^st^{2s+1}}{(2s+1)!}\right)
\left(\begin{array}{ccc}0
&-\ptl_z& \ptl_y\\ \ptl_z&0&-\ptl_x\\
-\ptl_y&\ptl_x&0\end{array}\right)\left(\begin{array}{c}\lmd_1e^{\pi(\vec
k^\dg\cdot \vec x)i}\\ \lmd_2e^{\pi(\vec k^\dg\cdot \vec x)i}\\
\lmd_3e^{\pi(\vec k^\dg\cdot \vec x)i}\end{array}\right)
\\&=&i\mbb{M}(\vec k,t)\left(\begin{array}{c}\lmd_1e^{\pi(\vec
k^\dg\cdot \vec x)i}\\ \lmd_2e^{\pi(\vec k^\dg\cdot \vec x)i}\\
\lmd_3e^{\pi(\vec k^\dg\cdot \vec
x)i}\end{array}\right)\hspace{9.3cm}(4.7.33)\end{eqnarray*} with
$$\mbb{M}(\vec k,t)=
\left(\begin{array}{ccc}0&-\frac{k_3^\dg\sin \pi t |\vec
k^\dg|}{|\vec k^\dg|}&\frac{k_2^\dg \sin
\pi t |\vec k^\dg|}{|\vec k^\dg|}\\
\frac{k_3^\dg\sin \pi t |\vec k^\dg|}{|\vec
k^\dg|}&0&-\frac{k_1^\dg\sin \pi t |\vec
k^\dg|}{|\vec k^\dg|}\\
-\frac{k_2^\dg\sin \pi t |\vec k^\dg|}{|\vec
k^\dg|}&\frac{k_1^\dg\sin \pi t |\vec k^\dg|}{|\vec
k^\dg|}&0\end{array}\right).\eqno(4.7.34)$$ Thus for $\vec
k\in\mbb{Z}^3$ and $\lmd_r\in\mbb{R}$ with $r\in\ol{1,6}$, the
vector-valued function
\begin{eqnarray*}\qquad& &\left(\begin{array}{cc}\cos t\mbb{D}&\sin t\mbb{D}\\ -\sin
t\mbb{D}&\cos
t\mbb{D}\end{array}\right)\left(\begin{array}{c}\lmd_1e^{\pi(\vec
k^\dg\cdot \vec x)i}\\\vdots\\
\lmd_6e^{\pi(\vec k^\dg\cdot \vec x)i}\end{array}\right)\\ &=&
\left(\begin{array}{cc}\mbb{K}(\vec k,t)&i\mbb{M}(\vec k,t)\\
-i\mbb{M}(\vec k,t)&\mbb{K}(\vec
k,t)\end{array}\right)\left(\begin{array}{c}\lmd_1e^{\pi(\vec
k^\dg\cdot \vec x)i}\\\vdots\\
\lmd_6e^{\pi(\vec k^\dg\cdot \vec
x)i}\end{array}\right)\hspace{5.25cm}(4.7.35)\end{eqnarray*} is a
complex solution of the equation (4.7.8).  Considering the real and
imaginary parts of (4.7.35), we get two real solutions of the
Maxwell equation (4.7.1):
$${\bf E}=\mbb{K}(\vec k,t)\left(\begin{array}{c}\lmd_1\cos\pi(\vec
k^\dg\cdot \vec x)\\ \lmd_2\cos\pi(\vec k^\dg\cdot \vec x)\\
\lmd_2\cos\pi(\vec k^\dg\cdot \vec x)\end{array}\right)-\mbb{M}(\vec
k,t)\left(\begin{array}{c}\lmd_4\sin\pi(\vec
k^\dg\cdot \vec x)\\ \lmd_5\sin\pi(\vec k^\dg\cdot \vec x)\\
\lmd_6\sin\pi(\vec k^\dg\cdot \vec x)\end{array}\right),
\eqno(4.7.36)$$
$${\bf B}=\mbb{M}(\vec k,t)\left(\begin{array}{c}\lmd_1\sin\pi(\vec
k^\dg\cdot \vec x)\\ \lmd_2\sin\pi(\vec k^\dg\cdot \vec x)\\
\lmd_3\sin\pi(\vec k^\dg\cdot \vec x)\end{array}\right)+\mbb{K}(\vec
k,t)\left(\begin{array}{c}\lmd_4\cos\pi(\vec
k^\dg\cdot \vec x)\\ \lmd_5\cos\pi(\vec k^\dg\cdot \vec x)\\
\lmd_6\cos\pi(\vec k^\dg\cdot \vec x)\end{array}\right)
\eqno(4.7.37)$$ and
$${\bf E}=\mbb{K}(\vec k,t)\left(\begin{array}{c}\mu_1\sin\pi(\vec
k^\dg\cdot \vec x)\\ \mu_2\sin\pi(\vec k^\dg\cdot \vec x)\\
\mu_3\sin\pi(\vec k^\dg\cdot \vec x)\end{array}\right)+\mbb{M}(\vec
k,t)\left(\begin{array}{c}\mu_4\cos\pi(\vec
k^\dg\cdot \vec x)\\ \mu_5\cos\pi(\vec k^\dg\cdot \vec x)\\
\mu_6\cos\pi(\vec k^\dg\cdot \vec x)\end{array}\right),
\eqno(4.7.38)$$
$${\bf B}=-\mbb{M}(\vec k,t)\left(\begin{array}{c}\mu_1\cos\pi(\vec
k^\dg\cdot \vec x)\\ \mu_2\cos\pi(\vec k^\dg\cdot \vec x)\\
\mu_3\cos\pi(\vec k^\dg\cdot \vec x)\end{array}\right)+\mbb{K}(\vec
k,t)\left(\begin{array}{c}\mu_4\sin\pi(\vec
k^\dg\cdot \vec x)\\ \mu_5\sin\pi(\vec k^\dg\cdot \vec x)\\
\mu_6\sin\pi(\vec k^\dg\cdot \vec x)\end{array}\right),
\eqno(4.7.39)$$ where $\lmd_r,\mu_r\in\mbb{R}$ for $r\in\ol{1,6}$.

Write
$$\lmd_r=b_r(\vec k),\;\;\mu_r=c_r(\vec k)\qquad\for\;\;r\in\ol{1,6}.\eqno(4.7.40)$$
By superposition principle,
\begin{eqnarray*}{\bf E}=\left(\begin{array}{c}E_1\\ E_2\\
E_3\end{array}\right)&=&\sum_{0\preceq \vec
k\in\mbb{Z}^{\:3}}[\mbb{K}(\vec k,t)\left(\begin{array}{c}b_1(\vec
k)\cos\pi(\vec k^\dg\cdot \vec x)+c_1(\vec k)\sin\pi(\vec k^\dg\cdot
\vec x)\\ b_2(\vec k)\cos\pi(\vec k^\dg\cdot \vec x)+c_2(\vec
k)\sin\pi(\vec
k^\dg\cdot \vec x)\\
b_3(\vec k)\cos\pi(\vec k^\dg\cdot \vec x)+c_3(\vec k)\sin\pi(\vec
k^\dg\cdot \vec x)\end{array}\right)\\ &&+\mbb{M}(\vec
k,t)\left(\begin{array}{c}c_4(\vec k)\cos\pi(\vec k^\dg\cdot \vec
x)-b_4(\vec k)\sin\pi(\vec k^\dg\cdot \vec x)\\ c_5(\vec
k)\cos\pi(\vec k^\dg\cdot \vec x)-b_5(\vec k)\sin\pi(\vec
k^\dg\cdot \vec x)\\
c_6(\vec k)\cos\pi(\vec k^\dg\cdot \vec x)-b_6(\vec k)\sin\pi(\vec
k^\dg\cdot \vec
x)\end{array}\right)]\hspace{1.6cm}(4.7.41)\end{eqnarray*} and
\begin{eqnarray*}{\bf B}=\left(\begin{array}{c}B_1\\ B_2\\
B_3\end{array}\right)&=&\sum_{0\preceq \vec
k\in\mbb{Z}^{\:3}}[\mbb{M}(\vec k,t)\left(\begin{array}{c}b_1(\vec
k)\sin\pi(\vec k^\dg\cdot \vec x)-c_1(\vec k)\cos\pi(\vec k^\dg\cdot
\vec x)\\ b_2(\vec k)\sin\pi(\vec k^\dg\cdot \vec x)-c_2(\vec
k)\cos\pi(\vec
k^\dg\cdot \vec x)\\
b_3(\vec k)\sin\pi(\vec k^\dg\cdot \vec x)-c_3(\vec k)\cos\pi(\vec
k^\dg\cdot \vec x)\end{array}\right)\\ & &+\mbb{K}(\vec
k,t)\left(\begin{array}{c}b_4(\vec k)\cos\pi(\vec k^\dg\cdot \vec
x)+c_4(\vec k)\sin\pi(\vec k^\dg\cdot \vec x)\\ b_5(\vec
k)\cos\pi(\vec k^\dg\cdot \vec x)+c_5(\vec k)\sin\pi(\vec
k^\dg\cdot \vec x)\\
b_6(\vec k)\cos\pi(\vec k^\dg\cdot \vec x)+c_6(\vec k)\sin\pi(\vec
k^\dg\cdot \vec
x)\end{array}\right)]\hspace{1.6cm}(4.7.42)\end{eqnarray*} is a
general solution of the equations in (4.7.1), where $\mbb{K}(\vec
k,t)$ is given in (4.7.32) and $\mbb{M}(\vec k,t)$ is given in
(4.7.34).

Note $\mbb{K}(\vec k,0)=I_3$ and $\mbb{M}(\vec k,0)=0_{3\times 3}$.
So
$${\bf E}(0,x_1,x_2,x_3)=\sum_{0\preceq \vec
k\in\mbb{Z}^{\:3}}\left(\begin{array}{c}b_1(\vec k)\cos\pi(\vec
k^\dg\cdot \vec x)+c_1(\vec k)\sin\pi(\vec k^\dg\cdot \vec x)\\
b_2(\vec k)\cos\pi(\vec k^\dg\cdot \vec x)+c_2(\vec k)\sin\pi(\vec
k^\dg\cdot \vec x)\\
b_3(\vec k)\cos\pi(\vec k^\dg\cdot \vec x)+c_3(\vec k)\sin\pi(\vec
k^\dg\cdot \vec x)\end{array}\right)\eqno(4.7.43)$$ and
$${\bf B}(0,x_1,x_2,x_3)=\sum_{0\preceq \vec
k\in\mbb{Z}^{\:3}}\left(\begin{array}{c}b_4(\vec k)\cos\pi(\vec
k^\dg\cdot \vec x)+c_4(\vec k)\sin\pi(\vec k^\dg\cdot \vec x)\\
b_5(\vec k)\cos\pi(\vec k^\dg\cdot \vec x)+c_5(\vec k)\sin\pi(\vec
k^\dg\cdot \vec x)\\
b_6(\vec k)\cos\pi(\vec k^\dg\cdot \vec x)+c_6(\vec k)\sin\pi(\vec
k^\dg\cdot \vec x)\end{array}\right).\eqno(4.7.44)$$
Write $$ {\bf E}_0=\left(\begin{array}{c}h_1(x,y,z)\\ h_2(x,y,z)\\
h_3(x,y,z)\end{array}\right),\qquad {\bf
B}_0=\left(\begin{array}{c}h_4(x,y,z)\\ h_5(x,y,z)\\
h_6(x,y,z)\end{array}\right),\eqno(4.7.45)$$ which must be of the
form (4.7.23) and (4.7.24).  By Fourier expansion and the
Kovalevskaya Theorem on the existence and uniqueness of the solution
of linear partial differential equations, we have:
 \psp

{\bf Theorem 4.7.1}. {\it The solution of the initial value problem
of the Maxwell equations (4.7.1)-(4.7.3) is given in (4.7.41) and
(4.7.42) with
$$b_r(\vec
k)=\frac{1}{2^{2+\dlt_{\vec k,\vec
0}}a_1a_2a_3}\int_{-a_1}^{a_1}\int_{-a_2}^{a_2}\int_{-a_3}^{a_3}
h_r(\vec x)\cos \pi (\vec k^\dg\cdot\vec x)\:dzdydx,\eqno(4.7.46)$$
$$c_r(\vec
k)=\frac{1}{4a_1a_2a_3}\int_{-a_1}^{a_1}\int_{-a_2}^{a_2}\int_{-a_3}^{a_3}
h_r(\vec x)\sin \pi (\vec k^\dg\cdot\vec x)\:dzdydx.\eqno(4.7.47)$$}
\psp

The above result is due to our work [X10]. Ciattonic, Crosignanic,
Di Porto and Yariv [CCDY] found the spatial Kerr solutions as exact
solutions of Maxwell equations. Fushchich  and  Revenko [FR]
obtained some exact solutions of the Lorentz-Maxwell equations.

\section{Dirac Equation and  Acoustic System}

The {\it classical free Dirac equation} is: \index{Dirac equation}
$$[\sum_{r=0}^3[\gm^rP_r-m]\psi=0\eqno(4.8.1)$$
with
$$P_0=i\ptl_t,\;\;P_1=i\ptl_x,\;\;P_2=i\ptl_y,\;\;
P_3=i\ptl_z,\eqno(4.8.2)$$ and the {\it Dirac matrices}:
$$\gm^0=\left(\begin{array}{rr}I_2&0\\
0&-I_2\end{array}\right),\qquad \gm^r=\left(\begin{array}{rr}0&\sgm_r\\
-\sgm_r&0\end{array}\right),\qquad r=1,2,3,\eqno(4.8.3)$$ where $m$
is a positive real constant, $I_2$ is the $2\times 2$ identity
matrix and
$$\sgm_1=\left(\begin{array}{rr}0&1\\
1&0\end{array}\right),\qquad\sgm_2=\left(\begin{array}{rr}0&-i\\
i&0\end{array}\right),\qquad\sgm_3=\left(\begin{array}{rr}1&0\\
0&-1\end{array}\right)\eqno(4.8.4)$$ are the {\it Pauli matrices}.
We want to solve the free Dirac equation (4.8.1) subject to the
initial condition:
$$\psi(0,x,y,z)=\psi_0(x,y,z)\qquad\for\;\;x\in[-a_1,a_1],\;y\in[-a_2,a_2],\;z\in[-a_3,a_3],
\eqno(4.8.5)$$ where $\psi_0(x,y,z)$ is a given continuous complex
vector-valued function.

The {\it Dirac matrices}:\index{Dirac matrices}
$$\gm^0=\left(\begin{array}{rrrr}1&0&0&0\\ 0&1&0&0\\ 0&0&-1&0\\
0&0&0&-1\end{array}\right),\qquad \gm^1=
\left(\begin{array}{rrrr}0&0&0&1\\ 0&0&1&0\\ 0&-1&0&0\\
-1&0&0&0\end{array}\right),\eqno(4.8.6)$$
$$\gm^2=\left(\begin{array}{rrrr}0&0&0&-i\\ 0&0&i&0\\ 0&i&0&0\\
-i&0&0&0\end{array}\right),\qquad \gm^3=
\left(\begin{array}{rrrr}0&0&1&0\\ 0&0&0&-1\\ -1&0&0&0\\
0&1&0&0\end{array}\right).\eqno(4.8.7)$$ Now free Dirac equation is
equivalent to: $\ptl_t(\psi)=\mbb{D}\psi$ with
$$\mbb{D}=\left(\begin{array}{cccc}mi& 0&
-\ptl_{z}&-\ptl_{x}+i\ptl_{y}\\
0&mi&-\ptl_{x}-i\ptl_{y}&\ptl_{z}\\
-\ptl_{z}&-\ptl_{x}+i\ptl_{y}&-mi&0\\
-\ptl_{x}-i\ptl_{y}&\ptl_{z}&0&-mi\end{array}\right).\eqno(4.8.8)$$
Observe
$$\mbb{D}^2=(\ptl_{x}^2+\ptl_{y}^2+\ptl_{z}^2-m^2)I_4,\eqno(4.8.9)$$
where $I_4$ is the $4\times 4$ identity matrix. Thus
$$e^{t\mbb{D}}=\left(\sum_{s=0}^\infty\frac{(\ptl_{x}^2+\ptl_{y}^2+\ptl_{z}^2-m^2)^st^{2s}}{(2s)!}\right)I_4+
\left(\sum_{s=0}^\infty\frac{(\ptl_{x}^2+\ptl_{y}^2+\ptl_{z}^2-m^2)^st^{2s+1}}{(2s+1)!}
\right)\mbb{D}.\eqno(4.8.10)$$ We take the settings
(4.7.25)-(4.7.30). Set
$$\la\vec k^\dg\ra=\sqrt{|\vec k^\dg|^2-m^2},\eqno(4.8.11)$$
$$\widehat{\mbb{D}}(\vec{k})=\left(\begin{array}{cccc}m& 0&
k_3^\dg i&k_1^\dg i+k_2^\dg\\
0&m&k_1^\dg i-k_2^\dg&-k_3^\dg i\\
k_3^\dg i&k_1^\dg i+k_2^\dg&-m&0\\
k_1^\dg i-k_2^\dg&-k_3^\dg i&0&-m\end{array}\right).\eqno(4.8.12)$$
Then
\begin{eqnarray*}\qquad & &e^{t\mbb{D}}\left(\begin{array}{c}a_1(\vec
k)e^{\pi(\vec
k^\dg\cdot \vec x)i}\\a_2(\vec k)e^{\pi(\vec k^\dg\cdot \vec x)i}\\
a_3(\vec k)e^{\pi(\vec k^\dg\cdot \vec x)i}\\
a_4(\vec k)e^{\pi(\vec k^\dg\cdot \vec x)i}\end{array}\right) =
[\left(\sum_{s=0}^\infty\frac{(\ptl_{x}^2+\ptl_{y}^2+\ptl_{z}^2-m^2)^st^{2s}}{(2s)!}\right)I_4\\
& &+
\left(\sum_{s=0}^\infty\frac{(\ptl_{x}^2+\ptl_{y}^2+\ptl_{z}^2-m^2)^st^{2s+1}}{(2s+1)!}
\right)\mbb{D}]\left(\begin{array}{c}a_1(\vec k)e^{\pi(\vec
k^\dg\cdot \vec x)i}\\a_2(\vec k)e^{\pi(\vec k^\dg\cdot \vec x)i}\\
a_3(\vec k)e^{\pi(\vec k^\dg\cdot \vec x)i}\\  a_4(\vec
k)e^{\pi(\vec k^\dg\cdot \vec x)i}\end{array}\right)
\\ &=&\left[\cos \pi\la\vec k^\dg\ra t\;I_4-\frac{\sin \pi \la\vec k^\dg\ra t}{\la\vec
k^\dg\ra}\widehat{\mbb{D}}(\vec{k})
\right]\left(\begin{array}{c}a_1(\vec k)e^{\pi(\vec
k^\dg\cdot \vec x)i}\\a_2(\vec k)e^{\pi(\vec k^\dg\cdot \vec x)i}\\
a_3(\vec k)e^{\pi(\vec k^\dg\cdot \vec x)i}\\  a_4(\vec
k)e^{\pi(\vec k^\dg\cdot \vec
x)i}\end{array}\right)\hspace{3.4cm}(4.8.13)\end{eqnarray*} is a
solution of the Dirac equation (4.8.1), where $a_r(\vec k)$ with
$r\in\ol{1,4}$ are complex constants.

We write
$$\psi_0(x,y,z)=\left(\begin{array}{c}f_1(x,y,z)\\ f_2(x,y,z)\\ f_3(x,y,z)\\
f_4(x,y,z)\end{array}\right)\eqno(4.8.14)$$ and take
$$a_r(\vec
k)=\frac{1}{8a_1a_2a_3}\int_{-a_1}^{a_1}\int_{-a_2}^{a_2}\int_{-a_3}^{a_3}
f_r(x,y,z)e^{-\pi(\vec k^\dg\cdot \vec x)i}\;dzdydx\eqno(4.8.15)$$
for $r\in\ol{1,3}$ and $\vec k\in\mbb{Z}^3$. By the theory of
Fourier expansion,
$$f_r(x,y,z)=\sum_{\vec
k\in\mbb{Z}^{\:3}}a_r(\vec k)e^{\pi(\vec k^\dg\cdot \vec
x)i}\;\;\for\;\;r\in\ol{1,4}.\eqno(4.8.16)$$ According to
superposition principle and the Kovalevskaya Theorem on the
existence and uniqueness of the solution of linear partial
differential equations,
 we obtain:\psp

{\bf Theorem 4.8.1}. {\it The solution of the initial value problem
of the free Dirac equation is:} $$\psi=\sum_{\vec
k\in\mbb{Z}^3}\left[\cos \pi\la\vec k^\dg\ra t\;I_4-\frac{\sin \pi
\la\vec k^\dg\ra t}{\la\vec
k^\dg\ra}\widehat{\mbb{D}}(\vec{k})\right]\left(\begin{array}{c}a_1(\vec
k)e^{\pi(\vec
k^\dg\cdot \vec x)i}\\a_2(\vec k)e^{\pi(\vec k^\dg\cdot \vec x)i}\\
a_3(\vec k)e^{\pi(\vec k^\dg\cdot \vec
x)i}\\
a_4(\vec k)e^{\pi(\vec k^\dg\cdot \vec
x)i}\end{array}\right).\eqno(4.8.17)$$
 \pse

The above result is taken from the author's work [X10]. Ibragimov
[In1] studied the invariance of Dirac equations. Fushchich, Shtelen
and Spichak [FSS] found  a connection between solutions of Dirac and
Maxwell equations.  Moreover, Hounkonnou and Mendy [HM] obtained
some exact solutions of Dirac equation for neutrinos in presence of
external fields. Furthermore, Inoue [Ia] constructed the fundamental
solution for the free Dirac equation by Hamiltonian path-integral
method. In addition, Moayedi and Darabi derived the exact solutions
of Dirac equation on 2D gravitational background.

The {\it $n$-dimensional generalized acoustic
system}\index{generalized acoustic system}
$$\lmd_t+\sum_{r=1}^n
u_{rx_r}=0,\;\;u_{pt}+\lmd_{x_p}=0,\;\;p\in\ol{1,n},\eqno(4.8.18)$$
comes from the linear approximation of the compressible Euler
equations in fluid dynamics. Denote
$$\vec u(t,x_1,...,x_n)=\left(\begin{array}{c}\lmd(t,x_1,...,x_n)\\ u_1(t,x_1,...,x_n)\\ \vdots
\\ u_n(t,x_1,...,x_n)\end{array}\right),\qquad \nabla=\left(\begin{array}{c}\ptl_{x_1}\\
\ptl_{x_2}\\ \vdots\\ \ptl_{x_n}\end{array}\right).\eqno(4.8.19)$$
Set
$$\mbb{A}=\left(\begin{array}{cc}0&\nabla^T\\ \nabla&0_{n\times
n}\end{array}\right),\eqno(4.8.20)$$ where the up-index ``$T$"
denotes the transpose of matrix and $0_{n\times n}$ denotes the
$n\times n$ matrix whose all entries are $0$. The system (4.8.18)
can be rewritten as
$$\vec u_t+\mbb{A}\vec u=0.\eqno(4.8.21)$$

We want to solve (4.8.21) for $t\in\mbb{R}$ and $x_r\in [-a_r, a_r]$
with $r\in\ol{1,n}$ subject to
$$\vec u(0,x_1,...,x_n)=\left(\begin{array}{c}\lmd(0,x_1,...,x_n)\\ u_1(0,x_1,...,x_n)\\ \vdots
\\ u_n(0,x_1,...,x_n)\end{array}\right)=\left(\begin{array}{c}f_0(x_1,...,x_n)\\
f_1(x_1,...,x_n)\\ \vdots\\
f_n(x_1,...,x_n)\end{array}\right).\eqno(4.8.22)$$ Recall the
Laplace operator
$$\Dlt=\ptl_{x_1}^2+\ptl_{x_2}^2+\cdots+\ptl_{x_n}^2=\nabla^T\nabla.\eqno(4.8.23)$$
We calculate
$$\mbb{A}^{2m+2}=\left(\begin{array}{cc}\Dlt^{m+1}&0\\
0&\Dlt^m\nabla\nabla^T\end{array}\right),\qquad \mbb{A}^{2m+1}=\left(\begin{array}{cc}0&\Dlt^m\nabla^T\\
\Dlt^m\nabla&0_{n\times n}\end{array}\right).\eqno(4.8.24)$$ Thus
\begin{eqnarray*}\qquad\qquad e^{-t\mbb{A}}&=&I_{n+1}+
\left(\sum_{m=0}^\infty\frac{t^{2m+2}\Dlt^m}{(2m+2)!}\right)\left(\begin{array}{cc}\Dlt&0\\
0&\nabla\nabla^T\end{array}\right)\\
& &-\left(\sum_{m=0}^\infty\frac{t^{2m+1}\Dlt^m}{(2m+1)!}\right)
\left(\begin{array}{cc}0&\nabla^T\\
\nabla&0_{n\times
n}\end{array}\right),\hspace{4.8cm}(4.8.25)\end{eqnarray*} where
$I_{n+1}$ is the $(n+1)\times(n+1)$ identity matrix.

For convenience, we  again denote
$$k^\dg_r=\frac{k_r}{a_r},\;\;\vec
k^\dg=(k^\dg_1,...,k_n^\dg),\;\;|\vec
k^\dg|=\sqrt{(k_1^\dg)^2+\cdots+(k_n^\dg)^2},\;\;\vec k^\dg\cdot\vec
x=\sum_{r=1}^n k^\dg_rx_r\eqno(4.8.26)$$
 for $\vec
k=(k_1,...,k_n)\in\mbb{Z}^{n}.$ Recall the equations in (4.7.25) and
take the convention (4.7.30). Let $\mu_r\in\mbb{R}$ with
$r\in\ol{0,n}$. Then
\begin{eqnarray*}\qquad&
&e^{-t\mbb{A}}\left(\begin{array}{c}\mu_0e^{\pi(\vec k^\dg\cdot
\vec x)i}\\\mu_1e^{\pi(\vec k^\dg\cdot \vec x)i}\\\vdots\\
\mu_ne^{\pi(\vec k^\dg\cdot \vec x)i}\end{array}\right)= [I_{n+1}+
\left(\sum_{m=0}^\infty\frac{t^{2m+2}\Dlt^m}{(2m+2)!}\right)\left(\begin{array}{cc}\Dlt&0\\
0&\nabla\nabla^T\end{array}\right)\\
& &-\left(\sum_{m=0}^\infty\frac{t^{2m+1}\Dlt^m}{(2m+1)!}\right)
\left(\begin{array}{cc}0&\nabla^T\\
\nabla&0_{n\times
n}\end{array}\right)]\left(\begin{array}{c}\mu_0e^{\pi(\vec
k^\dg\cdot
\vec x)i}\\\mu_1e^{\pi(\vec k^\dg\cdot \vec x)i}\\\vdots\\
\mu_ne^{\pi(\vec k^\dg\cdot \vec x)i}\end{array}\right)\\
&=&[\mbb{K}(\vec k,t)-i\mbb{M}(\vec
k,t)]\left(\begin{array}{c}\mu_0e^{\pi(\vec k^\dg\cdot
\vec x)i}\\\mu_1e^{\pi(\vec k^\dg\cdot \vec x)i}\\\vdots\\
\mu_ne^{\pi(\vec k^\dg\cdot \vec
x)i}\end{array}\right)\hspace{6.4cm}(4.8.27)\end{eqnarray*} is a
complex solution of the equation (4.8.21), where
$$\mbb{K}(\vec k,t)=\left(\begin{array}{cc}
\cos \pi|\vec k^\dg|t&0\\
0&I_n+|\vec k^\dg|^{-2}(\cos \pi|\vec k^\dg|t-1)(\vec k^\dg)^T\vec
k^\dg\end{array}\right)\eqno(4.8.28)$$ and
$$\mbb{M}(\vec k,t)=|\vec k^\dg|^{-1}\sin \pi|\vec k^\dg|t\;
\left(\begin{array}{cc}0&\vec k^\dg\\
(\vec k^\dg)^T&0_{n\times n}\end{array}\right).\eqno(4.8.29)$$

Considering the real and imaginary parts of (4.8.27), we get two
real solutions of the equation (4.8.21):
$$\vec u=\mbb{K}(\vec k,t)\left(\begin{array}{c}b_0(\vec k)\cos \pi(\vec
k^\dg\cdot
\vec x)
\\b_1(\vec k)\cos \pi(\vec
k^\dg\cdot
\vec x)\\\vdots\\
b_n(\vec k)\cos \pi(\vec k^\dg\cdot \vec x)\end{array}\right)
+\mbb{M}(\vec k,t)\left(\begin{array}{c}b_0(\vec k)\sin \pi(\vec
k^\dg\cdot \vec x)
\\ b_1(\vec k)\sin \pi(\vec
k^\dg\cdot
\vec x)\\\vdots\\
b_n(\vec k)\sin \pi(\vec k^\dg\cdot \vec
x)\end{array}\right)\eqno(4.8.30)$$ and
$$\vec u=\mbb{K}(\vec k,t)\left(\begin{array}{c}c_0(\vec k)\sin \pi(\vec
k^\dg\cdot \vec x)
\\c_1(\vec k)\sin \pi(\vec
k^\dg\cdot
\vec x)\\\vdots\\
c_n(\vec k)\sin \pi(\vec k^\dg\cdot \vec x)\end{array}\right)
-\mbb{M}(\vec k,t)\left(\begin{array}{c}b_0(\vec k)\cos \pi(\vec
k^\dg\cdot \vec x)
\\ b_1(\vec k)\cos \pi(\vec
k^\dg\cdot
\vec x)\\\vdots\\
b_n(\vec k)\cos \pi(\vec k^\dg\cdot \vec
x)\end{array}\right).\eqno(4.8.31)$$ We take
$$b_r(\vec
k)=\frac{1}{2^{n-1+\dlt_{\vec k,\vec
0}}\prod_{r=1}^na_r}\int_{-a_1}^{a_1}\int_{-a_2}^{a_2}\cdots\int_{-a_n}^{a_n}
f_r(\vec x)\cos \pi (\vec k^\dg\cdot\vec x)\:dx_1dx_2\cdots
dx_n,\eqno(4.8.32)$$
$$c_r(\vec
k)=\frac{1}{2^{n-1}\prod_{r=1}^na_r}\int_{-a_1}^{a_1}\int_{-a_2}^{a_2}\cdots\int_{-a_n}^{a_n}
f_r(\vec x)\sin \pi (\vec k^\dg\cdot\vec x)\:dx_1dx_2\cdots
dx_n\eqno(4.8.33)$$  (cf. (4.8.22)). Then we have the Fourier
expansions:
$$f_r(x_1,...,x_n)=\sum_{0\preceq \vec k\in\mbb{Z}^{\:n}}(b_r(\vec
k)\cos \pi(\vec k^\dg\cdot \vec x)+c_r(\vec k)\sin \pi(\vec
k^\dg\cdot \vec x)).\eqno(4.8.34)$$

Note $\mbb{K}(\vec k,0)=I_{(n+1)\times (n+1)}$ and $\mbb{M}(\vec
k,0)=0_{(n+1)\times (n+1)}$. According to superposition principle
and the Kovalevskaya Theorem on the existence and uniqueness of the
solution of linear partial differential equations,
 we obtain:\psp

{\bf Theorem 4.8.2}. {\it The solution of the $n$-dimensional
generalized acoustic system (4.8.18) subject to the initial
condition (4.8.22) is
\begin{eqnarray*}\qquad
\left(\begin{array}{c}\lmd\\ u_1\\
\vdots
\\ u_n\end{array}\right)&=&\sum_{0\preceq \vec k\in\mbb{Z}^n}[
\mbb{K}(\vec k,t)\left(\begin{array}{c}b_0(\vec k)\cos \pi(\vec
k^\dg\cdot \vec x)+c_0(\vec k)\sin \pi(\vec k^\dg\cdot \vec x)
\\b_1(\vec k)\cos \pi(\vec
k^\dg\cdot \vec x)+c_1(\vec k)\sin \pi(\vec
k^\dg\cdot \vec x)\\\vdots\\
b_n(\vec k)\cos \pi(\vec k^\dg\cdot \vec x)+c_n(\vec k)\sin \pi(\vec
k^\dg\cdot \vec x)\end{array}\right)
\\ & &+\mbb{M}(\vec k,t)\left(\begin{array}{c}b_0(\vec k)\sin \pi(\vec
k^\dg\cdot \vec x)-c_0(\vec k)\cos \pi(\vec k^\dg\cdot \vec x)
\\b_1(\vec k)\sin \pi(\vec
k^\dg\cdot \vec x)-c_1(\vec k)\cos \pi(\vec
k^\dg\cdot \vec x)\\\vdots\\
b_n(\vec k)\sin \pi(\vec k^\dg\cdot \vec x)-c_n(\vec k)\cos \pi(\vec
k^\dg\cdot \vec
x)\end{array}\right)\hspace{1.8cm}(4.8.35)\end{eqnarray*} with
$\mbb{K}(\vec k,t)$ given in (4.8.28) and $\mbb{M}(\vec k,t)$ given
(4.8.29).}\psp

The result in Theorem 4.8.2 was newly obtained. Cao [Cb1] determined
all the polynomial solutions of the Navier equation in elasticity
and their representation structure. Moreover, he solved the
initial-value problem of the Navier equation and the related
Lam\'{e} equation.

 \psp

{\bf Exercise 4.8}\psp

Solve the Lam\'{e} Equations
$$\vec u_{tt}=(\kappa\Dlt+\nabla\cdot\nabla ^T)(\vec u)$$
for $t\in\mbb{R}$ and $x_r\in[a_r,-a_r]$ with $r\in\ol{1,n}$ subject
to
$$\vec{u}(0,x_{1},\ldots,x_{n}) =
        \vec{g}_0(x_{1},\ldots,x_{n}),\;\;\vec{u}_t(0,x_{1},\ldots,x_{n})  =
        \vec{g}_1(x_{1},\ldots,x_{n}),$$
where $\kappa$ is a nonzero constant, $a_r$ are positive real
numbers and $g_1,g_2$ are continuous functions (cf. [Cb1]).

\chapter{Nonlinear Scalar Equations}

This chapter deals with nonlinear scalar (one dependent variable)
partial differential equations. First we do symmetry analysis on the
KdV equation, and obtain  the traveling-wave solutions of the KdV
equation in terms of the functions $\wp(z),\tan^2z,\coth^2z$ and
$\cn^2(z|m)$, respectively. In particular, the soliton solution is
obtained by taking $\lim_{m\rta 1}$ of a special case of the last
solution. Moreover, we derive the Hirota bilinear presentation of
the KdV equation and use it to find the two-soliton solution.

The KP equation can be viewed as an extension of the KdV equation.
Any solution of the KdV equation is obviously a solution of the KP
equation. In this chapter, we have done the symmetry analysis on the
KP equation and use the symmetry transformations to extend the
solutions of the KdV equation that are independent of $y$ to a more
sophisticated solution of the KP equation that depends on $y$.
Moreover, we solve the KP equation for solutions that are polynomial
in $x$, and obtain many solutions that can not be obtained from the
solutions of the KdV equation. Furthermore, we find the Hirota
bilinear presentation of the KP equation and obtain the ``lump"
solution. The above results  are well-known (e.g., cf. [AC]) and we
reformulate them here just for pedagogic purpose.

Lin, Reisner and Tsien [LRT] (1948) found the equation of transonic
gas flows. We derive  the symmetry
 transformations of the  equation.  Using the stable range of the
nonlinear term  and generalized power series method, we find a
family of singular solutions with seven arbitrary parameter
functions in $t$ and a family of analytic solutions with six
arbitrary parameter functions in $t$. Khristianovich and Rizhov [KR]
(1958) discovered the equations of short waves in connection with
the nonlinear reflection of weak shock waves. Khokhlov and
Zabolotskaya [KZ] (1969) found an equation for quasi-plane waves in
nonlinear acoustics of bounded bundles. The solutions of the above
equations similar to those of the LRT equation are derived. Kibel'
[Kt] (1954) introduced an equation for geopotential forecast on a
middle level. The  symmetry transformations and two new families of
exact solutions with multiple parameter functions of the equation
are derived.

\section{Kortweg and de Vries Equation}

Soliton phenomenon was first observed by J. Scott Russel in 1834
when he was riding on horseback beside the narrow Union Canal near
Edinburgh, Scotland. He described his observations as follows:

``I was observing the motion of a boat which was rapidly drawn along
a narrow channel by a pair of horse, when the boat suddenly
stopped---not so the mass of water in the channel which it had put
in motion; it accumulates round the prow of the vessel in a state of
violent agitation, then suddenly leaving it behind, rolled forward
with great velocity, assuming the form of a large solitary
elevation, a rounded, smooth and well-defined heap of water, which
continued its course along the channel apparently without change of
form or diminution of speed. I followed it on horseback, and
overtook it still rolling on at a rate of some eight or nine miles
an hour, preserving its original figure some thirty feet long and a
foot to a foot and a half in height. Its height gradually
diminished, and after a chase of one or two miles I lost it in the
windings of the channel. Such, in the month of August 1834, was my
first chance interview with that rare and beautiful phenomenon which
I called the Wave of Translation... ."

The phenomenon had been theoretically studied by Russel, Airy
(1845), Stokes (1847), Boussinesq (1871, 1872) and Rayleigh (1876).
Boussinesq's study lead him to discover the $(1+1)$-dimensional
Boussinesq equation. There had been an intensive discussion and
controversy on whether the inviscid equations of water wave would
posses such solitary wave solutions. The problem was finally solved
by Kortweg and de Vries (1895). They derived a nonlinear evolution
equation governing long one-dimensional, small amplitude, surface
gravity waves propagating in a shallow channel of water:
$$\frac{\ptl \eta}{\ptl
\tau}=\frac{3}{2}\sqrt{\frac{g}{h}}\frac{\ptl}{\ptl\xi}\left(\frac{1}{2}\eta^2+\frac{2}{3}\al\eta+\frac{1}{3}
\sgm\frac{\ptl^2\eta}{\ptl\xi^2}\right),\qquad\sgm
=\frac{1}{3}h^2-\frac{Th}{\rho g},\eqno(5.1.1)$$ where $\eta$ is the
surface elevation of the wave above the equilibrium level $h$, $\al$
is a small arbitrary constant related to the uniform motion of the
liquid, $g$ is the gravitational constant, $T$ is the surface
tension and $\rho$ is the density (the term ``long" and `` small"
are meant in comparison to the depth of the channel). By the
nondimensional transformation
$$t=\frac{1}{2}\sqrt{\frac{g}{h\sgm}}\;\tau,\qquad
x=-\frac{\xi}{\sqrt{\sgm}},\qquad
u=\frac{1}{2}\eta+\frac{1}{3}\al,\eqno(5.1.2)$$ the equation (5.1.1)
becomes
$$ u_t+6uu_x+u_{xxx}=0,\eqno(5.1.3)$$\index{KdV equation}the standard modern KdV equation.

A transformation is called a {\it symmetry}\index{symmetry} of a
partial differential equation if it maps the solution space of the
equation to itself. Since the equation (5.1.3) does not contain
variable coefficients, the translation
$$T_{a_1,a_2}(u(t,x))=u(t+a_1,x+a_2)\eqno(5.1.4)$$
leave (5.1.3) invariant, that is, it changes (5.1.3) to
$$ u_t(t+a_1,x+a_2)+6u(t+a_1,x+a_2)u_x(t+a_1,x+a_2)+u_{xxx}(t+a_1,x+a_2)=0,\eqno(5.1.5)$$
where $a_1,a_2\in\mbb{R}$ and the subindices denote the partial
derivatives with respect to the original independent variables. Thus
it maps a solution of (5.1.3) to another solution of (5.1.3).
Equivalently  $T_{a_1,a_2}$ is a symmetry of the KdV equation. Next
we want to find dilation (scaling) symmetry. We do the following
{\it degree analysis}.\index{degree analysis} Suppose that
$$\mbox{deg}\:t=\ell_1,\;\;\mbox{deg}\:x=\ell_2,\;\;\mbox{deg}\:u=\ell_3.\eqno(5.1.6)$$
We want to make all the terms in (5.1.3) having the same degree in
order to find invariant scaling transformation. Note
$$\mbox{deg}\;u_t=\ell_3-\ell_1,\;\;\mbox{deg}\;uu_x=2\ell_3-\ell_2,\;\;\mbox{deg}\;u_{xxx}=\ell_3-3\ell_2.
\eqno(5.1.7)$$ We impose
$$\ell_3-\ell_1=2\ell_3-\ell_2=\ell_3-3\ell_2.\eqno(5.1.8)$$
Thus
$$\ell_1=3\ell_2,\qquad \ell_3=-2\ell_2.\eqno(5.1.9)$$
Hence the scaling
$$S_b(u(t,x))=b^2u(b^3t,bx)\eqno(5.1.10)$$
with $0\neq b\in\mbb{R}$ keeps (5.1.3) invariant, that is, it
changes (5.1.3) to
$$ b^5[u_t(b^3t,bx)+6u(b^3t,bx)u_x(b^3t,bx)+u_{xxx}(b^3t,bx)]=0,\eqno(5.1.11)$$
where the subindices again denote the partial derivatives with
respect to the original independent variables; equivalently,
$$u_t(b^3t,bx)+6u(b^3t,bx)u_x(b^3t,bx)+u_{xxx}(b^3t,bx)=0.\eqno(5.1.12)$$
Thus $S_b$  maps a solution of (5.1.3) to another solution of
(5.1.3) because (5.1.12) implies (5.1.11). Observe that the
transformation $u(t,x)\mapsto u(t,x+ct)$ with $c\in\mbb{R}$ changes
(5.1.3) to
$$u_t(t,x+ct)+cu_x(t,x+ct)+6u(t,x+ct)u_x(t,x+ct)+u_{xxx}(t,x+ct)=0,\eqno(5.1.13)$$
where the subindices once again denote the partial derivatives with
respect to the original independent variables. On the other hand,
the transformation $u(t,x)\mapsto u(t,x)-c/6$ changes (5.1.3) to
$$u_t(t,x)-cu_x(t,x)+6u(t,x)u_x(t,x)+u_{xxx}(t,x)=0.\eqno(5.1.14)$$
 So (5.1.3) is invariant under the following {\it Galilean
 boost}\index{Galilean boost}
$$G_c(u(t,x))=u(t,x+ct)-\frac{c}{6}\eqno(5.1.15)$$
with the independent variable $x$ replaced by $x+ct$ and the same
meaning of the subindices.

A solution of (5.1.3) is called a {\it traveling-wave
solution}\index{traveling-wave solution} if it is of the form
$u=f(at+bx)$ with $a,b\in\mbb{R}$. To find such an interesting
solution, we can assume that $u=\xi(x)$ is independent of $t$;
otherwise, we replace $u$ by some $G_c(u)$ so that the ``$t$"
disappears. Under this assumption, (5.1.3) becomes
$${{\xi'}'}'+6\xi\xi'=0\sim {{\xi}'}'+3\xi^2=k.\eqno(5.1.16)$$
If we take $\mbox{deg}\:x=1$, we have to take $\mbox{deg}\:\xi=-2$
in order to make the two nonzero terms in the first equation in
(5.1.16) to have the same degree. This shows that we can try the
real function with a pole of order 2 when it is viewed as a complex
function. Note ${(x^{-2})'}'=6x^{-4}$. Assume $\xi=ax^{-2}$ is a
solution of (5.1.16). Then
$$6ax^{-4}+2a^2x^{-4}=k\lra a=-2.\eqno(5.1.17)$$
So $u=-2x^{-2}$ is a solution of  the KdV equation (5.1.3). Applying
$T_{0,a}$ in (5.1.4) and $G_c$ in (5.1.15)  , we get a more general
traveling-wave solution
$$u=-\frac{2}{(x+ct+a)^2}-\frac{c}{6}.\eqno(5.1.18)$$

Recall the Weierstrass's elliptic function $\wp(z)$ defined in
(3.4.9). Moreover, ${\wp'}'(z)=6\wp^2(z)-g_2/2$ with the $g_2$ given
in (3.4.29). In (3.4.9), we take $\omega_1\in\mbb{C}$ such that
$\mbox{Re}\:\w_1,\mbox{Im}\:\w_1\neq 0$ and $\w_2=\ol{\w_1}$. Then
$\wp(z)$ is real if $z\in\mbb{R}$ and  $g_2$ is a real number. Thus
$\xi=-2\wp(x)$ is a solution of (5.1.16). Applying the
transformation in (5.1.4) and (5.1.15), we get the following
traveling-wave solution of the KdV equation (5.1.3):
$$u=-2\wp(x+ct+a)-\frac{c}{6},\qquad a,b,c\in\mbb{R},\;b\neq 0.\eqno(5.1.19)$$

Note that for $a\in\mbb{R}$,
$${(f^2(x)+a)'}'={(f^2(x))'}'=2[f(x){f'}'(x)+(f'(x))^2].\eqno(5.1.20)$$
By (3.5.17), \begin{eqnarray*} \tan x\;{\tan'}'x+(\tan' x)^2&=&\tan
x\;(2\tan^3x+2\tan x)+(\tan^2x+1)^2\\ &=&3\tan^4
x+4\tan^2x+1=3(\tan^2x+2/3)^2-1/3.\hspace{0.9cm}(5.1.21)\end{eqnarray*}
Thus $\xi=-2(\tan^2x+2/3)$ is a solution of (5.1.16). Applying the
transformations in (5.1.4), (5.1.10) and (5.1.15), we find another
traveling-wave solution of the KdV equation (5.1.3):
$$u=-2b^2\tan^2(bx+cb^3t+a)-\frac{b^2(8+c)}{6},\qquad
a,b,c\in\mbb{R},\;b\neq 0.\eqno(5.1.22)$$ Taking $c=-8$, we get
$u=-b^2\tan^2(bx-8b^3t+a)$. According to (3.5.19),
\begin{eqnarray*} \coth x\;{\coth'}'x+(\coth' x)^2&=&\coth
x\;(2\coth^3 x-2\coth x)+(1-\coth^2 x)^2\\
&=&3(\coth^2x-2/3)^2-1/3.\hspace{4.7cm}(5.1.23)\end{eqnarray*} So we
have the following traveling-wave solution of the KdV equation
(5.1.3):
$$u=-2b^2\coth^2(bx+cb^3t+a)+\frac{b^2(8-c)}{6},\qquad
a,b,c\in\mbb{R},\;b\neq 0.\eqno(5.1.24)$$ Taking $c=8$, we get
$u=-2b^2\coth^2(bx+8b^3t+a)$.

Next (3.5.10), (3.5.13) and (3.5.14) imply
\begin{eqnarray*}& &
\sn(x|m)\;{\sn'}'(x|m)+(\sn'(x|m))^2\\&=&\sn(x|m)\;[2m^2\mbox{sn}^3(x|m)-(m^2+1)\sn(x|m)]+
\mbox{cn}^2 (x|m)\;\mbox{dn}^2(x|m)\\
&=&2m^2\mbox{sn}^4(x|m)-(m^2+1)\mbox{sn}^2(x|m)+(1-\mbox{sn}^2(x|m))(1-m^2\mbox{sn}^2(x|m))\\
&=&3m^2\mbox{sn}^4(x|m)-2(m^2+1)\mbox{sn}^2(x|m)+1
\\
&=&3m^2\left(\mbox{sn}^2(x|m)-\frac{m^2+1}{3m^2}\right)^2+\frac{m^2-m^4-1}{3m^2}.\hspace{5.5cm}(5.1.25)
\end{eqnarray*}
Thus
$$\xi=-2m^2\left(\mbox{sn}^2(x|m)-\frac{2m^2+2}{3m^2}\right)=2m^2\mbox{cn}^2(x|m)+\frac{2-4m^2}{3}\eqno(5.1.26)$$
is a solution of (5.1.16). Hence we have the following
traveling-wave solution of the KdV equation (5.1.3):
$$u=2b^2m^2\mbox{cn}^2(bx+cb^3t+a|m)+\frac{b^2(4-8m^2-c)}{6},\qquad
a,b,c\in\mbb{R},\;b\neq 0.\eqno(5.1.27)$$ Taking $c=4-8m^2$, we have
$u=2b^2m^2\mbox{cn}^2(bx+(4-8m^2)b^3t+a|m)$. Recall $\lim_{m\rta
1}\cn (x|m)\\ =\mbox{sech}\: x.$ Therefore, we have the soliton
solution\index{soliton solution!of KdV equation}
$$u=2b^2\mbox{sech}^2(bx-4b^3t+a),\eqno(5.1.28)$$
which describes the phenomenon observed by Russel in 1834.

There is another obvious solution $u=x/6t$ of the KdV equation
(5.1.3). Applying $T_{a_1,a_2}$ in (5.1.4), we get the following
traveling-wave solution of the KdV equation (5.1.3):
$$u=\frac{x-a_2}{6(t-a_1)},\qquad a_1,a_2\in\mbb{R}.\eqno(5.1.29)$$

Next we look for the solution of the KdV equation (5.1.3) in the
form
$$u=\rho\ptl_x^2\ln v(t,x),\eqno(5.1.30)$$
where $\rho$ is a nonzero constant to be determined when we try to
simplify the resulted equation.
 Then (5.1.3) becomes
$$\rho\ptl_x^2\ptl_t\ln v+3\rho^2\ptl_x(\ptl_x^2\ln v)^2+\rho\ptl_x^5\ln
v=0,\eqno(5.1.31)$$ equivalently,
$$\ptl_x\ptl_t\ln v+3\rho(\ptl_x^2\ln v)^2+\ptl_x^4\ln
v=\nu(t)\eqno(5.1.32)$$ for some function $\nu$ in $t$. Note
$$\ptl_x\ptl_t\ln v=\frac{vv_{tx}-v_tv_x}{v^2},\;\;\;\ptl_x^2\ln v=
\frac{vv_{xx}-v_x^2}{v^2},\eqno(5.1.33)$$ $$\ptl_x^3\ln v=
\frac{v^2v_{xxx}-3vv_xv_{xx} +2v_x^3}{v^3},\eqno(5.1.34)$$
$$\ptl_x^4\ln v=
\frac{v^3v_{xxxx}-4v^2v_xv_{xxx}-3v^2v_{xx}^2+12vv_x^2v_{xx}-6v_x^4}{v^4}.
\eqno(5.1.35)$$ Since
$$(\ptl_x^2\ln
v)^2=\frac{v^2v_{xx}^2-2vv_x^2v_{xx}+v_x^4}{v^4},\eqno(5.1.36)$$ we
take $\rho=2$, and (5.1.32) becomes
$$vv_{tx}-v_tv_x+vv_{xxxx}-4v_xv_{xxx}+3v_{xx}^2=\nu v^2.\eqno(5.1.37)$$

We assume
$$v=1+k_1e^{a_1t+b_1x}+k_2e^{a_2t+b_2x}+k_3e^{(a_1+a_2)t+(b_1+b_2)x},\qquad a_1,a_2,b_1,b_2,k_1,k_2,k_3
\in\mbb{R}.\eqno(5.1.38)$$ Then $$
v_t=a_1k_1e^{a_1t+b_1x}+a_2k_2e^{a_2t+b_2x}+(a_1+a_2)k_3e^{(a_1+a_2)t+(b_1+b_2)x},\eqno(5.1.39)$$
$$v_{tx}=a_1b_1k_1e^{a_1t+b_1x}+a_2b_2k_2e^{a_2t+b_2x}+(a_1+a_2)(b_1+b_2)k_3e^{(a_1+a_2)t+(b_1+b_2)x},\eqno(5.1.40)$$
$$\ptl_x^m(v)=b_1^mk_1e^{a_1t+b_1x}+b_2^mk_2e^{a_2t+b_2x}+(b_1+b_2)^mk_3e^{(a_1+a_2)t+(b_1+b_2)x}.\eqno(5.1.41)$$
Moreover,
\begin{eqnarray*}& &vv_{tx}-v_tv_x=v_{tx}
+ (k_1e^{a_1t+b_1x}+k_2e^{a_2t+b_2x}+k_3e^{(a_1+a_2)t+(b_1+b_2)x})\\
&
&\times(a_1b_1k_1e^{a_1t+b_1x}+a_2b_2k_2e^{a_2t+b_2x}+(a_1+a_2)(b_1+b_2)k_3e^{(a_1+a_2)t+(b_1+b_2)x})
\\ & &-(a_1k_1e^{a_1t+b_1x}+a_2k_2e^{a_2t+b_2x}+(a_1+a_2)k_3e^{(a_1+a_2)t+(b_1+b_2)x})\\
& &\times(b_1k_1e^{a_1t+b_1x}+b_2k_2e^{a_2t+b_2x}
+(b_1+b_2)k_3e^{(a_1+a_2)t+(b_1+b_2)x})\\&=&
a_1b_1(k_1e^{a_1t+b_1x}+k_2k_3e^{(a_1+2a_2)t+(b_1+2b_2)x})+(k_2e^{a_2t+b_2x}+k_1k_3e^{(2a_1+a_2)t+(2b_1+b_2)x})\\
& &\times a_2b_2+
[(a_1+a_2)(b_1+b_2)k_3+k_1k_2(a_1-a_2)(b_1-b_2)]e^{(a_1+a_2)t+(b_1+b_2)x},\hspace{0.75cm}(5.1.42)
\end{eqnarray*}
\begin{eqnarray*}& &vv_{xxxx}-4v_xv_{xxx}+3v_{xx}^2=v_{xxxx}+
(k_1e^{a_1t+b_1x}+k_2e^{a_2t+b_2x}+k_3e^{(a_1+a_2)t+(b_1+b_2)x})\\ &
&\times(b_1^4k_1e^{a_1t+b_1x}+b_2^4k_2e^{a_2t+b_2x}+(b_1+b_2)^4k_3e^{(a_1+a_2)t+(b_1+b_2)x})-4
(b_1k_1e^{a_1t+b_1x}+b_2k_2\\
& &\times e^{a_2t+b_2x}+(b_1+b_2)k_3e^{(a_1+a_2)t+(b_1+b_2)x})
(b_1^3k_1e^{a_1t+b_1x}+(b_1+b_2)^3k_3e^{(a_1+a_2)t+(b_1+b_2)x}\\&
&+b_2^3k_2e^{a_2t+b_2x})+
3(b_1^2k_1e^{a_1t+b_1x}+b_2^2k_2e^{a_2t+b_2x}+(b_1+b_2)^2k_3e^{(a_1+a_2)t+(b_1+b_2)x})^2
\\ &=&
[(b_1+b_2)^4k_3+k_1k_2(b_1-b_2)^4]e^{(a_1+a_2)t+(b_1+b_2)x}+k_2k_3b_1^4e^{(a_1+2a_2)t+(b_1+2b_2)x}
\\&
&+k_1k_3b_2^4e^{(2a_1+a_2)t+(2b_1+b_2)x}+k_1b_1^4e^{a_1t+b_1x}+k_2b_2^4e^{a_2t+b_2x}.\hspace{3.8cm}(5.1.43)
\end{eqnarray*}
Substituting the above expressions into (5.1.37) and taking
$\nu\equiv 0$, we find that (5.1.37) is equivalent to
$$a_1=-b_1^3,\;\;\;a_2=-b_2^3\eqno(5.1.44)$$
and
$$(a_1+a_2)(b_1+b_2)k_3+k_1k_2(a_1-a_2)(b_1-b_2)+(b_1+b_2)^4k_3+k_1k_2(b_1-b_2)^4=0,\eqno(5.1.45)$$
which is equivalent to
$$3b_1b_2(b_1+b_2)^2k_3=3b_1b_2(b_1-b_2)^2k_1k_2\lra
k_3=\left(\frac{b_1-b_2}{b_1+b_2}\right)^2k_1k_2.\eqno(5.1.46)$$
Hence we have a {\it two-solition solution}\index{two soliton
solution!of KdV equation}
$$u=2\ptl_x^2\ln\left(1+k_1e^{b_1x-b_1^3t}+k_2e^{b_2x-b_2^3t}+\left(\frac{b_1-b_2}{b_1+b_2}\right)^2
k_1k_2e^{(b_1+b_2)x-(b_1^3+b_2^3)t}\right)\eqno(5.1.47)$$ for the
KdV equation (5.1.3), where $0\neq b_1,b_2,k_1,k_2\in\mbb{R}$ and
$b_1+b_2\neq 0$. The above two-solition solution was discovered by
Hirota (1971) [Hr]. Hirota introduced a bilinear form (now called
{\it Hirota bilinear form}) as follows. For two functions
$f(x_1,...,x_n)$ and $g(x_1,...,x_n)$, we define the {\it Hirota
bilinear form}\index{Hirota bilinear form}
$$D_{x_{r_1}}^{k_1}D_{x_{r_2}}^{k_2}(f\cdot g)=\sum_{s_1=0}^{k_1}\sum_{s_2=0}^{k_2}{k_1\choose
s_1}{k_2\choose
s_2}(-1)^{s_1+s_2}\ptl_{x_{r_1}}^{k_1-s_1}\ptl_{x_{r_2}}^{k_2-s_2}(f)\ptl_{x_{r_1}}^{s_1}\ptl_{x_{r_2}}^{s_2}(g)
\eqno(5.1.48)$$for $r_1,r_2\in\ol{1,n}$ and $k_1,k_2\in\mbb{N}$. The
reason for the KdV equation to have the two-soliton solution
(5.1.47) is because the equation (5.1.37) with $\nu\equiv 0$ can be
written as
$$D_tD_x(v\cdot v)+D_x^4(v\cdot v)=0,\eqno(5.1.49)$$
which is called the {\it Hirota bilinear form
presentation}\index{Hirota bilinear form!for the KdV equation} of
the KdV equation.\psp

{\bf Exercise 5.1}\psp

Find exact solutions of the following {\it one-dimensional
Boussinesq equation}\index{ one-dimensional Boussinesq equation}
$$u_{tt}+uu_{xx}+(u_x)^2+u_{xxxx}=0$$
(Hint: prove that if $u=f(x)$ is a solution, then so is
$f(x+ct)-c^2$).

\section{Kadomtsev and Petviashvili Equation}

The Kadomtsev and Petviashvili (KP) equation
$$(u_t+6uu_x+u_{xxx})_x+3\es u_{yy}=0\eqno(5.2.1)$$\index{KP
equation}with $\es=\pm 1$ is used to describe the evolution of long
water waves of small amplitude if they are weakly two-dimensional
(cf. [KP]). The choice of $\es$ depends on the relevant magnitude of
gravity and surface tension. The equation has also been proposed as
a model for surface waves and internal waves in straits or channels
of varying depth and width.

Let $\al(t)$ be a differentiable function. Then the transformation
$u(t,x,y)\mapsto u(t,x+\al,y)$ changes the KP equation to
$$(u_t+\al' u_x+6uu_x+u_{xxx})_x+3\es u_{yy}=0,\eqno(5.2.2)$$
where the independent variable $x$ is replaced by $x+\al$ and the
subindices denote the partial derivatives with respect to the
original independent variables. Moreover, the transformation
$u\mapsto u-\al'/6$ changes the KP equation to
$$(u_t-\al' u_x+6uu_x+u_{xxx})_x+3\es u_{yy}=0.\eqno(5.2.3)$$
 So the transformation
$$T_{2,\al}(u(t,x,y))=u(t,x+\al,y)-\frac{\al'}{6}\eqno(5.2.4)$$
keeps the KP equation invariant with the independent variable $x$ is
replaced by $x+\al$; equivalently, $T_{2,\al}$ maps a solution of
the KP equation to another solution of the KP equation. Moreover,
the transformation $u(t,x,y)\mapsto u(t,x,y+\al)$ changes the KP
equation to
$$(u_t+\al' u_y+6uu_x+u_{xxx})_x+3\es u_{yy}=0\eqno(5.2.5)$$
with the independent variable $y$ replaced by $y+\al$, and the
transformation $u(t,x,y)\mapsto u(t,x+\be y,y)$ changes the KP
equation to
$$(u_t+\be'y u_x+6uu_x+u_{xxx})_x+3\es
u_{yy}+3\es \be^2u_{xx}+6\es\be u_{xy}=0\eqno(5.2.6)$$ with the
independent variable $x$ replaced by $x+\be y$;
 equivalently,
$$(u_t+(\be'y+3\es \be^2) u_x+6\es \be u_y+6uu_x+u_{xxx})_x+3\es
u_{yy}=0.\eqno(5.2.7)$$ Thus the transformation
$$T_{3,\al}(u(t,x,y))=u\left(t,x-\frac{\al'y}{6\es},y+\al\right)+\frac{2{\al'}'y-{\al'}^2}{72\es}\eqno(5.2.8)$$
leaves the KP equation invariant with the independent variable $y$
replaced by $y+\al$ and the variable $x$ replaced by $x-\es\al'y/6$.
Hence $T_{3,\al}$ maps a solution of the KP equation to another
solution of the KP equation.

From the degree analysis in (5.1.6)-(5.1.9), we can make the KP
equation homogeneous if we take
$\mbox{deg}\:y=2\mbox{deg}\:x=2\ell_2$. Hence the transformation
$$T_{a,b}(u(t,x,y))=b^2u(b^3t+a,bx,b^2y)\eqno(5.2.9)$$
keeps the KP equation invariant for $a,b\in\mbb{R}$ and $b\neq 0$.
Therefore, the transformation
$${\cal T}(u(t,x,y))=b^2
u(b^3t+a,b(x-\es
\al'y/6+\be),b^2(y+\al))+\frac{2{\al'}'y-{\al'}^2}{72\es}-\frac{\be'}{6}
\eqno(5.2.10)$$ maps a solution of the KP equation to another
solution for any functions $\al,\be$ in $t$ and $a,b\in\mbb{R}$ with
$b\neq 0$.

Note that any solution of the KdV equation is also a solution of the
KP equation. Using the above symmetry transformations in (5.2.4) and
(5.2.8), we can get more sophisticated solutions of the KP equation
from the solutions of the KdV equation in last section: (1)
$$u=-\frac{2}{(x-\es\al y/6+\be)^2}+\frac{2\al'y-\al^2}{72\es}-\frac{\be'}{6}
\eqno(5.2.11)$$ from the  solution $u=-2/x^2$ of the KdV equation;
 (2)
$$u=-2\wp(x-\es\al y/6+\be)+\frac{2\al'y-\al^2}{72\es}-\frac{\be'}{6}
\eqno(5.2.12)$$ from the solution $u=-2\wp(x)$ of the KdV equation;
(3)
$$u=-2b^2\tan^2b(x-\es\al y/6+\be)+\frac{2\al'y-\al^2}{72\es}-\frac{8b^2+\be'}{6}
\eqno(5.2.13)$$ from the solution $u=-2b^2(\tan^2bx+2/3)$ of the KdV
equation; (4)
$$u=-2b^2\coth^2(b(x-\es\al y/6+\be))+\frac{2\al' y-\al^2}{72\es}+\frac{8b^2-\be'}{6}
\eqno(5.2.14)$$from the solution $u=-2b^2(\coth^2bx-2/3)$ of the KdV
equation;
 (5)
$$u=2m^2b^2\mbox{cn}^2(b(x-\es\al y/6+\be)|m)+\frac{2\al' y-\al^2}{72\es}+\frac{(4-8m^2)b^2-\be'}{6}
\eqno(5.2.15)$$ from the solution
$u=2b^2m^2\mbox{cn}^2(bx|m)+(2-4m^2)b^2/3$ of the KdV equation,
which becomes a {\it line-soliton solution}\index{line-soliton
solution}
$$u=2b^2\mbox{sech}^2(b(x-\es c y-(3\es c^2+4b^2)t+a))\eqno(5.2.16)$$\index{soliton solution!of KP equation}
when we take $\al=6c,\;\be=-(3\es c^2+4b^2)t+a$ and let $m\rta 1$;
(6)
$$u=\frac{x-\es\al y/6+\be}{6(t+a)}+\frac{2\al'y-\al^2}{72\es}-\frac{\be'}{6}\eqno(5.2.17)$$
from the solution $u=x/6(t+a)$ of the KdV equation; (7)
\begin{eqnarray*}u&=&2\ptl_x^2\ln[1+k_1e^{b_1(x-\es\al y/6+\be)-b_1^3t}+k_2e^{b_2(x-\es\al y/6+\be)-b_2^3t}
\\ & &
+\left(\frac{b_1-b_2}{b_1+b_2}\right)^2 k_1k_2e^{(b_1+b_2)(x-\es\al
y/6+\be)-(b_1^3+b_2^3)t}]+\frac{2\al'y-\al^2}{72\es}-\frac{\be'}{6}
\hspace{2.25cm}(5.2.18)\end{eqnarray*} from the  solution (5.1.47)
of the KdV equation, which becomes a two-soliton solution\index{two
soliton solution!of KP equation}
\begin{eqnarray*}\qquad u&=&2\ptl_x^2\ln[1+k_1e^{b_1(x-\es c y-3\es c^2t)-b_1^3t}+k_2e^{b_2(x-\es c y-3\es c^2t)-b_2^3t}
\\ & &
+\left(\frac{b_1-b_2}{b_1+b_2}\right)^2 k_1k_2e^{(b_1+b_2)(x-\es c
y-3\es c^2t)-(b_1^3+b_2^3)t}] \hspace{4.6cm}(5.2.19)\end{eqnarray*}
 when we take $\al=6c$ and $\be=-3\es c^2t$.

Next we assume that
$$u=h(t,y)+g(t,y)x+f(t,y)x^2\eqno(5.2.20)$$
is a solution of the KP equation, where $h,g$ and $f$ are functions
in $t$ and $x$ to be determined. Then
$$g_t+3\es h_{yy}+6g^2+12fh+[2f_t+3\es g_{yy}+36fg]x+3(\es
f_{yy}+12f^2)x^2=0,\eqno(5.2.21)$$ equivalently,
$$\es
f_{yy}+12f^2=0,\eqno(5.2.22)$$
$$2f_t+3\es g_{yy}+36fg=0,\eqno(5.2.23)$$
$$g_t+3\es h_{yy}+6g^2+12fh=0.\eqno(5.2.24)$$
Recall the Weierstrass's elliptic function $\wp(z)$ defined in
(3.4.9). Moreover, ${\wp'}'(z)=6\wp^2(z)-g_2/2$ with the $g_2$ given
in (3.4.29). In (3.4.9), we take $\omega_1\in\mbb{C}$ such that
$\mbox{Re}\:\w_1,\mbox{Im}\:\w_1\neq 0$ and $\w_2=\ol{\w_1}$ for
which $g_2=0$. Then $\wp(z)$ is real if $z\in\mbb{R}$. An obvious
solution of the system (5.2.22)-(5.2.24) is $f=-\es \wp(y)/2$ and
$g=h=0$. So $u=-\es x^2\wp(y)/2$ is a solution of the KP equation.
Applying the transformations in (5.2.4) and (5.2.8), we get a more
sophisticated solution
$$u=-\frac{\es}{2} (x-\es
\al'y/6+\be)^2\wp(y+\al)+\frac{2{\al'}'y-{\al'}^2}{72\es}-\frac{\be'}{6}
\eqno(5.2.25)$$ for any differentiable functions $\al,\be$ in $t$.

Note that  $f=-\es/2(y-\al)^2$ is a solution of (5.2.22) for any
function $\al$ in $t$. Replacing $u$ by $T_{3,\al}(u)$ (cf.
(5.2.8)), we can assume $\al=0$, that is, $f=-\es/2y^2$.
Substituting it into (5.2.23), we get
 $g_{yy}=6g/y^2\sim
y^2g_{yy}=6g$. Assume
$$g=\sum_{m\in\mbb{Z}}a_m(t)y^m,\eqno(5.2.26)$$
where $a_m(t)$ are functions in $t$ to be determined. Then
$$\sum_{m\in\mbb{Z}}m(m-1)a_my^m=6\sum_{m\in\mbb{Z}}a_my^m\sim
[m(m-1)-6]a_m=0,\;\;m\in\mbb{Z}.\eqno(5.2.27)$$ Moreover,
$$[m(m-1)-6]a_m=0\sim (m-3)(m+2)a_m=0.\eqno(5.2.28)$$
So $a_m=0$ if $m\neq -2,3$. Hence
$$g=\frac{\be}{y^2}+\gm y^3,\eqno(5.2.29)$$
where $\be$ and $\gm$ are arbitrary functions in $t$.

Recall $u=fx^2+gx+h$ and observe
$$fx^2+gx=\frac{-\es x^2+2\be x}{2y^2}+\gm
y^3x=\frac{-\es(x-\es\be)^2+\es\be^2}{2y^2}+\gm y^3x.\eqno(5.2.30)$$
Replacing $u$ by $T_{2,\es\be}(u)$ (cf. (5.2.4)), we can assume
$\be=0$, that is, $g=\gm y^3$. Next (5.2.24) becomes
$$\gm'y^3+3\es
h_{yy}+6\gm^2y^6-\frac{6\es}{y^2}h=0,\eqno(5.2.31)$$
 equivalently,
$$y^2h_{yy}-2h=-\frac{\es\gm'}{3}y^5-2\es\gm^2y^8.\eqno(5.2.32)$$
 Suppose
$$h=\sum_{m\in\mbb{Z}}b_m(t)y^m,\eqno(5.2.33)$$
where $b_m(t)$ are functions in $t$ to be determined. Substituting
it into (5.2.32), we have
$$\sum_{m\in\mbb{Z}}[m(m-1)-2]b_my^m=-\frac{\es\gm'}{3}y^5-2\es\gm^2y^8.\eqno(5.2.34)$$
Thus
$$b_5=-\frac{\es\gm'}{54},\;\;b_8=-\frac{\es\gm^2}{27},\;\;[m(m-1)-2]b_m=(m-2)(m+1)b_m=0,\;\;m\neq
5,8.\eqno(5.2.35)$$ Hence
$$h=\frac{\vt}{y}+\nu
y^2-\frac{\es\gm'}{54}y^5-\frac{\es\gm^2}{27}y^8,\eqno(5.2.36)$$
where $\vt$ and $\nu$ are two arbitrary functions in $t$. Therefore,
we obtain following solution of the KP equation (5.2.1):
$$u=-\frac{\es x^2}{2y^2}+\gm xy^3+\frac{\vt}{y}+\nu
y^2-\frac{\es\gm'}{54}y^5-\frac{\es\gm^2}{27}y^8.\eqno(5.2.37)$$
Applying the transformations in (5.2.4) and (5.2.8), we have: \psp

{\bf Theorem 5.2.1}. {\it For any functions $\al,\be,\gm,\vt$ and
$\nu$ in $t$, the following is a solution of the KP equation
(5.2.1): \begin{eqnarray*}\qquad u&=&-\frac{\es
(x-\al'y/6\es+\be)^2}{2(y+\al)^2}+\gm
(x-\al'y/6\es+\be)(y+\al)^3+\frac{\vt}{y+\al}\\ & &+\nu
(y+\al)^2-\frac{\es\gm'}{54}(y+\al)^5-\frac{\es\gm^2}{27}(y+\al)^8+\frac{2{\al'}'y-{\al'}^2}{72\es}-\frac{\be'}{6}.
\hspace{1.6cm}(5.2.38)\end{eqnarray*}}\pse

Let $f=0$ in (5.2.22). Then (5.2.23) becomes $g_{yy}=0$. So $g=\al
y+\be$ for some functions $\al$ and $\be$ in $t$. Now (5.2.24)
yields
$$3\es h_{yy}
+6\al^2y^2+(\al'+12\al\be)y+6\be^2+\be'=0.\eqno(5.2.39)$$ Thus we
get the following solution of the KP equation
$$u=(\al y+\be)x-\frac{\es\al^2}{6}y^4-\frac{\es(\al'+12\al\be)}{18}y^3-\frac{\es(6\be^2+\be')}{6}y^2
+\gm y+\theta,\eqno(5.2.40)$$ where $\al,\be,\gm$ and $\theta$ are
arbitrary functions in $t$. Note that the solution (5.2.17) is a
special case of the above solution.

Changing variable $u=2\ptl_x^2\ln v$, we find the following
presentation of the KP equation in Hirota bilinear form\index{Hirota
bilinear form!of the KP equation}
$$D_tD_x(v\cdot v)+D_x^4(v\cdot v)+3\es D_y^2(v\cdot v)=0\eqno(5.2.41)$$
(cf. (5.1.37) and (5.1.49)). Suppose that
$$v=(x+a_0t)^2+by^2+c\eqno(5.2.42)$$
is a solution (5.2.41), where all the coefficients are constants to
be determined and $b\neq 0$. By (5.2.41),
$$2(a_0+3\es b)v-4a_0(x+a_0t)^2+12-12\es b^2y^2=0,\eqno(5.2.43)$$
equivalently,
$$ a_0=3\es b, \;\;c=-\frac{\es}{b^2}.\eqno(5.2.44)$$
So
$$u=2\ptl_x^2\ln v=2\ptl_x^2\ln((x+3\es bt)^2+by^2-\es/b^2)\eqno(5.2.45)$$
is a solution of the KP equation.  Applying the transformations in
(5.2.4) and (5.2.8), we obtain the following solution of the KP
equation:
$$u=2\ptl_x^2\ln((x-\es
\al'y/6+\be+3\es
bt+a)^2+b(y+\al)^2-\es/b^2)+\frac{2{\al'}'y-{\al'}^2}{72\es}-\frac{\be'}{6}.
\eqno(5.2.46)$$ Taking $\al=6\es t$ and $\be=-3\es c^2t$, we  get
the following {\it lump solution} of the KP equation:\index{lump
solution}
$$u=2\ptl_x^2\ln((x-cy+3\es(b-c^2)t+a)^2+b(y+6\es ct)^2-\es/b^2),\eqno(5.2.47)$$
where $a,b,c\in\mbb{R}$ and $b\neq 0$.\psp

Jimbo and Miwa [JM] found the $\tau$-function solutions of the KP
equation via the orbit of the vacuum vector for the fermionic
representation of the general linear group $GL(\infty)$ and the
Boson-Fermion correspondence in quantum field theory.  Kupershmidt
[Kb] found geometric-Hamiltonian form for the KP equation. Cao [Cb2]
found some algebraic approaches to the exact solutions of the
Jimbo-Miwa equation, which is the second equation in the KP
hierarchy.

\section{Equation of Transonic Gas Flows}

Lin, Reisner and Tsien  [LRT] (1948) found the  equation
$$2u_{tx}+u_xu_{xx}-u_{yy}=0\eqno(5.3.1)$$\index{equation of
transonic gas flows}for two-dimensional non-steady motion of a
slender body in a compressible fluid, which was later called the
``equation of
 transonic gas flows" (cf. [Me1]).

Mamontov [Me1]  (1969) obtained the Lie point symmetries of the
above equation  and solved the problem of existence of analytic
solutions in [Me2] (1972). Sevost'janov [Sg] (1977) found explicit
solutions of the equation (5.3.1), describing nonstationary
transonic flows in plane nozzles. Sukhinin [Sv] (1978) studied  the
group property and conservation laws of the equation. In this
section, we give the stable-range approach to the LRT equation
(5.3.1). The results are taken from our work [X8].

First we give an intuitive derivation of the symmetry
transformations of the LRT equation due to Mamontov [Me1].
  Suppose
 $$\mbox{deg}\;u=\ell_1,\qquad \mbox{deg}\;x=\ell_2.\eqno(5.3.2)$$
 To make each nonzero term having the same degree, we have to take
 $$\mbox{deg}\;t=2\ell_2-\ell_1,\qquad
 \mbox{deg}\;y=\frac{3}{2}\ell_2-\frac{1}{2}\ell_1.\eqno(5.3.3)$$
 Since the LRT equation (5..3.1) does not contain variable
 coefficients, it is translation invariant. Thus
 the transformation
$$T^{(a)}_{b_1,b_2}(u(t,x,y))=b_1^2u(b_1^2b_2^4t+a,b_2^2x,b_1b_2^3y)\eqno(5.3.4)$$
 keep the LRT equation invariant for $a,b_1,b_2\in\mbb{R}$ such that $b_1,b_2\neq 0$, with the independent variables $t$
 replaced by $b_1^2b_2^4t+a$, $x$ replaced by $b_2^2x$ and $y$
 replaced by $b_1b_2^3y$, where the
subindices denote the partial derivatives with respect to the
original independent variables. So  $T^{(a)}_{b_1,b_2}$ maps  a
solutions of the LRT equation to another solution.

Let $\al$  be differentiable functions in $t$.  Then the
transformation $u\mapsto u+\al$ keeps (5.3.1) invariant.
 Moreover, the transformation $u(t,x,y)\mapsto u(t,x+\al,y)$ changes
 the LRT equation to
 $$2u_{tx}+2\al'u_{xx}+u_xu_{xx}-u_{yy}=0\eqno(5.3.5)$$
with the independent variables $x$
 replaced by $x+\al$ and the
subindices denoting the partial derivatives with respect to the
original independent variables. Furthermore, the transformation
$u(t,x,y)\mapsto u(t,x,y)-2\al'x$ changes
 the LRT equation to
 $$-4{\al'}'+2u_{tx}-2\al'u_{xx}+u_xu_{xx}-u_{yy}=0.\eqno(5.3.6)$$
In addition, the transformation $u(t,x,y)\mapsto
u(t,x,y)-2{\al'}'y^2$ changes
 the LRT equation to
$$2u_{tx}+u_xu_{xx}-u_{yy}+4{\al'}'=0.\eqno(5.3.7)$$
Thus the transformation
$$T_{2,\al}(u(t,x,y))=u(t,x+\al,y)-2\al'x-2{\al'}'y^2\eqno(5.3.8)$$
keeps the LRT equation invariant with the independent variable $x$
 replaced by $x+\al$ and the
subindices denoting the partial derivatives with respect to the
original independent variables; equivalently, $T_{2,\al}$ maps  a
solutions of the LRT equation to another solution. Since $u=0$ is a
solution, $u=T_{2,\al}(0)=-2\al'x-2{\al'}'y^2$ is a nontrivial
solution of the LRT equation.

Given a  differentiable function $\be$ in $t$, the transformation
$u(t,x,y)\mapsto u(t,x,y+\be)$ changes
 the LRT equation to
 $$2u_{tx}+2\be'u_{xy}+u_xu_{xx}-u_{yy}=0\eqno(5.3.9)$$
with the independent variable $y$
 replaced by $y+\be$ and the
subindices denoting the partial derivatives with respect to the
original independent variables. Moreover, the transformation
$u(t,x,y)\mapsto u(t,x+\be'y,y)$ changes
 the LRT equation to
 $$2u_{tx}+2{\be'}'yu_{xx}+u_xu_{xx}-u_{yy}-2\be'u_{xy}-{\be'}^2u_{xx}=0\eqno(5.3.10)$$
with the independent variable $x$
 replaced by $x+\be'y$. Furthermore,
 the transformation
$u(t,x,y)\mapsto u(t,x,y)-2{\be'}'xy+{\be'}^2x$ changes
 the LRT equation to
$$4\be'{\be'}'-4{{\be'}'}'y+2u_{tx}-2{\be'}'yu_{xx}+{\be'}^2xu_{xx}+u_xu_{xx}-u_{yy}=0.\eqno(5.3.11)$$
In addition, the transformation $u(t,x,y)\mapsto
u(t,x,y)+2\be'{\be'}'y^2-2{{\be'}'}'y^3/3$ changes
 the LRT equation to
$$2u_{tx}+u_xu_{xx}-u_{yy}-4\be'{\be'}'+4{{\be'}'}'y=0.\eqno(5.3.12)$$
Therefore, the transformation
$$T_{3,\be}(u(t,x,y))=u(t,x+\be'y,y+\be)+{\be'}^2x+2\be'{\be'}'y^2-2
{\be'}'xy-\frac{2{{\be'}'}'}{3}y^3\eqno(5.3.13)$$  leave the
equation (5.3.1) invariant with the independent variables $x$
 replaced by $x+\be'y$ and $y$ replaced by $y+\be$, where the
subindices denote the partial derivatives with respect to the
original independent variables. In other words, $T_{3,\be}$ maps  a
solutions of the LRT equation to another solution. In particular,
$u=T_{3,\be}(0)={\be'}^2x+2\be'{\be'}'y^2-2
{\be'}'xy-\frac{2{{\be'}'}'}{3}y^3$ is a solution of the LRT
equation.

 In summary,  the transformation
\begin{eqnarray*}T_{b_1,b_2;\gm}^{(a;\al,\be)}(u(t,x,y))&=&b_1^2u(b_1^2b_2^4t+a,b_2^2(x+\be'y+\al),
b_1b_2^3(y+\be))+\gm\\ & &
+({\be'}^2-2\al')x+2(\be'{\be'}'-{\al'}')y^2-2
{\be'}'xy-\frac{2{{\be'}'}'}{3}y^3\hspace{1.4cm}(5.3.14)\end{eqnarray*}
maps a solutions of the LRT equation to another solution.

 Note that the maximal finite-dimensional subspace of $\mbb{R}[x]$ invariant under the transformation
$u\mapsto u_xu_{xx}$ is $\sum_{r=0}^3\mbb{R}x^r$ ({\it stable
range}). \index{stable range} We look for
  a solution  of the form:
 $$u=f(t,y)+g(t,y)x+h(t,y)x^2+\xi(t,y)x^3,\eqno(5.3.15)$$
where $f(t,y),\;g(t,y),\;h(t,y)$ and $\xi(t,y)$ are
suitably-differentiable functions to be determined. Note
$$u_x=g+2hx+3\xi x^2,\qquad u_{xx}=2h+6\xi x,\eqno(5.3.16)$$
$$u_{tx}=g_t+2h_tx+3\xi_t x^2,\qquad
u_{yy}=f_{yy}+g_{yy}x+h_{yy}x^2+\xi_{yy}x^3,\eqno(5.3.17)$$ Now
(5.3.1) becomes $$2(g_t+2h_tx+3\xi_t x^2)+(g+2hx+3\xi x^2)(2h+6\xi
x)-f_{yy}-g_{yy}x-h_{yy}x^2-\xi_{yy}x^3=0,\eqno(5.3.18)$$ which is
equivalent to the following system of partial differential
equations:
$$\xi_{yy}=18\xi^2,\eqno(5.3.19)$$
$$h_{yy}=6\xi_t+18\xi h,\eqno(5.3.20)$$
$$g_{yy}=4h_t+4h^2+6\xi g,\eqno(5.3.21)$$
$$f_{yy}=2g_t+2gh.\eqno(5.3.22)$$
Recall the Weierstrass's elliptic function $\wp(z)$ defined in
(3.4.9). Moreover, ${\wp'}'(z)=6\wp^2(z)-g_2/2$ with the $g_2$ given
in (3.4.29). In (3.4.9), we take $\omega_1\in\mbb{C}$ such that
$\mbox{Re}\:\w_1,\mbox{Im}\:\w_1\neq 0$ and $\w_2=\ol{\w_1}$ for
which $g_2=0$. Then $\wp(z)$ is real if $z\in\mbb{R}$.
 An obvious
solution of the equation (5.3.19)-(5.3.22) is $\xi=\wp(y)/3$ and
$f=g=h=0$. So $u=x^3\wp(y)/3$ is a solution of the LRT equation
(5.3.1). Applying the transformation $T_{1,1;\gm}^{(0;\al,\be)}$ in
(5.3.14), we get a more sophisticated solution
$$u=\frac{1}{3}(x+\be'y+\al)^3\wp(y+\be)+({\be'}^2-2\al')x+2(\be'{\be'}'-{\al'}')y^2-2
{\be'}'xy-\frac{2{{\be'}'}'}{3}y^3 +\gm\eqno(5.3.23)$$ of the LRT
equation (5.3.1).

Observe that
$$\xi=\frac{1}{3y^2}\eqno(5.3.24)$$ is a solution of
the equation (5.3.19). Substituting (5.3.24) into (5.3.20), we get
$$h_{yy}=\frac{6}{y^2}h.\eqno(5.3.25)$$
 Write
$$h(t,y)=\sum_{m\in\mbb{Z}}\frac{a_m(t)}{y^m}.\eqno(5.3.26)$$
Then (5.3.25) is equivalent to
$$[m(m+1)-6]a_m=0\sim
(m-2)(m+3)a_m=0\qquad\for\;\;m\in\mbb{Z}.\eqno(5.3.27)$$ Thus
$$h=\frac{\al}{y^2}+\gm y^3,\eqno(5.3.28)$$
where $\al$ and $\gm$ are arbitrary differentiable functions in $t$.

Note
$$\xi x^3+hx^2=\frac{x^3+3\al x^2}{3y^2}+\gm y^3x^2=\frac{(x+\al)^3-3\al^2 x-\al^3}{3y^2}+\gm y^3x^2.\eqno(5.3.29)$$
Replacing $u$ by $T_{2,-\al}(u)$ (cf. (5.3.8)), we can assume
$\al=0$, that is, $h=\gm y^3$. Now
$$h_t=\gm'y^3,\;\;h^2=\gm^2y^6.\eqno(5.3.30)$$
Substituting the above equation into (5.3.21), we have:
$$g_{yy}-\frac{2}{y^2}g
 =4\gm' y^3+4\gm^2y^6.\eqno(5.3.31)$$
Write
$$g(t,y)=\sum_{m\in\mbb{Z}}b_m(t)y^m.\eqno(5.3.32)$$
Then (5.3.31) is equivalent to
$$b_5=\frac{2\gm'}{9},\;\;b_8=\frac{2\gm^2}{27},\;\;(m+1)(m-2)a_m=0,\;\;m\neq
5,8.\eqno(5.3.33)$$ So
$$g=\frac{\vt}{y}+\Im y^2+\frac{2\gm'}{9}y^5+\frac{2\gm^2}{27}y^8,
\eqno(5.3.34)$$ where $\vt$ and $\Im$ are arbitrary differentiable
functions in $t$.

Observe that
$$g_t=\frac{\vt'}{y}+\Im'y^2+\frac{2{\gm'}'}{9}y^5+\frac{4\gm\gm'}{27}y^8,
\eqno(5.3.35)$$
$$gh=\gm\vt y^2+\gm\Im
y^5+\frac{2\gm\gm'}{9}y^8+\frac{2\gm^3}{27}y^{11}.\eqno(5.3.36)$$
Hence (5.3.22) becomes
$$f_{yy}=2\left[\frac{\vt'}{y}+(\Im'+\gm\vt)y^2+\frac{9\gm\Im+2{\gm'}'}{9}y^5
+\frac{10\gm\gm'}{27}y^8+\frac{2\gm^3}{27}y^{11}\right].\eqno(5.3.37)$$
Therefore,
$$f=2\vt'y(\ln
y-1)+\frac{\Im'+\gm\vt}{6}y^4+\frac{9\gm\Im+2{\gm'}'}{189}y^7
+\frac{2\gm\gm'}{243}y^{10}+\frac{\gm^3}{1053}y^{13}+\rho
y,\eqno(5.3.38)$$ where $\rho$ is any function in $t$.
 \psp

{\bf Theorem 5.3.1}. {\it Let $\al,\be,\gm,\vt,\Im,\rho,\mu$ be
arbitrary functions in $t$, which are differentiable up to need.
 We have the following solution of the LRT
equation (5.3.1):
\begin{eqnarray*}u&=&\vf=\frac{x^3}{3y^2}+\gm x^2y^3+
\left(\frac{\vt}{y}+\Im
y^2+\frac{2\gm'}{9}y^5+\frac{2\gm^2}{27}y^8\right)x+2\vt'y(\ln
y-1)\\ & &+\frac{\Im'+\gm\vt}{6}y^4+\frac{9\gm\Im+2{\gm'}'}{189}y^7
+\frac{2\gm\gm'}{243}y^{10}+\frac{\gm^3}{1053}y^{13}+\rho
y.\hspace{3.5cm}(5.3.39)\end{eqnarray*}  Moreover,
$u=T^{(0;\al,-\be)}_{1,1;\mu}(\vf)$ is solution of the LRT equation
(5.3.1) blowing up on the moving line $y=\be(t)$.}\psp

We remark that the solution $u=T^{(0,\al,-\be)}_{1,1;\mu}(\vf)$ may
reflect the phenomenon of abrupt high-speed wind. If we take
$\vf=x^3/3y^2$, then
$$u=\frac{(x+\be'y+\al)^3}{3(y-\be)^2}
+({\be'}^2-2\al')x+2(\be'{\be'}'-{\al'}')y^2-2
{\be'}'xy-\frac{2{{\be'}'}'}{3}y^3+\mu.\eqno(5.3.40)$$

Take the trivial solution $\xi=0$ of (5.3.19), which is the only
solution polynomial in $y$. Then (5.3.20) and (5.3.21) become
$$h_{yy}=0,\qquad g_{yy}=4h_t+4h^2.\eqno(5.3.41)$$
Replacing $u$ by $T_{3,\al}(u)$ for some proper function $\al$ in
$t$ if necessary (cf. (5.3.13)), we can  take $h=\be y$, where $\be$
is an arbitrary function in $t$. Hence
$$g_{yy}=4\be'y+4\be^2 y^2.\eqno(5.3.42)$$
So
$$g=\gm+\sgm y+\frac{2\be'}{3}y^3+\frac{\be^2}{3}
y^4,\eqno(5.3.43)$$ where $\gm$ and $\sgm$ are arbitrary functions
in $t$. Now (5.3.22) yields
$$f_{yy}=2\gm'+2(\be\gm+\sgm')y+2\be\sgm y^2+\frac{4{\be'}'}{3}y^3+\frac{8\be\be'}{3}y^4+ \frac{2}{3}\be^3
y^5.\eqno(5.3.44)$$ Replacing $u$ by some $T_{1,1;\al}^{(0;0,0)}(u)$
if necessary (cf. (5.3.14)), we have $$ f=\rho
y+\gm'y^2+\frac{\be\gm+\sgm'}{3}y^3 +\frac{\be\sgm}{6}y^4
+\frac{{\be'}'}{15}y^5+\frac{4\be\be'}{45}y^6+ \frac{\be^3}{63}
y^7.\eqno(5.3.45)$$ \pse

{\bf Theorem 5.3.2}. {\it The following is a solution of the
equation (5.3.1):
\begin{eqnarray*} \qquad u&=&\psi=\be x^2y+\left(\gm+\sgm y+\frac{2\be'}{3}y^3+\frac{\be^2}{3}
y^4\right)x+\rho y+\gm'y^2\\ & &+\frac{\be\gm+\sgm'}{3}y^3
+\frac{\be\sgm}{6}y^4+\frac{{\be'}'}{15}y^5+\frac{4\be\be'}{45}y^6+
\frac{\be^3}{63} y^7,\hspace{3.9cm}(5.3.46)\end{eqnarray*} where
$\be,\gm,\sgm$ and $\rho$  are arbitrary functions in $t$. Moreover,
any solution polynomial in $x$ and $y$ of (5.3.1) must be of the
form $u=T_{1,1;\vt}^{(0;0,\al)}(\psi)$, where $\al$ and $\vt$ are
another two arbitrary functions in $t$.}

{\it Proof}. We only need to prove the last statement. Suppose that
$u$ is a solution of (5.3.1) polynomial in $x$ and $y$. By comparing
the term with highest degree of $x$, $u$ must be of the form
(5.3.15) and (5.3.19)-(5.3.22) hold. Since $\xi$ is polynomial in
$y$, (5.3.19) forces $\xi=0$. Then the conclusion follows from the
 arguments (5.3.41)-(5.3.45). $\qquad\Box$

\section{Short  Wave Equation}

Khristianovich and Rizhov [KR] (1958) discovered the equations of
short waves
$$u_y-2v_t-2(v-x)v_x-2kv=0,\;\;v_y+u_x=0\eqno(5.4.1)$$\index{short
wave equation}
 in connection with the nonlinear reflection of weak
shock waves,  where $k$ is a real constant. Bagdoev and Petrosyan
[BP] (1985)  showed that the modulation equation of a gas-fluid
mixture coincides in main orders with the corresponding short-wave
equations. Kraenkel, Manna and Merle [KMM] (2000) studied nonlinear
short-wave propagation in ferrites and Ermakov  [Es] (2006)
investigated short-wave interaction in film slicks. By the second
equation in (5.4.1), there exist a potential function $w(t,x,y)$
such that $u=w_y$ and $v=-w_x$. Then the first equation becomes:
$$2w_{tx}-2(x+w_x)w_{xx}+w_{yy}+2kw_x=0.\eqno(5.4.2)$$
To solve the short wave equations (5.4.1) is equivalent to solve the
equation (5.4.2). The reader may find the other interesting results
in literatures such as [RRD, Kp]. In this section, we want to solve
the short wave equation by the stable-range approach. The results
come from our work [X13]

  The
symmetry group and conservation laws of (5.4.2) were first studied
by Kucharczyk [Kp] (1965) and later by Khamitova [Kr] (1982).   Let
$\al$ be a differentiable function in $t$. Note that the
transformation $w(t,x,y)\mapsto w(t,x+\al,y)$ changes the equation
(5.4.2) to
$$2\al'w_{xx}+2w_{tx}-2(x+\al+w_x)w_{xx}+w_{yy}+2kw_x=0\eqno(5.4.3)$$
with the independent variables $x$
 replaced by $x+\al$ and the
subindices denoting the partial derivatives with respect to the
original independent variables. Moreover,   the transformation
$w(t,x,y)\mapsto w(t,x,y)+(\al'-\al)x$ changes the equation (5.4.2)
to
$$2({\al'}'-\al')+2w_{tx}-2\al'w_{xx}-2(x-\al+w_x)w_{xx}+w_{yy}+2kw_x+2k(\al'-\al)=0.\eqno(5.4.4)$$
Furthermore, the transformation $w(t,x,y)\mapsto
w(t,x,y)+(k\al+(1-k)\al'-{\al'}')y^2$ changes the equation (5.4.2)
to
$$2w_{tx}-2(x+w_x)w_{xx}+w_{yy}+2(k\al+(1-k)\al'-{\al'}')+2kw_x=0.\eqno(5.4.5)$$
Thus the transformation
$$T_{2,\al}(w(t,x,y))=w(t,x+\al,y)+(\al'-\al)x+(k\al+(1-k)\al'-{\al'}')y^2\eqno(5.4.6)$$
keeps the equation (5.4.2) invariant with the independent variable
$x$ replaced by $x+\al$, that is, the transformation $T_{2,\al}$
maps a solution of (5.4.2) to another solution. In particular,
$T_{2,\al}(0)=(\al'-\al)x+(k\al+(1-k)\al'-{\al'}')y^2$ is a solution
of the equation (5.4.2).

Given a  differentiable function $\be$ in $t$, the transformation
$w(t,x,y)\mapsto w(t,x,y+\be)$ changes
 the  equation (5.4.2) to
$$2\be'w_{xy}+2w_{tx}-2(x+w_x)w_{xx}+w_{yy}+2kw_x=0\eqno(5.4.7)$$
with the independent variable $y$
 replaced by $y+\be$ and the
subindices denoting the partial derivatives with respect to the
original independent variables. Moreover, the transformation
$w(t,x,y)\mapsto w(t,x-\be'y,y)$ changes the  equation (5.4.2) to
$$-2{\be'}'yw_{xx}+2w_{tx}-2(x-\be'y+w_x)w_{xx}+w_{yy}-2\be'w_{xy}+{\be'}^2w_{xx}+2kw_x=0\eqno(5.4.8)$$
with the independent variable $x$
 replaced by $x-\be'y$. Furthermore,
 the transformation
$w(t,x,y)\mapsto w(t,x,y)+{\be'}^2x/2+(\be'-{\be'}')xy$ changes the
equation (5.4.2) to
\begin{eqnarray*}\qquad & &
2\be'{\be'}'+2({\be'}'-{{\be'}'}')y
+2w_{tx}-2(x+{\be'}^2/2+(\be'-{\be'}')y+w_x)w_{xx}\\
&
&+w_{yy}+2kw_x+k{\be'}^2+2k(\be'-{\be'}')y=0.\hspace{6cm}(5.4.9)\end{eqnarray*}
In addition, the transformation
$$w(t,x,y)\mapsto
w(t,x,y)-(\be'{\be'}'+k{\be'}^2/2)y^2+({{\be'}'}'+(k-1){\be'}'-k{\be'})y^3/3\eqno(5.4.10)$$
changes the equation (5.4.2) to
$$2w_{tx}-2(x+w_x)w_{xx}+w_{yy}-(2\be'{\be'}'+k{\be'}^2)+2({{\be'}'}'+(k-1){\be'}'-k{\be'})y
+2kw_x=0.\eqno(5.4.11)$$ Therefore, the transformation
\begin{eqnarray*} \qquad T_{3,\be}(w(t,x,y))&=&w(t,x-\be'y,y+\be)
+{\be'}^2x/2+(\be'-{\be'}')xy\\ &
&-(\be'{\be'}'+k{\be'}^2/2)y^2+({{\be'}'}'+(k-1){\be'}'-k{\be'})y^3/3\hspace{1.2cm}(5.4.12)\end{eqnarray*}
 leaves the
equation (5.4.2) invariant with the independent variables $x$
 replaced by $x-\be'y$ and $y$ replaced by $y+\be$, where the
subindices denote the partial derivatives with respect to the
original independent variables. In other words, $T_{3,\be}$ maps  a
solutions of the equation (5.4.2) to another solution. In
particular,
$$u=T_{3,\be}(0)={\be'}^2x/2+(\be'-{\be'}')xy-(\be'{\be'}'+k{\be'}^2/2)y^2+({{\be'}'}'
+(k-1){\be'}'-k{\be'})y^3/3\eqno(5.4.13)$$ is a solution of the
equation (5.4.2).

To make each term in (5.4.2) having the same degree, we take
$$\mbox{deg}\:w=2\:\mbox{deg}\:x=4\:\mbox{deg}\:y,\qquad\mbox{deg}\:t=0.\eqno(5.4.14)$$
Thus the transformation
$$T_{a,b}(w(t,x,y))=b^{-4}w(t+a,b^2x,by)\eqno(5.4.15)$$
keeps the equation (5.4.2) invariant with the independent variables
$t$ replaced by $t+a$, where $a,b\in\mbb{R}$ and $b\neq 0$. In
summary, the transformation
\begin{eqnarray*}& &
T_{a,b;\gm}^{(\al,\be)}(w(t,x,y))\\&=&b^{-4}w(t+a,b^2(x-\be'y+\al),b(y+\be))
+\gm+(\be'-{\be'}')xy+[k\al+(1-k)\al'-{\al'}'
\\ &
&-\be'{\be'}'-k{\be'}^2/2]y^2+({{\be'}'}'+(k-1){\be'}'-k{\be'})y^3/3+({\be'}^2/2+\al'-\al)x\hspace{1.1cm}(5.4.16)\end{eqnarray*}
 maps  a
solutions of the equation (5.4.2) to another solution, where
$\al,\be,\gm$ are functions in $t$ and $a,b\in\mbb{R}$ with $b\neq
0$.

In this section, we study solutions polynomial in $x$ for the short
wave equation (5.4.2).
 By comparing the terms of highest degree in $x$, we find that
 such a solution must be of the form:
 $$w=f(t,y)+g(t,y)x+h(t,y)x^2+\xi(t,y)x^3,\eqno(5.4.17)$$
where $f(t,y),\;g(t,y),\;h(t,y)$ and $\xi(t,y)$ are
suitably-differentiable functions to be determined. Note
$$w_x=g+2hx+3\xi x^2,\qquad w_{xx}=2h+6\xi x,\eqno(5.4.18)$$
$$w_{tx}=g_t+2h_tx+3\xi_t x^2,\qquad
w_{yy}=f_{yy}+g_{yy}x+h_{yy}x^2+\xi_{yy}x^3,\eqno(5.4.19)$$ Now
(5.4.2) becomes \begin{eqnarray*}\hspace{1cm}& & 2(g_t+2h_tx+3\xi_t
x^2)-2(g+(2h+1)x+3\xi x^2)(2h+6\xi x)\\ & &
+f_{yy}+g_{yy}x+h_{yy}x^2+\xi_{yy}x^3+2k(g+2hx+3\xi
x^2)=0,\hspace{2.9cm}(5.4.20)\end{eqnarray*} which is equivalent to
the following systems of partial differential equations:
$$\xi_{yy}=36\xi^2,\eqno(5.4.21)$$
$$h_{yy}=6\xi(6h+2-k)-6\xi_t,\eqno(5.4.22)$$
$$g_{yy}=8h^2+4(1-k)h+12\xi g-4h_t,\eqno(5.4.23)$$
$$f_{yy}=4gh-2g_t-2kg.\eqno(5.4.24)$$

First we observe that
$$\xi=\frac{1}{6y^2}\eqno(5.4.25)$$ is a solution of
the equation (5.4.21). Substituting (5.4.25) into (5.4.22), we get
$$h_{yy}=\frac{6h+2-k}{y^2}.\eqno(5.4.26)$$
Write
$$h(t,y)=\sum_{m\in\mbb{Z}}\frac{a_m(t)}{y^m},\eqno(5.4.27)$$
where $a_m(t)$ are functions in $t$ to be determined. Then (5.4.26)
is equivalent to
$$a_0=\frac{k-2}{6},\;\;[m(m+1)-6]a_m=(m+3)(m-2)a_m=0,\;\;m\neq 0.\eqno(5.4.28)$$
 Thus
$$h=\frac{\al}{y^2}+\frac{k-2}{6}
+\gm y^3,\eqno(5.4.29)$$ where $\al$ and $\gm$ are arbitrary
differentiable functions in $t$. Observe
$$\xi x^3+hx^2=\frac{x^3+6\al x^2}{6y^2}+\frac{k-2}{6}x^2+\gm x^2
y^3.\eqno(5.4.30)$$ Replacing $w$ by $T_{2,-2\al}(w)$, we can take
$\al=0$, that is,
$$h=\frac{k-2}{6}
+\gm y^3.\eqno(5.4.31)$$

Calculate
$$h^2=\frac{(k-2)^2}{36}+\frac{(k-2)\gm}{3}y^3 +\gm^2 y^6.\eqno(5.4.32)$$ Note (5.4.23)
becomes
$$g_{yy}-\frac{2g}{y^2}
 =\frac{2(k-2)(1-2k)}{9}-\frac{4[(k+1)\gm+3\gm']}{3}y^3 +8\gm^2
 y^6.\eqno(5.4.33)$$
 Write
$$g(t,y)=\sum_{m\in\mbb{Z}}b_m(t)y^m,\eqno(5.4.34)$$
where $b_m(t)$ are functions in $t$ to be determined. Now (5.4.33)
is equivalent to
$$(k-2)(1-2k)=0,\;\;b_5=-\frac{2(k+1)\gm}{27},\;\;b_8=\frac{4\gm^2}{27},\eqno(5.4.35)$$
$$m(m+3)b_{m+2}=0,\qquad m\neq 0,3,6.\eqno(5.4.36)$$
Thus $k=1/2,\;2$ and
$$g=\frac{\vt}{y}+\sgm y^2-\frac{2(k+1)\gm+6\gm'}{27}y^5+\frac{4\gm^2}{27}y^8,\eqno(5.4.37)$$
where $\vt$ and $\sgm$ are arbitrary differentiable  functions in
$t$.

Note
$$g_t=\frac{\vt'}{y}+\sgm'y^2-\frac{2(k+1)\gm'+6{\gm'}'}{27}y^5+\frac{8\gm\gm'}{27}y^8,\eqno(5.4.38)$$
\begin{eqnarray*}\qquad gh&=&\frac{(k-2)\vt}{6y}+\frac{(k-2)\sgm+6\gm\vt}{6}y^2+\frac{81\gm\sgm-[(k+1)\gm+3\gm'](k-2)}{81}y^5
\\ &
&-\frac{2\gm[(2k+5)\gm+9\gm']}{81}y^8+\frac{4\gm^3}{27}y^{11}.\hspace{6.2cm}(5.4.39)\end{eqnarray*}
Thus (5.4.24) becomes
\begin{eqnarray*}f_{yy}&=&-\frac{4(k+1)\vt+6\vt'}{3y}-\frac{4(k+1)\sgm+6\sgm'-12\gm\vt}{3}y^2
-\frac{40\gm[(k+1)\gm+3\gm']}{81}y^8\\ & &
+\frac{8(k+1)^2\gm+12(3k+2)\gm'-36{\gm'}'+244\gm\sgm}{81}y^5+\frac{16\gm^3}{27}y^{11}
.\hspace{2.6cm}(5.4.40)\end{eqnarray*} So
\begin{eqnarray*}f&=&\frac{4(k+1)\vt+6\vt'}{3}y(1-\ln
y)-\frac{2(k+1)\sgm+3\sgm'-6\gm\vt}{18}y^4
-\frac{4\gm[(k+1)\gm+3\gm']}{729}y^{10}\\ & &
+\frac{4(k+1)^2\gm+6(3k+2)\gm'-18{\gm'}'+122\gm\sgm}{1701}y^7+\frac{4\gm^3}{1053}y^{13}+\vs
y,\hspace{2.2cm}(5.4.41)\end{eqnarray*}
 where $\vs$ is an arbitrary functions in $t$. \psp

{\bf Theorem 5.4.1}. {\it Suppose $k=1/2,\;2$. We have the following
solution of the equation (5.4.2):
\begin{eqnarray*}
w&=&\psi=\frac{x^3}{6y^2}+\left(\frac{k-2}{6} +\gm
y^3\right)x^2+\left(\frac{\vt}{y}+\sgm
y^2-\frac{2(k+1)\gm+6\gm'}{27}y^5+\frac{4\gm^2}{27}y^8\right)x+\vs
y\\ & &+\frac{4(k+1)\vt+6\vt'}{3}y(1-\ln
y)-\frac{2(k+1)\sgm+3\sgm'-6\gm\vt}{18}y^4
-\frac{4\gm[(k+1)\gm+3\gm']}{729}y^{10}\\ & &
+\frac{4(k+1)^2\gm+6(3k+2)\gm'-18{\gm'}'+122\gm\sgm}{1701}y^7+\frac{4\gm^3}{1053}y^{13}+\vs
y,\hspace{2.1cm}(5.4.42)\end{eqnarray*}
 where
$\gm,\vt,\sgm$ and $\vs$ are arbitrary functions in $t$, whose
derivatives appeared in the above exist in a certain open set of
$\mbb{R}$. Moreover, $w=T_{0,1;\zeta}^{(\al,-\be)}(\psi)$ is
solution of the equation (5.4.2) blowing up on the moving line
$y=\be(t)$.}\psp

The simplest case is
$$\psi=\frac{x^3}{6y^2}+\frac{k-2}{6}x^2.\eqno(5.4.43)$$
So the simplest solution of the equation (5.4.2) blowing up on the
moving line $y=\be(t)$ is
\begin{eqnarray*} \qquad w&=&\frac{(x+\be'y)^3}{6(y-\be)^2}+\frac{k-2}{6}(x+\be'y)^2
+\frac{{\be'}^2x}{2}+({\be'}'-\be')xy\\ &
&-\left(\be'{\be'}'+\frac{k{\be'}^2}{2}\right)y^2-\frac{{{\be'}'}'+(k-1){\be'}'-k{\be'}}{3}y^3.\hspace{3.85cm}(5.4.44)\end{eqnarray*}

Take the trivial solution $\xi=0$ of (5.4.21), which is the only
solution polynomial in $y$. Then (5.4.22) and (5.4.23) become
$$h_{yy}=0,\qquad g_{yy}=8h^2+4(1-k)h-4h_t,\eqno(5.4.45)$$
Replacing $u$ by some $T_{3,\be}(u)$ if necessary (cf.(5.4.13)), we
have $h=\gm y$ for some function $\gm$ in $t$. Hence
$$g_{yy}=8\gm^2y^2+4(1-k)\gm y-4\gm' y.\eqno(5.4.46)$$
So
$$g=\frac{2\gm^2}{3}y^4+\frac{2[(1-k)\gm-\gm']}{3}y^3+\vt
y+\rho\eqno(5.4.47)$$ for some functions $\vt$ and $\rho$ in $t$.

Observe
$$gh=\frac{2\gm^3}{3}y^5+\frac{2\gm[(1-k)\gm-\gm']}{3}y^4+\gm\vt
y^2+\gm \rho y\eqno(5.4.48)$$ and
$$g_t=\frac{4\gm\gm'}{3}y^4+\frac{2[(1-k)\gm'-{\gm'}']}{3}y^3+\vt'
y+\rho'.\eqno(5.4.49)$$
 Now (5.4.24) yields
\begin{eqnarray*}\qquad f_{yy}&=&\frac{8\gm^3}{3}y^5
+\frac{4\gm[(2-3k)\gm-4\gm']}{3}y^4-
\frac{4[k(1-k)\gm-(2k-1)\gm'-{\gm'}']}{3}y^3
\\ & &+4\gm\vt
y^2+(4\gm\rho-2k\vt-2\vt')y-2k\rho-2\rho'.\hspace{4.6cm}(5.4.50)\end{eqnarray*}
Replacing $u$ by some $T_{0,1;\al}^{(0;0,0)}(u)$ if necessary (cf.
(5.4.16)), we have
\begin{eqnarray*} f&=&
\frac{4\gm^3}{63}y^7 +\frac{2\gm[(2-3k)\gm-4\gm']}{45}y^6-
\frac{k(1-k)\gm-(2k-1)\gm'-{\gm'}'}{15}y^5
\\ & &+\frac{\gm\vt}{3}
y^4+\frac{2\gm\rho-k\vt-\vt'}{3}y^3-(k\rho+\rho')y^2+\vs
y\hspace{5.2cm}(5.4.51)\end{eqnarray*} for some function $\vs$ of
$t$. \psp

{\bf Theorem 5.4.2}. {\it The following is a solution of the
equation (5.4.2):
\begin{eqnarray*}w &=&\vf=\gm x^2y+\left(\frac{2\gm^2}{3}y^4+\frac{2[(1-k)\gm-\gm']}{3}y^3+\vt
y+\rho\right)x\\ & &+\frac{4\gm^3}{63}y^7
+\frac{2\gm[(2-3k)\gm-4\gm']}{45}y^6-
\frac{k(1-k)\gm-(2k-1)\gm'-{\gm'}'}{15}y^5
\\ & &+\frac{\gm\vt}{3}
y^4+\frac{2\gm\rho-k\vt-\vt'}{3}y^3-(k\rho+\rho')y^2+\vs
y\hspace{4.9cm}(5.4.52)\end{eqnarray*}
 where
$\gm,\vt,\rho$ and $\vs$ are arbitrary functions in $t$, whose
derivatives  exist as they appear. Moreover, any solution polynomial
in $x$ and $y$ of (5.4.2) must be of the above form
$w=T_{0,1;\al}^{(0,\be)}(\vf)$, where $\al$ and $\be$ are another
arbitrary functions in $t$. }

\section{Khokhlov and Zabolotskaya  Equation}

 Khokhlov and Zabolotskaya [KZ] (1969) found the equation
$$2u_{tx}+(uu_x)_x-u_{yy}=0.\eqno(5.5.1)$$\index{Khokhlov-Zabolotskaya
equation}for quasi-plane waves in nonlinear acoustics of bounded
bundles. More specifically,  the equation describes the propagation
of a diffraction sound beam in a nonlinear medium. Kupershmidt [Kb]
(1994) constructed a geometric Hamiltonian form for the
Khokhlov-Zabolotskaya equation (5.5.1). Certain group-invariant
solutions of (5.1.1) were found by Korsunskii [Ks] (1991), and by
Lin and Zhang [LZ] (1995). Sanchez [Sd] (2005) studied long waves in
ferromagnetic media via Khokhlov-Zabolotskaya equation. There are
the other interesting results on the equation (e.g., cf. [Gj, KS,
KiPg, Mo, RN, Ra1, Ra2, Sf, Va]). In this section, we present the
stable-range approach to the equation (5.5.1) due to our work [X13].

Suppose
 $$\mbox{deg}\;u=\ell_1,\qquad \mbox{deg}\;x=\ell_2.\eqno(5.5.2)$$
 To make each nonzero term in (5.5.1) having the same degree, we have to take
 $$\mbox{deg}\;t=\ell_2-\ell_1,\qquad
 \mbox{deg}\;y=\ell_2-\frac{1}{2}\ell_1.\eqno(5.5.3)$$
 Since the Khokhlov-Zabolotskaya equation (5.5.1) does not contain variable
 coefficients, it is translation invariant. Thus
 the transformation
$$T^{(a)}_{b_1,b_2}(u(t,x,y))=b_1^2u(b_1^2b_2t+a,b_2x,b_1b_2y)\eqno(5.5.4)$$
 keeps the Khokhlov-Zabolotskaya equation invariant for $a,b_1,b_2\in\mbb{R}$ such that $b_1,b_2\neq 0$,  with the independent variables $t$
 replaced by $b_1^2b_2t+a$, $x$ replaced by $b_2x$ and $y$
 replaced by $b_1b_2y$, where the
subindices denote the partial derivatives with respect to the
original independent variables. So  $T^{(a)}_{b_1,b_2}$ maps  a
solutions of the  equation to another solution.

Let $\al$  be differentiable functions in $t$.  Then the
transformation $u(t,x,y)\mapsto u(t,x+\al,y)$ changes
 the Khokhlov-Zabolotskaya equation to
 $$2\al'u_{xx}+2u_{tx}+(uu_x)_x-u_{yy}=0.\eqno(5.5.5)$$
with the independent variables $x$
 replaced by $x+\al$ and the
subindices denoting the partial derivatives with respect to the
original independent variables. Furthermore, the transformation
$u(t,x,y)\mapsto u(t,x,y)-2\al'$ changes
 the Khokhlov-Zabolotskaya equation to
 $$2u_{tx}-2\al'u_{xx}+(uu_x)_x-u_{yy}=0.\eqno(5.5.6)$$
Thus the transformation
$$T_{2,\al}(u(t,x,y))=u(t,x+\al,y)-2\al'\eqno(5.5.7)$$
keep the Khokhlov-Zabolotskaya equation invariant with the
independent variables $x$
 replaced by $x+\al$ and the
subindices denoting the partial derivatives with respect to the
original independent variables; equivalently, $T_{2,\al}$ maps  a
solutions of the Khokhlov-Zabolotskaya equation to another solution.

Given a  differentiable function $\be$ in $t$, the transformation
$u(t,x,y)\mapsto u(t,x,y+\be)$ changes
 the Khokhlov-Zabolotskaya equation to
 $$2u_{tx}+2\be'u_{xy}+(uu_x)_x-u_{yy}=0\eqno(5.5.8)$$
with the independent variable $y$
 replaced by $y+\be$ and the
subindices denoting the partial derivatives with respect to the
original independent variables. Moreover, the transformation
$u(t,x,y)\mapsto u(t,x+\be'y,y)$ changes
 the Khokhlov-Zabolotskaya equation to
 $$2u_{tx}+2{\be'}'yu_{xx}+(uu_x)_x-u_{yy}-2\be'u_{xy}-{\be'}^2u_{xx}=0\eqno(5.5.9)$$
with the independent variable $x$
 replaced by $x+\be'y$. Furthermore,
 the transformation
$u(t,x,y)\mapsto u(t,x,y)-2{\be'}'y+{\be'}^2$ changes
 the Khokhlov-Zabolotskaya equation to
$$2u_{tx}-2{\be'}'yu_{xx}+{\be'}^2u_{xx}+(uu_x)_x-u_{yy}=0.\eqno(5.5.10)$$
Therefore, the transformation
$$T_{3,\be}(u(t,x,y))=u(t,x+\be'y,y+\be)+{\be'}^2-2
{\be'}'y\eqno(5.5.11)$$  leaves the equation (5.5.1) invariant with
the independent variables $x$
 replaced by $x+\be'y$ and $y$ replaced by $y+\be$, where the
subindices denote the partial derivatives with respect to the
original independent variables. In other words, $T_{3,\be}$ maps  a
solutions of the Khokhlov-Zabolotskaya  equation to another
solution.

In summary,  the transformation
$$T_{a;b_1,b_2}^{(\al,\be)}(u(t,x,y))=b_1^2u(b_1^2b_2t+a,b_2(x+\be'y+\al),
b_1b_2(y+\be))-2\al'+{\be'}^2-2{\be'}'y\eqno(5.5.12)$$
 maps a solutions of the Khokhlov-Zabolotskaya  equation
to another solution.

Comparing the terms with highest degree of $x$, we find that the
solution of the equation (5.5.1) polynomial in $x$ must be of the
form
$$u=f(t,y)+g(t,y)x+\xi(t,y)x^2.\eqno(5.5.13)$$
Then
$$u_x=g+2\xi x,\qquad u_{tx}=g_t+2\xi_t x,\qquad
u_{yy}=f_{yy}+g_{yy}x+\xi_{yy}x^2,\eqno(5.5.14)$$
$$(uu_x)_x=\ptl_x(fg+(g^2+2f\xi)x+3g\xi x^2+2\xi^2 x^3)=g^2+2f\xi+6g\xi x+6\xi^2
x^2.\eqno(5.5.15)$$ Substituting them into (5.5.1), we get
$$2(g_t+2\xi_t x)+g^2+2f\xi+6g\xi
x+6\xi^2x^2-f_{yy}-g_{yy}x-\xi_{yy}x^2=0,\eqno(5.5.16)$$
equivalently,
$$\xi_{yy}=6\xi^2,\eqno(5.5.17)$$
$$g_{yy}-6g\xi=4\xi_t,\eqno(5.5.18)$$
$$f_{yy}-2f\xi=2g_t+g^2.\eqno(5.5.19)$$

Recall the Weierstrass's elliptic function $\wp(z)$ defined in
(3.4.9). Moreover, ${\wp'}'(z)=6\wp^2(z)-g_2/2$ with the $g_2$ given
in (3.4.29). In (3.4.9), we take $\omega_1\in\mbb{C}$ such that
$\mbox{Re}\:\w_1,\mbox{Im}\:\w_1\neq 0$ and $\w_2=\ol{\w_1}$ for
which $g_2=0$. Then $\wp(z)$ is real if $z\in\mbb{R}$.
 An obvious
solution of the equation (5.5.17)-(5.5.19) is $\xi=\wp(y)$ and
$g=f=0$. Applying $T_{0;1,1}^{(\al,\be)}$, we obtain a more
sophisticated solution
$$u=(x+\be'
y+\al)^2\wp(y+\be)-2\al'+{\be'}^2-2{\be'}'y.\eqno(5.5.20)$$

Observe that $\xi=1/y^2$ is a solution of the equation (5.5.17).
Substituting it into (5.5.18), we obtain
$$g_{yy}-\frac{6g}{y^2}=0.\eqno(5.5.21)$$
Write
$$g(t,y)=\sum_{m\in\mbb{Z}}a_m(t)y^m.\eqno(5.5.22)$$
Then (5.5.21) becomes
$$\sum_{m\in\mbb{Z}}[(m+2)(m+1)-6]a_{m+2}(t)y^m=0\sim
(m+4)(m-1)a_{m+2}=0\;\;\for\;\;m\in\mbb{Z} \eqno(5.5.23)$$ Hence
$$g=\frac{\al}{y^2}+\be y^3,\eqno(5.5.24)$$
 where $\al$ and $\be$ are arbitrary  differentiable
functions in $t$. Note
$$x^2\xi+xg=\frac{x^2+\al x}{y^2}+\be xy^3.\eqno(5.5.25)$$
Replacing $u$ by $T_{2,-\al/2}(u)$, we can take $\al=0$. That is,
 $g=\be y^3$.

  We can write (5.5.19) as
 $$f_{yy}-\frac{2}{y^2}f=2\be'y^3+\be^2y^6.\eqno(5.5.26)$$
Suppose
$$f(t,y)=\sum_{m\in\mbb{Z}}b_m(t)y^m.\eqno(5.5.27)$$
Then (5.5.26) becomes
$$\sum_{m\in\mbb{Z}}[(m+2)(m+1)-2]a_{m+2}(t)y^m=2\be'y^3+\be^2y^6,\eqno(5.5.28)$$
equivalently,
$$18a_5=2\be',\;\;54a_8=\be^2,\;\;(m+3)ma_{m+2}=0,\;\;m\neq
3,6.\eqno(5.5.29)$$ Thus
$$f=\frac{\gm}{y}+\vt
y^2+\frac{\be'}{9}y^5+\frac{\be^2}{54}y^8,\eqno(5.5.30)$$ where
$\gm$ and $\vt$ are arbitrary functions in $t$. \psp

{\bf Theorem 5.5.1}. {\it We have the following solution of the
equation:
$$u=\vf=\frac{x^2}{y^2}+\be xy^3+\frac{\gm}{y}+\vt
y^2+\frac{\be'}{9}y^5+\frac{\be^2}{54}y^8,\eqno(5.5.31)$$  where
$\be, \gm$ and $\vt$ are arbitrary functions in $t$. Moreover,
$u=T^{(\al,-\sgm)}_{0;1,1}(\vf)$ is solution of the
Khokhlov-Zabolotskaya equation (5.5.1) blowing up on the moving line
$y=\sgm(t)$.}\psp

The simplest solution of the Khokhlov-Zabolotskaya equation (5.5.1)
blowing up on the moving line $y=\sgm(t)$:
$$u=\frac{(x-\sgm' y)^2}{(y-\sgm)^2}+{\sgm'}^2-2
{\sgm'}'y\eqno(5.5.32)$$

Suppose that $\xi$  is polynomial in  $y$, then $\xi=0$ by comparing
the terms with highest degree of $y$ in (5.5.17).
  Then (5.5.18) and (5.5.19) become
$$g_{yy}=0,\qquad f_{yy}=2g_t+g^2.\eqno(5.5.33)$$
Replacing $u$ by some $T_{3,\al}(u)$ (cf. (5.5.11)), we have $g=\be
y$ for some function $\be$ in $t$.  Hence
$$f_{yy}=2\be' y+\be^2 y^2.\eqno(5.5.34)$$
So
$$f=\gm+\sgm
y+\frac{\be'}{3}y^3+\frac{\be^2}{12} y^4,\eqno(5.5.35)$$ where $\gm$
and $\sgm$ are arbitrary functions in $t$.\psp

{\bf Theorem 5.5.2}. {\it The following is a solution of the
Khokhlov-Zabolotskaya equation (5.5.1):
$$u=\psi=\be xy+\gm+\sgm
y+\frac{\be'}{3}y^3+\frac{\be^2}{12} y^4,\eqno(5.5.36)$$ where
$\be,\gm$ and $\sgm$ are arbitrary functions in $t$. Moreover, any
solution polynomial in $x$ and $y$ of (5.5.1) must be of the form $u
=T_{3,\al}(\psi)$. }\psp

\section{Equation of Geopotential  Forecast}

In a book on short term weather forecast,  Kibel' [Kt] (1954) used
the partial differential equation\index{equation of geopotential
forecast}
$$(H_{xx}+H_{yy})_t+H_x(H_{xx}+H_{yy})_y-H_y(H_{xx}+H_{yy})_x=k
H_x\eqno(5.6.1)$$ for geopotential forecast on a middle level in
earth sciences, where $k$ is a real constant. Moreover, Kibel' [Kt]
found the Gaurvitz solution of the above equation. Syono [Ss] (1958)
got another special solution. The other known solutions are related
to the physical backgrounds such as configuration of type of narrow
gullies and crests, flows of type of isolate whirlwinds, stream
flow, springs and drains, hyperbolic points, and cyclone formation.
Katkov [Kv1, Kv2] (1965, 1966) determined the Lie point symmetries
and obtained certain invariant solutions of the above equation. In
this section, we give new approaches to the equation (5.6.1).

To make the nonzero terms in (5.6.1) having the same degree, we
suppose
$$\mbox{deg}\:x=\mbox{deg}\:y=\ell_1,\;\;\mbox{deg}\:H=\ell_2.\eqno(5.6.2)$$
Then
$$\ell_2-2\ell_1-\mbox{deg}\:t=2\ell_2-4\ell_1=\ell_2-\ell_1\sim
\ell_2=3\ell_1,\;\; \mbox{deg}\:t=-\ell_1.\eqno(5.6.3)$$ Since
(5.6.1) dose not contain variable coefficients, it is translation
invariant. Thus the transformation
$$T_{a,b;c}(H)=c^{-3}H(c^{-1}t+a,cx,cy+b)\eqno(5.6.4)$$
keeps the  equation (5.6.1) invariant for $a,b,c\in\mbb{R}$ and
$c\neq 0$ with the independent variables $t$
 replaced by $c^{-1}t+a$, $x$ replaced by $cx$ and $y$
 replaced by $cy+b$, where the
subindices denote the partial derivatives with respect to the
original independent variables. So $T_{a,b;c}$ maps a solution of
the geopotential equation (5.6.1) to another solution.

Let $\al$ and $\be$ be two differentiable functions in $t$. The
transformation $H(t,x,y)\mapsto H(t,x+\al,y)$ changes the equation
(5.6.1) to
$$\al'(H_{xx}+H_{yy})_x+ (H_{xx}+H_{yy})_t+H_x(H_{xx}+H_{yy})_y-H_y(H_{xx}+H_{yy})_x=k
H_x,\eqno(5.6.5)$$
 with the independent variables  $x$ replaced by $x+\al$, where the
subindices denote the partial derivatives with respect to the
original independent variables. Moreover, the transformation
$H(t,x,y)\mapsto H(t,x,y)+\al' y$ changes the equation (5.6.1) to
$$(H_{xx}+H_{yy})_t+H_x(H_{xx}+H_{yy})_y-(H_y+\al')(H_{xx}+H_{yy})_x=k
H_x.\eqno(5.6.6)$$ Hence the transformation
$$T_{\al,\be}(H)=H(t,x+\al,y)+\al'y+\be\eqno(5.6.7)$$
 leaves the  equation (5.6.1) invariant with the independent variables  $x$ replaced by $x+\al$ , where the
subindices denote the partial derivatives with respect to the
original independent variables. Thus $T_{\al,\be}$ maps a solution
of the geopotential equation (5.6.1) to another solution.

In summary, the transformation
$$T_{a,b;c}^{(\al,\be)}(H(t,x,y))=c^{-3}H(c^{-1}t+a,c(x+\al),cy+b)+\al'y+\be\eqno(5.6.8)$$
maps a solution of the geopotential equation (5.6.1) to another
solution.

Fix two functions $\al$ and $\be$ in $t$. Denote
$$\varpi=\al x+\be y.\eqno(5.6.9)$$
Assume
$$H=\phi(t,\varpi)+\mu y^2+\tau x+\nu
y,\eqno(5.6.10)$$
 where $\phi$ is a two-variable function and $\tau,\mu,\nu$ are
functions in $t$. Note
$$H_x=\al\phi_\varpi+\tau,\qquad H_y=\be\phi_\varpi+2\mu
y+\nu,\qquad
H_{xx}+H_{yy}=2\mu+(\al^2+\be^2)\phi_{\varpi\varpi},\eqno(5.6.11)$$
$$
(H_{xx}+H_{yy})_t=2\mu'+(\al^2+\be^2)'\phi_{\varpi\varpi}+
(\al^2+\be^2)[\phi_{t\varpi\varpi}+(\al'x+\be'y)\phi_{\varpi\varpi\varpi}],\eqno(5.6.12)$$
$$(H_{xx}+H_{yy})_x=(\al^2+\be^2)\al\phi_{\varpi\varpi\varpi},
\qquad(H_{xx}+H_{yy})_y=(\al^2+\be^2)\be\phi_{\varpi\varpi\varpi}.
\eqno(5.6.13)$$ Thus (5.6.1) becomes
\begin{eqnarray*}\hspace{2cm} & &2\mu'+(\al^2+\be^2)'\phi_{\varpi\varpi}+
(\al^2+\be^2)\phi_{t\varpi\varpi}-k(\al\phi_\varpi+\tau)\\ & & +
(\al^2+\be^2)[\al'x+(\be'-2\al\mu)y+\be\tau-\al\nu]
\phi_{\varpi\varpi\varpi}=0.\hspace{2.6cm}(5.6.14)\end{eqnarray*}

In order to solve the above equation, we assume
$$2\mu'=k\tau,\qquad \tau=\al{\vt'}',\qquad\nu=
\be{\vt'}',\eqno(5.6.15)$$ for some function $\vt$ in $t$, and
$$\al'x+(\be'-2\al\mu)y=0.\eqno(5.6.16)$$
Note that (5.6.16) is equivalent to the following system of ordinary
differential equations:
$$\al'=0,\qquad \be'-2\al\mu=0.\eqno(5.6.17)$$
By the first equation and replacing $H$ by $T_{0,0;c}(H)$ (cf.
(5.6.8)) if necessary, we have $\al=1$. So $\tau={\vt'}'$ according
to the second equation in (5.6.15). Moreover,
 the first equation in (5.6.15) yields
$$ \mu=\frac{k\vt'+c_0}{2},\qquad c_0\in\mbb{R}.\eqno(5.6.18)$$
Hence the second equation in (5.6.17) becomes
$$\be'-(k\vt'+c_0)=0.\eqno(5.6.19)$$
Therefore,
$$\be=k\vt+c_0t+d,\qquad d\in\mbb{R}.\eqno(5.6.20)$$
According to the third equation in (5.6.15),
$$\nu=(k\vt+c_0t+d){\vt'}'.\eqno(5.6.21)$$

 Now (5.6.14) becomes
$$(\al^2+\be^2)'\phi_{\varpi\varpi}+
(\al^2+\be^2)\phi_{t\varpi\varpi}-k \phi_\varpi=0. \eqno(5.6.22)$$
Replacing $H$ by some $T^{(0,\vs)}_{0,0;1}(H)$ if necessary, we
have:
$$(\al^2+\be^2)'\phi_\varpi+
(\al^2+\be^2)\phi_{t\varpi} -k\phi=0. \eqno(5.6.23)$$ The above
equation can written as
$$[(\al^2+\be^2)\phi_\varpi]_t-k\phi=0.\eqno(5.6.24)$$
So we take the form
$$\phi=\frac{\hat\phi(t,\varpi)}{\al^2+\be^2}=\frac{\hat\phi(t,\varpi)}
{1+(k\vt+c_0t+d)^2}.\eqno(5.6.25)$$ Then (5.6.23) becomes
$$\hat\phi_{\varpi t}=\frac{k\hat\phi}{1+(k\vt+c_0t+d)^2}.\eqno(5.6.26)$$

We use the separation of variables
$$\hat\phi=\xi(\varpi)\eta(t),\eqno(5.6.27)$$
where $\xi$ and $\eta$ are one-variable functions. Then (5.6.26)
becomes
$$\frac{\xi'(\varpi)}{k\xi(\varpi)}=\frac{\eta(t)}{(1+(k\vt+c_0t+d)^2)\eta'(t)},\eqno(5.6.28)$$
which must be a constant. To find more solutions, we assume
$$\frac{\xi'(\varpi)}{k\xi(\varpi)}=\frac{\eta(t)}{(1+(k\vt+c_0t+d)^2)\eta'(t)}=a+bi\neq 0\eqno(5.6.29)$$
for some $a,b\in\mbb{R}$. Thus $\xi'=(a+bi)\xi$ and
$$\eta'=\frac{\eta}{(a+bi)(1+(k\vt+c_0t+d)^2)}=\frac{(a-bi)\eta}
{(a^2+b^2)(1+(k\vt+c_0t+d)^2)}.\eqno(5.6.30)$$ We have
$$\xi=e^{k(a+bi)\varpi},\;\;\eta=\exp\left(\frac{a-bi}{a^2+b^2}\int\frac{dt}{1+(k\vt+c_0t+d)^2}\right),\eqno(5.6.31)$$
that is,
$$\hat\phi=e^{k(a+bi)\varpi}\exp\left(\frac{a-bi}{a^2+b^2}\int\frac{dt}{1+(k\vt+c_0t+d)^2}\right)\eqno(5.6.32)$$
is a complex solution (5.6.26). Since (5.6.26) is a linear equation
with real coefficients, the real part
 \begin{eqnarray*}\qquad\zeta_1&=&\exp\left(ka\varpi+\frac{a}{a^2+b^2}\int\frac{dt}{1+(k\vt+c_0t+d)^2}\right)
\\ &
&\times\cos\left(kb\varpi-\frac{b}{a^2+b^2}\int\frac{dt}{1+(k\vt+c_0t+d)^2}\right)\hspace{4.1cm}(5.6.33)\end{eqnarray*}
and
 the imaginary  part
 \begin{eqnarray*}\qquad\zeta_2&=&\exp\left(ka\varpi+\frac{a}{a^2+b^2}\int\frac{dt}{1+(k\vt+c_0t+d)^2}\right)
\\ &
&\times\sin\left(kb\varpi-\frac{b}{a^2+b^2}\int\frac{dt}{1+(k\vt+c_0t+d)^2}\right)\hspace{4.1cm}(5.6.34)\end{eqnarray*}
are real solutions of (5.6.26). For any $c\in\mbb{R}$,
\begin{eqnarray*}\zeta_1\sin c+\zeta_2\cos c&=&\exp\left(ka\varpi+\frac{a}{a^2+b^2}\int\frac{dt}{1+(k\vt+c_0t+d)^2}\right)
\\ &
&\times\sin\left(c+kb\varpi-\frac{b}{a^2+b^2}\int\frac{dt}{1+(k\vt+c_0t+d)^2}\right)\hspace{1.6cm}(5.6.35)\end{eqnarray*}
is a solution of (5.6.26) by the additivity of solutions for linear
equation. Applying the additivity again, we have more general
solution
\begin{eqnarray*}\hspace{2cm}\hat\phi&=&\sum_{r=1}^md_r\exp\left(ka_r\varpi+\frac{a_r}{a_r^2+b_r^2}
\int\frac{dt}{1+(k\vt+c_0t+d)^2}\right)\\ & &\times\sin
\left(kb_r\varpi+c_r-\frac{b_r}{a_r^2+b_r^2}
\int\frac{dt}{1+(k\vt+c_0t+d)^2}\right),\hspace{1.8cm}
(5.6.36)\end{eqnarray*} where $a_r,b_r,c_r,d_r$ are real constants
such that $(a_r,b_r)\neq (0,0)$. By (5.6.10), (5.6.18), (5.6.20),
(5.6.21) and (5.6.25), we have:\psp

{\bf Theorem 5.6.1}. {\it Let $\vt$ be any function in $t$ and let
$a_r,b_r,c_r,d_r,c_0,d$ for $r=1,...,m$ be real constants such that
$(c,d),(a_r,b_r)\neq (0,0)$. We have the following solution of the
geopotential forecast equation (5.6.1):
\begin{eqnarray*}
H&=&\frac{k\vt'+c_0}{2}y^2+{\vt'}'[x+(k\vt+c_0t+d)y]+
\frac{1} {1+(k\vt+c_0t+d)^2}\\
&
&\times\sum_{r=1}^md_r\exp\left(ka_r[x+(k\vt+c_0t+d)y]+\frac{a_r}{a_r^2+b_r^2}
\int\frac{dt}{1+(k\vt+c_0t+d)^2}\right)\\
& &\times\sin
\left(kb_r[x+(k\vt+c_0t+d)y]+c_r-\frac{b_r}{a_r^2+b_r^2}
\int\frac{dt}{1+(k\vt+c_0t+d)^2}\right).\hspace{0.4cm}
(5.6.37)\end{eqnarray*}}\pse

Applying the transformation $T_{0,b;c}^{(\al,\be)}$ in (5.6.8) to
the above solution, we will get a more general solution the
geopotential forecast equation (5.6.1).

Next we set
$$\varpi=x^2+y^2.\eqno(5.6.38)$$
Assume
$$H=\xi(\varpi)-y\eqno(5.6.39)$$
where  $\xi$ is a one-variable function. Note
$$H_x=2x\xi',\qquad H_y=2y\xi'-1,\qquad
H_{xx}+H_{yy}=4(\xi'+\varpi{\xi'}'),\eqno(5.6.40)$$
$$(H_{xx}+H_{yy})_x=8x(2{\xi'}'+\varpi{{\xi'}'}'),\qquad
(H_{xx}+H_{yy})_y=8y(2{\xi'}'+\varpi{{\xi'}'}').\eqno(5.6.41)$$ Then
(5.6.1) is equivalent to:
$$4(2{\xi'}'+\varpi{{\xi'}'}')=k\xi'.\eqno(5.6.42)$$
Replacing $H$ by some $T^{(0,\vs)}_{0,0;1}(H)$ if necessary, we
have:
$$\xi'+\varpi{\xi'}'=\frac{k}{4}\xi.\eqno(5.6.43)$$
To solve the above ordinary differential equation, we assume
$$\xi=\sum_{s=0}^\infty \varpi^s(a_s+b_s\ln\varpi),\qquad
a_s,b_s\in\mbb{R}.\eqno(5.6.44)$$ Observe
$$\xi'=\sum_{s=0}^\infty
\varpi^{s-1}(sa_s+b_s+sb_s\ln\varpi),\eqno(5.6.45)$$
$${\xi'}'=\sum_{s=0}^\infty
\varpi^{s-2}(s(s-1)a_s+(2s-1)b_s+s(s-1)b_s\ln\varpi).\eqno(5.6.46)$$
So (5.6.43) becomes
$$\sum_{s=0}^\infty
\varpi^{s-1}(s^2a_s+2sb_s+s^2b_s\ln\varpi)=\frac{k}{4}\sum_{s=0}^\infty
\varpi^s(a_s+b_s\ln\varpi),\eqno(5.6.47)$$ equivalently,
$$(s+1)^2a_{s+1}+2(s+1)b_{s+1}=\frac{k}{4}a_s,\qquad
(s+1)^2b_{s+1}=\frac{k}{4}b_s.\eqno(5.6.48)$$ Hence
$$b_s=\frac{b_0k^s}{(s!)^24^s},\qquad
a_s=\frac{a_0k^s}{(s!)^24^s}-\frac{2b_0k^s}{(s!)^24^s}\sum_{r=1}^s\frac{1}{r}\qquad\for\;\;s>0.
\eqno(5.6.49)$$ Thus
$$\xi=a_0\sum_{s=0}^{\infty}\frac{k^s\varpi^s}{(s!)^24^s}
+b_0[\ln \varpi+\sum_{j=1}^{\infty}\frac{(k\varpi)^j}
{(j!)^24^j}(\ln\varpi-2\sum_{r=1}^jr^{-1})]. \eqno(5.6.50)$$ \pse

{\bf Theorem 5.6.2}. {\it Let $b$ and $c$ be any real constants. We
have the following steady solution of the geopotential forecast
equation (5.6.1):
\begin{eqnarray*}\qquad\qquad H&=&-y+b\sum_{s=0}^{\infty}\frac{k^s(x^2+y^2)^s}{(s!)^24^s}
+c[\ln (x^2+y^2)\\ & &+\sum_{j=1}^{\infty}\frac{(k(x^2+y^2))^j}
{(j!)^24^j}(\ln(x^2+y^2)-2\sum_{r=1}^jr^{-1})].
\hspace{3.3cm}(5.6.51)\end{eqnarray*}}\psp

{\bf Remark 5.6.3}. Although the above solution is time independent,
we apply $T_{0,0;1}^{(\al,\be)}$ to it and  obtain the following
time-dependent solution:
\begin{eqnarray*}\hspace{1cm}H&=&(\al'-1)y+\be+b\sum_{s=0}^{\infty}\frac{k^s((x+\al)^2+y^2)^s}{(s!)^24^s}
+c[\ln ((x+\al)^2+y^2)
\\ & &+\sum_{j=1}^{\infty}\frac{(k((x+\al)^2+y^2))^j}
{(j!)^24^j}(\ln((x+\al)^2+y^2)-2\sum_{r=1}^jr^{-1})],
\hspace{1.9cm}(5.6.52)\end{eqnarray*}
 where $\al$ and $\be$ are arbitrary functions in $t$.

\chapter{Nonlinear Schr\"{o}dinger and DS  Equations}

 The two-dimensional cubic nonlinear
Schr\"{o}dinger equation is used to describe the propagation of an
intense laser beam through a medium with Kerr nonlinearity. The
coupled two-dimensional cubic nonlinear  Schr\"{o}dinger equations
 are used to describe interaction of electromagnetic
waves with different polarizations in nonlinear optics. In this
chapter, we solve the above equations by imposing a quadratic
condition on the related argument functions and using their symmetry
transformations. More complete families of exact solutions of such
type are obtained. Many of them are the periodic, quasi-periodic,
aperiodic and singular solutions that may have practical
significance.

The Davey-Stewartson  equations are used to describe the long time
evolution of  three-dimensional packets of surface waves. Assuming
that the argument functions are quadratic in spacial variables, we
find in this chapter various
 exact solutions  for the Davey-Stewartson
 equations.

\section{Nonlinear  Schr\"{o}dinger Equation}

The two-dimensional cubic nonlinear  Schr\"{o}dinger
equation\index{nonlinear Schr\"{o}dinger equation}
$$ i\psi_t+\kappa(\psi_{xx}+\psi_{yy})+\ves|\psi|^2\psi=0\eqno(6.1.1)$$
is used to describe the propagation of an intense laser beam through
a medium with Kerr nonlinearity, where $t$ is the distance in the
direction of propagation, $x$ and $y$ are the transverse spacial
coordinates,  $\psi$ is a complex valued function in $t,x,y$
standing for electric field amplitude, and $\kappa,\ves$ are nonzero
real constants. Akhnediev, Eleonskii and Kulagin [AEK] (1987) found
certain exact solutions of (6.1.1) whose real and imaginary parts
are linearly dependent over the functions in $t$. Moreover, Gagnon
and Winternitz [GW] (1989) found exact solutions of the cubic and
quintic nonlinear Schr\"{o}dinger equation for a cylindrical
geometry. Mihalache and Panoin [MP] (1992) used the method of
Akhnediev, Eleonskii and Kulagin to obtain new solutions which
describe the propagation of dark envelope soliton light pulses in
optical fibers under the normal group velocity dispersion regime.
Furthermore, Saied, EI-Rahman and Ghonamy [SEG] (2003) used various
similarity variables to reduce the above equation to certain
ordinary differential equations and obtain some exact solutions.
However, many of their solutions are equivalent to each other under
the action of the known symmetry transformations of the above
equation. There are the other interesting results on the equation
(6.1.1) (e.g., cf. [AP, Pa, Sy]).

The objective of this section is to give a direct more systematical
study on the exact solutions of the nonlinear Schr\"{o}dinger
equation. We solve them by imposing the quadratic condition on the
argument functions and using their symmetry transformations.
  More complete families  of explicit exact
solutions of this type with multiple parameter functions are
obtained. Many of them are the periodic, quasi-periodic, aperiodic
and singular solutions that physicists and engineers expect to know.
 For instance, soliton solutions are sitting in our families. The
 results are from our work [X14].

To make the nonzero terms in (6.1.1) to have the same degree, we
have to take
$$\mbox{deg}\;x=\mbox{deg}\;y=-\mbox{deg}\;\psi=\frac{1}{2}\mbox{deg}\;t.\eqno(6.1.2)$$
Moreover, the Laplace operator $\ptl_x^2+\ptl_y^2$ is invariant
under rotations and (6.1.1) is translation invariant because it does
not contain variable coefficients.
 Thus the transformation
$$T_{a;b;\sta}^{(a_1,a_2,a_3)}(\psi)=be^{ai}
\psi(b^2(t+a_1),b(x\cos \sta+y\sin\sta+a_2),b(-x\sin \sta+y\cos
\sta+a_3))\eqno(6.1.3)$$ maps a solution of the Schr\"{o}dinger
equation (6.1.1) to another solution, where $a,a_1,a_2,a_3,\\
b,\sta\in\mbb{R}$ and $b\neq 0$.

Fix $a_1,a_2\in\mbb{R}$. Note that the transformation
$\psi(t,x,y)\mapsto \psi(t,x-2\kappa a_1t,y-2\kappa a_2t)$ changes
the equation (6.1.1) to
$$-2\kappa
i(a_1\psi_x+a_2\psi_y)+i\psi_t+\kappa(\psi_{xx}+\psi_{yy})+\ves|\psi|^2\psi=0
\eqno(6.1.4)$$ with the independent variables  $x$ replaced by
$x-2\kappa a_1t$ and $y$ replaced by $y-2\kappa a_2t$ , where the
subindices denote the partial derivatives with respect to the
original independent variables. Moreover, the transformation
$\psi\mapsto e^{[(a_1x+a_2y)-\kappa(a_1^2+a_2^2)t]i}\psi$ changes
the equation (6.1.1) to
$$e^{[(a_1x+a_2y)-\kappa(a_1^2+a_2^2)t]i}[i\psi_t+2\kappa
i(a_1\psi_x+a_2\psi_y)+\kappa(\psi_{xx}+\psi_{yy})+\ves|\psi|^2\psi]=0.
\eqno(6.1.5)$$ Hence the transformation
$$S_{a_1,a_2}(\psi(t,x,y))=e^{[(a_1x+a_2y)-\kappa(a_1^2+a_2^2)t]i}\psi(t,x-2\kappa
a_1t,y-2\kappa a_2t)\eqno(6.1.6)$$ changes the equation (6.1.1) to
$$e^{(a_1x+a_2y)i}[i\psi_t+\kappa(\psi_{xx}+\psi_{yy})+\ves|\psi|^2\psi]=0,\eqno(6.1.7)$$
equivalently, (6.1.1) holds with the independent variables  $x$
replaced by $x-2\kappa a_1t$ and $y$ replaced by $y-2\kappa a_2t$ ,
where the subindices denote the partial derivatives with respect to
the original independent variables. Therefore, $S_{a_1,a_2}$ maps a
solution of the Schr\"{o}dinger equation (6.1.1) to another
solution.

 Write
$$\psi=\xi(t,x,y)e^{i\phi(t,x,y)},\eqno(6.1.8)$$
where $\xi$ and $\phi$ are real functions in $t,x,y$. Note
$$\psi_t=(\xi_t+i\xi\phi_t)e^{i\phi},\qquad
\psi_x=(\xi_x+i\xi\phi_x)e^{i\phi},\qquad
\psi_y=(\xi_y+i\xi\phi_y)e^{i\phi},\eqno(6.1.9)$$
$$\psi_{xx}=(\xi_{xx}-\xi\phi_x^2+i(2\xi_x\phi_x
+\xi\phi_{xx}))e^{i\phi},\;\;
\psi_{yy}=(\xi_{yy}-\xi\phi_y^2+i(2\xi_y\phi_y
+\xi\phi_{yy}))e^{i\phi}.\eqno(6.1.10)$$ So the equation (6.1.1)
becomes
\begin{eqnarray*}\hspace{2cm}&
&i\xi_t-\phi_t\xi+\ves\xi^3
+\kappa[\xi_{xx}+\xi_{yy}-\xi(\phi_x^2+\phi_y^2)\\
& &+i(2\xi_x\phi_x+2\xi_y
\phi_y+\xi(\phi_{xx}+\phi_{yy}))]=0,\hspace{4.9cm}(6.1.11)\end{eqnarray*}
equivalently,
$$\xi_t+\kappa(2\xi_x \phi_x+2\xi_y
\phi_y+\xi(\phi_{xx}+\phi_{yy}))=0,\eqno(6.1.12)$$
$$-\xi[\phi_t+\kappa(\phi_x^2+\phi_y^2)]
+\kappa(\xi_{xx}+\xi_{yy})+\ves\xi^3=0.\eqno(6.1.13)$$

Note that it is very difficult to solve the above system without
pre-assumptions.  From the algebraic characteristics of the above
system of partial differential equations, it is most affective to
assume that $\phi$ is quadratic in $x$ and $y$. After sorting case
by case, we only have the following four cases that lead us to exact
solutions of (6.1.12) and (6.1.13).\psp

{\it Case 1}. $\phi=\be(t)$ is a function in $t$.\psp

According to (6.1.12), $\xi_t=0$. Moreover, (6.1.13) becomes
$$-\be'\xi
+\kappa(\xi_{xx}+\xi_{yy})+\ves\xi^3=0.\eqno(6.1.14)$$ Replacing
$\psi$ by some $T_{a;1;0}^{(0,0,0)}(\psi)$, we have
$$\be=bt,\qquad b\in\mbb{R}.\eqno(6.1.15)$$
Then (6.1.14) becomes
$$-b\xi
+\kappa(\xi_{xx}+\xi_{yy})+\ves\xi^3=0.\eqno(6.1.16)$$ First we
assume $\xi_y=0$. The above equation becomes an ordinary
differential equation:
$$-b\xi
+\kappa{\xi'}'+\ves\xi^3=0.\eqno(6.1.17)$$

Recall
$${\left(\frac{1}{x}\right)'}'=2\left(\frac{1}{x}\right)^3,\eqno(6.1.18)$$
$${(\tan z)'}'=2(\tan^3 z+\tan
z), \qquad{(\sec z)'}'=2\sec^3 z-\sec z\eqno(6.1.19)$$ (cf. (3.5.17)
and (3.5.18)),
$${(\coth z)'}'=2(\coth^3z-\coth z),\qquad
{(\csch z)'}'=2\csch^3 z+\csch z\eqno(6.1.20)$$ (cf. (3.5.19) and
(3.5.20)),
$${\sn'}'(z|m)=2m^2\mbox{sn}^3(z|m)-(m^2+1)\sn(z|m),\eqno(6.1.21)$$
$${\cn'}'(z|m)=-2m^2\mbox{cn}^3(z|m)+(2m^2-1)\cn(z|m),\eqno(6.1.22)$$
$${\dn'}'(z|m)=-2\mbox{dn}^3(z|m)+(2-m^2)\dn(z|m)\eqno(6.1.23)$$
(cf. (3.5.14)-(3.5.16)).

Substituting  $\xi=kf(x)$ to (6.1.17) with $k\in\mbb{R}$ and $f=1/x,
\tan x, \sec x,\coth x,\csch x,\\ \sn(x|m),\cn(x|m),\dn(x|m)$, we
find the following solutions: if $\kappa\ves<0$,
$$\xi=\frac{1}{x}\sqrt{-\frac{2\kappa}{\ves}},\qquad
b=0;\eqno(6.1.24)$$
$$\xi=\sqrt{-\frac{2\kappa}{\ves}}\:\tan x,\qquad b=2\kappa;\eqno(6.1.25)$$
$$\xi=\sqrt{-\frac{2\kappa}{\ves}}\:\sec x,\qquad b=-\kappa;\eqno(6.1.26)$$
$$\xi=\sqrt{-\frac{2\kappa}{\ves}}\:\coth x,\qquad b=-2\kappa;\eqno(6.1.27)$$
$$\xi=\sqrt{-\frac{2\kappa}{\ves}}\:\csch x,\qquad b=\kappa;\eqno(6.1.28)$$
$$\xi=m\sqrt{-\frac{2\kappa}{\ves}}\:\sn(x|m),\qquad b=-(1+m^2)\kappa.\eqno(6.1.29)$$
When $\kappa\ves>0$, we get the following solutions:
$$\xi=m\sqrt{\frac{2\kappa}{\ves}}\:\cn (x|m),\qquad b=(2m^2-1)\kappa,\eqno(6.1.30)$$
$$\xi=\sqrt{\frac{2\kappa}{\ves}}\:\dn (x|m),\qquad b=(2-m^2)\kappa.\eqno(6.1.31)$$

Observe that
$$(\ptl_x^2+\ptl_y^2)\left(\sqrt{\frac{1}{x^2+y^2}}\right)=\left(\sqrt{\frac{1}{x^2+y^2}}\right)^3.
\eqno(6.1.32)$$ Thus we have solution
$$\xi=\sqrt{-\frac{\kappa}{\ves(x^2+y^2)}},\qquad b=0\eqno(6.1.33)$$
 if $\kappa\ves<0$.\psp

{\bf Theorem 6.1.1}. {\it Let $m\in\mbb{R}$ such that $0<m<1$. The
following functions are solutions $\psi$ of the two-dimensional
cubic nonlinear cubic nonlinear Schr\"{o}dinger equation (6.1.1): if
$\ves\kappa<0$,
$$\sqrt{-\frac{2\kappa}{\ves}}\:\frac{1}{x},\;
 \;\sqrt{-\frac{\kappa}{\ves(x^2+y^2)}},\;
  e^{2\kappa ti}\sqrt{-\frac{2\kappa}{\ves}}\:\tan x,\;
   e^{-\kappa ti}\sqrt{-\frac{2\kappa}{\ves}}\sec x,
  \eqno(6.1.34)$$}
$$e^{-2\kappa ti}\sqrt{-\frac{2\kappa}{\ves}}\:\coth
x,\; e^{\kappa ti}\sqrt{-\frac{2\kappa}{\ves}}\:\csch x,\;
me^{-(1+m^2)\kappa ti}\sqrt{-\frac{2\kappa}{\ves}}\sn(
x|m);\eqno(6.1.35)$$ {\it when} $\ves\kappa>0$,
$$me^{(2m^2-1)\kappa ti}\sqrt{\frac{2\kappa}{\ves}}\:\cn (x|m),\qquad e^{(2-m^2)\kappa ti}
\sqrt{\frac{2\kappa}{\ves}}\:\dn (x|m).\eqno(6.1.36)$$ \pse

{\bf Remark 6.1.2}. Recall $\lim_{m\rta 1}\cn (x|m)=\sech x$. Thus
we have the solution
$$\psi=\lim_{m\rta 1}me^{(2m^2-1)\kappa
ti}\sqrt{\frac{2\kappa}{\ves}}\:\cn (x|m)=e^{\kappa
ti}\sqrt{\frac{2\kappa}{\ves}}\:\sech x.\eqno(6.1.37)$$ Applying the
transformation $T_{c;b;\sta}^{(0,a,0)}$ (cf. (6.1.3)) and $S_{d,0}$
(cf. (6.1.6)), we get a soliton solution\index{soliton solution!of
 nonlinear Schr\"{o}dinger equations}
$$\psi=b\sqrt{\frac{2\kappa}{\ves}}\:e^{(b^2\kappa(1-d^2)
t+bd(x\cos \sta+y\sin\sta+a)+c)i}\sech b(x\cos
\sta+y\sin\sta-2bd\kappa t+a).\eqno(6.1.38)$$ We can also apply the
transformations (6.1.3) and (6.1.6) to the other solutions in the
above theorem and obtain more general solutions.\psp

{\it Case 2}. $\phi=x^2/4\kappa t+\be$ for some function $\be$ of
$t$.\psp

In this case, (6.1.12) becomes
$$\xi_t+\frac{x}{t}\xi_x +\frac{1}{2t}\xi=0.\eqno(6.1.39)$$
Thus
$$\xi=\frac{1}{\sqrt{t}}\zeta(u,y),\qquad u=\frac{x}{t},\eqno(6.1.40)$$
for some two-variable function $\zeta$. Now (6.1.13) becomes
(6.1.14). Note
$$
\xi_{xx}=t^{-5/2}\zeta_{uu},\qquad\xi_{yy}=t^{-1/2}\zeta_{yy},\qquad\xi^3=t^{-3/2}\zeta^3.\eqno(6.1.41)$$
So (6.1.14) become
$$-\frac{\be'}{\sqrt{t}}\zeta+\kappa(t^{-5/2}\zeta_{uu}+t^{-1/2}\zeta_{yy})+\ves t^{-3/2}\zeta^3
=0,\eqno(6.1.42)$$
 whose coefficients of $t^{-3/2}$ force
 us to take
$$\xi=\frac{b}{\sqrt{t}},\qquad b\in\mbb{R}.\eqno(6.1.43)$$
Now (6.1.14) becomes
$$-\be'+\frac{\ves b^2}{t}=0\lra\be=\ves b^2\ln t\eqno(6.1.44)$$
because otherwise we can replace $\psi$ by some
$T_{a;1;0}^{(0,0,0)}(\psi)$.\psp

{\it Case 3}. $\phi=x^2/4\kappa t+y^2/4\kappa(t-d)+\be$ for some
function $\be$ in $t$ with $0\neq d\in\mbb{R}$. \psp

In this case, (6.1.12) becomes
$$\xi_t+\frac{x}{t}\xi_x +\frac{y}{t-d}\xi_y+
\left(\frac{1}{2t}+\frac{1}{2(t-d)}\right)\xi=0.\eqno(6.1.45)$$
Hence we have:
$$\xi=\frac{1}{\sqrt{t(t-d)}}\zeta(u,v),\qquad
u=\frac{x}{t},\;v=\frac{y}{t-d}, \eqno(6.1.46)$$ for some
two-variable function $\zeta$.
 Again (6.1.13) becomes (6.1.14). Note
 $$\xi_{xx}=t^{-5/2}(t-d)^{-1/2}\zeta_{uu},\;\xi_{yy}=
 t^{-1/2}(t-d)^{-5/2}\zeta_{vv},\;\xi^3=t^{-3/2}(t-d)^{-3/2}\zeta^3.
 \eqno(6.1.47)$$
So (6.1.14) becomes
$$-\frac{\be'}{\sqrt{t(t-d)}}\zeta+\kappa(t^{-5/2}(t-d)^{-1/2}\zeta_{uu}+t^{-1/2}(t-d)^{-5/2}\zeta_{vv})
+\ves t^{-3/2}(t-d)^{-3/2}\zeta^3=0,\eqno(6.1.48)$$ whose
coefficients of $t^{-3/2}(t-d)^{-3/2}$ force
 us to take
$$\xi=\frac{b}{\sqrt{t(t-d)}},\qquad b\in\mbb{R}.\eqno(6.1.49)$$
Now (6.1.14) becomes
$$-\be'+\frac{\ves b^2}{t(t-d)}=0\lra\be=\frac{\ves b^2}{d}\ln
\frac{t-d}{t}\eqno(6.1.50)$$ because otherwise we can replace $\psi$
by some $T_{a;1;0}^{(0,0,0)}(\psi)$.
 \psp

 {\bf Theorem 6.1.3}. {\it Let $b,d\in\mbb{R}$ with $d\neq 0$. The following functions are solutions
$\psi$ of the two-dimensional cubic nonlinear cubic nonlinear
Schr\"{o}dinger equation:
$$bt^{\ves b^2i-1/2}e^{x^2i/4\kappa t},\qquad
bt^{-\ves b^2i/d-1/2}(t-d)^{\ves b^2i/d-1/2}e^{x^2i/4\kappa
t+y^2i/4\kappa(t-d)} .\eqno(6.1.51)$$}\pse

{\bf Remark 6.1.4}. Applying (6.1.3) to the above first solution, we
get another solution
$$\psi=b(t+a)^{a\kappa b^2i-1/2}\exp\left(\frac{
(x\cos\sta+y\sin\sta+a_0)^2}{4\kappa(t+a)}+d\right)i,\eqno(6.1.52)$$
for $a,a_0,b,d,\sta\in\mbb{R}$.  Moreover, we obtain a more
sophisticated solution:
\begin{eqnarray*}\hspace{1cm}\psi&=&b(t+a)^{\kappa b^2i-1/2}e^{(a_1x+a_2y-\kappa(a_1^2+a_2^2)t+d)i}\\ & &
\times\exp\frac{ ((x-2\kappa a_1t)\cos\sta+(y-2\kappa
a_2t)\sin\sta+a_0)^2i}{4\kappa(t+a)}
 \hspace{3.3cm}(6.1.53)\end{eqnarray*} by applying the
transformation (6.1.6) to (6.1.52), where $a_1,a_2\in\mbb{R}$.\psp

{\it Case 4}. $\phi=(x^2+y^2)/4\kappa t+\be$ for some function $\be$
in $t$.\psp

Under our assumption, (6.1.12) becomes
$$\xi_t+\frac{x}{t}\xi_x +\frac{y}{t}\xi_y+
\frac{1}{t}\xi=0.\eqno(6.1.54)$$ Thus we have:
$$\xi=\frac{1}{t}\zeta(u,v),\qquad
u=\frac{x}{t},\;v=\frac{y}{t}, \eqno(6.1.55)$$ for some two-variable
function $\zeta$. Moreover, (6.1.13) becomes
$$-\be'\zeta+\frac{\kappa}{t^2}(\zeta_{uu}+\zeta_{vv})
+\frac{\ves}{t^2}\zeta^3=0.\eqno(6.1.56)$$ An obvious solution is
$$\zeta=d,\qquad\be=-\frac{\ves d^2}{t},\qquad
d\in\mbb{R}.\eqno(6.1.57)$$ If $\ves\kappa<0$, we have the simple
following solutions with $\be=0$:
$$\zeta=\frac{1}{u}\sqrt{-\frac{2\kappa}{\ves}}
\qquad\mbox{or}\qquad\sqrt{-\frac{\kappa}{\ves(u^2+v^2)}}.\eqno(6.1.58)$$

Next we take
$$\be'=\frac{b}{t^2}\lra \be=-\frac{b}{t},\eqno(6.1.59)$$
 where $b$ is a real constant to
be determined. Then (6.1.58) is equivalent to
$$-b\zeta+\kappa(\zeta_{uu}+\zeta_{vv})
+\ves\zeta^3=0,\eqno(6.1.60)$$ which is the equation of the type
(6.1.16). By Theorem 6.1.1, we have:\psp

{\bf Theorem 6.1.5}. {\it Let $m\in\mbb{R}$ such that $0<m<1$. The
following functions are solutions $\psi$ of the two-dimensional
cubic nonlinear cubic nonlinear Schr\"{o}dinger equation (6.1.1): if
$\ves\kappa<0$,
$$\frac{d}{t}e^{(x^2+y^2-4\kappa \ves d^2)i/4\kappa
t}\sqrt{-\frac{2\kappa}{\ves}},\;\;e^{(x^2+y^2)i/4\kappa
t}\sqrt{-\frac{2\kappa}{\ves}}\:\frac{1}{x},\;\;e^{(x^2+y^2)i/4\kappa
t}\sqrt{-\frac{\kappa}{\ves(x^2+y^2)}},\eqno(6.1.61)$$
$$\frac{e^{(x^2+y^2-8\kappa^2)i/4\kappa
t}}{t}\sqrt{-\frac{2\kappa}{\ves}}\:\tan \frac{x}{t},\;
   \frac{e^{(x^2+y^2+4\kappa^2)i/4\kappa
t}}{t}\sqrt{-\frac{2\kappa}{\ves}}\sec \frac{x}{t},
  \eqno(6.1.62)$$}
$$
\frac{e^{(x^2+y^2+8\kappa^2)i/4\kappa
t}}{t}\sqrt{-\frac{2\kappa}{\ves}}\:\coth
\frac{x}{t},\;\;\frac{e^{(x^2+y^2-4\kappa^2)i/4\kappa
t}}{t}\sqrt{-\frac{2\kappa}{\ves}}\:\csch
\frac{x}{t},\eqno(6.1.63)$$
$$
\frac{me^{(x^2+y^2+4(1+m^2)\kappa^2)i/4\kappa t}}{t}
\sqrt{-\frac{2\kappa}{\ves}}\sn\left(\frac{x}{t}|m\right);\eqno(6.1.64)$$
{\it when} $\ves\kappa>0$,
$$\frac{me^{(x^2+y^2+4(1-2m^2)\kappa^2)i/4\kappa t}}{t}\sqrt{\frac{2\kappa}{\ves}}\:\cn \left(\frac{x}{t}|m\right),
\eqno(6.1.65)$$ $$\frac{e^{(x^2+y^2+4(m^2-2)\kappa^2)i/4\kappa
t}}{t}
\sqrt{\frac{2\kappa}{\ves}}\:\dn\left(\frac{x}{t}|m\right).\eqno(6.1.66)$$
\pse

{\bf Remark 6.1.6}. Recall $\lim_{m\rta 1}\cn (x|m)=\sech x$. Thus
we have the solution
$$\psi=\frac{e^{(x^2+y^2-4\kappa^2)i/4\kappa t}}{t}\sqrt{\frac{2\kappa}{\ves}}\:\sech \frac{x}{t}
.\eqno(6.1.67)$$ Applying the transformations
$T_{0;b;\sta}^{(a,a_2,0)}$ (cf. (6.1.3)) and $S_{a_1,0}$ (cf.
(6.1.6)), we get a more general soliton-like
solution\index{soliton-like solution!of nonlinear Schr\"{o}dinger
equation}
\begin{eqnarray*}\qquad\qquad\psi&=& \sqrt{\frac{2\kappa}{\ves}}\;\frac{e^{((x-2a_1\kappa t)^2+y^2-4\kappa^2/b^2)i/4\kappa(t-a)
+a_1(x-a_1\kappa t)i}}{b(t-a)}\\ & &\times \:\sech
\frac{(x-2a_1\kappa t)\cos
\sta+y\sin\sta}{b(t-a)}.\hspace{5.6cm}(6.1.68)\end{eqnarray*} Of
course, applying the general forms of the transformations in (6.1.3)
and (6.1.6) to the solutions in the above theorem, we will get more
solutions of the Schr\"{o}dinger equation.

\section{Coupled Schr\"{o}dinger Equations}

The coupled two-dimensional cubic nonlinear  Schr\"{o}dinger
equations\index{coupled  nonlinear Schr\"{o}dinger equations}
$$ i\psi_t+\kappa_1(\psi_{xx}+\psi_{yy})+(\ves_1|\psi|^2+\es_1|\vf|^2)\psi=0,
\eqno(6.2.1)$$
$$ i\vf_t+\kappa_2(\vf_{xx}+\vf_{yy})+(\ves_2|\psi|^2+\es_2|\vf|^2)\vf=0
\eqno(6.2.2)$$ are used to describe interaction of electromagnetic
waves with different polarizations in nonlinear optics, where
$\kappa_1,\kappa_2,\ves_1,\ves_2,\es_1$ and $\es_2$ are real
constants. Radhakrishnan and Lakshmanan [RL1] (1995) used
Painlev\'{e} analysis to find a Hirota bilinearization of the above
system of partial differential equations and obtained bright and
dark multiple soliton soutions. They [RL2] (1995) also generalized
their results to the coupled nonlinear Schr\"{o}dinger equations
with higher-order effects. Gr\'{e}bert and Guillot [GG] (1996)
construcetd periodic solutions of coupled one-dimensional nonlinear
Schr\"{o}dinger equations with periodic boundary conditions in some
resonance situations. Moreover, Hioe and Salter [HS] (2002) found a
connections between Lam\'{e} functions and solutions of the above
coupled equations. In this section, we want to apply the
quadratic-argument approach to the coupled nonlinear Schr\"{o}dinger
equations. Results are due to our work [X14].

As (6.1.3), we have the following symmetric transformations of the
coupled equations (6.2.1) and (6.2.2):
$$T_{a,a_0;b;\sta}^{(a_1,a_2,a_3)}(\psi)=be^{ai}
\psi(b^2(t+a_1),b(x\cos\sta+y\sin\sta+a_2),b(-x\sin\sta+y\cos
\sta+a_3)),\eqno(6.2.3)$$
$$T_{a,a_0;b;\sta}^{(a_1,a_2,a_3)}(\vf)=be^{a_0i}
\vf(b^2(t+a_1),b(x\cos\sta+y\sin\sta+a_2),b(-x\sin\sta+y\cos
\sta+a_3)).\eqno(6.2.4)$$ Moreover, (6.1.6) implies the following
symmetry
$$S_{a_1,a_2}(\psi(t,x,y))=e^{[(a_1x+a_2y)-(a_1^2+a_2^2)t]i/\kappa_1}\psi(t,x-2
a_1t,y-2a_2t),\eqno(6.2.5)$$
$$S_{a_1,a_2}(\vf(t,x,y))=e^{[(a_1x+a_2y)-(a_1^2+a_2^2)t]i/\kappa_2}\vf(t,x-2
a_1t,y-2 a_2t)\eqno(6.2.6)$$ of the coupled equations. In addition
to the above
 symmetries, we
 also solve the coupled equations modulo the following symmetry:
 $$(\psi,\kappa_1,\ves_1,\es_1)\leftrightarrow
(\vf,\kappa_2,\ves_2,\es_2).\eqno(6.2.7)$$

 Write
$$\psi=\xi(t,x,y)e^{i\phi(t,x,y)},\qquad \vf=\eta(t,x,y)e^{i\mu(t,x,y)}\eqno(6.2.8)$$
where $\xi,\phi,\eta$ and $\mu$ are real functions in $t,x,y$. As
the arguments in (6.1.8)-(6.1.13), the system (6.2.1) and (6.2.2) is
equivalent to the following system for real functions:
$$\xi_t+\kappa_1(2\xi_x \phi_x+2\xi_y
\phi_y+\xi(\phi_{xx}+\phi_{yy}))=0,\eqno(6.2.9)$$
$$-\xi[\phi_t+\kappa_1(\phi_x^2+\phi_y^2)]
+\kappa_1(\xi_{xx}+\xi_{yy})+(\ves_1\xi^2+\es_1\eta^2)\xi=0,\eqno(6.2.10)$$
$$\eta_t+\kappa_2(2\eta_x \mu_x+2\eta_y
\mu_y+\eta(\mu_{xx}+\mu_{yy}))=0,\eqno(6.2.11)$$
$$-\eta[\mu_t+\kappa_2(\mu_x^2+\mu_y^2)]
+\kappa_2(\eta_{xx}+\eta_{yy})+(\ves_2\xi^2+\es_2\eta^2)\eta=0.\eqno(6.2.12)$$
Based on our experience in last section, we will solve the above
system according to the following cases. For the convenience, we
always assume the conditions on the constants involved in an
expression such that it make sense. For instance, when we use
$\sqrt{d_1-d_2}$, we naturally assume $d_1\geq d_2$. \psp

{\it Case 1}. $(\phi,\mu)=(0,0)$ and $\ves_1\es_2-\ves_2\es_1\neq
0$.\psp

In this case, $\xi_t=\eta_t=0$ by (6.2.9) and (6.2.11). Moreover,
(6.2.10) and (6.2.12) become
$$\kappa_1(\xi_{xx}+\xi_{yy})+(\ves_1\xi^2+\es_1\eta^2)\xi=0,\qquad
\kappa_2(\eta_{xx}+\eta_{yy})+(\ves_2\xi^2+\es_2\eta^2)\eta=0,\eqno(6.2.13)$$
where $\iota_1$ and $\iota_2$ are constants to be determined. Assume
$$\xi=\frac{\iota_1}{x},\qquad\eta=\frac{\iota_2}{x}.\eqno(6.2.14)$$
Then (6.2.13) is equivalent to:
$$\ves_1\iota_1^2+\es_1\iota_2^2+2\kappa_1=0,\qquad
\ves_2\iota_1^2+\es_2\iota_2^2+2\kappa_2=0.\eqno(6.2.15)$$ Solving
the above linear algebraic equations for $\iota_1^2$ and
$\iota_2^2$, we have:
$$\iota_1^2=\frac{2(\es_1\kappa_2-\es_2\kappa_1)}{\ves_1\es_2-\ves_2\es_1},\qquad\iota_2^2=
\frac{2(\ves_2\kappa_1-\ves_1\kappa_2)}{\ves_1\es_2-\ves_2\es_1}.\eqno(6.2.16)$$
Thus we have the following solution
$$\xi=\frac{\sgm_1}{x}\sqrt{\frac{2(\es_1\kappa_2-\es_2\kappa_1)}{\ves_1\es_2-\ves_2\es_1}},\qquad
\eta=\frac{\sgm_2}{x}\sqrt{\frac{2(\ves_2\kappa_1-\ves_1\kappa_2)}{\ves_1\es_2-\ves_2\es_1}}
\eqno(6.2.17)$$ for $\sgm_1,\sgm_2\in\{1,-1\}$. Similarly, we have
the solution:
$$\xi=\sgm_1\sqrt{\frac{\es_1\kappa_2-\es_2\kappa_1}{(\ves_1\es_2-\ves_2\es_1)(x^2+y^2)}},\qquad
\eta=\sgm_2\sqrt{\frac{\ves_2\kappa_1-\ves_1\kappa_2}{(\ves_1\es_2-\ves_2\es_1)(x^2+y^2)}}.\eqno(6.2.18)$$
\psp

{\it Case 2}. $(\phi,\mu)=(k_1t,k_2t)$ with $k_1,k_2\in\mbb{R}$.
\psp

Again we have $\xi_t=\eta_t=0$ by (6.2.9) and (6.2.11). Moreover,
(6.2.10) and (6.2.12) become
$$-k_1\xi+\kappa_1(\xi_{xx}+\xi_{yy})+(\ves_1\xi^2+\es_1\eta^2)\xi=0,\;\;
-k_2\eta+\kappa_2(\eta_{xx}+\eta_{yy})+(\ves_2\xi^2+\es_2\eta^2)\eta=0.\eqno(6.2.19)$$

First we assume $\ves_1\es_2-\ves_2\es_1\neq 0$ and
$$\xi=\iota_1\Im(x),\qquad \eta=\iota_2\Im(x),\eqno(6.2.20)$$
where $\iota_1$ and $\iota_2$ are constants to be determined. Then
(6.2.19) becomes
$$-k_1\Im+\kappa_1{\Im'}'+(\ves_1\iota_1^2+\es_1\iota_2^2)\Im^3=0,\qquad
-k_2\Im+\kappa_2{\Im'}'+(\ves_2\iota_1^2+\es_2\iota_2^2)\Im^3=0.\eqno(6.2.21)$$
 According to (3.5.17)-(3.5.20), when $\Im=\tan
x,\;\sec x,\;\coth x$ and $\csch x$, we always have
$$\ves_1\iota_1^2+\es_1\iota_2^2+2\kappa_1=0,\qquad
\ves_2\iota_1^2+\es_2\iota_2^2+2\kappa_2=0.\eqno(6.2.22)$$  Thus for
$\sgm_1,\sgm_2\in\{1,-1\}$, we have the following solutions:
$$\xi=\sgm_1\sqrt{\frac{2(\es_1\kappa_2-\es_2\kappa_1)}{\ves_1\es_2-\ves_2\es_1}}\:\tan
x,\;\;
\eta=\sgm_2\sqrt{\frac{2(\ves_2\kappa_1-\ves_1\kappa_2)}{\ves_1\es_2-\ves_2\es_1}}\:\tan
x \eqno(6.2.23)$$ with $(k_1,k_2)=2(\kappa_1,\kappa_2)$;
$$\xi=\sgm_1\sqrt{\frac{2(\es_1\kappa_2-\es_2\kappa_1)}{\ves_1\es_2-\ves_2\es_1}}\:\sec x,\;\;
\eta=\sgm_2\sqrt{\frac{2(\ves_2\kappa_1-\ves_1\kappa_2)}{\ves_1\es_2-\ves_2\es_1}}\:\sec
x\eqno(6.2.24)$$ with $(k_1,k_2)=-(\kappa_1,\kappa_2)$;
$$\xi=\sgm_1\sqrt{\frac{2(\es_1\kappa_2-\es_2\kappa_1)}{\ves_1\es_2-\ves_2\es_1}}\:\coth
x,\;\;
\eta=\sgm_2\sqrt{\frac{2(\ves_2\kappa_1-\ves_1\kappa_2)}{\ves_1\es_2-\ves_2\es_1}}\:\coth
x\eqno(6.2.25)$$ with $(k_1,k_2)=-2(\kappa_1,\kappa_2);$
$$\xi=\sgm_1\sqrt{\frac{2(\es_1\kappa_2-\es_2\kappa_1)}{\ves_1\es_2-\ves_2\es_1}}\:\csch x,\;\;
\eta=\sgm_2\sqrt{\frac{2(\ves_2\kappa_1-\ves_1\kappa_2)}{\ves_1\es_2-\ves_2\es_1}}\:\csch
x\eqno(6.2.26)$$ with $(k_1,k_2)=(\kappa_1,\kappa_2).$ Similarly,
(3.5.14)-(3.5.16) give us the following solutions:
$$\xi=m\sgm_1\sqrt{\frac{2(\es_1\kappa_2-\es_2\kappa_1)}{\ves_1\es_2-\ves_2\es_1}}\;\sn(x|m),\qquad
\eta=m\sgm_2\sqrt{\frac{2(\ves_2\kappa_1-\ves_1\kappa_2)}{\ves_1\es_2-\ves_2\es_1}}\;\sn
(x|m) \eqno(6.2.27)$$ with $(k_1,k_2)=-(1+m^2)(\kappa_1,\kappa_2);$
$$\xi=m\sgm_1\sqrt{\frac{2(\es_2\kappa_1-\es_1\kappa_2)}{\ves_1\es_2-\ves_2\es_1}}\;\cn (x|m)\qquad
\eta=m\sgm_2\sqrt{\frac{2(\ves_1\kappa_2-\ves_2\kappa_1)}{\ves_1\es_2-\ves_2\es_1}}\;\cn
(x|m) \eqno(6.2.28)$$ with $(k_1,k_2)=(2m^2-1)(\kappa_1,\kappa_2);$
$$\xi=\sgm_1\sqrt{\frac{2(\es_2\kappa_1-\es_1\kappa_2)}{\ves_1\es_2-\ves_2\es_1}}\;\dn
(x|m),\qquad
\eta=\sgm_2\sqrt{\frac{2(\ves_1\kappa_2-\ves_2\kappa_1)}{\ves_1\es_2-\ves_2\es_1}}\;\dn
(x|m)\eqno(6.2.29)$$ with $(k_1,k_2)=(2-m^2)(\kappa_1,\kappa_2).$

If $(\ves_1,\es_1)= \ves_1(1,d^2)$ and
$(\ves_2,\es_2)=\ves_2(1,d^2)$ with $d\in\mbb{R}$, then (6.2.19)
becomes
$$-k_1\xi+\kappa_1(\xi_{xx}+\xi_{yy})+\ves_1(\xi^2+d^2\eta^2)\xi=0,\;\;
-k_2\eta+\kappa_2(\eta_{xx}+\eta_{yy})+\ves_2(\xi^2+d^2\eta^2)\eta=0.\eqno(6.2.30)$$
The sum of squares and $\sin^2x+\cos^2x=1$ motivate us to try
$$\xi=d\ell\sin x,\qquad \eta=\ell\cos x\eqno(6.2.31)$$
for any $0\neq \ell\in\mbb{R}$. Substitute them into (6.2.30), we
have
$$-k_1-\kappa_1+d^2\ell^2\ves_1=0,\;\;
-k_2-\kappa_2+d^2\ell^2\ves_2=0.\eqno(6.2.32)$$ So
$$
(k_1,k_2)=(d^2\ell^2\ves_1-\kappa_1,d^2\ell^2\ves_2-\kappa_2).\eqno(6.2.33)$$

 When
$(\ves_1,\es_1)= \ves_1(1,-d^2)$ and $(\ves_2,\es_2)=\ves_2(1,-d^2)$
with $d\in\mbb{R}$, then (6.2.19) becomes
$$-k_1\xi+\kappa_1(\xi_{xx}+\xi_{yy})+\ves_1(\xi^2-d^2\eta^2)\xi=0,\;\;
-k_2\eta+\kappa_2(\eta_{xx}+\eta_{yy})+\ves_2(\xi^2-d^2\eta^2)\eta=0.\eqno(6.2.34)$$
The difference of squares and $\cosh^2x-\sinh^2x=1$ motivate us to
try
$$\xi=d\ell\cosh x,\qquad \eta=\ell\sinh x\eqno(6.2.35)$$
for any $0\neq \ell\in\mbb{R}$. Substitute them into (6.2.34), we
have
$$-k_1+\kappa_1+d^2\ell^2\ves_1=0,\;\;
-k_2+\kappa_2+d^2\ell^2\ves_2=0.\eqno(6.2.36)$$ Hence
$$
(k_1,k_2)=(d^2\ell^2\ves_1+\kappa_1,d^2\ell^2\ves_2+\kappa_2).\eqno(6.2.37)$$

In summary, we have the following theorem. \psp

{\bf Theorem 6.2.1}. {\it Let $d,\ell,m\in\mbb{R}$ with $0<m<1$ and
let $\sgm_1,\sgm_2\in\{1,-1\}$. If $a_1\es_2-\ves_2\es_1\neq 0$, we
have the following solutions of the coupled two-dimensional cubic
nonlinear Schr\"{o}dinger equations (6.2.1) and (6.2.2):
$$\psi=\frac{\sgm_1}{x}\sqrt{\frac{2(\es_1\kappa_2-\es_2\kappa_1)}{\ves_1\es_2-\ves_2\es_1}},\qquad
\vf=\frac{\sgm_2}{x}\sqrt{\frac{2(\ves_2\kappa_1-\ves_1\kappa_2)}{\ves_1\es_2-\ves_2\es_1}};
\eqno(6.2.38)$$
$$\psi=\sgm_1\sqrt{\frac{\es_1\kappa_2-\es_2\kappa_1}{(\ves_1\es_2-\ves_2\es_1)(x^2+y^2)}},\qquad
\vf=\sgm_2\sqrt{\frac{\ves_2\kappa_1-\ves_1\kappa_2}{(\ves_1\es_2-\ves_2\es_1)(x^2+y^2)}};\eqno(6.2.39)$$
$$\psi=\sgm_1\sqrt{\frac{2(\es_1\kappa_2-\es_2\kappa_1)}{\ves_1\es_2-\ves_2\es_1}}\:e^{2\kappa_1ti}\tan x,\;\;
\vf=\sgm_2\sqrt{\frac{2(\ves_2\kappa_1-\ves_1\kappa_2)}{\ves_1\es_2-\ves_2\es_1}}\:e^{2\kappa_2ti}\tan
x;\eqno(6.2.40)$$
$$\psi=\sgm_1\sqrt{\frac{2(\es_1\kappa_2-\es_2\kappa_1)}{\ves_1\es_2-\ves_2\es_1}}\:e^{-\kappa_1ti}\sec x,\;\;
\vf=\sgm_2\sqrt{\frac{2(\ves_2\kappa_1-\ves_1\kappa_2)}{\ves_1\es_2-\ves_2\es_1}}\:e^{-\kappa_2ti}\sec
x;\eqno(6.2.41)$$
$$\psi=\sgm_1\sqrt{\frac{2(\es_1\kappa_2-\es_2\kappa_1)}{\ves_1\es_2-\ves_2\es_1}}\:e^{-2\kappa_1ti}\coth x,\;\;
\vf=\sgm_2\sqrt{\frac{2(\ves_2\kappa_1-\ves_1\kappa_2)}{\ves_1\es_2-\ves_2\es_1}}\:e^{-2\kappa_2ti}\coth
x;\eqno(6.2.42)$$}
$$\psi=\sgm_1\sqrt{\frac{2(\es_1\kappa_2-\es_2\kappa_1)}{\ves_1\es_2-\ves_2\es_1}}\:e^{\kappa_1ti}\csch
x,\qquad
\vf=\sgm_2\sqrt{\frac{2(\ves_2\kappa_1-\ves_1\kappa_2)}{\ves_1\es_2-\ves_2\es_1}}\:e^{\kappa_2ti}\csch
x;\eqno(6.2.43)$$
$$\psi=m\sgm_1\sqrt{\frac{2(\es_1\kappa_2-\es_2\kappa_1)}{\ves_1\es_2-\ves_2\es_1}}\;e^{-(1+m^2)\kappa_1ti}
\sn (x|m),\eqno(6.2.44)$$
$$\vf=m\sgm_2\sqrt{\frac{2(\ves_2\kappa_1-\ves_1\kappa_2)}{\ves_1\es_2-\ves_2\es_1}}\;e^{-(1+m^2)\kappa_2ti}\sn
(x|m); \eqno(6.2.45)$$
$$\psi=m\sgm_1\sqrt{\frac{2(\es_2\kappa_1-\es_1\kappa_2)}{\ves_1\es_2-\ves_2\es_1}}\;e^{(2m^2-1)\kappa_1ti}
\cn (x|m),\eqno(6.2.46)$$
$$\vf=m\sgm_2\sqrt{\frac{2(\ves_1\kappa_2-\ves_2\kappa_1)}{\ves_1\es_2-\ves_2\es_1}}\;e^{(2m^2-1)\kappa_2ti}\cn
(x|m);\eqno(6.2.47)$$
$$\psi=\sgm_1\sqrt{\frac{2(\es_2\kappa_1-\es_1\kappa_2)}{\ves_1\es_2-\ves_2\es_1}}\;e^{(2-m^2)\kappa_1ti}
\dn (x|m),\eqno(6.2.48)$$
$$\vf=\sgm_2\sqrt{\frac{2(\ves_1\kappa_2-\ves_2\kappa_1)}{\ves_1\es_2-\ves_2\es_1}}\;e^{(2-m^2)\kappa_1ti}\dn
(x|m).\eqno(6.2.49)$$
 {\it If $(\ves_1,\es_1)= \ves_1(1,d^2)$ and
$(\ves_2,\es_2)=\ves_2(1,d^2)$},
$$\psi=d\ell e^{(d^2\ell^2\ves_1-\kappa_1)ti}\sin x,\qquad\vf=\ell e^{(d^2\ell^2\ves_2
-\kappa_2)ti} \cos x.\eqno(6.2.50)$$ {\it When $(\ves_1,\es_1)=
\ves_1(1,-d^2)$ and $(\ves_2,\es_2)=\ves_2(1,-d^2)$},
$$\psi=d\ell e^{(d^2\ell^2\ves_1+\kappa_1)ti}\cosh x,\qquad
\eta=\ell e^{(d^2\ell^2\ves_2+\kappa_2)ti}\sinh x.\eqno(6.2.51)$$
\pse

{\bf Remark 6.2.2}. Applying the symmetric transformations
(6.2.3)-(6.2.6) to the above solutions, we can get more
sophisticated ones. For instance, by (6.2.38), we get the following
traveling-wave solution
$$\psi=\frac{\sgm_1e^{ai+a_1(x\cos\sta+y\sin
\sta+a_2-a_1t)i/\kappa_1}}{x\cos\sta+y\sin\sta-2a_1t+a_2}
\sqrt{\frac{2(\es_1\kappa_2-\es_2\kappa_1)}{\ves_1\es_2-\ves_2\es_1}},\eqno(6.2.52)$$
$$\vf=\frac{\sgm_2e^{a_0i+a_1(x\cos\sta+y\sin
\sta+a_2-a_1t)i/\kappa_2}}{x\cos\sta+y\sin
\sta-2a_1t+a_2}\sqrt{\frac{2(\ves_2\kappa_1-\ves_1\kappa_2)}{\ves_1\es_2-\ves_2\es_1}}.
\eqno(6.2.53)$$ Since $\lim_{m\rta 1}cn(x|m)=\sech x$, (6.2.46) and
(6.2.47) yield the solution
$$\psi=\sgm_1\sqrt{\frac{2(\es_2\kappa_1-\es_1\kappa_2)}{\ves_1\es_2-\ves_2\es_1}}\;e^{\kappa_1ti}
\sech
x,\;\;\vf=\sgm_2\sqrt{\frac{2(\ves_1\kappa_2-\ves_2\kappa_1)}{\ves_1\es_2-\ves_2\es_1}}\;e^{\kappa_2ti}\sech
x.\eqno(6.2.54)$$ The symmetric transformations (6.2.3)-(6.2.6) give
us the following soliton solution\index{soliton solution!of coupled
 Schr\"{o}dinger equations}
\begin{eqnarray*}\qquad\qquad\psi&=&b\sgm_1\sqrt{\frac{2(\es_2\kappa_1-\es_1\kappa_2)}{\ves_1\es_2-\ves_2\es_1}}
e^{(b^2\kappa_1t+a)i+a_1b(x\cos\sta+y\sin
\sta+a_2-a_1bt)i/\kappa_1}\\& &\times \sech
b(x\cos\sta+y\sin\sta-2a_1bt+a_2),\hspace{4.8cm}(6.2.55)\end{eqnarray*}
\begin{eqnarray*}\qquad\qquad\vf&=&b\sgm_2\sqrt{\frac{2(\ves_1\kappa_2-\ves_2\kappa_1)}{\ves_1\es_2-\ves_2\es_1}}
\;e^{(b^2\kappa_2t+a_0)i+a_1b(x\cos\sta+y\sin
\sta+a_2-a_1bt)i/\kappa_2}\\& &\times\sech
b(x\cos\sta+y\sin\sta-2a_1bt+a_2).\hspace{4.9cm}(6.2.56)\end{eqnarray*}
If $(\ves_1,\es_1)= \ves_1(1,d^2)$ and
$(\ves_2,\es_2)=\ves_2(1,d^2)$,  (6.2.3)-(6.2.6) and (6.2.50) yield
the following wave solution
\begin{eqnarray*}\qquad\qquad\psi&=&bd\ell
e^{[b^2(d^2\ell^2\ves_1-\kappa_1)t+a]i+a_1b(x\cos\sta+y\sin
\sta+a_2-a_1bt)i/\kappa_1}\\ & &\times\sin
b(x\cos\sta+y\sin\sta-2a_1bt+a_2),\hspace{4.9cm}(6.2.57)\end{eqnarray*}
\begin{eqnarray*}\qquad\qquad\vf&=&b\ell e^{[b^2(d^2\ell^2\ves_2
-\kappa_2)t+a_0]i+a_1b(x\cos\sta+y\sin \sta+a_2-a_1bt)i/\kappa_2}
\\ & &\times\cos b(x\cos\sta+y\sin\sta-2a_1bt+a_2).\hspace{5cm}(6.2.58)\end{eqnarray*}\pse

{\it Case 3}. $\phi=x^2/4\kappa_1t+\be_1$ and
$\mu=(x-d)^2/4\kappa_2(t-\ell)+\be_2$ or
$\mu=y^2/4\kappa_2(t-\ell)+\be_2$
 for some functions $\be_1$ and $\be_2$ in $t$ and real constants
 $d$ and $\ell$.\psp

First we assume $\mu=(x-d)^2/4\kappa_2(t-\ell)+\be_2$. Then (6.2.9)
and (6.2.11) become
$$\xi_t+\frac{x}{t}\xi_x+\frac{1}{2t}\xi=0,\qquad
 \eta_t+\frac{x-d}{t-\ell}\eta_x+\frac{1}{2(t-\ell)}\eta=0.
\eqno(6.2.59)$$ Thus
$$\xi=\frac{1}{\sqrt{t}}\hat\xi(t^{-1}x,y),\qquad\eta=
\frac{1}{\sqrt{t-\ell}}\hat\eta((t-\ell)^{-1}(x-d),y)\eqno(6.2.60)$$
for some two-variable functions $\hat\xi$ and $\hat\eta$. On the
other hand, (6.2.10) and (6.2.12) become
$$-\be_1'\xi+\kappa_1(\xi_{xx}+\xi_{yy})+(\ves_1\xi^2+\es_1\eta^2)\xi=0,\eqno(6.2.61)$$
$$-\be_2'\eta+\kappa_2(\eta_{xx}+\eta_{yy})+(\ves_2\xi^2+\es_2\eta^2)\eta=0.
\eqno(6.2.62)$$ As (6.1.40)-(6.1.43), the above two equations force
us to take
$$\xi=\frac{c_1}{\sqrt{t}},\qquad\eta=
\frac{c_2}{\sqrt{t-\ell}}.\eqno(6.2.63)$$ So (6.2.61) and (6.2.62)
are implied by the equations:
$$\be_1'=\frac{c_1^2\ves_1}{t}+\frac{c_2^2\es_1}{t-\ell},\qquad
\be_2'=\frac{c_1^2\ves_2}{t}+\frac{c_2^2\es_2}{t-\ell}.\eqno(6.2.64)$$
For simplicity, we take
$$\be_1=c_1^2\ves_1\ln t+c_2^2\es_1\ln(t-\ell),\qquad
\be_2=c_1^2\ves_2\ln t+c_2^2\es_2\ln(t-\ell).\eqno(6.2.65)$$ Exact
same approach holds for $\mu=y^2/4\kappa_2(t-\ell)+\be_2$.\psp

{\bf Theorem 6.2.3}. {\it Let $c_1,c_2,d,\ell\in\mbb{R}$. We have
the following solutions of the coupled two-dimensional cubic
nonlinear Schr\"{o}dinger equations (6.2.1) and (6.2.2)}:
$$\psi=c_1t^{c_1^2\ves_1i-1/2}(t-\ell)^{c_2^2\es_1i}e^{x^2i/2\kappa_1t},\qquad
\vf=c_2t^{c_1^2\ves_2i}(t-\ell)^{c_2^2\es_2i-1/2}e^{(x-d)^2i/2\kappa_2(t-\ell)};
\eqno(6.2.66)$$
$$\psi=c_1t^{c_1^2\ves_1i-1/2}(t-\ell)^{c_2^2\es_1i}e^{x^2i/2\kappa_1t},\qquad
\vf=c_2t^{c_1^2\ves_2i}(t-\ell)^{c_2^2\es_2i-1/2}e^{y^2i/2\kappa_2(t-\ell)}.
\eqno(6.2.67)$$\pse

{\it Case 4}. $\phi=x^2/4\kappa_1t+\be_1$ and
$\mu=(x-d)^2/4\kappa_2(t-\ell_1)+y^2/4\kappa_2(t-\ell_2)+\be_2$
 for some functions $\be_1$ and $\be_2$ in $t$ and real constants
 $d,\ell_1$ and $\ell_2$.
 \psp

In this case,  (6.2.9) and (6.2.11) become
$$\xi_t+\frac{x}{t}\xi_x+\frac{1}{2t}\xi=0,\;\;\eta_t+\frac{x-d}{t-\ell_1}\eta_x +\frac{y}{t-\ell_2}\eta_y+
\left(\frac{1}{2(t-\ell_1)}+\frac{1}{2(t-\ell_2)}\right)\xi=0.
\eqno(6.2.68)$$ Thus
$$\xi=\frac{1}{\sqrt{t}}\hat\xi(t^{-1}x,y),\qquad\eta=
\frac{1}{\sqrt{(t-\ell_1)(t-\ell_2)}}\hat\eta((t-\ell_1)^{-1}(x-d),
(t-\ell_2)^{-1}y)\eqno(6.2.69)$$ for some two-variable functions
$\hat\xi$ and $\hat\eta$ by the method of characteristic lines in
Section 4.1. Again (6.2.10) and (6.2.12) become (6.2.61) and
(6.2.62), respectively. Moreover, they force us to take
$$\xi=\frac{c_1}{\sqrt{t}},\qquad\eta=
\frac{c_2}{\sqrt{(t-\ell_1)(t-\ell_2)}}.\eqno(6.2.70)$$ So (6.2.10)
and (6.2.12) are implied by the equations:
$$\be_1'=\frac{c_1^2\ves_1}{t}+\frac{c_2^2\es_1}{(t-\ell_1)(t-\ell_2)},\qquad
\be_2'=\frac{c_1^2\ves_2}{t}+\frac{c_2^2\es_2}{(t-\ell_1)(t-\ell_2)}.
\eqno(6.2.71)$$ For simplicity, we get
$$\be_1=c_1^2\ves_1\ln t+\frac{c_2^2\es_1}{\ell_2-\ell_1}\ln\frac
{t-\ell_1}{t-\ell_2},\qquad\be_2=\ves_2c_1^2\ln
t+\frac{\es_2c_2^2}{\ell_2-\ell_1}\ln\frac
{t-\ell_1}{t-\ell_2}\eqno(6.2.72)$$ if $\ell_1\neq \ell_2$, and
$$\be_1=c_1^2\ves_1\ln t-\frac{c_2^2\es_1}{t-\ell_1},
,\qquad\be_2=c_1^2\ves_2\ln
t-\frac{c_2^2\es_2}{t-\ell_1}\eqno(6.2.73)$$ when
$\ell_1=\ell_2$.\psp

{\bf Theorem 6.2.4}. {\it Let $c_1,c_2,\ell_1,\ell_2\in\mbb{R}$ such
that $\ell_1\neq \ell_2$. We have the following solutions of the
coupled two-dimensional cubic nonlinear Schr\"{o}dinger equations
(6.2.1) and (6.2.2)}:
$$\psi=c_1t^{c_1^2\ves_1i-1/2}(t-\ell_1)^{c_2^2\es_1(\ell_2-\ell_1)^{-1}i}
(t-\ell_2)^{-c_2^2\es_1(\ell_2-\ell_1)^{-1}i}e^{x^2i/4\kappa_1t},\eqno(6.2.74)$$
\begin{eqnarray*}\hspace{2cm}\vf&=&c_2t^{c_1^2\ves_2i}(t-\ell_1)^{c_2^2\es_2(\ell_2-\ell_1)^{-1}i-1/2}
(t-\ell_2)^{-c_2^2\es_2(\ell_2-\ell_1)^{-1}i-1/2}\\
& &\times\exp\left(\frac{(x-d)^2i} {4\kappa_2(t-\ell_1)}+\frac{y^2i}
{4\kappa_2(t-\ell_1)}\right);\hspace{4.7cm}(6.2.75)\end{eqnarray*}
$$\psi=c_1t^{c_1^2\ves_1i-1/2}\exp\left(\frac{x^2i} {4\kappa_1t}
-\frac{c_2^2\es_1i}{t-\ell_1}\right),\eqno(6.2.76)$$
$$\vf=\frac{c_2t^{c_1^2\ves_2i}}{t-\ell_1}
\exp\frac{((x-d)^2+y^2-4c_2^2\kappa_2\es_2)i}
{4\kappa_2(t-\ell_1)}.\eqno(6.2.77)$$ \pse

{\it Case 5}. For $\ell_1,\ell_2,\ell,d_1,d_2\in\mbb{R}$ and
functions $\be_1,\be_2$ in $t$,
$$\phi=\frac{x^2}{4\kappa_1t}+\frac{y^2}{4\kappa_1(t-\ell)}+\be_1,\qquad
\mu=\frac{(x-d_1)^2}{4\kappa_2(t-\ell_1)}+\frac{(y-d_2)^2}{4\kappa_1(t-\ell_2)}+\be_2.
\eqno(6.2.78)$$\pse

As  the above case, we get
$$\xi=\frac{c_1}{\sqrt{t(t-\ell)}},\qquad\eta=
\frac{c_2}{\sqrt{(t-\ell_1)(t-\ell_2)}}.\eqno(6.2.79)$$ So (6.2.10)
and (6.2.12) are implied by the equations:
$$\be_1'=\frac{c_1^2\ves_1}{t(t-\ell)}+\frac{c_2^2\es_1}{(t-\ell_1)(t-\ell_2)},\qquad
\be_2'=\frac{c_1^2\ves_2}{t}+\frac{c_2^2\es_2}{(t-\ell_1)(t-\ell_2)}.
\eqno(6.2.80)$$ For simplicity, we have
$$\be_1=\frac{c_1^2\ves_1}{\ell}\ln \frac{t-\ell}{t}+\frac{c_2^2\es_1}{\ell_2-\ell_1}\ln\frac
{t-\ell_1}{t-\ell_2},\qquad\be_2=\frac{c_1^2\ves_2}{\ell}\ln
\frac{t-\ell}{t}+\frac{c_2^2\es_2}{\ell_2-\ell_1}\ln\frac
{t-\ell_1}{t-\ell_2}\eqno(6.2.81)$$ if $\ell\neq 0$ and
$\ell_1\neq\ell_2$;
$$\be_1=-\frac{c_1^2\ves_1}{t}+\frac{c_2^2\es_1}{\ell_2-\ell_1}\ln\frac
{t-\ell_1}{t-\ell_2},\qquad\be_2=-\frac{c_1^2\ves_2}{t}+\frac{c_2^2\es_2}{\ell_2-\ell_1}\ln\frac
{t-\ell_1}{t-\ell_2}\eqno(6.2.82)$$ when $\ell=0$  and
$\ell_1\neq\ell_2$;
$$\be_1=\frac{c_1^2\ves_1}{t}-\frac{c_2^2\es_1}{t-\ell_1},
,\qquad\be_2=\frac{c_1^2\ves_2}{t}
t-\frac{c_2^2\es_2}{t-\ell_1}\eqno(6.2.83)$$ if $\ell=0$ and
$\ell_1=\ell_2$. Therefore, we obtain:\psp

{\bf Theorem 6.2.5}. {\it Let
$c_1,c_2,\ell,d_1,d_2,\ell_1,\ell_2\in\mbb{R}$ such that $\ell\neq
0$ and $\ell_1\neq\ell_2$. We have the following solutions of the
coupled two-dimensional cubic nonlinear Schr\"{o}dinger equations
(6.2.1) and (6.2.2)}:
$$\psi=\frac{c_1}{t}
\exp\left(\frac{(x^2+y^2-4c_1^2\kappa_1\ves_1)i}
{4\kappa_1t}-\frac{c_2^2\es_1i}{t-\ell_1}\right),\eqno(6.2.84)$$
$$\vf=\frac{c_2}{t-\ell_1}
\exp\left(\frac{((x-d_1)^2+(y-d_2)^2-4c_2^2\kappa_2\es_2)i}
{4\kappa_2(t-\ell_1)}-\frac{c_1^2\ves_2i}{t}\right);\eqno(6.2.85)$$
$$\psi=\frac{c_1(t-\ell_1)^{c_2^2\es_1i/(\ell_2-\ell_1)}
(t-\ell_2)^{-c_2^2\es_1i/(\ell_2-\ell_1)}}{t}
\exp\frac{(x^2+y^2-4c_1^2\kappa_1\ves_1)i}
{4\kappa_1t},\eqno(6.2.86)$$
\begin{eqnarray*}\hspace{2cm}\vf&=&c_2(t-\ell_1)^{c_2^2\es_2i/(\ell_2-\ell_1)-1/2}
(t-\ell_2)^{-c_2^2\es_2i/(\ell_2-\ell_1)-1/2}
\\ & &\times\exp\left(\frac{(x-d_1)^2i}{4\kappa_2(t-\ell_1)}+\frac{(y-d_2)^2i}{4\kappa_2(t-\ell_2)}
-\frac{c_1^2\ves_2i}{t}\right);\hspace{3.3cm}(6.2.87)\end{eqnarray*}
\begin{eqnarray*}\hspace{1.9cm}\psi&=&c_1t^{-c_1^2\ves_1i/\ell-1/2}
(t-\ell)^{c_1^2\ves_1i/\ell-1/2}
(t-\ell_1)^{c_2^2\es_1i/(\ell_2-\ell_1)}\\ & &\times
(t-\ell_2)^{-c_2^2\es_1i/(\ell_2-\ell_1)}
\exp\left(\frac{x^2i}{4\kappa_1t}+\frac{y^2i}{4\kappa_1(t-\ell)}
\right),\hspace{2.7cm}(6.2.88)\end{eqnarray*}
\begin{eqnarray*}\hspace{1cm}\vf&=&c_2t^{-c_1^2\ves_2i/\ell}
(t-\ell)^{c_1^2\ves_2i/\ell}
(t-\ell_1)^{c_2^2\es_2i/(\ell_2-\ell_1)-1/2}\\ & &\times
(t-\ell_2)^{-c_2^2\es_2i/(\ell_2-\ell_1)-1/2}
\exp\left(\frac{(x-d_1)^2i}{4\kappa_2(t-\ell_1)}+
\frac{(y-d_2)^2i}{4\kappa_2(t-\ell_2)}
\right).\hspace{1.7cm}(6.2.89)\end{eqnarray*} \pse

{\it Case 6}. For  two functions $\be_1,\be_2$ in $t$,
$$\phi=\frac{x^2+y^2}{4\kappa_1t}+\be_1,\qquad
\mu=\frac{x^2+y^2}{4\kappa_2t}+\be_2. \eqno(6.2.90)$$\pse

As  Case 4, (6.2.9) and (6.2.11) imply
$$\xi=\frac{1}{t}\hat\xi(u,v),\qquad
\eta=\frac{1}{t}\hat\eta(u,v),\qquad
u=\frac{x}{t},\;v=\frac{y}{t}.\eqno(6.2.91)$$ Moreover, (6.2.10) and
(6.2.12) become
$$-\be_1'\hat\xi
+\frac{\kappa_1}{t^2}(\hat\xi_{uu}+\hat\xi_{vv})+
\frac{1}{t^2}(\ves_1\hat\xi^2+\es_1\hat\eta^2)\hat\xi=0,\eqno(6.2.92)$$
$$-\be_2'\hat\eta
+\frac{\kappa_2}{t^2}(\hat\eta_{uu}+\hat\eta_{vv})+
\frac{1}{t^2}(\ves_2\hat\xi^2+\es_2\hat\eta^2)\hat\eta=0.\eqno(6.2.93)$$
To solve the above system, we assume
$$\be_1=-\frac{c_1}{t},\qquad \be_2=-\frac{c_2}{t},\qquad
c_1,c_2\in\mbb{R}.\eqno(6.2.94)$$ Then (6.2.92) and (6.2.93) are
equivalent to:
$$-c_1\hat\xi
+\kappa_1(\hat\xi_{uu}+\hat\xi_{vv})+
(\ves_1\hat\xi^2+\es_1\hat\eta^2)\hat\xi=0,\eqno(6.2.95)$$
$$-c_2\hat\eta
+\kappa_2(\hat\eta_{uu}+\hat\eta_{vv})+(\ves_2\hat\xi^2+\es_2\hat\eta^2)
\hat\eta=0.\eqno(6.2.96)$$

For simplicity, we assume $\hat\xi$ and $\hat\eta$ are independent
of $v$. If $(\ves_1,\es_1)=\ves_1(1,d^2)$ and
$(\ves_2,\es_2)=\ves_2(1,d^2)$ with $d\in\mbb{R}$, we have the
following solution:
$$\hat\xi=d\ell\sin u,\;\; \hat\eta=\ell\cos u,\;\;
(c_1,c_2)=(d^2\ell^2\ves_1-\kappa_1,d^2\ell^2\ves_2-
\kappa_2)\eqno(6.2.97)$$ for $\ell\in\mbb{R}$.
 When
$(\ves_1,\es_1)=\ves_1(1,-d^2)$ and $(\ves_2,\es_2)=\ves_2(1,-d^2)$
with $d\in\mbb{R}$, we get the solution:
$$\hat\xi=d\ell\cosh\varpi,\;\; \hat\eta=\ell\sinh\varpi,\;\;
(c_1,c_2)=(d^2\ell^2\ves_1+\kappa_1,d^2\ell^2\ves_2+
\kappa_2)\eqno(6.2.98)$$ for $\ell\in\mbb{R}$.\psp

{\bf Theorem 6.2.6}. {\it For $d,\ell\in\mbb{R}$, we have the
following solutions of the coupled two-dimensional cubic nonlinear
Schr\"{o}dinger equations (6.2.1) and (6.2.2)}:
$$\psi=\frac{d\ell\sin (x/t)}
{t}\exp\left(\frac{x^2+y^2}{4\kappa_1t}+\frac{\kappa_1
-d^2\ell^2\ves_1}{t}\right)i,\eqno(6.2.99)$$
$$\vf=\frac{\ell\cos (x/t)}
{t}\exp\left(\frac{x^2+y^2}{4\kappa_2t}+\frac{\kappa_2
-d^2\ell^2\ves_2}{t}\right)i\eqno(6.2.100)$$ {\it if
$(\ves_1,\es_1)= \ves_1(1,d^2)$ and $(\ves_2,\es_2)=\ves_2(1,d^2)$};
$$\psi=\frac{d\ell\cosh(x/t)}
{t}\exp\left(\frac{x^2+y^2}{4\kappa_1t}-\frac{\kappa_1
+d^2\ell^2\ves_1}{t}\right)i,\eqno(6.2.101)$$
$$\vf=\frac{\ell\sinh(x/t)}
{t}\exp\left(\frac{x^2+y^2}{4\kappa_2t}-\frac{\kappa_2
+d^2\ell^2\ves_2}{t}\right)i\eqno(6.2.102)$$ {\it when
$(\ves_1,\es_1)= \ves_1(1,-d^2)$ and
$(\ves_2,\es_2)=\ves_2(1,-d^2)$}.\psp

{\bf Remark 6.2.7}. Applying the transformation in (6.2.3) and
(6.2.4) with $a=a_0=a_2=a_3=0$ to (6.2.99) and (6.2.100), we get a
more general wave-like solution:
$$\psi=\frac{d\ell\sin [(x\cos\sta+y\sin\sta)/(b(t-a_1))]}
{b(t-a_1)}\exp\left(\frac{x^2+y^2}{4\kappa_1(t-a_1)}+\frac{\kappa_1
-d^2\ell^2\ves_1}{b^2(t-a_1)}\right)i,\eqno(6.2.103)$$
$$\vf=\frac{\ell\cos [(x\cos\sta+y\sin\sta)/(b(t-a_1))]}
{b(t-a_1)}\exp\left(\frac{x^2+y^2}{4\kappa_2(t-a_1)}+\frac{\kappa_2
-d^2\ell^2\ves_2}{b^2(t-a_1)}\right)i\eqno(6.2.104)$$ if
$(\ves_1,\es_1)= \ves_1(1,d^2)$ and $(\ves_2,\es_2)=\ves_2(1,d^2)$,
where $a_1,b,\sta\in\mbb{R}$ with $b\neq 0$. We can get more
sophisticated wave-like solution if we apply the general forms of
the transformations in (6.2.3)-(6.2.6).\psp

Finally, we assume $\ves_1\es_2-\ves_2\es_1\neq 0$. Again we assume
that $\hat\xi$ and $\hat\eta$ are independent of $v$. By the
arguments in (6.2.19)-(6.2.30), we have:\psp

{\bf Theorem 6.2.8}. {\it Let $d,\ell,m\in\mbb{R}$ with $0<m<1$ and
let $\sgm_1,\sgm_2\in\{1,-1\}$. If $\ves_1\es_2-\ves_2\es_1\neq 0$,
we have the following solutions of the coupled two-dimensional cubic
nonlinear Schr\"{o}dinger equations (6.2.1) and (6.2.2):
$$\psi=\frac{\sgm_1}{x}\sqrt{\frac{2(\sgm_1\kappa_2-\es_2\kappa_1)}{\ves_1\es_2-\ves_2\es_1}}\;
\exp\frac{(x^2+y^2)i}{4\kappa_1t},\eqno(6.2.105)$$
$$\vf=\frac{\sgm_2}{x}\sqrt{\frac{2(\ves_2\kappa_1-\ves_1\kappa_2)}{\ves_1\es_2-\ves_2\es_1}}\;
\exp\frac{(x^2+y^2)i}{4\kappa_2t}; \eqno(6.2.106)$$
$$\psi=\sgm_1\sqrt{\frac{\es_1\kappa_2-\es_2\kappa_1}{(\ves_1\es_2-\ves_2\es_1)(x^2+y^2)}}\;
\exp\frac{(x^2+y^2)i}{4\kappa_1t},\eqno(6.2.107)$$
$$
\vf=\sgm_2\sqrt{\frac{\ves_2\kappa_1-\ves_1\kappa_2}{(\ves_1\es_2-\ves_2\es_1)(x^2+y^2)}}\;
\exp\frac{(x^2+y^2)i}{4\kappa_2t};\eqno(6.2.108)$$
$$\psi=\frac{\sgm_1}{t}\sqrt{\frac{2(\es_1\kappa_2-\es_2\kappa_1)}{\ves_1\es_2-\ves_2\es_1}}\tan\frac{x}{t}\:
\exp\left(\frac{x^2+y^2}{4\kappa_1t}-\frac{2\kappa_1}{t}\right)i,\eqno(6.2.109)$$
$$
\vf=\frac{\sgm_2}{t}\sqrt{\frac{2(\ves_2\kappa_1-\ves_1\kappa_2)}{\ves_1\es_2-\ves_2\es_1}}\tan
\frac{x}{t}\:
\exp\left(\frac{x^2+y^2}{4\kappa_2t}-\frac{2\kappa_2}{t}\right)i;\eqno(6.2.110)$$
$$\psi=\frac{\sgm_1}{t}\sqrt{\frac{2(\es_1\kappa_2-\es_2\kappa_1)}{\ves_1\es_2-\ves_2\es_1}}\:\sec \frac{x}{t}
\:
\exp\left(\frac{x^2+y^2}{4\kappa_1t}+\frac{\kappa_1}{t}\right)i,\eqno(6.2.111)$$
$$
\vf=\frac{\sgm_2}{t}\sqrt{\frac{2(\ves_2\kappa_1-\ves_1\kappa_2)}{\ves_1\es_2-\ves_2\es_1}}\:\sec
\frac{x}{t}\:
\exp\left(\frac{x^2+y^2}{4\kappa_2t}+\frac{\kappa_2}{t}\right)i;\eqno(6.2.112)$$
$$\psi=\frac{\sgm_1}{t}\sqrt{\frac{2(\es_1\kappa_2-\es_2\kappa_1)}{\ves_1\es_2-\ves_2\es_1}}\:\coth \frac{x}{t},
\:
\exp\left(\frac{x^2+y^2}{4\kappa_1t}+\frac{2\kappa_1}{t}\right)i,\eqno(6.2.113)$$
$$
\vf=\frac{\sgm_2}{t}\sqrt{\frac{2(\ves_2\kappa_1-\ves_1\kappa_2)}{\ves_1\es_2-\ves_2\es_1}}\:\coth
\frac{x}{t}\:
\exp\left(\frac{x^2+y^2}{4\kappa_2t}+\frac{2\kappa_2}{t}\right)i;\eqno(6.2.114)$$}
$$\psi=\frac{\sgm_1}{t}\sqrt{\frac{2(\es_1\kappa_2-\es_2\kappa_1)}{\ves_1\es_2-\ves_2\es_1}}\:\csch
\frac{x}{t}\:
\exp\left(\frac{x^2+y^2}{4\kappa_1t}-\frac{\kappa_1}{t}\right)i,\eqno(6.2.115)$$
$$
\vf=\frac{\sgm_2}{t}\sqrt{\frac{2(\ves_2\kappa_1-\ves_1\kappa_2)}{\ves_1\es_2-\ves_2\es_1}}\:\csch
\frac{x}{t}\:
\exp\left(\frac{x^2+y^2}{4\kappa_2t}-\frac{\kappa_2}{t}\right)i;\eqno(6.2.116)$$
$$\psi=\frac{m\sgm_1}{t}\sqrt{\frac{2(\es_1\kappa_2-\es_2\kappa_1)}{\ves_1\es_2-\ves_2\es_1}}\:
\sn (x/t|m)\:
\exp\left(\frac{x^2+y^2}{4\kappa_1t}+\frac{(1+m^2)\kappa_1}{t}\right)i
,\eqno(6.2.117)$$
$$\vf=\frac{m\sgm_2}{t}\sqrt{\frac{2(\ves_2\kappa_1-\ves_1\kappa_2)}{\ves_1\es_2-\ves_2\es_1}}\:\sn
(x/t|m)\:
\exp\left(\frac{x^2+y^2}{4\kappa_2t}+\frac{(1+m^2)\kappa_2}{t}\right)i;
\eqno(6.2.118)$$
$$\psi=\frac{m\sgm_1}{t}\sqrt{\frac{2(\es_2\kappa_1-\es_1\kappa_2)}{\ves_1\es_2-\ves_2\es_1}}\:
\cn (x/t|m)\:
\exp\left(\frac{x^2+y^2}{4\kappa_1t}+\frac{(1-2m^2)\kappa_1}{t}\right)i,\eqno(6.2.119)$$
$$\vf=\frac{m\sgm_2}{t}\sqrt{\frac{2(\ves_1\kappa_2-\ves_2\kappa_1)}{\ves_1\es_2-\ves_2\es_1}}\:\cn
(x/t|m)\:
\exp\left(\frac{x^2+y^2}{4\kappa_2t}+\frac{(1-2m^2)\kappa_2}{t}\right)i;\eqno(6.2.120)$$
$$\psi=\frac{\sgm_1}{t}\sqrt{\frac{2(\es_2\kappa_1-\es_1\kappa_2)}{\ves_1\es_2-\ves_2\es_1}}
\:\dn (x/t|m)\:
\exp\left(\frac{x^2+y^2}{4\kappa_1t}+\frac{(m^2-2)\kappa_1}{t}\right)i,\eqno(6.2.121)$$
$$\vf=\frac{\sgm_2}{t}\sqrt{\frac{2(\ves_1\kappa_2-\ves_2\kappa_1)}{\ves_1\es_2-\ves_2\es_1}}\:\dn
(x/t|m)\:
\exp\left(\frac{x^2+y^2}{4\kappa_1t}+\frac{(m^2-2)\kappa_1}{t}\right)i.\eqno(6.2.122)$$\pse

{\bf Remark 6.2.9}.  Since $\lim_{m\rta 1}cn(x|m)=\sech x$,
(6.2.119) and (6.2.120) yield the solution
$$\psi=\frac{\sgm_1}{t}\sqrt{\frac{2(\es_2\kappa_1-\es_1\kappa_2)}{\ves_1\es_2-\ves_2\es_1}}\;
\sech\frac{x}{t}\:
\exp\left(\frac{x^2+y^2}{4\kappa_1t}-\frac{\kappa_1}{t}\right)i,\eqno(6.2.123)$$
$$\vf=\frac{\sgm_2}{t}\sqrt{\frac{2(\ves_1\kappa_2-\ves_2\kappa_1)}{\ves_1\es_2-\ves_2\es_1}}\;\sech\frac{x}{t}
\:
\exp\left(\frac{x^2+y^2}{4\kappa_2t}-\frac{\kappa_2}{t}\right)i;\eqno(6.2.124)$$
 Applying the transformation in (6.2.3)-(6.2.4) with $a=a_0=a_2=a_3=0$ and the transformation $S_{c,0}$
 in (6.2.5)-(6.2.6), we get a more general
soliton-like solution:\index{soliton-like solution!of coupled
nonlinear Schr\"{o}dinger equations}
\begin{eqnarray*} \hspace{0.6cm}\psi&=&\frac{\sgm_1}{b^2(t-a_1)}\sqrt{\frac{2(\es_2\kappa_1-\es_1\kappa_2)}
{\ves_1\es_2-\ves_2\es_1}}\;
\sech\frac{(x-2ct)\cos\sta+y\sin\sta}{b(t-a_1)}\:
\\ & &\times
\exp\left(\frac{(x-2ct)^2+y^2}{4\kappa_1(t-a_1)}-\frac{\kappa_1}{b^2(t-a_1)}+\frac{c(x-ct)}{\kappa_1}+a\right)i,
\hspace{2.4cm}(6.2.125)\end{eqnarray*}
\begin{eqnarray*}\qquad \hspace{0.6cm}\vf&=&\frac{\sgm_2}{b^2(t-a_1)}\sqrt{\frac{2(\es_2\kappa_1-\es_1\kappa_2)}
{\ves_1\es_2-\ves_2\es_1}}\;
\sech\frac{(x-2ct)\cos\sta+y\sin\sta}{b(t-a_1)}\:
\\ & &\times
\exp\left(\frac{(x-2ct)^2+y^2}{4\kappa_2(t-a_1)}-\frac{\kappa_2}{b^2(t-a_1)}+\frac{c(x-ct)}{\kappa_2}+a_0\right)i,
\hspace{1.4cm}(6.2.126)\end{eqnarray*} where
$a,a_0,a_1,b,c,\sta\in\mbb{R}$ with $b\neq 0$. We can get more
sophisticated soliton-like solution if we apply the general forms of
the transformations in (6.2.3)-(6.2.6).

\section{Davey and Stewartson  Equations}

 Davey and Stewartson [DS] (1974) used the method of multiple scales
 to derive the following system of nonlinear partial differential
 equations\index{Davey-Stewartson equations}
$$2iu_t+ \es_1u_{xx}+u_{yy}-2\es_2|u|^2u-2uv=0,\eqno(6.3.1)$$
$$v_{xx}-\es_1(v_{yy}+2(|u|^2)_{xx})=0\eqno(6.3.2)$$
that describe the long time evolution of  three-dimensional packets
of surface waves, where $u$ is a complex-valued function, $v$ is a
real valued function and $\es_1,\es_2=\pm 1$. The equations are
called the {\it Davey-Stewartson I equations} if $\es_1=1$, and the
{\it Davey-Stewartson II equations} when $\es_1=-1$. They were used
to study the stability of the uniform Stokes wave train with respect
to small disturbance. The soliton solutions of the Davey-Stewartson
equations were first studied by Anker and Freeman [AF] (1978). Kirby
and Dalrymple [KD] (1983) obtained oblique envelope solutions of the
equations in intermediate water depth. Omote [Om] (1988) found
infinite-dimensional symmetry algebras and an infinite number of
conserved quantities for the equations.

Arkadiev,  Pogrebkov and  Polivanov [APP1] (1989) studied the
solutions of the Davey-Stewartson II equations whose singularities
form closed lines with string-like behavior. They  [APP2] (1989)
also applied the inverse scattering transform method to the
Davey-Stewartson II equations. Gilson and Nimmo [GN] (1991) found
dromion solutions and Malanyuk [Mt1, Mt2] (1991, 1994) obtained
finite-gap solutions of the equations. van de Linden (1992) studied
the solutions under a certain boundary condition. Clarkson and Hood
[CH] (1994) obtained certain symmetry reductions of the equations to
ordinary differential equations with no intervening steps and
provided new exact solutions which are not obtainable by the Lie
group approach. Guil and Manas [GM]
 (1995) found certain solutions of the Davey-Stewartson I
equations by deforming dromion. Manas and Santini [MS] (1997)
studied a large class of solutions of the Davey-Stewartson II
equations by a Wronskian scheme. There are the other interesting
works on solutions of the Davey-Stewartson equations (e.g., cf.
[Vj]). It is obvious that the some of above solutions are equivalent
to each other under the known symmetric transformations. It is time
to study solutions of the Davey-Stewartson equations modulo the
known symmetric transformations.

 In this section, we use the
quadratic-argument approach to study exact solutions of  the
Davey-Stewartson equations modulo the most known symmetry
transformations. This is a revision of our earlier preprint [X18].

By (6.1.2), (6.3.1) and (6.3.2), we take
$$\mbox{deg}\;x=\mbox{deg}\;y=-\mbox{deg}\;u=\frac{1}{2}\mbox{deg}\;t=-\frac{1}{2}\mbox{deg}\;v\eqno(6.3.3)$$
in order to make the nonzero terms in (6.3.1) and (6.3.2) having the
same degree. Moreover, the equation (6.3.1) and (6.3.2) are
translation invariant because they do not contain variable
coefficients. Thus the transformation
$$T_{a,b}(u(t,x,y))=bu(b^2t+a,bx,by),\qquad T_{a,b}(v(t,x,y))=b^2v(b^2t+a,bx,by)\eqno(6.3.4)$$
maps a solution of the  Davey-Stewartson equations (6.3.1) and
(6.3.2) to another solution, where $a,b\in\mbb{R}$ and $b\neq 0$.
Let $\al,\be$ and $\gm$ be functions in $t$.  The transformation
$u(t,x,y)\mapsto u(t,x+\al,y+\be)$ and $v(t,x,y)\mapsto
v(t,x+\al,y+\be)$ changes (6.3.1) to
$$2i(\al'u_x+\be'u_y+u_t)+ \es_1u_{xx}+u_{yy}-2\es_2|u|^2u-2uv=0\eqno(6.3.5)$$
and leaves (6.3.2) invariant, where the independent variables $x$
 is replaced by $x+\al$, the independent variables $y$
 is replaced by $y+\be$ and  the
subindices denote the partial derivatives with respect to the
original independent variables. Moreover, the transformation
$u\mapsto e^{-(\es_1\al'x+\be'y+\gm)i}u$ and $v\mapsto v$ changes
(6.3.1) to
\begin{eqnarray*}\qquad \qquad&
&2[((\es_1{\al'}'x+{\be'}'y)+\gm')u+iu_t]-(\es_1{\al'}^2+{\be'}^2)u
-2\al'iu_x-2\be'i u_y\\
&
&+\es_1u_{xx}+u_{yy}-2\es_2|u|^2u-2uv=0\hspace{6.2cm}(6.3.6)\end{eqnarray*}
and leaves (6.3.2) invariant. Furthermore, the transformation
$$u\mapsto u\;\;\mbox{and}\;\; v\mapsto
v+\es_1{\al'}'x+{\be'}'y-\frac{\es_1{\al'}^2+{\be'}^2}{2}+\gm'\eqno(6.3.7)$$
changes (6.3.1) to \begin{eqnarray*}\qquad \qquad& &2iu_t+
\es_1u_{xx}+u_{yy}-2\es_2|u|^2u-2uv\\
& &+[\es_1{\al'}^2+{\be'}^2+2\gm'-2(\es_1{\al'}'x+{\be'}'y
)]u=0\hspace{4.7cm}(6.3.8)\end{eqnarray*} and keeps (6.3.2)
invariant. Thus the transformation
$$S_{\al,\be,\gm}(u(t,x,y))=e^{-(\es_1\al'x+\be'y+\gm)i}u(t,x+\al,y+\be),\eqno(6.3.9)$$
$$S_{\al,\be,\gm}(v(t,x,y))=v(t,x+\al,y+\be)+\es_1{\al'}'x+{\be'}'y-\frac{\es_1{\al'}^2+{\be'}^2}{2}+\gm'\eqno(6.3.10)$$
maps a solution of the  Davey-Stewartson equations (6.3.1) and
(6.3.2) to another solution.

 Write
$$u=\xi(t,x,y)e^{i\phi(t,x,y)},\eqno(6.3.11)$$
where $\xi$ and $\phi$ are real functions in $t,x,y$. Note
$$u_t=(\xi_t+i\xi\phi_t)e^{i\phi},\qquad
u_x=(\xi_x+i\xi\phi_x)e^{i\phi},\qquad
u_y=(\xi_y+i\xi\phi_y)e^{i\phi},\eqno(6.3.12)$$
$$u_{xx}=(\xi_{xx}-\xi\phi_x^2+i(2\xi_x\phi_x
+\xi\phi_{xx}))e^{i\phi},\;\;
u_{yy}=(\xi_{yy}-\xi\phi_y^2+i(2\xi_y\phi_y
+\xi\phi_{yy}))e^{i\phi}.\eqno(6.3.13)$$ Then (6.3.1) is equivalent
to
\begin{eqnarray*}\hspace{2cm}& &2i\xi_t-2\xi\phi_t+\es_1(\xi_{xx}-\xi\phi_x^2+i(2\xi_x\phi_x
+\xi\phi_{xx}))\\ &&+\xi_{yy}-\xi\phi_y^2+i(2\xi_y\phi_y
+\xi\phi_{yy})-2\es_2\xi^3-2\xi
v=0,\hspace{3.2cm}(6.3.14)\end{eqnarray*} equivalently,
$$2\xi_t+2(\es_1\xi_x\phi_x+\xi_y\phi_y)
+\xi(\es_1\phi_{xx}+\phi_{yy})=0,\eqno(6.3.15)$$
$$\xi(2\phi_t+\es_1\phi_x^2+\phi_y^2)-\es_1\xi_{xx}-\xi_{yy}+2\es_2\xi^3+2\xi
v=0.\eqno(6.3.16)$$ Moreover, (6.3.2) becomes
$$v_{xx}-\es_1(v_{yy}+2(\xi^2)_{xx})=0.\eqno(6.3.17)$$

{\it Case 1}. $\phi=0$. \psp

In this case, (6.3.15) becomes $\xi_t=0$. Moreover, (6.3.16) gives
$$-\es_1\xi_{xx}-\xi_{yy}+2\es_2\xi^3+2\xi
v=0.\eqno(6.3.18)$$ Fixing $\ell_1,\ell_2\in\mbb{R}$ , we denote
$$\varpi=\ell_1 x+\ell_2y.\eqno(6.3.19)$$
Assume $\xi=f(\varpi)$ and $v=g(\varpi)$ for some one-variable
functions $f$ and $g$. Then (6.3.17) and (6.3.18) become
$$(\ell_1^2-\es_1\ell_2^2){g'}'-2\es_1\ell_1^2{(f^2)'}'=0,\eqno(6.3.20)$$
$$-(\es_1\ell_1^2+\ell_2^2){f'}'+2\es_2f^3+2fg=0.\eqno(6.3.21)$$
Suppose
$$\ell_1^2-\es_1\ell_2^2\neq 0\;\;\mbox{and}\;\; \es_1\ell_1^2+\ell_2^2\neq
0\sim \ell_1^4\neq \ell_2^4.\eqno(6.3.22)$$ Then
$$g=\frac{2\ell_1^2f^2}{\es_1\ell_1^2-\ell_2^2}+c(\es_1\ell_1^2+\ell_2^2)\eqno(6.3.23)$$
is a solution of (6.3.20) with $c\in\mbb{R}$.

Substituting (6.3.23) into (6.3.21), we get
$$-(\es_1\ell_1^2+\ell_2^2){f'}'+2
\frac{(2+\es_1\es_2)\ell_1^2-\es_2\ell_2^2}{\es_1\ell_1^2-\ell_2^2}
f^3+2c(\es_1\ell_1^2+\ell_2^2)f =0,\eqno(6.3.24)$$ equivalently,
$${f'}'+2
\frac{(2+\es_1\es_2)\ell_1^2-\es_2\ell_2^2}{\ell_2^4-\ell_1^4}
f^3-2cf =0.\eqno(6.3.25)$$ If
$$\es_2=1\;\;\mbox{and}\;\;\ell_2=\pm\sqrt{2+\es_1}\:\ell_1,\eqno(6.3.26)$$
then (6.3.25) becomes ${f'}'=2cf$. Assuming $c=2c_1^2$ with
$c_1\in\mbb{R}$, we have the solution
$$f=a_1e^{2c_1\varpi}+a_2e^{-2c_1\varpi}\;\;\mbox{and}\;\;g=
-f^2+8c_1^2\ell_1^2.\eqno(6.3.27)$$ Letting $c=-2c_1^2$ with
$c_1\in\mbb{R}$, we obtain another solution
$$f=a_1\sin 2c_1\varpi\;\;\mbox{and}\;\;g=
-f^2-8c_1^2\ell_1^2.\eqno(6.3.28)$$ Since
$\varpi=\ell_1x+\ell_2y=\ell_1(x\pm\sqrt{2+\es_1}\:y)$, we can take
$2c_1\ell_1=1$ if we replace $u$ by $T_{0,(2c_1\ell_1)^{-1}}(u)$ and
$v$ by $T_{0,(2c_1\ell_1)^{-1}}(v)$. Thus have
$$f=a_1e^{x\pm\sqrt{2+\es_1}\:y}+a_2e^{-x\mp\sqrt{2+\es_1}\:y}\;\;\mbox{and}\;\;
g= -f^2+2;\eqno(6.3.29)$$
$$f=a_1\sin(x\pm\sqrt{2+\es_1}\:y) \;\;\mbox{and}\;\;g=
-f^2-2.\eqno(6.3.30)$$

Next we assume
$$(2+\es_1\es_2)\ell_1^2-\es_2\ell_2^2\neq 0.\eqno(6.3.31)$$
Recall (6.1.18)-(6.1.23). Substituting  $\xi=f=k\vf(x)$ to (6.3.25)
with $k\in\mbb{R}$ and $\vf=1/x, \tan x, \sec x,\coth x,\csch x,
\sn(x|m),\cn(x|m),\dn(x|m)$, we find the following solutions:
$$f=\frac{1}{\varpi}\sqrt{\frac{\ell_1^4-\ell_2^4}{(2+\es_1\es_2)\ell_1^2-\es_2\ell_2^2}},\;\;
\;g=\frac{2\ell_1^2f^2}{\es_1\ell_1^2-\ell_2^2};\eqno(6.3.32)$$
$$f=\sqrt{\frac{\ell_1^4-\ell_2^4}{(2+\es_1\es_2)\ell_1^2-\es_2\ell_2^2}}\;\tan\varpi,\;\;
\;g=\frac{2\ell_1^2f^2}{\es_1\ell_1^2-\ell_2^2}+\es_1\ell_1^2+\ell_2^2;\eqno(6.3.33)$$
$$f=\sqrt{\frac{\ell_1^4-\ell_2^4}{(2+\es_1\es_2)\ell_1^2-\es_2\ell_2^2}}\;\sec\varpi,\;\;
\;g=\frac{2\ell_1^2f^2}{\es_1\ell_1^2-\ell_2^2}-\frac{\es_1\ell_1^2+\ell_2^2}{2};\eqno(6.3.34)$$
$$f=\sqrt{\frac{\ell_1^4-\ell_2^4}{(2+\es_1\es_2)\ell_1^2-\es_2\ell_2^2}}\;\coth\varpi,\;\;
\;g=\frac{2\ell_1^2f^2}{\es_1\ell_1^2-\ell_2^2}-\es_1\ell_1^2-\ell_2^2;\eqno(6.3.35)$$
$$f=\sqrt{\frac{\ell_1^4-\ell_2^4}{(2+\es_1\es_2)\ell_1^2-\es_2\ell_2^2}}\;\csch\varpi,\;\;
\;g=\frac{2\ell_1^2f^2}{\es_1\ell_1^2-\ell_2^2}+\frac{\es_1\ell_1^2+\ell_2^2}{2};\eqno(6.3.36)$$
$$f=m\sqrt{\frac{\ell_1^4-\ell_2^4}{(2+\es_1\es_2)\ell_1^2-\es_2\ell_2^2}}\;\sn(\varpi|m),\;\;
\;g=\frac{2\ell_1^2f^2}{\es_1\ell_1^2-\ell_2^2}-\frac{(m^2+1)(\es_1\ell_1^2+\ell_2^2)}{2};\eqno(6.3.37)$$
$$f=m\sqrt{\frac{\ell_2^4-\ell_1^4}{(2+\es_1\es_2)\ell_1^2-\es_2\ell_2^2}}\;\cn(\varpi|m),\;
\;\;g=\frac{2\ell_1^2f^2}{\es_1\ell_1^2-\ell_2^2}+\frac{(2m^2-1)(\es_1\ell_1^2+\ell_2^2)}{2};\eqno(6.3.38)$$
$$f=\sqrt{\frac{\ell_2^4-\ell_1^4}{(2+\es_1\es_2)\ell_1^2-\es_2\ell_2^2}}\;\dn(\varpi|m),\;\;
\;g=\frac{2\ell_1^2f^2}{\es_1\ell_1^2-\ell_2^2}+\frac{(2-m^2)(\es_1\ell_1^2+\ell_2^2)}{2}.\eqno(6.3.39)$$
In summary, we have:\psp

{\bf Theorem 6.3.1}. {\it If $\es_2=1$, we have the following
solutions of the   Davey-Stewartson equations (6.3.1) and (6.3.2):
for $a_1,a_2\in\mbb{R}$ and $a_1\neq 0$,
$$u=a_1e^{x\pm\sqrt{2+\es_1}\:y}+a_2e^{-x\mp\sqrt{2+\es_1}\:y}\;\;\mbox{and}\;\;
v= -u^2+2;\eqno(6.3.40)$$
$$u=a_1\sin(x\pm\sqrt{2+\es_1}\:y) \;\;\mbox{and}\;\;v=
-u^2-2.\eqno(6.3.41)$$ Let $\ell_1,\ell_2\in\mbb{R}$ such that
$$\ell_1^4\neq\ell_2^4\;\;\mbox{and}\;\;(2+\es_1\es_2)\ell_1^2\neq
\es_2\ell_2^2.\eqno(6.3.42)$$ Then we the following solutions of the
Davey-Stewartson equations (6.3.1) and (6.3.2):
$$u=\frac{1}{\ell_1 x+\ell_2y}\sqrt{\frac{\ell_1^4-\ell_2^4}{(2+\es_1\es_2)\ell_1^2-\es_2\ell_2^2}},\;\;
\;v=\frac{2\ell_1^2u^2}{\es_1\ell_1^2-\ell_2^2};\eqno(6.3.43)$$
$$u=\sqrt{\frac{\ell_1^4-\ell_2^4}{(2+\es_1\es_2)\ell_1^2-\es_2\ell_2^2}}\;\tan(\ell_1 x+\ell_2y),\;\;
\;v=\frac{2\ell_1^2u^2}{\es_1\ell_1^2-\ell_2^2}+\es_1\ell_1^2+\ell_2^2;\eqno(6.3.44)$$
$$u=\sqrt{\frac{\ell_1^4-\ell_2^4}{(2+\es_1\es_2)\ell_1^2-\es_2\ell_2^2}}\;\sec(\ell_1 x+\ell_2y),\;\;
\;v=\frac{2\ell_1^2u^2}{\es_1\ell_1^2-\ell_2^2}-\frac{\es_1\ell_1^2+\ell_2^2}{2};\eqno(6.3.45)$$}
$$u=\sqrt{\frac{\ell_1^4-\ell_2^4}{(2+\es_1\es_2)\ell_1^2-\es_2\ell_2^2}}\;\coth(\ell_1 x+\ell_2y),\;\;
\;v=\frac{2\ell_1^2u^2}{\es_1\ell_1^2-\ell_2^2}-\es_1\ell_1^2-\ell_2^2;\eqno(6.3.46)$$
$$u=\sqrt{\frac{\ell_1^4-\ell_2^4}{(2+\es_1\es_2)\ell_1^2-\es_2\ell_2^2}}\;\csch(\ell_1 x+\ell_2y),\;\;
\;v=\frac{2\ell_1^2u^2}{\es_1\ell_1^2-\ell_2^2}+\frac{\es_1\ell_1^2+\ell_2^2}{2};\eqno(6.3.47)$$
$$u=m\sqrt{\frac{\ell_1^4-\ell_2^4}{(2+\es_1\es_2)\ell_1^2-\es_2\ell_2^2}}\;\sn(\ell_1
x+\ell_2y|m),\eqno(6.3.48)$$
$$v=\frac{2\ell_1^2u^2}{\es_1\ell_1^2-\ell_2^2}-\frac{(m^2+1)(\es_1\ell_1^2+\ell_2^2)}{2};\eqno(6.3.49)$$
$$u=m\sqrt{\frac{\ell_2^4-\ell_1^4}{(2+\es_1\es_2)\ell_1^2-\es_2\ell_2^2}}\;\cn(\ell_1 x+\ell_2y|m),\eqno(6.3.50)$$
$$v=\frac{2\ell_1^2u^2}{\es_1\ell_1^2-\ell_2^2}+\frac{(2m^2-1)(\es_1\ell_1^2+\ell_2^2)}{2};\eqno(6.3.51)$$
$$u=\sqrt{\frac{\ell_2^4-\ell_1^4}{(2+\es_1\es_2)\ell_1^2-\es_2\ell_2^2}}\;\dn(\ell_1 x+\ell_2y|m),
\;v=\frac{2\ell_1^2u^2}{\es_1\ell_1^2-\ell_2^2}+\frac{(2-m^2)(\es_1\ell_1^2+\ell_2^2)}{2}.\eqno(6.3.52)$$
\pse

{\bf Remark 6.3.2}.  Since $\lim_{m\rta 1}\dn(x|m)=\sech x$,
(6.3.52)
 yields the solution
$$u=\sqrt{\frac{\ell_2^4-\ell_1^4}{(2+\es_1\es_2)\ell_1^2-\es_2\ell_2^2}}\;\sech(\ell_1 x+\ell_2y),
\;v=\frac{2\ell_1^2u^2}{\es_1\ell_1^2-\ell_2^2}+\frac{\es_1\ell_1^2+\ell_2^2}{2}.\eqno(6.3.53)$$
Applying $S_{\al,\be,\gm}$ in (6.3.9) and (6.3.10), we get a more
general solution
$$u=\sqrt{\frac{\ell_2^4-\ell_1^4}{(2+\es_1\es_2)\ell_1^2-\es_2\ell_2^2}}\;e^{-(\es_1\al'x+\be'y+\gm)i}
\sech(\ell_1 (x+\al)+\ell_2(y+\be)),\eqno(6.3.54)$$
\begin{eqnarray*}\qquad\;\;
v&=&\frac{2\ell_1^2(\es_1\ell_1^2+\ell_2^2)}{\es_2\ell_2^2-(2+\es_1\es_2)\ell_1^2}\;\sech^2(\ell_1
(x+\al)+\ell_2(y+\be))\\ &
&+\frac{\es_1\ell_1^2+\ell_2^2}{2}+\es_1{\al'}'x+{\be'}'y-\frac{\es_1(\al')^2+(\be')^2}{2}+\gm',\hspace{3.5cm}(6.3.55)\end{eqnarray*}
 where $\al,\be$ and $\gm$ are arbitrary functions of
$t$. Taking $\al=a_1t,\;\be=a_2t$ and $\gm=(\es_1a_1^2+a_2^2)t/2$,
we have a solition solution\index{soliton solution!of
Davey-Stewartson equations}
$$u=\sqrt{\frac{\ell_2^4-\ell_1^4}{(2+\es_1\es_2)\ell_1^2-\es_2\ell_2^2}}\;e^{-(\es_1a_1x+a_2y+(\es_1a_1^2+a_2^2)t/2)i}
\;\sech(\ell_1x+\ell_2y+(a_1\ell_1+a_2\ell_2)t),\eqno(6.3.56)$$
$$v=\frac{2\ell_1^2(\es_1\ell_1^2+\ell_2^2)}{\es_2\ell_2^2-(2+\es_1\es_2)\ell_1^2}\;\sech^2(\ell_1x+\ell_2y+(a_1\ell_1+a_2\ell_2)t)
+\frac{\es_1\ell_1^2+\ell_2^2}{2}, \eqno(6.3.57)$$ where
$a_1,a_2\in\mbb{R}$.\psp

{\it Case 2}. $\phi=\es_1x^2/2t$ or $y^2/2t$.\psp

Suppose $\phi=\es_1x^2/2t$. Then (6.3.15) and (6.3.16) become
$$\xi_t+\frac{x}{t}\xi_x+\frac{1}{2t}\xi=0,\eqno(6.3.58)$$
$$-\es_1\xi_{xx}-\xi_{yy}+2\es_2\xi^3+2\xi
v=0.\eqno(6.3.59)$$ By (6.1.40)-(6.1.43),  we have
$$\xi=\frac{a}{\sqrt{t}},\;\;v=-\frac{\es_2a^2}{t},\qquad a\in\mbb{R},\eqno(6.3.60)$$
which satisfies (6.3.17). Moreover, (6.3.60) also holds when
$\phi=y^2/2t$. \psp

{\it Case 3}. $\phi=\es_1x^2/2t+y^2/2(t-d)$  with $0\neq
d\in\mbb{R}$. \psp

In this case, (6.1.45) and (6.3.59) hold. By (6.1.46)-(6.1.49),
$$\xi=\frac{a}{\sqrt{t(t-d)}},\;\;v=-\frac{\es_2a^2}{t(t-d)},\qquad a\in\mbb{R}.\eqno(6.3.61)$$
\psp

In summary, we have: \psp

{\bf Theorem 6.3.3}. {\it For $a,d\in\mbb{R}$ with $d\neq 0$, we
have the following solutions of the  Davey-Stewartson equations
(6.3.1) and (6.3.2):
$$u=\frac{ae^{\es_1x^2i/2t}}{\sqrt{t}},\qquad
v=-\frac{\es_2a^2}{t};\eqno(6.3.62)$$
$$u=\frac{ae^{y^2i/2t}}{\sqrt{t}},\qquad
v=-\frac{\es_2a^2}{t};\eqno(6.3.63)$$
$$u=\frac{ae^{(\es_1x^2/2t+y^2/2(t-d))i}}{\sqrt{t(t-d)}},\qquad
v=-\frac{\es_2a^2}{t(t-d)}.\eqno(6.3.64)$$}

{\it Case 4}. $\phi=(\es_1x^2+y^2)/2t$.\psp

In this case, (6.3.15) becomes (6.1.54). So

$$\xi=\frac{1}{t}\zeta(z,s),\qquad
z=\frac{x}{t},\;s=\frac{y}{t}, \eqno(6.3.65)$$ for some two-variable
function $\zeta$ by (6.1.55). Moreover, (6.3.16) becomes
$$-\frac{\es_1\zeta_{zz}+\zeta_{ss}}{t^2}+2\frac{\es_2\zeta^3}{t^2}+2\zeta v=0.\eqno(6.3.66)$$
Assume
$$v=\frac{\eta(z,s)}{t^2}\eqno(6.3.67)$$
for some two-variable functions $\eta$. Then (6.3.66) becomes
$$-\es_1\zeta_{zz}-\zeta_{ss}+2\es_2\zeta^3+2\zeta\eta=0\eqno(6.3.68)$$
and (6.3.17) becomes
$$\eta_{zz}-\es_1(\eta_{ss}+2(\zeta^2)_{zz})=0.\eqno(6.3.69)$$

By the arguments in (6.3.17)-(6.3.39), we obtain:\psp

{\bf Theorem 6.3.4}. {\it If $\es_2=1$, we have the following
solutions of the   Davey-Stewartson equations (6.3.1) and (6.3.2):
for $a_1,a_2\in\mbb{R}$ and $a_1\neq 0$,
$$u=\frac{e^{(\es_1x^2+y^2)i/2t}}{t}(a_1e^{(x\pm\sqrt{2+\es_1}\:y)/t}+a_2e^{(-x\mp\sqrt{2+\es_1}\:y)/t}),\eqno(6.3.70)$$
$$v=\frac{1}{t^2}[2-(a_1e^{(x\pm\sqrt{2+\es_1}\:y)/t}+a_2e^{(-x\mp\sqrt{2+\es_1}\:y)/t})^2];\eqno(6.3.71)$$
$$u=\frac{a_1e^{(\es_1x^2+y^2)i/2t}}{t}\sin\frac{x\pm\sqrt{2+\es_1}\:y}{t},\;\;v=-
\frac{1}{t^2}\left(2+a_1^2\sin^2\frac{x\pm\sqrt{2+\es_1}\:y}{t}\right);\eqno(6.3.72)$$
Let $\ell_1,\ell_2\in\mbb{R}$ such that
$$\ell_1^4\neq\ell_2^4\;\;\mbox{and}\;\;(2+\es_1\es_2)\ell_1^2\neq
\es_2\ell_2^2.\eqno(6.3.73)$$ Then we the following solutions of the
Davey-Stewartson equations (6.3.1) and (6.3.2):
$$u=\frac{e^{(\es_1x^2+y^2)i/2t}}{\ell_1 x+\ell_2y}\sqrt{\frac{\ell_1^4-\ell_2^4}{(2+\es_1\es_2)\ell_1^2-\es_2\ell_2^2}},\;\;
\;v=\frac{2\ell_1^2(\es_1\ell_1^2+\ell_2^2)}{((2+\es_1\es_2)\ell_1^2-\es_2\ell_2^2)(\ell_1
x+\ell_2y)^2};\eqno(6.3.74)$$
$$u=\sqrt{\frac{\ell_1^4-\ell_2^4}{(2+\es_1\es_2)\ell_1^2-\es_2\ell_2^2}}\;\frac{e^{(\es_1x^2+y^2)i/2t}}{t}
\tan\frac{\ell_1 x+\ell_2y}{t},\eqno(6.3.75)$$
$$
v=\frac{2\ell_1^2(\es_1\ell_1^2+\ell_2^2)}{((2+\es_1\es_2)\ell_1^2-\es_2\ell_2^2)t^2}\tan^2\frac{\ell_1
x+\ell_2y}{t}+\frac{\es_1\ell_1^2+\ell_2^2}{t^2};\eqno(6.3.76)$$
$$u=\sqrt{\frac{\ell_1^4-\ell_2^4}{(2+\es_1\es_2)\ell_1^2-\es_2\ell_2^2}}\;\frac{e^{(\es_1x^2+y^2)i/2t}}{t}
\sec\frac{\ell_1 x+\ell_2y}{t},\eqno(6.3.77)$$
$$
v=\frac{2\ell_1^2(\es_1\ell_1^2+\ell_2^2)}{((2+\es_1\es_2)\ell_1^2-\es_2\ell_2^2)t^2}\sec^2\frac{\ell_1
x+\ell_2y}{t}-\frac{\es_1\ell_1^2+\ell_2^2}{2t^2};\eqno(6.3.78)$$
$$u=\sqrt{\frac{\ell_1^4-\ell_2^4}{(2+\es_1\es_2)\ell_1^2-\es_2\ell_2^2}}\;\frac{e^{(\es_1x^2+y^2)i/2t}}{t}
\coth\frac{\ell_1 x+\ell_2y}{t},\eqno(6.3.79)$$
$$
v=\frac{2\ell_1^2(\es_1\ell_1^2+\ell_2^2)}{((2+\es_1\es_2)\ell_1^2-\es_2\ell_2^2)t^2}\coth^2\frac{\ell_1
x+\ell_2y}{t}-\frac{\es_1\ell_1^2+\ell_2^2}{t^2};\eqno(6.3.80)$$}
$$u=\sqrt{\frac{\ell_1^4-\ell_2^4}{(2+\es_1\es_2)\ell_1^2-\es_2\ell_2^2}}\;\frac{e^{(\es_1x^2+y^2)i/2t}}{t}
\csch\frac{\ell_1 x+\ell_2y}{t},\eqno(6.3.81)$$
$$
v=\frac{2\ell_1^2(\es_1\ell_1^2+\ell_2^2)}{((2+\es_1\es_2)\ell_1^2-\es_2\ell_2^2)t^2}\mbox{csch}^2\frac{\ell_1
x+\ell_2y}{t}+\frac{\es_1\ell_1^2+\ell_2^2}{2t^2};\eqno(6.3.82)$$
$$u=m\sqrt{\frac{\ell_1^4-\ell_2^4}{(2+\es_1\es_2)\ell_1^2-\es_2\ell_2^2}}\;\frac{e^{(\es_1x^2+y^2)i/2t}}{t}
\sn\left(\frac{\ell_1 x+\ell_2y}{t}|m\right),\eqno(6.3.83)$$
$$
v=\frac{2\ell_1^2(\es_1\ell_1^2+\ell_2^2)}{((2+\es_1\es_2)\ell_1^2-\es_2\ell_2^2)t^2}\mbox{sn}^2\left(\frac{\ell_1
x+\ell_2y}{t}|m\right)-\frac{(m^2+1)(\es_1\ell_1^2+\ell_2^2)}{2t^2};\eqno(6.3.84)$$
$$u=m\sqrt{\frac{\ell_2^4-\ell_1^4}{(2+\es_1\es_2)\ell_1^2-\es_2\ell_2^2}}\;\frac{e^{(\es_1x^2+y^2)i/2t}}{t}
\cn\left(\frac{\ell_1 x+\ell_2y}{t}|m\right),\eqno(6.3.85)$$
$$
v=-\frac{2\ell_1^2(\es_1\ell_1^2+\ell_2^2)}{((2+\es_1\es_2)\ell_1^2-\es_2\ell_2^2)t^2}\mbox{cn}^2\left(\frac{\ell_1
x+\ell_2y}{t}|m\right)+\frac{(2m^2-1)(\es_1\ell_1^2+\ell_2^2)}{2t^2};\eqno(6.3.86)$$
$$u=\sqrt{\frac{\ell_2^4-\ell_1^4}{(2+\es_1\es_2)\ell_1^2-\es_2\ell_2^2}}\;\frac{e^{(\es_1x^2+y^2)i/2t}}{t}
\dn\left(\frac{\ell_1 x+\ell_2y}{t}|m\right),\eqno(6.3.87)$$
$$
v=-\frac{2\ell_1^2(\es_1\ell_1^2+\ell_2^2)}{((2+\es_1\es_2)\ell_1^2-\es_2\ell_2^2)t^2}\mbox{dn}^2\left(\frac{\ell_1
x+\ell_2y}{t}|m\right)+\frac{(2-m^2)(\es_1\ell_1^2+\ell_2^2)}{2t^2}.\eqno(6.3.88)$$
\pse

{\bf Remark 6.3.5}.  Since $\lim_{m\rta 1}\dn(x|m)=\sech x$,
(6.3.87) and (6.3.88)
 yield the solution
$$u=\sqrt{\frac{\ell_2^4-\ell_1^4}{(2+\es_1\es_2)\ell_1^2-\es_2\ell_2^2}}\;\frac{e^{(\es_1x^2+y^2)i/2t}}{t}
\sech\frac{\ell_1 x+\ell_2y}{t},\eqno(6.3.89)$$
$$
v=-\frac{2\ell_1^2(\es_1\ell_1^2+\ell_2^2)}{((2+\es_1\es_2)\ell_1^2-\es_2\ell_2^2)t^2}\mbox{sech}^2\frac{\ell_1
x+\ell_2y}{t}+\frac{\es_1\ell_1^2+\ell_2^2}{2t^2}.\eqno(6.3.90)$$
Applying $S_{a_1t,a_2t,(\es_1a_1^2+a_2^2)t/2}$ in (6.3.9) and
(6.3.10), we get a solition-like solution\index{soliton-like
solution!of Davey-Stewartson equations}
\begin{eqnarray*}\qquad\qquad u&=&
\sqrt{\frac{\ell_2^4-\ell_1^4}{(2+\es_1\es_2)\ell_1^2-\es_2\ell_2^2}}\;\frac{e^{(\es_1x^2+y^2)i/2t-(\es_1a_1x+a_2y+(\es_1a_1^2+a_2^2)t/2)i}}{t}
\\ & &\times \sech\frac{\ell_1
x+\ell_2y+(a_1\ell_1+a_2\ell_2)t}{t},\hspace{5.7cm}(6.3.91)\end{eqnarray*}
$$
v=-\frac{2\ell_1^2(\es_1\ell_1^2+\ell_2^2)}{((2+\es_1\es_2)\ell_1^2-\es_2\ell_2^2)t^2}\mbox{sech}^2\frac{\ell_1
x+\ell_2y+(a_1\ell_1+a_2\ell_2)t}{t}+\frac{\es_1\ell_1^2+\ell_2^2}{2t^2}.\eqno(6.3.92)$$

\chapter{Dynamic Convection in a Sea}

The rotation of the earth influences both the atmospheric and
oceanic flows. In fact, the fast rotation and small aspect ratio are
two main characteristics of the large scale atmospheric and oceanic
flows. The small aspect ratio characteristic leads to the primitive
equations, and the fast rotation leads to the quasi-geostropic
equations (e.g., cf. [GC, LTW1, LTW2, Pj]). A main objective in
climate dynamics and in geophysical fluid dynamics is to understand
and predict the periodic, quasi-periodic, aperiodic, and fully
turbulent characteristics of the large scale atmospheric and oceanic
flows (e.g., cf. [HMW, Le]). The general model of atmospheric and
oceanic flows is very complicated. In this chapter, we study a
simplified model of dynamic convection in a sea due to Ovsiannikov
(1967) (e.g., cf. Page 203 in [In3]).

In Section 7.1, we present the equations for dynamic convection in a
sea and the symmetry analysis on them. In Section 7.2, we use a new
variable of moving line to solve the equations. An approach of using
the product of cylindrical invariant function with $z$ is introduced
in Section 7.3. In Section 7.4, we reduce the three-dimensional
(spacial) equations  into a two-dimensional problem and then solve
it with three different ansatzes (assumptions). This chapter is a
revision of our earlier preprint [X17].

\section{Equations and Symmetries}

 The following  equations
$$u_x+v_y+w_z=0,\qquad \rho=p_z,\eqno(7.1.1)$$
$$\rho_t+u\rho_x+v\rho_y+w\rho_z=0,\eqno(7.1.2)$$
$$u_t+uu_x+vu_y+wu_z+v=-\frac{1}{\rho}p_x,\eqno(7.1.3)$$
$$v_t+uv_x+vv_y+wv_z-u=-\frac{1}{\rho}p_y\eqno(7.1.4)$$\index{equation
of dynamic convection}are used to describe the dynamic convection of
a sea in geophysics, where $u,\:v$ and $w$ are components of
velocity vector of relative motion of fluid in Cartesian coordinates
$(x,y,z)$, $\rho=\rho(x,y,z,t)$ is the density of fluid and $p$ is
the pressure (e.g., cf. Page 203 in [In3]). Ovsiannikov  determined
the Lie point symmetries of the above equations and found two very
special solutions.

Let us first do degree analysis. Denote
$$\mbox{deg}\:u=\ell,\;\;\mbox{deg}\:x=\ell_1,\;\;\mbox{deg}\:y=\ell_2,\;\;
\mbox{deg}\:x=\ell_3.\eqno(7.1.5)$$ To make the nonzero terms in
(7.1.1)-(7.1.4) to have the same degree, we have to take
$$\mbox{deg}\:u_x=\mbox{deg}\:v_y\lra
\mbox{deg}\:v=\ell+\ell_2-\ell_1,\eqno(7.1.6)$$
$$\mbox{deg}\:u_x=\mbox{deg}\:w_z\lra
\mbox{deg}\:w=\ell+\ell_3-\ell_1,\eqno(7.1.7)$$
$$\mbox{deg}\:u_t=\mbox{deg}\:uu_x\lra
\mbox{deg}\:t=\ell_1-\ell,\eqno(7.1.8)$$
$$\mbox{deg}\:u_t=\mbox{deg}\:v\sim
2\ell-\ell_1=\ell+\ell_2-\ell_1\lra \ell_2=\ell,\eqno(7.1.9)$$
$$\mbox{deg}\:v_t=\mbox{deg}\:u\sim 2\ell+\ell_2-2\ell_1=\ell\lra
\ell_1=\ell,\eqno(7.1.10)$$
$$\mbox{deg}\:\rho=\mbox{deg}\:p_z\lra
\mbox{deg}\:\rho=\mbox{deg}\:p-\ell_3,\eqno(7.1.11)$$
$$\mbox{deg}\:u=\mbox{deg}\:\frac{1}{\rho}p_y\sim\ell=\mbox{deg}\:p
-\mbox{deg}\:\rho-\ell_2\lra\ell=\ell_3-\ell_2\lra\ell_3=2\ell.\eqno(7.1.12)$$

In summary,
$$\mbox{deg}\:u=\mbox{deg}\:v=\mbox{deg}\:x=\mbox{deg}\:y=\ell,\eqno(7.1.13)$$
$$\mbox{deg}\:w=\mbox{deg}\:z=2\ell=\mbox{deg}\:p
-\mbox{deg}\:\rho,\;\;\mbox{deg}\:t=0.\eqno(7.1.14)$$ Moreover, the
equations (7.1.1)-(7.1.4) are translation invariant because they do
not contain variable coefficients. Thus the transformation
$$T_{a;b_1,b_2}(u(t,x,y,z))=b_1^{-1}u(t+a,b_1x,b_1y,b_1^2z),\eqno(7.1.15)$$
$$T_{a;b_1,b_2}(v(t,x,y,z))=b_1^{-1}v(t+a,b_1x,b_1y,b_1^2z),\eqno(7.1.16)$$
$$T_{a;b_1,b_2}(w(t,x,y,z))=b_1^{-2}w(t+a,b_1x,b_1y,b_1^2z),\eqno(7.1.17)$$
$$T_{a;b_1,b_2}(\rho(t,x,y,z))=b_2\rho(t+a,b_1x,b_1y,b_1^2z),\eqno(7.1.18)$$
$$T_{a;b_1,b_2}(p(t,x,y,z))=b_1^{-2}b_2p(t+a,b_1x,b_1y,b_1^2z)\eqno(7.1.19)$$
is a symmetry of the equations (7.1.1)-(7.1.4).

Let $\al$ be a function in $t$. Note that the transformation
$$F(t,x,y,z)\mapsto
F(t,x+\al,y,z)\;\;\mbox{with}\;\;F=u,v,w,p,\rho\eqno(7.1.20)$$
leaves (7.1.1) invariant and changes (7.1.2)-(7.1.4) to:
$$\al'\rho_x+\rho_t+u\rho_x+v\rho_y+w\rho_z=0,\eqno(7.1.21)$$
$$\al'u_x+u_t+uu_x+vu_y+wu_z+v=-\frac{1}{\rho}p_x,\eqno(7.1.22)$$
$$\al'v_x+v_t+uv_x+vv_y+wv_z-u=-\frac{1}{\rho}p_y,\eqno(7.1.23)$$
 where the independent variable $x$ is
replaced by $x+\al$ and the partial derivatives are with respect to
the original variables. Thus the transformation
$$F(t,x,y,z)\mapsto
F(t,x+\al,y,z)-\dlt_{u,F}\al'\;\;\mbox{with}\;\;F=u,v,w,p,\rho\eqno(7.1.24)$$
leaves (7.1.1) and  (7.1.2)  invariant, and changes (7.1.3) and
(7.1.4) to
$$-{\al'}'+u_t+uu_x+vu_y+wu_z+v=-\frac{1}{\rho}p_x\eqno(7.1.25)$$
$$v_t+uv_x+vv_y+wv_z-u+\al'=-\frac{1}{\rho}p_y.\eqno(7.1.26)$$

On the other hand, the transformation
$$F(t,x,y,z)\mapsto
F(t,x,y,z+{\al'}'x-\al'y)\;\;\mbox{with}\;\;F=u,v,w,p,\rho\eqno(7.1.27)$$
leaves the second equation in (7.1.1) invariant and changes the
first equation in (7.1.1), and (7.1.2)-(7.1.4) to:
$${\al'}'u_z+u_x-\al'v_z+v_y+w_z=0,\eqno(7.1.28)$$
$$({{\al'}'}'x-{\al'}'y)\rho_z+\rho_t+{\al'}'u\rho_z+u\rho_x-\al'v\rho_z+v\rho_y+w\rho_z=0,\eqno(7.1.29)$$
$$({{\al'}'}'x-{\al'}'y)u_z+u_t+{\al'}'uu_z
+uu_x-\al'vu_z+vu_y+wu_z+v=-\frac{1}{\rho}p_x-{\al'}',\eqno(7.1.30)$$
$$({{\al'}'}'x-{\al'}'y)v_z+v_t+{\al'}'uv_z+uv_x-\al'vv_z+vv_y+wv_z-u=-\frac{1}{\rho}p_y+\al'.\eqno(7.1.31)$$
Thus we have the following symmetry transformation of
(7.1.1)-(7.1.4):
$$S_{1,\al}(F(t,x,y,z))=F(t,x+\al,y,z+{\al'}'x-\al'y)-\dlt_{u,F}\al'\;\;\mbox{with}\;\;F=u,v,p,\rho\eqno(7.1.32)$$
and
$$S_{1,\al}(w(t,x,y,z))=w(t,x+\al,y,z+{\al'}'x-\al'y)-{\al'}'u+\al'v-{{\al'}'}'x+{\al'}'y.\eqno(7.1.33)$$

Similarly, we have the symmetry transformation of (7.1.1)-(7.1.4):
$$S_{2,\al}(F(t,x,y,z))=F(t,x,y+\al,z+\al'x+{\al'}'y)-\dlt_{v,F}\al'\;\;\mbox{with}\;\;F=u,v,p,\rho\eqno(7.1.34)$$
and
$$S_{2,\al}(w(t,x,y,z))=w(t,x+\al,y,z+\al'x+{\al'}'y)-\al'u-{\al'}'v-{\al'}'x-{{\al'}'}'y.\eqno(7.1.35)$$
Let $\be$ be another function in $t$. We have the following symmetry
transformation of (7.1.1)-(7.1.4):
$$S_{\al,\be}(F(t,x,y,z))=F(t,x,y,z+\al)-\dlt_{w,F}\al'+\dlt_{p,F}\be\;\;\mbox{with}\;\;F=u,v,w,p,\rho.\eqno(7.1.36)$$
 The above
transformations transform one solution of the equations
(7.1.1)-(7.1.4) into another solution. Applying the above
transformations to any solution found in this chapter will yield
another solution with four extra parameter functions.

\section{Moving-Line Approach}

Let $\al$ and $\be$ be given functions in $t$. Denote the variable
of {\it moving line}\index{moving line}
$$\varpi=\al'x+\be'y+z.\eqno(7.2.1)$$
Suppose that $f,g,h$ are functions in $t,x,y,z$ that are linear in
$x,y,z$ such that
$$f_x+g_y+h_z=0.\eqno(7.2.2)$$
 We assume
$$u=\phi(t,\varpi)+f,\qquad v=\psi(t,\varpi)+g,\eqno(7.2.3)$$
$$w=h-\al'\phi(t,\varpi)- \be'\psi(t,\varpi),\qquad
p=\zeta(t,\varpi),\eqno(7.2.4)$$ where $\phi,\psi,\zeta$ are
two-variable functions to be determined. Note that the first
equation in (7.1.1) naturally holds and  $\rho=p_z=\zeta_\varpi$ by
the second equation in (7.1.1). Moreover, (7.1.2)-(7.1.4) become
$$\zeta_{\varpi
t}+\zeta_{\varpi\varpi}({\al'}'x+{\be'}'y+\al'f+\be'g+h)=0,\eqno(7.2.5)$$
\begin{eqnarray*}\hspace{1cm}& &f_t+g+ff_x+gf_y+hf_z+\al'+\phi_t
+(f_x-\al'f_z)\phi +(f_y-\be'f_z+1)\psi\\ &
&+\phi_\varpi({\al'}'x+{\be'}'y+\al'f+\be'g+h)
=0,\hspace{6.2cm}(7.2.6)\end{eqnarray*}
\begin{eqnarray*}\hspace{1cm}& &g_t-f+fg_x+gg_y+hg_z+\be'+\psi_t
+(g_x-\al'g_z-1)\phi +(g_y-\be'g_z)\psi\\ &
&+\psi_\varpi({\al'}'x+{\be'}'y+\al'f+\be'g+h)
=0.\hspace{6.2cm}(7.2.7)\end{eqnarray*}

In order to solve the above system of partial differential
equations, we assume
$${\al'}'x+{\be'}'y+\al'f+\be'g+h=-\gm'\varpi=-\gm'(\al'x+\be'y+z)
\eqno(7.2.8)$$ for some function $\gm$ in $t$, and
$$f_t+g+ff_x+gf_y+hf_z+\al'=0,\eqno(7.2.9)$$
$$g_t-f+fg_x+gg_y+hg_z+\be'=0.\eqno(7.2.10)$$
Then (7.2.5)-(7.2.7) become
$$\zeta_{\varpi
t}-\gm'\varpi\zeta_{\varpi\varpi}=0,\eqno(7.2.11)$$
$$\phi_t
+(f_x-\al'f_z)\phi
+(f_y-\be'f_z+1)\psi-\gm'\varpi\phi_\varpi=0,\eqno(7.2.12)$$
$$\psi_t+(g_x-\al'g_z-1)\phi
+(g_y-\be'g_z)\psi-\gm'\varpi\psi_\varpi=0.\eqno(7.2.13)$$

According to (7.2.8),
$$h=-{\al'}'x-{\be'}'y-\al'f-\be'g-\gm'\varpi.\eqno(7.2.14)$$
Substituting the above equation into (7.2.9) and (7.2.10), we have:
$$f_t+f(f_x-\al'f_z)+g(f_y-\be'f_z+1)-f_z({\al'}'x+{\be'}'y+\gm'\varpi)
+\al'=0,\eqno(7.2.15)$$
$$g_t+f(g_x-\al'g_z-1)+g(g_y-\be'g_z)-g_z({\al'}'x+{\be'}'y+\gm'\varpi)
+\be'=0.\eqno(7.2.16)$$ Our linearity assumption implies that
$$A=\left(\begin{array}{cc}f_x-\al'f_z&f_y-\be'f_z+1\\
g_x-\al'g_z-1&g_y-\be'g_z\end{array}\right)\eqno(7.2.17)$$ is a
matrix function in $t$. In order to solve the system (7.2.12) and
(7.2.13), and the system (7.2.15) and (7.2.16), we need the
commutativity of $A$ with $dA/dt$. For simplicity, we assume
$$f_y-\be'f_z+1=g_x-\al'g_z-1=0.\eqno(7.2.18)$$
So
$$f_y=\be'f_z-1,\qquad g_x=\al'g_z+1.\eqno(7.2.19)$$
Moreover, (7.2.15) and (7.2.16) become
$$f_t+f(f_x-\al'f_z)-f_z({\al'}'x+{\be'}'y+\gm'\varpi)
+\al'=0,\eqno(7.2.20)$$
$$g_t+g(g_y-\be'g_z)-g_z({\al'}'x+{\be'}'y+\gm'\varpi)
+\be'=0.\eqno(7.2.21)$$ Write
$$f=\al_1x+(\be'\al_2-1)y+\al_2 z+\al_3,\eqno(7.2.22)$$
$$g=(\al'\be_2+1)x+\be_1y+\be_2 z+\be_3\eqno(7.2.23)$$
by our linearity assumption and (7.2.19), where $\al_l$ and $\be_j$
are functions in $t$.

 Now (7.2.20) is equivalent to the following
system of ordinary differential equations:
$$\al_1'+\al_1(\al_1-\al'\al_2)-\al_2({\al'}'+\gm'\al')=0,\eqno(7.2.24)$$
$$(\be'\al_2)'+(\be'\al_2-1)(\al_1-\al'\al_2)
-\al_2({\be'}'+\gm'\be')=0,\eqno(7.2.25)$$
$$\al_2'+\al_2(\al_1-\al'\al_2-\gm')=0,\eqno(7.2.26)$$
$$\al_3'+\al_3(\al_1-\al'\al_2)+\al'=0.\eqno(7.2.27)$$
Observe that $(7.2.25)-\be'\times(7.2.26)$ becomes
$$-\al_1+\al'\al_2=0.\eqno(7.2.28)$$
So (7.2.26) becomes
$$\al_2'-\gm'\al_2=0\lra \al_2=b_1e^\gm,\qquad
b_1\in\mbb{R}.\eqno(7.2.29)$$ According to (7.2.28),
$$\al_1=b_1\al'e^\gm.\eqno(7.2.30)$$
With the data (7.2.29) and (7.2.30), (7.2.24) naturally holds. By
(7.2.27), we take
$$\al_3=-\al.\eqno(7.2.31)$$

Note that (7.2.21) is equivalent to the following system of ordinary
differential equations:
$$\al'\be_2'+(\al'\be_2+1)(\be_1-\be'\be_2)-
\al'\be_2\gm'=0,\eqno(7.2.32)$$
$$\be_1'+\be_1(\be_1-\be'\be_2)-\be_2({\be'}'+\be'\gm')=0,\eqno(7.2.33)$$
$$\be_2'+\be_2(\be_1-\be'\be_2-\gm')=0,\eqno(7.2.34)$$
$$\be_3'+\be_3(\be_1-\be'\be_2)+\be'=0.\eqno(7.2.35)$$
Similarly, we have:
$$\be_1=b_2\be'e^\gm,\qquad\be_2=b_2e^\gm,\qquad
\be_3=\be\eqno(7.2.36)$$ with $b_2\in\mbb{R}$. Moreover, (7.2.2)
gives $\gm'=0$ by (7.2.14), (7.2.28) and (7.2.36). We take $\gm=0$.
Therefore, $\phi=\Im(\varpi)$ and $\psi=\iota(\varpi)$ by (7.2.12)
and (7.2.13) for some one-variable functions $\Im$ and $\iota$.
Furthermore, we take $\zeta=\sgm(\varpi)$ by (7.2.11) for another
one-variable function $\sgm$. In summary, we have:\psp

{\bf Theorem 7.2.1}. {\it Let $\al,\be$ be functions in $t$ and let
$b_1,b_2\in\mbb{R}$. Suppose that $\Im,\;\iota$ and $\sgm$ are
arbitrary one-variable functions. The following is a solution of the
equations (7.1.1)-(7.1.4) of  dynamic convection in a sea:
$$u=b_1\al'x+(b_1\be'-1)y+b_1
z-\al+\Im(\al'x+\be'y+z),\eqno(7.2.37)$$
$$v=(b_2\al'+1)x+b_2\be' y+b_2 z+\be+
\iota(\al'x+\be'y+z),\eqno(7.2.38)$$
\begin{eqnarray*}w&=&-({\al'}'+b_1{\al'}^2+(b_2\al'+1)\be')x
-({\be'}'+\al'(b_1\be'-1)+b_2{\be'}^2)y-(b_1\al'+b_2\be')z\\
& &+\al\al'-\be\be'-\al'\Im(\al'x+\be'y+z)-\be'\iota(\al'x+\be'y+z),
\hspace{3.4cm}(7.2.39)\end{eqnarray*}
$$p=\sgm(\al'x+\be'y+z),\qquad\rho=\sgm'(\al'x+\be'y+z).\eqno(7.2.40)$$
}

\section{Approach of Cylindrical Product}

Let $\sgm$ be a fixed one-variable function and set the variable of
{\it cylindrical product}:\index{cylindrical product}
$$\varpi=z\sgm(x^2+y^2).\eqno(7.3.1)$$
Suppose that $f$ and $g$  are functions in $t,x,z$ that are linear
homogeneous in $x,y$ and
$$h=\frac{\gm}{\sgm}-z(f_x+g_y),
\eqno(7.3.2)$$ where $\gm$ is a function in $t$. Assume
$$u=f+y\psi(t,\varpi),\qquad v=g-x\psi(t,\varpi),\qquad
w=h,\qquad p=\phi(t,\varpi)\eqno(7.3.3)$$ where $\psi$ and $\phi$
are two-variable functions. Note
$$u_t=f_t+y\psi_t,\qquad u_x=f_x+2xyz\sgm'\psi_\varpi,\eqno(7.3.4)$$
$$u_y=f_y+\psi+2y^2z\sgm'\psi_\varpi,\qquad
u_z=f_z+y\sgm\psi_\varpi,\eqno(7.3.5)$$
$$v_t=g_t-x\psi_t,\qquad v_x=g_x-\psi-2x^2z\sgm'
\psi_\varpi,\eqno(7.3.6)$$
$$v_y=g_y-2xyz\sgm'\psi_\varpi,\qquad
v_z=g_z-x\sgm\psi_\varpi.\eqno(7.3.7)$$ Hence (7.1.3) becomes
\begin{eqnarray*}&&u_t+uu_x+vu_y+wu_z+v=f_t+y\psi_t+(f+y\psi)
(f_x+2xyz\sgm'\psi_\varpi)\\ &
&+(g-x\psi)(f_y+1+\psi+2y^2z\sgm'\psi_\varpi)+y\sgm h\psi_\varpi
\\ &=&f_t+ff_x+g(1+f_y)+x(g_x-f_y-1)\psi-x\psi^2\\ & &+y[\psi_t
+(f_x+g_y)\psi+(2(xf+yg)\sgm'z+h\sgm)
\psi_\varpi]=-\frac{2xz\sgm'}{\sgm}\hspace{3.4cm}(7.3.8)
\end{eqnarray*}
and (7.1.4) gives
\begin{eqnarray*}&&v_t+uv_x+vv_y+wv_z-u=g_t-x\psi_t+(f+y\psi)
(g_x-1-\psi-2x^2z\sgm'\psi_\varpi)\\ &
&+(g-x\psi)(g_y-2xyz\sgm'\psi_\varpi)-x\sgm h\psi_\varpi
\\ &=&g_t+f(g_x-1)+gg_y-y(1+f_y-g_x)\psi-y\psi^2\\ & &-x[\psi_t+(f_x+g_y)\psi
+(2(xf+yg)\sgm'z+h\sgm)
\psi_\varpi]=-\frac{2yz\sgm'}{\sgm}.\hspace{3.3cm}(7.3.9)
\end{eqnarray*}

In order to solve the above system of differential equations, we
assume
$$f=\al' x-\frac{y}{2},\qquad g= \frac{x}{2}+\al' y,\qquad
\sgm(x^2+y^2)=\frac{1}{x^2+y^2}\eqno(7.3.10)$$ for some function
$\al$ in $t$. According to (7.3.2),
$$h=\frac{\gm}{\sgm}-2\al' z.
\eqno(7.3.11)$$ Now (7.3.8) becomes
$$({\al'}'+{\al'}^2+4^{-1}-\psi^2)x+y[\psi_t+2\al'\psi+(\gm-4\al'\varpi)
\psi_\varpi]=2x\varpi\eqno(7.3.12)$$ and (7.3.9) yields
$$({\al'}'+{\al'}^2+4^{-1}-\psi^2)y-x[\psi_t+2\al'\psi+(\gm-4\al'\varpi)
\psi_\varpi]=2y\varpi.\eqno(7.3.13)$$ The above system is equivalent
to
$${\al'}'+{\al'}^2+4^{-1}-\psi^2=2\varpi,\eqno(7.3.14)$$
$$\psi_t+2\al'\psi+(\gm-4\al'\varpi)\psi_\varpi=0.\eqno(7.3.15)$$
By (7.3.14), we take
$$\psi=\sqrt{{\al'}'+{\al'}^2+4^{-1}-2\varpi},\eqno(7.3.16)$$
due to the skew-symmetry of $(u,x)$ and $(v,y)$. Substituting
(7.3.16) into (7.3.15), we get
$${{\al'}'}'+2\al'{\al'}'+4\al'({\al'}'+{\al'}^2+4^{-1}-2\varpi)
-2(\gm-4\al'\varpi)=0, \eqno(7.3.17)$$ equivalently,
$$\gm=2{\al'}^3+3\al'{\al'}'+\frac{{{\al'}'}'+\al'}{2}.\eqno(7.3.18)$$

According to the second equation in (7.1.1), we have
$\rho=\sgm\phi_\varpi$. Note
$$\rho_t=\sgm\phi_{\varpi
t},\qquad\rho_x=2x\sgm'(\phi_\varpi+\varpi\phi_{\varpi\varpi}),
\eqno(7.3.19)$$
$$\rho_y=2y\sgm'(\phi_\varpi+\varpi\phi_{\varpi\varpi}),\qquad
\rho_z=\sgm^2\phi_{\varpi\varpi}.\eqno(7.3.20)$$ So (7.1.2) becomes
$$\phi_{\varpi t}-2\al'\phi_\varpi
+(\gm-4\al'\varpi)\phi_{\varpi\varpi}=0.\eqno(7.3.21)$$ Modulo some
$S_{0,\be}$ in (7.1.36), the above equation is equivalent to:
$$\phi_ t+2\al'\phi+(\gm-4\al'\varpi)\phi_\varpi=0.\eqno(7.3.22)$$
Set
$$\td\psi=e^{2\al}\psi,\qquad\td\phi=e^{2\al}\phi.\eqno(7.3.23)$$
Then (7.3.15) and (7.3.22) are equivalent to the equations:
$$\td\psi_t+(\gm-4\al'\varpi)\td\psi_\varpi=0,\qquad
\td\phi_t+(\gm-4\al'\varpi)\td\phi_\varpi=0,\eqno(7.3.24)$$
respectively. So we have the solution
$$\td\phi=\Im(\td\psi)\lra \phi=e^{-2\al}\Im\left(e^{2\al}\sqrt{{\al'}'+{\al'}^2+4^{-1}
-2\varpi}\right)\eqno(7.3.25)$$ for some one-variable function
$\Im$. Thus we have:\psp

{\bf Theorem 7.3.1}. {\it Let $\al$ be any function in $t$ and let
$\Im$ be arbitrary one-variable function.
 The following is a solution of the
equations (7.1.1)-(7.1.4) of  dynamic convection in a sea:
$$u=\al'
x-\frac{y}{2}+y\sqrt{{\al'}'+{\al'}^2+\frac{1}{4}-\frac{2z}{x^2+y^2}},
\eqno(7.3.26)$$
$$v=\al'y+\frac{x}{2}-x\sqrt{{\al'}'+{\al'}^2+\frac{1}{4}-\frac{2z}{x^2+y^2}},
\eqno(7.3.27)$$
$$w=\left(2{\al'}^3+3\al'{\al'}'+\frac{{{\al'}'}'+\al'}{2}\right)(x^2+y^2)
-2\al' z,\eqno(7.3.28)$$
$$p=e^{-2\al}\Im\left(e^{2\al}\sqrt{{\al'}'+{\al'}^2
+\frac{1}{4}-\frac{2z}{x^2+y^2}}\right),\eqno(7.3.29)$$
$$\rho
=-\frac{\Im'\left(e^{2\al}\sqrt{{\al'}'+{\al'}^2
+\frac{1}{4}-\frac{2z}{x^2+y^2}}\right)}{(x^2+y^2)\sqrt{\al'+\al^2+\frac{1}{4}-\frac{2z}{x^2+y^2}}}
.\eqno(7.3.30)$$ }\pse

{\bf Remark 7.3.2}. Let $\be_1,\be_2,\be_3$ and $\gm$ be functions
in $t$. Applying $S_{1,\be_1}$ in (7.1.32)-(7.1.33), $S_{2,\be_2}$
in (7.1.34)-(7.1.35) and $S_{\be_3,\gm}$ in (7.1.36) to the above
solution,  we get a more general solution:
\begin{eqnarray*}u&=&(y+\be_2)\sqrt{{\al'}'+{\al'}^2+\frac{1}{4}
-\frac{2(z+(({\be_1'}'+\be_2')x+({\be_2'}'-\be_1')y+\be_3)}{(x+\be_1)^2+(y+\be_2)^2}}\\&
& +\al' (x+\be_1)-\frac{y+\be_2}{2}-\be_1',
\hspace{8.3cm}(7.3.31)\end{eqnarray*}
\begin{eqnarray*}v&=&-(x+\be_2)\sqrt{{\al'}'+{\al'}^2+\frac{1}{4}
-\frac{2(z+(({\be_1'}'+\be_2')x+({\be_2'}'-\be_1')y+\be_3)}{(x+\be_1)^2+(y+\be_2)^2}}\\&
& +\al' (y+\be_2)+\frac{x+\be_1}{2}-\be_2',
\hspace{8.3cm}(7.3.32)\end{eqnarray*}
\begin{eqnarray*}w&=&\left(2{\al'}^3+3\al'{\al'}'+\frac{{{\al'}'}'+\al'}{2}\right)((x+\be_1)^2+(y+\be_2)^2)
\\ & &-2\al' (z+({\be_1'}'+\be_2')x+({\be_2'}'-\be_1')y+\be_3)-\be_3'
\\ &
&-({\be_1'}'+\be_2')u+(\be_1'-{\be_2'}')v-({{\be_1'}'}'+{\be_2'}')x+({\be_1'}'-{{\be_2'}'}')y,
\hspace{3cm}(7.3.33)\end{eqnarray*}
$$p=e^{-2\al}\Im\left(e^{2\al}\sqrt{{\al'}'+{\al'}^2+\frac{1}{4}
-\frac{2(z+(({\be_1'}'+\be_2')x+({\be_2'}'-\be_1')y+\be_3)}{(x+\be_1)^2+(y+\be_2)^2}}\right)+\gm,\eqno(7.3.34)$$
$$\rho
=-\frac{\Im'\left(e^{2\al}\sqrt{{\al'}'+{\al'}^2+\frac{1}{4}
-\frac{2(z+(({\be_1'}'+\be_2')x+({\be_2'}'-\be_1')y+\be_3)}{(x+\be_1)^2+(y+\be_2)^2}}\right)}
{[(x+\be_1)^2+(y+\be_2)^2]\sqrt{{\al'}'+{\al'}^2+\frac{1}{4}
-\frac{2(z+(({\be_1'}'+\be_2')x+({\be_2'}'-\be_1')y+\be_3)}{(x+\be_1)^2+(y+\be_2)^2}}}
.\eqno(7.3.35)$$

\section{Dimensional Reduction}

\index{dimensional reduction}

 Suppose that $u,v,\zeta$ and $\eta$ are functions in $t,x,y$.
Assume
$$w=\zeta-(u_x+v_y)z,\qquad p=z+\eta,\qquad \rho=1.\eqno(7.4.1)$$
 Then the equations (7.1.1)-(7.1.4) are equivalent to the
following two-dimensional problem:
$$u_t+uu_x+vu_y+v=-\eta_x,\eqno(7.4.2)$$
$$v_t+uv_x+vv_y-u=-\eta_y.\eqno(7.4.3)$$
The compatibility $\eta_{xy}=\eta_{yx}$ gives
$$(u_y-v_x)_t+u(u_y-v_x)_x+v(u_y-v_x)_y
+(u_x+v_y)(u_y-v_x+1)=0.\eqno(7.4.4)$$

Let $\vt$ be a function in $t,x,y$ that is harmonic in $x$ and $y$,
i.e.
$$\vt_{xx}+\vt_{yy}=0.\eqno(7.4.5)$$
 We assume
$$u=\vt_{xx},\qquad v=\vt_{xy}.\eqno(7.4.6)$$
Then (7.4.4) naturally holds. Indeed,
$$u_t+uu_x+vu_y+v=
\left(\vt_{xt}+2^{-1}(\vt_{xx}^2+\vt_{xy}^2)+\vt_y\right)_x,\eqno(7.4.7)$$
$$v_t+uv_x+vv_y-u=\left(\vt_{xt}+2^{-1}(\vt_{xx}^2+\vt_{xy}^2)+\vt_y\right)_y.
\eqno(7.4.8)$$ By (7.4.2) and (7.4.3), we take
$$\eta=-\vt_{xt}-\vt_y-\frac{1}{2}(\vt_{xx}^2+\vt_{xy}^2).\eqno(7.4.9)$$
Hence we have the following easy result:\psp

{\bf Proposition 7.4.1}. {\it Let $\vt$ and $\zeta$ be functions in
$t,x,y$ such that (7.4.5) holds. The following is a solution of the
equations (7.1.1)-(7.1.4) of  dynamic convection in a sea:
$$u=\vt_{xx},\qquad v=\vt_{xy},\qquad w=\zeta,\eqno(7.4.10)$$
$$\rho=1,\qquad p=z-\vt_{xt}-\vt_y-\frac{1}{2}(\vt_{xx}^2+\vt_{xy}^2).
\eqno(7.4.11)$$}

The above approach is the well-known rotation-free approach. We are
more interested in the approaches that the rotation may not be zero.
Let $f$ and $g$ be functions in $t,x,y$ that are linear in $x,y$.
Denote
$$\varpi=x^2+y^2.\eqno(7.4.12)$$
Consider
$$u=f+y\phi(t,\varpi),\qquad v=g
-x\phi(t,\varpi),\eqno(7.4.13)$$ where $\phi$ is a two-variable
function to be determined. Then
$$u_x=f_x+2xy\phi_\varpi,\qquad
u_y=f_y+\phi+2y^2\phi_\varpi,\eqno(7.4.14)$$
$$v_x=g_x-\phi-2x^2\phi_\varpi,\qquad
u_y=g_y-2xy\phi_\varpi.\eqno(7.4.15)$$ Thus
$$u_x+v_y=f_x+g_y, \qquad
u_y-v_x=f_y-g_x+2(\varpi\phi)_\varpi.\eqno(7.4.16)$$ For simplicity,
we assume
$$f=-\frac{\al'x}{2\al}-\frac{y}{2},\qquad g=\frac{x}{2}
-\frac{\al'y}{2\al}\eqno(7.4.17)$$ for some functions $\al$ and
$\be$ in $t$. Then (7.4.4) becomes
$$(\varpi\phi)_{\varpi t}-\frac{\al'}{\al}
\varpi(\varpi\phi)_{\varpi\varpi}-\frac{\al'}{\al}
(\varpi\phi)_\varpi=0.\eqno(7.4.18)$$
 Hence
$$\phi=\frac{\gm+\Im(\al\varpi)}{\varpi}\eqno(7.4.19)$$
for some function $\gm$ in $t$ and one-variable function $\Im$.

Now (7.4.12), (7.4.17) and (7.4.19) imply
$$u=-\frac{\al'x}{2\al}-\frac{y}{2}
+\frac{(\gm+\Im(\al\varpi))y}{\varpi},\eqno(7.4.20)$$
$$v=\frac{x}{2}-\frac{\al'y}{2\al}-\frac{(\gm+\Im(\al\varpi))x}{\varpi}.
\eqno(7.4.21)$$ By (7.4.13) and (7.4.17), we calculate
\begin{eqnarray*}& &u_t+uu_x+vu_y+v\\
&=&f_t+y\phi_t+(f+y\phi)(f_x+2xy\phi_\varpi)+(g-x\phi)(f_y+\phi+2y^2\phi_\varpi)+g-x\phi
\\&=&
f_t+ff_x+g(f_y+1)+y\phi_t+(f_xy-f_yx+g-x)\phi+2(fx+gy)y\phi_\varpi-x\phi^2
\\&=&\left(\frac{3{\al'}^2-2\al{\al'}'}{4\al^2}+\frac{1}{4}
\right)x-x\phi^2+y\left(\phi_t-\frac{\al'}{\al}(\varpi\phi)_\varpi)\right),\hspace{4cm}(7.4.22)
\end{eqnarray*}
\begin{eqnarray*}& &v_t+uv_x+vv_y-u\\
&=&g_t-x\phi_t+(f+y\phi)(g_x-\phi-2x^2\phi_\varpi)+(g-x\phi)(g_y-2xy\phi_\varpi)-f-y\phi
\\&=&
g_t+f(g_x-1)+gg_y-x\phi_t+(g_xy-g_yx-f-y)\phi-2(fx+gy)x\phi_\varpi-y\phi^2
\\&=&\left(\frac{3{\al'}^2-2\al{\al'}'}{4\al^2}+\frac{1}{4}
\right)y-x\phi^2-x\left(\phi_t-\frac{\al'}{\al}(\varpi\phi)_\varpi)\right).\hspace{4.1cm}(7.4.23)
\end{eqnarray*}

On the other hand, (7.4.19) says that
$$\phi_t-\frac{\al'}{\al}(\varpi\phi)_\varpi=\frac{\gm'+\al'\varpi\Im'(\al\varpi)}{\varpi}-\al'\Im'(\al\varpi)=
\frac{\gm'}{\varpi}.\eqno(7.4.24)$$ Thus (7.4.2) and (7.4.3) yield
$$\left(\frac{3{\al'}^2-2\al{\al'}'}{4\al^2}+\frac{1}{4}
\right)x +\frac{\gm' y}{\varpi}-x\phi^2 =-\eta_x,\eqno(7.4.25)$$
$$\left(\frac{3{\al'}^2-2\al{\al'}'}{4\al^2}+\frac{1}{4}
\right)y-\frac{\gm' x}{\varpi}-y\phi^2 =-\eta_y\eqno(7.4.26)$$ by
(7.4.22) and (7.4.23). Hence
$$\eta=\frac{1}{2}\int
\frac{(\gm+\Im(\al\varpi))^2d\varpi}{\varpi^2}
-\frac{1}{2}\left(\frac{3{\al'}^2-2\al{\al'}'}{4\al^2}+\frac{1}{4}
\right)\varpi+\gm' \arctan\frac{y}{x}.\eqno(7.4.27)$$\pse

{\bf Theorem 7.4.2}. {\it Let $\al,\gm$ be any functions in $t$.
Suppose that $\Im$ is an arbitrary one-variable function and $\zeta$
is any function in $t,x,y$. The following is a solution of the
equations (7.1.1)-(7.1.4) of  dynamic convection in a sea:
$$u=-\frac{\al'x}{2\al}-\frac{y}{2}
+\frac{(\gm+\Im((x^2+y^2)\al))y}{x^2+y^2},\eqno(7.4.28)$$
$$v=\frac{x}{2}-\frac{\al'y}{2\al}-
\frac{(\gm+\Im((x^2+y^2)\al))x}{x^2+y^2},\eqno(7.4.29)$$
$$w=\frac{\al'}{\al}z+\zeta,\qquad
\rho=1,\eqno(7.4.30)$$
$$p=z+\frac{1}{2}\int
\frac{(\gm+\Im(\al\varpi))^2d\varpi}{\varpi^2}
-\frac{(3{\al'}^2-2\al{\al'}')\al^{-2}+1}{8}(x^2+y^2)+\gm'
\arctan\frac{y}{x}\eqno(7.4.31)$$ with $\varpi=x^2+y^2$.} \psp

Next we assume
$$u=\ves(t,x),\qquad v=\phi(t,x)+\psi(t,x)y,\eqno(7.4.32)$$
where $\ves,\;\phi$ and $\psi$ are functions in $t,x$ to be
determined. Substituting (7.4.32) into (7.4.4), we get
$$\phi_{tx}+\psi_{tx}y+\ves(\phi_{xx}+\psi_{xx}y)+(\phi+\psi
y)\psi_x+(\ves_x+\psi)(\phi_x+\psi_xy-1)=0,\eqno(7.4.33)$$
equivalently,
$$(\phi_t+\ves\phi_x+\phi\psi-\ves)_x-\psi=0,
\eqno(7.4.34)$$
$$(\psi_t+\ves\psi_x+\psi^2)_x=0.\eqno(7.4.35)$$
For simplicity, we take
$$\psi=-\al',\eqno(7.4.36)$$
a function in $t$.

Denote
$$\phi=\hat\phi+x.\eqno(7.4.37)$$
Then (7.4.34) becomes
$$(\hat\phi_t+\ves\hat\phi_x-\al'\hat\phi)_x=0.\eqno(7.4.38)$$
To solve the above equation, we assume
$$\ves=\frac{\be}{\hat\phi_x}-\frac{\vt_t(t,x)}{\vt_x(t,x)}\eqno(7.4.39)$$
for some functions $\be$ in $t$, and $\vt$ in $t$ and $x$.  We have
the following solution of (7.4.38):
$$\hat\phi=e^\al\Im(\vt)\lra\phi=e^\al\Im(\vt)+x\lra
v=e^\al\Im(\vt)+x-\al'y \eqno(7.4.40)$$ for another one-variable
function $\Im$. Moreover,
$$\ves=\frac{\be e^{-\al}}{\vt_x\Im'(\vt)}-
\frac{\vt_t}{\vt_x}.\eqno(7.4.41)$$ Note
\begin{eqnarray*} & &u_t+uu_x+vu_y+v
=\frac{(\be e^{-\al})'}{\vt_x\Im'(\vt)}- \frac{\be
e^{-\al}(\vt_{xt}\Im'(\vt)+\vt_t\vt_x{\Im'}'(\vt))
}{(\vt_x\Im'(\vt))^2}-
\frac{\vt_{tt}\vt_x-\vt_t\vt_{xt}}{\vt_x^2}\\
& &+\left(\frac{\be e^{-\al}}{\vt_x\Im'(\vt)}-
\frac{\vt_t}{\vt_x}\right)\left(\frac{\be e^{-\al}}{\vt_x\Im'(\vt)}-
\frac{\vt_t}{\vt_x}\right)_x+
e^\al\Im(\vt)+x-\al'y.\hspace{3.3cm}(7.4.42)
\end{eqnarray*}

By (7.4.36), (7.4.40) and (7.4.41),
\begin{eqnarray*}& &\phi_t+\ves\phi_x+\psi\phi-\ves
\\&=&e^\al(\al'\Im(\vt)+\vt_t\Im'(\vt))+\left(\frac{\be
e^{-\al}}{\vt_x\Im'(\vt)}-
\frac{\vt_t}{\vt_x}\right)e^\al\vt_x\Im'(\vt)-\al'(e^\al\Im(\vt)+x)
\\ &=&\be-\al'x.\hspace{12.1cm}(7.4.43)
\end{eqnarray*}
Thus (7.4.32) (7.4.36) and (7.4.43) yield
\begin{eqnarray*} v_t+uv_x+vv_y-u&=&\phi_t+\psi_ty+\ves(\phi_x+\psi_xy-1)+
(\phi+\psi y)\psi\\
&=&\phi_t+\ves\phi_x+\psi\phi-\ves+(\psi_t+\ves\psi_x+\psi^2)y
\\
&=&\be-\al' x+(-{\al'}'+\al^2)y.\hspace{5.9cm}(7.4.44)
\end{eqnarray*}
According to (7.4.2) and (7.4.3),
\begin{eqnarray*}\eta&=&\int\left(\frac{\be
e^{-\al}(\vt_{xt}\Im'(\vt)+\vt_t\vt_x{\Im'}'(\vt))
}{(\vt_x\Im'(\vt))^2}+\frac{\vt_{tt}\vt_x-\vt_t\vt_{xt}}{\vt_x^2}
-\frac{(\be e^{-\al})'}{\vt_x\Im'(\vt)}-e^\al\Im(\vt)\right)dx
\\ &  &+\al' xy-\be
y+\frac{({\al'}'-{\al'}^2)y^2-x^2}{2}-\frac{1}{2}\left(\frac{\be
e^{-\al}}{\vt_x\Im'(\vt)}-
\frac{\vt_t}{\vt_x}\right)^2.\hspace{2.9cm}(7.4.45)\end{eqnarray*}\pse

{\bf Theorem 7.4.3}. {\it Let $\al,\be$ be functions in $t$ and let
$\Im$ be a one-variable function. Suppose that $\vt$ is a function
in $t,x$, and $\zeta$ is function in $t,x,y$. The following is a
solution of the equations (7.1.1)-(7.1.4) of  dynamic convection in
a sea:
$$u=\frac{\be e^{-\al}}{\vt_x\Im'(\vt)}-
\frac{\vt_t}{\vt_x},\qquad v=e^\al\Im(\vt)+x-\al'y,\eqno(7.4.46)$$
$$w=\left(\al'+\frac{\be e^{-\al}(\vt_{xx}\Im'(\vt)+\vt_x^2{\Im'}'(\vt)
)}{(\vt_x\Im'(\vt))^2}+\frac{\vt_{xt}\vt_x-\vt_t\vt_{xx}}{\vt_x^2}\right)z+
\zeta,\qquad \rho=1,\eqno(7.4.47)$$
\begin{eqnarray*}p&=&z+\int\left(\frac{\be
e^{-\al}(\vt_{xt}\Im'(\vt)+\vt_t\vt_x{\Im'}'(\vt))
}{(\vt_x\Im'(\vt))^2}+\frac{\vt_{tt}\vt_x-\vt_t\vt_{xt}}{\vt_x^2}
-\frac{(\be e^{-\al})'}{\vt_x\Im'(\vt)}-e^\al\Im(\vt)\right)dx
\\ &  &+\al' xy-\be
y+\frac{({\al'}'-{\al'}^2)y^2-x^2}{2}-\frac{1}{2}\left(\frac{\be
e^{-\al}}{\vt_x\Im'(\vt)}-
\frac{\vt_t}{\vt_x}\right)^2.\hspace{3.1cm}(7.4.48)\end{eqnarray*} }
\psp

Finally, we suppose that $\al,\be$ are functions in $t$ and $f,g$
are functions in $t,x,y$ that are linear homogeneous in $x$ and $y$.
Denote $\varpi=\al x+\be y$. Assume
$$u=f+\be \phi(t,\varpi),\qquad
v=g-\al\phi(t,\varpi).\eqno(7.4.49)$$ Then
$$u_y-v_x=f_y-g_x+(\al^2+\be^2)\phi_\varpi,\qquad
u_x+v_y=f_x+g_y.\eqno(7.4.50)$$ Now (7.4.4) becomes
\begin{eqnarray*}\hspace{0.5cm}& &f_{yt}-g_{xt}+(\al^2+\be^2)'\phi_\varpi+
(\al^2+\be^2)(\phi_{\varpi
t}+(\al'x+\be'y+\al f+\be g)\phi_{\varpi\varpi})\\
&&+(f_x+g_y)(f_y-g_x+1+(\al^2+\be^2)\phi_\varpi)=0.
\hspace{5.6cm}(7.4.51)\end{eqnarray*} In order to solve the above
equation, we assume
$$g_x=\vf,\qquad f_y=\vf-1,\eqno(7.4.52)$$
$$\al'x+\be'y+\al f+\be g=0\eqno(7.4.53)$$ for some function $\vf$ in $t$. The
equation (7.4.52) is equivalent to:
$$\al'+\al f_x+\vf\be=0\lra
f_x=-\frac{\al'+\vf\be}{\al},\eqno(7.4.54)$$
$$\be'+\be g_y+\al(\vf-1)=0\lra
g_y=-\frac{\be'+\al(\vf-1)}{\be}.\eqno(7.4.55)$$ Thus
$$f=-\frac{\al'+\vf\be}{\al}x+(\vf-1)y,\qquad g=\vf
x-\frac{\be'+\al(\vf-1)}{\be}y.\eqno(7.4.56)$$

Now (7.4.51) becomes
$$\phi_{\varpi
t}-\left(\frac{\al'+\vf\be}{\al}
+\frac{\be'+\al(\vf-1)}{\be}-\frac{(\al^2+\be^2)'}{\al^2+\be^2}\right)
\phi_\varpi=0.\eqno(7.4.57)$$ Thus we have the following solution
$$\phi=\frac{\al\be}{\al^2+\be^2}e^{\int(\al\be^{-1}(\vf-1)+\al^{-1}\be\vf)dt}
\Im'(\varpi),\eqno(7.4.58)$$ where $\Im$ is an arbitrary
one-variable function. Note that (7.4.56) and (7.4.58) give
\begin{eqnarray*}&&
u_t+uu_x+vu_y+v\\ &=&f_t+\be'
\phi+\be\phi_t+(\al'x+\be'y)\be\phi_\varpi+(f+\be\phi)(f_x+\al\be\phi_\varpi)\\
& &+ (g-\al\phi)(f_y+1+\be^2\phi_\varpi)\\&=&
f_t+ff_x+g(f_y+1)+(\be'+\be
f_x-\al (f_y+1))\phi +\be\phi_t\\
& &+(\al'x+\be'y+\al f+\be g)\be\phi_\varpi
\\&=&\frac{\al^2\be}{\al^2+\be^2}
\left(\frac{2(\al\be'-\al'\be)}{\al^2+b^2}-1\right)e^{\int(\al\be^{-1}(\vf-1)+\al^{-1}\be\vf)dt}
\Im'(\varpi)
\\& &+\left(\frac{2{\al'}^2+(\vf\be)^2+3\al'\be\vf-\al(\vf\be)'
-\al{\al'}'}{\al^2}+\vf^2\right)x\\ & & + \left(\vf'-
\frac{(\vf-1)(\al'+\vf\be)}{\al}
-\frac{\vf(\be'+\al(\vf-1))}{\be}\right)y,\hspace{4.3cm}(7.4.59)
\end{eqnarray*}
\begin{eqnarray*}& &v_t+uv_x+vv_y-u\\&=&g_t-\al'
\phi-\al\phi_t-(\al'x+\be'y)\al\phi_\varpi+(f+\be\phi)(g_x-1-2\al^2\phi_\varpi)\\
& &+ (g-\al\phi)(g_y-2\al\be\phi_\varpi)\\&=&
g_t+f(g_x-1)+gg_y-(\al'+\al
g_y-\be (g_x-1))\phi -\al\phi_t\\
& &-(\al'x+\be'y+\al f+\be g)\al\phi_\varpi
\\&=&\frac{\al\be^2}{\al^2+\be^2}
\left(\frac{2(\al\be'-\al'\be)}{\al^2+b^2}-1\right)e^{\int(\al\be^{-1}(\vf-1)+\al^{-1}\be\vf)dt}
\Im'(\varpi)+[(\vf-1)^2\\ &&-\frac{\be((\vf-1)\al)'
+\be{\be'}'-2{\be'}^2-((\vf-1)\al)^2-3\al\be'(\vf-1)}{\be^2}]y\\
& &+ \left(\vf'- \frac{(\vf-1)(\al'+\vf\be)}{\al}
-\frac{\vf(\be'+\al(\vf-1))}{\be}\right)x.\hspace{4.4cm}(7.4.60)
\end{eqnarray*}
By (7.4.2) and (7.4.3),
\begin{eqnarray*}\eta&=&\frac{y^2}{2}\left(\frac{\be((\vf-1)\al)'
+\be{\be'}'-2{\be'}^2-((\vf-1)\al)^2-3\al\be'(\vf-1)}{\be^2}-(\vf-1)^2
\right)\\ & &-
\frac{x^2}{2}\left(\frac{2{\al'}^2+(\vf\be)^2+3\al'\be\vf-\al(\vf\be)'
-\al{\al'}'}{\al^2}+\vf^2\right)\\ & &+ [
\frac{(\vf-1)(\al'+\vf\be)}{\al}-\vf'
+\frac{\vf(\be'+\al(\vf-1))}{\be}]xy \\ &
&+\frac{\al\be}{\al^2+\be^2}
\left(1-\frac{2(\al\be'-\al'\be)}{\al^2+b^2}\right)e^{\int(\al\be^{-1}(\vf-1)+\al^{-1}\be\vf)dt}
\Im(\varpi).\hspace{3.1cm}(7.4.61)
\end{eqnarray*}
\pse

{\bf Theorem 7.4.4}. {\it Let $\al,\be,\vf$ be functions in $t$ and
let $\Im$ be a one-variable function. Suppose that $\zeta$ is a
function in $t,x,y$. The following is a solution of the equations
(7.1.1)-(7.1.4) of  dynamic convection in a sea:
$$u=(\vf-1)y-\frac{(\al'+\vf\be)x}{\al}+
\frac{\al\be^2}{\al^2+\be^2}
e^{\int(\al\be^{-1}(\vf-1)+\al^{-1}\be\vf)dt} \Im'(\al x+\be
y),\eqno(7.4.62)$$
$$v=\vf x-\frac{(\be'+(\vf-1)\al)y}{\be}-
\frac{\al^2\be}{\al^2+\be^2}e^{\int(\al\be^{-1}(\vf-1)+\al^{-1}\be\vf)dt}
\Im'(\al x+\be y),\eqno(7.4.63)$$
$$w=\left(\frac{\al'+\vf\be}{\al}
+\frac{\be'+(\vf-1)\al}{\be}\right)z+\zeta,\qquad\rho=1,\eqno(7.4.64)$$
\begin{eqnarray*}& &p=z+\frac{y^2}{2}\left(\frac{\be((\vf-1)\al)'
+\be{\be'}'-2{\be'}^2-((\vf-1)\al)^2-3\al\be'(\vf-1)}{\be^2}-(\vf-1)^2
\right)\\ & &-
\frac{x^2}{2}\left(\frac{2{\al'}^2+(\vf\be)^2+3\al'\be\vf-\al(\vf\be)'
-\al{\al'}'}{\al^2}+\vf^2\right)\\ & &+xy [
\frac{(\vf-1)(\al'+\vf\be)}{\al}
-\vf'+\frac{\vf(\be'+\al(\vf-1))}{\be}]
\\ & &+\frac{\al\be}{\al^2+\be^2}
\left(1-\frac{2(\al\be'-\al'\be)}{\al^2+b^2}\right)e^{\int(\al\be^{-1}(\vf-1)+\al^{-1}\be\vf)dt}
\Im(\al x+\be y).\hspace{2.4cm}(7.4.65)
\end{eqnarray*}}

\chapter{ Boussinesq Equations in Geophysics}

 Boussinesq systems of nonlinear partial
differential equations are fundamental equations in geophysical
fluid dynamics. In this chapter, we  use asymmetric ideas  and
moving frames  to solve the two-dimensional Boussinesq equations
with partial viscosity terms and the three-dimensional stratified
rotating Boussinesq equations. We obtain new families of explicit
exact solutions with multiple parameter functions. Many of them are
the periodic, quasi-periodic, aperiodic solutions that may have
practical significance. By Fourier expansion and some of our
solutions, one can obtain discontinuous solutions. In addition, the
symmetries of these equations are used to simplify our arguments.

In Section 8.1, we solve the two-dimensional Boussinesq equations
and obtain four families of explicit exact solutions. In Section
8.2, we give the symmetry analysis on the three-dimensional
stratified rotating Boussinesq equations. In Section 8.3, we find
the solutions of the three-dimensional equations that are linear in
$x$ and $y$. In Section 8.4, we obtain two families of explicit
exact solutions under certain conditions on the variable $z$. In
Section 8.5, we obtain  a family of explicit exact solutions of the
three-dimensional equations that are independent of $x$. The status
can be changed by applying symmetry transformations. This chapter is
a revision of our preprint [X16].

\section{Two-Dimensional Equations}

The  Boussinesq system for the incompressible fluid in $\mbb{R}^2$
is
$$u_t+uu_x+vu_y-\nu\Dlt u=-p_x,\qquad v_t+uv_x+vv_y-\nu\Dlt
v-\sta=-p_y,\eqno(8.1.1)$$\index{Boussinesq
equations!two-dimensional}
$$\sta_t+u\sta_x+v\sta_y-\kappa \Dlt\sta=0,\qquad
u_x+v_y=0,\eqno(8.1.2)$$ where $(u,v)$ is the velocity vector field,
$p$ is the scalar pressure, $\sta$ is the scalar temperature,
$\nu\geq 0$ is the viscosity and $\kappa\geq 0$ is the thermal
diffusivity. The above system is a simple model in atmospheric
sciences (e.g., cf. [Ma]). Chae [Cd] proved the global regularity,
and Hou and Li [HL] obtained the well-posedness of the above system.

Let us do the degree analysis. Note that $\Dlt=\ptl_x^2+\ptl_y^2$ in
this case. To make the nonzero terms to have the same degree, we
have to take
$$\deg x=\deg y=\ell\;\;\mbox{and}\;\;\deg uu_x=\deg u_{xx}\lra \deg
u=-\ell,\eqno(8.1.3)$$
$$\deg vv_y=\deg v_{yy}\lra \deg v=-\ell,\;\;\deg u_t=\deg
u_{xx}\lra\deg t=2\ell,\eqno(8.1.4)$$
$$\deg p_x=\deg uu_x\lra\deg p=-2\ell,\;\;\deg\sta=\deg
v_t=-3\ell.\eqno(8.1.5)$$ Moreover, (8.1.1) and (8.1.2) are
translation invariant because they do not contain variable
coefficients. Thus the transformation
$$T_{a,b}(u(t,x,y))=bu(b^2t+a,bx,by),\;\;T_{a,b}(v(t,x,y))=bv(b^2t+a,bx,by),\eqno(8.1.6)$$
$$T_{a,b}(p(t,x,y))=b^2(b^2t+a,bx,by),\;\;T_{a,b}(\sta(t,x,y))=b^3\sta(b^2t+a,bx,by)\eqno(8.1.7)$$
is a symmetry of the equations (8.1.1) and (8.1.2), where
$a,b\in\mbb{R}$ with $b\neq 0$. By the arguments in
(7.1.20)-(7.1.24), we have the following symmetry of the equations
(8.1.1) and (8.1.2):
$$S_{\al,\be;\gm}(u(t,x,y))=u(t,x+\al,y+\be)-\al',\;\;S_{\al,\be;\gm}(\sta(t,x,y))=\sta(t,x+\al,y+\be),\eqno(8.1.8)$$
$$S_{\al,\be;\gm}(v(t,x,y))=v(t,x+\al,y+\be)-\be',\eqno(8.1.9)$$
$$S_{\al,\be;\gm}(p(t,x,y))=p(t,x+\al,y+\be)+{\al'}'x+{\be'}'y+\gm,\eqno(8.1.10)$$ where $\al,\be$ and $\gm$ are arbitrary functions in
$t$.

According to the second equation in (8.1.2), we take the potential
form:
$$u=\xi_y,\qquad v=-\xi_x\eqno(8.1.11)$$
for some functions $\xi$ in $t,x,y$. Then the two-dimensional
Boussinesq equations become
$$\xi_{yt}+\xi_y\xi_{xy}-\xi_x\xi_{yy}-\nu\Dlt \xi_y=-p_x,\qquad \xi_{xt}+\xi_y\xi_{xx}-\xi_x\xi_{xy}-\nu\Dlt
\xi_x+\sta=p_y,\eqno(8.1.12)$$
$$\sta_t+\xi_y\sta_x-\xi_x\sta_y-\kappa \Dlt\sta=0.\eqno(8.1.13)$$
By our assumption $p_{xy}=p_{yx}$, the compatible condition of the
equations in (8.1.12) is
$$(\Dlt \xi)_t+\xi_y(\Dlt \xi)_x-\xi_x(\Dlt
\xi)_y-\nu\Dlt^2\xi+\sta_x=0.\eqno(8.1.14)$$ Now we first solve the
system (8.1.13) and (8.1.14). To do this, we impose some asymmetric
conditions.

Firs we assume
$$\sta=\ves(t,y),\qquad\xi=\phi(t,y)+x\psi(t,y)\eqno(8.1.15)$$
for some functions $\ves,\phi$ and $\psi$ in $t,y$. Then (8.1.13)
becomes
$$\ves_t-\psi\ves_y-\kappa\ves_{yy}=0.\eqno(8.1.16)$$
Moreover, (8.1.14) becomes
$$\phi_{yyt}+x\psi_{yyt}+(\phi_y+x\psi_y)\psi_{yy}-\psi(\phi_{yyy}+x\psi_{yyy})-\nu(\phi_{yyyy}+x\psi_{yyyy})=0,
\eqno(8.1.17)$$ equivalently,
$$\phi_{yyt}+\phi_y\psi_{yy}-\psi\phi_{yyy}-\nu\phi_{yyyy}=0,
\eqno(8.1.18)$$
$$\psi_{yyt}+\psi_y\psi_{yy}-\psi\psi_{yyy}-\nu\psi_{yyyy}=0.
\eqno(8.1.19)$$ The above two equations are equivalent to:
$$\phi_{yt}+\phi_y\psi_y-\psi\phi_{yy}-\nu\phi_{yyy}=\al_1,
\eqno(8.1.20)$$
$$\psi_{yt}+\psi_y^2-\psi\psi_{yy}-\nu\psi_{yyy}=\al_2
\eqno(8.1.21)$$ for some functions $\al_1$ and $\al_2$ in $t$ to be
determined.

Observe that
$$\psi=6\nu y^{-1}\eqno(8.1.22)$$
is a solution of (8.1.21) with $\al_2=0$. In order to solve
(8.1.20), we assume
$$\phi=\sum_{m=1}^\infty\gm_my^m,\eqno(8.1.23)$$
 where $\gm_m$ are
functions in $t$ to be determined. Now (8.1.20) becomes
$$\sum_{m=1}^\infty[m\gm_m'-\nu(m+2)(m+3)(m+4)\gm_{m+2}]y^{m-1}-6\nu\gm_1y^{-2}
-18\nu\gm_2y^{-1}=\al_1,\eqno(8.1.24)$$ equivalently,
$$\gm_1=\gm_2=0,\qquad \al_1=-60\nu\gm_3,\eqno(8.1.25)$$
$$m\gm_m'-\nu(m+2)(m+3)(m+4)\gm_{m+2}=0,\qquad
m> 1.\eqno(8.1.26)$$ Thus
$$\gm_{2m+2}=\frac{2m\gm_{2m}'}{\nu(2m+2)(2m+3)(2m+4)}=0,\qquad
m\geq 1,\eqno(8.1.27)$$
$$\gm_{2m+3}=\frac{(2m+1)\gm_{2m+1}'}{\nu(2m+3)(2m+4)(2m+5)}=\frac{360\gm_3^{(m)}}{\nu^m(2m+3)(2m+5)!},\qquad
m\geq 1.\eqno(8.1.28)$$ For simplicity, we redenote $\al=\gm_3$.
Then
$$\phi=360\sum_{m=0}^\infty
\frac{\al^{(m)}y^{2m+3}}{\nu^m(2m+3)(2m+5)!}.\eqno(8.1.29)$$

To solve (8.1.16), we also assume
$$\ves=\sum_{n=0}^\infty\be_n y^n,\eqno(8.1.30)$$
where $\be_n$ are functions in $t$ to be determined. Then (8.1.16)
becomes
$$6\nu\be_1y^{-1}+\sum_{n=0}^\infty[\be_n'-(n+2)(6\nu+(n+1)\kappa)\be_{n+2}]y^n=0,\eqno(8.1.31)$$
that is, $\be_1=0$ and
$$\be_n'-(n+2)(6\nu+(n+1)\kappa)\be_{n+2}=0,\qquad n\geq 0.\eqno(8.1.32)$$
Hence
$$\sta=\be+\sum_{n=1}^\infty\frac{\be^{(n)}y^{2n}}{2^nn!\prod_{r=1}^n(6\nu+(2r-1)\kappa)},\eqno(8.1.33)$$
where $\be$ is an arbitrary function in $t$. Moreover, (8.1.11),
(8.1.20), (8.1.21) and (8.1.25) lead to
\begin{eqnarray*}\qquad& &u_t+uu_x+vu_y-\nu\Dlt
u\\
&=&\phi_{yt}+x\psi_{yt}+(\phi_y+x\psi_y)\psi_y-\psi(\phi_{yy}+x\psi_{yy})-\nu
(\phi_{yyy}+x\psi_{yyy})\\
&=&\phi_{yt}+\phi_y^2-\psi\phi_{yy}-\nu\phi_{yyy}+(\psi_{yt}+\psi_y\psi_y-\psi\psi_{yy}-\nu\psi_{yyy})x
\\&=&\al_2x+\al_1=-60\nu\al.
\hspace{9.1cm}(8.1.34)\end{eqnarray*} Furthermore, (8.1.22) and
(8.1.33) give
\begin{eqnarray*}\qquad& &v_t+uv_x+vv_y-\nu\Dlt(v)-\sta
=-\psi_t+\psi\psi_y+\nu\psi_{yy}-\sta
\\&=&-24\nu^2y^{-3}-\be-\sum_{n=1}^\infty\frac{\be^{(n)}y^{2n}}{2^nn!\prod_{r=1}^n(6\nu+(2r-1)\kappa)}
.\hspace{4cm}(8.1.35)
\end{eqnarray*}
By (8.1.15), (8.1.22) and (8.1.29),
$$\xi=6\nu xy^{-1}+360\sum_{m=0}^\infty
\frac{\al^{(m)}y^{2m+3}}{\nu^m(2m+3)(2m+5)!}.\eqno(8.1.36)$$

 According (8.1.1) and (8.1.11), we have:\psp

{\bf Theorem 8.1.1}. {\it  The following is a solution of the
two-dimensional Boussinesq equations (8.1.1)-(8.1.2):
$$u=360\sum_{m=0}^\infty
\frac{\al^{(m)}y^{2m+2}}{\nu^m(2n+5)!}-6\nu xy^{-2},\qquad v=-6\nu
y^{-1},\eqno(8.1.37)$$
$$p=60\nu\al x+12\nu^2 y^{-2}+\be y+
\sum_{n=1}^\infty\frac{\be^{(n)}y^{2n+1}}{2^nn!(2n+1)\prod_{r=1}^n(6\nu+(2r-1)\kappa)}\eqno(8.1.38)$$
and $\sta$ is given in (8.1.33), where $\al $ and $\be$ are
arbitrary functions in $t$. }\psp

{\bf Remark 8.1.2}. Let $\gm,\gm_1,\gm_2$ be arbitrary functions in
$t$. Applying the symmetry transformation $S_{\gm_1,\gm_2;\gm}$ in
(8.1.8)-(8.1.10) to the above solution, we get a more general
solution of the two-dimensional Boussinesq equations
(8.1.1)-(8.1.2):
$$u=360\sum_{m=0}^\infty
\frac{\al^{(m)}(y+\gm_2)^{2m+2}}{\nu^m(2n+5)!}-6\nu
(x+\gm_1)(y+\gm_2)^{-2}-\gm_1',\eqno(8.1.39)$$ $$ v=-6\nu
(y+\gm_2)^{-1}-\gm_2',\eqno(8.1.40)$$
$$\sta=\be+\sum_{n=1}^\infty\frac{\be^{(n)}(y+\gm_2)^{2n}}{2^nn!\prod_{r=1}^n(6\nu+(2r-1)\kappa)},\eqno(8.1.41)$$
\begin{eqnarray*}\qquad p&=&60\nu\al(x+\gm_1)+12\nu^2
(y+\gm_2)^{-2}+\be(y+\gm_2)+{\gm_1'}'x+{\gm_2'}'y+\gm\\ &&+
\sum_{n=1}^\infty\frac{\be^{(n)}(y+\gm_2)^{2n+1}}{2^nn!(2n+1)\prod_{r=1}^n(6\nu+(2r-1)\kappa)}.\hspace{5.1cm}(8.1.42)\end{eqnarray*}
\pse

Let $c$ be a fixed real constant and let $\gm$ be a fixed function
in $t$. We define
$$\zeta_1(y)=\frac{e^{\gm y}-ce^{-\gm y}}{2},\qquad \eta_1(y)=\frac{e^{\gm y}+ce^{-\gm
y}}{2},\eqno(8.1.43)$$
$$\zeta_0(y)=\sin\gm y,\qquad \eta_0(y)=\cos\gm y.\eqno(8.1.44)$$
Then
$$\eta_r^2(y)+(-1)^r\zeta_r^2(y)=c^r,\eqno(8.1.45)$$
$$\ptl_y(\zeta_r(y))=\gm\eta_r(y),\qquad
\ptl_y(\eta_r(y))=-(-1)^r\gm\zeta_r(y)\eqno(8.1.46)$$ and
$$\ptl_y(\zeta_r(y))=\gm'y\eta_r(y),\qquad
\ptl_t(\eta_r(y))=-(-1)^r\gm'y\zeta_r(y),\eqno(8.1.47)$$
 where we treat
$0^0=1$ when $c=r=0$.

 First we assume
$$\psi=\be_1y+\be_2\zeta_r(y)\eqno(8.1.48)$$
for some functions $\be_1$ and $\be_2$ in $t$, where $r=0,1$. Then
(8.1.21) becomes
\begin{eqnarray*}\qquad&
&\be_1'+(\be_2\gm)'\eta_r-(-1)^r\be_2\gm\gm'y\zeta_r+(\be_1+\be_2\gm\eta_r)^2\\
& &+(-1)^r\be_2\gm^2(\be_1y+\be_2\zeta_r)\zeta_r
+(-1)^r\nu\be_2\gm^3\eta_r\\&=&
\be_1'+c^r\be_2^2\gm^2+\be_1^2+[(\be_2\gm)'+(-1)^r\nu\be_2\gm^3+2\be_1\be_2\gm]
\eta_r
\\ & &+(-1)^r\be_2\gm(\be_1\gm-\gm')y\zeta_r=\al_2,\hspace{7.2cm}
(8.1.49)\end{eqnarray*} which is implied by the following equations:
$$\be_1'+c^r\be_2^2\gm^2+\be_1^2=\al_2,\qquad\be_1\gm-\gm'=0,\eqno(8.1.50)$$
$$(\be_2\gm)'+(-1)^r\nu\be_2\gm^3+2\be_1\be_2\gm=0.\eqno(8.1.51)$$
For convenience, we assume
$$\gm=\sqrt{\al'}\eqno(8.1.52)$$ for some increasing
function $\al$ in $t$. Thus we have
$$\be_1=\frac{\gm'}{\gm}=\frac{{\al'}'}{2\al'}\eqno(8.1.53)$$
by the second equation in (8.1.50). Now (8.1.51) becomes
$$(\be_2\gm)'+\left((-1)^r\nu\al'+\frac{{\al'}'}{\al'}\right)\be_2\gm=0.\eqno(8.1.54)$$
Hence
$$\be_2\gm=\frac{b_1e^{-(-1)^r\nu\al}}{\al'}\lra
 \be_2=\frac{b_1e^{-(-1)^r\nu\al}}{\sqrt{(\al')^3}},\qquad
b_1\in\mbb{R}.\eqno(8.1.55)$$

To solve (8.1.20), we assume
$$\phi=\be_3\eta_r(y)\eqno(8.1.56)$$
 for some function $\be_3$.
 Now (8.1.20) becomes
\begin{eqnarray*}&
&-(-1)^r[(\be_3\gm)'\zeta_r+\be_3\gm\gm'y\eta_r+\be_3\gm\zeta_r(\be_1+\be_2\gm\eta_r)-
\be_3\gm^2(\be_1y+\be_2\zeta)\eta_r]-\nu\be_3\gm^3\zeta
\\&=&-[(-1)^r((\be_3\gm)'+\be_1\be_3\gm)+\nu\be_3\gm^3]\zeta_r(y)=\al_1\hspace{5.7cm}
(8.1.57)\end{eqnarray*} by (8.1.46), (8.1.47) and the second
equation in (8.1.50), equivalently, $\al_1=0$ and
$$(-1)^r((\be_3\gm)'+\be_1\be_3\gm)+\nu\be_3\gm^3=0.\eqno(8.1.58)$$
According to (8.1.52) and (8.1.53),
$$(\be_3\gm)'+\left(\frac{{\al'}'}{2\al'}+(-1)^r\al'\right)\be_3\gm=0.\eqno(8.1.59)$$
Thus
$$\be_3\gm=\frac{b_2e^{-(-1)^r\nu\al}}{\sqrt{\al'}}\lra
\be_3=\frac{b_2e^{-(-1)^r\nu\al}}{\al'},\eqno(8.1.60)$$
 where $b_2$ is a real constant.

In order to solve (8.1.16), we assume
$$\ves=be^{\gm_1\eta_r(y)},\eqno(8.1.61)$$ where $b$ is a real constant and $\gm_1$ is a function in $t$.
Then (8.1.16) changes to
$$b(\gm_1'\eta_r-(-1)^r\gm_1\gm'y\zeta_r)+(-1)^rb\gm_1\gm(\be_1y+\be_2\zeta_r)\zeta_r
-b\kappa\gm_1\gm^2(-(-1)^r\eta_r+\gm_1\zeta_r^2)=0,\eqno(8.1.62)$$
which is implied by
$$\gm_1'+(-1)^r\kappa\gm^2\gm_1=0,\qquad(-1)^r\be_2-\kappa\gm\gm_1=0.\eqno(8.1.63)$$
Then the first  equation  and (8.1.52) imply
$$\gm_1=b_3e^{-(-1)^r\kappa\al}\eqno(8.1.64)$$
for some constant $b_3$. By the second equations in (8.1.63) and
(8.1.55), we have:
$$(-1)^r\frac{b_1e^{-(-1)^r\nu\al}}{\sqrt{(\al')^3}}=b_3\kappa\sqrt{\al'}e^{-(-1)^r\kappa\al}.\eqno(8.1.65)$$
For convenience, we take
$$b_1=(-1)^r\kappa b_3.\eqno(8.1.66)$$
Then (8.1.65) is implied by
$$\al'e^{(-1)^r(\nu-\kappa)\al/2}=1.\eqno(8.1.67)$$
If $\nu=\kappa$, (8.1.65) is implied by $\al=t$.
 When $\nu\neq
\kappa$, (8.1.65) becomes
$$\left(\frac{2e^{(-1)^r(\nu-\kappa)\al/2}}{\nu-\kappa}\right)'=(-1)^r.\eqno(8.1.68)$$
Thus
$$\al=\frac{2(-1)^r}{\nu-\kappa}\ln[(-1)^r(\nu-\kappa)t/2+c_0],\qquad c_0\in\mbb{R}.\eqno(8.1.69)$$

Suppose $\nu=\kappa$. Then $\gm=\sqrt{\al'}=1$ and $\be_1=0$. By
(8.1.48), (8.1.55), (8.1.56) and (8.1.60),
$$\phi=b_2e^{-(-1)^r\nu t}\eta_r(y),\qquad\psi=(-1)^rb_3\nu e^{-(-1)^r\nu
t}\zeta_r(y)\eqno(8.1.70)$$ Moreover, (8.1.15), (8.1.61) and
(8.1.64) yield
$$\sta=b\exp(b_3e^{-(-1)^r\nu t}\eta_r(y)).\eqno(8.1.71)$$
Furthermore, (8.1.15) and (8.1.66) give
$$\xi=b_2e^{-(-1)^r\nu
t}\eta_r(y)+(-1)^rb_3\nu e^{-(-1)^r\nu t}x\zeta_r(y).\eqno(8.1.72)$$
 According to (8.1.11),
$$u=\xi_y=(-1)^r[-b_2e^{-(-1)^r\nu t}\zeta_r(y)+b_3\nu e^{-(-1)^r\nu
t}x\eta_r(y)],\eqno(8.1.73)$$ $$ v=-\xi_x=-(-1)^rb_3\nu
e^{-(-1)^r\nu t}\zeta_r(y).\eqno(8.1.74)$$ Note
$$u_t+uu_x+vu_y-\nu\Dlt u= b_3^2\nu^2c^r e^{-(-1)^r2\nu
t}x,\eqno(8.1.75)$$
$$v_t+uv_x+vv_y-\nu\Dlt v-\sta
=vv_y-b\exp(b_3e^{-(-1)^r\nu t}\eta_r(y)).\eqno(8.1.76)$$ By
(8.1.1), we have
$$p=
b\int\exp(b_3e^{-(-1)^r\nu t}\eta_r(y))dy-\frac{1}{2}b_3^2\nu^2
e^{-(-1)^r2\nu t}(c^rx^2+\zeta_r^2(y)).\eqno(8.1.77)$$\pse

{\bf Theorem 8.1.3}. {\it Suppose $\kappa=\nu$. For
$b,b_2,b_3,c\in\mbb{R}$, we have the following solutions of  the
two-dimensional Boussinesq equations (8.1.1)-(8.1.2): (1)
$$u=\frac{e^{\nu t}}{2}[b_2(e^y-ce^{-y})-b_3\nu x(e^y+ce^{-y})],\eqno(8.1.78)$$
$$ v=\frac{1}{2}b_3\nu
e^{\nu t}(e^y-ce^{-y}),\eqno(8.1.79)$$
$$\sta=b\exp(b_3e^{\nu t}(e^y+ce^{-y})/2)\eqno(8.1.80)$$
and
$$p=
b\int\exp(b_3e^{\nu t}(e^y+ce^{-y})/2)dy-\frac{1}{2}b_3^2\nu^2
e^{2\nu t}(cx^2+(e^y-ce^{-y})^2/4);\eqno(8.1.81)$$ (2)
$$u=e^{-\nu t}[-b_2\sin y+b_3\nu x\cos y],\;\; v=-b_3\nu
e^{-\nu t}\sin y,\eqno(8.1.82)$$
$$\sta=b\exp(b_3e^{-\nu t}\cos y)\eqno(8.1.83)$$
and
$$p=
b\int\exp(b_3e^{-\nu t}\cos y)dy-\frac{1}{2}b_3^2\nu^2 e^{-2\nu
t}(x^2+\cos^2y).\eqno(8.1.84)$$
 }\psp

Applying the symmetry transformations in (8.1.6)-(8.1.10) to the
above solutions, we can get more general solutions the
two-dimensional Boussinesq equations (8.1.1)-(8.1.2).

Consider the case $\nu\neq \kappa$. Then
$$\gm=\sqrt{\al'}=\frac{1}{\sqrt{(-1)^r(\nu-\kappa)t/2+c_0}}\eqno(8.1.85)$$
by (8.1.69). Moreover,
$$\be_1=\frac{\gm'}{\gm}=\frac{(-1)^r(\kappa-\nu)}{4[(-1)^r(\nu-\kappa)t/2+c_0]}
\eqno(8.1.86)$$ by (8.1.53),
 $$\be_2=\frac{b_1e^{-(-1)^r\nu\al}}{\sqrt{(\al')^3}}=(-1)^rb_3\kappa
[(-1)^r(\nu-\kappa)t/2+c_0]^{2\nu/(\kappa-\nu)+3/2}\eqno(8.1.87)$$
according to (8.1.55), (8.1.66) and (8.1.69),
$$\be_3=\frac{b_2e^{-(-1)^r\nu\al}}{\al'}=b_2[(-1)^r(\nu-\kappa)t/2+c_0]^{2\nu/(\kappa-\nu)+1}
\eqno(8.1.88)$$ by (8.1.60), and
$$\gm_1=b_3e^{-(-1)^r\kappa\al}=b_3[(-1)^r(\nu-\kappa)t/2+c_0]^{2\kappa/(\kappa-\nu)}
\eqno(8.1.89)$$ by (8.1.64). Thus (8.1.56) and (8.1.88) yield
$$\phi=b_2[(-1)^r(\nu-\kappa)t/2+c_0]^{2\nu/(\kappa-\nu)+1}\eta_r(y).\eqno(8.1.90)$$
 Furthermore,
$$\psi=\frac{(-1)^r(\kappa-\nu)y}{4[(-1)^r(\nu-\kappa)t/2+c_0]}+
(-1)^rb_3\kappa
[(-1)^r(\nu-\kappa)t/2+c_0]^{2\nu/(\kappa-\nu)+3/2}\zeta_r(y)\eqno(8.1.91)$$
by (8.1.48), (8.1.86) and (8.1.87).

According to (8.1.15), (8.1.61) and (8.1.89),
$$\sta=b\exp\big(b_3[(-1)^r(\nu-\kappa)t/2+c_0]^{2\kappa/(\kappa-\nu)}\eta_r(y)\big).\eqno(8.1.92)$$
By (8.1.15),
\begin{eqnarray*}\hspace{1cm}\xi&=&\frac{(-1)^r(\kappa-\nu)xy}{4[(-1)^r(\nu-\kappa)t/2+c_0]}+
(-1)^rb_3\kappa
[(-1)^r(\nu-\kappa)t/2+c_0]^{2\nu/(\kappa-\nu)+3/2}x\zeta_r(y)\\
&&+b_2[(-1)^r(\nu-\kappa)t/2+c_0]^{2\nu/(\kappa-\nu)+1}\eta_r(y).\hspace{4.9cm}(8.1.93)\end{eqnarray*}
Then (8.1.11) and (8.1.93) say that
\begin{eqnarray*}u&=&\frac{(-1)^r(\kappa-\nu)x}{4[(-1)^r(\nu-\kappa)t/2+c_0]}+
(-1)^rb_3\kappa
[(-1)^r(\nu-\kappa)t/2+c_0]^{2\nu/(\kappa-\nu)+1}x\eta_r(y)\\ &&
-(-1)^rb_2[(-1)^r(\nu-\kappa)t/2+c_0]^{2\nu/(\kappa-\nu)+1/2}\zeta_r(y),
\hspace{4.6cm}(8.1.94)\end{eqnarray*}
$$v=\frac{(-1)^r(\nu-\kappa)y}{4[(-1)^r(\nu-\kappa)t/2+c_0]}-
(-1)^rb_3\kappa
[(-1)^r(\nu-\kappa)t/2+c_0]^{2\nu/(\kappa-\nu)+3/2}\zeta_r(y),\eqno(8.1.95)$$

By (8.1.20) with $\al_1=0$ and (8.1.21) with $\al_2$ given in
(8.1.49), we have
\begin{eqnarray*}\qquad& &u_t+uu_x+vu_y-\nu\Dlt
u\\
&=&\phi_{yt}+x\psi_{yt}+(\phi_y+x\psi_y)\psi_y-\psi(\phi_{yy}+x\psi_{yy})-\nu
(\phi_{yyy}+x\psi_{yyy})\\
&=&\phi_{yt}+\phi_y^2-\psi\phi_{yy}-\nu\phi_{yyy}+(\psi_{yt}+\psi_y\psi_y-\psi\psi_{yy}-\nu\psi_{yyy})x
\\&=&(\be_1'+c^r\be_2^2\gm^2+\be_1^2)x=
b_3^2c^r\kappa^2 [(-1)^r(\nu-\kappa)t/2+c_0]^{4\nu/(\kappa-\nu)+2}x
\\ & &+\frac{3(\nu-\kappa)^2x}{16[(-1)^r(\nu-\kappa)t/2+c_0]^2}.
\hspace{7.8cm}(8.1.96)\end{eqnarray*} Moreover, (8.1.48), the second
equation in (8.1.50) and (8.1.85)-(8.1.87) yield
\begin{eqnarray*}& &v_t+uv_x+vv_y-\nu\Dlt(v)-\sta
=-\psi_t+\psi\psi_y+\nu\psi_{yy}-\sta
\\&=&-(\be_1'y+\be_2'\zeta_r+\be_2\gm'y\eta_r)+(\be_1y+\be_2\zeta_r)(\be_1+\be_2\gm\eta_r)
-(-1)^r\nu\be_2\gm^2\zeta_r-\sta
\\&=&(\be_1^2-\be_1')y+(\be_1\be_2-\be_2'-(-1)^r\nu\be_2\gm^2)\zeta_r+\be_2(\be_1\gm-\gm')y\eta_r
+\frac{\be_2^2}{2}\ptl_y(\zeta_r^2)-\sta
\\&=&\frac{3(\nu-\kappa)^2y}{16[(-1)^r(\nu-\kappa)t/2+c_0]^2}
-be^{b_3[(-1)^r(\nu-\kappa)t/2+c_0]^{2\kappa/(\kappa-\nu)}\eta_r(y)}
\\ & &+b_3\kappa(\kappa-\nu)
[(-1)^r(\nu-\kappa)t/2+c_0]^{2\nu/(\kappa-\nu)+1/2}\zeta_r(y)
\\ & &+
\frac{b_3^2}{2}\kappa^2
[(-1)^r(\nu-\kappa)t/2+c_0]^{4\nu/(\kappa-\nu)+3}\ptl_y\zeta_r^2(y).\hspace{5.4cm}(8.1.97)
\end{eqnarray*}
According to (8.1.11), we have
\begin{eqnarray*}p&=&b\int
e^{b_3[(-1)^r(\nu-\kappa)t/2+c_0]^{2\kappa/(\kappa-\nu)}\eta_r(y)}dy
-\frac{b_3^2}{2}c^r\kappa^2
[(-1)^r(\nu-\kappa)t/2+c_0]^{4\nu/(\kappa-\nu)+2}x^2
\\ & &-\frac{3(\nu-\kappa)^2(x^2+y^2)}{32[(-1)^r(\nu-\kappa)t/2+c_0]^2}
-\frac{b_3^2}{2}\kappa^2
[(-1)^r(\nu-\kappa)t/2+c_0]^{4\nu/(\kappa-\nu)+3}\zeta_r^2(y)
\\
& &+(-1)^rb_3\kappa(\kappa-\nu)
[(-1)^r(\nu-\kappa)t/2+c_0]^{2\nu/(\kappa-\nu)+1}\eta_r(y).\hspace{3.6cm}(8.1.98)\end{eqnarray*}
\pse

{\bf Theorem 8.1.4}. {\it Suppose $\kappa\neq\nu$. For
$b,b_2,b_3,c,c_0\in\mbb{R}$, we have the following solutions of  the
two-dimensional Boussinesq equations (8.1.1)-(8.1.2): (1)
\begin{eqnarray*}u&=&
-\frac{b_3}{2}\kappa [(\kappa-\nu)t/2+c_0]^{2\nu/(\kappa-\nu)+1}x
(e^{y/\sqrt{(\kappa-\nu)t/2+c_0}}+ce^{y/\sqrt{(\kappa-\nu)t/2+c_0}})
\\
&&+\frac{b_2}{2}[(\kappa-\nu)t/2+c_0]^{2\nu/(\kappa-\nu)+1/2}(e^{y/\sqrt{(\kappa-\nu)t/2+c_0}}-ce^{y/
\sqrt{(\kappa-\nu)t/2+c_0}})
\\
& &+\frac{(\nu-\kappa)x}{4[(\kappa-\nu)t/2+c_0]},
\hspace{9.6cm}(8.1.99)\end{eqnarray*}
\begin{eqnarray*}v&=&
\frac{b_3}{2}\kappa [(\kappa-\nu)t/2+c_0]^{2\nu/(\kappa-\nu)+3/2}x
(e^{y/\sqrt{(\kappa-\nu)t/2+c_0}}-ce^{y/\sqrt{(\kappa-\nu)t/2+c_0}})
\\& &+\frac{(\kappa-\nu)y}{4[(\kappa-\nu)t/2+c_0]},
\hspace{9.4cm}(8.1.100)\end{eqnarray*}
$$\sta=b\exp\big(2^{-1}b_3[(\kappa-\nu)t/2+c_0]^{2\kappa/(\kappa-\nu)}
(e^{y/\sqrt{(\kappa-\nu)t/2+c_0}}+ce^{y/\sqrt{(\kappa-\nu)t/2+c_0}})
\big)\eqno(8.1.101)$$ and
\begin{eqnarray*}p&=&b\int
\exp\big(2^{-1}b_3[(\kappa-\nu)t/2+c_0]^{2\kappa/(\kappa-\nu)}
(e^{y/\sqrt{(\kappa-\nu)t/2+c_0}}+ce^{y/\sqrt{(\kappa-\nu)t/2+c_0}})
\big)dy
\\ & &-\frac{b_3^2}{8}\kappa^2
[(\kappa-\nu)t/2+c_0]^{4\nu/(\kappa-\nu)+3}(e^{y/\sqrt{(\kappa-\nu)t/2+c_0}}-ce^{y/\sqrt{(\kappa-\nu)t/2+c_0}})
\\
& &-\frac{b_3}{2}\kappa(\kappa-\nu)
[(\kappa-\nu)t/2+c_0]^{2\nu/(\kappa-\nu)+1}(e^{y/\sqrt{(\kappa-\nu)t/2+c_0}}+ce^{y/\sqrt{(\kappa-\nu)t/2+c_0}})
\\ & &-\frac{b_3^2}{2}c\kappa^2
[(\kappa-\nu)t/2+c_0]^{4\nu/(\kappa-\nu)+2}x^2
-\frac{3(\nu-\kappa)^2(x^2+y^2)}{32[(\kappa-\nu)t/2+c_0]^2};\hspace{2.4cm}(8.1.102)\end{eqnarray*}
(2)
\begin{eqnarray*}u&=&
b_3\kappa [(\nu-\kappa)t/2+c_0]^{2\nu/(\kappa-\nu)+1}x\cos
\frac{y}{\sqrt{(\nu-\kappa)t/2+c_0}}+\frac{(\kappa-\nu)x}{4[(\nu-\kappa)t/2+c_0]}\\
& & -b_2[(\nu-\kappa)t/2+c_0]^{2\nu/(\kappa-\nu)+1/2}\sin
\frac{y}{\sqrt{(\nu-\kappa)t/2+c_0}},
\hspace{3.4cm}(8.1.103)\end{eqnarray*}
$$v=\frac{(\nu-\kappa)y}{4[(\nu-\kappa)t/2+c_0]}-
b_3\kappa [(\nu-\kappa)t/2+c_0]^{2\nu/(\kappa-\nu)+3/2}\sin
\frac{y}{\sqrt{(\nu-\kappa)t/2+c_0}},\eqno(8.1.104)$$
$$\sta=b\exp\big(b_3[(\nu-\kappa)t/2+c_0]^{2\kappa/(\kappa-\nu)}\cos
\frac{y}{\sqrt{(\nu-\kappa)t/2+c_0}}\big),\eqno(8.1.105)$$
\begin{eqnarray*}p&=&b\int
\exp\big(b_3[(\nu-\kappa)t/2+c_0]^{2\kappa/(\kappa-\nu)}\cos
\frac{y}{\sqrt{(\nu-\kappa)t/2+c_0}}\big)dy
\\ & &-\frac{b_3^2}{2}\kappa^2
[(\nu-\kappa)t/2+c_0]^{4\nu/(\kappa-\nu)+2}x^2
-\frac{3(\nu-\kappa)^2(x^2+y^2)}{32[(\nu-\kappa)t/2+c_0]^2}
\\ & &-\frac{b_3^2}{2}\kappa^2
[(\nu-\kappa)t/2+c_0]^{4\nu/(\kappa-\nu)+3}\sin^2
\frac{y}{\sqrt{(\nu-\kappa)t/2+c_0}}
\\
& &+b_3\kappa(\kappa-\nu)
[(\nu-\kappa)t/2+c_0]^{2\nu/(\kappa-\nu)+1}\cos
\frac{y}{\sqrt{(\nu-\kappa)t/2+c_0}}.\hspace{2.2cm}(8.1.106)\end{eqnarray*}
 }\pse

Applying the symmetry transformations in (8.1.6)-(8.1.10) to the
above solutions, we can get more general solutions the
two-dimensional Boussinesq equations (8.1.1)-(8.1.2).

 Let $\gm$ be a function in $t$.
Denote the  {\it moving frame} \index{moving frame!for
two-dimensional Boussinesq equations}
$$\X=x\cos\gm+y\sin\gm,\qquad \Y=y\cos\gm-x\sin\gm.\eqno(8.1.107)$$
Then
$$\ptl_t(\X)=\gm'\Y,\qquad
\ptl_t(\Y)=-\gm'\X.\eqno(8.1.108)$$ By the chain rule of taking
partial derivatives,
$$\ptl_x=\cos\gm\;\ptl_\X-\sin\gm\;\ptl_\Y,\qquad
\ptl_y=\sin\gm\;\ptl_\X+\cos\gm\;\ptl_\Y.\eqno(8.1.109)$$
 Solving the above system, we get
$$\ptl_{\X}=\cos\gm\:\ptl_x+\sin\gm\:\ptl_y,\qquad
\ptl_{\Y}=-\sin\gm\:\ptl_x+\cos\gm\:\ptl_y.\eqno(8.1.110)$$
Moreover, (8.1.107) and (8.1.110) imply
$$\ptl_\X(\Y)=0,\qquad\ptl_\Y(\X)=0.\eqno(8.1.111)$$
 In
particular,
$$\Dlt=\ptl_x^2+\ptl_y^2=\ptl_{\X}^2+\ptl_{\Y}^2,\;\;x^2+y^2=\X^2+\Y^2.\eqno(8.1.112)$$

We assume
$$\xi=\phi(t,\X)-\frac{\gm'}{2}(x^2+y^2)
,\qquad\sta=\psi(t,\X),\eqno(8.1.113)$$ where $\phi$ and $\psi$ are
functions in $t,\X$. Note
$$\xi_y\ptl_x-\xi_x\ptl_y=(\X-\phi_\X)\ptl_\Y-\gm'\Y\ptl_\X.\eqno(8.1.114)$$
Then (8.1.13) becomes
$$\psi_t-\kappa\psi_{\X\X}=0\eqno(8.1.115)$$  and (8.1.14) becomes
$$-2{\gm'}'+\phi_{t\X\X}
-\nu\phi_{\X\X\X\X}+\psi_{\X}\cos\gm =0\eqno(8.1.116)$$ by (8.1.111)
and (8.1.114). Modulo the transformation in (8.1.8)-(8.1.11), the
above equation is equivalent to $$-2{\gm'}'\X+\phi_{t\X}
-\nu\phi_{\X\X\X}+\psi\cos\gm =0.\eqno(8.1.117)$$

Note that (8.1.115) is a heat conduction equation. Assume
$\nu=\kappa$. We take its solution
$$\psi=\sum_{r=1}^ma_rd_re^{a_r^2\kappa t\cos2b_rt+a_r\X\cos
b_r}\sin(a_r^2\kappa t\sin2b_r +a_r\X\sin
b_r+b_r+c_r),\eqno(8.1.118)$$ where $a_r,b_r,c_r,d_r\in\mbb{R}$ with
$(a_r,b_r)\neq (0,0)$ and $d_r\neq 0$. Then
$$\psi=\ptl_\X[\sum_{r=1}^md_re^{a_r^2\kappa t\cos2b_rt+a_r\X\cos
b_r}\sin(a_r^2\kappa t\sin2b_r +a_r\X\sin b_r+c_r)].\eqno(8.1.119)$$
 Moreover,  (8.1.117) is implied by the following
equation:
\begin{eqnarray*}\hspace{1.5cm}& &2\nu\gm'-{\gm'}'\X^2+\phi_t
-\nu\phi_{\X\X}+[\sum_{r=1}^md_re^{a_r^2\kappa t\cos 2b_r+a_r\X\cos
b_r}\\ & &\times\sin(a_r^2\kappa t\sin 2b_r+a_r\X\sin
b_r+c_r)]\cos\gm=0\hspace{4.1cm}(8.1.120)\end{eqnarray*} by
(8.1.119). Thus we have the following solution of (8.1.117):
\begin{eqnarray*}\phi&=&-[\sum_{r=1}^rd_re^{a_r^2\kappa
t\cos 2b_r+a_r\X\cos b_r}\sin(a_r^2\kappa t\sin 2b_r+a_r\X\sin
b_r+c_r)]\int \cos\gm\:dt\\ & &+\gm'\X^2+\sum_{s=1}^n\hat d_se^{\hat
a_s^2\kappa t\cos 2\hat b_s+\hat a_s\X\cos \hat b_s}\sin(\hat
a_s^2\kappa t\sin 2\hat b_s+\hat a_s\X\sin \hat b_s+\hat
c_s),\hspace{0.9cm}(8.1.121)\end{eqnarray*} where $\hat a_s,\hat
b_s,\hat c_s,\hat d_s$ are real numbers.

Suppose $\nu\neq \kappa$. To make (8.1.117) solvable, we choose the
following solution of (8.1.115):
$$\psi=\sum_{r=1}^m a_rd_re^{a_r^2\kappa t+a_r\X}.\eqno(8.1.122)$$
Now  (8.1.117) is implied by the following equation:
$$2\nu\gm'-{\gm'}'\X^2+\phi_t
-\nu\phi_{\X\X}+\sum_{r=1}^md_re^{a_r^2\kappa
t+a_r\X}\cos\gm=0.\eqno(8.1.123)$$  We obtain the following solution
of (8.1.117):
\begin{eqnarray*}\hspace{1cm}\phi&=&\gm'\X^2+\sum_{s=1}^n\hat d_se^{\hat
a_s^2\kappa t\cos 2\hat b_s+\hat a_s\X\cos \hat b_s}\sin(\hat
a_s^2\kappa t\sin 2\hat b_s+\hat a_s\X\sin \hat b_s+\hat c_s)\\
& &-\sum_{r=1}^md_re^{a_r^2\nu t+a_r\X}\int
e^{a_r^2(\kappa-\nu)t}\cos\gm\:dt.\hspace{5.6cm}(8.1.124)\end{eqnarray*}

Note
$$u=\phi_\X\sin\gm-\gm'y,\qquad v=\gm'x
-\phi_\X\cos\gm.\eqno(8.1.125)$$ Moreover,
$$u\ptl_x+v\ptl_y=-\phi_\X\ptl_\Y+\gm'(x\ptl_y-y\ptl_x).\eqno(8.1.126)$$
By (8.1.117) and (8.1.126), we find
\begin{eqnarray*} & & u_t+uu_x+vu_y-\nu\Dlt u\\ &=&
\gm'\phi_\X\cos\gm+\phi_{\X
t}\sin\gm+\gm'\Y\phi_{\X\X}\sin\gm-{\gm'}'y+\gm'\phi_\X\ptl_\Y(y)\\
&
&+\gm'(x\ptl_y-y\ptl_x)(\phi_\X)\sin\gm-{\gm'}^2x-\nu\phi_{\X\X\X}\sin\gm
\\&=&(\phi_{\X
t}-\nu\phi_{\X\X\X})\sin\gm+2\gm'\phi_\X\cos\gm-\gm'^2x-{\gm'}'y
\\
&=&(2{\gm'}'\X-\psi\cos\gm)\sin\gm+2\gm'\phi_\X\cos\gm-\gm'^2x-{\gm'}'y,
\\ &=&{\gm'}'(x\sin 2\gm-y\cos 2\gm)
+(2\gm'\phi_\X-\psi\sin\gm) \cos\gm-\gm'^2x, \hspace{3.5cm}(8.1.127)
\end{eqnarray*}
\begin{eqnarray*} \hspace{1cm}& &v_t+uv_x+vv_y-\nu\Dlt
v-\sta\\ &=& \gm'\phi_\X\sin\gm-\phi_{\X
t}\cos\gm-\gm'\Y\phi_{\X\X}\cos\gm+{\gm'}'x-\gm'\phi_\X\ptl_\Y(x)\\
&
&-\gm'(x\ptl_y-y\ptl_x)(\phi_\X)\cos\gm-{\gm'}^2y+\nu\phi_{\X\X\X}\cos\gm
-\psi
\\ &=&(\nu\phi_{\X\X\X}-\phi_{\X
t})\cos\gm+2\gm'\phi_\X\sin\gm-\gm'^2y+{\gm'}'x -\psi\\
&=&(\psi\cos\gm-2{\gm'}'\X)\cos\gm+2\gm'\phi_\X\sin\gm-\gm'^2y+{\gm'}'x
-\psi\\ &=&-{\gm'}'(x\cos 2\gm+y\sin
2\gm)+(2\gm'\phi_\X-\psi\sin\gm)\sin\gm-\gm'^2y.\hspace{2.3cm}(8.1.128)
\end{eqnarray*}
 According to (8.1.1),
$$p=\frac{{\gm'}^2-{\gm'}'\sin 2\gm}{2}x^2+\frac{{\gm'}^2+{\gm'}'\sin
2\gm}{2}y^2+{\gm'}'xy\cos2\gm+\int\psi d\X\:\sin\gm-2\gm'\phi.
\eqno(8.1.129)$$\psp

{\bf Theorem 8.1.5}. {\it Let $\gm$ be any function in $t$ and
denote $\X=x\cos\gm+y\sin\gm$. Take
$$\{a_r,b_r,c_r,d_r,\hat a_s,\hat
b_s,\hat c_s,\hat d_s\mid
r=1,...,m;s=1,...,n\}\subset\mbb{R}.\eqno(8.1.130)$$ If
$\nu=\kappa$, we have the following solutions of the two-dimensional
Boussinesq equations (8.1.1)-(8.1.2):
\begin{eqnarray*}u&=&\{\sum_{s=1}^n\hat a_s\hat d_se^{\hat a_s^2\kappa
t\cos 2\hat b_s+\hat a_s\X\cos \hat b_s}\sin(\hat a_s^2\kappa
t\sin 2\hat b_s+\hat a_s\X\sin \hat b_s+\hat b_s+\hat c_s)\\
& &-[\sum_{r=1}^m a_rd_re^{a_r^2\kappa t\cos 2b_r+a_r\X\cos
b_r}\sin(a_r^2\kappa t\sin 2b_r+b_r+a_r\X\sin b_r+c_r)]\\ &
&\times\int \cos\gm\:dt+2\gm'\X\}\sin\gm-\gm'
y,\hspace{7.2cm}(8.1.131)\end{eqnarray*}
\begin{eqnarray*}v&=&-\{\sum_{s=1}^n\hat a_s\hat d_se^{\hat a_s^2\kappa
t\cos 2\hat b_s+\hat a_s\X\cos \hat b_s}\sin(\hat a_s^2\kappa t\sin
2\hat b_s+\hat a_s\X\sin \hat b_s+\hat b_s+\hat c_s)\\
& &-[\sum_{r=1}^m a_rd_re^{a_r^2\kappa t\cos 2b_r+a_r\X\cos
b_r}\sin(a_r^2\kappa t\sin 2b_r+a_r\X\sin b_r+b_r+c_r)]\\ &
&\times\int
\cos\gm\:dt+2\gm'\X\}\cos\gm+\gm'x,\hspace{7.1cm}(8.1.132)\end{eqnarray*}
$\sta=\psi$ in (8.1.118), and
\begin{eqnarray*}
p&=&(\sin\gm+2\gm'\int\cos\gm)[\sum_{r=1}^md_re^{a_r^2\kappa t\cos
2b_r+a_r\X\cos b_r}\sin(a_r^2\kappa t\sin 2b_r+a_r\X\sin b_r+c_r)]\\
& &+\frac{{\gm'}^2-2{\gm'}'\sin
2\gm}{2}x^2+\frac{{\gm'}^2+{\gm'}'\sin
2\gm}{2}y^2+{\gm'}'xy\cos2\gm-\gm'^2\X^2\\ & &-2\gm'\sum_{s=1}^n\hat
d_se^{\hat a_s^2\kappa t\cos 2\hat b_s+\hat a_s\X\cos \hat
b_s}\sin(\hat a_s^2\kappa t\sin 2\hat b_s+\hat a_s\X\sin \hat
b_s+\hat c_s).\hspace{1.8cm}(8.1.133)\end{eqnarray*}

When $\nu\neq\kappa$,  we have the following solutions of the
two-dimensional Boussinesq equations (8.1.1)-(8.1.2):
\begin{eqnarray*}\hspace{1cm}u&=&\{\sum_{s=1}^n\hat a_s\hat
d_se^{\hat a_s^2\kappa t\cos 2\hat b_s+\hat a_s\X\cos \hat
b_s}\sin(\hat a_s^2\kappa t\sin 2\hat b_s+\hat a_s\X\sin \hat
b_s+\hat b_s+\hat c_s)\\
& &+2\gm'\X-\sum_{r=1}^ma_rd_re^{a_r^2\nu t+a_r\X}\int
e^{a_r^2(\kappa-\nu)t}\cos\gm\:dt
\}\sin\gm-\gm'y,\hspace{1.7cm}(8.1.134)\end{eqnarray*}
\begin{eqnarray*}\hspace{1cm}v&=&-\{\sum_{s=1}^n\hat a_s\hat
d_se^{\hat a_s^2\kappa t\cos 2\hat b_s+\hat a_s\X\cos \hat
b_s}\sin(\hat a_s^2\kappa t\sin 2\hat b_s+\hat a_s\X\sin \hat
b_s+\hat b_s+\hat c_s)\\
& &+2\gm'\X-\sum_{r=1}^m a_rd_re^{a_r^2\nu t+a_r\X}\int
e^{a_r^2(\kappa-\nu)t}\cos\gm\:dt
\}\cos\gm+\gm'x,\hspace{1.7cm}(8.1.135)\end{eqnarray*} $\sta=\psi$
in (8.1.122), and
\begin{eqnarray*}
p&=&\frac{{\gm'}^2-{\gm'}'\sin
2\gm}{2}x^2+\frac{{\gm'}^2+{\gm'}'\sin
2\gm}{2}y^2+{\gm'}'xy\cos2\gm-2\gm'^2\X^2
\\&&-2\gm'\sum_{s=1}^n\hat
d_se^{\hat a_s^2\kappa t\cos 2\hat b_s+\hat a_s\X\cos \hat
b_s}\sin(\hat a_s^2\kappa t\sin 2\hat b_s+\hat a_s\X\sin \hat
b_s+\hat c_s)\\ & &+\sum_{r=1}^m d_re^{a_r^2\nu t+a_r\X}(2\gm'\int
e^{a_r^2(\kappa-\nu)t}\cos\gm\:dt+\sin\gm).\hspace{4.3cm}(8.1.136)\end{eqnarray*}
}\pse

{\bf Remark 8.1.6}. By Fourier expansion, we can use the above
solution to obtain the one depending on two  piecewise continuous
functions of $\X$. Applying the symmetry transformations in
(8.1.6)-(8.1.10) to the above solution, we can get more general
solutions of the two-dimensional Boussinesq equations
(8.1.1)-(8.1.2).

\section{Three-Dimensional Equations and Symmetry}

Another slightly simplified version of the system of  primitive
equations in geophysics is the three-dimensional stratified rotating
Boussinesq system (e.g., cf. [LTW], [Pj]):
$$u_t+uu_x+vu_y+wu_z-\frac{1}{R_0}v=\sgm(\Dlt u-p_x),\eqno(8.2.1)$$
$$v_t+uv_x+vv_y+wv_z+\frac{1}{R_0}u=\sgm(\Dlt v-p_y),\eqno(8.2.2)$$
$$w_t+uw_x+vw_y+ww_z-\sgm R T=\sgm(\Dlt w-p_z),\eqno(8.2.3)$$
$$T_t+uT_x+vT_y+wT_z=\Dlt T+w,\eqno(8.2.4)$$
$$ u_x+v_y+w_z=0,\eqno(8.2.5)$$\index{Boussinesq equations!three-dimensional}where
$(u,v,w)$ is the velocity vector filed, $T$ is the temperature
function, $p$ is the pressure function, $\sgm$ is the Prandtle
number, $R$ is the thermal Rayleigh number and $R_0$ is the Rossby
number. Moreover, the vector $(1/R_0)(-v,u,0)$ represents the
Coriolis force and the term $w$ in (8.2.4) is derived using
stratification. So the above equations are the extensions of
Navier-Stokes equations by adding the Coriolis force and the
stratified temperature equation. Due to the Coriolis force, the
two-dimensional system (8.1.1) and (8.1.2) is not a special case of
the above three-dimensional system. Hsia, Ma and Wang [HMW] studied
the bifurcation and periodic solutions of the above system
(8.2.1)-(8.2.5).

After the degree analysis, we find that  the three-dimensional
stratified rotating Boussinesq system
 is not dilation invariant. It is translation invariant. Let $\al$
 be a function in $t$. The transformation
 $$F(t,x,y,z)\mapsto
 F(t,x+\al,y,z)-\dlt_{u,F}\al'\qquad\for\;\;F=u,v,w,T,p\eqno(8.2.6)$$
 leaves (8.2.3)-(8.2.5) invariant and changes (8.2.1) and (8.2.2)
 to
 $$-{\al'}'+u_t+uu_x+vu_y+wu_z-\frac{1}{R_0}v=\sgm(\Dlt u-p_x),\eqno(8.2.7)$$
and
$$v_t+uv_x+vv_y+wv_z+\frac{1}{R_0}u-\frac{\al'}{R_0}=\sgm(\Dlt v-p_y),\eqno(8.2.8)$$
where the independent variable $x$ is replaced by $x+\al$ and the
partial derivatives are with respect to the original variables. Thus
the transformation
$$S_{1,\al}(F(t,x,y,z))=
 F(t,x+\al,y,z)-\dlt_{u,F}\al'+\dlt_{p,F}\sgm^{-1}({\al'}'x+\al'y/R_0)
 \eqno(8.2.9)$$
for $F=u,v,w,T,p$, is a symmetry of the system (8.2.1)-(8.2.5).
Similarly, we have the following symmetry of the system
(8.2.1)-(8.2.5):
$$S_{2,\al}(F(t,x,y,z))=
 F(t,x,y+\al,z)-\dlt_{v,F}\al'+\dlt_{p,F}\sgm^{-1}({\al'}'y-\al'x/R_0)
 \eqno(8.2.10)$$
for $F=u,v,w,T,p$.

Note that the transformation
 $$F(t,x,y,z)\mapsto
 F(t,x,y,z+\al)-\dlt_{w,F}\al'\qquad\for\;\;F=u,v,w,T,p\eqno(8.2.11)$$
leaves (8.2.1), (8.2.2) and (8.2.5) invariant, and changes (8.2.3)
and (8.2.4) to
$$-{\al'}'+w_t+uw_x+vw_y+ww_z-\sgm R T=\sgm(\Dlt w-p_z),\eqno(8.2.12)$$
and
$$T_t+uT_x+vT_y+wT_z=\Dlt T+w-\al',\eqno(8.2.13)$$
where the independent variable $x$ is replaced by $x+\al$ and the
partial derivatives are with respect to the original variables.
Hence the transformation
$$S_{3,\al}(F(t,x,y,z))=
 F(t,x,y,z+\al)-\dlt_{w,F}\al'+\dlt_{p,F}(\sgm^{-1}{\al'}'-\al
 /R)z-\dlt_{T,F}\al
 \eqno(8.2.14)$$
for $F=u,v,w,T,p$, is a symmetry of the system (8.2.1)-(8.2.5) .
Obviously, the transformation
 $$S_{4,\al}(F(t,x,y,z))=
 F(t,x,y,z)+\dlt_{p,F}\al'
 \eqno(8.2.15)$$
for $F=u,v,w,T,p$, is a symmetry of the system.

For convenience of computation, we denote
$$\Phi_1=u_t+uu_x+vu_y+wu_z-\frac{1}{R_0}v-\sgm(u_{xx}+u_{yy}+u_{zz}),\eqno(8.2.16)$$
$$\Phi_2=v_t+uv_x+vv_y+wv_z+\frac{1}{R_0}u-\sgm(v_{xx}+v_{yy}+v_{zz}),\eqno(8.2.17)$$
$$\Phi_3=w_t+uw_x+vw_y+ww_z-\sgm R T-\sgm(w_{xx}+w_{yy}+w_{zz}).\eqno(8.2.18)$$
Then the equations (8.2.1)-(8.2.3) become
 $$\Phi_1+\sgm p_x=0,\qquad
 \Phi_2+\sgm p_y=0,\qquad
\Phi_3+\sgm p_z=0.
 \eqno(8.2.19)$$
 Our strategy is  to solve the following
 compatibility conditions:
 $$\ptl_y(\Phi_1)=\ptl_x(\Phi_2),\qquad
 \ptl_z(\Phi_1)=\ptl_x(\Phi_3),\qquad\ptl_z(\Phi_2)=\ptl_y(\Phi_3).
 \eqno(8.2.20)$$

\section{Asymmetric Approach I}

Starting from this section, we use  asymmetric methods to solve the
stratified rotating Boussinesq equations (8.2.1)-(8.2.5).

\index{asymmetric approach I!for 3-D Boussinesq equations}

First we assume
$$u=\phi_z(t,z) x+\vs(t,z) y+\mu(t,z),\qquad v=\tau(t,z)
x+\psi_z(t,z) y+\ves(t,z),\eqno(8.3.1)$$
$$ w=-\phi(t,z)-\psi(t,z),\qquad T=\vt(t,z)+z,\eqno(8.3.2)$$ where $\phi,\vt,\vs,\mu,\tau,$ and
$\ves$ are functions of $t,z$ to be determined. Then
\begin{eqnarray*}\Phi_1&=&\phi_{tz}x+\vs_t y+\mu_t+
\phi_z(\phi_z x+\vs y+\mu)+(\vs-1/R_0)(\tau x+\psi_zy+\ves)\\ &
&-(\phi+\psi)(\phi_{zz}x+\vs_z y+\mu_z)
-\sgm(\phi_{zzz}x+\vs_{zz} y+\mu_{zz})\\
&=&[\phi_{tz}+\phi_z^2+\tau(\vs-1/R_0)-\phi_{zz}(\phi+\psi)-\sgm\phi_{zzz}]x\\
& &+[\vs_t+\vs\phi_z+\psi_z(\vs-1/R_0)-\vs_z(\phi+\psi)-\sgm \vs_{zz}]y\\
& &+\mu_t+ \mu\phi_z+(\vs-1/R_0)\ves-\mu_z(\phi+\psi)-\sgm\mu_{zz},
\hspace{5cm}(8.3.3)\end{eqnarray*}
\begin{eqnarray*}\Phi_2&=&\tau_tx+\psi_{tz}y+\ves_t+\psi_z(\tau x+\psi_zy+\ves)+
(\tau+1/R_0)(\phi_zx+\vs y+\mu)\\
& &-(\phi+\psi)(\tau_zx+\psi_{zz}y+\ves_z)
-\sgm(\tau_{zz}x+\psi_{zzz}y+\ves_{zz})\\
&=&[\psi_{tz}+\psi_z^2+\vs(\tau+1/R_0)-(\phi+\psi)\psi_{zz}-\sgm\psi_{zzz}]y\\
& &+[\tau_t+\tau\psi_z+(\tau+1/R_0)\phi_z-(\phi+\psi)\tau_z-\sgm
\tau_{zz}]x\\ & &+\ves_t+
\ves\psi_z+(\tau+1/R_0)\mu-(\phi+\psi)\ves_z-\sgm\ves_{zz},
\hspace{5.1cm}(8.3.4)\end{eqnarray*}
$$\Phi_3=-\phi_t-\psi_t+(\phi+\psi)(\phi_z+\psi_z)-\sgm
R(\vt+z)+\sgm(\phi_{zz}+\psi_{zz}).\eqno(8.3.5)
$$
Thus (8.2.20) is equivalent to the following system of partial
differential equations:
$$\phi_{tz}+\phi_z^2+\tau(\vs-1/R_0)-\phi_{zz}(\phi+\psi)-\sgm\phi_{zzz}=\al_1,\eqno(8.3.6)$$
$$\vs_t+\vs\phi_z+\psi_z(\vs-1/R_0)-\vs_z(\phi+\psi)-\sgm
\vs_{zz}=\al,\eqno(8.3.7)$$
$$\mu_t+
\mu\phi_z+(\vs-1/R_0)\ves-\mu_z(\phi+\psi)-\sgm\mu_{zz}=\al_2,\eqno(8.3.8)$$
$$\psi_{tz}+\psi_z^2+\vs(\tau+1/R_0)-(\phi+\psi)\psi_{zz}-\sgm\psi_{zzz}=\be_1,\eqno(8.3.9)$$
$$\tau_t+\tau\psi_z+(\tau+1/R_0)\phi_z-(\phi+\psi)\tau_z-\sgm
\tau_{zz}=\al,\eqno(8.3.10)$$
$$\ves_t+\ves\psi_z+(\tau+1/R_0)\mu-(\phi+\psi)\ves_z-\sgm\ves_{zz}=\be_2\eqno(8.3.11)$$
for some $\al,\al_1,\al_2,\be_1,\be_2$ are functions in $t$.

Let $0\neq b$ and $c$ be fixed real constants. We define
$$\zeta_1(z)=\frac{e^{bz}-ce^{-bz}}{2},\qquad \eta_1(z)=\frac{e^{b
z}+ce^{-b z}}{2},\eqno(8.3.12)$$
$$\zeta_0(z)=\sin bz,\qquad \eta_0(z)=\cos bz.\eqno(8.3.13)$$
Then
$$\eta_r^2(z)+(-1)^r\zeta_r^2(z)=c^r.\eqno(8.3.14)$$
 We assume
$$\phi=b^{-1}\gm_1\zeta_r(z),\qquad
\psi=b^{-1}(\gm_2\zeta_r(z)+\gm_3\eta_r(z)),\eqno(8.3.15)$$
$$\vs=\gm_4(\gm_2\eta_r(z)-(-1)^r\gm_3\zeta_r(z)),\qquad\tau=\gm_5\gm_1\eta_r(z),\qquad\gm_4\gm_5=1,\eqno(8.3.16)$$
where $\gm_j$ are functions in $t$ to be determined. Moreover,
(8.3.6) becomes
$$(\gm_1'+(-1)^rb^2\sgm\gm_1-\gm_1\gm_5/R_0)\eta_r(z)+(\gm_1+\gm_2)\gm_1c^r
=\al_1,\eqno(8.3.17)$$ which is implied by
$$\al_1=(\gm_1+\gm_2)\gm_1c^r,\eqno(8.3.18)$$
$$\gm_1'+(-1)^rb^2\sgm\gm_1-\gm_1\gm_5/R_0=0.\eqno(8.3.19)$$

On the other hand, (8.3.10) becomes
$$[(\gm_1\gm_5)'+ \gm_1/R_0+(-1)^rb^2\sgm\gm_1\gm_5]\eta_r+
\gm_1\gm_5(\gm_1+\gm_2)c^r=\al,\eqno(8.3.20)$$ which gives
$$\al=\gm_1\gm_5(\gm_1+\gm_2)c^r,\eqno(8.3.21)$$
$$(\gm_1\gm_5)'+(-1)^rb^2\sgm\gm_1\gm_5+ \gm_1/R_0=0.\eqno(8.3.22)$$ Solving (8.3.19) and (8.3.22) for $\gm_1$ and
$\gm_1\gm_5$, we get
$$\gm_1=b_1e^{-(-1)^rb^2\sgm t}\sin\frac{t}{R_0},\qquad\gm_1\gm_5=
b_1e^{-(-1)^rb^2\sgm t}\cos\frac{t}{R_0},\eqno(8.3.23)$$ where $b_1$
is a real constant. In particular, we take
$$\gm_5=\cot\frac{t}{R_0}.\eqno(8.3.24)$$
Observe that (8.3.7) becomes
\begin{eqnarray*}\hspace{1cm}& &[(\gm_2\gm_4)'+(-1)^rb^2\sgm\gm_2\gm_4-\gm_2/R_0]\eta_r
+\gm_4(\gm_1\gm_2+\gm_2^2+(-1)^r\gm_3^2)c^r\\
&&-(-1)^r[(\gm_3\gm_4)'+(-1)^rb^2\sgm\gm_2\gm_4-\gm_3/R_0]\zeta_r=\al
\hspace{4.4cm}(8.3.25)\end{eqnarray*} and (8.3.9) becomes
\begin{eqnarray*}\hspace{1cm}& &[\gm_2'+(-1)^rb^2\sgm\gm_2+\gm_2\gm_4/R_0]\eta_r
+(\gm_1\gm_2+\gm_2^2+(-1)^r\gm_3^2)c^r\\
&&-(-1)^r[\gm_3'+(-1)^rb^2\sgm\gm_3+\gm_3\gm_4/R_0]\zeta_r=\be_1,
\hspace{5cm}(8.3.26)\end{eqnarray*} equivalently,
$$\al=\gm_4(\gm_1\gm_2+\gm_2^2+(-1)^r\gm_3^2)c^r,\eqno(8.3.27)$$
$$\be_1=(\gm_1\gm_2+\gm_2^2+(-1)^r\gm_3^2)c^r,\eqno(8.3.28)$$
$$(\gm_2\gm_4)'+(-1)^rb^2\sgm\gm_2\gm_4-\gm_2/R_0=0,\eqno(8.3.29)$$
$$\gm_2'+(-1)^rb^2\sgm\gm_2+\gm_2\gm_4/R_0=0,\eqno(8.3.30)$$
$$(\gm_3\gm_4)'+(-1)^rb^2\sgm\gm_2\gm_4-\gm_3/R_0=0,\eqno(8.3.31)$$
$$\gm_3'+(-1)^rb^2\sgm\gm_3+\gm_3\gm_4/R_0=0.\eqno(8.3.32)$$
Solving (8.3.29)-(8.3.32) under the assumption $\gm_4\gm_5=1$, we
obtain
$$\gm_2\gm_4=b_2e^{-(-1)^rb^2\sgm t}\sin\frac{t}{R_0},\qquad\gm_2=
b_2e^{-(-1)^rb^2\sgm t}\cos\frac{t}{R_0},\eqno(8.3.33)$$
$$\gm_3\gm_4=b_3e^{-(-1)^rb^2\sgm t}\sin\frac{t}{R_0},\qquad\gm_3=
b_3e^{-(-1)^rb^2\sgm t}\cos\frac{t}{R_0}.\eqno(8.3.34)$$ In
particular, we have:
$$\gm_4=\tan\frac{t}{R_0}.\eqno(8.3.35)$$

According to (8.3.21) and (8.3.27),
$$\gm_1\gm_5(\gm_1+\gm_2)c^r=\gm_4(\gm_1\gm_2+\gm_2^2+(-1)^r\gm_3^2)c^r.
\eqno(8.3.36)$$ Multiplying $\gm_4$ to the above equation and
dividing by $c^r$ , we have
$$
\gm_1(\gm_1+\gm_2)=\gm_1\gm_4(\gm_2\gm_4)+
(\gm_2\gm_4)^2+(-1)^r(\gm_3\gm_4)^2.\eqno(8.3.37)$$ By (8.3.23) and
(8.3.33)-(8.3.35), the above equation is equivalent to
$$b_1^2\sin^2\frac{t}{R_0}+\frac{b_1b_2}{2}\sin\frac{2t}{R_0}
=b_1b_2\tan\frac{t}{R_0}\;\sin^2\frac{t}{R_0}+(b_2^2+(-1)^rb_3^2)\sin^2\frac{t}{R_0},\eqno(8.3.38)$$
which can be rewritten as
$$-b_1b_2\cos\frac{2t}{R_0}\;\tan\frac{t}{R_0}+(b_2^2-b_1^2+(-1)^rb_3^2)\sin\frac{2t}{R_0}
=0.\eqno(8.3.39)$$ Thus
$$b_1b_2=0,\qquad b_2^2-b_1^2+(-1)^rb_3^2=0.\eqno(8.3.40)$$
So
$$ r=0,\qquad b_2=0,\qquad b_1=b_3\eqno(8.3.41)$$
or
$$ r=1,\qquad b_1=0,\qquad b_2=b_3.\eqno(8.3.42)$$

Assume $r=0$ and $b_1\neq 0$. Then
$$\phi=b^{-1}b_1e^{-b^2\sgm t}\sin bz\:\sin\frac{t}{R_0},\qquad \psi=
b^{-1}b_1e^{-b^2\sgm t}\cos bz\:\cos\frac{t}{R_0},\eqno(8.3.43)$$
$$\vs=-b_1e^{-b^2\sgm t}\sin bz\:\sin\frac{t}{R_0},\qquad\tau=b_1e^{-b^2\sgm
t}\cos bz\:\cos\frac{t}{R_0}.\eqno(8.3.44)$$ Moreover, we take
$\mu=\ves=\vt=0$. So (8.2.4), (8.3.8) and (8.3.11) naturally hold.
Observe
$$\Phi_1=\gm_1^2(x+\gm_5y)=b_1^2e^{-2b^2\sgm
t}\left(x\sin\frac{t}{R_0}+y\cos\frac{t}{R_0}\right)\sin\frac{t}{R_0}\eqno(8.3.45)$$
by (8.3.3), (8.3.6)-(8.3.8), (8.3.18) and (8.3.21). Similarly
$$\Phi_2=b_1^2e^{-2b^2\sgm
t}\left(x\sin\frac{t}{R_0}+y\cos\frac{t}{R_0}\right)\cos\frac{t}{R_0}.\eqno(8.3.46)$$

According to (8.3.5),
$$\Phi_3=\left[b^{-1}R_0^{-1}b_1e^{-b^2\sgm
t}-b^{-1}b_1^2e^{-2b^2\sgm t}\cos
\left(bz-\frac{t}{R_0}\right)\right]\sin\left(bz-\frac{t}{R_0}\right)-R\sgm
z.\eqno(8.3.47)$$ By (8.2.19), we have
\begin{eqnarray*}\hspace{1cm}p&=&\frac{Rz^2}{2}+\frac{b_1e^{-b^2\sgm
t}}{b^2\sgm R_0}\cos
\left(bz-\frac{t}{R_0}\right)-\frac{b_1^2e^{-2b^2\sgm
t}}{2\sgm b^2}\cos ^2\left(bz-\frac{t}{R_0}\right)\\
& &-\frac{b_1^2e^{-2b^2\sgm
t}}{2\sgm}\left(y^2\cos^2\frac{t}{R_0}+x^2\sin^2\frac{t}{R_0}+xy\sin\frac{2t}{R_0}
\right).\hspace{3.3cm}(8.3.48)\end{eqnarray*}

Suppose  $r=1$ and $b_2\neq 0$. Then
 $$\phi=\tau=\mu=\ves=\vt=0,\;\;\psi=b^{-1}b_2e^{bz+b^2\sgm t}\cos\frac{t}{R_0},
\qquad\vs=b_2e^{bz+b^2\sgm t}\sin\frac{t}{R_0}.\eqno(8.3.49)$$
Moreover,
$$\Phi_1=\Phi_2=0,\;\;\Phi_3=b^{-1}b_2R_0^{-1}e^{bz+b^2\sgm
t}\sin\frac{t}{R_0}+b^{-1}b_2^2e^{2(bz+b^2\sgm
t)}\cos^2\frac{t}{R_0}-R\sgm z.\eqno(8.3.50)$$ According to
(8.2.19),
$$p=\frac{Rz^2}{2}-\frac{b_2e^{bz+b^2\sgm
t}}{b^2\sgm R_0}\sin\frac{t}{R_0}-\frac{b_2^2e^{2(bz+b^2\sgm
t)}}{2b^2\sgm}\cos^2\frac{t}{R_0}.\eqno(8.3.51)$$  by (8.3.1) and
(8.3.2), we get:\psp

{\bf Theorem 8.3.1}. {\it Let $b,b_1,b_2\in\mbb{R}$ with $b\neq 0$.
We have the following solutions of the three-dimensional stratified
rotating Boussinesq equations (8.2.1)-(8.2.5): (1)
$$u=b_1e^{-b^2\sgm t}(x\cos bz-y\sin bz)\sin\frac{t}{R_0},\;\;v=
b_1e^{-b^2\sgm t}(x\cos bz-y\sin
bz)\cos\frac{t}{R_0},\eqno(8.3.52)$$
$$w=-b^{-1}b_1e^{-b^2\sgm t}\cos\left(bz-\frac{t}{R_0}\right),\qquad
T=z\eqno(8.3.53)$$ and $p$ is given in (8.3.48); (2)
$$u=b_2e^{bz+b^2\sgm t}y\sin\frac{t}{R_0},\qquad v=b_2e^{bz+b^2\sgm
t}y\cos\frac{t}{R_0},\eqno(8.3.54)$$
$$w=-b^{-1}b_2e^{bz+b^2\sgm
t}\cos\frac{t}{R_0}\qquad T=z\eqno(8.3.55)$$ and $p$ is given in
(8.3.51).} \psp

Next we assume $\phi=\vs=\psi=\tau=0$. Then
$$\mu_t-\frac{1}{R_0}\ves-\sgm\mu_{zz}=\al_2,\;\;
\ves_t+\frac{1}{R_0}\nu-\sgm\ves_{zz}=\be_2,\;\;\vt_t-\vt_{zz}=0.\eqno(8.3.56)$$
Solving them, we get:\psp

{\bf Theorem 8.3.2}. {\it Let $a_s,b_s,c_s,d_s,\hat a_r,\hat
b_r,\hat c_r,\hat d_r,\td a_j,\td b_j,\td c_j,\td d_j$ be real
numbers.
 We have the following solutions of the
three-dimensional stratified rotating Boussinesq equations
(8.2.1)-(8.2.5):
\begin{eqnarray*}u&=&\cos\frac{t}{R_0}\;\sum_{s=1}^md_se^{a_s^2\sgm t\cos
2b_s+ a_sz\cos b_s}\sin(a_s^2\sgm t\sin 2b_s+a_sz\sin b_s+c_s)\\
& &+\sin\frac{t}{R_0}\;\sum_{r=1}^n \hat d_re^{\hat a_r^2\sgm t\cos
2\hat b_r+a_rz\cos\hat b_r}\sin(\hat a_r^2\sgm t\sin 2\hat b_r+\hat
a_rz\sin \hat b_r+\hat c_r),\hspace{1.4cm}(8.3.57)\end{eqnarray*}
\begin{eqnarray*}v&=&-\sin\frac{t}{R_0}\;\sum_{s=1}^md_se^{a_s^2\sgm t\cos
2b_s+ a_sz\cos b_s}\sin(a_s^2\sgm t\sin 2b_s+a_sz\sin b_s+c_s)\\
& &+\cos\frac{t}{R_0}\;\sum_{r=1}^n \hat d_re^{\hat a_r^2\sgm t\cos
2\hat b_r+a_rz\cos\hat b_r}\sin(\hat a_r^2\sgm t\sin 2\hat b_r+\hat
a_rz\sin \hat b_r+\hat c_r),\hspace{1.3cm}(8.3.58)\end{eqnarray*}
$$w=0,\;\;T=z+\sum_{j=1}^k\td a_j\td d_j e^{\td a_j^2 t\cos
2\td b_j+ \td a_jz\cos \td b_j}\sin(\td a_j^2 t\sin 2\td b_j+\td
a_jz\sin \td b_j+\td b_j+\td c_j),\eqno(8.3.59)$$
$$p=\frac{R z^2}{2}+R\sum_{j=1}^k\td d_j e^{\td a_j^2 t\cos
2\td b_j+ \td a_jz\cos \td b_j}\sin(\td a_j^2 t\sin 2\td b_j+\td
a_jz\sin \td b_j+\td c_j).\eqno(8.3.60)$$ }\pse

{\bf Remark 8.3.3}. By Fourier expansion, we can use the above
solution to obtain the one depending on three arbitrary piecewise
continuous functions of $z$.

\section{Asymmetric Approach II}

In this section, we solve the stratified rotating Boussinesq
equations (8.2.1)-(8.2.5) under the assumption
$$u_z=v_z=w_{zz}=T_{zz}=0.\eqno(8.4.1)$$

\index{asymmetric approach II!for 3-D Boussinesq equations}

Let $\gm$ be a function in $t$ and we use the moving frame  in
(8.1.107). Assume
$$u=f(t,\X)\sin\gm-\gm'y,\qquad
v=-f(t,\X)\cos\gm+\gm'x,\eqno(8.4.2)$$
$$w=\phi(t,\X),\qquad
T=\psi(t,\X)+z,\eqno(8.4.3)$$ for some functions $f,\;\phi$ and
$\psi$ in $t$ and $\X$.

Using (8.1.108)-(8.1.112) and (8.2.16)-(8.2.18), we get
$$u\ptl_x+v\ptl_y=-f\ptl_\Y+\gm'(x\ptl_y-y\ptl_x)\eqno(8.4.4)$$
and
$$\Phi_1=-(\gm'^2+\gm'/R_0)x-{\gm'}'y+f_t\sin\gm+(2\gm'+1/R_0)f\cos\gm-\sgm
f_{\X\X}\sin\gm,\eqno(8.4.5)$$
$$\Phi_2=-(\gm'^2+\gm'/R_0)y+{\gm'}'x-f_t\cos\gm+(2\gm'+1/R_0)f\sin\gm+\sgm
f_{\X\X}\cos\gm,\eqno(8.4.6)$$
$$\Phi_3=\phi_t-\sgm\phi_{\X\X}-\sgm
R(\psi+z).\eqno(8.4.7)$$ By (8.2.20), we have
$$-2{\gm'}'+f_{\X t}-\sgm
f_{\X\X\X}=0,\eqno(8.4.8)$$
$$\phi_t-\sgm\phi_{\X\X}-\sgm R\psi=0.\eqno(8.4.9)$$
Moreover, (8.2.4) becomes
$$\psi_t-\psi_{\X\X}=0.\eqno(8.4.10)$$

Solving (8.4.8), we have:
$$f=2\gm'\X+\sum_{j=1}^m a_jd_je^{a_j^2\kappa t\cos 2b_j+a_j\X\cos
b_j}\sin(a_j^2\kappa t\sin 2b_j+a_j\X\sin
b_j+b_j+c_j),\eqno(8.4.11)$$
 where $a_j,b_j,c_j,d_j$ are arbitrary  real numbers.
Moreover, (8.4.9) and (8.4.10) yield
$$\phi=\sum_{r=1}^n \hat d_re^{\hat a_r^2 t\cos 2\hat b_r+\hat a_r\X\cos
\hat b_r}\sin(\hat a_r^2 t\sin 2\hat b_r+\hat a_r\X\sin \hat
b_r+\hat c_r)+\sgm Rt\psi,\eqno(8.4.12)$$
$$\psi=\sum_{s=1}^k\td d_se^{\td a_s^2t\cos 2\td
b_s+\td a_s\X\cos \td b_s}\sin(\td a_s^2 t\sin 2\td b_s+\td
a_s\X\sin \td b_s+\td c_s)\eqno(8.4.13)$$ if $\sgm=1$, and
\begin{eqnarray*}\phi&=&\sum_{r=1}^n \hat d_re^{\hat a_r^2\sgm t\cos 2\hat b_r+\hat a_r\X\cos
\hat b_r}\sin(\hat a_r^2\sgm t\sin 2\hat b_r+\hat a_r\X\sin \hat
b_r+\hat c_r)\\ & &+\frac{\sgm R}{1-\sgm}\sum_{s=1}^k\td d_se^{\td
a_s^2t\cos 2\td b_s+\td a_s\X\cos \td b_s}\sin(\td a_s^2 t\sin 2\td
b_s+\td a_s\X\sin \td b_s+\td
c_s),\hspace{1.8cm}(8.4.14)\end{eqnarray*}
$$\psi=\sum_{s=1}^k\td a_s^2\td d_se^{\td a_s^2t\cos 2\td b_s+\td a_s\X\cos
\td b_s}\sin(\td a_s^2 t\sin 2\td b_s+\td a_s\X\sin \td b_s+2\td
b_s+\td c_s)\eqno(8.4.15)$$ when $\sgm\neq 1$, where $\hat a_r,\hat
b_r,\hat c_r,\hat d_r,\td a_s,\td b_s,\td c_s,\td d_s$ are arbitrary
real numbers.

Now
$$\Phi_1=({\gm'}'\sin2\gm-\gm'^2-\gm'/R_0)x-{\gm'}'y\cos2\gm+(2\gm'+1/R_0)f\cos\gm,\eqno(8.4.16)$$
$$\Phi_2=-({\gm'}'\sin2\gm+\gm'^2+\gm'/R_0)y-{\gm'}'x\cos2\gm+(2\gm'+1/R_0)f\sin\gm\eqno(8.4.17)$$
and $\Phi_3=-\sgm Rz$. Thanks to (8.2.19), we have
\begin{eqnarray*}p&=&
-\frac{2\gm'+1/R_0}{\sgm}[\gm'\X^2+\sum_{j=1}^m d_je^{a_j^2\kappa
t\cos 2b_j+a_j\X\cos b_j}\sin(a_j^2\kappa
t\sin 2b_j+a_j\X\sin b_j+c_j)]\\
&&+\frac{R}{2}z^2+\frac{(\gm'^2+\gm'/R_0)(x^2+y^2)+{\gm'}'(y^2-x^2)\sin2\gm}
{2\sgm}+\frac{{\gm'}'}{\sgm}xy\cos2\gm.\hspace{1.4cm}(8.4.18)\end{eqnarray*}

\psp

{\bf Theorem 8.4.1}. {\it Let $a_j,b_j,c_j,d_j,\hat a_r,\hat
b_r,\hat c_r,\hat d_r,\td a_s,\td b_s,\td c_s,\td d_s$ be real
numbers and let $\gm$ be any function in $t$. Denote
$\X=x\cos\gm+y\sin\gm$.
 We have the following solutions of the
three-dimensional stratified rotating Boussinesq equations
(8.2.1)-(8.2.5): \begin{eqnarray*}u&=&[\sum_{j=1}^m
a_jd_je^{a_j^2\kappa t\cos 2b_j+a_j\X\cos b_j}\sin(a_j^2\kappa t\sin
2b_j+a_j\X\sin b_j+b_j+c_j)\\ & &+2\gm'\X]\sin\gm-\gm'
y,\hspace{9.8cm}(8.4.19)\end{eqnarray*}
\begin{eqnarray*}v&=&[-\sum_{j=1}^m a_jd_je^{a_j^2\kappa
t\cos 2b_j+a_j\X\cos b_j}\sin(a_j^2\kappa t\sin 2b_j+a_j\X\sin
b_j+b_j+c_j)\\ & &+2\gm'\X]\cos\gm+\gm'
x,\hspace{9.8cm}(8.4.20)\end{eqnarray*} $p$ is given in (8.4.18);
\begin{eqnarray*}w&=&\sum_{r=1}^n \hat d_re^{\hat a_r^2 t\cos 2\hat b_r+\hat a_r\X\cos
\hat b_r}\sin(\hat a_r^2 t\sin 2\hat b_r+\hat a_r\X\sin \hat
b_r+\hat c_r)\\ & &+\sgm Rt\sum_{s=1}^k\td d_se^{\td a_s^2t\cos 2\td
b_s+\td a_s\X\cos \td b_s}\sin(\td a_s^2 t\sin 2\td b_s+\td
a_s\X\sin \td b_s+\td c_s),\hspace{2.2cm}(8.4.21)\end{eqnarray*}
$$T=z+\sum_{s=1}^k\td d_se^{\td a_s^2t\cos 2\td
b_s+\td a_s\X\cos \td b_s}\sin(\td a_s^2 t\sin 2\td b_s+\td
a_s\X\sin \td b_s+\td c_s)\eqno(8.4.22)$$ if $\sgm=1$, and
\begin{eqnarray*}w&=&\sum_{r=1}^n \hat d_re^{\hat a_r^2\sgm t\cos 2\hat b_r+\hat a_r\X\cos
\hat b_r}\sin(\hat a_r^2\sgm t\sin 2\hat b_r+\hat a_r\X\sin \hat
b_r+\hat c_r)\\ & &+\frac{\sgm R}{1-\sgm}\sum_{s=1}^k\td d_se^{\td
a_s^2t\cos 2\td b_s+\td a_s\X\cos \td b_s}\sin(\td a_s^2 t\sin 2\td
b_s+\td a_s\X\sin \td b_s+\td
c_s),\hspace{1.8cm}(8.4.23)\end{eqnarray*}
$$T=z+\sum_{s=1}^k\td a_s^2\td d_se^{\td a_s^2t\cos 2\td b_s+\td a_s\X\cos
\td b_s}\sin(\td a_s^2 t\sin 2\td b_s+\td a_s\X\sin \td b_s+2\td
b_s+\td c_s)\eqno(8.4.24)$$ when $\sgm\neq 1$. }\psp

{\bf Remark 8.4.2}. By Fourier expansion, we can use the above
solution to obtain the one depending on three arbitrary piecewise
continuous functions of $\X$.\psp

Next we  set
$$\varpi=x^2+y^2.\eqno(8.4.25)$$
We assume
$$u=y\phi(t,\varpi),\qquad
v=-x\phi(t,\varpi),\eqno(8.4.26)$$
$$w=\psi(t,\varpi),\qquad T=\vt(t,\varpi)+z\eqno(8.4.27)$$
where $\phi,\psi$ and $\vt$ are functions in $t,\varpi$. Note that
(8.2.16)-(8.2.18) give
$$\Phi_1=y\phi_t+\frac{x}{R_0}
\phi-x\phi^2-4\sgm y(\varpi\phi)_{\varpi \varpi},\eqno(8.4.28)$$
$$\Phi_2=-x\phi_t+\frac{y}{R_0}
\phi-y\phi^2+4\sgm x(\varpi\phi)_{\varpi \varpi},\eqno(8.4.29)$$
$$\Phi_3=\psi_t-\sgm R(\vt+z)-4\sgm(\varpi\psi_\varpi)_\varpi.\eqno(8.4.30)$$

According to (8.2.20),
$$[\varpi(\phi_t-4\sgm (\varpi\phi)_{\varpi
\varpi})]_\varpi=0,\eqno(8.4.31)$$
$$\ptl_x[\psi_t-\sgm R\vt-4\sgm(\varpi\psi_\varpi)_\varpi]=
\ptl_y[\psi_t-\sgm
R\vt-4\sgm(\varpi\psi_\varpi)_\varpi]=0,\eqno(8.4.32)$$
$$\phi_t-4\sgm (\varpi\phi)_{\varpi
\varpi}=\frac{\al'}{\varpi},\eqno(8.4.33)$$
$$\psi_t-\sgm
R\vt-4\sgm(\varpi\psi_\varpi)_\varpi=\be'\eqno(8.4.34)$$ for some
functions $\al$ and $\be$ in $t$.

Write
$$\phi=\sum_{j=-1}^\infty\al_j\varpi^j,\eqno(8.4.35)$$
where $\al_j$ are functions in $t$ to be determined. Then (8.4.33)
becomes
$$\sum_{j=-1}^\infty(\al_j'-4\sgm(j+2)(j+1)\al_{j+1})
\varpi^j=\frac{\al'}{\varpi},\eqno(8.4.36)$$ equivalently,
$$\al_{-1}'=\al',\;\;4\sgm(j+2)(j+1)\al_{j+1}=\al_j'\;\;\for\;\;j\geq
0.\eqno(8.4.37)$$ We take $\al_{-1}=\al$ and redenote $\al_0=\gm$.
The above second equation implies
$$\al_j=\frac{\gm^{(j)}}{j!(j+1)!(4\sgm)^j}\qquad \for\;\;j\geq
0.\eqno(8.4.38)$$So
$$\phi=\frac{\al}{\varpi}+\sum_{j=0}^\infty
\frac{\gm^{(j)}\varpi^j}{j!(j+1)!(4\sgm)^j}.\eqno(8.4.39)$$

Observe that (8.2.4) becomes
$$\vt_t-4(\varpi\vt_{\varpi})_{\varpi}=0\eqno(8.4.40)$$
by (8.4.25)-(8.4.27). The arguments in the above show
$$\vt=\sum_{r=0}^\infty
\frac{\gm_1^{(r)}\varpi^r}{r!(r+1)!(4\sgm)^r},\eqno(8.4.41)$$ where
$\gm_1$ is an arbitrary function in $t$. Substituting (8.4.41) into
(8.4.34), we get
 $$\psi_t-4\sgm(\varpi\psi_\varpi)_\varpi=\be'+4\sgm R\sum_{r=0}^\infty
\frac{\gm_1^{(r)}\varpi^r}{r!(r+1)!(4\sgm)^r}.\eqno(8.4.42)$$ Write
$$\psi=\sum_{r=1}^\infty\be_r\varpi^r,\eqno(8.4.43)$$
where $\be_r$ are functions in $t$ to be determined. Then (8.4.42)
becomes
$$\sum_{r=0}^\infty(\be_r'-4\sgm(r+2)(r+1)\be_{r+1})
\varpi^r=\be'+4\sgm R\sum_{r=0}^\infty
\frac{\gm_1^{(r)}\varpi^r}{r!(r+1)!(4\sgm)^r},\eqno(8.4.44)$$
equivalently,
$$8\sgm\be_1=\be_0'-\be'-4\sgm R\gm,\eqno(8.4.45)$$
$$\be_{r+1}
=\frac{\be_r'}{4\sgm(r+2)(r+1)}-
\frac{R\gm_1^{(r)}}{(r+2)!(r+1)!(4\sgm)^r}\qquad\for\;\;r\geq
1.\eqno(8.4.46)$$ Thus
$$\be_r=\frac{\be_0^{(r)}-\be^{(r)}}{r!(r+1)!(4\sgm)^r}-
\frac{R\gm_1^{(r-1)}}{(r+1)!(r-1)!(4\sgm)^{r-1}}\qquad\for\;\;r\geq
1.\eqno(8.4.47)$$ So
$$\psi=
\be_0+\sum_{r=1}^\infty\frac{(\be_0^{(r)}-\be^{(r)}-4\sgm
R\gm_1^{(r-1)})\varpi^r}{r!(r+1)!(4\sgm)^r}.\eqno(8.4.48)$$

Now (8.4.28), (8.4.29) and (8.4.33) give
$$\Phi_1=
\frac{\al' y}{\varpi}+\frac{x}{R_0}\phi-x\phi^2,\eqno(8.4.49)$$
$$\Phi_2=
-\frac{\al' x}{\varpi}+\frac{y}{R_0}\phi-y\phi^2.\eqno(8.4.50)$$
Moreover,
$$\Phi_3=\be'-\sgm Rz\eqno(8.4.51)$$
by (8.4.30) and (8.4.34).
 Thanks to (8.2.19), we have
\begin{eqnarray*}p&=&\frac{Rz^2}{2} +\frac{\al'}{\sgm}\arctan\frac{y}{x}
-\frac{\be'}{\sgm}z-\frac{\al\ln(x^2+y^2)}{2\sgm R_0}-\frac{1}{\sgm
R_0}\sum_{j=0}^\infty
\frac{\gm^{(j)}(x^2+y^2)^{j+1}}{[(j+1)!]^2(4\sgm)^j}\\ &
&-\frac{\al^2}{2\sgm(x^2+y^2)}-\frac{\al\gm\ln(x^2+y^2)}{\sgm}
+\frac{\al}{\sgm}\sum_{j=1}^\infty
\frac{\gm^{(j)}(x^2+y^2)^j}{jj!(j+1)!(4\sgm)^j}\\ &
&+\frac{1}{2\sgm}\sum_{j_1,j_2=0}^\infty\frac{\gm^{(j_1)}\gm^{(j_2)}
(x^2+y^2)^{j_1+j_2+1}}{(j_1+j_2+1)j_1!j_2!(j_1+1)!(j_2+1)!(4\sgm)^{j_1+j_2}}.
 \hspace{3.4cm}(8.4.52)
\end{eqnarray*}

By (8.4.25)-(8.4.27), (8.4.39), (8.4.41), and (8.4.48), we have:\psp

{\bf Theorem 8.4.3} {\it Let $\al,\be,\be_0,\gm,\gm_1$ be any
functions in $t$. We have the following solutions of the
three-dimensional stratified rotating Boussinesq equations
(8.2.1)-(8.2.5):
$$u=\frac{\al y}{x^2+y^2}+y\sum_{j=0}^\infty
\frac{\gm^{(j)}(x^2+y^2)^j}{j!(j+1)!(4\sgm)^j},\eqno(8.4.53)$$
$$v=-\frac{\al x}{x^2+y^2}-x\sum_{j=0}^\infty
\frac{\gm^{(j)}(x^2+y^2)^j}{j!(j+1)!(4\sgm)^j},\eqno(8.4.54)$$
$$w=\be_0+\sum_{r=1}^\infty\frac{(\be_0^{(r)}-\be^{(r)}-4\sgm
R\gm_1^{(r-1)})(x^2+y^2)^r}{r!(r+1)!(4\sgm)^r},\eqno(8.4.55)$$
$$T=z+\sum_{r=0}^\infty
\frac{\gm_1^{(r)}(x^2+y^2)^r}{r!(r+1)!(4\sgm)^r}\eqno(8.4.56)$$
 and $p$ is given in (8.4.52).}

\section{Asymmetric Approach III }

In this section, we solve (8.2.1)-(8.2.5) under the assumption
$v_x=w_x=T_x=0$.

\index{asymmetric approach III!for 3-D Boussinesq equations}

Let $c$ be a real constant. Set
$$\varpi=y\cos c+z\sin c.\eqno(8.5.1)$$
Suppose
$$u=f(t,\varpi),\qquad
v=\phi(t,\varpi)\sin c,\eqno(8.5.2)$$
$$w=-\phi(t,\varpi)\cos c,\qquad
T=\psi(t,\varpi)+z,\eqno(8.5.3)$$ where $f,\;\phi$ and $\psi$ are
functions in $t$ and $\varpi$. According to (8.2.16)-(8.2.18),
$$\Phi_1=f_t-\sgm
f_{\varpi\varpi}-\frac{\sin c}{R_0}\phi,\eqno(8.5.4)$$
$$\Phi_2=(\phi_t-\sgm\phi_{\varpi\varpi})\sin c+\frac{1}{R_0}f,\eqno(8.5.5)$$
$$\Phi_3=(\sgm\phi_{\varpi\varpi}-\phi_t)\cos c-\sgm R(\psi+z).\eqno(8.5.6)$$ By (8.2.20),
$$f_{\varpi t}-\sgm
f_{\varpi\varpi\varpi}-\frac{\sin
c}{R_0}\phi_{\varpi}=0,\eqno(8.5.7)$$
$$(\phi_t-\sgm\phi_{\varpi\varpi})_{\varpi}+\frac{\sin c}{R_0}f_{\varpi}+\sgm
R\psi_{\varpi}\cos c=0.\eqno(8.5.8)$$ For simplicity, we take
$$f_t-\sgm
f_{\varpi\varpi}-\frac{\sin c}{R_0}\phi=0,\eqno(8.5.9)$$
$$\phi_t-\sgm\phi_{\varpi\varpi}+\frac{\sin c}{R_0}f+\sgm
R\psi\cos c=0.\eqno(8.5.10)$$

Denote
$$\left(\begin{array}{c}\hat
f\\\hat\phi\end{array}\right)=\left(\begin{array}{cc}\cos\frac{t\sin
c}{R_0}&-\sin\frac{t\sin c}{R_0}\\ \sin\frac{t\sin
c}{R_0}&\cos\frac{t\sin
c}{R_0}\end{array}\right)\left(\begin{array}{c}
f\\\phi\end{array}\right).\eqno(8.5.11)$$ Then (8.5.9) and (8.5.10)
become
$$\hat f_t-\sgm\hat f_{\varpi\varpi}-\sgm R\psi\cos c\;\sin\frac{t\sin
c}{R_0}=0,\eqno(8.5.12)$$
$$\hat\phi_t-\sgm\hat\phi_{\varpi\varpi}+\sgm R\psi\cos c\;\cos\frac{t\sin
c}{R_0}=0.\eqno(8.5.13)$$ On the other hand, (8.2.4) becomes
$$\psi_t-\psi_{\varpi\varpi}=0.\eqno(8.5.14)$$

Assume $\sgm=1$. We have the following solution:
$$\psi=\sum_{j=1}^m a_jd_je^{a_j^2t\cos 2b_j+a_j\varpi\cos
b_j}\sin(a_j^2t\sin 2b_j+a_j\varpi\sin b_j+b_j+c_j),\eqno(8.5.15)$$
\begin{eqnarray*}\hat f\!\!\!&=&\!\!\!- RR_0\cot c\;\cos\frac{t\sin
c}{R_0}\;\sum_{j=1}^m a_jd_je^{a_j^2 t\cos 2b_j+a_j\varpi\cos
b_j}\sin(a_j^2t\sin 2b_j+a_j\varpi\sin b_j+b_j+c_j)\\ &
&+\sum_{r=1}^n \hat a_r\hat d_re^{\hat a_r^2 t\cos 2\hat b_r+\hat
a_r\varpi\cos \hat b_r}\sin(\hat a_r^2t\sin 2\hat b_r+\hat
a_r\varpi\sin \hat b_r+\hat b_r+\hat
c_r),\hspace{1.8cm}(8.5.16)\end{eqnarray*}
\begin{eqnarray*}\hat \phi\!\!\!&=&\!\!\!- RR_0\cot c\;\sin\frac{t\sin
c}{R_0}\;\sum_{j=1}^m a_jd_je^{a_j^2t\cos 2b_j+a_j\varpi\cos
b_j}\sin(a_j^2t\sin 2b_j+a_j\varpi\sin b_j+b_j+c_j)\\ &
&+\sum_{s=1}^k\td a_s\td d_se^{\td a_s^2t\cos 2\td b_s+\td
a_s\varpi\cos \td b_s}\sin(\td a_s^2 t\sin 2\td b_s+\td
a_s\varpi\sin \td b_s+\td b_s+\td
c_s),\hspace{1.8cm}(8.5.17)\end{eqnarray*} where $a_j,b_j,c_j,\hat
a_r,\hat b_r,\hat c_r,\hat d_r, \td a_s,\td b_s,\td c_s,\td d_s$
 are arbitrary  real numbers. According to (8.5.11),
\begin{eqnarray*}\!\!f\!\!\!&=&\!\!\!-RR_0\cot c\;\cos\frac{2t\sin
c}{R_0}\;\sum_{j=1}^m a_jd_je^{a_j^2t\cos 2b_j+a_j\varpi\cos
b_j}\sin(a_j^2t\sin 2b_j+a_j\varpi\sin b_j+b_j+c_j)
\\ \!\!\!& & \!\!\!+\cos\frac{t\sin c}{R_0}\;\sum_{r=1}^n \hat a_r\hat d_re^{\hat a_r^2 t\cos 2\hat b_r+\hat
a_r\varpi\cos \hat b_r}\sin(\hat a_r^2t\sin 2\hat b_r+\hat
a_r\varpi\sin \hat b_r+\hat b_r+\hat c_r) \\ \!\!\!& &\!\!\!
+\sin\frac{t\sin c}{R_0}\;\sum_{s=1}^k\td a_s\td d_se^{\td
a_s^2t\cos 2\td b_s+\td a_s\varpi\cos \td b_s}\sin(\td a_s^2 t\sin
2\td b_s+\td a_s\varpi\sin \td b_s+\td b_s+\td
c_s),\hspace{0.5cm}(8.5.18)\end{eqnarray*}
\begin{eqnarray*}& &\phi=-\sin\frac{t\sin c}{R_0}\;\sum_{r=1}^n \hat a_r\hat d_re^{\hat a_r^2 t\cos 2\hat b_r+\hat
a_r\varpi\cos \hat b_r}\sin(\hat a_r^2t\sin 2\hat b_r+\hat
a_r\varpi\sin \hat b_r+\hat b_r+\hat c_r)\\ & &+\cos\frac{t\sin
c}{R_0}\;\sum_{s=1}^k\td a_s\td d_se^{\td a_s^2t\cos 2\td b_s+\td
a_s\varpi\cos \td b_s}\sin(\td a_s^2 t\sin 2\td b_s+\td
a_s\varpi\sin \td b_s+\td b_s+\td
c_s).\hspace{0.5cm}(8.5.19)\end{eqnarray*}

Suppose $\sgm\neq 1$. We take the following solution of (8.5.14):
$$\psi=\sum_{j=1}^m a_jd_je^{a_j^2t+a_j\varpi},\eqno(8.5.20)$$
where $a_j,d_j$ are real constants. Substituting
$$\hat
f=\al_je^{a_j^2t+a_j\varpi},\;\;\hat\phi=\be_je^{a_j^2t+a_j\varpi},\;\;\psi=a_jd_je^{a_j^2t+a_j\varpi}\eqno(8.5.21)$$
into (8.5.12) and (8.5.13), we get
$$\al_j'+a_j^2(1-\sgm)\al_j-\sgm Ra_jd_j\cos c\;\sin\frac{t\sin
c}{R_0}=0,\eqno(8.5.22)$$
$$\be_j'+a_j^2(1-\sgm)\be_j+\sgm Ra_jd_j\cos c\;\cos\frac{t\sin
c}{R_0}=0.\eqno(8.5.23)$$ We have the solutions
$$\al_j=\sgm Ra_jd_j\cos c\;\frac{a_j^2(1-\sgm)\sin\frac{t\sin c}{R_0}-R_0^{-1}\sin
c\:\cos\frac{t\sin
c}{R_0}}{a_j^4(1-\sgm)^2+R_0^{-2}\sin^2c},\eqno(8.5.24)$$
$$\be_j=-\sgm Ra_jd_j\cos c\;\frac{a_j^2(1-\sgm)\cos\frac{t\sin c}{R_0}+R_0^{-1}\sin
c\:\sin\frac{t\sin
c}{R_0}}{a_j^4(1-\sgm)^2+R_0^{-2}\sin^2c}.\eqno(8.5.25)$$ Thus we
have the following solutions of (8.5.12) and (8.5.13):

\begin{eqnarray*}\hat f&=&\sgm R\sum_{j=1}^m
a_jd_je^{a_j^2t+a_j\varpi}\frac{\cos
c\:\left[a_j^2(1-\sgm)\sin\frac{t\sin c}{R_0}-R_0^{-1}\sin
c\:\cos\frac{t\sin c}{R_0}\right]}{a_j^4(1-\sgm)^2+R_0^{-2}\sin^2c}
\\ & &+\sum_{r=1}^n \hat a_r\hat d_re^{\hat a_r^2\sgm t\cos
2\hat b_r+\hat a_r\varpi\cos \hat b_r}\sin(\hat a_r^2\sgm t\sin
2\hat b_r+\hat a_r\varpi\sin \hat b_r+\hat b_r+\hat
c_r),\hspace{1.1cm}(8.5.26)\end{eqnarray*}
\begin{eqnarray*}\hat \phi&=&\sgm R\sum_{j=1}^m
a_jd_je^{a_j^2t+a_j\varpi}\frac{\cos
c\:\left[a_j^2(\sgm-1)\cos\frac{t\sin c}{R_0}-R_0^{-1}\sin
c\:\sin\frac{t\sin c}{R_0}\right]}{a_j^4(1-\sgm)^2+R_0^{-2}\sin^2c}
\\ &
&+\sum_{s=1}^k\td a_s\td d_se^{\td a_s^2\sgm t\cos 2\td b_s+\td
a_s\varpi\cos \td b_s}\sin(\td a_s^2\sgm t\sin 2\td b_s+\td
a_s\varpi\sin \td b_s+\td b_s+\td
c_s),\hspace{1.1cm}(8.5.27)\end{eqnarray*} where $\hat a_r,\hat
b_r,\hat c_r,\hat d_r, \td a_s,\td b_s,\td c_s,\td d_s$
 are arbitrary  real numbers.

  According to (8.5.11),
\begin{eqnarray*}f&=&\cos\frac{t\sin c}{R_0}\;\sum_{r=1}^n \hat a_r\hat d_re^{\hat a_r^2\sgm t\cos 2\hat b_r+\hat
a_r\varpi\cos \hat b_r}\sin(\hat a_r^2\sgm t\sin 2\hat b_r+\hat
a_r\varpi\sin \hat b_r+\hat b_r+\hat c_r) \\&& +\sin\frac{t\sin
c}{R_0}\;\sum_{s=1}^k\td a_s\td d_se^{\td a_s^2\sgm t\cos 2\td
b_s+\td a_s\varpi\cos \td b_s}\sin(\td a_s^2\sgm t\sin 2\td b_s+\td
a_s\varpi\sin \td b_s+\td b_s+\td c_s)\\
& &-\sgm R\sum_{j=1}^m\frac{a_jd_je^{a_j^2t+a_j\varpi}\sin
2c}{2R_0(a_j^4(1-\sgm)^2+R_0^{-2}\sin^2c)}
,\hspace{6.3cm}(8.5.28)\end{eqnarray*}
\begin{eqnarray*}\phi&=&-\sin\frac{t\sin c}{R_0}\;\sum_{r=1}^n \hat a_r\hat d_re^{\hat a_r^2\sgm t\cos 2\hat b_r+\hat
a_r\varpi\cos \hat b_r}\sin(\hat a_r^2\sgm t\sin 2\hat b_r+\hat
a_r\varpi\sin \hat b_r+\hat b_r+\hat c_r)\\ & &+\cos\frac{t\sin
c}{R_0}\;\sum_{s=1}^k\td a_s\td d_se^{\td a_s^2\sgm t\cos 2\td
b_s+\td a_s\varpi\cos \td b_s}\sin(\td a_s^2\sgm t\sin 2\td b_s+\td
a_s\varpi\sin \td b_s+\td b_s+\td
c_s)\\
& &+\sgm R\sum_{j=1}^m\frac{a_j^3d_j(\sgm-1)e^{a_j^2t+a_j\varpi}\cos
c}{a_j^4(1-\sgm)^2+R_0^{-2}\sin^2c}
.\hspace{7.2cm}(8.5.29)\end{eqnarray*}

By (8.5.4)-(8.5.6), (8.5.9) and (8.5.10), $\Phi_1=0$,
$$\Phi_2=\left(\frac{\cos c}{R_0}f-\sgm R\psi\sin c\right)\cos
c,\eqno(8.5.30)$$
$$\Phi_3=\left(\frac{\cos c}{R_0}f-\sgm R\psi\sin c\right)\sin
c-\sgm Rz.\eqno(8.5.31)$$ Thanks to (8.2.19),
\begin{eqnarray*}p&=&\frac{R\cos^2 c}{\sin c}\cos\frac{2t\sin
c}{R_0}\;\sum_{j=1}^m d_je^{a_j^2t\cos 2b_j+a_j\varpi\cos
b_j}\sin(a_j^2t\sin 2b_j+a_j\varpi\sin b_j+c_j)
\\ & &-\frac{\cos c}{ R_0}\cos\frac{t\sin c}{R_0}\;\sum_{r=1}^n \hat d_re^{\hat a_r^2 t\cos 2\hat b_r+\hat
a_r\varpi\cos \hat b_r}\sin(\hat a_r^2t\sin 2\hat b_r+\hat
a_r\varpi\sin \hat b_r+\hat c_r) \\&& -\frac{\cos c}{
R_0}\sin\frac{t\sin c}{R_0}\;\sum_{s=1}^k\td d_se^{\td a_s^2t\cos
2\td b_s+\td a_s\varpi\cos \td b_s}\sin(\td a_s^2 t\sin 2\td b_s+\td
a_s\varpi\sin \td b_s+\td c_s)\\ & &+R\sin c\:\sum_{j=1}^m
d_je^{a_j^2t\cos 2b_j+a_j\varpi\cos b_j}\sin(a_j^2t\sin
2b_j+a_j\varpi\sin
b_j+c_j)+\frac{R}{2}z^2\hspace{0.7cm}(8.5.32)\end{eqnarray*}
 if $\sgm=1$, and
\begin{eqnarray*}p&=&-\frac{\cos c}{\sgm R_0}\cos\frac{t\sin c}{R_0}\;\sum_{r=1}^n \hat d_re^{\hat a_r^2\sgm t\cos 2\hat b_r+\hat
a_r\varpi\cos \hat b_r}\sin(\hat a_r^2\sgm t\sin 2\hat b_r+\hat
a_r\varpi\sin \hat b_r+\hat c_r) \\&& -\frac{\cos c}{\sgm
R_0}\sin\frac{t\sin c}{R_0}\;\sum_{s=1}^k\td d_se^{\td a_s^2\sgm
t\cos 2\td b_s+\td a_s\varpi\cos \td b_s}\sin(\td a_s^2\sgm t\sin
2\td b_s+\td
a_s\varpi\sin \td b_s+\td c_s)\\
& &+ \sum_{j=1}^m\frac{d_jRe^{a_j^2t+a_j\varpi}\sin 2c\;\cos
c}{2R_0^2(a_j^4(1-\sgm)^2+R_0^{-2}\sin^2c)}+R\sin c\;\sum_{j=1}^m
d_je^{a_j^2t+a_j\varpi}+\frac{R}{2}z^2
\hspace{1.6cm}(8.5.33)\end{eqnarray*} when $\sgm\neq 1$.

In summary, we have:\psp

{\bf Theorem 8.5.1}. {\it Let $a_j,b_j,c_j,\hat a_r,\hat b_r,\hat
c_r,\hat d_r, \td a_s,\td b_s,\td c_s,\td d_s,c$ be arbitrary real
numbers. Denote $\varpi=y\cos x+z\sin c$. We have the following
solutions of the three-dimensional stratified rotating Boussinesq
equations (8.2.1)-(8.2.5):
$$u=f,\qquad
v=\phi\sin c,\qquad w=-\phi\cos c,\qquad T=\psi+z,\eqno(8.5.34)$$
where (1) $\sgm=1$, $f$ is given in (8.5.18), $\phi$ is given in
(8.5.19), $\psi$ is given in (8.5.15) and $p$ is given in (8.5.32);
(2)  $\sgm\neq 1$, $f$ is given in (8.5.28), $\phi$ is given in
(8.5.29), $\psi$ is given in (8.5.20) and $p$ is given in (8.5.33).}
\psp

{\bf Remark 8.5.2}. By Fourier expansion, we can use the above
solution to obtain the one depending on three arbitrary piecewise
continuous functions of $\varpi$.

\chapter{Navier-Stokes Equations}

In this chapter, we introduce a method of imposing asymmetric
conditions on the velocity vector with respect to independent
spacial variables and a method of moving frame for solving the three
dimensional Navier-Stokes equations. Seven families of non-steady
rotating asymmetric solutions with various parameters are obtained.
In particular, one family of solutions blow up  on a moving plane,
which may be used to study abrupt high-speed rotating flows. Using
Fourier expansion and two families of our solutions, one can obtain
discontinuous solutions that may be useful in study of shock waves.
Another family of solutions are partially cylindrical invariant,
containing two parameter functions in $t$, which may be used to
describe incompressible fluid in a nozzle. Most of our solutions are
globally analytic with respect to spacial variables. The results are
due to our work [X12]. Cao [Cb3] applied our approaches to the
magnetohydrodynamic  equations of incompressible viscous fluids with
finite electrical conductivity, which describe the motion of viscous
electrically conducting fluids in a magnetic field.

\section{Background and Symmetry}

The most fundamental differential equations in the motion of
incompressible viscous fluid are the Navier-Stokes
equations:\index{Navier-Stokes equations}
$$u_t+uu_x+vu_y+wu_z+\frac{1}{\rho}p_x=\nu (u_{xx}+u_{yy}+u_{zz})
 ,\eqno(9.1.1)$$
$$v_t+uv_x+vv_y+wv_z+\frac{1}{\rho}p_y=\nu (v_{xx}+v_{yy}+v_{zz})
 ,\eqno(9.1.2)$$
$$w_t+uw_x+vw_y+ww_z+\frac{1}{\rho}p_z=\nu (w_{xx}+w_{yy}+w_{zz})
 ,\eqno(9.1.3)$$
$$u_x+v_y+w_z=0,\eqno(9.1.4)$$
 where
$(u,v,w)$ stands for the velocity vector of the fluid, $p$ stands
for the pressure of the fluid, $\rho$ is the density constant and
$\nu$ is the coefficient constant of the kinematic viscosity.

Assuming nullity of certain components of the tensor of momentum
flow density, Landau [Ll] (1944) found an exact solution of the
Navier-Stokes equations (9.1.1)-(9.1.4), which describes axially
symmetrical jet discharging from a thin pipe into unbounded space.
Moreover, Kapitanskii [Kl] (1978) found certain cylindrical
invariant solutions of the equations and Yakimov [Y] (1984) obtained
exact solutions with a singularity of the type of a vortex filament
situated on a half line. Furthermore, Gryn [Gv] (1991) obtained
certain exact solution describing flows between porous walls in the
presence of injection and suction at identical rates. Brutyan and
Karapivskii [BK] (1992) got exact solutions describing the evolution
of a vortex structure in a generalized shear flow. Leipnik [Lr]
(1996) obtained exact solutions by recursive series of diffusive
quotients. In addition, Polyanin [Pa] (2001) used the method of
generalized separation of variables to find certain exact solutions
and Vyskrebtsov [Vv] (2001) studied self-similar solutions for an
axisymmetric flow of a viscous incompressible flow. There also are
other works on exact solutions on the Navier-Stokes equations (e.g.,
cf. [Bv, Pv, Sh1, Sh2]).

A $3\times 3$ real matrix $A$ is called {\it
orthogonal}\index{orthogonal matrix} if $A^TA=AA^T=I_3$, where the
up-index ``$T$" denotes the transpose of matrix. To show that the
Navier-Stokes  are invariant under the orthogonal transformation, we
need to rewrite the Navier-Stokes equations in terms of matrices and
column  vectors (which are also viewed as special matrices). Denote
$$\vec u=\left(\begin{array}{c}u\\ v\\ w\end{array}\right),\;\;\;\vec
x=\left(\begin{array}{c}x\\ y\\
z\end{array}\right)=\left(\begin{array}{c}x_1\\ x_2\\
x_3\end{array}\right),\eqno(9.1.5)$$
$$\nabla=\left(\begin{array}{c}\ptl_x\\ \ptl_y\\
\ptl_z\end{array}\right),\;\;\Dlt=\nabla^T\nabla=\ptl_x^2+\ptl_y^2+\ptl_z^2.\eqno(9.1.6)$$
Note $\vec u^T\nabla=u\ptl_x+v\ptl_y+w\ptl_w$. Then (9.1.1)-(9.1.3)
become
$$\vec u_t+(\vec u^T\nabla)(\vec u)
+\frac{1}{\rho}\nabla(p)=\nu\Dlt(\vec u)\eqno(9.1.7)$$ and (9.1.4)
changes to
$$\nabla^T\vec u=0.\eqno(9.1.8)$$
For a  $3\times 3$ orthogonal matrix $A=(a_{r,s})_{3\times 3}$, we
define
$$T_A(\vec u(t,\vec x^T))=A\vec u(t,\vec x^TA),\;\;T_A(p(t,\vec
x^T))=p(t,\vec x^TA).\eqno(9.1.9)$$ Note that for any function
$f(t,\vec x)$ in $t,x,y,z$,
$$\nabla(f(t,\vec x^TA))=
A\left(\begin{array}{c}f_x(t,\vec x^TA)\\ f_y(t,\vec x^TA)\\
f_z(t,\vec x^TA)\end{array}\right)=A\nabla(f)(t,\vec
x^TA),\eqno(9.1.10)$$ equivalently,
$$\ptl_{x_r}(f(t,\vec x^TA))=\sum_{s=1}^sa_{r,s}f_{x_s}(t,\vec
x^TA)\qquad\for\;\;r\in\ol{1,3}.\eqno(9.1.11)$$
\begin{eqnarray*}& &\Dlt(f(t,\vec
x^TA))\\ &=&(\nabla^T\nabla)(f(t,\vec
x^TA))=\nabla^T[\nabla(f(t,\vec
x^TA))]=\nabla^T[(A\nabla(f))(t,\vec x^TA)]\\
&=&[\nabla^TA^T(A\nabla(f))](t,\vec x^TA)
=[\nabla^T\nabla(f)](t,\vec x^TA)=\Dlt(f)(t,\vec
x^TA).\hspace{2.4cm}(9.1.12)\end{eqnarray*} Now
\begin{eqnarray*}& &\ptl_t(T_A(\vec u))+(T_A(\vec
u))^T\nabla(T_A(\vec u)))+\frac{1}{\rho}\nabla(T_A(p))
\\ &=&\ptl_t(A\vec u(t,\vec x^TA))+\vec u^T(t,\vec
x^TA)A^T\nabla(A\vec u(t,\vec x^TA)) +\frac{1}{\rho}\nabla(p(t,\vec
x^TA))
\\&=&A\vec u_t(t,\vec x^TA)+[(\vec u^T(t,\vec
x^TA)A^TA\nabla)(A\vec u)](t,\vec x^TA)
+\frac{1}{\rho}A\nabla(p)(t,\vec x^TA)
\\&=&A\vec u_t(t,\vec x^TA)+[(\vec u^T(t,\vec
x^TA)\nabla)(A\vec u)](t,\vec x^TA))
+\frac{1}{\rho}A\nabla(p)(t,\vec x^TA)
\\&=&A\vec u_t(t,\vec x^TA)+A(\vec u^T(t,\vec
x^TA)\nabla)(\vec u)(t,\vec x^TA) +\frac{1}{\rho}A\nabla(p)(t,\vec
x^TA)
\\&=&A\left[\vec u_t(t,\vec x^TA)+(\vec u^T(t,\vec
x^TA)\nabla)(\vec u)(t,\vec x^TA) +\frac{1}{\rho}\nabla(p)(t,\vec
x^TA)\right],\hspace{2.5cm}(9.1.13)
\end{eqnarray*}
$$\nu\Dlt((T_A(\vec u)))=\nu\Dlt(A\vec u(t,\vec x^TA)) =\nu
A\Dlt(\vec u(t,\vec x^TA)=A[\nu\Dlt(\vec u)(t,\vec
x^TA))]\eqno(9.1.14)$$ by (9.1.12), and
\begin{eqnarray*}\qquad\qquad\nabla^T(T_A(\vec u))&=&\nabla^T(A\vec u(t,\vec x^TA))=A\nabla^T(\vec
u(t,\vec x^TA))\\ &=&AA^T(\nabla^T\vec u)(t,\vec x^TA)=(\nabla^T\vec
u)(t,\vec x^TA).\hspace{3.3cm}(9.1.15)
\end{eqnarray*}

If $[\vec u(t,x,y,z),p(t,x,y,z)]$ is a solution of the Navier-Stokes
equations (9.1.1)-(9.1.4), then $$\vec u_t(t,\vec x^TA)+ (\vec
u^T(t,\vec x^TA)\nabla)(\vec u)(t,\vec
x^TA)+\frac{1}{\rho}\nabla(p)(t,\vec x^TA)=\nu\Dlt(\vec u)(t,\vec
x^TA)\eqno(9.1.16)$$ and
$$(\nabla^T\vec u)(t,\vec x^TA)=0.\eqno(9.1.17)$$ Thus
$$\ptl_t(T_A(\vec u))+(T_A(\vec
u))^T\nabla(T_A(\vec
u)))+\frac{1}{\rho}\nabla(T_A(p))=\nu\Dlt((T_A(\vec u)))
\eqno(9.1.18)$$ by (9.1.13) and (9.1.14). Moreover, (9.1.15) implies
$$\nabla^T(T_A(\vec u))=0.\eqno(9.1.19)$$
Therefore, $[T_A(\vec u),T_A(p)]$ is also a solution of the the
Navier-Stokes equations (9.1.1)-(9.1.4), that is, $T_A$ is a
symmetry of the equations.

Let us do the degree analysis. Due to the term $\Dlt(u)$ in (9.1.1),
we assume
$$\mbox{deg}\:x=\mbox{deg}\:y=\mbox{deg}\:z=\ell_1.\eqno(9.1.20)$$
Moreover, to make the nonzero terms in (9.1.4) to have the same
degree, we have to take
$$\mbox{deg}\:u=\mbox{deg}\:v=\mbox{deg}\:w=\ell_2.\eqno(9.1.21)$$
Note that in (9.1.1),
$$\mbox{deg}\:u_t=\mbox{deg}\;uu_x=\mbox{deg}\;p_x=\mbox{deg}\:\Dlt(u).\eqno(9.1.22)$$
Thus
$$\mbox{deg}\;t=2\ell_1=-\mbox{deg}\;p,\qquad\ell_2=-\ell_1.\eqno(9.1.23)$$
Moreover, the Navier-Stokes equations are translation invariant
because they do not contain variable coefficients. Hence the
transformation
$$T_{a,b}(\vec u(t,x,y,z))=b\vec u(b^2t+a,bx,by,bz),\eqno(9.1.24)$$
$$T_{a,b}(p(t,x,y,z))=b^2p(b^2t+a,bx,by,bz)\eqno(9.1.25)$$ keeps the
Navier-Stokes equations invariant for $a,b\in\mbb{R}$ with $b\neq
0$, that is, $T_{a,b}$ maps a solution of (9.1.1)-(9.1.4) to another
solution.

Let $\al$ be a function in $t$. Note that the transformation
$$\vec u(t,x,y,z)\mapsto \vec u(t,x+\al,y,z),\;\;p(t,x,y,z)\mapsto
p(t,x+\al,y,z)\eqno(9.1.26)$$ changes the equation (9.1.7) to
$$\vec u_t^T+\al'\vec u_x^T+\vec u^T(\nabla(u),\nabla(v),\nabla(w))
+\frac{1}{\rho}\nabla^T(p)=\nu\Dlt(\vec u^T)\eqno(9.1.27)$$ and
keeps (9.1.4) invariant, where the independent variable $x$ is
replaced by $x+\al$ and the partial derivatives are with respect to
the original variables.
 On the other
hand, the transformation
$$\vec u^T(t,x,y,z)\mapsto \vec u^T(t,x,y,z)-(\al',0,0),\;\;p(t,x,y,z)\mapsto
p(t,x,y,z)+\rho{\al'}'x\eqno(9.1.28)$$ changes the equation (9.1.7)
to
$$\vec u_t^T+\vec u^T(\nabla(u),\nabla(v),\nabla(w))-\al'\vec u_x^T
+\frac{1}{\rho}\nabla^T(p)=\nu\Dlt(\vec u^T)\eqno(9.1.29)$$  by
(9.1.1)-(9.1.3) and keeps (9.1.4) invariant. Thus the transformation
$$T_{1,\al}(\vec u^T(t,x,y,z))=\vec
u^T(t,x+\al,y,z)-(\al',0,0),\eqno(9.1.30)$$ $$
T_{1,\al}(p(t,x,y,z))= p(t,x+\al,y,z)+\rho{\al'}'x\eqno(9.1.31)$$ is
a symmetry of the Navier-Stokes equations. Symmetrically, we have
that the transformation
$$T_{\al_1,\al_2,\al_3;\be}(\vec u^T(t,x,y,z))=\vec
u^T(t,x+\al_1,y+\al_2,z+\al_3)-(\al_1',\al_2',\al_3'),\eqno(9.1.32)$$
$$T_{\al_1,\al_2,\al_3;\be}(p(t,x,y,z))
=p(t,x+\al_1,y+\al_2,z+\al_3)+\rho({\al_1'}'x+{\al_2'}'y+{\al_3'}'z)+\be\eqno(9.1.33)$$
is a symmetry of the Navier-Stokes equations for any functions
$\al_1,\al_2,\al_3$ and $\be$ in $t$.

\section{Asymmetric Approaches }

In this section, we will solve the incompressible Navier-Stokes
equations (9.1.1)-(9.1.4) by imposing asymmetric assumptions on
$u,\;v$ and $w$.

\index{asymmetric approach!for Navier-Stokes equations}

For convenience of computation, we denote
$$\Phi_1=u_t+uu_x+vu_y+wu_z-\nu
(u_{xx}+u_{yy}+u_{zz}),\eqno(9.2.1)$$
$$\Phi_2=v_t+uv_x+vv_y+wv_z-\nu
(v_{xx}+v_{yy}+v_{zz}),\eqno(9.2.2)$$
$$\Phi_3=w_t+uw_x+vw_y+ww_z-\nu (w_{xx}+w_{yy}+w_{zz})
.\eqno(9.2.3)$$ Then the Navier-Stokes equations become
 $$\Phi_1+\frac{1}{\rho}p_x=0,\qquad
 \Phi_2+\frac{1}{\rho}p_y=0,\qquad
\Phi_3+\frac{1}{\rho}p_z=0
 \eqno(9.2.4)$$
 and $u_x+v_y+w_z=0.$ Our strategy is first to solve the following
 compatibility conditions:
 $$\ptl_y(\Phi_1)=\ptl_x(\Phi_2),\qquad
 \ptl_z(\Phi_1)=\ptl_x(\Phi_3),\qquad\ptl_z(\Phi_2)=\ptl_y(\Phi_3)
 \eqno(9.2.5)$$
and then find $p$ via (9.2.4).

Let us first look for simplest non-steady solutions of the
Navier-Stokes equations (indeed, the corresponding Euler equations)
that are not rotation free. This will help the reader to better
understand our later approaches.  Assume
$$u=\gm_1x-\al_1y-\al_2z,\;\;v=\al_1x+\gm_2y-\al_3z,\;\;w=\al_2x+\al_3y
+\gm_3z,\eqno(9.2.6)$$ where $\al_j$ and $\gm_j$ are functions in
$t$ such that $\gm_1+\gm_2+\gm_3=0$.  Then
$$\Phi_1=(\gm_1'+\gm_1^2-\al_1^2-\al_2^2)x-(\al_1'-\al_1\gm_3
+\al_2\al_3)y+(\al_1\al_3-\al_2'+\al_2\gm_2)z,\eqno(9.2.7)$$
$$\Phi_2=(\al_1'-\al_1\gm_3-\al_2\al_3)x+(\gm_2'+\gm_2^2-\al_1^2
-\al_3^2)y-(\al_3'+\al_1\al_2-\al_3\gm_1)z,\eqno(9.2.8)$$
$$\Phi_3=(\al_2'+\al_1\al_3-\al_2\gm_2)x
+(\al_3'-\al_1\al_2-\al_3\gm_1)y+(\gm_3'+\gm_3^2-\al_2^2
-\al_3^2)z.\eqno(9.2.9)$$ Furthermore,
$$\ptl_y(\Phi_1)=\ptl_x(\Phi_2)\lra\gm_3=\frac{\al_1'}{\al_1},\eqno(9.2.10)$$
$$\ptl_z(\Phi_1)=\ptl_x(\Phi_3)\lra
\gm_2=\frac{\al_2'}{\al_2},\eqno(9.2.11)$$
$$\ptl_z(\Phi_2)=\ptl_y(\Phi_3)\lra
\gm_1=\frac{\al_3'}{\al_3}.\eqno(9.2.12)$$

Note
$$\gm_1+\gm_2+\gm_3=0\sim \frac{\al_1'}{\al_1}+
\frac{\al_2'}{\al_2}+\frac{\al_3'}{\al_3}=0\sim
\al_1\al_2\al_3=c\eqno(9.2.13)$$ for some real constant. Moreover,
$$\Phi_1=({\al_3'}'\al_3^{-1}
-\al_1^2-\al_2^2)x-\al_2\al_3y+\al_1\al_3z,\eqno(9.2.14)$$
$$\Phi_2=-\al_2\al_3x+({\al_2'}'\al_2^{-1}-\al_1^2
-\al_3^2)y-\al_1\al_2z,\eqno(9.2.15)$$
$$\Phi_3=\al_1\al_3x
-\al_1\al_2y+({\al_1'}'\al^{-1}_1-\al_2^2 -\al_3^2)z.\eqno(9.2.16)$$
By (9.2.4),
\begin{eqnarray*}p&=&\frac{\rho}{2}[
(\al_1^2+\al_2^2-{\al_3'}'\al_3^{-1})x^2+
(\al_1^2+\al_3^2-{\al_2'}'\al_2^{-1})y^2+
(\al_2^2+\al_3^2-{\al_1'}'\al_1^{-1})z^2]\\ &
&+\rho(\al_2\al_3xy-\al_1\al_3xz+\al_1\al_2yz),\hspace{7.4cm}(9.2.17)\end{eqnarray*}
after replacing $p$ by some $T_{0,0,0;\be}(p)$ if necessary (cf.
(9.1.32) and (9.1.33)).\psp

{\bf Proposition 9.2.1}. {\it Let $\al_1,\;\al_2$ and $\al_3$ be
functions in $t$ such that $\al_1\al_2\al_3=c$ for some real
constant $c$. Then we have the following solution of the
Navier-Stokes equations (9.1.1)-(9.1.4):
$$u=\frac{{\al_3}'}{\al_3}x-\al_1y-\al_2z,\;\;v=\al_1x
+\frac{{\al_2}'}{\al_2}y -\al_3z,\;\;w=\al_2x+\al_3y
+\frac{{\al_1}'}{\al_1}z\eqno(9.2.18)$$ and $p$ is given in
(9.2.17).}\psp

 Next we assume
$$v=-\frac{{\be'}'}{2\be'} y,\qquad w=\psi(t,z),\eqno(9.2.19)$$
where $\be$ is a function in $t$, $\psi$ is a function of $t,z$ and
$v$ is so written just for computational convenience by our earlier
experience. According to (9.1.4),
$$u=f(t,y,z)+\left(\frac{{\be'}'}{2\be'}-\psi_z\right)x\eqno(9.2.20)$$
for some function $f$ of $t,y,z$. Then
\begin{eqnarray*}\hspace{1cm}\Phi_1&=&
f_t+f\left(\frac{{\be'}'}{2\be'}-\psi_z\right)-\frac{{\be'}'}{2\be'}
y f_y+\psi f_z-\nu(f_{yy}+f_{zz})
\\ & &+\left[\left(\frac{{\be'}'}{2\be'}-\psi_z\right)^2
+\frac{\be'{{\be'}'}'-{{\be'}'}^2}{2{\be'}^2}
-\psi_{zt}-\psi\psi_{zz}+\nu\psi_{zzz}\right]x,
\hspace{1.8cm}(9.2.21)\end{eqnarray*}
$$\Phi_2=\frac{(3{{\be'}'}^2-2\be'{{\be'}'}')y}{4{\be'}^2},\qquad
\Phi_3=\psi_t+\psi\psi_z-\nu\psi_{zz}.\eqno(9.2.22)$$ Thus (9.2.5)
is equivalent to the following equations:
$${\cal T}\left[f_t+f\left(\frac{{\be'}'}{2\be'}-\psi_z\right)-\frac{{\be'}'}{2\be'}
y f_y+\psi f_z-\nu(f_{yy}+f_{zz})\right]=0,\eqno(9.2.23)$$
$${\cal T}\left[\psi_z^2-\frac{{\be'}'}{\be'}\psi_z
-\psi_{zt}-\psi\psi_{zz}+\nu\psi_{zzz}\right]=0\eqno(9.2.24)$$ with
${\cal T}=\ptl_y,\;\ptl_z$. The above two equations are equivalent
to
$$f_t+f\left(\frac{{\be'}'}{2\be'}-\psi_z\right)-\frac{{\be'}'}{2\be'}
y f_y+\psi f_z-\nu(f_{yy}+f_{zz})=\tau_1,\eqno(9.2.25)$$
$$\psi_z^2-\frac{{\be'}'}{\be'}\psi_z
-\psi_{zt}-\psi\psi_{zz}+\nu\psi_{zzz}=\tau_2\eqno(9.2.26)$$ for
some functions $\tau_1$ and $\tau_2$ in $t$.

We solve (9.2.26) first. The idea is to linearize it. Note  that
$$\psi=e^{\nu\gm\pm\sqrt{\gm'}z},\;e^{-\nu\gm}\sin\sqrt{\gm'}z,\;e^{-\gm}\cos\sqrt{\gm'}z\eqno(9.2.27)$$
can simplify the expression
$$-\psi_{zt}+\nu\psi_{zzz}\eqno(9.2.28)$$
for any increasing function $\gm$ in $t$ such that $\gm'\not\equiv
0$ . The nonlinear term $\psi_z^2-\psi\psi_{zz}$ hints us to use
$$\xi_0=e^{\nu\gm}(\es_1e^{\sqrt{\gm'}z}-\es_2e^{-\sqrt{\gm'}z}),\qquad\xi_1=
e^{-\nu\gm}\sin(\sqrt{\gm'}z),\eqno(9.2.29)$$
$$\zeta_0=e^{\nu\gm}(\es_1e^{\sqrt{\gm'}z}+\es_2e^{-\sqrt{\gm'}z}),\qquad\zeta_1=
e^{-\nu\gm}\cos(\sqrt{\gm'}z),\eqno(9.2.30)$$ where
$\es_1,\es_2\in\mbb{R}$. In fact,
$$\zeta_0^2-\xi_0^2=4\es_1\es_2e^{2\nu\gm},\qquad
\xi_1^2+\zeta_1^2=e^{-2\nu\gm}.\eqno(9.2.31)$$

Assume
$$\psi=\lmd\xi_r+\mu z,\eqno(9.2.32)$$
where $r=0,1$ and $\lmd,\mu$ are functions in $t$ to be determined.
We calculate
$$\psi_z=\lmd\sqrt{\gm'}\zeta_r+\mu,\;\;\psi_{zz}=(-1)^r\lmd\gm'\xi_r,\;\;
\psi_{zzz}=(-1)^r\lmd{\gm'}^{3/2}\zeta_r,\eqno(9.2.33)$$
$$\psi_{tz}=(-1)^r\lmd\sqrt{\gm'}(\nu\gm'\zeta_r+{\gm'}'z\xi_r/2\sqrt{\gm'})+
(\lmd'\sqrt{\gm'}+\lmd{\gm'}'/2\sqrt{\gm'})\zeta_r.\eqno(9.2.34)$$
Substituting (9.2.33) and (9.2.34) into (9.2.26), we find
\begin{eqnarray*}
&
&\lmd^2\gm'(\zeta_r^2-(-1)^r\xi_r^2)+2\lmd\mu\sqrt{\gm'}\zeta_r+\mu^2-(-1)^r\lmd\mu\gm'z\xi_r
-{\be'}'\lmd\sqrt{\gm'}\zeta_r/\be'-{\be'}'\mu/\be'\\ & &
-(-1)^r\lmd{\gm'}'z\xi_r/2-
(\lmd'\sqrt{\gm'}+\lmd{\gm'}'/2\sqrt{\gm'})\zeta_r=\tau_2,\hspace{5.1cm}(9.2.35)\end{eqnarray*}
equivalently
$$\lmd^2\gm'(\zeta_r^2-(-1)^r\xi_r^2)+\mu^2-{\be'}'\mu/\be'=\tau_2\eqno(9.2.36)$$
by the terms that are independent of spacial variables,
$$-(-1)^r\lmd\mu\gm'-(-1)^r\lmd{\gm'}'/2=0\eqno(9.2.37)$$
by the coefficients of $z\xi_r$ and
$$2\lmd\mu\sqrt{\gm'}
-{\be'}'\lmd\sqrt{\gm'}/\be'-
(\lmd'\sqrt{\gm'}+\lmd{\gm'}'/2\sqrt{\gm'})=0\eqno(9.2.38)$$ by the
coefficients of $\zeta_r$. According to (9.2.37),
$$\mu=-\frac{{\gm'}'}{2\gm'}.\eqno(9.2.39)$$
Substituting it into (9.2.38), we get $$
-{\be'}'\lmd\sqrt{\gm'}/\be'-
\lmd'\sqrt{\gm'}-3\lmd{\gm'}'/2\sqrt{\gm'}=0\lra\lmd=\frac{1}{\be'\sqrt{{\gm'}^3}}.\eqno(9.2.40)$$
So
$$\psi=\frac{\xi_r}{\be'\sqrt{{\gm'}^3}}-\frac{{\gm'}'z}{2\gm'}\eqno(9.2.41)$$
and
$$\tau_2=\frac{4\es_1\es_2
e^{2\nu\gm}\dlt_{r,0}+e^{-2\nu\gm}\dlt_{r,1}}{(\be'\gm')^2}+\frac{{{\gm'}'}^2}{4{\gm'}^2}
+\frac{{\be'}'{\gm'}'}{2\be'\gm'}\eqno(9.2.42)$$ by (9.2.36).
According to (9.2.21), (9.2.25) and (9.2.26),
$$\Phi_1=\tau_1+\left[\frac{2\be'{{\be'}'}'-{{\be'}'}^2}{4{\be'}^2}+
\frac{4\es_1\es_2
e^{2\nu\gm}\dlt_{r,0}+e^{-2\nu\gm}\dlt_{r,1}}{(\be'\gm')^2}+\frac{{{\gm'}'}^2}{4{\gm'}^2}
+\frac{{\be'}'{\gm'}'}{2\be'\gm'}\right]x.\eqno(9.2.43)$$

Substituting (9.2.41) into (9.2.25), we find
$$f_t+\left(\frac{{\be'}'}{2\be'}+\frac{{\gm'}'}{2\gm'}
\right)f+\frac{\xi_rf_z-\sqrt{\gm'}\zeta_rf}{\be'\sqrt{{\gm'}^3}}
-\frac{{\be'}'}{2\be'} y f_y-\frac{{\gm'}'}{2\gm'}zf_z
-\nu(f_{yy}+f_{zz})=\tau_1.\eqno(9.2.44)$$ We assume
$$f=\frac{g(t,\varpi)\zeta_r}{\sqrt{\be'\gm'}},\qquad
\varpi=\sqrt{\be'}y,\eqno(9.2.45)$$ where $g(t,\varpi)$ is a
two-variable function to be determined. We calculate
$$f_t=\frac{g_t\zeta_r}{\sqrt{\be'\gm'}}-\left(\frac{{\be'}'}{2\be'}+\frac{{\gm'}'}{2\gm'}
\right)f+(-1)^r\frac{{\gm'}'zg\xi_r}{2\gm'\sqrt{\be'}}+\frac{(-1)^r\nu\gm'g\zeta_r}{\sqrt{\be'\gm'}}
+\frac{{\be'}'yg_\varpi\zeta_r}{2\be'\sqrt{\gm'}},\eqno(9.2.46)$$
$$f_y=\frac{g_\varpi\zeta_r}{\sqrt{\gm'}},\qquad
f_{yy}=\frac{\sqrt{\be'}g_{\varpi\varpi}\zeta_r}{\sqrt{\gm'}},\eqno(9.2.47)$$
$$f_z=\frac{(-1)^rg\xi_r}{\sqrt{\be'}},\qquad
f_{zz}=\frac{(-1)^r\sqrt{\gm'}g\zeta_r}{\sqrt{\be'}}.\eqno(9.2.48)$$
Substituting (9.2.46)-(9.2.48) into (9.2.44), we get
$$\frac{g_t\zeta_r}{\sqrt{\be'\gm'}}+\frac{((-1)^r\xi_r^2-\zeta_r^2)g}{\sqrt{(\be'\gm')^3}}
 -\frac{\nu\sqrt{\be'}g_{\varpi\varpi}\zeta_r}{\sqrt{\gm'}}=\tau_1.\eqno(9.2.49)$$
\pse

{\it Case 1}. $g=a\in\mbb{R}$. \psp

 In this case
$$f=\frac{a\zeta_r}{\sqrt{\be'\gm'}},\;\;\tau_1=\frac{((-1)^r\xi_r^2-\zeta_r^2)g}{\sqrt{(\be'\gm')^3}}
=-\frac{a(4\es_1\es_2
e^{2\nu\gm}\dlt_{r,0}+e^{-2\nu\gm}\dlt_{r,1})}{\sqrt{(\be'\gm')^3}}\eqno(9.2.50)$$
by (9.2.49) \pse

{\it Case 2}. $r=0=\es_2$ and $\es_1=1$.\psp

In this case, $\tau_1=0$ and
$$g_t-\nu\be'g_{\varpi\varpi}=0\eqno(9.2.51)$$
by (6.2.49). So
$$g=e^{\nu((a+ci)^2\be)+(a+ci)\varpi}\eqno(9.2.52)$$is
a complex solution of (9.2.51) for any $a,c\in\mbb{R}$. Thus we have
real solutions
$$e^{\nu(a^2-c^2)\be+a\varpi}\sin(2ac\nu\be+c\varpi),\;\;e^{\nu(a^2-c^2)\be+a\varpi}\cos(2ac\nu\be+c\varpi).\eqno(9.2.53)$$
In particular, any linear combination
\begin{eqnarray*}\;\;\;\qquad & &e^{\nu(a^2-c^2)\be+a\varpi}(C_1\sin(2ac\nu\be+c\varpi)+C_2\cos(2ac\nu\be+c\varpi))
\\&=&
be^{\nu(a^2-c^2)\be+a\varpi}\sin(2ac\nu\be+c\varpi+\sta)\hspace{6.2cm}(9.2.54)\end{eqnarray*}
 of them is a solution of (9.2.51), where $C_1,C_2\in\mbb{R}$ and
 $b=\sqrt{C_1^2+C_2^2},\;C_1/b=\cos\sta$. By superposition
 principle, we have more general solution:
 $$g=\sum_{s=1}^nb_se^{\nu(a_s^2-c_s^2)\be+a_s\varpi}\sin(2a_sc_s\nu\be+c_s\varpi+\sta_s)\eqno(9.2.55)$$
 for $a_s,b_s,c_s,\sta_s\in\mbb{R}$ such that $b_s\neq
 0,\;(a_s,c_s)\neq (0,0)$. Recall $\varpi=\sqrt{\be'}y$. Thanks to
 (9.2.45),
$$f=\frac{\zeta_r}{\sqrt{\be'\gm'}}\sum_{s=1}^nb_se^{\nu(a_s^2-c_s^2)\be+a_s\sqrt{\be'}y}
\sin(2a_sc_s\nu\be+c_s\sqrt{\be'}y+\sta_s).\eqno(9.2.56)$$

Next we calculate the pressure $p$ via (9.2.4). First we assume
$g=a$ and $r=1$. In this case,
$$\psi=\frac{e^{-\nu\gm}\sin(\sqrt{\gm'}z)}{\be'\sqrt{{\gm'}^3}}-\frac{{\gm'}'z}{2\gm'}.\eqno(9.2.57)$$
Denote
$$\hat\psi=-\frac{e^{-\nu\gm}\cos(\sqrt{\gm'}z)}{\be'\gm'}-\frac{{\gm'}'z^2}{4\gm'}.\eqno(9.2.58)$$
Then $\hat\psi_z=\psi$. According to (9.2.4), (9.2.22), (9.2.43) and
(9.2.50),
\begin{eqnarray*}
\qquad
p&=&\rho\left(\nu\psi_z-\frac{\psi^2}{2}-\hat\psi_t+\frac{e^{-2\nu\gm}x}{\sqrt{(\be'\gm')^3}}
-\frac{(3{{\be'}'}^2-2\be'{{\be'}'}')y^2}{8{\be'}^2}\right)\\
& &-\frac{\rho
x^2}{2}\left[\frac{2\be'{{\be'}'}'-{{\be'}'}^2}{4{\be'}^2}+
\frac{e^{-2\nu\gm}}{(\be'\gm')^2}+\frac{{{\gm'}'}^2}{4{\gm'}^2}
+\frac{{\be'}'{\gm'}'}{2\be'\gm'}\right].\hspace{3.9cm}(9.2.59)\end{eqnarray*}

Consider the case $g=a$ and $r=0$. We have
$$\psi=\frac{e^{\nu\gm}(\es_1e^{\sqrt{\gm'}z}-\es_2e^{-\sqrt{\gm'}z})}{\be'\sqrt{{\gm'}^3}}
-\frac{{\gm'}'z}{2\gm'}.\eqno(9.2.60)$$ Denote
$$\hat\psi=\frac{e^{\nu\gm}(\es_1e^{\sqrt{\gm'}z}+\es_2e^{-\sqrt{\gm'}z})}{\be'\gm'}
-\frac{{\gm'}'z^2}{4\gm'}.\eqno(9.2.61)$$ According to (9.2.4),
(9.2.22), (9.2.43) and (9.2.56),
\begin{eqnarray*}
\qquad
p&=&\rho\left(\nu\psi_z-\frac{\psi^2}{2}-\hat\psi_t+\frac{4\es_1\es_2e^{2\nu\gm}x}{\sqrt{(\be'\gm')^3}}
-\frac{(3{{\be'}'}^2-2\be'{{\be'}'}')y^2}{8{\be'}^2}\right)\\
& &-\frac{\rho
x^2}{2}\left[\frac{2\be'{{\be'}'}'-{{\be'}'}^2}{4{\be'}^2}+
\frac{4\es_1\es_2e^{2\nu\gm}}{(\be'\gm')^2}+\frac{{{\gm'}'}^2}{4{\gm'}^2}
+\frac{{\be'}'{\gm'}'}{2\be'\gm'}\right].\hspace{3.6cm}(9.2.62)\end{eqnarray*}

Suppose $r=0=\es_2$ and $\es_1=1$. Then the pressure is the
corresponding special case of (9.2.62):
\begin{eqnarray*}
\qquad\qquad p&=&\rho\left(\nu\psi_z-\frac{\psi^2}{2}-\hat\psi_t
-\frac{(3{{\be'}'}^2-2\be'{{\be'}'}')y^2}{8{\be'}^2}\right)\\
& &-\frac{\rho
x^2}{2}\left[\frac{2\be'{{\be'}'}'-{{\be'}'}^2}{4{\be'}^2}+\frac{{{\gm'}'}^2}{4{\gm'}^2}
+\frac{{\be'}'{\gm'}'}{2\be'\gm'}\right]\hspace{5cm}(9.2.63)\end{eqnarray*}
with
$$\psi=\frac{e^{\nu\gm}e^{\sqrt{\gm'}z}}{\be'\sqrt{{\gm'}^3}}
-\frac{{\gm'}'z}{2\gm'},\;\;\hat\psi=\frac{e^{\nu\gm}e^{\sqrt{\gm'}z}}{\be'\gm'}
-\frac{{\gm'}'z^2}{4\gm'}.\eqno(9.2.64)$$\pse

{\bf Theorem 9.2.2}. {\it Let $\al, \be$ and $\gm$ be any functions
in $t$. For any  $0\neq a,\es_1,\es_2\in\mbb{R}$, we have the
following solutions of the Navier Stokes equations (9.1.1)-(9.1.4):
$$u=\frac{ae^{-\nu\gm}\cos(\sqrt{\gm'}z)}{\sqrt{\be'\gm'}}+\left(\frac{{\be'}'}{2\be'}+\frac{{\gm'}'}{2\gm'}
-\frac{e^{-\nu\gm}\cos(\sqrt{\gm'}z)}{\be'\gm}\right)x,\eqno(9.2.65)$$
$$v=-\frac{{\be'}'}{2\be'} y,\;\;\;
w=\frac{e^{-\nu\gm}\sin(\sqrt{\gm'}z)}{\be'\sqrt{{\gm'}^3}}-\frac{{\gm'}'z}{2\gm'},\eqno(9.2.66)$$
and $p$ is given in (9.2.59);
$$u=\frac{ae^{\nu\gm}(\es_1e^{\sqrt{\gm'}z}+\es_2e^{-\sqrt{\gm'}z})
}{\sqrt{\be'\gm'}}+\left(\frac{{\be'}'}{2\be'}+\frac{{\gm'}'}{2\gm'}
-\frac{e^{\nu\gm}(\es_1e^{\sqrt{\gm'}z}+\es_2e^{-\sqrt{\gm'}z})}{\be'\gm}\right)x,\eqno(9.2.67)$$
$$v=-\frac{{\be'}'}{2\be'} y,\;\;\;
w=\frac{e^{\nu\gm}(\es_1e^{\sqrt{\gm'}z}-\es_2e^{-\sqrt{\gm'}z})}{\be'\sqrt{{\gm'}^3}}
-\frac{{\gm'}'z}{2\gm'},\eqno(9.2.68)$$ and $p$ is given in
(9.2.62).

For $a_s,b_s,c_s,\sta_s\in\mbb{R}$ with $s\in\ol{1,n}$ such that
$b_s\neq
 0,\;(a_s,c_s)\neq (0,0)$, we have
the following solutions of the Navier Stokes equations
(9.1.1)-(9.1.4):
\begin{eqnarray*}\qquad u&=&\frac{e^{\nu \gm+\sqrt{\gm'}z}}{\sqrt{\be'\gm'}}\sum_{s=1}^nb_se^{\nu(a_s^2-c_s^2)\be+a_s\sqrt{\be'}y}
\sin(2a_sc_s\nu\be+c_s\sqrt{\be'}y+\sta_s)
\\ & &+\left(\frac{{\be'}'}{2\be'}+\frac{{\gm'}'}{2\gm'}
-\frac{e^{\nu\gm}e^{\sqrt{\gm'}z}}{\be'\gm}\right)x,\hspace{7.1cm}(9.2.69)\end{eqnarray*}
$$v=-\frac{{\be'}'}{2\be'} y,\;\;\;
w=\frac{e^{\nu\gm+\sqrt{\gm'}z}}{\be'\sqrt{{\gm'}^3}}
-\frac{{\gm'}'z}{2\gm'}\eqno(9.2.70)$$ and $p$ is given in
(9.2.63).}
 \psp

{\bf Remark 9.2.3}. We can use Fourier expansion to solve the system
(9.2.51)  for $g(t,\sqrt{\be'}y)$ with given $g(0,\sqrt{\be'(0)}y)$.
In this way, we can obtain discontinuous solutions of the
Navier-Stokes equations (9.1.1)-(9.1.4), which may be useful in
studying shock waves.\psp

For $\sta\in\mbb{R}$, we denote the rotation
$$A=\left(\begin{array}{ccc}1&0&0\\ 0&\cos\sta&\sin\sta\\
0&-\sin\sta&\cos\sta\end{array}\right),\qquad A^T=\left(\begin{array}{ccc}1&0&0\\ 0&\cos\sta&-\sin\sta\\
0&\sin\sta&\cos\sta\end{array}\right).\eqno(9.2.71)$$ Applying $T_A$
in (9.1.9) to the above first solution, we get
\begin{eqnarray*} \qquad
u&=&\frac{ae^{-\nu\gm}\cos(\sqrt{\gm'}(y\sin\sta+z\cos\sta))}{\sqrt{\be'\gm'}}
\\ &&+\left(\frac{{\be'}'}{2\be'}+\frac{{\gm'}'}{2\gm'}
-\frac{e^{-\nu\gm}\cos(\sqrt{\gm'}(y\sin\sta+z\cos\sta)))}{\be'\gm}\right)x,\hspace{3.1cm}(9.2.72)\end{eqnarray*}
\begin{eqnarray*} \qquad
v&=&\left(\frac{e^{-\nu\gm}\sin(\sqrt{\gm'}(y\sin\sta+z\cos\sta))}{\be'\sqrt{{\gm'}^3}}
-\frac{{\gm'}'(y\sin\sta+z\cos\sta)}{2\gm'}\right)\sin\sta\\ & & -
\frac{{\be'}'}{2\be'}(y\cos\sta-z\sin\sta)\cos\sta,\hspace{7.5cm}(9.2.73)\end{eqnarray*}
\begin{eqnarray*} \qquad
w&=&\left(\frac{e^{-\nu\gm}\sin(\sqrt{\gm'}(y\sin\sta+z\cos\sta))}{\be'\sqrt{{\gm'}^3}}
-\frac{{\gm'}'(y\sin\sta+z\cos\sta)}{2\gm'}\right)\cos\sta\\ & & +
\frac{{\be'}'}{2\be'}(y\cos\sta-z\sin\sta)\sin\sta\hspace{7.6cm}(9.2.74)\end{eqnarray*}
and
\begin{eqnarray*}
\qquad p&=&\rho\big[\nu\psi_z(t,y\sin\sta+z\cos\sta)
-\frac{\psi^2(t,y\sin\sta+z\cos\sta)}{2}-\hat\psi_t(t,y\sin\sta+z\cos\sta)\\
& &+\frac{e^{-2\nu\gm}x}{\sqrt{(\be'\gm')^3}}
-\frac{(3{{\be'}'}^2-2\be'{{\be'}'}')(y\cos\sta-z\sin\sta)^2}{8{\be'}^2}\big]\\
& &-\frac{\rho
x^2}{2}\left[\frac{2\be'{{\be'}'}'-{{\be'}'}^2}{4{\be'}^2}+
\frac{e^{-2\nu\gm}}{(\be'\gm')^2}+\frac{{{\gm'}'}^2}{4{\gm'}^2}
+\frac{{\be'}'{\gm'}'}{2\be'\gm'}\right].\hspace{3.9cm}(9.2.75)\end{eqnarray*}

Set
$$\varpi=x^2+y^2.\eqno(9.2.76)$$
Consider
$$u=y\phi(t,\varpi),\qquad v=
-x\phi(t,\varpi),\qquad w=\psi(t,\varpi),\eqno(9.2.77)$$ where
$\phi$ and $\psi$ are functions in $t,\varpi$. Then (9.2.1)-(9.2.3)
give
$$\Phi_1=y\phi_t
-x\phi^2- 4y\nu(\varpi\phi)_{\varpi\varpi},\eqno(9.2.78)$$
$$\Phi_2=-x\phi_t
-y\phi^2+ 4x\nu(\varpi\phi)_{\varpi\varpi},\eqno(9.2.79)$$
$$\Phi_3=\psi_t-4\nu(\psi_\varpi+\varpi\psi_{\varpi\varpi}).\eqno(9.2.80)$$

Note that $\ptl_y(\Phi_1)=\ptl_x(\Phi_2)$ becomes
$$(\varpi\phi)_{\varpi t}
-4\nu((\varpi\phi)_{\varpi\varpi}+
\varpi(\varpi\phi)_{\varpi\varpi\varpi})=0.\eqno(9.2.81)$$ Set
$$\hat\phi=(\varpi\phi)_\varpi.\eqno(9.2.82)$$
Then (9.2.81) becomes
$$\hat\phi_t-
4\nu(\hat\phi_\varpi+\varpi\hat\phi_{\varpi\varpi})=0.\eqno(9.2.83)$$
Suppose that
$$\hat\phi=\sum_{m=0}^\infty a_m(t)\varpi^m,\eqno(9.2.84)$$
where $a_m(t)$ are functions in $t$ to be determined. Then (9.2.83)
becomes
$$\sum_{m=0}^\infty a_m'\varpi^m=4\nu\sum_{m=0}^\infty
m^2a_m\varpi^{m-1},\eqno(9.2.85)$$ equivalently,
$$a_m=\frac{ a_0^{(m)}}{(4\nu)^m(m!)^2}\qquad\for\;\;m\in\mbb{N}.\eqno(9.2.86)$$
Write $\al(t)=a_0(t)$. We have
$$\hat\phi=\sum_{m=0}^{\infty}\frac{
\al^{(m)}\varpi^m}{(4\nu)^m(m!)^2}.\eqno(9.2.87)$$ By (9.2.82), we
get
$$\phi=\be\varpi^{-1}+\sum_{m=0}^\infty\frac{
\al^{(m)}\varpi^m}{(4\nu)^mm!(m+1)!} \eqno(9.2.88)$$ for a function
$\be$ in $t$.

Note
$$\phi_t=\be'\varpi^{-1}+\sum_{m=0}^\infty\frac{
\al^{(m+1)}\varpi^m}{(4\nu)^mm!(m+1)!},\eqno(9.2.89)$$
$$4\nu(\varpi\phi)_{\varpi\varpi}=4\nu\hat\phi_\varpi=
\sum_{m=1}^{\infty}\frac{
\al^{(m)}\varpi^{m-1}}{(4\nu)^{m-1}(m-1)!m!}.\eqno(9.2.90)$$ Thus
$$\phi_t-
4\nu(\varpi\phi)_{\varpi\varpi}=\be'\varpi^{-1}.\eqno(9.2.91)$$
Therefore,
$$\Phi_1=\frac{\be'y}{x^2+y^2}-x\phi^2\eqno(9.2.92)$$ and
$$\Phi_2=-\frac{\be'x}{x^2+y^2}-y\phi^2.\eqno(9.2.93)$$

On the other hand, Equations $\ptl_z(\Phi_1)=\ptl_x(\Phi_3)$ and
$\ptl_z(\Phi_2)=\ptl_y(\Phi_3)$ are implied by the following
differential equation:
$$\psi_t-4\nu(\psi_\varpi+\varpi\psi_{\varpi\varpi})=0\eqno(9.2.94)$$ (cf.
(9.2.80)). Similarly, we have the solution:
$$\psi=\sum_{n=0}^{\infty}\frac{
\gm^{(n)}\varpi^n}{(4\nu)^n(n!)^2},\eqno(9.2.95)$$ where $\gm$ is a
smooth function in $t$. With this $\psi$, $\Phi_3=0$. By (9.2.4),
(9.2.76), (9.2.77),  (9.2.88), (9.2.92), (9.2.93) and (9.2.95), we
obtain:\psp

{\bf Theorem 9.2.4}. {\it Let $\al,\gm$ be any smooth functions in
$t$ and let $\be$ be any differentiable function in $t$. We have the
following solution of the Navier-Stokes equations (9.1.1)-(9.1.4):
$$u=\frac{\be y}{x^2+y^2}+y\sum_{m=0}^\infty\frac{
\al^{(m)}(x^2+y^2)^m}{(4\nu)^mm!(m+1)!},\eqno(9.2.96)$$
$$v=-\frac{\be x}{x^2+y^2}-x\sum_{m=0}^\infty\frac{
\al^{(m)}(x^2+y^2)^m}{(4\nu)^mm!(m+1)!},\eqno(9.2.97)$$
$$w=\sum_{n=0}^{\infty}\frac{
\gm^{(n)}(x^2+y^2)^n}{(4\nu)^n(n!)^2},\eqno(9.2.98)$$
$$p=\rho\be'\arctan\frac{y}{x}
+\rho\sum_{m,n=0}^\infty\frac{\al^{(m)}\al^{(n)}(x^2+y^2)^{m+n+1}}
{2(m+n+1)m!(m+1)!n!(n+1)!(4\nu)^{m+n}}.\eqno(9.2.99)$$}

{\bf Remark 9.2.5}. When $\al$ and $\gm$ are polynomials in $t$, the
summations in the above theorem are finite. Let $\gm_1,\gm_2,\gm_3$
and $\vt$ be functions in $t$. For $\sta\in\mbb{R}$, we the matrices
in (9.2.71). Recall the transformations in (9.1.9) and
(9.1.32)-(9.1.33). Applying $T_AT_{\gm_1,\gm_2,\gm_3;\vt}$ to the
above solution,  we get the following solution of the Navier-Stokes
equations with six parameter functions in $t$:

\begin{eqnarray*}
u&=&(y\cos\sta-z\sin\sta+\gm_2)\sum_{m=0}^\infty\frac{
\al^{(m)}[(x+\gm_1)^2+(y\cos\sta-z\sin\sta+\gm_2)^2]^m}{(4\nu)^mm!(m+1)!}
\\ & &+\frac{\be
(y\cos\sta-z\sin\sta+\gm_2)}{(x+\gm_1)^2+(y\cos\sta-z\sin\sta+\gm_2)^2}-\gm_1'
,\hspace{5.5cm}(9.2.100)\end{eqnarray*}
\begin{eqnarray*}
\qquad v&=&-\big[\sum_{m=0}^\infty\frac{
\al^{(m)}[(x+\gm_1)^2+(y\cos\sta-z\sin\sta+\gm_2)^2]^m}{(4\nu)^mm!(m+1)!}
\\ & &+\frac{\be x
}{(x+\gm_1)^2+(y\cos\sta-z\sin\sta+\gm_2)^2}-\gm_2'\big]\cos\sta-\gm_2'
\\ & &+\sin\sta\;\sum_{n=0}^{\infty}\frac{
\gm^{(n)}[(x+\gm_1)^2+(y\cos\sta-z\sin\sta+\gm_2)^2]^n}{(4\nu)^n(n!)^2}
,\hspace{2.8cm}(9.2.101)\end{eqnarray*}
\begin{eqnarray*}
\qquad w&=&\big[\sum_{m=0}^\infty\frac{
\al^{(m)}[(x+\gm_1)^2+(y\cos\sta-z\sin\sta+\gm_2)^2]^m}{(4\nu)^mm!(m+1)!}
\\ & &+\frac{\be x
}{(x+\gm_1)^2+(y\cos\sta-z\sin\sta+\gm_2)^2}-\gm_2'\big]\sin\sta-\gm_3'
\\ & &+\cos\sta\;\sum_{n=0}^{\infty}\frac{
\gm^{(n)}[(x+\gm_1)^2+(y\cos\sta-z\sin\sta+\gm_2)^2]^n}{(4\nu)^n(n!)^2}
,\hspace{2.7cm}(9.2.102)\end{eqnarray*}
\begin{eqnarray*}
\qquad
p&=&\rho\sum_{m,n=0}^\infty\frac{\al^{(m)}\al^{(n)}[(x+\gm_1)^2+(y\cos\sta-z\sin\sta+\gm_2)^2]^{m+n+1}}
{2(m+n+1)m!(m+1)!n!(n+1)!(4\nu)^{m+n}}\\ &&
+\rho\be'\arctan\frac{y\cos\sta-z\sin\sta+\gm_2}{x}+\rho({\gm_1'}'x+{\gm_2'}'y+{\gm_3'}'z)
+\vt.\hspace{1.6cm}(9.2.103)\end{eqnarray*}

\section{Moving-Frame Approach I}

Let $\al,\be$ be given differentiable functions in $t$. Denote
$$\Upsilon=\left(\begin{array}{ccc}\cos\al&\sin\al\;\cos\be&\sin\al\;\sin\be\\
-\sin\al&\cos\al\;\cos\be&\cos\al\;\sin\be\\ 0&
-\sin\be&\cos\be\end{array}\right)\eqno(9.3.1)$$ and
$$Q=\left(\begin{array}{ccc}0&\al'&\be'\sin\al\\
-\al'&0&\be'\cos\al\\-\be'\sin\al&-\be'\cos\al&0
\end{array}\right).\eqno(9.3.2)$$
Then
$$\Upsilon^{-1}=\Upsilon^T=\left(\begin{array}{ccc}\cos\al&-\sin\al&0\\
\sin\al\;\cos\be &\cos\al\;\cos\be&-\sin\be
\\\sin\al\;\sin\be&\cos\al\;\sin\be&\cos\be\end{array}\right)\eqno(9.3.3)$$
and
$$\frac{d}{dt}(\Upsilon)=Q\Upsilon.\eqno(9.3.4)$$
Define the moving frames:\index{moving-frame!for Navier-Stokes
equations}
$$\vec{\cal U}=\left(\begin{array}{c}{\cal U}\\{\cal V}\\ {\cal
W}\end{array}\right)=\Upsilon\left(\begin{array}{c}u(t,x,y,z)\\ v(t,x,y,z)\\
w(t,x,y,z)\end{array}\right),\qquad \vec{\cal
X}=\left(\begin{array}{c}{\cal X}\\{\cal Y}\\ {\cal
Z}\end{array}\right)=\Upsilon\left(\begin{array}{c}x\\ y\\
z\end{array}\right).\eqno(9.3.5)$$ Set
$$\widetilde{\nabla}^T=(\ptl_{\cal X},\ptl_{\cal Y},\ptl_{\cal
Z}).\eqno(9.3.6)$$ Then
$$\nabla=\Upsilon^T\widetilde\nabla.\eqno(9.3.7)$$
Thus
$$\Dlt=\ptl_x^2+\ptl_y^2+\ptl_z^2=\nabla^T\nabla=(\widetilde\nabla^T\Upsilon)(\Upsilon^T\widetilde\nabla)=
\widetilde\nabla^T\widetilde\nabla=\ptl_{\cal X}^2+\ptl_{\cal Y}^2+
\ptl_{\cal Z}^2,\eqno(9.3.8)$$ Recall the notion in (9.1.5). The
equation (9.3.7) yields
$$u_x+v_y+w_z=\nabla^T\vec
u=(\widetilde\nabla^T\Upsilon)(\Upsilon^{-1}\vec{\cal
U})=\widetilde\nabla^T\vec{\cal U}=\U_\X+\V_\Y+\W_\Z\eqno(9.3.9)$$
and
$$\vec u^T\nabla=(\Upsilon^T\vec {\cal U})^T(\Upsilon^T\widetilde\nabla)=
\vec {\cal U}^T\Upsilon\Upsilon^T\widetilde\nabla=\vec {\cal
U}^T\widetilde\nabla.\eqno(9.3.10)$$ According to (9.3.3) and
(9.3.5),
$$\vec {\cal U}=\Upsilon\vec u(t,\vec x^T)=\Upsilon\vec u(t,\vec {\cal
X}^T\Upsilon),\;\;p(t,\vec x)=p(t,\vec {\cal
X}^T\Upsilon).\eqno(9.3.11)$$ By (9.3.3) and (9.3.4), we get
$$\ptl_t(\vec{\cal X})=\frac{d}{dt}(\Upsilon)\vec x=Q\Upsilon\vec
x=Q\vec{\cal X},\eqno(9.3.12)$$
$$\ptl_t(\vec{\cal U})=\frac{d}{dt}(\Upsilon)\vec u+\Upsilon\vec
u_t=Q\Upsilon\vec u+\Upsilon\vec u_t=Q\vec{\cal U}+\Upsilon\vec u_t
.\eqno(9.3.13)$$ On the other hand,
$$\ptl_t(\vec{\cal U})=\vec{\cal U}_t+(\ptl_t({\cal
X}^T)\widetilde\nabla)(\vec{\cal U})=\vec{\cal U}_t+({\cal
X}^TQ^T\widetilde\nabla)(\vec{\cal U}).\eqno(9.3.14)$$ Thus
$$\Upsilon\vec u_t=\vec{\cal U}_t+({\cal
X}^TQ^T\widetilde\nabla)(\vec{\cal U})-Q\vec{\cal U}.\eqno(9.3.15)$$

Multiplying $\Upsilon$ to (9.1.7) from the left side, we get
$$\Upsilon\vec u_t+(\vec u^T\nabla)(\Upsilon\vec u)
+\frac{1}{\rho}\Upsilon\nabla(p)=\nu\Dlt(\Upsilon\vec
u),\eqno(9.3.16)$$which is equivalent to
$$\vec{\cal U}_t+({\cal
X}^TQ^T\widetilde\nabla)(\vec{\cal U})-Q\vec{\cal U} +(\vec {\cal
U}^T\widetilde\nabla)(\vec{\cal
U})+\frac{1}{\rho}\widetilde\nabla(p)=\nu\Dlt(\vec {\cal U})
\eqno(9.3.17)$$
 by (9.3.7)-(9.3.9) and (9.3.15). Moreover, (9.1.8), (9.3.5) and
 (9.3.7) imply
$$(\widetilde\nabla^T\Upsilon)(\Upsilon^{-1}\vec{\cal U})=0\sim
\widetilde\nabla^T\vec{\cal U}=0 .\eqno(9.3.18)$$

Next we want to find the analogue of (9.2.4). According to (9.3.2),
(9.3.8) and (9.3.17), we denote
\begin{eqnarray*}& &R_1=\U_t+\al'(\Y\U_\X-\X\U_\Y
-\V)+\be'(\Z\U_\X-\X\U_\Z-\W)\sin\al\\
&
&+\be'(\Z\U_\Y-\Y\U_\Z)\cos\al+\U\U_\X+\V\U_\Y+\W\U_\Z-\nu\Dlt(\U),
\hspace{3.6cm}(9.3.19)\end{eqnarray*}
\begin{eqnarray*}& &R_2=\V_t+\al'(\Y\V_\X-\X\V_\Y
+\U)+\be'(\Z\V_\X-\X\V_\Z)\sin\al\\
&&+\be'(\Z\V_\Y-\Y\V_\Z-\W)\cos\al+\U\V_\X+\V\V_\Y+\W\V_\Z-\nu\Dlt(\V),
\hspace{2.7cm}(9.3.20)\end{eqnarray*}
\begin{eqnarray*}& &R_3=\W_t+\al'(\Y\W_\X-\X\W_\Y
)+\be'(\Z\W_\X-\X\W_\Z+\U)\sin\al\\&&+\be'(\Z\W_\Y-\Y\W_\Z+\V)\cos\al
+\U\W_\X+\V\W_\Y+\W\W_\Z-\nu\Dlt(\W),\hspace{1.9cm}(9.3.21)\end{eqnarray*}
Then the Navier-Stokes equations (9.1.1)-(9.1.4) become
$$R_1+\frac{1}{\rho}p_{_\X}=0,\qquad R_2+
\frac{1}{\rho}p_{_\Y}=0,\qquad R_3+
\frac{1}{\rho}p_{_\Z}=0,\eqno(9.3.22)$$
$$\U_\X+\V_\Y+\W_\Z=0\eqno(9.3.23)$$
by (9.3.17) and (9.3.18). Instead of solving the equations in
(9.3.21), we will first solve the following compatibility equations:
$$\ptl_\Y(R_1)=\ptl_\X(R_2),\qquad\ptl_\Z(R_1)=\ptl_\X(R_3),
\qquad \ptl_\Z(R_2)=\ptl_\Y(R_3)\eqno(9.3.24)$$ for $\U,\V,\W$, and
then find $p$ from the equations via (9.3.22).

Let $f,g,h$ be functions of $t,\X,\Y,\Z$ that are linear in
$\X,\Y,\Z$ and $f_\X+g_{_\Y}+h_\Z=0$.  Assume
$$\U=f+6\nu\X^{-1},\qquad\V=g+6\nu\Y\X^{-2},\qquad\W=h.\eqno(9.3.25)$$
Then (9.3.19)-(9.3.21) become
\begin{eqnarray*}\hspace{1.1cm}R_1&=&f_t+ff_\X+f_\Y g+f_\Z h
+6\nu f_\X\X^{-1}+\al'(\Y f_\X-\X f_\Y-g)
\\
& &+\be'(\Z f_\X-\X f_\Z-h)\sin\al
+\be'(\Z f_\Y-\Y f_\Z)\cos\al \\
& &-6\nu (f-\Y
f_\Y+2\al'\Y+\be'\Z\sin\al)\X^{-2}-48\nu^2\X^{-3},\hspace{2.8cm}(9.3.26)\end{eqnarray*}
\begin{eqnarray*}\hspace{0.5cm}R_2&=&g_t+fg_{_\X}+gg_{_\Y}+g_{_\Z}h+
\al'(\Y g_{_\X}-\X g_{_\Y} +f)+\be'(\Z g_{_\X}-\X g_{_\Z})\sin\al\\
& &-6\nu g_{_\X}\X^{-1} +6\nu(g+\be'\Z\cos\al+\Y g_{_\Y})\X^{-2}\\ &
&+\be'(\Z g_{_\Y}-g_{_\Z}\Y-h)\cos\al
-12\nu\Y(f+\al'\Y+\be'\Z\sin\al)\X^{-3},
\hspace{1.5cm}(9.3.27)\end{eqnarray*}
\begin{eqnarray*}\qquad R_3&=&h_t+fh_\X+gh_\Y+hh_\Z
+\al'(\Y h_\X-\X h_\Y )\\ & &+\be'(\Z h_\X-\X h_\Z+f)\sin\al+\be'(\Z
h_\Y-\Y h_\Z+g)\cos\al \\&&+6\nu(h_x+\be'\sin\al)\X^{-1}
+6\nu(h_\Y+\be'\cos\al) \Y\X^{-2}.
\hspace{3.4cm}(9.3.28)\end{eqnarray*} By the coefficients of
$\X^{-4}$ in $\ptl_\Y(R_1)=\ptl_\X(R_2)$, we take
$$f=\gm\X-\al'\Y-\be'\Z\sin\al,\eqno(9.3.29)$$
where $\gm$ is a functions in $t$. Moreover, the coefficients of
$\X^{-3}$ and the coefficients of $\X^{-2}$ in
$\ptl_\Y(R_1)=\ptl_\X(R_2)$ imply
$$g=\al'\X+\gm\Y-\be'\Z\cos\al.\eqno(9.3.30)$$
Furthermore, $\ptl_\Y(R_1)=\ptl_\X(R_2)$ does not contain $\X^{-1}$.

According to the coefficients of $\X^{-2}$ in
$\ptl_\Z(R_2)=\ptl_\Y(R_3)$, we find $h_\Y=-\be'\cos\al$. Moreover,
the coefficients of $\X^{-2}$ in $\ptl_\Z(R_1)=\ptl_\X(R_3)$ force
$h_x=-\be'\sin\al$. The condition $f_\X+g_\Y+h_\Z=0$ implies
$h_\Z=-2\gm$. For simplicity, we take
$$h=-(\be'\X\sin\al+\be'\Y\cos\al+2\gm\Z).\eqno(9.3.31)$$
With the above $f,g$ and $h$, we have:
\begin{eqnarray*}\hspace{1.1cm}R_1&=&(\gm'+\gm^2-{\al'}^2+
3{\be'}^2\sin^2\al)\X +12\nu \al'\Y\X^{-2}-48\nu^2\X^{-3}
\\ & &+(3{\be'}^2\sin\al\;\cos \al-{\al'}'-2\al'\gm)\Y
+(4\be'\gm-{\be'}')\Z\sin\al, \hspace{2cm}(9.3.32)\end{eqnarray*}
\begin{eqnarray*}\hspace{1.1cm}R_2&=&(\gm'+\gm^2-{\al'}^2+
3{\be'}^2\cos^2\al)\Y-12\nu \al'\X^{-1}\\&
&+({\al'}'+2\al'\gm+3{\be'}^2\sin\al\;\cos \al)\X
+(4\be'\gm-{\be'}')\Z\cos\al,\hspace{1.9cm}(9.3.33)\end{eqnarray*}
$$R_3=(4\gm^2-2\gm'-{\be'}^2)\Z+(4\be'\gm-{\be'}')
(\X\sin\al+\Y\cos\al).\eqno(9.3.34)$$

Thanks to (9.3.32)-(9.3.34), (9.3.24) is now equivalent to
$$-{\al'}'-2\al'\gm
={\al'}'+2\al'\gm\lra\gm=-\frac{{\al'}'}{2\al'}.\eqno(9.3.35)$$ Thus
$$\U=-\frac{{\al'}'}{2\al'}\X-\al'\Y-\be'\Z\sin\al+6\nu\X^{-1},\eqno(9.3.36)$$
$$
\V=\al'\X-\frac{{\al'}'}{2\al'}\Y-\be'\Z\cos\al+6\nu\Y\X^{-2},\eqno(9.3.37)$$
$$\W=\frac{{\al'}'}{\al'}\Z-\be'\X\sin\al-\be'\Y\cos\al\eqno(9.3.38)$$
by (9.3.25), (9.3.29)-(9.3.31) and (9.3.35). Moreover, (9.3.24) and
(9.3.32)-(9.3.34) imply
\begin{eqnarray*}p&=&\rho\{
\frac{(2\al'{{\al'}'}'+4{\al'}^4-3{{\al'}'}^2)(\X^2+\Y^2)}{8{\al'}^2}+({\be'}'+2{\al'}'\be'/\al')\Z(\X\sin\al+\Y\cos\al)\\
& &- \frac{3{\be'}^2(\X\sin\al+\Y\cos\al)^2}{2}
+12\nu(3\nu\X^{-2}-\al'\Y\X^{-1}) +
\frac{(\al'{\be'}^2-{{\al'}'}')\Z^2}{2\al'}
\}.\hspace{0.4cm}(9.3.39)\end{eqnarray*}

Note $\vec u=\Upsilon^{-1}\vec{\cal U}$ by (9.3.5). Thus (9.3.3)
yields
$$u=\left(\frac{{\al'}'}{2\al'}-6\nu\Y\X^{-2}\right)(\Y\sin\al-\X\cos\al)
-\al'(\X\sin\al+\Y\cos\al),\eqno(9.3.40)$$
\begin{eqnarray*}\hspace{0.6cm}v&=&\left(6\nu\Y\X^{-2}-\frac{{\al'}'}
{2\al'}\right)(\X\sin\al+\Y\cos\al)\cos\be
+\al'(\X\cos\al-\Y\sin\al)\cos\be\\ &&-\be'\Z\cos\be
+\left(\be'\X\sin\al+\be'\Y\cos\al-\frac{{\al'}'}{\al'}\Z\right)\sin\be,
\hspace{3.4cm}(9.3.41)\end{eqnarray*}
\begin{eqnarray*}\hspace{0.6cm}w&=&\left(6\nu\Y\X^{-2}-\frac{{\al'}'}
{2\al'}\right)(\X\sin\al+\Y\cos\al)\sin\be
+\al'(\X\cos\al-\Y\sin\al)\sin\be\\ &&-\be'\Z\sin\be
+\left(\frac{{\al'}'}{\al'}\Z-\be'\X\sin\al-\be'\Y\cos\al\right)\cos\be.
\hspace{3.4cm}(9.3.42)\end{eqnarray*}

According to (9.3.5), $\vec {\cal X}=\Upsilon\vec u$. So (9.3.1)
gives
$$\Y\sin\al-\X\cos\al=-x,\qquad
\X\sin\al+\Y\cos\al=y\cos\be+z\sin\be,\eqno(9.3.43)$$
$$\X^2+\Y^2=x^2+(y\cos\be+z\sin\be)^2.\eqno(9.3.44)$$
Therefore, we have the following theorem:\psp

 {\bf Theorem 9.3.1}. {\it Let
$\al$ and $\be$ be functions in $t$ with $\al'\neq 0$. We have the
following solution of the Navier-Stokes equations (9.1.1)-(9.1.4):
$$u=\frac{6\nu
x[(y\cos\be+z\sin\be)\cos\al-x\sin\al]}{[(y\cos\be+z\sin\be)\sin\al+x\cos\al]^2}
-\frac{{\al'}'x}{2\al'}-\al'(y\cos\be+z\sin\be),\eqno(9.3.45)$$
\begin{eqnarray*}v&=&\frac{6\nu
[(y\cos\be+z\sin\be)\cos\al-x\sin\al](y\cos\be+z\sin\be)\cos\be
}{[(y\cos\be+z\sin\be)\sin\al+x\cos\al]^2}+\al' x\cos\be\\
& &\!\!\!+[\be'\sin2\be+({\al'}'/2\al')(\sin^2\be-\cos
2\be)]y-[\be'\cos2\be+(3{\al'}'/4\al')\sin2\be]z,
\hspace{0.3cm}(9.3.46)\end{eqnarray*}
\begin{eqnarray*}w&=&\frac{6\nu
[(y\cos\be+z\sin\be)\cos\al-x\sin\al](y\cos\be+z\sin\be)\sin\be
}{[(y\cos\be+z\sin\be)\sin\al+x\cos\al]^2}+\al' x\sin\be\\
&
&\!\!\!-[\be'\cos2\be+(3{\al'}'/4\al')\sin2\be]y+[({\al'}'/2\al')(\cos^2\be+\cos
2\be)-\be'\sin2\be]z \hspace{0.3cm}(9.3.47)\end{eqnarray*} and
\begin{eqnarray*}p&=&\rho\{\frac{12\nu[6\nu+\al'[(x^2-(y\cos\be+z\sin\be)^2)\sin2\al-2x(y\cos\be+z\sin\be)\cos2\al]]}
{2[(y\cos\be+z\sin\be)\sin\al+x\cos\al]^2}\\ & &+
\frac{(2\al'{{\al'}'}'+4{\al'}^4-3{{\al'}'}^2)[x^2+(y\cos\be+z\sin\be)^2]}{8{\al'}^2}\\
& &+({\be'}'/2+{\al'}'\be'/\al')[(z^2-y^2)\sin2\be+2yz\cos2\be]
\\
& &- \frac{3{\be'}^2(y\cos\be+z\sin\be)^2}{2}
 +
\frac{(\al'{\be'}^2-{{\al'}'}')(z\cos\be-y\sin\be)^2}{2\al'}
\}.\hspace{2.5cm}(9.3.48)\end{eqnarray*} The above solution blows up
on the following rotating plane:}
 $$\{(x,y,z)\in\mbb{R}^3\mid(y\cos\be+z\sin\be)\sin\al+x\cos\al=0\}.\eqno(9.3.49)$$
\pse

Applying the symmetry transformation in (9.1.32) and (9.1.33) to the
above solution, we can get a solutions with six parameter functions
and blowing up on a more general moving plane. Next let $f$ be a
function in $t,\Y,\Z$ such that $\ptl_\Y^2(f)=\ptl_\Z^2(f)=0$, and
let $\phi,\psi$ be functions in $t,\X$. Suppose that $\gm$ is a
function in $t$. Assume
$$\U=f-2\gm'\X,\qquad\V=\phi+\gm'\Y,\qquad\W=\psi+\gm'\Z
.\eqno(9.3.50)$$ Then
\begin{eqnarray*}\hspace{0.2cm}& &R_1=f_t-2{\gm'}'\X-\al'(3\gm'\Y+\X f_\Y+\phi
)-\be'(3\gm'\Z+\X f_\Z+\psi)\sin\al \\ & &+\be'(\Z f_\Y-\Y
f_\Z)\cos\al-2\gm'(f-2\gm'\X)+f_\Y(\phi+\gm'\Y)+f_\Z(\psi+\gm'\Z)
,\hspace{1cm}(9.3.51)\end{eqnarray*}
\begin{eqnarray*}\hspace{1cm}R_2&=&\phi_t+{\gm'}'\Y+\al'(\Y\phi_\X-3\gm'\X +f)
+\be'\Z\phi_\X\sin\al-\be'\psi\cos\al\\
&&+(f-2\gm'\X)\phi_\X+\gm'\phi+{\gm'}^2\Y
-\nu\phi_{\X\X},\hspace{5.1cm}(9.3.52)\end{eqnarray*}
\begin{eqnarray*}\hspace{1cm}R_3&=&\psi_t+{\gm'}'\Z+\al'\Y\psi_\X
+\be'(\Z\psi_\X-3\gm'\X+f)\sin\al-\nu\psi_{\X\X}\\&&
+\be'\phi\cos\al
+(f-2\gm'\X)\psi_\X+\gm'(\psi+\gm'\Z).\hspace{4.4cm}(9.3.53)\end{eqnarray*}
Now (9.3.24) becomes
\begin{eqnarray*}\hspace{1.2cm}& &
\phi_{t\X}+(\al'\Y
+\be'\Z\sin\al+f)\phi_{\X\X}-\be'\psi_\X\cos\al-2\gm'(\X\phi_\X)_\X
\\ & &+\gm'\phi_\X -\nu\phi_{\X\X\X}=f_{t\Y}-\be'
f_\Z\cos\al-\gm'f_\Y,\hspace{4.8cm}(9.3.54)\end{eqnarray*}
\newpage

\begin{eqnarray*}\hspace{1.2cm}& &\psi_{t\X}+(\al'\Y
+\be'\Z\sin\al+f)\psi_{\X\X}-\nu\psi_{\X\X\X} +\be'\phi_\X\cos\al
\\&&-2(\gm'\X\psi_\X)_\X+\gm'\psi_\X=
f_{t\Z}+\be'
f_\Y\cos\al-\gm'f_\Z,\hspace{4cm}(9.3.55)\end{eqnarray*}
$$\al'f_\Z +(\be'\sin\al+f_\Z)\phi_\X=
(\al'+f_\Y)\psi_\X +\be'f_\Y\sin\al.\eqno(9.3.56)$$

By (9.3.54) and (9.3.55), we take
$$f=-\al'\Y-\be'\Z\sin\al.\eqno(9.3.57)$$
Note that (9.3.56) is implied by (9.3.57). Integrating (9.3.54) and
(9.3.55), we obtain
$$\phi_t-2\gm'\X\phi_\X +\gm'\phi -\nu\phi_{\X\X}-\be'\psi\cos\al
=[{\be'}^2
\sin\al\;\cos\al+\al'\gm'-{\al'}']\X+\be_1,\eqno(9.3.58)$$
\begin{eqnarray*}\hspace{2cm}& &\psi_t-2\gm'\X\psi_\X
+\gm'\psi-\nu\psi_{\X\X} +\be'\phi\cos\al \\ &=& -[(\be'\sin\al)'
+\al'\be'\cos\al-\gm'\be'\sin\al]\X+\be_2,
\hspace{3.6cm}(9.3.59)\end{eqnarray*} where $\be_1$ and $\be_2$ are
arbitrary functions in $t$. To solve the above problem, we write
$$\be'=\frac{\vf'}{\cos\al},\qquad \gm=\frac{1}{4}\ln\mu'\eqno(9.3.60)$$
and set
$$\left(\begin{array}{c}\hat\phi\\ \hat\psi\end{array}\right)
=\sqrt[4]{\mu'}\left(\begin{array}{cc}\cos\vf&-\sin\vf\\
\sin\vf&\cos\vf\end{array}\right) \left(\begin{array}{c}\phi\\
\psi\end{array}\right),\eqno(9.3.61)$$
$$\left(\begin{array}{c}\gm_1\\ \gm_2\end{array}\right)
=\int\frac{1}{\sqrt[4]{\mu'}}\left(\begin{array}{cc}\cos\vf&
-\sin\vf\\
\sin\vf&\cos\vf\end{array}\right) \left(\begin{array}{c}{\vf'}^2
\tan\al+\frac{\al'{\mu'}'}{4\mu'}-{\al'}'
\\ -(\vf'\tan\al)'-\al'\vf'+\frac{{\mu'}'\vf'}{4\mu'}\tan\al
\end{array}\right)dt.\eqno(9.3.62)$$
Then (9.3.58) and (9.3.59) are equivalent to:
$$\hat\phi_t-\frac{{\mu'}'}{2\mu'}\X\hat\phi_\X
-\nu\hat\phi_{\X\X}=\gm_1'\sqrt{\mu'}\X+\vf_1',\eqno(9.3.63)$$
$$\hat\psi_t-\frac{{\mu'}'}{2\mu'}\X\hat\psi_\X
-\nu\hat\psi_{\X\X}=\gm_2'\sqrt{\mu'}\X+\vf_2',\eqno(9.3.64)$$ where
$\vf_1$ and $\vf_2$ are arbitrary functions in $t$. Note the first
two terms in the above equations motivate us to write
$$\hat\phi=\td\phi(t,\varpi)+\gm_1\varpi+\vf_1
,\;\;\hat\psi=\td\psi(t,\varpi)+\gm_2\varpi+\vf_2,\;\;\varpi=\sqrt{\mu'}\X.\eqno(9.3.65)$$
Then the above equations become equations:
$$\td\phi_t-\nu\mu'\td\phi_{\varpi\varpi}=0,\qquad\td\psi_t-\nu\mu'\td\psi_{\varpi\varpi}=0.
\eqno(9.3.66)$$ Thus we have the following solution:
$$\td\phi=\sum_{r=1}^m
a_rd_re^{a_r^2\nu\mu\cos 2b_r+a_r\varpi\cos b_r}\sin(a_r^2\nu
\mu\sin 2b_r+a_r\varpi\sin b_r+b_r+c_r),\eqno(9.3.67)$$
$$\td\psi=\sum_{s=1}^n\hat a_s\hat d_se^{\hat a_s^2\nu\mu\cos 2\hat b_s+\hat a_s\varpi\cos \hat b_s}\sin(\hat
a_s^2\nu \mu\sin 2\hat b_s+\hat a_s\varpi\sin \hat b_s+\hat b_s+\hat
c_s),\eqno(9.3.68)$$ where $a_r,\hat a_s,b_r,\hat b_s,c_r,\hat
c_s,d_r$ and $\hat d_s$ are real constants. Therefore,
\begin{eqnarray*}\hat\phi&=&\sum_{r=1}^m
a_rd_re^{a_r^2\nu\mu\cos 2b_r+a_r\sqrt{\mu'}\X\cos b_r}\sin(a_r^2\nu
\mu\sin 2b_r+a_r\sqrt{\mu'}\X\sin b_r+b_r+c_r)\\ & &+
\gm_1\sqrt{\mu'}\X+\vf_1,\hspace{10.4cm}(9.3.69)\end{eqnarray*}
\begin{eqnarray*}\hat\psi&=&\sum_{s=1}^n\hat a_s\hat d_se^{\hat a_s^2\nu\mu\cos 2\hat b_s+\hat a_s\sqrt{\mu'}\X
\cos \hat b_s}\sin(\hat a_s^2\nu \mu\sin 2\hat b_s+\hat
a_s\sqrt{\mu'}\X\sin \hat b_s+\hat b_s+\hat c_s)
\\ & &+
\gm_2\sqrt{\mu'}\X+\vf_2.\hspace{10.4cm}(9.3.70)\end{eqnarray*}

According to (9.3.61), we have
\begin{eqnarray*}\phi&=&\frac{\cos\vf}{\sqrt[4]{\mu'}}\sum_{r=1}^m
a_rd_re^{a_r^2\nu\mu\cos 2b_r+a_r\sqrt{\mu'}\X\cos b_r}
\sin(a_r^2\nu \mu\sin 2b_r+a_r\sqrt{\mu'}\X\sin b_r+b_r+c_r)\\ & &+
\frac{\sin\vf}{\sqrt[4]{\mu'}}\sum_{s=1}^n\hat a_s\hat d_se^{\hat
a_s^2\nu\mu\cos 2\hat b_s+\hat a_s\sqrt{\mu'}\X \cos \hat
b_s}\sin(\hat a_s^2\nu \mu\sin 2\hat b_s+\hat a_s\sqrt{\mu'}\X\sin
\hat b_s+\hat b_s+\hat c_s)\\
& &+\sqrt[4]{\mu'}(\gm_1\cos\vf+\gm_2\sin\vf) \X+\sgm_1,
\hspace{7.3cm}(9.3.71)\end{eqnarray*}
\begin{eqnarray*}\psi&=&-\frac{\sin\vf}{\sqrt[4]{\mu'}}\sum_{r=1}^m
a_rd_re^{a_r^2\nu\mu\cos 2b_r+a_r\sqrt{\mu'}\X\cos b_r}\sin(a_r^2\nu
\mu\sin 2b_r+a_r\sqrt{\mu'}\X\sin b_r+b_r+c_r)\\ & &+
\frac{\cos\vf}{\sqrt[4]{\mu'}}\sum_{s=1}^n\hat a_s\hat d_se^{\hat
a_s^2\nu\mu\cos 2\hat b_s+\hat a_s\sqrt{\mu'}\X \cos \hat
b_s}\sin(\hat a_s^2\nu \mu\sin 2\hat b_s+\hat a_s\sqrt{\mu'}\X\sin
\hat b_s+\hat b_s+\hat c_s)\\ &
&+\sqrt[4]{\mu'}(\gm_2\cos\vf-\gm_1\sin\vf) \X+\sgm_2,
\hspace{7.3cm}(9.3.72)\end{eqnarray*} where $\sgm_1$ and $\sgm_2$
are arbitrary functions in $t$. By (9.3.50), (9.3.57), (9.3.60),
(9.3.71) and (9.3.72),
$$\U=-\al'\Y-\vf'\Z\tan\al-\frac{{\mu'}'\X}{2\mu'},\eqno(9.3.73)$$
\begin{eqnarray*}\V&=&\frac{\cos\vf}{\sqrt[4]{\mu'}}\sum_{r=1}^m
a_rd_re^{a_r^2\nu\mu\cos 2b_r+a_r\sqrt{\mu'}\X\cos b_r}
\sin(a_r^2\nu \mu\sin 2b_r+a_r\sqrt{\mu'}\X\sin b_r+b_r+c_r)\\ & &+
\frac{\sin\vf}{\sqrt[4]{\mu'}}\sum_{s=1}^n\hat a_s\hat d_se^{\hat
a_s^2\nu\mu\cos 2\hat b_s+\hat a_s\sqrt{\mu'}\X \cos \hat
b_s}\sin(\hat a_s^2\nu \mu\sin 2\hat b_s+\hat a_s\sqrt{\mu'}\X\sin
\hat b_s+\hat b_s+\hat c_s)\\
& &+\sqrt[4]{\mu'}(\gm_1\cos\vf+\gm_2\sin\vf)
\X+\frac{{\mu'}'\Y}{4\mu'}+\sgm_1,
\hspace{5.9cm}(9.3.74)\end{eqnarray*}
\begin{eqnarray*}\W&=&-\frac{\sin\vf}{\sqrt[4]{\mu'}}\sum_{r=1}^m
a_rd_re^{a_r^2\nu\mu\cos 2b_r+a_r\sqrt{\mu'}\X\cos b_r}\sin(a_r^2\nu
\mu\sin 2b_r+a_r\sqrt{\mu'}\X\sin b_r+b_r+c_r)\\ & &+
\frac{\cos\vf}{\sqrt[4]{\mu'}}\sum_{s=1}^n\hat a_s\hat d_se^{\hat
a_s^2\nu\mu\cos 2\hat b_s+\hat a_s\sqrt{\mu'}\X \cos \hat
b_s}\sin(\hat a_s^2\nu \mu\sin 2\hat b_s+\hat a_s\sqrt{\mu'}\X\sin
\hat b_s+\hat b_s+\hat c_s)\\ &
&+\sqrt[4]{\mu'}(\gm_2\cos\vf-\gm_1\sin\vf)
\X++\frac{{\mu'}'\Z}{4\mu'}+\sgm_2.
\hspace{5.5cm}(9.3.75)\end{eqnarray*}

To find the pressure $p$, we recalculate
\begin{eqnarray*}R_1&=&({\vf'}^2\Y-2\vf'\psi)\tan\al-2\al'\phi
-{\al'}'\Y-({\vf'}'\tan\al+\al'\vf'(1+\sec^2\al))\Z
\\ & &-\frac{{\mu'}'(\al'\Y+\vf'\Z\tan\al)}{2\mu'}+({\al'}^2+{\vf'}^2
\tan^2\al) \X+\frac{(3{{\mu'}'}^2-2\mu'{{\mu'}'}')\X}{4{\mu'}^2}
,\hspace{1.2cm}(9.3.76)\end{eqnarray*}
$$R_2=\frac{(4\mu'{{\mu'}'}'-3{{\mu'}'}^2)\Y}{16{\mu'}^2}
-{\al'}^2\Y+ \sgm_1'+(\vf'\X-\al'\Z)\vf'\tan\al-
\frac{(\al'{\mu'}'+2{\al'}'\mu')\X}{2\mu'},\eqno(9.3.77)$$
\begin{eqnarray*}\hspace{2cm}R_3&=&\frac{(4{{\mu'}'}'-3{{\mu'}'}^2)\Z}{16{\mu'}^2}
-\frac{{\mu'}'\X+2\al'\mu'\Y}{2\mu'}\vf'\tan\al+ \sgm_2'\\&
&-{\vf'}^2\Z\tan^2\al-({\vf'}'\tan\al+\al'\vf'(1+\sec^2\al))\X
\hspace{2.8cm}(9.3.78)
\end{eqnarray*}
by (9.3.51)-(9.3.53),  (9.3.57)-(9.3.60), (9.3.71) and (9..72).
Thanks to (9.3.22), we have
\begin{eqnarray*}p&=&\rho\{
({\al'}'\Y+({\vf'}'\tan\al+\al'\vf'(1+\sec^2\al))\Z)\X
+\frac{{\mu'}'(\al'\Y+\vf'\Z\tan\al)\X}{2\mu'}\\
& &-\sgm_1'\Y-\sgm_2'\Z+\frac{\X^2}{2}
\left(\frac{(2\mu'{{\mu'}'}'-3{{\mu'}'}^2)}{4{\mu'}^2}-{\al'}^2-{\vf'}^2\tan^2\vf\right)
+\frac{{\al'}^2\Y^2+{\gm'}^2\Z^2\tan^2\al}{2}
\\ &
&+\frac{(3{{\mu'}'}^2-4\mu'{{\mu'}'}')(\Y^2+\Z^2)}{32{\mu'}^2}+(\al'\Z-\vf'\X)\vf'\Y\tan\al
+2\frac{\al'\cos\vf-\vf'\sin\vf\:\tan\al}{\sqrt[4]{{\mu'}^3}}\\ &
&\times\sum_{r=1}^m d_re^{a_r^2\nu\mu\cos 2b_r+a_r\sqrt{\mu'}\X\cos
b_r} \sin(a_r^2\nu \mu\sin 2b_r+a_r\sqrt{\mu'}\X\sin b_r+c_r)\\ &
&+2(\al'\sgm_1+\vf'\sgm_2\tan\al)\X+2
\frac{\al'\sin\vf+\vf'\cos\vf\:\tan\al}{\sqrt[4]{{\mu'}^3}}\\
& &\times\sum_{s=1}^n\hat d_se^{\hat a_s^2\nu\mu\cos 2\hat b_s+\hat
a_s\sqrt{\mu'}\X \cos \hat b_s}\sin(\hat a_s^2\nu \mu\sin 2\hat
b_s+\hat a_s\sqrt{\mu'}\X\sin
\hat b_s+\hat c_s)\\
& &+\sqrt[4]{\mu'}[\gm_1(\al'\cos\vf-\vf'\sin\vf\:\tan\al)
+\gm_2(\al'\sin\vf+\vf'\cos\vf\:\tan\al)] \X^2.\}
\hspace{0.9cm}(9.3.79)\end{eqnarray*}

By (9.3.3), (9.3.5) and (9.3.73)-(9.3.75), we get: \psp

{\bf Theorem 9.3.2}. {\it Let $\al,\vf,\mu,\sgm_1,\sgm_2$ be
functions in $t$ with $\mu'> 0$. Take real constants $\{r,\hat
a_s,b_r,\hat b_s,c_r,\hat c_s,d_r,\hat d_s\mid
i=1,...,m;s=1,...,n\}$. Denote $\be=\int\vf'\sec\al\;dt$ and define
$\gm_1,\gm_2$ by (9.3.62). Take the notations $\X,\Y,\Z$ given in
(9.3.1) and (9.3.5). We have the following solution of the
Navier-Stokes equations (9.1.1)-(9.1.4):
\begin{eqnarray*}& &u=-\left(\frac{{\mu'}'\X}{2\mu'}
+\al'\Y+\vf'\Z\tan\al\right)\cos\al
-\big[\frac{\cos\vf}{\sqrt[4]{\mu'}}\sum_{r=1}^m
a_rd_re^{a_r^2\nu\mu\cos 2b_r+a_r\sqrt{\mu'}\X\cos b_r} \\&
&\times\sin(a_r^2\nu \mu\sin 2b_r+a_r\sqrt{\mu'}\X\sin b_r+b_r+c_r)+
\frac{\sin\vf}{\sqrt[4]{\mu'}}\sum_{s=1}^n\hat a_s\hat d_se^{\hat
a_s^2\nu\mu\cos 2\hat b_s+\hat a_s\sqrt{\mu'}\X \cos \hat b_s}\\&
&\times\sin(\hat a_s^2\nu \mu\sin 2\hat b_s+\hat
a_s\sqrt{\mu'}\X\sin
\hat b_s+\hat b_s+\hat c_s)\\
& &+\sqrt[4]{\mu'}(\gm_1\cos\vf+\gm_2\sin\vf)
\X+\frac{{\mu'}'\Y}{4\mu'}+\sgm_1\big]\sin\al,
\hspace{5.4cm}(9.3.80)\end{eqnarray*}
\begin{eqnarray*}& &v=\left(\frac{{\mu'}'\X}{2\mu'}
-\al'\Y-\vf'\Z\tan\al\right)\sin\al\;\cos\be+\frac{{\mu'}'(\Y\cos\al\;\cos\be-\Z\sin\be)}{4\mu'}
\\ &&+\frac{\cos\vf\;\cos\al\;\cos\be+\sin\vf\;\sin\be}{\sqrt[4]{\mu'}}\sum_{r=1}^m
a_rd_re^{a_r^2\nu\mu\cos 2b_r+a_r\sqrt{\mu'}\X\cos b_r} \\&
&\times\sin(a_r^2\nu \mu\sin 2b_r+a_r\sqrt{\mu'}\X\sin b_r+b_r+c_r)+
\frac{\sin\vf\;\cos\al\;\cos\be-\cos\vf\;\sin\be}{\sqrt[4]{\mu'}}\\
& &\times\sum_{s=1}^n\hat a_s\hat d_se^{\hat a_s^2\nu\mu\cos 2\hat
b_s+\hat a_s\sqrt{\mu'}\X \cos \hat b_s}\sin(\hat a_s^2\nu \mu\sin
2\hat b_s+\hat a_s\sqrt{\mu'}\X\sin \hat b_s+\hat b_s+\hat c_s)\\
&&+\sqrt[4]{\mu'}[\gm_1(\cos\vf\;\cos\al\;\cos\be+\sin\vf\;\sin\be)
 +\gm_2(\sin\vf\;\cos\al\;\cos\be-\cos\vf\;\sin\be)]\X
\\ & &+\sgm_1\cos\al\;\cos\be- \sgm_2\sin\be,
\hspace{8.9cm}(9.3.81)\end{eqnarray*}
\begin{eqnarray*}& &w=\left(\frac{{\mu'}'\X}{2\mu'}
-\al'\Y-\vf'\Z\tan\al\right)\sin\al\;\sin\be+\frac{{\mu'}'(\Y\cos\al\;\sin\be+\Z\cos\be)}{4\mu'}
\\ &&+\frac{\cos\vf\;\cos\al\;\sin\be-\sin\vf\;\cos\be}{\sqrt[4]{\mu'}}\sum_{r=1}^m
a_rd_re^{a_r^2\nu\mu\cos 2b_r+a_r\sqrt{\mu'}\X\cos b_r} \\&
&\times\sin(a_r^2\nu \mu\sin 2b_r+a_r\sqrt{\mu'}\X\sin b_r+b_r+c_r)+
\frac{\sin\vf\;\cos\al\;\sin\be+\cos\vf\;\cos\be}{\sqrt[4]{\mu'}}\\
& &\times\sum_{s=1}^n\hat a_s\hat d_se^{\hat a_s^2\nu\mu\cos 2\hat
b_s+\hat a_s\sqrt{\mu'}\X \cos \hat b_s}\sin(\hat a_s^2\nu \mu\sin
2\hat b_s+\hat a_s\sqrt{\mu'}\X\sin
\hat b_s+\hat b_s+\hat c_s)\\
&&+\sqrt[4]{\mu'}[\gm_1(\cos\vf\;\cos\al\;\sin\be-\sin\vf\;\cos\be)
 +\gm_2(\sin\vf\;\cos\al\;\sin\be+\cos\vf\;\cos\be)]\X
\\ & &+\sgm_1\cos\al\;\sin\be+\sgm_2\cos\be,
\hspace{8.9cm}(9.3.82)\end{eqnarray*} and $p$ is given in
(9.3.79).}\psp

{\bf Remark 9.3.3}. We can use Fourier expansion to solve the system
(9.3.66) for $\td\phi(t,\sqrt{\mu'}\X)$ and
$\td\psi(t,\sqrt{\mu'}\X)$ with given $\td\phi(0,\sqrt{\mu'(0)}\X)$
and $\td\psi(0,\sqrt{\mu'(0)}\X)$. In this way, we can obtain
discontinuous solutions of the Navier-Stokes equations
(9.1.1)-(9.1.4), which may be useful in studying shock waves.

\section{Moving-Frame Approach II}

Motivated from the first solution in Theorem 9.2.2, we will solve
the equations (9.3.17) and (9.3.18) by sin, cos, sinh and cosh
functions.

First we rewrite (9.3.19)-(9.3.21):
\begin{eqnarray*}& &R_1=\U_t+(\al'\Y+\be'\Z\sin\al+\U)\U_\X
+(\V-\al'\X+\be'\Z\cos\al)\U_\Y \\
&&+(\W-\be'(\X\sin\al+\Y\cos\al))\U_\Z-\al'\V-\be'\W\sin\al-\nu\Dlt(\U),
\hspace{3cm}(9.4.1)\end{eqnarray*}
\begin{eqnarray*}& &R_2=\V_t+(\al'\Y+\be'\Z\sin\al+\U)\V_\X
+(\V-\al'\X+\be'\Z\cos\al)\V_\Y \\
&&+(\W-\be'(\X\sin\al+\Y\cos\al))\V_\Z+\al'\U-\be'\W\cos\al-\nu\Dlt(\V),
\hspace{2.9cm}(9.4.2)\end{eqnarray*}
\begin{eqnarray*}& &R_3=\W_t+(\al'\Y+\be'\Z\sin\al+\U)\W_\X
+(\V-\al'\X+\be'\Z\cos\al)\W_\Y \\
&&+(\W-\be'(\X\sin\al+\Y\cos\al))\W_\Z+
\be'(\U\sin\al+\V\cos\al)-\nu\Dlt(\W).
\hspace{2cm}(9.4.3)\end{eqnarray*} Let $\al_1,\be_1,\gm$ be
functions in $t$. Set
$$\xi_0=\sinh(\al_1\Y+\be_1\Z),\;\;\zeta_0=
\cosh(\al_1\Y+\be_1\Z),\;\;\phi_0=\sinh\gm\X,\eqno(9.4.4)$$
$$\psi_0=\cosh\gm\X,\;\;\xi_1=\sin(\al_1\Y+\be_1\Z),\;\;\zeta_1=\cos(\al_1\Y+\be_1\Z),\eqno(9.4.5)$$
 $$\phi_1=\sin\gm\X,\;\;
\psi_1=\cos\gm\X, \qquad\Dlt_1=\ptl_\Y^2+\ptl_\Z^2.\eqno(9.4.6)$$

Suppose that $f$ and $h$ are functions in $t,\Y,\Z$. Moreover,
$\sgm$ and $\tau$ are functions in $t$. According to
(9.3.29)-(9.3.31), we assume
$$\U=-\al'\Y-\be'\Z\sin\al-(f_\Y+h_\Z)\X
-(\al_1\sgm+\be_1\tau)\zeta_r\phi_s,\eqno(9.4.7)$$
$$\V=\al'\X-\be'\Z\cos\al+f+\sgm\gm\xi_r\psi_s,\qquad
\W=\be'(\X\sin\al+\Y\cos\al)+h+\tau\gm\xi_r\psi_s.\eqno(9.4.8)$$ By
(9.4.1)-(9.4.3), we have
\begin{eqnarray*}&
&R_1= -(\al_1\sgm+\be_1\tau)'\zeta_r\phi_s-
(\al_1\sgm+\be_1\tau)[(-1)^r(\al_1'\Y+\be_1'\Z)\xi_r\phi_s+\gm'\X\zeta_r\psi_s]
\\ & &-(f_{\Y t}+h_{\Z t})\X+((f_\Y+h_\Z)\X+(\al_1\sgm+\be_1\tau)\zeta_r\phi_s)(
f_\Y+h_\Z+\gm(\al_1\sgm+\be_1\tau)\zeta_r\psi_s)
\\ & &-\al'(f+\al'\X-\be'\Z\cos\al+\gm\sgm\xi_r\psi_s)
-\be'(\be'(\X\sin\al+\Y\cos\al)+h+\gm\tau\xi_r\psi_s)\sin\al
\\ &&-(f+\gm\sgm\xi_r\psi_s)(\al'+(f_{\Y\Y}+h_{\Y\Z})\X+(-1)^r\al_1(\al_1\sgm+\be_1\tau)\xi_r\phi_s)
-(h+\gm\tau\xi_r\psi_s)\\ & &\times
(\be'\sin\al+(f_{\Y\Z}+h_{\Z\Z})\X+(-1)^r\be_1(\al_1\sgm+\be_1\tau)\xi_r\phi_s)
+\nu\{\Dlt_1(f_\Y+h_\Z)\X
\\ & &+(\al_1\sgm+\be_1\tau)[(-1)^r(\al_1^2+\be_1^2)+(-1)^s\gm^2]
\zeta_r\phi_s\}-{\al'}'\Y-(\be'\sin\al)'\Z\\&
&=\{(\gm(f_\Y+h_\Z)-\gm')\X\zeta_r\psi_s-
(-1)^r(\al_1'\Y+\be_1'\Z+\al_1f+\be_1h
)\xi_r\phi_s\}(\al_1\sgm+\be_1\tau)
\\ & &+\{(\al_1\sgm+\be_1\tau)[(-1)^s\nu\gm^2
+(-1)^r\nu(\al_1^2+\be_1^2)+f_\Y+h_\Z]-(\al_1\sgm+\tau\be_1)'\}\zeta_r\phi_s
\\ &&-\gm\{2(\sgm\al'+\tau\be'\sin\al)
+[\sgm(f_{\Y\Y}+h_{\Y\Z})+\tau(f_{\Y\Z}+h_{\Z\Z})]\X\} \xi_r\psi_s
-(f_{\Y t}+h_{\Z t})\X\\
&&+(f_\Y+h_\Z)^2\X-f(\al'+(f_{\Y\Y}+h_{\Y\Z})\X)
-h(\be'\sin\al+(f_{\Y\Z}+h_{\Z\Z})\X)
\\ & &-\al'(f+\al'\X-\be'\Z\cos\al)
-\be'(\be'(\X\sin\al+\Y\cos\al)+h)\sin\al -{\al'}'\Y\\ &
&-(\be'\sin\al)'\Z+\nu\Dlt_1(f_\Y+h_\Z)\X+\gm(\al_1\sgm+\be_1\tau)^2\phi_s\psi_s
, \hspace{4.2cm}(9.4.9)\end{eqnarray*}
\begin{eqnarray*}&
&R_2={\al'}'\X-(\be'\cos\al)'\Z+
f_t+(\gm\sgm)'\xi_r\psi_s+\gm\sgm((\al_1'\Y+\be_1'\Z)\zeta_r\psi_s
+(-1)^s\gm'\X\xi_r\phi_s)\\ &
&-\al'[\al'\Y+\be'\Z\sin\al+(f_\Y+h_\Z)\X+(\al_1\sgm+\be_1\tau)\zeta_r\phi_s]
-\be'[\be'(\X\sin\al+\Y\cos\al)\\ & &+ h+\gm\tau\xi_r\psi_s]\cos\al
-[(f_\Y+h_\Z)\X+(\al_1\sgm+\be_1\tau)\zeta_r\phi_s]
(\al'+(-1)^s\gm^2\sgm\xi_r\phi_s)
\\ & &+(f+\gm\sgm\xi_r\psi_s)(f_\Y+
\al_1\gm\sgm\zeta_r\psi_s)+(h+\gm\tau\xi_r\psi_s)(f_\Z-\be'\cos\al
+\be_1\gm\sgm\zeta_r\psi_s)\\& &
-\nu[\Dlt_1(f)+\gm\sgm((-1)^r(\al_1^2+\be_1^2)+(-1)^s\gm^2)]\xi_r\psi_s\\&&={\al'}'\X+f_t+\gm\sgm(\al_1'\Y+\be_1'\Z+\al_1f+\be_1h)
\zeta_r\psi_s +(-1)^s\gm\sgm(\gm'-\gm(f_\Y+h_\Z))\X \xi_r\phi_s \\ &
&+\{\gm[\sgm f_\Y+\tau f_\Z
-\nu\sgm[(-1)^r(\al_1^2+\be_1^2)+(-1)^s\gm^2]
-2\tau\be'\cos\al]+(\gm\sgm)'\}\xi_r\psi_s
\\ & &-2\al'(\al_1\sgm+\be_1\tau)\zeta_r\phi_s-(\be'\cos\al)'\Z
-\al'[\al'\Y+\be'\Z\sin\al+2(f_\Y+h_\Z)\X]
 \\ &
 &-\be'[\be'(\X\sin\al+\Y\cos\al)+h]\cos\al+ff_\Y+h(f_\Z-\be'\cos\al)
-\nu\Dlt_1(f)\\ & & +\gm^2\sgm(\al_1\sgm+\be_1\tau)\xi_r\zeta_r ,
\hspace{9.9cm}(9.4.10)\end{eqnarray*}
\begin{eqnarray*}& &R_3=(\be'\sin\al)'\X+(\be'\cos\al)'\Y+h_t
+(\tau\gm)'\xi_r\psi_s+ \tau\gm[(\al_1'\Y+\be_1'\Z) \zeta_r\psi_s\\
& &+(-1)^s\gm'\X\xi_r\phi_s] -[(f_\Y+h_\Z)\X+
(\al_1\sgm+\be_1\tau)\zeta_r\phi_s](\be'\sin\al+(-1)^s
\tau\gm^2\xi_r\phi_s)\\ & &+(f+\sgm\gm\xi_r\psi_s)
(\be'\cos\al+h_\Y+\al_1\tau\gm\zeta_r\psi_s)+(h+\tau\gm\xi_r\psi_s)
(h_\Z+\be_1\tau\gm\zeta_r\psi_s)\\ &
&-\be'(\al'\Y+\be'\Z\sin\al+(f_\Y+h_\Z)\X+(\al_1\sgm+\be_1\tau)
\zeta_r\phi_s)\sin\al +\be'(\al'\X-\be'\Z\cos\al\\ &
&+f+\sgm\gm\xi_r\psi_s)\cos\al
-\nu[\Dlt_1(h)+\gm\tau((-1)^r(\al_1^2+\be_1^2)
+(-1)^s\gm^2)]\xi_r\psi_s\\
&&=\gm\tau(\al_1'\Y+\be_1'\Z+\al_1f+\be_1h)\zeta_r\psi_s
+\{(\tau\gm)'-\nu\gm\tau[(-1)^r(\al_1^2+\be_1^2) +(-1)^s\gm^2]\\
&&+\gm(2\be'\sgm\cos\al+\sgm h_\Y+\tau h_\Z) \}\xi_r\psi_s
-2\be'(\al_1\sgm+\be_1\tau)\zeta_r\phi_s\sin\al+
(-1)^s\gm\tau(\gm'\\ & &-\gm(f_\Y+h_\Z))\X\xi_r\phi_s
+(\be'\sin\al)'\X+(\be'\cos\al)'\Y+h_t-\be'(f_\Y+h_\Z)\X\sin\al
\\ &&+f(\be'\cos\al+h_\Y)+hh_\Z-
\be'(\al'\Y+\be'\Z\sin\al+(f_\Y+h_\Z)\X)\sin\al+\be'(\al'\X\\ &
&-\be'\Z\cos\al+f)\cos\al -\nu\Dlt_1(h)
+\gm^2\tau(\al_1\sgm+\be_1\tau)\xi_r\zeta_r.
\hspace{4.2cm}(9.4.11)\end{eqnarray*}

By the coefficients of $\xi_r\psi_s$ in the equation
$\ptl_\Y(R_1)=\ptl_\X(R_2)$, we have
$$\gm^2\sgm=(-1)^{r+s+1}\al_1(\al_1\sgm+\be_1\tau),\qquad
[\sgm(f_{\Y\Y}+h_{\Y\Z})+\tau(f_{\Y\Z}+h_{\Z\Z})]_\Y=0.\eqno(9.4.12)$$
Moreover, the coefficients of $\zeta_r\phi_s$ in the equation
$\ptl_\Y(R_1)=\ptl_\X(R_2)$ suggest
$$(f_\Y+h_\Z)_\Y=0,\eqno(9.4.13)$$
which implies the second equation in (9.4.12). According the
coefficients of $\xi_r\phi_s$ in the equation
$\ptl_\Y(R_1)=\ptl_\X(R_2)$, we get
$$\sgm\be_1h_\Y =\tau\al_1( f_\Z -2\be'\cos\al).\eqno(9.4.14)$$
Furthermore, the coefficients of $\zeta_r\psi_s$ in the equation
$\ptl_\Y(R_1)=\ptl_\X(R_2)$ yield
$$\al_1\be'\sin\al=\al'\be_1.\eqno(9.4.15)$$
Symmetrically, we have (9.4.15),
$$\gm^2\tau=(-1)^{r+s+1}\be_1(\al_1\sgm+\be_1\tau),\qquad
(f_\Y+h_\Z)_\Z=0\eqno(9.4.16)$$ and
$$\tau\al_1 f_\Z
= \sgm\be_1(h_\Y+2\be'\cos\al)\eqno(9.4.17)$$ (cf. (9.4.7) and
(9.4.8)). By the first equation in (9.4.12) and (9.4.16), we have
$$\sgm\be_1=\tau\al_1.\eqno(9.4.18)$$
Then (9.4.14) is implied by (9.4.17) and (9.4.18). Note that the
equations of  the coefficients
$\xi_r\psi_s,\;\zeta_r\psi_s,\;\xi_r\phi_s$ and $\zeta_r\phi_s$ in
$\ptl_\Z(R_2)=\ptl_\Y(R_3)$ are implied by (9.4.15), (9.4.17) and
(9.4.18).

According to (9.4.13) and the second equation in (9.4.16),
$$f_\Y+h_\Z=\gm_1,\eqno(9.4.19)$$
a function in $t$. Under the conditions in (9.4.15), the first
equation in (9.4.16), and (9.4.17)-(9.4.19),
$\ptl_\Y(R_1)=\ptl_\X(R_2)$ becomes
$$ \al'h_\Z-\be'h_\Y\sin\al
={\al'}',\eqno(9.4.20)$$ $\ptl_\Z(R_1)=\ptl_\X(R_3)$ is equivalent
to
$$\be'h_\Z\sin\al+\al'
h_y=\be'\gm_1\sin\al-(\be'\sin\al)'-2\al'\be'\cos\al \eqno(9.4.21)$$
and $\ptl_\Z(R_2)=\ptl_\Y(R_3)$ says
$$(ff_\Y+hf_\Z)_\Z
=(fh_\Y+hh_\Z)_\Y +2\be'\gm_1\cos\al.\eqno(9.4.22)$$ By (9.4.17) and
(9.4.19)-(9.4.21), we assume that $f_\Y,\;f_\Z,\;h_\Y$ and $h_\Z$
are functions in $t$. Then (9.4.22) can be written as
$$(f_\Y+h_\Z)f_\Z=(f_\Y+h_\Z)h_\Y+2\be'\gm_1\cos\al,\eqno(9.4.23)$$
which is implied by (9.4.17) and (9.4.19). Solving (9.4.20) and
(9.4.21), we get
$$h_\Y=\frac{\al'\be'\gm_1\sin\al-(\al'\be'\sin\al)'
-2{\al'}^2\be'\cos\al}{{\al'}^2+{\be'}^2\sin^2\al},\eqno(9.4.24)$$
$$h_\Z= \frac{\al'{\al'}'+{\be'}^2\gm_1\sin^2\al-
(\be'\sin\al)(\be'\sin\al)'-\al'{\be'}^2\sin
2\al}{{\al'}^2+{\be'}^2\sin^2\al}.\eqno(9.4.25)$$ Moreover,
$$f_\Y=\frac{\gm_1{\al'}^2-\al'{\al'}'+
(\be'\sin\al)(\be'\sin\al)'+\al'{\be'}^2\sin
2\al}{{\al'}^2+{\be'}^2\sin^2\al}\eqno(9.4.26)$$ by (9.4.19) and
(9.4.25), and
$$f_\Z=\frac{\al'\be'\gm_1\sin\al-(\al'\be'\sin\al)'
+2{\be'}^2\sin^2\al\;\cos\al}{{\al'}^2+{\be'}^2\sin^2\al}
\eqno(9.4.27)$$ by (9.4.17) and (9.4.24). With the above data, we
take
$$f=f_\Y\Y+f_\Z\Z,\qquad h=h_\Y\Y+h_\Z\Z.\eqno(9.4.28)$$
 Furthermore,  (9.4.18)
and the first equation in (9.4.16) yield $r+s+1\in 2\mbb{Z}$,
$$\al_1=\vf\al',\qquad \gm=\pm\vf\sqrt{{\al'}^2+{\be'}^2\sin^2\al},
\eqno(9.4.29)$$
$$\be_1=\vf\be'\sin\al,\qquad\sgm=\mu\al',\qquad
\tau=\mu\be'\sin\al.\eqno(9.4.30)$$ In particular,
$\al,\be,\gm_1,\vf$ and $\mu$ are arbitrary functions in $t$.
 Thanks to (9.3.22) and (9.4.9)-(9.4.11), the pressure
\begin{eqnarray*}&
&p=\rho \{\gm\mu\vf^{-1}[(\gm'-\gm\gm_1)\X\zeta_r\phi_s
-((\vf\al')'\Y+(\vf\be'\sin\al)'\Z+\vf(\al'f+\be'h\sin\al) )\xi_r\psi_s] \\
& & +(-1)^s\vf^{-1}[(\gm\mu)'-\gm\mu\vf'\vf^{-1}]\zeta_r\psi_s
+2\mu({\al'}^2+{\be'}^2\sin^2\al)\xi_r\phi_s+2(\al'f+\be'h\sin\al)\X\\
& & +\frac{{\al'}^2+{\be'}^2\sin^2\al+\gm_1'-\gm_1^2}{2}\X^2
  +[(\be'\sin\al)'-\al'\be'\cos\al]\X\Z-
\frac{1}{2}\gm^4\mu^2\vf^{-2}(\phi_s^2 +\xi_r^2)
\\ & & +\left(\frac{{\be'}^2}{2}\sin2\al+{\al'}'\right)\X\Y
+[(\be'\cos\al)'+\al'\be'\sin\al-f_{\Z t}-f_\Y f_\Z-h_\Y h_\Z]\Y\Z\\
& &+\frac{{\be'}^2-h_{\Z t}-f_\Z^2-h_\Z^2}{2}\Z^2 +
\frac{{\al'}^2+{\be'}^2\cos\al-f_{\Y t}-f_\Y^2-h_\Y^2}{2}\Y^2
\}.\hspace{2.4cm}(9.4.31)
\end{eqnarray*}
 By (9.3.3) and (9.3.5), we have the
following theorem: \psp

{\bf Theorem 9.4.1}. {\it Let $\al,\be,\gm_1,\vf$ and $\mu$ be
arbitrary functions in $t$ such that $\vf\neq 0$ and
${\al'}^2+{\be'}^2\sin^2\al\neq 0$. The notations $\X,\Y$ and $\Z$
are defined in (9.3.5) via (9.3.1), and $\al_1,\be_1$ and $\gm$ are
given in (9.4.29) and (9.4.30). Moreover, $f_\Y,f_\Z,h_\Y,h_\Z$ and
$f, h$ are given in (9.4.24)-(9.4.28). We have the following
solution of the Navier-Stokes equations (9.1.1)-(9.1.4): (1)
\begin{eqnarray*} u&=&-\al'(\X\sin\al+\Y\cos\al)
-(f+\mu\al'\gm\sinh(\al_1\Y+\be_1\Z)\;\cos\gm\X)\sin\al\\ & &
-(\gm_1\X+ \vf
\mu({\al'}^2+{\be'}^2\sin^2\al)\cosh(\al_1\Y+\be_1\Z)\;\sin\gm\X)\cos\al,\hspace{2.3cm}(9.4.32)
\end{eqnarray*}
\begin{eqnarray*}v&=&(f\cos\al-\be'\Z)\cos\be
-(\al'\sin\al\;\cos\be+\be'\cos\al\;\sin\be)\Y
\\ & &-(\gm_1\X+ \vf
\mu({\al'}^2+{\be'}^2\sin^2\al)\cosh(\al_1\Y+\be_1\Z)\;\sin\gm\X)\sin\al\;\cos\be -h\sin\be\\
& &+(\al'\cos\al\;\cos\be-
\be'\sin\al\;\sin\be)(\X+\gm\mu\sinh(\al_1\Y+\be_1\Z)\;\cos\gm\X),\hspace{1.1cm}(9.4.33)\end{eqnarray*}
\begin{eqnarray*}w&=&(\be'\cos\al\;\cos\be-
\al'\sin\al\;\sin\be)\Y +(f\cos\al-\be'\Z)\sin\be\\& & -(\gm_1\X+
\vf
\mu({\al'}^2+{\be'}^2\sin^2\al)\cosh(\al_1\Y+\be_1\Z)\;\sin\gm\X)\sin\al\;\sin\be
+h\cos\be\\&&+(\al'\cos\al\;\sin\be+\be'\sin\al\;\cos\be)(\X+\gm\mu\sinh(\al_1\Y+\be_1\Z)\;\cos\gm\X)
\hspace{1.1cm}(9.4.34)\end{eqnarray*}
\begin{eqnarray*}&
&p=\rho
\{\gm\mu\vf^{-1}[(\gm'-\gm\gm_1)\X\cosh(\al_1\Y+\be_1\Z)\;\sin\gm\X
-((\vf\al')'\Y+(\vf\be'\sin\al)'\Z\\ & &+\vf(\al'f+\be'h\sin\al)
)\sinh(\al_1\Y+\be_1\Z)\;\cos\gm\X]
-\vf^{-1}[(\gm\mu)'-\gm\mu\vf'\vf^{-1}]\\
& &\times\cosh(\al_1\Y+\be_1\Z)\;\cos\gm\X
+2\mu({\al'}^2+{\be'}^2\sin^2\al)\sinh(\al_1\Y+\be_1\Z)\;\sin\gm\X\\
& &+2(\al'f+\be'h\sin\al)\X+[(\be'\cos\al)'+\al'\be'\sin\al-f_{\Z
t}-f_\Y f_\Z-h_\Y h_\Z]\Y\Z
 \\ & &+\frac{{\al'}^2+{\be'}^2\sin^2\al+\gm_1'-\gm_1^2}{2}\X^2
 +[(\be'\sin\al)'-\al'\be'\cos\al]\X\Z
\\ & & +\left(\frac{{\be'}^2}{2}\sin2\al+{\al'}'\right)\X\Y-
\frac{1}{2}\gm^4\mu^2\vf^{-2}(\sin^2\gm\X +\sinh^2(\al_1\Y+\be_1\Z))
\\
& &+\frac{{\be'}^2-h_{\Z t}-f_\Z^2-h_\Z^2}{2}\Z^2 +
\frac{{\al'}^2+{\be'}^2\cos\al-f_{\Y t}-f_\Y^2-h_\Y^2}{2}\Y^2
\};\hspace{2.3cm}(9.4.35)
\end{eqnarray*}
(2)
\begin{eqnarray*} u&=&-\al'(\X\sin\al+\Y\cos\al)
-(f+\mu\al'\gm\sin(\al_1\Y+\be_1\Z)\;\cosh\gm\X)\sin\al\\ & &
-(\gm_1\X+ \vf
\mu({\al'}^2+{\be'}^2\sin^2\al)\cos(\al_1\Y+\be_1\Z)\;\sinh\gm\X)\cos\al,\hspace{2.3cm}(9.4.36)
\end{eqnarray*}
\begin{eqnarray*}v&=&(f\cos\al-\be'\Z)\cos\be
-(\al'\sin\al\;\cos\be+\be'\cos\al\;\sin\be)\Y
\\ & &-(\gm_1\X+ \vf
\mu({\al'}^2+{\be'}^2\sin^2\al)\cos(\al_1\Y+\be_1\Z)\;\sinh\gm\X)\sin\al\;\cos\be -h\sin\be\\
& &+(\al'\cos\al\;\cos\be-
\be'\sin\al\;\sin\be)(\X+\gm\mu\sin(\al_1\Y+\be_1\Z)\;\cosh\gm\X),\hspace{1.1cm}(9.4.37)\end{eqnarray*}
\begin{eqnarray*}w&=&(\be'\cos\al\;\cos\be-
\al'\sin\al\;\sin\be)\Y +(f\cos\al-\be'\Z)\sin\be\\& & -(\gm_1\X+
\vf
\mu({\al'}^2+{\be'}^2\sin^2\al)\cos(\al_1\Y+\be_1\Z)\;\sinh\gm\X)\sin\al\;\sin\be
+h\cos\be\\&&+(\al'\cos\al\;\sin\be+\be'\sin\al\;\cos\be)(\X+\gm\mu\sin(\al_1\Y+\be_1\Z)\;\cosh\gm\X)
\hspace{1.1cm}(9.4.38)\end{eqnarray*}
\begin{eqnarray*}&
&p=\rho
\{\gm\mu\vf^{-1}[(\gm'-\gm\gm_1)\X\cos(\al_1\Y+\be_1\Z)\;\sinh\gm\X
-((\vf\al')'\Y+(\vf\be'\sin\al)'\Z\\ & &+\vf(\al'f+\be'h\sin\al)
)\sin(\al_1\Y+\be_1\Z)\;\cosh\gm\X]
+\vf^{-1}[(\gm\mu)'-\gm\mu\vf'\vf^{-1}]\\
& &\times\cos(\al_1\Y+\be_1\Z)\;\cosh\gm\X
+2\mu({\al'}^2+{\be'}^2\sin^2\al)\sin(\al_1\Y+\be_1\Z)\;\sinh\gm\X\\
& &+2(\al'f+\be'h\sin\al)\X+[(\be'\cos\al)'+\al'\be'\sin\al-f_{\Z
t}-f_\Y f_\Z-h_\Y h_\Z]\Y\Z
 \\ & &+\frac{{\al'}^2+{\be'}^2\sin^2\al+\gm_1'-\gm_1^2}{2}\X^2
 +[(\be'\sin\al)'-\al'\be'\cos\al]\X\Z
\\ & & +\left(\frac{{\be'}^2}{2}\sin2\al+{\al'}'\right)\X\Y-
\frac{1}{2}\gm^4\mu^2\vf^{-2}(\sinh^2\gm\X +\sin^2(\al_1\Y+\be_1\Z))
\\
& &+\frac{{\be'}^2-h_{\Z t}-f_\Z^2-h_\Z^2}{2}\Z^2 +
\frac{{\al'}^2+{\be'}^2\cos\al-f_{\Y t}-f_\Y^2-h_\Y^2}{2}\Y^2
\}.\hspace{2.4cm}(9.4.39)
\end{eqnarray*}}

Let $\gm_1,\gm_2$ be functions in $t$ and let $a,b,c$ be real
numbers. Denote
$$\phi_0=
e^{\gm_1\Y+\gm_2\Z}-ae^{-\gm_1\Y-\gm_2\Z},
\qquad\phi_1=\sin(\gm_1\Y+\gm_2\Z),\eqno(9.4.40)$$
$$\psi_0=
e^{\gm_1\Y+\gm_2\Z}+ae^{-\gm_1\Y-\gm_2\Z},
\qquad\psi_1=\cos(\gm_1\Y+\gm_2\Z),\eqno(9.4.41)$$
$$\xi_0=
be^{\gm_1\Y+\gm_2\Z}-ce^{-\gm_1\Y-\gm_2\Z},
\qquad\xi_1=c\sin(\gm_1\Y+\gm_2\Z+b),\eqno(9.4.42)$$
$$\zeta_0=
be^{\gm_1\Y+\gm_2\Z}+ce^{-\gm_1\Y-\gm_2\Z},
\qquad\zeta_1=c\cos(\gm_1\Y+\gm_2\Z+b).\eqno(9.4.43)$$ Suppose that
$\sgm,\tau$ are functions in $t$ and $f,k,h$ are functions in
$t,\X,\Y,\Z$ such that $h$ and $g$ are linear in $\X,Y,\Z$ and
$$f_\X+k_\Y+h_\Z=0.\eqno(9.4.44)$$
Motivated from the above solution, we consider the solution of the
form:
$$\U=-\al'\Y-\be'\Z\sin\al+f-(\gm_1^2+\gm_2^2)
(\tau\zeta_r\X+\sgm\psi_r\X^2),\eqno(9.4.45)$$
$$\V=\al'\X-\be'\Z\cos\al+k+\gm_1(\tau\xi_r+2\sgm\phi_r\X),\eqno(9.4.46)$$
$$\W=\be'(\X\sin\al+\Y\cos\al)+h+\gm_2(\tau\xi_r+2\sgm\phi_r\X)
.\eqno(9.4.47)$$

For convenience of computation, we denote
$$\gm=\gm_1^2+\gm_2^2,\qquad f^\ast=f-f_x\X
\qquad\Dlt_1=\ptl_\Y^2+\ptl_\Z^2.\eqno(9.4.48)$$ Now (9.4.1) becomes
\begin{eqnarray*}&
&R_1=-{\al'}'\Y-(\be'\sin\al)'\Z+f_t-(-1)^r\gm(\gm_1'\Y+\gm_2'\Z)(\tau\xi_r\X+\sgm\phi_r\X^2)\\
& & +((-1)^r\nu \gm^2\tau-(\gm\tau)')\zeta_r\X +
(f-\gm(\tau\zeta_r\X+\sgm\psi_r\X^2))(f_\X-\gm(\tau\zeta_r+2\sgm\psi_r\X))
\\ & &+(k+\gm_1(\tau\xi_r+2\sgm\phi_r\X)) [f_\Y-2\al'-(-1)^r\gm\gm_1
(\tau\xi_r\X+\sgm\phi_r\X^2)]-\nu\Dlt_1(f)\\ & &
+(h+\gm_2(\tau\xi_r+2\sgm\phi_r\X))[f_\Z-2\be'\sin\al-(-1)^r\gm\gm_2
(\tau\xi_r\X+\sgm\phi_r\X^2)]+2\nu\gm\sgm\psi_r\\&&-\al'(\al'\X-\be'\Z\cos\al)
-{\be'}^2(\X\sin\al+\Y\cos\al)\sin\al+((-1)^r\nu
\gm^2\sgm-(\gm\sgm)')\psi_r\X^2
\\&&=-({\al'}^2+{\be'}^2\sin^2\al)\X-({\al'}'+2^{-1}{\be'}^2\sin2\al)\Y
+(\al'\be'\cos\al-(\be'\sin\al)')\Z\\ &
&+\gm^2[\tau^2(4b\dlt_{0,r}+c\dlt_{1,r})c\X+3\sgm\tau(2\dlt_{0,r}(ab+c)+
\dlt_{1,r}c\cos b)\X^2+2\sgm^2 (4a\dlt_{0,r}+\dlt_{1,r})\X^3]
\\ & &-(-1)^r\gm(\gm_1'\Y+\gm_2'\Z+k\gm_1+h\gm_2)
(\tau\xi_r\X+\sgm\phi_r\X^2)+ff_\X+k(f_\Y-2\al')\\ &
&+h(f_\Z-2\be'\sin\al)+((-1)^r\nu \gm^2\sgm-(\gm\sgm)'-3\gm\sgm
f_\X)\psi_r\X^2+\nu(2\gm\sgm\psi_r-\Dlt_1(f))\\ & &-\gm\tau
f^\ast\zeta_r-[((\gm\tau)'+2\gm\tau f_\X-(-1)^r\nu
\gm^2\tau)\zeta_r+2\gm\sgm
f^\ast\psi_r]\X+f_t\\
&& +(\gm_1 (f_\Y-2\al') +\gm_2(f_\Z-2\be'\sin\al))
(\tau\xi_r+2\sgm\phi_r\X).\hspace{4.6cm}(9.4.49)
\end{eqnarray*}

To solve (9.3.24), we assume
$$\gm_1'\Y+\gm_2'\Z+k\gm_1+h\gm_2=0\eqno(9.4.50)$$
and
$$
(-1)^r\nu\gm^2\sgm-(\gm\sgm)'-3\gm\sgm f_\X=0,\eqno(9.4.51)$$
 Moreover, (9.4.2) and
(9.4.3) become
\begin{eqnarray*}& &R_2={\al'}'\X-(\be'\cos\al)'\Z+
((\gm_1\tau)'-(-1)^r\nu\gm\gm_1\tau)\xi_r+2((\gm_1\sgm)'-
(-1)^r\nu\gm\gm_1\sgm)\phi_r\X\\ &
&+k_t+(\gm_1'\Y+\gm_2'\Z)\gm_1(\tau\zeta_r+2\sgm\psi_r\X)
+(f-\gm(\tau\zeta_r\X+\sgm\psi_r\X^2))(2\al'+k_\X+2\gm_1\sgm\phi_r)
\\ & &+(k+\gm_1(\tau\xi_r+2\sgm\phi_r\X))(k_\Y+\gm_1^2(\tau\zeta_r
+2\sgm\psi_r\X))-{\be'}^2(\X\sin\al+\Y\cos\al)\cos\al\\ & &
-\al'(\al'\Y+\be'\Z\sin\al)+(h+\gm_2(\tau\xi_r+2\sgm\phi_r\X))(k_\Z
-2\be'\cos\al+\gm_1\gm_2
(\tau\zeta_r+2\sgm\psi_r\X))\\
&&=({\al'}'-2^{-1}{\be'}^2\sin2\al+f_\X(2\al'+k_\X))\X
-({\al'}^2+{\be'}^2\cos^2\al)\Y+k_t+kk_\Y\\ & & +
[\tau(\gm_1k_\Y+\gm_2(k_\Z-2\be'\cos\al))+(\gm_1\tau)'-(-1)^r\nu\gm\gm_1\tau]\xi_r
-((\be'\cos\al)'+\al'\be'\sin\al)\Z\\&&+\gm\sgm(2\sgm\gm_1\phi_r-2\al'
-k_\X)\psi_r\X^2+f^\ast(2\al'+k_\X+2\gm_1\sgm\phi_r)+\gm\gm_1\tau^2\xi_r\zeta_r\\
& &+h(k_\Z-2\be'\cos\al) + \{2\gm\gm_1\sgm\tau\xi_r\psi_r
+2[(\gm_1\sgm)'-\sgm\gm_1(h_\Z+ (-1)^r\nu\gm)\\ &
&+\gm_2\sgm(k_\Z-2\be'\cos\al))]\phi_r
-\gm\tau(2\al'+k_\X)\zeta_r\}\X,\hspace{5.5cm}(9.4.52)\end{eqnarray*}
\begin{eqnarray*}& &R_3=(\be'\sin\al)'\X+(\be'\cos\al)'\Y+
 (\gm_2\tau)'\xi_r+2(\gm_2\sgm)'\phi_r\X-(-1)^r\nu\gm_2\gm(\tau\xi_r+2\sgm\phi_r\X)\\ &
&+(\gm_1'\Y+\gm_2'\Z)\gm_2(\tau\zeta_r+2\sgm\psi_r\X)
+(f-\gm(\tau\zeta_r\X+\sgm\psi_r\X^2))(2\be'\sin\al+h_\X+2\gm_2\sgm\phi_r)
\\ & &+(k+\gm_1(\tau\xi_r+2\sgm\phi_r\X))(2\be'\cos\al+
h_\Y+\gm_1\gm_2(\tau\zeta_r +2\sgm\psi_r\X)-{\be'}^2\Z+h_t\\ &
&+\al'\be'(\X\cos\al-\Y\sin\al)
+(h+\gm_2(\tau\xi_r+2\sgm\phi_r\X))(h_\Z
+\gm_2^2(\tau\zeta_r+2\sgm\psi_r\X))\\ &
&=[(\be'\sin\al)'+\al'\be'\cos\al+f_\X(2\be'\sin\al+h_\X)]\X+
[(\be'\cos\al)'-\al'\be'\sin\al]\Y
\\ & & +[(\gm_2\tau)'+(\gm_1(2\be'\cos\al+
h_\Y)+\gm_2h_\Z-(-1)^r\nu\gm\gm_2)\tau]\xi_r+\{
2\gm\gm_2\tau\sgm\xi_r\psi_r +2[(\gm_2\sgm)'\\ &
&-\gm_2\sgm(k_\Y+(-1)^r\nu\gm)+\gm_1\sgm
(2\be'\cos\al+h_\Y)]\phi_r-\gm\tau(2\be'\sin\al+ h_\X)\zeta_r\}\X\\
& &+f^\ast(2\be'\sin\al+h_\X+2\gm_2\sgm\phi_r) +k(2\be'\cos\al+
h_\Y)+h_t+hh_\Z+\gm\gm_2\tau^2\xi_r\zeta_r\\ &
&+\gm\sgm(2\gm_2\sgm\phi_r-2\be'\sin\al-h_\X)\psi_r\X^2-{\be'}^2\Z
\hspace{6.1cm}(9.4.53)
\end{eqnarray*}
by (9.4.50).

Thanks to the coefficients of $\X^2$ in $\ptl_\Z(R_2)=\ptl_\Y(R_3)$,
we have:
$$\gm_2(2\al'+k_\X)=\gm_1(2\be'\sin\al+h_\X).\eqno(9.4.54)$$
According to (9.4.50),
$$k_\X\gm_1+h_\X\gm_2=0,\;\;\gm_1'+\gm_1k_\Y+\gm_2h_\Y=0,\;\;
\gm_2'+\gm_1k_\Z+\gm_2h_\Z=0.\eqno(9.4.55)$$ Solving (9.4.54) and
the first equation in (9.4.55), we obtain
$$k_\X=2\gm^{-1}\gm_2(\be'\gm_1\sin\al-\al'\gm_2),\qquad h_\X=
-2\gm^{-1}\gm_1(\be'\gm_1\sin\al-\al'\gm_2).\eqno(9.4.56)$$
Moreover, the coefficients of $\X$ in $\ptl_\Z(R_2)=\ptl_\Y(R_3)$
give
$$\gm_1'\gm_2-\gm_1\gm_2'+\gm_1\gm_2(k_\Y-h_\Z
)+\gm_2^2k_\Z-\gm_1^2h_\Y-2\gm\be'\cos\al=0 \eqno(9.4.57)$$ by
(9.4.50). According to (9.4.55), the above equation can be rewritten
as
$$k_\Z-h_\Y=2\be'\cos\al.
\eqno(9.4.58)$$ Furthermore,  (9.4.54) and the coefficients of
$\X^0$ in $\ptl_\Z(R_2)=\ptl_\Y(R_3)$ show that $f$ is a function of
$t$ and $\gm_1\Y+\gm_2\Z$ by the method of characteristics in
Section 4.1.
 According to the coefficients of $\X$ in
$\ptl_\Y(R_1)=\ptl_\X(R_2)$ and $\ptl_\Z(R_1)=\ptl_\X(R_3)$, we take
$$f^\ast=\vf\vt_r+\sgm\td\varpi\phi_r+\al_1
,\eqno(9.4.59)$$ where $\vf$ and $\al_1$ are functions in $t$, and
$$\td\varpi=\gm_1\Y+\gm_2\Z,\qquad
\vt_0=b_1e^{\td\varpi}-c_1e^{-\td\varpi},\qquad
\vt_1=c_1\sin(\td\varpi+b_1)\eqno(9.4.60)$$ for $b_1,c_1\in\mbb{R}$.

Note $$2\sgm(\gm_1f_\Y+\gm_2f_\Z)\phi_r=2\gm\sgm
f^\ast_{\td\varpi}\phi_r.\eqno(9.4.61)$$ Denote
$$\hat\vt_0=b_1e^{\td\varpi}+c_1e^{-\td\varpi},\qquad
\hat\vt_1=c_1\cos(\td\varpi+b_1).\eqno(9.4.62)$$ Then
$$f^\ast_{\td\varpi}=\vf\hat\vt_r+\sgm(\phi_r+\td\varpi\psi_r),\;\;f^\ast_{\td\varpi\td\varpi}=
(-1)^r(\vf\vt_r+\sgm\td\varpi\phi_r)+2\sgm\psi_r.\eqno(9.4.63)$$
Moreover. \begin{eqnarray*}\qquad & &\ptl_\Y(2\gm\sgm
f^\ast_{\td\varpi}\phi_r-2\gm\sgm f^\ast\psi_r)
\\&=&2\gm\gm_1\sgm[f^\ast_{\td\varpi\td\varpi}\phi_r+f^\ast_{\td\varpi}\psi_r
-(f^\ast_{\td\varpi}\psi_r+(-1)^rf^\ast\phi_r)]\\ &=& 2\gm\gm_1\sgm[
((-1)^r(\vf\vt_r+\sgm\td\varpi\phi_r)+2\sgm\psi_r)\phi_r-(-1)^r(\vf\vt_r+\sgm\td\varpi\phi_r+\al_1)\phi_r]
\\ &=&4\gm\gm_1\sgm^2\phi_r\psi_r-(-1)^r2\al_1\gm\gm_1\sgm\phi_r.\hspace{7.1cm}(9.4.64)
\end{eqnarray*}
Similarly,
$$\ptl_\Z(2\gm\sgm
f^\ast_{\td\varpi}\phi_r-2\gm\sgm
f^\ast\psi_r)=4\gm\gm_2\sgm^2\phi_r\psi_r-(-1)^r2\al_1\gm\gm_2\sgm\phi_r.\eqno(9.4.65)$$
 Now the coefficients of $\X$ in $\ptl_\Y(R_1)=\ptl_\X(R_2)$ give
\begin{eqnarray*}\qquad& &-(-1)^r\gm_1[((\gm\tau)'+2\gm\tau f_\X-(-1)^r\nu
\gm^2\tau)\xi_r+2\al_1\gm\sgm\phi_r]\\ & & -4\gm_1\sgm(\gm_1\al'
+\gm_2\be'\sin\al)\psi_r
=-2\gm\sgm(2\al'+k_\X)\psi_r\hspace{4.4cm}(9.4.66)
\end{eqnarray*}
by (9.4.49), (9.4.52) and (9.4.64). According to (9.4.49), (9.4.53)
and (9.4.65), the coefficients of $\X$ in
$\ptl_\X(R_1)=\ptl_\X(R_3)$ imply
\begin{eqnarray*}\qquad& &-(-1)^r\gm_2[((\gm\tau)'+2\gm\tau f_\X-(-1)^r\nu
\gm^2\tau)\xi_r+2\al_1\gm\sgm\phi_r]\\ & & -4\gm_2\sgm(\gm_1\al'
+\gm_2\be'\sin\al)\psi_r
=-2\gm\sgm(2\be'\sin\al+h_\X)\psi_r.\hspace{3.3cm}(9.4.67)
\end{eqnarray*}

Observe that (9.4.56) yields
$$\gm(2\al'+k_\X)=2\al'\gm+2\gm_2(\be'\gm_1\sin\al-\al'\gm_2)=2\gm_1(\gm_1\al'+\gm_2\be'\sin\al),\eqno(9.4.68)$$
$$\gm(2\be'\sin\al+h_\X)=2\be'\gm\sin\al-2\gm_1(\be'\gm_1\sin\al-\al'\gm_2)=2\gm_2(\gm_2\be'\sin\al+\gm_1\al').
\eqno(9.4.69)$$ Thus (9.4.66) and (9.4.67) are  implied by
$$((\gm\tau)'+2\gm\tau f_\X-(-1)^r\nu
\gm^2\tau)\xi_r+2\al_1\gm\sgm\phi_r=0\eqno(9.4.70)$$

As (9.4.64) and (9.4.65), Expressions (9.4.40)-(9.4.43) and
(9.4.59)-(9.4.62) give
$$\gm\tau\ptl_\Y(f^\ast_{\td\varpi}\xi_r-
f^\ast\zeta_r)=\gm\gm_1\tau(2\sgm\xi_r\psi_r-(-1)^r\al_1\xi_r+\hat
c_r\sgm),\eqno(9.4.71)$$
$$\gm\tau\ptl_\Z(f^\ast_{\td\varpi}\xi_r-
f^\ast\zeta_r)=\gm\gm_2\tau(2\sgm\xi_r\psi_r-(-1)^r\al_1\xi_r+\hat
c_r\sgm),\eqno(9.4.72)$$ where
$$\hat c_0=\xi_0\psi_0-\zeta_0\phi_0=2(ab-c),\qquad\hat
c_1=\xi_1\psi_1-\\zeta_1\phi_1=c\sin b.\eqno(9.4.73)$$

Moreover,
$$kf_\Y+hf_\Z=(\gm_1k+\gm_2h)f^\ast_{\td\varpi}=-(\gm_1'\Y+\gm_2'\Z)f^\ast_{\td\varpi}=-\ptl_t(\td\varpi)
f^\ast_{\td\varpi}\eqno(9.4.74)$$ by (6.4.55). On the other hand,
$$\ptl_t(f^\ast)=f^\ast_t+\ptl_t(\td\varpi)
f^\ast_{\td\varpi}.\eqno(9.4.75)$$ Thus the coefficients of $\X^0$
in $\ptl_\Y(R_1)=\ptl_\X(R_2)$ give
\begin{eqnarray*}&&[(f_\X-(-1)^r\gm\nu)\vf+\vf')\vt_r
+((f_\X-(-1)^r\nu\gm)\sgm+\sgm')\td\varpi\phi_r
-\al_1\gm\tau\zeta_r]_\Y\\
 &=&2{\al'}'-(2\al'+k_\X)h_\Z+k_{\X
t}+(h_\X+2\be'\sin\al)h_\Y-\hat c_r\gm_1\gm\sgm\tau\\ &
&+2[(\gm_1\sgm)'-\gm_1\sgm (h_\Z+ (-1)^r\nu\gm)+\gm_2\sgm
h_\Y]\phi_r\hspace{5.8cm}(9.4.76)
\end{eqnarray*}
and the coefficients of $\X^0$ in $\ptl_\X(R_1)=\ptl_\X(R_3)$ yield
\begin{eqnarray*}&&
[(f_\X-(-1)^r\gm\nu)\vf+\vf')\vt_r
+((f_\X-(-1)^r\nu\gm)\sgm+\sgm')\td\varpi\phi_r
-\al_1\gm\tau\zeta_r]_\Z
\\ &=&2(\be'\sin\al)'+h_{\X t}-
(h_\X+2\be'\sin\al)k_\Y
+(k_\X+2\al') k_\Z -\hat c_r\gm_2\gm\sgm\tau\\
& &+2[(\gm_2\sgm)'-\gm_2\sgm (k_\Y+ (-1)^r\nu\gm)+\gm_1\sgm k_\Z
]\phi_r \hspace{5.8cm}(9.4.77)
\end{eqnarray*}
by (9.4.44), (9.4.49), (9.4.52), (9.4.53), (9.4.58),  (9.4.68),
(9.4.69) and (9.4.71)-(9.4.74). Thus we have:
$$2{\al'}'-(2\al'+k_\X)h_\Z+k_{\X
t}+(h_\X+2\be'\sin\al)h_\Y-\hat c_r\gm_1\gm\sgm\tau=0\eqno(9.4.78)$$
and
$$2(\be'\sin\al)'+h_{\X t}-
(h_\X+2\be'\sin\al)k_\Y +(k_\X+2\al') k_\Z-\hat
c_r\gm_2\gm\sgm\tau=0.\eqno(9.4.79)$$

For simplicity, we only consider two special cases a follows. \psp

{\it Case 1}. $\vt_r=\zeta_r,\;\sgm=0,\;\gm_1=\al'\mu$ and
$\gm_2=\be'\mu\sin\al$, where $\mu$ is a function in $t$.\psp

In this case,
$$ k_\X=h_\X=0\eqno(9.4.80)$$
by (9.4.56).  Moreover, (9.4.78) and (9.4.79) becomes
$$\al'h_\Z-\be'\sin\al\;h_\Y={\al'}',\qquad\al'k_\Z-\be'\sin\al\;k_\Y=-(\be'\sin\al)'.\eqno(9.4.81)$$
Furthermore, (9.4.55) becomes
$$\al'k_\Y+\be'\sin\al\;h_\Y=-{\al'}'-\al'\frac{\mu'}{\mu},\eqno(9.4.82)$$
$$\al'k_\Z+\be'\sin\al\;h_\Z=-(\be'\sin\al)'
-\frac{\be'\mu'}{\mu}\sin\al.\eqno(9.4.83)$$ Adding (9.4.82) to the
first equation in (9.4.81), we get
$$\al'(k_\Y+h_\Z)=-\al'\frac{\mu'}{\mu}\sim\al'f_\X=\al'\frac{\mu'}{\mu}
\lra f_\X=\frac{\mu'}{\mu}\eqno(9.4.84)$$ by (9.4.44). Note
$$h_\Z=-f_\X-h_\Y=-\frac{\mu'}{\mu}-k_\Y.\eqno(9.4.85)$$
Substituting (9.4.85) into the first equation (9.4.81), we have
$$h_\Y=-\frac{\mu(\al'k_\Y+{\al'}')+\al'\mu'}{\be'\mu\sin\al}.\eqno(9.4.86)$$
In addition, the second equation in (9.4.81) yields
$$k_\Z=\frac{\be'\sin\al\;k_\Y-(\be'\sin\al)'}{\al'}.\eqno(9.4.87)$$
Note that (9.4.85)-(9.4.87) satisfy (9.4.82) and (9.4.83).

According to (9.4.58),
$$\frac{\be'\sin\al\;k_\Y-(\be'\sin\al)'}{\al'}
+\frac{\mu(\al'k_\Y+{\al'}')+\al'\mu'}{\be'\mu\sin\al}=2\be'\cos\al.
\eqno(9.4.88)$$ Thus
$$\mu({\al'}^2+{\be'}^2\sin^2\al)k_\Y-\mu(\be'\sin\al)'\be'\sin\al+\mu\al'{\al'}'+{\al'}^2\mu'=
\al'{\be'}^2\mu\sin2\al\eqno(9.4.89)$$
$$\lra
k_\Y=\frac{\mu[\al'({\be'}^2\sin2\al-{\al'}')+(\be'\sin\al)'\be'\sin\al]-{\al'}^2\mu'}{\mu({\al'}^2+{\be'}^2\sin^2\al)}
.\eqno(9.4.90)$$ By (9.4.87),
$$k_\Z=\frac{[\mu({\be'}^2\sin2\al-{\al'}')-\al'\mu']\be'\sin\al-\mu\al'(\be'\sin\al)'}{\mu({\al'}^2+{\be'}^2\sin^2\al)}.
\eqno(9.4.91)$$ Moreover,
$$h_\Y=k_\Z-2\be'\cos\al=-\frac{\be'[(\mu{\al'}'+\al'\mu')\sin\al+2\mu{\al'}^2\cos\al]+\mu\al'(\be'\sin\al)'}
{\mu({\al'}^2+{\be'}^2\sin^2\al)}.\eqno(9.4.92)$$ In addition,
(9.4.85) gives
$$h_\Z=-\frac{\al'({\be'}^2\sin2\al-{\al'}')+(\be'\sin\al)'\be'\sin\al+{\be'}^2\sin^2\al}{{\al'}^2+{\be'}^2\sin^2\al}
.\eqno(9.4.93)$$ In particular, $k=k_\Y\Y+k_\Z\Z$ and
$h=h_\Y\Y+h_\Z\Z$ are determined by (9.4.90)-(9.4.93).

Now (9.4.70) is equivalent to
$$(\gm\tau)'+2\gm\tau f_\X-(-1)^r\nu
\gm^2\tau=0.\eqno(9.4.94)$$ According to (9.4.84), the above
equation can be written as
$$(\gm\tau)'+\frac{2\mu'}{\mu}(\gm\tau)-(-1)^r\nu
\gm(\gm\tau)=0.\eqno(9.4.95)$$ So
$$\gm\tau=\frac{1}{\mu^2}\exp((-1)^r\nu\int
\gm
dt)=\frac{1}{\mu^2}\exp[(-1)^r\nu\int\mu^2({\al'}^2+{\be'}^2\sin^2\al)dt].\eqno(9.4.96)$$
Hence
$$\tau=\frac{\exp[(-1)^r\nu\int\mu^2({\al'}^2+{\be'}^2\sin^2\al)dt]}{\mu^4({\al'}^2+{\be'}^2\sin^2\al)}.\eqno(9.4.97)$$
 Note that (9.4.76) and (9.4.77) are implied by
$$(f_\X-(-1)^r\gm\nu)\vf+\vf'-\al_1\gm\tau=0\eqno(9.4.98)$$
$$\lra
\al_1=\frac{[\mu\mu'-(-1)^r\mu^4({\al'}^2+{\be'}^2\sin^2\al)]\vf+\mu^2\vf'}{
\exp[(-1)^r\nu\int\mu^2({\al'}^2+{\be'}^2\sin^2\al)dt]}.\eqno(9.4.99)$$
It can be verified that the equation for the coefficients of $\X^0$
in $\ptl_\Z(R_2)=\ptl_\Y(R_3)$ is implied by (9.4.55), (9.4.58) and
the assumption that $\sgm=0,\;\gm_1=\al'\mu$ and
$\gm_2=\be'\mu\sin\al$.

According to (9.4.45)-(9.4.47), (9.4.59), (9.4.90)-(9.4.93),
(9.4.97) and (9.4.99), we have
\begin{eqnarray*}\U&=&\frac{\mu'}{\mu}\X-\al'\Y-\be'\Z\sin\al
+[\vf-\mu^{-2}\exp[(-1)^r\nu\int\mu^2({\al'}^2+{\be'}^2\sin^2\al)dt]]\zeta_r\\
&
&+\frac{[\mu\mu'-(-1)^r\mu^4({\al'}^2+{\be'}^2\sin^2\al)]\vf+\mu^2\vf'}{
\exp[(-1)^r\nu\int\mu^2({\al'}^2+{\be'}^2\sin^2\al)dt]}
,\hspace{5.2cm}(9.4.100)\end{eqnarray*}
$$\V=\al'\X-\be'\Z\cos\al+k+\frac{\al'\xi_r\exp[(-1)^r\nu\int\mu^2({\al'}^2
+{\be'}^2\sin^2\al)dt]}{\mu^3({\al'}^2+{\be'}^2\sin^2\al)},
\eqno(9.4.101)$$
$$\W=\be'(\X\sin\al+\Y\cos\al)+h
+\frac{\be'\xi_r\sin\al\;\exp[(-1)^r\nu\int\mu^2({\al'}^2+{\be'}^2\sin^2\al)dt]}{\mu^3({\al'}^2+{\be'}^2\sin^2\al)}.
\eqno(9.4.102)$$

Observe that $f^\ast=\vf\zeta_r+\al_1$ and so
$$f^\ast_{\td\varpi}\xi_r-f^\ast\zeta_r=\vf((-1)^r\xi_r^2-\zeta_r^2)-\al_1\zeta_r=
-(4b\dlt_{0,r}+c\dlt_{1,r})c\vf-\al_1\zeta_r.\eqno(9.4.103)$$ Hence
\begin{eqnarray*}
R_1&=&(\mu'/\mu-{\al'}^2-{\be'}^2\sin^2\al)\X+({\al'}'+2\al'\mu'/\mu-2^{-1}{\be'}^2\sin2\al)\Y
\\ & &+(\al'\be'\cos\al+(\be'\sin\al)'+2\be'\mu'\sin\al/\mu)\Z\\ & &+
\gm\tau(4b\dlt_{0,r}+c\dlt_{1,r})c(\gm\tau\X-\vf) -2\gm\tau\xi_r/\mu
\hspace{5.5cm}(9.4.104)
\end{eqnarray*}
by (9.4.49), (9.4.55), (9.4.74), (9.4.75) and (9.4.103). Moreover,
(9.4.52), (9.4.53), (9.4.55) and (9.4.58) yield
\begin{eqnarray*}R_2&=&({\al'}'-2^{-1}{\be'}^2\sin2\al+2\al'\mu'/\mu)\X
+(k_{\Y t}-{\al'}^2-{\be'}^2\cos^2\al)\Y+\\ & & +
\gm_1(\tau'-(-1)^r\nu\gm\tau)\xi_r
+((k_\Z-\be'\cos\al)'-\al'\be'\sin\al)\Z\\&&+2\al'f^\ast+\gm\gm_1\tau^2\xi_r\zeta_r
-2\al'\gm\tau\zeta_r\X+kk_\Y+hh_\Y,\hspace{4.4cm}(9.4.105)\end{eqnarray*}
\begin{eqnarray*}R_3&=&[(\be'\sin\al)'+\al'\be'\cos\al+2\be'\mu'\sin\al/\mu]\X+
[(h_\Y+\be'\cos\al)'-\al'\be'\sin\al]\Y
\\ & & +\gm_2(\tau'-(-1)^r\nu\gm\tau)\xi_r-2\be'\gm\tau\sin\al\;\zeta_r\X+(h_{\Z t}-{\be'}^2)\Z\\
& &+2\be'\sin\al\;f^\ast +kk_\Z+hh_\Z+\gm\gm_2\tau^2\xi_r\zeta_r.
\hspace{5.7cm}(9.4.106)
\end{eqnarray*}
\pse

By (9.3.3), (9.3.5), (9.3.22), (9.4.100)-(9.4.102) and
(9.4.104)-(9.4.106),  we have the following theorem: \psp

{\bf Theorem 9.4.2}. {\it Let $\al,\be,\vf,\mu$ be arbitrary
functions in $t$ such that $\mu({\al'}^2+{\be'}^2\sin^2\al)\neq 0$,
and let $b,c$ be arbitrary real constants. Define the moving frame
$\X,\;\Y$ and $\Z$ by (9.3.1) and (9.3.5), and
$$\xi_0=be^{\mu(\al'\Y+\be'\Z\sin\al)}-ce^{-\mu(\al'\Y+\be'\Z\sin\al)},\qquad
\xi_1=c\sin[\mu(\al'\Y+\be'\Z\sin\al)+b],\eqno(9.4.107)$$
$$\zeta_0=be^{\mu(\al'\Y+\be'\Z\sin\al)}+ce^{-\mu(\al'\Y+\be'\Z\sin\al)},\qquad
\zeta_1=c\cos[\mu(\al'\Y+\be'\Z\sin\al)+b].\eqno(9.4.108)$$
Moreover, $k=k_\Y\Y+k_\Z\Z$ and $h=h_\Y\Y+h_\Z\Z$ are defined by
(9.4.90)-(9.4.93). For $r=0,1$, we have the following solution of
the Navier-Stokes equations (9.1.1)-(9.1.4):
\begin{eqnarray*}u&=&\left(\frac{\mu'}{\mu}\X-\al'\Y+[\vf-\mu^{-2}\exp[(-1)^r\nu\int\mu^2({\al'}^2+{\be'}^2\sin^2\al)dt]]\zeta_r\right)\cos\al
\\& &
-\left[\al'\X+k+\frac{\al'\xi_r\exp[(-1)^r\nu\int\mu^2({\al'}^2+{\be'}^2\sin^2\al)dt]}{\mu^3({\al'}^2
+{\be'}^2\sin^2\al)}\right]\sin\al
\\
&&+\frac{[\mu\mu'-(-1)^r\mu^4({\al'}^2+{\be'}^2\sin^2\al)]\vf+\mu^2\vf'}{
\exp[(-1)^r\nu\int\mu^2({\al'}^2+{\be'}^2\sin^2\al)dt]}\cos\al
,\hspace{4.3cm}(9.4.109)\end{eqnarray*}
\begin{eqnarray*}v&=&
\left(\frac{\mu'}{\mu}\X-\al'\Y+[\vf-\mu^{-2}\exp[(-1)^r\nu\int\mu^2({\al'}^2+{\be'}^2\sin^2\al)dt]]\zeta_r\right)\sin\al\;
\cos\be
\\& &
+[(\al'\X+k)\cos\al-\be'\Z]\cos\be-[\be'(\X\sin\al+\Y\cos\al)+h]\sin\be
\\ & &+\frac{(\al'\cos\al\;\cos\be-\be'\sin\al\;\sin\be)
\xi_r\exp[(-1)^r\nu\int\mu^2({\al'}^2+{\be'}^2\sin^2\al)dt]}{\mu^3({\al'}^2
+{\be'}^2\sin^2\al)}
\\&&+\frac{[\mu\mu'-(-1)^r\mu^4({\al'}^2+{\be'}^2\sin^2\al)]\vf+\mu^2\vf'}{
\exp[(-1)^r\nu\int\mu^2({\al'}^2+{\be'}^2\sin^2\al)dt]}\sin\al\;\cos\be
,\hspace{3.3cm}(9.4.110)\end{eqnarray*}
\begin{eqnarray*}w&=&
\left(\frac{\mu'}{\mu}\X-\al'\Y+[\vf-\mu^{-2}\exp[(-1)^r\nu\int\mu^2({\al'}^2+{\be'}^2\sin^2\al)dt]]\zeta_r\right)
\sin\al\; \sin\be
\\& &
+[(\al'\X+k)\cos\al-\be'\Z]\sin\be+[\be'(\X\sin\al+\Y\cos\al)+h]\cos\be
\\ & &+\frac{(\al'\cos\al\;\sin\be+\be'\sin\al\;\cos\be)
\xi_r\exp[(-1)^r\nu\int\mu^2({\al'}^2+{\be'}^2\sin^2\al)dt]}{\mu^3({\al'}^2
+{\be'}^2\sin^2\al)}
\\&&+\frac{[\mu\mu'-(-1)^r\mu^4({\al'}^2+{\be'}^2\sin^2\al)]\vf+\mu^2\vf'}{
\exp[(-1)^r\nu\int\mu^2({\al'}^2+{\be'}^2\sin^2\al)dt]}\sin\al\;\sin\be
,\hspace{3.3cm}(9.4.111)\end{eqnarray*}
\begin{eqnarray*}p&=&\rho\{({\al'}^2+{\be'}^2\sin^2\al-\mu'/\mu)\X^2/2
+(2^{-1}{\be'}^2\sin2\al-{\al'}'-2\al'\mu'/\mu)\X\Y\\ & &
-(\al'\be'\cos\al+(\be'\sin\al)'+2\be'\mu'\sin\al/\mu)\X\Z-2\gm\tau\xi_r\X/\mu
\\ &
&+2^{-1}[({\al'}^2+{\be'}^2\cos^2\al-k_{\Y t})\Y^2+({\be'}^2-h_{\Z
t})\Z^2-k^2-h^2]
\\ & &+(4b\dlt_{0,r}+c\dlt_{1,r})c\mu^{-2}\exp[(-1)^r\nu\int\mu^2({\al'}^2+{\be'}^2\sin^2\al)dt]
\\ & &\times(\vf-2^{-1}\mu^{-2}\exp[(-1)^r\nu\int\mu^2({\al'}^2+{\be'}^2\sin^2\al)dt]\X)
\\& &+(\al'\be'\sin\al-(k_\Z-\be'\cos\al)')\Y\Z-
\frac{\xi^2\exp[(-1)^r2\nu\int\mu^2({\al'}^2+{\be'}^2\sin^2\al)dt]}{2\mu^8({\al'}^2+{\be'}^2\sin^2\al)}
\\ & &-2\mu^{-1}\vf\xi_r-\frac{2(\al'\Y+\be'\sin\al\;\Z)[\mu\mu'-(-1)^r\mu^4({\al'}^2+{\be'}^2\sin^2\al)]\vf+\mu^2\vf'}{
\exp[(-1)^r\nu\int\mu^2({\al'}^2+{\be'}^2\sin^2\al)dt]}
\\ & &-(-1)^r
\frac{\zeta_r[\mu^4({\al'}^2+{\be'}^2\sin^2\al)]'\exp[(-1)^r\nu\int\mu^2({\al'}^2+{\be'}^2\sin^2\al)dt]}
{\mu^4({\al'}^2+{\be'}^2\sin^2\al)}.\hspace{1.2cm}(9.4.112)
\end{eqnarray*}}
\pse

{\it Case 2}. $\gm_2=\al_1=\tau=0$ and $\gm_1\neq 0$.\psp

Under the assumption, (9.4.70) naturally holds. According to
(9.4.55) and (9.4.58),
$$k_\Y=-\frac{\gm_1'}{\gm_1},\qquad k_\X=k_\Z=0,\qquad
h_\Y=-2\be'\cos\al.\eqno(9.4.113)$$ Note $\gm=\gm_1^2$. Moreover,
(9.4.56) says
$$h_\X=-2\be'\sin\al.\eqno(9.4.114)$$
Furthermore, (9.4.78)  becomes
$$2{\al'}'-2\al'h_\Z=0\lra h_Z=\frac{{\al'}'}{\al'}\eqno(9.4.115)$$ and
(9.4.79) is satisfied naturally. Equation (9.4.44) yields
$$f_\X=\frac{\gm_1'}{\gm_1}-\frac{{\al'}'}{\al'}.\eqno(9.4.116)$$

Now (9.4.76) and (9.4.77) are equivalent to
$$\frac{\vf'}{\vf}=\frac{\sgm'}{\sgm}=(-1)^r\gm\nu-f_\X
=(-1)^r\nu\gm_1^2-\frac{\gm_1'}{\gm_1}+\frac{{\al'}'}{\al'},\eqno(9.4.117)$$
$$\frac{(\gm_1\sgm)'}{(\gm_1\sgm)}=h_\Z+(-1)^r\nu\gm=(-1)^r\nu\gm_1^2+\frac{{\al'}'}{\al'}.\eqno(9.4.118)$$
According to (9.4.117),
$$\vf=b_2\al'\gm_1^{-1}e^{(-1)^r\nu\int\gm_1^2dt},\qquad\sgm=b_3\al'\gm_1^{-1}e^{(-1)^r\nu\int\gm_1^2dt}\eqno(9.4.119)$$
with $b_2,b_3\in\mbb{R}$. Moreover, (9.4.118) is satisfied by the
above $\sgm$.

Next
$$\gm\sgm=b_2\al'\gm_1e^{(-1)^r\nu\int\gm_1^2dt}\lra
\frac{(\gm\sgm)'}{(\gm\sgm)}=(-1)^r\nu\gm_1^2+\frac{{\al'}'}{\al'}+\frac{\gm_1'}{\gm}.\eqno(9.4.120)$$
On the other hand, (9.4.51) implies
$$\frac{(\gm\sgm)'}{(\gm\sgm)}=(-1)^r\nu\gm-3f_\X=(-1)^r\nu\gm_1^2-\frac{3\gm_1'}{\gm_1}+\frac{3{\al'}'}{\al'}.
\eqno(9.4.121)$$ So
$$\frac{{\al'}'}{\al'}=2\frac{\gm_1'}{\gm}\lra\gm_1=c_2\sqrt{\al'},\qquad
0\neq c_2\in\mbb{R}.\eqno(9.4.122)$$ Thus
$$\vf=b_2c_2^{-1}\sqrt{\al'}e^{(-1)^rc_2^2\nu\al},\;\;\sgm=b_3c_2^{-1}\sqrt{\al'}e^{(-1)^rc_2^2\nu\al},\;\;
f_\X=k_\Y=-\frac{{\al'}'}{2\al'}.\eqno(9.4.123)$$

Observe
\begin{eqnarray*}\qquad\qquad& &\gm_1f_\Y\phi_r-\gm
f^\ast\psi_r\\
&=&\gm[(\vf\hat\vt_r+\sgm\phi_r+\sgm\td\varpi\psi_r)\phi_r-(\vf\vt_r+\sgm\td\varpi\phi_r)\psi_r]
\\
&=&\gm[\vf(\hat\vt_r\phi_r-\vt_r\psi_r)+\sgm\phi_r\xi_r]
\\ &=&[2(c-ab_1)\dlt_{r,0}-c_1\sin
b_1\;\dlt_{r,1}]\gm\vf+\gm\sgm\phi_r^2\hspace{4.7cm}(9.4.124)
\end{eqnarray*}
by (9.4.60) and (9.4.62). According to (9.4.49), (9.4.52), (9.4.53),
(9.4.113) and (9.4.114),  we have
\begin{eqnarray*}
R_1&=&[f_{\X t}
-{\al'}^2-{\be'}^2\sin^2\al+2\gm\vf\sgm[2(c-ab_1)\dlt_{r,0}-c_1\sin
b_1\;\dlt_{r,1}]]\X
\\ & &
-({\al'}'+2^{-1}{\be'}^2\sin2\al)\Y
+(\al'\be'\cos\al-(\be'\sin\al)')\Z+2\gm^2\sgm^2
(4a\dlt_{0,r}+\dlt_{1,r})\X^3
\\ & &-2\al'k-2\be'h\sin\al+2\sgm(\gm\sgm\phi_r-2\gm_1\al')
\phi_r\X,\hspace{4.9cm}(9.4.125)
\end{eqnarray*}
\begin{eqnarray*}R_2&=&({\al'}'-2^{-1}{\be'}^2\sin2\al+2\al'f_\X)\X
-({\al'}^2+{\be'}^2\cos^2\al)\Y+k_t+kk_\Y\\ & &
-((\be'\cos\al)'+\al'\be'\sin\al)\Z-2\be'h\cos\al
\\&&+\gm\sgm(2\sgm\gm_1\phi_r-2\al'
)\psi_r\X^2+f^\ast(2\al'+2\gm_1\sgm\phi_r),\hspace{4.7cm}(9.4.126)\end{eqnarray*}
$$R_3=[(\be'\sin\al)'+\al'\be'\cos\al]\X+
[(\be'\cos\al)'-\al'\be'\sin\al]\Y  +h_t+hh_\Z-{\be'}^2\Z.
\eqno(9.4.127)$$ In particular, (9.3.24) holds by (9.4.113),
(9.4.114) and (9.4.123).

Expressions (9.4.45)-(9.4.47) become
$$\U=-\frac{{\al'}'}{2\al'}\X-\al'\Y-\be'\Z\sin\al+
[b_2c_2^{-1}\sqrt{\al'}\vt_r+
b_3\al'(\Y\phi_r-c_2\sqrt{\al'}\psi_r\X^2)]e^{(-1)^rc_2^2\nu\al}.\eqno(9.4.128)$$
$$\V=\al'\X-\frac{{\al'}'}{2\al'}\Y-\be'\Z\cos\al+
2b_3c_2^2{\al'}^2e^{(-1)^rc_2^2\nu\al}\phi_r\X,\eqno(9.4.129)$$
$$\W=-\be'(\X\sin\al+\Y\cos\al)+\frac{{\al'}'}{\al'}\Z.\eqno(9.4.130)$$
By (9.3.3), (9.3.5), (9.3.22), (9.4.113), (9.4.114), (9.4.123) and
(9.4.125)-(9.4.130), we have the following theorem: \psp

{\bf Theorem 9.4.3}. {\it Let $\al,\be$ be arbitrary functions in
$t$ and let $a,b_1,b_2,c_2$ be real constants. Define the moving
frame $\X,\;\Y$ and $\Z$ by (9.3.1) and (9.3.5), and
$$\phi_0=
e^{c_2\sqrt{\al'}\Y}-ae^{-c_2\sqrt{\al'}\Y},\;\;\phi_1=\sin(c_2\sqrt{\al'}\Y),\;\;\psi_0=
e^{c_2\sqrt{\al'}\Y}+ae^{-c_2\sqrt{\al'}\Y},\eqno(9.4.131)$$ $$
\psi_1=\cos(c_2\sqrt{\al'}\Y), \;\;\vt_0=
b_1e^{c_2\sqrt{\al'}\Y}-c_1e^{-c_2\sqrt{\al'}\Y},\;\;\vt_1=c_1\sin(c_2\sqrt{\al'}\Y+b_1).\eqno(9.4.132)$$
For $r=0,1$, we have the following solution of  the Navier-Stokes
equations (9.1.1)-(9.1.4):
\begin{eqnarray*}u&=&
[-{\al'}'\X/(2\al')-\al'\Y+ [b_2c_2^{-1}\sqrt{\al'}\vt_r+
b_3\al'(\Y\phi_r-c_2\sqrt{\al'}\psi_r\X^2)]e^{(-1)^rc_2^2\nu\al}]\cos\al
\\ & &-[\al'\X-{\al'}'\Y/(2\al')+
2b_3c_2^2{\al'}^2e^{(-1)^rc_2^2\nu\al}\phi_r\X]\sin\al,\hspace{4.1cm}(9.4.133)\end{eqnarray*}
\begin{eqnarray*}v\!\!\!&=&\!\!\!
[-{\al'}'\X/(2\al')-\al'\Y+ [b_2c_2^{-1}\sqrt{\al'}\vt_r+
b_3\al'(\Y\phi_r-c_2\sqrt{\al'}\psi_r\X^2)]e^{(-1)^rc_2^2\nu\al}]\sin\al\:\cos\be
\\ \!\!\!& &\!\!\!+[[\al'\X-{\al'}'\Y/(2\al')+
2b_3c_2^2{\al'}^2e^{(-1)^rc_2^2\nu\al}\phi_r\X]\cos\al-\be'\Z]\cos\be
\\ \!\!\!&&\!\!\!+[\be'(\X\sin\al+\Y\cos\al)-{\al'}'\Z/\al']\sin\be
,\hspace{6.4cm}(9.4.134)\end{eqnarray*}
\begin{eqnarray*}w\!\!\!&=&\!\!\!
[-{\al'}'\X/(2\al')-\al'\Y+ [b_2c_2^{-1}\sqrt{\al'}\vt_r+
b_3\al'(\Y\phi_r-c_2\sqrt{\al'}\psi_r\X^2)]e^{(-1)^rc_2^2\nu\al}]\sin\al\:\sin\be
\\ \!\!\!& &\!\!\!+[[\al'\X-{\al'}'\Y/(2\al')+
2b_3c_2^2{\al'}^2e^{(-1)^rc_2^2\nu\al}\phi_r\X]\cos\al-\be'\Z]\sin\be
\\ \!\!\!&&\!\!\!-[\be'(\X\sin\al+\Y\cos\al)-{\al'}'\Z/\al']\cos\be
,\hspace{6.3cm}(9.4.135)\end{eqnarray*}
\begin{eqnarray*}
p&=&\frac{\rho}{2}\{[\frac{\al'{{\al'}'}'-{{\al'}'}^2}{2{\al'}^2}
+{\al'}^2-3{\be'}^2\sin^2\al-2b_2b_3[2(c-ab_1)\dlt_{r,0}-c_1\sin
b_1\;\dlt_{r,1}]\\ & &\times {\al'}^2e^{(-1)^r2c_2^2\nu\al}]\X^2
-3{\be'}^2\X\Y\sin2\al -(b_3c_2)^2
(4a\dlt_{0,r}+\dlt_{1,r}){\al'}^3e^{(-1)^r2c_2^2\nu\al}\X^4\\ &
&+2\frac{(\al'{\be'}'+2{\al'}'\be')\sin\al}{\al'}\X\Z
+2b_3c_2^{-1}{\al'}^2e^{(-1)^rc_2^2\nu\al}
(2-b_3c_2e^{(-1)^rc_2^2\nu\al}\phi_r) \phi_r\X^2\\
&&-4c_2^{-1}\sqrt{{\al'}^3}e^{(-1)^rc_2^2\nu\al}\int(b_2\vt_r+b_3c_2\sqrt{\al'}\Y\phi_r)
  (1+b_3e^{(-1)^rc_2^2\nu\al}\phi_r)d\Y\\& &+2\frac{({\be'}'-2{\al'}'\be')\cos\al}{\al'}\Y\Z
  +\frac{({\al'}^2{\be'}^2-\al'{{\al'}'}'+{{\al'}'}^2)\Z^2}{{\al'}^2}
  \\&&+[\frac{2\al'{{\al'}'}'-3{{\al'}'}^2}{4{\al'}^2}+{\al'}^2-3{\be'}^2\cos^2\al]\Y^2
  \}.\hspace{5.9cm}(9.4.136)
\end{eqnarray*}
}

\addcontentsline{toc}{chapter}{\numberline{}Index}

xxx

\printindex

\end{document}